\documentclass[12pt,a4paper]{book}
\usepackage[whole]{bxcjkjatype}
\pdfoutput=1

\usepackage{amsmath,amssymb,amsbsy,bm,bbm}
\usepackage[mathscr]{eucal}
\usepackage{color}

\usepackage{makeidx}
\makeindex

\usepackage[vcentermath]{youngtab}

\usepackage{graphicx}
\usepackage[unicode,
 pdftitle={Instanton Counting, Quantum Geometry and Algebra},
 pdfauthor={Taro Kimura},
 hidelinks
]{hyperref}

\usepackage{tikz}
\usetikzlibrary{arrows}
\usetikzlibrary{calc}
\usetikzlibrary{cd}
\usetikzlibrary{decorations.pathreplacing} 
\usetikzlibrary{decorations.markings}
\usetikzlibrary{knots}
\usetikzlibrary{shadows,fadings}

\usepackage{dynkin-diagrams}
\usetikzlibrary{backgrounds}

\tikzset{->-/.style={decoration={
  markings,
  mark=at position .55 with {\arrow{stealth}}},postaction={decorate}}}

\tikzset{-<-/.style={decoration={
  markings,
  mark=at position .55 with {\arrow[>=stealth]{<}}},postaction={decorate}}}  

\newcommand{\bZ}{\mathbb{Z}}
\newcommand{\bC}{\mathbb{C}}
\newcommand{\bR}{\mathbb{R}}

\newcommand{\cS}{\mathcal{S}}

\newcommand{\rO}{\mathrm{O}}

\newcommand{\rU}{\mathrm{U}}

\newcommand{\SU}{\mathrm{SU}}
\newcommand{\SO}{\mathrm{SO}}
\newcommand{\Sp}{\mathrm{Sp}}
\newcommand{\SL}{\mathrm{SL}}
\newcommand{\GL}{\mathrm{GL}}
\newcommand{\Spin}{\mathrm{Spin}}
\newcommand{\End}{\operatorname{End}}
\newcommand{\Hom}{\operatorname{Hom}}
\newcommand{\sdim}{\operatorname{sdim}}
\newcommand{\Tr}{\operatorname{Tr}}
\newcommand{\tr}{\operatorname{tr}}

\newcommand{\str}{\operatorname{str}}
\newcommand{\sdet}{\operatorname{sdet}}

\newcommand{\id}{\mathbbm{1}}

\newcommand{\np}{\mathrm{e}}
\newcommand{\im}{\mathrm{i}}
\newcommand{\imag}{\operatorname{Im}}
\newcommand{\real}{\operatorname{Re}}
\newcommand{\ch}{\operatorname{ch}}
\newcommand{\sch}{\operatorname{sch}}
\def\diag{\mathop{\mathrm{diag}}}

\def\res{\mathop{\mathrm{Res}}}
\def\rk{\mathop{\mathrm{rk}}}

\newcommand{\cvev}[1]{\left( \, #1 \, \right)}
\newcommand{\vev}[1]{\langle \, #1 \, \rangle}
\newcommand{\VEV}[1]{\Big< \, #1 \, \Big>}
\newcommand{\bra}[1]{\left< \, #1 \, \right|}
\newcommand{\ket}[1]{\left| \, #1 \, \right>}

\def\Xint#1{\mathchoice
   {\XXint\displaystyle\textstyle{#1}}%
   {\XXint\textstyle\scriptstyle{#1}}%
   {\XXint\scriptstyle\scriptscriptstyle{#1}}%
   {\XXint\scriptscriptstyle\scriptscriptstyle{#1}}%
   \!\int}
\def\XXint#1#2#3{{\setbox0=\hbox{$#1{#2#3}{\int}$}
     \vcenter{\hbox{$#2#3$}}\kern-.5\wd0}}

\def\dashint{\Xint-}

\usepackage[framemethod=TikZ]{mdframed}

\newenvironment{itembox}[1]{\begin{mdframed}[
  roundcorner=5pt,
  frametitleaboveskip=\dimexpr-0.7\baselineskip,
  innertopmargin=\dimexpr-0.25\baselineskip,
  innerbottommargin=\dimexpr0.5\baselineskip,
  frametitle={\tikz{\node[anchor=base,rectangle,fill=white]{\strut #1};}}]
  }
  {
   \end{mdframed}}

\makeatletter

\@addtoreset{equation}{section}
\makeatother

\date{\today}

\begin{document}

\pagenumbering{roman}

\begin{titlepage}

\renewcommand{\thefootnote}{\fnsymbol{footnote}}


 \vspace*{-5em}
 
 \begin{center}
  \includegraphics[height=4em]{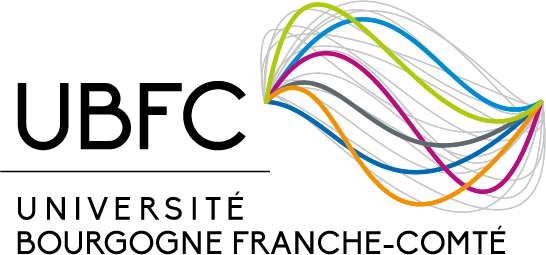} \hspace{5em}
  \includegraphics[height=4em]{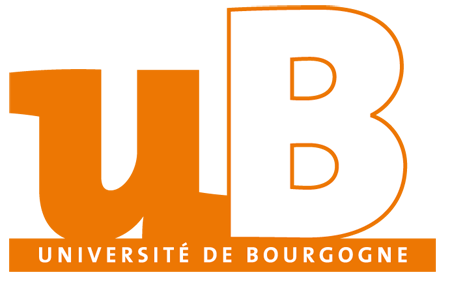} \hspace{5em}
  \includegraphics[height=4em]{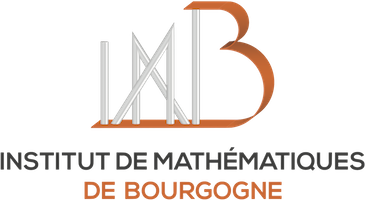}  
 \end{center}

 \vspace{1em}

 \begin{center}
  \textsc{%
  {\large 
  Universit\'e Bourgogne Franche-Comt\'e
  } \\[1.5em]
  \'Ecole Doctorale Carnot--Pasteur (ED 553) \\
  Institut de Math\'ematiques de Bourgogne (UMR 5584)
  }

  \vspace{1.5em}
  
  M\'emoire pr\'esent\'e pour l'obtention du dipl\^{o}me\\[1em]
  
  \textbf{\large
  D'Habilitation \`a Diriger des Recherches
  }\\[1em]

  Discipline : Math\'ematiques\\[1em]
  
  \vspace{1em}
  
  {\LARGE \textbf{%
  Instanton Counting,\\[.7em] 
  Quantum Geometry and Algebra    
  }}\\[1.5em]
  
  \textit{par}\\[1.5em]
  
  \setcounter{footnote}{1}

  {\large \textsc{Taro Kimura}}\\[.5em]  
  {\large 木村太郎}%

  \vspace{1.5em}
  

  Soutenance le 16 d\'ecembre 2020
  \`a Institut de Math\'ematiques de Bourgogne,
  Universit\'e Bourgogne Franche-Comt\'e,
  devant le jury compos\'e de\\[2em]
 \begin{tabular*}{.9\textwidth}{@{\extracolsep{\fill}}lll}
  \textsc{Giuseppe Dito} & Universit\'e de Bourgogne & Examinateur \\
  \textsc{Daniele Faenzi} & Universit\'e de Bourgogne & Pr\'esident du Jury\\
  \textsc{Stefan Hohenegger} & Universit\'e Claude Bernard Lyon I & Examinateur \\
  \textsc{Kenji Iohara} & Universit\'e Claude Bernard Lyon I & Examinateur \\  
  \textsc{Marcos Mari{\~n}o} & Universit\'e de Gen\`eve & Examinateur \\
  \textsc{Boris Pioline} & CNRS \& Sorbonne Universit\'e & Examinateur \\
  \textsc{Alessandro Tanzini} & SISSA 
      & Rapporteur \\
  \textsc{Maxim Zabzine} & Uppsala University & Rapporteur
 \end{tabular*}
 \vspace{1em}

 Rapporteurs du memoire :\qquad
  \begin{tabular*}{.6\textwidth}{@{\extracolsep{\fill}}lc}
   \textsc{Hiraku Nakajima} & University of Tokyo  \\
   \textsc{Alessandro Tanzini} & SISSA \\
   \textsc{Maxim Zabzine} & Uppsala University 
  \end{tabular*}
  
 \end{center}

\end{titlepage}

\setcounter{footnote}{0}

\newpage
\setcounter{page}{0}
\vspace*{\stretch{1}}
\thispagestyle{empty}
\newpage
\vspace*{\stretch{.8}}
\begin{center}
 \textbf{R\'esum\'e}
\end{center}
\addcontentsline{toc}{chapter}{R\'esum\'e}
\noindent
Le but de ce m\'emoire d'Habilitation \`a Diriger des Recherches est de pr\'esenter les aspects g\'eom\'etriques et alg\'ebriques quantiques des th\'eories de jauge supersym\'etriques, qui \'emergent de la nature non perturbative de la structure du vide, induite par les instantons.
Nous commen\c{c}ons par un bref r\'esum\'e de la localisation \'equivariante de l'espace des modules d'instantons, et montrons comment obtenir la fonction de partition d'instantons et sa g\'en\'eralisation aux th\'eories de jauge de carquois et de supergroupes de trois mani\`eres, par : la formule d'indice \'equivariant, la formule de l'int\'egrale de contour et la formule combinatoire.
Nous explorons ensuite la description g\'eom\'etrique de la th\'eorie de jauge $\mathcal{N} = 2$ bas\'ee sur la g\'eom\'etrie de Seiberg--Witten du point de vue de la th\'eorie des cordes et de la M-th\'eorie.
Avec sa relation aux systèmes int\'egrables, nous montrons comment quantifier une telle structure g\'eom\'etrique via la $\Omega$-d\'eformation de la th\'eorie de jauge.
Nous discutons \'egalement de la structure alg\'ebrique quantique sous-jacente aux vides supersym\'etriques.
Nous introduisons la notion de W-alg\`ebre de carquois construite par double quantification de la g\'eom\'etrie de Seiberg--Witten, et montrons ses sp\'ecificit\'es : les W-alg\`ebres de carquois affines, les W-alg\`ebres de carquois fractionnaires et leurs d\'eformations elliptiques.\\[1em]
\textbf{Mots-cl\'es} :
Instanton,
Localization \'equivariante,
Th\'eorie de jauje de carquois,
Th\'eorie de jauje de supergroupe,
G\'eom\'etrie de Seiberg--Witten,
Syst\`emes int\'egrables,
$qq$-caract\`ere,
W-alg\`ebre de carquois,
Alg\`ebre vertex.
\vspace{\stretch{1}}
\newpage
\vspace*{\stretch{1}}
\thispagestyle{empty}
\newpage
\vspace*{\stretch{.75}}
\begin{center}
 \textbf{Abstract}
\end{center}
\addcontentsline{toc}{chapter}{Abstract}

\noindent
The aim of this memoir for ``Habilitation \`a Diriger des Recherches'' is to present quantum geometric and algebraic aspects of supersymmetric gauge theory, which emerge from non-perturbative nature of the vacuum structure induced by instantons.
We start with a brief summary of the equivariant localization of the instanton moduli space, and show how to obtain the instanton partition function and its generalization to quiver gauge theory and supergroup gauge theory in three ways: the equivariant index formula, the contour integral formula, and the combinatorial formula.
We then explore the geometric description of $\mathcal{N} = 2$ gauge theory based on Seiberg--Witten geometry together with its string/M-theory perspective.
Through its relation to integrable systems, we show how to quantize such a geometric structure via the $\Omega$-deformation of gauge theory.
We also discuss the underlying quantum algebraic structure arising from the supersymmetric vacua.
We introduce the notion of quiver W-algebra constructed through double quantization of Seiberg--Witten geometry, and show its specific features: affine quiver W-algebras, fractional quiver W-algebras, and their elliptic deformations.\\[1em]
\vspace{\stretch{1}}
\newpage
\vspace*{\stretch{1}}
\thispagestyle{empty}
\newpage
\vspace*{\stretch{.8}}
\begin{center}
 \textbf{Acknowledgements}
\end{center}
\addcontentsline{toc}{chapter}{Acknowledgements}
\noindent

First of all, I'd like to express my sincere gratitude to my collaborators involved in my research, in particular, presented in this memoir:
Heng-Yu Chen,
Bertrand Eynard,
Toshiaki Fujimori,
Koji Hashimoto,
Norton Lee,
Hironori Mori,
Fabrizio Nieri,
Muneto Nitta,
Keisuke Ohashi,
Vasily Pestun,
Yuji Sugimoto,
Peng Zhao,
Rui-Dong Zhu.
In addition, I'm also grateful to
Yalong Cao,
Martijn Kool,
Sergej Monavari,
Kantaro Ohmori.
I could not have materialized this memoir without stimulating discussions and conversations with them, which provide a lot of clear insight for my understanding of the universe of physics and mathematics.

I'm deeply grateful to Hiraku Nakajima, Alessandro Tanzini, Maxim Zabzine,
 who kindly accepted to be the reporter of the memoir, Daniele Faenzi, who took the role of the president of the jury committee, and Giuseppe Dito, Kenji Iohara, Stefan Hohenegger, Marcos Mari{\~{n}}o, Boris Pioline, who participated the committee as the examinator.
 I would appreciate their kind help provided even in the difficult pandemic period.

I'd like to thank Institut de Math\'ematiques de Bourgogne,
Universit\'e de Bourgogne/Universit\'e Bourgogne Franche-Comt\'e, and the support from ``Investissements d'Avenir'' program, project ISITE-BFC (No.~ANR-15-IDEX-0003) and EIPHI Graduate School (No.~ANR-17-EURE-0002), for providing a chance to be committed to ``Habilitation \`a Diriger des Recherches'' with a warm environment to carry out my research there.

Finally, last but not least, it is my great pleasure to express my gratitude to my family, 
佳菜, 太智, 祥太, 紗弥子,
for eternal encouragement, and for making my life enjoyable and invaluable.

\vspace{\stretch{1}}
\newpage
\vspace*{\stretch{1}}
\thispagestyle{empty}
\newpage
\begin{center}
 \textbf{List of works presented in this memoir}
\end{center}
  \begin{description}
   \item[\cite{Kimura:2019hnw}]
	      T.~Kimura,\\
	      {\em Integrating over quiver variety and BPS/CFT correspondence},\\ \qquad
	      \href{https://doi.org/10.1007/s11005-020-01261-5}{Lett. Math. Phys. \textbf{110} (2020) 1237--1255},
	      \href{https://arxiv.org/abs/1910.03247}{\texttt{arXiv:1910.03247 [hep-th]}}.\\[-.5em]

   \item[\cite{Chen:2019vvt}]
	      H.-Y. Chen, T.~Kimura, and N.~Lee,\\
	      \emph{{Quantum Elliptic Calogero-Moser Systems from Gauge Origami}},\\
	      \href{https://doi.org/10.1007/JHEP02(2020)108}{JHEP \textbf{02} (2020)}, 108,
	      \href{https://arxiv.org/abs/1908.04928}{{\ttfamily arXiv:1908.04928 [hep-th]}}.\\[-.5em]

   \item[\cite{Chen:2020rxu}]
	      H.-Y. Chen, T.~Kimura, and N.~Lee,\\
	      \emph{{Quantum Integrable Systems from Supergroup Gauge Theories}},\\
  \href{https://doi.org/10.1007/JHEP09(2020)104}{JHEP \textbf{2009} (2020)}, 104, \href{https://arxiv.org/abs/2003.13514}{{\ttfamily arXiv:2003.13514 [hep-th]}}.\\[-.5em]
	      
   \item[\cite{Kimura:2020lmc}]
	      T.~Kimura and Y.~Sugimoto,\\ 
\emph{{Topological Vertex/anti-Vertex and Supergroup Gauge Theory}},\\ 
	      \href{https://doi.org/10.1007/JHEP04(2020)081}{JHEP \textbf{04} (2020)}, 081, \href{https://arxiv.org/abs/2001.05735}{{\ttfamily arXiv:2001.05735 [hep-th]}}.\\[-.5em]

   \item[\cite{Kimura:2019msw}]
	      T. Kimura and V. Pestun,\\
	      {\em Super instanton counting and localization},\\
	      Preprint, \href{https://arxiv.org/abs/1905.01513}{\texttt{arXiv:1905.01513 [hep-th]}}.\\[-.5em]

   \item[\cite{Kimura:2019xzj}]
	   T. Kimura and V. Pestun,\\
	   \emph{{Twisted reduction of quiver W-algebras}},\\
	   Preprint, \href{https://arxiv.org/abs/1905.03865}{{\ttfamily arXiv:1905.03865 [hep-th]}}.\\[-.5em]

   \item[\cite{Kimura:2019gon}]
	      T.~Kimura and R.-D. Zhu,\\ 
	      \emph{{Web Construction of ABCDEFG and Affine Quiver Gauge Theories}},\\ 
	      \href{https://doi.org/10.1007/JHEP09(2019)025}{JHEP \textbf{09} (2019)}, 025, \href{https://arxiv.org/abs/1907.02382}{{\ttfamily arXiv:1907.02382 [hep-th]}}.\\[-.5em]
	      
   \item[\cite{Chen:2018ntf}]
	      H.-Y. Chen and T.~Kimura,\\ 
	      \emph{{Quantum integrability from non-simply laced quiver gauge theory}},\\ 
	      \href{https://doi.org/10.1007/JHEP06(2018)165}{JHEP \textbf{06} (2018)}, 165, \href{https://arxiv.org/abs/1805.01308}{{\ttfamily
	      arXiv:1805.01308 [hep-th]}}.\\[-.5em]

   \item[\cite{Kimura:2017hez}]
	      T. Kimura and V. Pestun,\\
	      {\em Fractional quiver W-algebras},\\
	      \href{https://doi.org/10.1007/s11005-018-1087-7}{Lett. Math. Phys. \textbf{108} (2018) 2425--2451},
	      \href{https://arxiv.org/abs/1705.04410}{\texttt{arXiv:1705.04410 [hep-th]}}.\\[-.5em]
	      
   \item[\cite{Kimura:2016dys}]
	      T. Kimura and V. Pestun,\\
	      {\em Quiver elliptic W-algebras,}\\
	      \href{https://doi.org/10.1007/s11005-018-1073-0}{Lett. Math. Phys. \textbf{108} (2018) 1383--1405},
	      \href{https://arxiv.org/abs/1608.04651}{\texttt{arXiv:1608.04651 [hep-th]}}.\\[-.5em]
	      
   \item[\cite{Kimura:2015rgi}]
	      T. Kimura and V. Pestun,\\
	      {\em Quiver W-algebras,}\\
	      \href{https://doi.org/10.1007/s11005-018-1072-1}{Lett. Math. Phys. \textbf{108} (2018) 1351--1381},
	      \href{https://arxiv.org/abs/1512.08533}{\texttt{arXiv:1512.08533 [hep-th]}}.\\[-.5em]	      
	      
   \item[\cite{Kimura:2016ebq}]
	      T. Kimura,\\
	      {\em Double quantization of Seiberg--Witten geometry and W-algebras},\\
	      Topological Recursion and its Influence in Analysis, Geometry, and Topology,
	      \href{https://dx.doi.org/10.1090/pspum/100/01762}{Proc. Symp. Pure Math. \textbf{100} (2018) 405--431},
	      \href{https://arxiv.org/abs/1612.07590}{\texttt{arXiv:1612.07590 [hep-th]}}.\\[-.5em]

   \item[\cite{Kimura:2017auj}]
	   T.~Kimura, H.~Mori, and Y.~Sugimoto,\\ 
	      \emph{{Refined geometric transition and $qq$-characters}},\\ 
	   \href{https://doi.org/10.1007/JHEP01(2018)025}{JHEP \textbf{01} (2018)}, 025, \href{https://arxiv.org/abs/1705.03467}{{\ttfamily arXiv:1705.03467 [hep-th]}}.\\[-.5em]
	      
   \item[\cite{Hashimoto:2015dla}]
	      K. Hashimoto and T. Kimura,\\
	      {\em Band spectrum is D-brane},\\
	      \href{https://doi.org/10.1093/ptep/ptv181}{PTEP \textbf{2016} (2016) 013B04},
	      \href{https://arxiv.org/abs/1509.04676}{\texttt{arXiv:1509.04676 [hep-th]}}.\\[-.5em]

   \item[\cite{Hashimoto:2016dtm}]
	      K. Hashimoto and T. Kimura,\\
	      {\em Topological Number of Edge States},\\
	      \href{https://doi.org/10.1103/PhysRevB.93.195166}{Phys. Rev. \textbf{B93} (2016) 195166},
	      \href{https://arxiv.org/abs/1602.05577}{\texttt{arXiv:1602.05577 [cond-mat.mes-hall]}}.\\[-.5em]

   \item[\cite{Fujimori:2015zaa}]
	      T. Fujimori, T. Kimura, M. Nitta, and K. Ohashi,\\
	      {\em 2d partition function in $\Omega$-background and vortex/instanton correspondence},\\
	      \href{https://doi.org/10.1007/JHEP12(2015)110}{JHEP \textbf{12} (2015) 110},
	      \href{https://arxiv.org/abs/1509.08630}{\texttt{arXiv:1509.08630 [hep-th]}}.\\[-.5em]

   \item[\cite{Fujimori:2012ab}]
	      T.~Fujimori, T.~Kimura, M.~Nitta, and K.~Ohashi,\\
	      \emph{{Vortex counting from field theory}},\\
	      \href{https://doi.org/10.1007/JHEP06(2012)028}{JHEP \textbf{1206} (2012)}, 028,
	      \href{https://arxiv.org/abs/1204.1968}{{\ttfamily arXiv:1204.1968 [hep-th]}}.\\[-.5em]
	      
\item[\cite{Kimura:2011gq}]
	   T.~Kimura,\\ 	  
	   \emph{{$\beta$-ensembles for toric orbifold partition function}},\\
  \href{https://doi.org/10.1143/PTP.127.271}{Prog. Theor. Phys. \textbf{127}
  (2012)}, 271--285, \href{https://arxiv.org/abs/1109.0004}{{\ttfamily
	   arXiv:1109.0004 [hep-th]}}.\\[-.5em]

\item[\cite{{Kimura:2011wh}}]
	   T.~Kimura and M.~Nitta,\\ \emph{{Vortices on Orbifolds}}, \\
  \href{https://doi.org/10.1007/JHEP09(2011)118}{JHEP \textbf{1109} (2011)},
  118, \href{https://arxiv.org/abs/1108.3563}{{\ttfamily arXiv:1108.3563
  [hep-th]}}.\\[-.5em]
	   
\item[\cite{Kimura:2011zf}]
	   T.~Kimura,\\ 
	   \emph{{Matrix model from $\mathcal{N} = 2$ orbifold partition function}},\\ 
	   \href{https://doi.org/10.1007/JHEP09(2011)015}{JHEP \textbf{1109} (2011)}, 015, \href{https://arxiv.org/abs/1105.6091}{{\ttfamily arXiv:1105.6091 [hep-th]}}.
  \end{description}

\tableofcontents
\newpage
\pagenumbering{arabic}
\setcounter{chapter}{-1}
\chapter{Introduction}

\subsubsection{Gauge theory in physics and mathematics}

Since Yang--Mills' proposal to extend gauge symmetry to non-Abelian symmetry~\cite{Yang:1954ek}, gauge theory has been playing a crucial role in theoretical physics as a ubiquitous framework to describe fundamental interactions: electroweak interaction~\cite{Glashow:1961tr,Weinberg:1967tq}, quantum chromodynamics (QCD)~\cite{Nakano:1953zz,Nishijima:1955gxk,Gell-Mann:1956iqa,GellMann:1961ky,Greenberg:1964pe,Han:1965pf,Fritzsch:1973pi}, and gravity~\cite{Utiyama:1956sy}.
In addition to the significant role in theoretical physics, the influence of gauge theory is not restricted to physics, but also extended to wide-ranging fields of mathematics:
The study of self-duality equations in four-dimensions~\cite{Atiyah:1977bu,Atiyah:1978wi}, which leads to the so-called Atiyah--Drinfeld--Hitchin--Manin (ADHM) construction of the instantons~\cite{Atiyah:1978ri};
Morse theory~\cite{Atiyah:1982fa,Witten:1982im} in the relation to algebraic geometry;
Donaldson invariants of four-manifolds~\cite{Donaldson:1983wm};
Topological invariants of knots, known as Jones polynomial~\cite{Jones:1985dw}, from Chern--Simons gauge theory~\cite{Witten:1988hf};
Seiberg--Witten invariant~\cite{Witten:1994cg} motivated by Seiberg--Witten theory of $\mathcal{N} = 2$ supersymmetric gauge theory~\cite{Seiberg:1994rs,Seiberg:1994aj}.
In fact, these developments have been motivating various interplay between physics and mathematics up to now.
The aim of this memoir is to present new mathematical concepts emerging from such intersections of physics and mathematics.


\subsubsection{Universality of QFT}

In general, Quantum Field Theory (QFT) is a universal methodology to describe many-body interacting systems, which involves quite broad applications to particle physics, nuclear physics, astrophysics and cosmology, condensed-matter physics, and more.
In order to discuss the origin of its universality, one cannot say anything without mentioning the role of symmetry on the low energy behavior in the vicinity of the vacuum/ground state of the system, e.g., spacetime/internal symmetry, global/local symmetry, and non-local symmetry.

One may obtain several constraints on the spectrum, and also the conservation law from the symmetry argument, which provide useful information to discuss the effective description of the low energy behavior.
However, it is not straightforward to understand the vacuum structure of QFT, since it would be strongly coupled in many cases in the low energy regime, due to the so-called asymptotic freedom~\cite{Gross:1973id,Politzer:1973fx,Polyakov:1975rr}, and one cannot apply the systematic approach based on the perturbation theory with respect to a small coupling constant as in the weakly coupled regime.
In order to overcome this difficulty, it would be plausible to incorporate additional symmetry, i.e., supersymmetry, which provides further analytic framework for the study of QFT.
In fact, supersymmetric extension of gauge theory, which we mainly explore in this memoir, shows a lot of geometric and algebraic properties in the low energy regime.

\subsubsection{$\mathcal{N} = 2$ supersymmetry}

In this memoir, we mainly focus on $\mathcal{N} = 2$ supersymmetric gauge theory in four dimensions, and explore the associated geometric and algebraic structure emerging from the moduli space of the supersymmetric vacua.
$\mathcal{N} = 2$ theory has two sets of supersymmetries, which provide powerful tools to study its dynamics rather than non-supersymmetric and $\mathcal{N} = 1$ theories.
At the same time, it still shows various dynamical behaviors, e.g., the asymptotic freedom and the dynamical mass generation.

Actually, the instanton plays a crucial role to explore the vacuum structure of $\mathcal{N} = 2$ theory as well.
Since the instanton provides a solution to the classical equation of motion in the Yang--Mills theory, one may consider the perturbative expansion around the instanton configuration~\cite{tHooft:1976snw}.
Although it is still hard to control this expansion, we can apply the so-called topological twist to localize the path integral on the instanton configuration, if there exists $\mathcal{N} = 2$ supersymmetry~\cite{Witten:1988ze} (\S\ref{sec:instanton_sum}).
This drastically simplifies the analysis of gauge theory path integral, and one can deal with the gauge theory path integral as a statistical model of the instantons.
What remains is to evaluate the configuration space of the instantons, a.k.a., the instanton moduli space.

\subsubsection{Instanton counting}

From this point of view, we will provide the instanton counting argument with detailed study of the instanton moduli space.
We are in particular interested in the volume of the instanton moduli space, which gives rise to important contributions to the partition function based on the path integral formalism.
Since the naively defined moduli space is non-compact and singular, we should instead define a regularized version of the moduli space, and then apply the equivariant localization scheme to evaluate the volume of the moduli space.
The gauge theory partition function obtained by the equivariant integral over the instanton moduli space is called the {\em instanton partition function}~\cite{Losev:1997tp,Moore:1997dj,Lossev:1997bz} and also the {\em Nekrasov partition function}~\cite{Nekrasov:2002qd,Nekrasov:2003rj}, which will be one of the main objects in this memoir (\S\ref{sec:eq_ch_formula}).

The instanton partition function provides a lot of suggestive insights in the relation to various branches of mathematics:
Combinatorics of (2d and also higher dimensional) partitions;
Geometric representation theory;
$\tau$-function and integrable systems;
Vertex operator algebra and conformal field theory, and more.
The latter part of this memoir is devoted to the study of quantum geometric and algebraic aspects of $\mathcal{N} = 2$ gauge theory based on such interesting connections between the instanton partition function and various illuminating notions in mathematical physics.

\subsubsection{Seiberg--Witten theory}

A striking application of the instanton counting is the Seiberg--Witten theory for $\mathcal{N} = 2$ gauge theory in four dimensions~\cite{Seiberg:1994rs,Seiberg:1994aj}, which provides an algebraic geometric description for the low energy effective theory of $\mathcal{N} = 2$ theory (\S\ref{sec:SW_th}).
A remarkable property of $\mathcal{N} = 2$ theory is the one-to-one correspondence between the Lagrangian and the holomorphic function, known as the {\em prepotential}~\cite{Seiberg:1988ur}.
Seiberg--Witten theory provides a geometric characterization of the low energy effective prepotential based on the auxiliary algebraic curve, called the Seiberg--Witten curve.

The instanton partition function depends on the equivariant parameters associated with the spacetime rotation symmetry denoted by $(\epsilon_1,\epsilon_2) \in \mathbb{C}^2$ (also called the $\Omega$-background/deformation parameters).
The partition function diverges if we naively take the limit $\epsilon_{1,2} \to 0$.
In fact, Nekrasov's proposal was that the asymptotic expansion of the instanton partition function in the limit $\epsilon_{1,2} \to 0$ reproduces Seiberg--Witten's prepotential (\S\ref{sec:classical_lim}).
This proposal has been confirmed by Nekrasov--Okounkov~\cite{Nekrasov:2003rj}, Nakajima--Yoshioka~\cite{Nakajima:2003pg}, and Braverman--Etingof~\cite{Braverman:2004cr}, based on different approaches.

\subsubsection{Relation to integrable system}

Seiberg--Witten's geometric description implies a possible connection between $\mathcal{N} = 2$ gauge theory and classical integrable systems.
In fact, the Coulomb branch of the moduli space of the supersymmetric vacua of $\mathcal{N} = 2$ theory in four dimensions is identified with the base of the phase space of the algebraic integrable system~\cite{Gorsky:1995zq,Martinec:1995by,Donagi:1995cf,Seiberg:1996nz}.
This correspondence is based on the identification of the Seiberg--Witten curve with the spectral curve of the corresponding classical integrable system.
A primary example of the integrable system is the closed $n$-particle Toda chain ($\widehat{A}_{n-1}$ Toda chain), corresponding to $\mathcal{N} = 2$ $\SU(n)$ Yang--Mills theory.
One can also obtain the spin chain model from $\mathcal{N} = 2$ theory with the fundamental matters, a.k.a, $\mathcal{N} = 2$ supersymmetric QCD (SQCD).
In this context, the gauge symmetry (and the flavor symmetry) is not reflected in the symmetry of the integrable system, whereas the quiver structure does affect the symmetry algebra on the integrable system side.
These integrable systems are in general associated with the moduli space of periodic monopole~\cite{Nekrasov:2012xe}, obtained through a duality chain on the gauge theory side (\S\ref{sec:SW_curve_quiver}; \S\ref{sec:5d6d}).
In addition, imposing additional periodicity, we will obtain the trigonometric/elliptic integrable systems, corresponding to 5d $\mathcal{N} = 1$ on a circle $S^1$ and 6d $\mathcal{N} = (1,0)$ theory on a torus $T^2$, respectively.

\subsubsection{Quantization of geometry}

Once the correspondence to the classical integrable system is established, it is natural to ask:
Is it possible to see a quantum version of the correspondence?
If yes, how to quantize this relation?
Nekrasov--Shatashvili's proposal was to use the $\Omega$-background parameter, which was originally introduced as a regularization parameter to localize the path integral~\cite{Nekrasov:2009rc}.
See also~\cite{Nekrasov:2009zz,Nekrasov:2009uh,Nekrasov:2009ui,Nekrasov:2013xda}.
In particular, the limit $(\epsilon_1,\epsilon_2) \to (\hbar,0)$ is called the Nekrasov--Shatashvili (NS) limit, in which we can see the quantization of the cycle integral over the Seiberg--Witten curve, namely Bohr--Sommerfeld's quantization condition~(\S\ref{sec:BS_quantum}).
In this situation, the spectral curve is promoted to the quantum curve, which is now discussed in various research fields:
matrix model~\cite{Eynard:2007kz}%
\footnote{%
See Appendix~\ref{chap:matrix_model}.
}; 
topological string~\cite{Aganagic:2003qj,Dijkgraaf:2007sw,Dijkgraaf:2008fh}, knot invariant (AJ conjecture)~\cite{Garoufalidis:2008,Garoufalidis:2004GTM}, etc.
In the context of gauge theory, the quantum curve is identified with the TQ-relation of the quantum integrable system.
Similarly, the saddle point equation obtained from the instanton partition function is identified with the Bethe equation of the quantum integrable system~(\S\ref{sec:NS_integrability}).

\subsubsection{Quantum algebraic structure}

Quantum integrable systems in principle have infinitely many conserved Hamiltonians, which are constructed from the underlying infinite dimensional quantum algebra.
Then, the correspondence between gauge theory and integrable systems implies existence of such a quantum algebraic structure on the gauge theory side.
Furthermore, since the correspondence to the quantum integrable system is discussed in the NS limit, it is expected to obtain a doubly quantum algebra with generic $\Omega$-background parameters $(\epsilon_{1,2})$.
In fact, such a quantum algebra is then identified with {\em Virasoro/W-algebra}, which is an infinite dimensional (non-linear) symmetry algebra of conformal field theory (CFT)~\cite{Bouwknegt:1992wg,Bouwknegt:1994hmh,DiFrancesco:1997nk,Ribault:2014hia}.
From this point of view, the quantum integrability is described by the Poisson algebra obtained in the classical limit of W-algebras.

The algebraic correspondence between gauge theory and CFT is in general dubbed as {\em BPS/CFT correspondence}~\cite{Nekrasov:2015wsu,Nekrasov:2016qym,Nekrasov:2016ydq,Nekrasov:2017rqy,Nekrasov:2017gzb}, with a lot of examples explored so far.
The primary example is the {\em Alday--Gaiotto--Tachikawa (AGT) relation}~\cite{Alday:2009aq,Wyllard:2009hg}, which states the equivalence between the instanton partition function of $G$-YM theory and the conformal block of W$(G)$-algebra.
This relation is generalized to various situations:\index{AGT relation}
5d $\mathcal{N} = 1$ theory and $q$-CFT~\cite{Awata:2009ur};
The surface operator and the degenerate field insertion~\cite{Alday:2009fs}, and also the affine Lie algebra~\cite{Alday:2010vg};
The instanton partition function on the orbifold and the super/para-CFT~\cite{Belavin:2011pp,Nishioka:2011jk,Bonelli:2011jx,Bonelli:2011kv}.
See also review articles on the topic~\cite{Tachikawa:2014dja,Tachikawa:2016kfc,LeFloch:2020uop}.
Another important example is the {\em chiral algebra}~\cite{Gadde:2011ik,Beem:2013sza}, which is the correspondence between a class of the operators in $\mathcal{N} = 2$ theory and a two-dimensional chiral algebra (vertex operator algebra).
From this point of view, the superconformal index on the gauge theory side is identified with the character of the corresponding module on the chiral algebra side.
See also a recent review article~\cite{Lemos:2020pqv}.
In fact, these two relations are motivated by the class $\mathcal{S}$ description (compactification of 6d $\mathcal{N} = (2,0)$ theory with a generic Riemann surface) of $\mathcal{N} = 2$ gauge theory~\cite{Gaiotto:2009we}.

\subsubsection{Quiver W-algebra}

Regarding the correspondence to the quantum integrable system, the symmetry algebra is related to the quiver structure on the gauge theory side.
From this point of view, we may discuss a quantum algebraic structure from quiver gauge theory with generic $\Omega$-background parameters.
The {\em quiver W-algebra} W$_{q_{1,2}}(\Gamma)$ (or simply W$(\Gamma)$) is a doubly quantum algebra constructed from $\Gamma$-quiver gauge theory, and its algebraic structure is associated with the quiver structure $\Gamma$ of gauge theory~\cite{Kimura:2015rgi} (Chapter~\ref{chap:quiv_W}).
See also \cite{Kimura:2016ebq}.
In fact, this quiver W-algebra is linked to the AGT relation through the duality.

The formalism of quiver W-algebra exhibits several specific features.
Starting with the finite-type Dynkin-quiver, quiver W-algebra reproduces Frenkel--Reshetikhin's construction of $q$-deformation of W-algebras for $\Gamma = ADE$~\cite{Frenkel:1997}.
Quiver W-algebra is also applicable to affine quivers, and in that case, it gives rise to a new family of W-algebras (\S\ref{sec:affine_quiv_W}).
In order to extend this formalism to arbitrary quiver, including non-simply-laced quivers, we should consider the fractional quiver gauge theory, which partially breaks the symmetry of $\epsilon_1 \leftrightarrow \epsilon_2$~\cite{Kimura:2017hez}~(\S\ref{sec:frac_W}).
Applying this formalism to 6d $\mathcal{N} = (1,0)$ gauge theory, we obtain an elliptic deformation of W-algebras~\cite{Kimura:2016dys} (Chapter~\ref{chap:eW}).
This algebra has one more parameter corresponding to the modulus of the torus, on which the gauge theory is compactified.

\subsection*{Organization of the memoir}
\addcontentsline{toc}{section}{Organization of the memoir}

This memoir is organized as follows.

\subsection*{Part~\ref{partI}: Instanton Counting}

Chapter~\ref{chap:YM_instanton} is devoted to describe basic aspects of instanton counting.
We start with a brief review on gauge theory, in particular, with emphasis on the role of the instantons in the path integral.
We then discuss the ADHM construction of the instantons and the associated moduli space.
After introducing basic ideas on the equivariant localization, we apply it to the instanton moduli space to evaluate the equivariant volume, which will be identified with the instanton partition function.
In particular, we discuss how to characterize the fixed point under the equivariant action, on which the path integral will be localized.
We show equivalent expressions for the instanton partition function in the threefold way, the equivariant index formula, the contour integral formula, and the combinatorial formula.

In Chapter~\ref{chap:quiver_gauge_theory}, we generalize the instanton counting description to quiver gauge theory, which consists of multiple gauge field degrees of freedom.
We derive the instanton partition function in the threefold way via the equivariant localization for quiver gauge theory.
We will also introduce the quiver variety and its relation to the moduli space of instantons on the ALE space.
We then discuss the fractional quiver gauge theory, which is applied to describe a non-simply-laced quiver.
We will similarly obtain the instanton partition partition function for fractional quiver with the corresponding instanton moduli space.

In Chapter~\ref{sec:super_instanton}, we study the gauge theory with supergroup gauge symmetry.
Although it is inevitably non-unitary due to violation of the spin-statistics theorem, we will be able to explore its non-perturbative aspects of supergroup gauge theory.
We will show that supergroup gauge theory has several realizations in terms of non-supergroup gauge theory in the unphysical parameter regime.
We then discuss the ADHM construction of instantons in supergroup gauge theory together with its string theory perspective.
Based on this construction, we provide the threefold way derivation of the instanton partition function for supergroup gauge theory by applying the equivariant localization scheme.

\subsection*{Part~\ref{part2}: Quantum Geometry}

Chapter~\ref{chap:SW_theory} is devoted to geometric description of $\mathcal{N} = 2$ gauge theory, a.k.a, Seiberg--Witten theory.
Starting with basic properties of $\mathcal{N} = 2$ theory in four dimensions, we explain how the low energy behavior of $\mathcal{N} = 2$ theory is described in terms of algebraic geometry.
We also show its generalization to quiver gauge theory, and the Seiberg--Witten curve for quiver theory is in general given as the spectral curve of the corresponding algebraic integrable system.
We then discuss the string/M-theory perspective of the Seiberg--Witten geometry based on the brane description of $\mathcal{N} = 2$ gauge theory in four dimensions.
In particular, we see how the Coulomb branch description turns to the Higgs branch with emphasis on the role of the vortices in the brane configuration.
We show that one can similarly discuss the Seiberg--Witten geometry in 5d $\mathcal{N} = 1$ theory and 6d $\mathcal{N} = (1,0)$ theory compactified on a circle and a torus, respectively.
We will discuss several features specific to higher dimensional theories.

In Chapter~\ref{chap:geometry}, we discuss quantization of Seiberg--Witten geometry based on the $\Omega$-deformation of gauge theory.
In particular, we show that a doubly quantum deformation of the Seiberg--Witten curve is described by the $qq$-character generated by the adding/removing-instanton operation.
We then discuss the classical limit, which reproduces the ordinary Seiberg--Witten curve discussed earlier.
If one takes a partial classical limit (NS limit), it would be reduced to the $q$-character, which is identified with the transfer matrix of the corresponding quantum integrable system.
We show that the TQ-relation is interpreted as quantization of the Seiberg--Witten curve, and the Bethe equation is obtained as the saddle point equation obtained from the instanton partition function in the NS limit.

\subsection*{Part~\ref{part:algebra}: Quantum Algebra}

In Chapter~\ref{chap:operator}, we explain the operator formalism of $\mathcal{N} = 2$ gauge theory based on the holomorphic deformation of the prepotential to discuss the algebraic structure of gauge theory.
In this formalism, there are infinitely many harmonic oscillators, which generate the Fock space, and one can discuss the free field realization of the quantum algebra on it with the vertex operators.
From this point of view, we introduce the notion of the $Z$-state through the operator/state correspondence on the Fock space, and show that the instanton partition function is obtained as a chiral correlator of the corresponding vertex operator algebra.
We also clarify how to describe the pole cancellation mechanism in terms of the operator formalism.

In Chapter~\ref{chap:quiv_W}, we introduce the quiver W-algebra based on the operator formalism of gauge theory.
We show that the operator analog of the $qq$-character, called the $\mathsf{T}$-operator, plays a role of the generating current of the algebra W$_{q_{1,2}}(\Gamma)$ with $q_{1,2} = \np^{\epsilon_{1,2}} \in \mathbb{C}^\times$, and it is reduced to the commuting transfer matrix in the NS limit (the generating function of the conserved Hamiltonians).
We extend this formalism to fractional quiver gauge theory, and show that one can construct $q$-deformation of W-algebras associated with non-simply-laced algebras.
We then apply this formalism to affine quivers, and demonstrate that a new family of W-algebras are generated by the corresponding $qq$-character, which is an infinite series of the vertex operators.
We also discuss applications of the operator formalism to the contour integral formula of the instanton partition function.
We show that the partition function again has a chiral correlator realization, and the summation over the topological sectors is concisely described.

In Chapter~\ref{chap:eW}, we discuss an elliptic deformation of W-algebra constructed from 6d $\mathcal{N} = (1,0)$ gauge theory compactified on a torus.
In this case, the instanton partition function is given as an elliptic chiral correlator expressed in two ways:
The first is based on the elliptic deformation of the vertex operators, and the second is as a torus correlation function of the vertex operator without elliptic deformation.
The relation between these two descriptions is naturally understood from the duality.
Applying the formalism of quiver W-algebra to this elliptic theory, we can construct the holomorphic generating current of the elliptic W-algebra denoted by W$_{q_{1,2},p}(\Gamma)$, depending on the elliptic parameter $p \in \mathbb{C}^\times$.

\part{Instanton Counting}\label{partI}
\chapter{Instanton counting and localization}
\label{chap:YM_instanton}

The aim of this Chapter is to introduce the Yang--Mills (YM) theory, and explain how the specific solution, called the {\em instanton}, plays an important role in four-dimensional gauge theory.%
\footnote{%
See~\cite{tHooft:1999cgx} for introductory review on this topic,
and also~\cite{Freed:1991,Donaldson:1997} for mathematical description of gauge theory.
}
We will explain a systematic method to describe the instanton solution, a.k.a. ADHM construction~\cite{Atiyah:1978ri}, and discuss how the moduli space of the instanton plays a role in the path integral formalism of the YM theory.
In particular, volume of the instanton moduli space is an important quantity, but we should regularize it due to the singular behavior of the moduli space.
We will then consider the equivariant action on the instanton moduli space, and apply the equivariant localization scheme to evaluate the volume of the moduli space~\cite{Duistermaat:1982vw,Berline:1982,Atiyah:1984px}, which gives rise to the instanton partition function~~\cite{Losev:1997tp,Moore:1997dj,Lossev:1997bz,Nekrasov:2002qd,Nekrasov:2003rj}.

\section{Yang--Mills theory}\label{sec:YM_theory}

Let us briefly review the basics of gauge theory.
Gauge theory is mathematically formulated as a principal bundle with the structure group $G$ ($G$-bundle for short), which is also called the gauge group in the physicists' terminology.
Let $\mathcal{S}$ be the $\mathsf{d}$-dimensional Riemannian manifold as a base of the $G$-bundle, then the Lie algebra valued one-form, called the {\em connection}, is the fundamental ingredient of gauge theory,
$A : \mathcal{S} \longrightarrow T^* \mathcal{S} \otimes \mathbb{C}[\mathfrak{g}]$, with $\mathfrak{g} = \operatorname{Lie} G$.
The curvature two-form is given as
\begin{align}
 F = dA + A \wedge A \in \Omega^2(\mathcal{S}) \otimes \mathbb{C}[\mathfrak{g}]
\end{align}
where 
$\Omega^p(\mathcal{S}) \otimes \mathbb{C}[\mathfrak{g}] = \wedge^p T^* \mathcal{S} \otimes \mathbb{C}[\mathfrak{g}]$ is a set of $\mathfrak{g}$-valued $p$-forms on $\mathcal{S}$.
Under the $G$-gauge transformation $A \longmapsto g A g^{-1} + g d g^{-1}$ for $g \in G$, the curvature behaves as $F \longmapsto g F g^{-1} = \operatorname{ad}_g (F)$.

The Yang--Mills theory with the gauge group $G$ ($G$-YM theory) is described with the YM action functional:\index{Yang--Mills action}
\begin{align}
 S_\text{YM}[A] = \frac{1}{g^2} \int_\mathcal{S} d\text{vol} \, \left|F\right|^2 = - \frac{1}{g^2} \int_\mathcal{S} \tr \left( F \wedge \star F \right)
 \label{eq:YM-action}
\end{align}
where $d\text{vol}$ is the volume form, $\star$~is the Hodge star operator on $\mathcal{S}$, and the inner product is defined as $\left< A, B \right> = - \tr (AB)$, where the trace is with respect to the defining representation of $G$.%
\footnote{%
We apply the convention s.t., the Lie algebra generators are anti-Hermitian.
}
The gauge coupling constant is denoted by~$g$.
This YM action is invariant under the $G$-gauge transformation.

We are interested in a specific configuration minimizing the YM action~\eqref{eq:YM-action}, which is a solution to the equation of motion (e.o.m.; also referred to as the YM equation): 
\begin{align}
 \frac{\delta S_\text{YM}[A]}{\delta A} = 0
 \quad \implies \quad
 D \star F = 0
 \label{eq:eom_YM}
\end{align}
where we define the covariant derivative
\begin{align}
 D = d + A
 \, .
\end{align}
The e.o.m.~\eqref{eq:eom_YM} is a second order non-linear PDE, which is difficult to solve in general.
Hence, instead of finding a general solution, we will deal with a class of more tractable solutions.

For example, a naive solution is $F = 0$ (zero curvature), and the corresponding connection is called the {\em flat connection}.
However, the flat connection is not a good solution in higher dimensions in the following sense:
The two-form curvature $F \in \Omega^2(\mathcal{S}) \otimes \mathbb{C}[\mathfrak{g}]$ has $\displaystyle {\mathsf{d} \choose {2}} = \frac{\mathsf{d}(\mathsf{d}-1)}{2}$ components, while the connection $A \in \Omega^1(\mathcal{S}) \otimes \mathbb{C}[\mathfrak{g}]$ has $(\mathsf{d} - 1)$ components after gauge fixing.
Therefore, the zero curvature condition $F = 0$ overdetermines the connection $A$ except for $\mathsf{d} = 2$, so that it is difficult to discuss the corresponding moduli space.

  \section{Instanton}

From this point of view, we should find an alternative class of solutions in higher dimensions.
In the case of $\mathsf{d} = 4$, there is a special property of the Hodge star operator, which behaves as an endmorphism of the bundle of two-forms, $\star: \Omega^2(\mathcal{S}) \to \Omega^2(\mathcal{S})$.
In this case, since $\star^2 = 1$, we can decompose it into the self-dual (SD) and anti-self-dual (ASD) parts, $\Omega^2(\mathcal{S}) = \Omega^2_+(\mathcal{S}) \oplus \Omega^2_-(\mathcal{S})$, with respect to the eigenvalues of the Hodge star operator:
\begin{align}
 F_\pm := \frac{1}{2} \left( F \pm \star F \right) \in \Omega^2_\pm(\mathcal{S}) \otimes \mathbb{C}[\mathfrak{g}]
 \quad \implies \quad
 \star F_\pm = \pm F_\pm
\end{align}
where $F_\pm$ is called the (A)SD part of the curvature.
If the SD (ASD) part is vanishing, the curvature becomes ASD (SD):
\begin{align}
 F_\pm = 0
 \quad \iff \quad
 \star F = \mp F
 \, .
 \label{eq:ASDYM_eq}
\end{align}
The vanishing SD (ASD) condition is called the {\em ASD (SD) YM equation},\index{(A)SD YM!---equation} and the connection solving (A)SD YM equation is then called the {\em (A)SD YM connection}. \index{(A)SD YM!---connection}
In fact, the (A)SD YM connection turns out to be a solution to the e.o.m. via the Bianchi identity\index{Bianchi identity}
\begin{align}
 D \star F \stackrel{\text{(A)SD}}{=} \mp D F \stackrel{\text{Bianchi}}{=} 0
 \, .
\end{align}
Under this decomposition, the number of components of two-form correspondingly splits into $6 = 3 + 3$, so that the (A)SD condition \eqref{eq:ASDYM_eq} seems a good solution of $\mathsf{d} = 4$ YM theory.

Furthermore, the YM action is bounded by the topological term as follows:
\begin{align}
 S_\text{YM}[A]
 & = - \frac{1}{2 g^2} \int_\mathcal{S} \tr
 \left( F \pm \star F \right) \wedge \star \left( F \pm \star F \right)
 \pm \frac{1}{g^2} \int_\mathcal{S} \tr F \wedge F
 \nonumber \\
 & = \frac{2}{g^2} \int_\mathcal{S} d\text{vol} \left| F_\pm \right|^2 \pm \frac{8 \pi^2 k}{g^2}
 \ge \frac{8 \pi^2 |k|}{g^2}
\end{align}
where $k \in \mathbb{Z}$ is the topological charge, called the {\em instanton number}, given by integrating the second Chern class over the four-manifold $\mathcal{S}$,
\begin{align}
 k := c_2[\mathcal{S}] = \frac{1}{8\pi^2} \int_\mathcal{S} \tr F \wedge F
 \, .
\end{align}
Here a solution with positive $k$ is called the {\em instanton}, and with negative $k$ is the {\em anti-instanton}, respectively.
The properties of instanton and anti-instanton are summarized in Tab.~\ref{tab:SDYM}.%
\footnote{%
We will discuss more refined version of the classification in \S\ref{sec:super_ADHM} (See Tab.~\ref{tab:super_SDYM}).
}

\begin{table}[t]
 \begin{center}
  \begin{tabular*}{.6\textwidth}{@{\extracolsep{\fill}}c|cc}
    \hline\hline
   & instanton & anti-instanton \\ \hline
   SD or ASD & $\star F = - F $ & $\star F = + F$ \\ 
   topological \# & $k>0$ & $k<0$ \\
   \hline\hline
   \end{tabular*}
 \end{center}
 \caption{Properties of instanton and anti-instanton.}
 \label{tab:SDYM}
\end{table}

\section{Summing up instantons}\label{sec:instanton_sum}

\subsection{$\theta$-term}

In addition to the YM action, one can incorporate the gauge invariant term in $\mathsf{d} = 4$ theory, which is called the {\em $\theta$-term}:
\begin{align}
 S_\theta[A] = - \frac{\im \theta}{8 \pi^2} \int_\mathcal{S} \tr F \wedge F
 \, .
\end{align}
This term does not contribute to the e.o.m. since it is a topological term, which is invariant under the infinitesimal variation.
This means that the (anti-)instanton is still a solution, providing a local minimum of the action.
Then, expanding the total action around the $k$-instanton configuration, $A = A_\text{inst}^{(k)} + \delta A$, we obtain
\begin{align}
 S_\text{tot}[A]
 & := S_\text{YM}[A] + S_\theta[A]
 \nonumber \\
 & = 
 \frac{2}{g^2} \int_\mathcal{S} d\text{vol} \left|F_\pm\right|^2
 + \left( \pm \frac{1}{g^2} - \frac{\im \theta}{8 \pi^2} \right) \int_\mathcal{S} \tr F \wedge F
 \nonumber \\
 & =
 \frac{8 \pi^2}{g^2} |k| - \im \theta k + S_\text{fluc}[\delta A] 
 \label{eq:YM_fluctuation}
\end{align}
where we introduce the fluctuation term $S_\text{fluc}[\delta A]$ schematically.
Since the instanton is a solution to the e.o.m., there is no linear term in $\delta A$, so that $S_\text{fluc}[\delta A]$ starts with quadratic term $O(\delta A^2)$.
Although computation of the fluctuation is in general difficult~\cite{tHooft:1976snw}, in some cases equipped with supersymmetry, the higher order terms are suppressed, and we could manage the computation.
See \S{\ref{sec:topological_twist}}.

In the path integral formalism, this is rephrased as follows:
\begin{align}
 Z & = \int [DA] \, \np^{- S_\text{tot}[A]} 
 \nonumber \\
 & = \sum_k \mathfrak{q}^k \int [DA_\text{inst}^{(k)}] 
 \int [D\delta A] \, \np^{- S_\text{fluc}[\delta A]}
 \label{eq:path_integral_instanton_sum}
\end{align}
where $[DA]$ is the path integral measure of the gauge field $A$, which splits into the instanton part $[DA_\text{inst}^{(k)}]$ and the fluctuation part $[D\delta A]$.
The instanton counting parameter (fugacity) is defined as \index{fugacity (instanton)}
\begin{align}
 \mathfrak{q} = \exp \left( 2 \pi \im \tau \right) = \exp \left( - \frac{8 \pi^2}{g^2} + \im \theta \right) \in \mathbb{C}^\times
 \label{eq:inst_fugacity}
\end{align}
with the complexified coupling constant
\begin{align}
 \tau = \frac{4\pi \im}{g^2} + \frac{\theta}{2\pi}
 \, .
 \label{eq:complex_coupling}
\end{align}
In principle, the path integral~\eqref{eq:path_integral_instanton_sum} also involves the anti-instantons contributions counted with the complex conjugate $\bar{\mathfrak{q}} = \exp \left( - 8 \pi^2/g^2 - \im \theta \right)$, but we focus on the instanton contributions which will be relevant in the context of $\mathcal{N} = 2$ gauge theory.
See \S\ref{sec:topological_twist} and \S\ref{sec:N=2} for details.
As seen in the expression of the fugacity~\eqref{eq:inst_fugacity}, it is invariant under the coupling shift, $\tau \to \tau + 1$, corresponding to the shift of the $\theta$-angle, $\theta \to \theta + 2\pi$.
This is analogous to the modular $\mathsf{T}$-transformation. 
In fact, this complexified coupling constant would be related to the complex structure of the Riemann surface associated with gauge theory (Seiberg--Witten curve), and the $\mathsf{S}$-transformation corresponds to the strong-weak duality of $\mathcal{N} = 2$ gauge theory.
See \S\ref{sec:SW_th} for details.

Let us then look at the path integral with the measure $[DA_\text{inst}^{(k)}]$.
This is interpreted as integral over all the possible $k$-instanton configuration in $G$-YM theory, namely the configuration space of the $k$-instanton solution, a.k.a., the instanton moduli space denoted by $\mathfrak{M}_{G,k}$.
We will discuss details of the moduli space in \S{\ref{sec:ADHM_mod_sp}}.

\subsection{Topological twist}
\label{sec:topological_twist}
\index{topological twist}

The gauge field path integral is schematically given as an infinite series over the different topological sectors as shown in~\eqref{eq:path_integral_instanton_sum}.
We can deal with this argument more rigorously in supersymmetric gauge theory.

\begin{table}[t]
 \begin{center}
  \begingroup
  \renewcommand{\arraystretch}{1.3}
  \begin{minipage}[c]{.4\textwidth}
   \begin{tabular}{cccc} \hline\hline
    & $\SU(2)_L$ & $\SU(2)_R$ & $\SU(2)_I$ \\ \hline
    $Q^i_\alpha$ & $\mathbf{2}$ & $\mathbf{1}$ & $\mathbf{2}$ \\
    $Q^i_{\dot{\alpha}}$ & $\mathbf{1}$ & $\mathbf{2}$ & $\mathbf{2}$ \\ \hline  
    $A_\mu$ & $\mathbf{2}$ & $\mathbf{2}$ & $\mathbf{1}$ \\
    $\lambda^i_\alpha$ & $\mathbf{2}$ & $\mathbf{1}$ & $\mathbf{2}$ \\
    $\tilde\lambda^i_{\dot{\alpha}}$ & $\mathbf{1}$ & $\mathbf{2}$ & $\mathbf{2}$ \\  
    $\phi$ & $\mathbf{1}$ & $\mathbf{1}$ & $\mathbf{1}$ \\
    \hline \hline
   \end{tabular}    
  \end{minipage}
  \quad $\stackrel{\text{twist}}{\implies}$ \quad
  \begin{minipage}[c]{.35\textwidth}
   \begin{tabular}{ccc} \hline\hline
    & $\SU(2)_L$ & $\SU(2)_d$ \\ \hline
    $G_\mu$ & $\mathbf{2}$ & $\mathbf{2}$ \\ 
    $(\mathcal{Q}, \mathcal{Q}_{\mu\nu}^+)$ & $\mathbf{1}$ & $\mathbf{1} \oplus \mathbf{3}$ \\ \hline
    $A_\mu$ & $\mathbf{2}$ & $\mathbf{2}$ \\ 
    $\lambda_\mu$ & $\mathbf{2}$ & $\mathbf{2}$ \\ 
    $(\eta, \chi_{\mu\nu}^+)$ & $\mathbf{1}$ & $\mathbf{1} \oplus \mathbf{3}$ \\
    $\phi$ & $\mathbf{1}$ & $\mathbf{1}$ \\
    \hline \hline
   \end{tabular}    
  \end{minipage}   
  \endgroup
 \end{center}
 \caption{$\SU(2)$ representations under the topological twist.}
 \label{tab:topological_twist}
\end{table}

$\mathcal{N} = 2$ supersymmetric gauge theory on $\mathbb{R}^4 = \mathbb{C}^2$ has global symmetries: Lorentz symmetry, $\operatorname{Spin}(4) = \SU(2)_L \times \SU(2)_R$, and R-symmetry, $\rU(2)_I = \SU(2)_I \times \rU(1)_I$.%
\footnote{%
We will review more details of 4d $\mathcal{N} = 2$ gauge theory in \S\ref{sec:N=2}.
}
We denote the supercharges of this theory by $(Q^i_\alpha, Q^i_{\dot{\alpha}})$, where the index $(\alpha,\dot{\alpha},i)$ is for $(\SU(2)_L, \SU(2)_R, \SU(2)_I)$, so that there are 8 supercharges in total.
The {\em topological twist} is the procedure to mix these $\SU(2)$ symmetries~\cite{Witten:1988ze}.
In particular, we consider the diagonal subgroup $\SU(2)_d \subset \SU(2)_R \times \SU(2)_I$.
Under this twist, the representations of $\SU(2)$'s are given for the supercharges and the vector multiplet $(A_\mu, \lambda^i_\alpha, \tilde\lambda^i_{\dot{\alpha}}, \phi)$ as in Tab.~\ref{tab:topological_twist}.
 For example, we denote the two-dimensional representation of $\SU(2)$ by $\mathbf{2}$.
 The scalar supercharge $\mathcal{Q}$ is to be identified with the BRST charge, which would be combined with the vector supercharge $G_\mu$ to construct a new (equivariant) BRST charge~\cite{Nekrasov:2002qd}.
 The tensor supercharge and the fermionic field obey the SD condition, $\star (\mathcal{Q}_{\mu\nu}^+, \chi_{\mu\nu}^+) = (\mathcal{Q}_{\mu\nu}^+, \chi_{\mu\nu}^+)$.
 In addition, we introduce the auxiliary field $H_{\mu\nu}^+$ corresponding to the tensor fermionic field~$\chi_{\mu\nu}^+$, which plays a role of the Lagrange multiplier implementing the constraint:
 \begin{align}
  F_{\mu\nu}^+ = 0
  \, ,
  \label{eq:twist_constraint_SD}
 \end{align}
 which is the ASD YM equation~\eqref{eq:ASDYM_eq}. \index{(A)SD YM!---equation}
 Hence the path integral localizes on the ASD YM configuration, and thus is reduced to the integral over the instanton moduli space.

 In the presence of the hypermultiplet in the fundamental representation, the constraint \eqref{eq:twist_constraint_SD} is replaced with the monopole equation, together with the additional Weyl equation:
 \begin{subequations}
  \begin{align}
   F_{\mu\nu}^+ + \im \psi^\dag \bar{\sigma}_{\mu\nu} \psi & = 0
   \, , \\
  D\hspace{-.67em}\slash \hspace{.25em} \psi_\alpha & = 0
   \, ,
   \label{eq:Weyl_zero_mode}
  \end{align}
 \end{subequations}
 where we denote the Weyl fermion in the hypermultiplet by~$\psi_\alpha$.
 Therefore, the localization locus of the path integral is the monopole configuration with the Weyl zero mode.%
 \footnote{%
 We can also deform the localization locus of the path integral to that on the instanton configuration.
 See, for example, \cite{Shadchin:2005mx} for details.
 }
 From the moduli space of this monopole equation, one can construct the Seiberg--Witten invariant of the four-manifold.
 See \cite{Morgan:1995,Moore:1996} for details.
 
 \section{ADHM construction of instantons}\label{sec:ADHM_construction}
 \index{ADHM construction}
 
 In this Section, let us introduce the systematic construction of instantons, a.k.a. the {\em ADHM construction}~\cite{Atiyah:1978ri}.%
 \footnote{%
 See also extensive review articles~\cite{Dorey:2002ik,Nakajima:2003uh} on this topic.
 }
 Although the original ADHM construction is for instantons on the four-sphere $S^4$, we apply it to the case on $\bR^4 = \bC^2$ by conformal compactification:
 Since we are interested in the instanton with finite action, the curvature should vanish at infinity, $F \to 0$, which implies $A \to g d g^{-1}$ at infinity. 
 One can rotate the group element $g \in G$ using the global part of the gauge symmetry, which is called the {\em framing} of the connection at infinity. \index{framing}

 \subsection{ADHM equation}
 
 We consider the unitary gauge group $G = \rU(n)$. 
 Then, in order to construct the $k$-instanton solution in $\rU(n)$-YM theory, we define the complex vector spaces:
 \begin{align}
  N = \bC^n
  \, , \qquad
  K = \bC^k
  \, .
  \label{eq:ADHM_vect_sp}
 \end{align}
 We define the base manifold
 \begin{align}
  X = \Hom(K,K) \oplus \Hom(K,K) \oplus \Hom(N,K) \oplus \Hom(K,N)
  \, .
  \label{eq:ADHM_var_space}
 \end{align}  
 The ADHM variables $(B_{1,2}, I, J) \in X$ are the linear maps associated with these vector spaces:
 \begin{align}
  B_{1,2} \in \Hom(K,K) = \End(K)
  \, , \quad
  I \in \Hom(N,K)
  \, , \quad
  J \in \Hom(K,N)
  \, ,
  \label{eq:ADHM_data}
 \end{align}
 which satisfy the ADHM equation
 \begin{align}
  (\mu_\bR, \mu_\bC) = (0,0)
  \, ,
  \label{eq:ADHM_eq}
 \end{align}
 with the moment maps $(\mu_\bR,\mu_\bC): X \to \bR^3 \otimes \mathfrak{u}_k^*$
 \begin{subequations}\label{eq:moment_maps}
  \begin{align}
   \mu_\bR & = [B_1, B_1^\dag] + [B_2, B_2^\dag] + II^\dag - J^\dag J \, ,
   \\
   \mu_\bC & = [B_1, B_2] + IJ \, .
  \end{align}
 \end{subequations}
  Let $G^\vee = \rU(K) = \rU(k)$, which is called the dual gauge group in the ADHM construction.
  These ADHM variables are defined modulo $G^\vee$-action:
  \begin{align}
   (g) \cdot (B_{1,2}, I, J) = (g B_{1,2} g^{-1}, g I , J g^{-1})
  \, , \qquad
   g \in \rU(K)
   \, .
   \label{eq:ADHM_gauge_transform}
  \end{align}

 There is an additional group action on these variables, $G = \rU(N) = \rU(n)$, which corresponds to the gauge transformation of $\rU(n)$-YM theory at infinity (the framing):
  \begin{align}
  (\nu) \cdot (B_{1,2}, I, J) = (B_{1,2}, I \nu^{-1}, \nu J)
  \, , \qquad
  \nu \in \rU(N)
   \, .
   \label{eq:ADHM_flavor_transform}
  \end{align}
  This global symmetry plays a role of the flavor symmetry for the ADHM variables.

  \subsection{Constructing instanton}
  
 Given the ADHM variables \eqref{eq:ADHM_data}, the ASD connection is constructed as follows.
 Let $(z_1, z_2) \in \mathcal{S} = \mathbb{C}^2$ be a spacetime coordinate.
 Then, define the {\em dual Dirac operator}\index{dual Dirac operator}%
 \begin{align}
  D^\dag =
  \begin{pmatrix}
   B_1 - z_1 & B_2 - z_2 & I \\
   - B_2^\dag + \bar{z}_2 & B_1^\dag - \bar{z}_1 & - J^\dag
  \end{pmatrix}
  \, ,
  \label{eq:dual_Dirac_op}
 \end{align}
 which is a map $D^\dag: K \otimes \cS \oplus N \to K \otimes \cS$.
 Due to the ADHM equation \eqref{eq:ADHM_eq}, we obtain 
 \begin{align}
  D^\dag D
  = \Delta \otimes \id_\cS
 \end{align}
 where 
\begin{subequations}
 \begin{align}
  \Delta
  & = (B_1 - z_1)(B_1^\dag - \bar{z}_1) + (B_2 - z_2)(B_2^\dag - \bar{z}_2) + II^\dag
  \, , \\
  & =
  (B_1^\dag - \bar{z}_1)(B_1 - z_1) + (B_2^\dag - \bar{z}_2)(B_2 - z_2) + J^\dag J
  \, .
 \end{align}
 \end{subequations}
 This is a map $\Delta: K \to K$, which behaves $\Delta \to |z|^2 \id_K$ at $z \to \infty$.

 Let $\Psi: N \to K \otimes \cS \oplus N$ be a zero mode of the dual Dirac operator, $D^\dag \Psi = 0$ ($\Psi \in \operatorname{Ker} D^\dag$) with the normalization condition $\Psi^\dag \Psi = \id_N$.
 This zero mode defines the projector from $K \otimes \cS \oplus N$ onto $N$:
 \begin{align}
  P := \Psi \Psi^\dag = \id_{K \otimes \cS \oplus N} - D (\Delta^{-1} \otimes \id_\cS) D^\dag
  \, ,
 \end{align}
 with $P D = \Psi \Psi^\dag D = 0$.
 This follows from the completeness condition of the vector space $K \otimes \cS \oplus N$, namely $\id_{K \otimes \cS \oplus N} = \Psi \Psi^\dag + D (D^\dag D)^{-1} D^\dag$. 
 Then, the ASD connection is constructed as\index{(A)SD YM!---connection}
 \begin{align}
  A = \Psi^\dag d \Psi = - \Psi d \Psi^\dag
  \, ,
  \label{eq:ADHM_connection}
 \end{align}
 which takes a value in $\mathfrak{u}_n = \operatorname{Lie} \rU(n)$ since it is anti-hermitian $A^\dag = - A$ with rank $n$.
 Let us show the curvature constructed with this connection is ASD:
 \begin{align}
  F & = dA + A \wedge A
  = d \Psi^\dag (\id - \Psi \Psi^\dag) d \Psi
  = \Psi^\dag (d D) (\Delta \otimes \id_\cS)^{-1} (d D^\dag) \Psi
  \, .
 \end{align}
 Recalling the definition of the Dirac operator \eqref{eq:dual_Dirac_op}, $dD^\dag$ and $dD$ give rise to the quaternion basis,
 $\sigma$ and $\bar{\sigma}$,
  respectively.
  Thus the curvature is proportional to $\bar{\sigma} \wedge \sigma$, which is ASD.
  Changing the orientation of the four-manifold, we will instead obtain the SD curvature proportional to $\sigma \wedge \bar{\sigma}$.

  
  Let us then compute the instanton charge based on the ADHM variables.
  We can use Osborn's formula~\cite{Osborn:1981yf} to rewrite the integral,
  \begin{align}
   \frac{1}{8\pi^2} \int_\mathcal{S} \tr F \wedge F
   = \frac{1}{16 \pi^2} \int d^4 x \, \partial^2 \partial^2 \tr_K \log \Delta^{-1}
   = \tr_K \id_K
   = k
   \, ,
   \label{eq:Osborn_formula}
  \end{align}
  where the Laplacian is denoted by $\partial^2$.
  The second equality is due to the asymptotic behavior of $\Delta \to |z|^2 \id_K$ at $z \to \infty$.
  This confirms that the ASD connection~\eqref{eq:ADHM_connection} gives rise to $k$-instanton solution.

  \subsection{Dirac zero mode}

  The Atiyah--Singer index theorem claims that there exist $k$ Dirac (Weyl) zero modes in the $k$-instanton background, which contribute to the path integral (See \S{\ref{sec:topological_twist}}).
  We construct such a zero mode with the ADHM variables~\cite{Osborn:1978rn}.

  First we decompose the ADHM zero mode $\Psi^\dag = (v_1^\dag \ v_2^\dag \ u^\dag)$, where $v_{1,2} : N \to K$, and $u : N \to N$.
  Then each component of the Weyl zero mode is given by $\psi_\alpha = v_\alpha^\dag \Delta^{-1}$.
  Since this is a bosonic solution to the Weyl equation, we define a fermionic map $\lambda : M \to \Pi K$ to construct the fermionic zero modes with the (fundamental) flavor space
  \begin{align}
   M = \mathbb{C}^{n^\text{f}}
   \, .
  \end{align}
  Then, $n^\text{f}$ flavor fermionic zero modes are given by $\psi_\alpha = v_\alpha^\dag \Delta^{-1} \lambda$ for $\alpha = 1,2$.
  The antifundamental zero modes are similarly constructed by $\bar{\lambda} : \Pi K \to \widetilde{M}$ with the antifundamental flavor space
  \begin{align}
   \widetilde{M} = \mathbb{C}^{n^\text{af}}
   \, .
  \end{align}

  \subsection{String theory perspective}\label{sec:ADHM_brane}

  Let us comment on the string theory perspective of the ADHM construction~\cite{Witten:1995gx,Douglas:1995bn}.
   
  The standard realization of gauge theory in string theory is to consider a stack of D-branes.
  The $k$-instanton configuration in four-dimensional $\rU(n)$-YM theory, in which we are interested, is realized as a bound state of $k$ D0 and $n$ D4 branes.%
  \footnote{%
  In general, we may consider $k$ D($4-p$) and $n$ D$p$ branes.
  }
  In this setup, the Chan--Paton vector spaces are $\bC^k$ and $\bC^n$, which would be identified with the vector spaces defined in the ADHM construction~\eqref{eq:ADHM_vect_sp}.
  The ADHM variables $(B_{1,2}, I, J)$ are then identified with open string degrees of freedom connecting D0-D0, D4-D0, D0-D4, respectively, and the ADHM equation is identified with the Bogomolnyi--Prasad--Sommerfield (BPS) equation, the supersymmetry preserving condition for this brane configuration.
  The $\rU(k)$ symmetry of the ADHM variables is interpreted as gauge symmetry of D0 brane world-volume theory, and the $\rU(n)$ symmetry is the flavor symmetry associated with the flavor D4 branes in this perspective.

  This dual description is interpreted as a consequence of T-duality, which exchanges D0 and D4 branes on a four-torus $T^4$ and its dual $\check{T}^4$:  
  \begin{align}
   \begin{tikzpicture}[baseline=(current  bounding  box.center)]
    \node [rectangle, draw, fill=white,
    text width=20em, text centered, rounded corners,
    minimum height=2.5em, 
    ]
    (k-inst) at (0,1.5) {$k$-instanton in $\rU(n)$-YM theory on $T^4$};
    \draw[blue,ultra thick,latex-latex,text=black] (0,-.7) -- ++(0,.7) node [right] {T-dual} -- ++(0,.7);
    \node [rectangle, draw, fill=white,
    text width=20em, text centered, rounded corners,
    minimum height=2.5em, 
    ]
   (n-inst) at (0,-1.5) {$n$-instanton in $\rU(k)$-YM theory on $\check{T}^4$};   \end{tikzpicture}
   \label{eq:ADHM_Nahm_dual}
  \end{align}
  Let $R$ be the radius of $T^4$, and the radius of the dual torus $\check{T}^4$ is given by $\check{R} = 1/R$.
  Then, in the limit $R \to \infty$, we have $T^4 \to \mathbb{R}^4$ and $\check{T}^4 \to $ pt.
  This zero dimensional description gives rise to the ADHM construction, which does not involve the derivative terms.
  
  In addition to the correspondence shown above, the construction of the ASD connection~\eqref{eq:ADHM_connection} also has an interpretation in the context of the tachyon condensation~\cite{Hashimoto:2005qh}.
  Furthermore, through the identification of the ASD connection as the Berry connection, we could establish the {\em Band/Brane correspondence}~\cite{Hashimoto:2015dla,Hashimoto:2016dtm}, which claims the equivalence of the band spectrum and the D-brane shape.

 \section{Instanton moduli space}\label{sec:ADHM_mod_sp}

 We have seen that the instanton solutions are systematically obtained using the ADHM construction, where the ADHM variables parametrize the configuration of the instantons.
 Since it is defined modulo $\rU(K)$ action~\eqref{eq:ADHM_gauge_transform}, we define the ADHM instanton moduli space as a hyper-K\"ahler quotient:
 \begin{align}
  \mathfrak{M}_{n,k}
  & = \{(B_{1,2}, I, J) \mid (\mu_\bR, \mu_\bC) = (0,0) \} /\!\!/\!\!/ \rU(k)
  \nonumber \\
  & = \mu_\bR^{-1}(0) \cap \mu_\bC^{-1}(0) /\!\!/\!\!/ \rU(k)
  \, .
 \end{align}
 The reason why we call this hyper-K\"ahler is that the quotient is taken with three conditions, $(\mu_{\mathbb{R}}^{-1}(0), \real \mu_{\mathbb{C}}^{-1}(0), \imag\mu_{\mathbb{C}}^{-1}(0) )$.
  The (complex) dimension of the moduli space is given by
 \begin{align}
  \dim_\mathbb{C} \mathfrak{M}_{n,k}
  & = 2 \dim \Hom(K,K) + \dim \Hom(N,K) + \dim \Hom(K,N)
  \nonumber \\
  & \quad
 - \frac{3}{2} \dim \Hom(K,K) - \dim \rU(K)
  \nonumber \\
  & = 2 nk
  \, .
  \label{eq:ADHM_dim}
 \end{align}
 We remark the dimensions of the ADHM equations are given by $\frac{3}{2} \dim \Hom(K,K) = \frac{3}{2} k^2$.
 The moduli space naively defined here is non-compact and singular, and thus we should regularize this for the later purpose.

 \subsection{Compactification and resolution}

 The first step is to compactify the moduli space by adding the point-like instanton, a.k.a. the Uhlenbeck compactification:\index{Uhlenbeck compactification}
  \begin{align}
   \bigsqcup_{k'=0}^k \mathfrak{M}_{n,k-k'} \times \operatorname{Sym}^{k'} \bC^2
  \end{align}
  where $\operatorname{Sym}^{k'} \bC^2$ is the $k'$-th symmetric product of $\bC^2$, corresponding to the point-like instanton.
  This is analogous to the compactification of $\bC^2$ to $S^4$ by adding $\{\infty\}$.
  This point-like instanton gives rise to the so-called small instanton singularity, and the next step is to resolve such a singularity in the moduli space.
 The resolution is then done by modifying the ADHM equation~\eqref{eq:ADHM_eq} as follows~\cite{Nakajima:1993jg}: 
 \begin{align}
  \mathfrak{M}^\zeta_{n,k}
  & = {\mu}^{-1}(\zeta) /\!\!/\!\!/ \rU(K)
  \, ,
  \label{eq:ADHM_mod_sp_res}
 \end{align}
 where we use the vector notation, ${\mu} = (\mu_\bR, \real \mu_\bC, \imag \mu_\bC) \ ( = \vec{\mu})$ and $\zeta = (\zeta \id_K, 0, 0) \ ( = \vec{\zeta})$.%
 \footnote{%
 In general, we may consider $\zeta = (\zeta_\bR, \real \zeta_\bC, \imag \zeta_\bC) \in \mathbb{R}^3 \otimes \mathfrak{u}_k^*$ with $\operatorname{ad}_g^* (\zeta_i) = \zeta_i$ for any $g \in \rU(K)$.
 From this point of view, we should consider $\zeta \to \im \zeta$ to interpret $\zeta \in \mathbb{R}$, since $\vec{\zeta} \in \mathbb{R}^3 \otimes \mathfrak{u}_k^*$.
 }
 The topological structure of the moduli space does not depend on the deformation parameter as long as $\zeta \neq 0$~\cite{Nakajima:1999}.
 The deformed moduli space $\mathfrak{M}^\zeta_{n,k}$ was also introduced in the study of instantons on the noncommutative $\bR^4 = \bC^2$~\cite{Nekrasov:1998ss}:
 There are two noncommutative parameters, $[z_i,\bar{z_i}] = \zeta_i$ ($i = 1,2$), and identified with $\zeta = \zeta_1 + \zeta_2$.
 In this case, one can construct the $\rU(1)$-instanton without the small instanton singularity, which is inevitable on the ordinary commutative space.
 Another (but slightly different) situation, where the deformation parameter is incorporated, is the moduli space of instantons on the asymptotically locally Euclidean (ALE) space~\cite{Kronheimer:1990MA}.
 The ALE space is obtained by a blow-up resolution of the orbifold singularity, $\bC^2/\Gamma$, where $\Gamma$ is a finite subgroup of $\SU(2)$, classified into the ADE type.
 See also \S\ref{sec:ADHM_ALE}.
 Physically, this blow-up is interpreted as insertion of magnetic fluxes, which also gives rise to noncommutativity locally~\cite{Seiberg:1999vs}.

 We introduce another moduli space, the framed moduli space of torsion-free sheaves on $\mathbb{P}^2$, which is isomorphic to the resolved moduli space $\mathfrak{M}^\zeta_{n,k}$.
 It is given by the quotient in the sense of geometric invariant theory (GIT)~\cite{Nakajima:1999}:
\begin{align}
 \widetilde{\mathfrak{M}}_{n,k} = \mu_\bC^{-1}(0) /\!\!/ \GL(K)
 \, ,
\end{align}
together with the {\em stability condition}\index{stability condition}
\begin{align}
 K =
 \begin{cases}
  \bC [ B_1, B_2] \, I(N) & (\zeta > 0) \\
  \bC [ B_1^\dag, B_2^\dag] \, J^\dag(N) & (\zeta < 0)
 \end{cases}
 \, .
 \label{eq:stab_cond}
\end{align}
 The condition for $\zeta < 0$ is also called the {\em co-stability condition}.
 See~\cite{Lee:2020hfu} for a related discussion.
 Here $\GL(K)$ denotes the general linear group associated with the vector space $K = \bC^k$, which is the complexification of $\rU(K)$, 
 and the condition $\mu_\bR = \zeta_{\neq 0} \id_K$ is replaced with the (co-)stability condition.
 Hence, the (complex) dimension of the moduli space is given by
 \begin{align}
  \dim_\mathbb{C} \widetilde{\mathfrak{M}}_{n,k} = \underbrace{2k^2}_{B_{1,2}} + \underbrace{2 nk}_{I, J} - \underbrace{k^2}_{\mu_\mathbb{C} = 0} - \underbrace{k^2}_{\GL(K)} = 2nk
  \, ,
 \end{align}
 which is consistent with the previous case~\eqref{eq:ADHM_dim}.
 We shall apply either the stability or the co-stability condition to evaluate the equivariant volume of the instanton moduli space, while for the instanton moduli space associated with supergroup gauge theory, both of them have to be simultaneously taken into account (Chapter~{\ref{sec:super_instanton}}).

 \subsection{Stability condition}\label{sec:stab_cond}

 Derivation of the stability condition~\eqref{eq:stab_cond} from the ADHM equation is as follows:
 We denote
 \begin{align}
  K' = \bC [ B_1, B_2] \, I(N) \subseteq K
  \, , \qquad
  K'' = \bC [ B_1^\dag, B_2^\dag] \, J^\dag(N) \subseteq K
  \, ,
 \end{align}
 and define
 \begin{align}
  K'_\bot = K - K'
  \, , \qquad
  K''_\bot = K - K''
  \, .
 \end{align}
 The projection operator, $P^\bullet: K \to K^\bullet_\bot$, obeys $[P^\bullet, B_{1,2}] = 0$ and $P' I = P'' J^\dag = 0$.
 Then, from the condition $\mu_\bR = \zeta_{\neq 0} \id_K$, we obtain
 \begin{align}
  \zeta \, \id_{K_\bot^\bullet} = P^{\bullet} \, \mu_\mathbb{R} \, P^\bullet
  =
  \begin{cases}
   P' \left( [B_1,B_1^\dag] + [B_2,B_2^\dag] - J^\dag J \right) P' \\[.5em]
   P'' \left( [B_1,B_1^\dag] + [B_2,B_2^\dag] + II^\dag \right) P''    
  \end{cases}
  \, .
 \end{align}
 Taking a trace yields
 \begin{subequations}
 \begin{align}
  0 \le \tr \zeta_{>0} \id_{K_\bot'} & = - \tr P' J^\dag J P' \le 0 \, , \\
  0 \ge \tr \zeta_{<0} \id_{K_\bot''} & = \tr P'' II^\dag P'' \ge 0 \, ,
 \end{align}
 \end{subequations}
 which implies $K_\bot' = 0$ for $\zeta > 0$ and $K_\bot'' = 0$ for $\zeta < 0$.
 This proves the (co-)stability condition~\eqref{eq:stab_cond}.

\section{Equivariant localization of instanton moduli space}

 As mentioned in \S{\ref{sec:instanton_sum}}, we are interested in the volume of the instanton moduli space, which gives rise to an important contribution to the gauge theory path integral.
 If we naively consider the moduli space, however, we may have a diverging volume due to the small instanton singularity and non-compactness of the moduli space.

 After the regularization as discussed before, we can now consider the {\em equivariant integral}, which utilizes the equivariant group action on the manifold, together with the {\em localization formula}, claiming that the integral localizes on fixed point loci under the equivariant action.
 Duistermaat--Heckman's formula~\cite{Duistermaat:1982vw} is the primary example of the localization formula for a symplectic compact manifold equipped with $\rU(1)$ action associated with the moment map.
 This is then generalized by Berline--Vergne and Atiyah--Bott to the case for a generic compact manifold with $\rU(1)$ equivariant action~\cite{Berline:1982,Atiyah:1984px}.
 Although the localization formula is originally formulated for a compact manifold, we would formally apply it to a non-compact and infinite dimensional integral as well, as long as the equivariant fixed point is compact (equivariantly compact).
 \index{equivariantly compact}

 There have been a lot of applications of the localization formula to the path integral~\cite{Witten:1982im,Witten:1988ze,Witten:1988xj}, and it turns out to be applicable to the instanton moduli space~\cite{Losev:1997tp,Moore:1997dj,Lossev:1997bz,Moore:1998et,Bruzzo:2002xf}, which leads to the instanton partition function~\cite{Nekrasov:2002qd}.
 The localization method is then applied to the path integral of the supersymmetric gauge theory on a curved compact manifold~\cite{Pestun:2007rz}.
 See review articles on this topic~\cite{Szabo:1996md,Pestun:2016zxk} for more details. 


 \subsection{Equivariant cohomology}
 \index{equivariant!---cohomology}
 
 Let us briefly review the equivariant cohomology.
 See the textbook on this topic~\cite{Berline:2003} for more details.
 
 Let $X$ be a manifold equipped with a free group action $G$.%
 \footnote{%
 If the $G$-action is not free on $X$, $X/G$ is not an ordinary manifold.
 Thus, in this case, the $G$-equivariant cohomology is defined with the universal bundle $EG$, on which the group $G$ freely acts:
 \begin{align}
  H_G^\bullet(X) = H^\bullet(X \times_G EG) = H^\bullet((X \times EG)/G)
  \, .
 \end{align}
 In particular, for $X = \text{pt}$, it becomes $H_G^\bullet(\text{pt}) = H^\bullet(BG)$, where $BG = EG/G$ is the classifying space.
 See, for example, \cite{Pestun:2016qko} for details.
 }
 Then the $G$-equivariant cohomologies of $X$ are isomorphic to the de Rham cohomologies of $X/G$:
 \begin{align}
  H_G^\bullet(X) \cong H^\bullet (X/G) 
  \, .
 \end{align}
 This is constructed as follows.
 Let $\mathfrak{g} = \operatorname{Lie}(G)$, and we define the $G$-equivariant differential forms (See also \S\ref{sec:YM_theory}):\index{equivariant!---form} 
 \begin{align}
  \Omega^\bullet_G(X) = \left( \Omega^\bullet (X) \otimes \bC[\mathfrak{g}] \right)^G
  \, ,
 \end{align}
 where the induced action on $\alpha(\mathsf{x}) \in \Omega_G^\bullet(X)$, $\mathsf{x} \in \mathfrak{g}$, is given by the pullback:
 \begin{align}
  g^* \alpha(\mathsf{x}) = 
  \alpha(\operatorname{ad}(g) \mathsf{x})
  \qquad \text{for} \qquad
  ^\forall g \in G
  \, .
 \end{align}
 Denoting the vector field representing a Lie algebra element on $X$ by $V(x)$, we define the Lie derivative $\mathcal{L}_V : \Omega^\bullet(X) \to \Omega^\bullet(X)$ with respect to the vector field $V$:
 \begin{align}
  \mathcal{L}_V = d \iota_V + \iota_V d
 \end{align}
 with the nilpotent interior multiplication (also called the contraction) $\iota_V : \Omega^\bullet(X) \to \Omega^{\bullet - 1}(X)$.
 In addition, we define the equivariant exterior derivative $d_\mathfrak{g} : \Omega_G^\bullet(X) \to \Omega_G^{\bullet + 1}(X)$ with the interior multiplication $\iota_V$:
 \begin{align}
  d_\mathfrak{g} = d + \iota_V
  \, ,
 \end{align}
 which leads to
 \begin{align}
  d_\mathfrak{g}^2 = \mathcal{L}_V
  \, .
  \label{eq:d_to_Lie_der}
 \end{align}
 Since the Lie derivative vanishes on the equivariant forms $\alpha(\mathsf{x}) \in \Omega_G^\bullet(X)$:
\begin{align}
 \mathcal{L}_V \alpha(\mathsf{x}) = 0
 \, ,
\end{align}
we define the equivariant cohomology based on the nilpotent equivariant derivative:
\begin{align}
 H_G^\bullet (X) = \operatorname{Ker} d_\mathfrak{g} / \operatorname{Im} d_\mathfrak{g}
 \, .
\end{align}
In the following, we will consider the integral of the equivariant form, and show the equivariant localization formula based on this argument.
 
 \subsection{Equivariant localization}\label{sec:equiv_int}
 \index{equivariant!---localization}

 We consider the integral of the $G$-equivariant closed form $\alpha(x,dx)$, s.t., $d_\mathfrak{g} \alpha(x,dx) = 0$, on the manifold $X$:
 \begin{align}
  \int_X \alpha(x, dx)
  \, .
 \end{align}
 Let us evaluate this integral based on a physical argument in the following.
 
 Let $(x^\mu)_{\mu = 1,\ldots,n}$ be the coordinates of the manifold $X$ with $n = \dim X$, and define the fermionic (anti-commuting) coordinates $(\psi^\mu = d x^\mu)_{\mu = 1,\ldots, n}$ for the fibers of the odd (parity flipped) tangent bundle $\Pi TX$ (See also \S\ref{sec:sup_vec_sp}).
 In the QFT context, we identify the nilpotent exterior derivative as the BRST charge, $d_\mathfrak{g} \to \mathcal{Q}$, and the corresponding BRST transformation is given as follows:
 \begin{align}
  \mathcal{Q} x^\mu  = \psi^\mu
  \, , \qquad
  \mathcal{Q} \psi^\mu = V^\mu(x) = T^a V_a^\mu(x)
  \label{eq:equivariant_Q-trans}
 \end{align}
 where $(T^a)_{a = 1,\ldots,\dim \mathfrak{g}}$ are the generators of the $G$-action on $X$, and $V_a^\mu(x)$ is the corresponding component of the vector field $V^\mu(x)$.
 Namely, the operator $\mathcal{Q}^2$ generates the infinitesimal $G$-transformation, which corresponds to the relation \eqref{eq:d_to_Lie_der}. 
 The integral is rewritten as
 \begin{align}
  \int_X \alpha(x, dx) \, dx^1 \wedge \cdots \wedge dx^n
  & = \int_X \tilde{\alpha}(x, \psi) \, d^n x \, d^n \psi
  \, ,
 \end{align}
 where the new observable is given by
 \begin{align}
  \tilde{\alpha}(x, \psi) = \alpha(x,\psi) \psi^1 \cdots \psi^n
 \end{align}
 which is again supposed to be closed $\mathcal{Q} \tilde{\alpha}(x, \psi) = 0$.
 Let us then deform the integral with the $\mathcal{Q}$-exact term as follows:
 \begin{align}
  \int_X \tilde{\alpha}(x,\psi) \, \np^{- t \mathcal{Q} W} \, d^n x \, d^n \psi
  \, ,
  \label{eq:eq_int_t-def}
 \end{align}
 where $t$ is the deformation parameter, and the potential $W$ is an arbitrary fermionic function, so that $\mathcal{Q}W$ becomes a bosonic term.
 Recalling $\mathcal{Q} \tilde{\alpha}(x, \psi) = 0$, $\mathcal{Q}^2 W = 0$, we obtain
 \begin{align}
  \frac{d}{dt} \int_X \tilde{\alpha}(x,\psi) \, \np^{- t \mathcal{Q} W} \, d^n x \, d^n \psi
  & = - t \int_X \mathcal{Q} \left( \tilde{\alpha}(x, \psi) \, W \, \np^{-t\mathcal{Q}W} \right) d^n x \, d^n \psi
  \, .
 \end{align}
 Since it is now written as the total derivative form, it turns out that the integral~\eqref{eq:eq_int_t-def} is independent of the deformation parameter $t$, and we can evaluate the integral with arbitrary $t$.

 A typical choice is $t \to \infty$, which allows us to apply the semiclassical analysis.
 We set the potential $W = V_\mu(x) \psi^\mu$, and the corresponding contribution to the action is given by
 \begin{align}
  \mathcal{Q}W = \partial_\mu V_\nu(x) \psi^\mu \psi^\nu +  V_\mu(x) V^\mu(x)
  \, .
  \label{eq:QW}
 \end{align}
 Hence, in the limit $t \to \infty$, these quadratic terms are dominating in the action.
 The corresponding critical point $(x_\text{c}, \psi_\text{c})$ is then given by $V_\mu(x_\text{c}) = 0$ with $\psi_\text{c} = 0$, which is the fixed point under the $G$-action on $X$.
 Actually this critical point equation is generic as explained as follows:
 Because the potential $W$ should be fermionic, it contains odd number $\psi$ variables.
 In order to obtain a Lorentz scalar, at least one of them should be contracted with the vector field $V_\mu(x)$ as $V_\mu(x) \psi^\mu$.
 From this point of view, our choice $W = V_\mu(x) \psi^\mu$ is the minimal one.

 In order to apply the semiclassical analysis, we expand the $(x,\psi)$ variables around the critical point:
 \begin{align}
  x = x_\text{c} + t^{-\frac{1}{2}} \xi
  \, , \qquad
  \psi = \underbrace{\psi_\text{c}}_{= \, 0} + \ t^{-\frac{1}{2}} \eta
  \, ,
 \end{align}
 and the potential term~\eqref{eq:QW} is given by
 \begin{align}
  t \mathcal{Q}W = \partial_\mu V_\nu(x_\text{c}) \eta^\mu \eta^\nu + \partial_\mu V_\rho(x_\text{c}) \partial_\nu V^\rho(x_\text{c}) \xi^\mu \xi^\nu + O(t^{-1})
  \, .
 \end{align}
 We remark that, in this case, the non-linear terms in the BRST transformation~\eqref{eq:equivariant_Q-trans} are suppressed in the limit $t \to \infty$:
 \begin{align}
  \mathcal{Q} \xi^\mu = \eta^\mu
  \, , \qquad
  \mathcal{Q} \eta^\mu = \partial_\nu V^\mu(x_\text{c}) \xi^\nu
  \, .
 \end{align}
 We then perform the Gaussian integral to obtain the contribution associated with the critical point $(x_\text{c}, \psi_\text{c})$ with the measure given by
 \begin{align}
  d^n x \, d^n \psi
  = \left( t^{-\frac{n}{2}} d^n \xi \right) \left( t^{+\frac{n}{2}} d^n \eta \right)
  = d^n \xi \, d^n \eta
  \, .
 \end{align} 
 Summing up all the fixed point contributions, we obtain the equivariant localization formula:
 \begin{itembox}{Equivariant localization formula (Berline--Vergne--Atiyah--Bott formula)}
  Let $V_\mu(x)$ be the vector field associated with the $G$-action on the manifold $X$.
  Then, the integral of the $G$-closed form $\alpha(x,dx)$ over $X$ localizes on the critical configurations denoted by $\{x_\text{c}\}$,
  \begin{align}
   \int_X \alpha(x,dx)
  = \sum_{x_\text{c}}
   \frac{\alpha(x_\text{c}, 0)}{\sqrt{\det \partial_\mu V_\nu(x_\text{c})}}
   \, ,
   \label{eq:BVAB_formula}
  \end{align}
  with the critical/fixed point equation
  \begin{align}
   V_\mu(x_\text{c}) = 0
   \, .
  \end{align}  
 \end{itembox}
  Although we have implicitly assumed that $X$ is compact so far, we could formally apply the localization formula to non-compact manifolds, in particular, if the critical point is isolated and compact.
  Such a situation is called equivariantly compact.
  In many examples in physics, we would like to deal with non-compact, infinite dimensional manifolds, but still equivariantly compact.
  We could apply the equivariant localization formula to obtain exact results for such a case, which is used to discuss non-perturbative aspects of QFT.
  \index{equivariantly compact}

 \subsection{Equivariant action and fixed point analysis}\label{sec:eq_fixed_point}
 \index{equivariant!---action}
 
 In order to apply the equivariant localization formalism to the ADHM moduli space, let us specify the equivariant action and the corresponding fixed point under it.

 \subsubsection{Spacetime rotation}
 
 As shown in \S\ref{sec:ADHM_construction}, there are $G^\vee = \rU(k)$ and $G = \rU(n)$ actions on the ADHM variables, \eqref{eq:ADHM_gauge_transform} and \eqref{eq:ADHM_flavor_transform}.
 In addition to these group actions, there is another action corresponding to the spacetime rotation of gauge theory on $\mathcal{S} = \bC^2$:%
 \footnote{%
 This convention is chosen to be consistent with the equivariant character formula shown in \S\ref{sec:eq_ch_formula}.
 }
 \begin{align}
  (q_1,q_2) \cdot (B_1, B_2, I, J)
  = (q_1^{-1} B_1, q_2^{-1} B_2, I, q^{-1} J)
 \end{align}
 where
 \begin{align}
  (q_1, q_2) = (\np^{\epsilon_1}, \np^{\epsilon_2})
  \in \mathsf{T}_\mathbf{Q} := \rU(1)^2 \subset \operatorname{Spin}(4)
  \label{eq:q12_torus}
 \end{align}
 and
 \begin{align}
  q := q_1 q_2 = \np^{\epsilon_{12}}
  \, , \qquad
  \epsilon_{12} = \epsilon_1 + \epsilon_2 
  \, .
 \end{align}
 The parameters $(q_1,q_2)$ are the equivariant parameters for $\operatorname{Spin}(4)$, 
 which are the exponentiated (multiplicative) version of the $\Omega$-background parameters.
 
 \subsubsection{Fixed point analysis}

 In order to apply the localization formula, we should specify the fixed point under the equivariant action.
 Let us analyse the fixed point in the ADHM moduli space explicitly.
 
 We parametrize elements of $\rU(k)$ and $\rU(n)$ groups with the corresponding Lie algebras: 
 \begin{align}
  g = \np^{\phi}
  \, , \qquad
  \nu = \np^{\mathsf{a}}
 \end{align}
 where $\phi \in \mathfrak{u}_k$, $\mathsf{a} \in \mathfrak{u}_n$.
Then the fixed point equations are given as follows: 
\begin{subequations}\label{eq:fixed_pt1}
 \begin{align}
  [\phi, B_{1,2}] - \epsilon_{1,2} B_{1,2} & = 0
  \label{eq:fixed_pt_B}
  \, , \\
  \phi I - I \mathsf{a} & = 0
  \, , \\
  - J \phi + \mathsf{a} J - \epsilon_{12} J & = 0
  \, .
 \end{align}
\end{subequations}
Namely, they are invariant under the $\rU(n) \times \rU(1)^2$ action modulo $\rU(k)$ symmetry at the fixed point.
Here we can assume that the element $\mathsf{a} \in \mathfrak{u}_n$ is diagonal without loss of generality, so that it is an element of the Cartan subalgebra of $\mathfrak{u}_n$.
This is because, under $\rU(n)$ transformation, $\mathsf{a} \mapsto h \mathsf{a} h^{-1}$ $(h \in \rU(n))$, we have $\phi I - I (h \mathsf{a} h^{-1}) = 0 \iff \phi I' - I' \mathsf{a} = 0$ with $I' = I h$.
Similarly, we have $- J \phi + h \mathsf{a} h^{-1} J - \epsilon_{12} J = 0 \iff - J' \phi + \mathsf{a} J' - \epsilon_{12}  J' = 0$ with $J' = h^{-1} J$.
Hence, we can choose the basis which diagonalizes $\mathsf{a} \in \mathfrak{u}_n$.
We decompose it into one-dimensional elements:
 \begin{align}
  \mathsf{a} = \bigoplus_{\alpha=1}^n \mathsf{a}_\alpha
  \, .
  \label{eq:Coulomb_diagonal}
 \end{align}
 We may focus on the maximal torus of $G = \rU(n)$, $\mathsf{T}_N =  \rU(1)^n \subset \rU(n)$, and also decompose $(I,J)$ with respect to the $\mathsf{T}_N$ torus action:
 \begin{align}
  I = \bigoplus_{\alpha = 1}^n I_\alpha
  \, , \qquad
  J = \bigoplus_{\alpha = 1}^n J_\alpha
  \, .
 \end{align}
 The fixed point equations under the full torus action $\mathsf{T}_N \times \mathsf{T}_\mathbf{Q} = \rU(1)^n \times \rU(1)^2$ are given by
\begin{subequations}
\begin{align}
 \phi I_\alpha & = \mathsf{a}_\alpha I_\alpha
 \, , \\
 J_\alpha \phi & = ( \mathsf{a}_\alpha - \epsilon_{12} ) J_\alpha
 \, .
\end{align}
\end{subequations}
Now $(I_\alpha)_{\alpha = 1,\ldots,n}$ and $(J_\alpha)_{\alpha = 1,\ldots,n}$ are the left and right eigenvectors of $\phi$.
Since the corresponding eigenvalues do not coincide with each other for generic $(\epsilon_1,\epsilon_2)$, $\mathsf{a}_\alpha \neq \mathsf{a}_{\alpha'} + \epsilon_{12}$ for $\alpha,\alpha' \in (1,\ldots,n)$, we have
\begin{align}
 J_\alpha I_{\alpha'} = 0
 \, .
\end{align}
Therefore, together with the ADHM equation $\mu_\bC = 0$, $B_1$ and $B_2$ become commutative at the fixed point:
\begin{align}
 [B_1, B_2] = 0
 \, .
 \label{eq:B12_commuting}
\end{align}

Using the fixed point equation~\eqref{eq:fixed_pt_B} for $m = 1, 2$, we have
\begin{subequations}
 \begin{align}
  \phi B_m I_\alpha & = B_m \phi I_\alpha + \epsilon_m B_m I_\alpha = (\mathsf{a}_\alpha + \epsilon_m) B_m I_\alpha
  \, , \\
  J_\alpha B_m \phi & = J_\alpha \phi B_m - \epsilon_m J_{\alpha} B_m = J_\alpha B_m (\mathsf{a}_\alpha - \epsilon_{12} - \epsilon_m)
  \, .
 \end{align}
\end{subequations}
Hence we can generate the left and right eigenvectors by applying the matrices $B_{1,2}$ to $(I_\alpha,J_\alpha)_{\alpha = 1,\ldots,n}$.
Applying the same argument recursively, we obtain
\begin{subequations}\label{eq:phi_ev}
 \begin{align}
  \phi \left( B_1^{s_1 - 1} B_2^{s_2 - 1} I_\alpha \right) & = (\mathsf{a}_\alpha + (s_1 - 1) \epsilon_1 + (s_2 - 1) \epsilon_2) \left( B_1^{s_1 - 1} B_2^{s_2 - 1} I_\alpha \right)
  \, ,
  \label{eq:phi_ev_I} \\
  \left( J_\alpha B_1^{s_1 - 1} B_2^{s_2 - 1} \right) \phi & = (\mathsf{a}_\alpha - s_1 \epsilon_1 - s_2 \epsilon_2) \left( J_\alpha B_1^{s_1 - 1} B_2^{s_2 - 1} \right)
  \, ,
  \label{eq:phi_ev_J}
 \end{align}
\end{subequations}
for $s_{1,2} \in (1,\ldots,\infty)$.
We remark that $B_1$ and $B_2$ are commutative at the fixed point~\eqref{eq:B12_commuting}, so that the order of $B_{1,2}$-multiplication does not matter in the eigenvectors.
Although we obtain infinitely many eigenvectors formally, which are linearly independent with different eigenvalues, there should be only $k$ independent eigenvectors, since $\operatorname{rk} \phi = k$.
This is actually consistent with the stability condition \eqref{eq:stab_cond} discussed in \S\ref{sec:ADHM_mod_sp} (and the co-stability condition as well).
\index{stability condition}

We denote the discrete set of the equivariant $\mathsf{T}$-fixed points in the moduli space by $\mathfrak{M}^\mathsf{T}$.
Then, the fixed point is parametrized by the $n$-tuple partition $\lambda = (\lambda_{\alpha})_{\alpha = 1,\ldots,n}$ with $|\lambda| = \sum_{\alpha = 1}^n |\lambda_\alpha| = k$, and each $\lambda_\alpha$ is a partition $\lambda_\alpha = (\lambda_{\alpha,1} \ge \lambda_{\alpha,2} \ge \cdots \ge 0)$~\cite{Nakajima:1999,Nekrasov:2002qd,Nekrasov:2003rj}.
Each box $s = (s_1, s_2) \in \lambda_{\alpha}$%
 \footnote{%
 This terminology makes sense based on the graphical representation of the partition with the Young diagram.
 We may abuse some terminologies through this identification.
 }
 is associated with a monomial $z_1^{s_1 - 1} z_2^{s_2 - 1}$ with $s_1 \in (1,\ldots,\infty)$, $s_2 \in (1,\ldots,\lambda_{\alpha,s_1})$.
 Let $\mathbf{I}_{\lambda_\alpha} \subset \bC[z_1,z_2] = \mathbf{I}_\emptyset$ be the ideal generated by all monomials outside the partition, $z_1^{s_1 - 1} z_2^{s_2 - 1}$ with $s \not\in \lambda_\alpha$, while $K_{\lambda_\alpha} = \mathbf{I}_\emptyset/\mathbf{I}_{\lambda_\alpha}$ is generated by those inside the partition. 
 Then, we obtain 
 \begin{align}
  K = \bigoplus_{\alpha = 1}^n K_{\lambda_\alpha}
  \, ,
 \end{align}
 and the eigenvalue of $\phi$ associated to each box $s \in \lambda_\alpha$~\eqref{eq:phi_ev_I}, denoted by $\phi_s$, is given by
\begin{align}
 \phi_{s} = \mathsf{a}_\alpha + (s_1 - 1) \epsilon_1 + (s_2 - 1) \epsilon_2
 \, .
  \label{eq:phi_ev_I2}
\end{align}
This is called the $\mathsf{a}$-shifted content of the box $(s_1, s_2) \in \lambda_\alpha$ in the partition.
We can similarly formulate with the co-stability condition.
In this case, we instead assign the eigenvalue of $\phi$~\eqref{eq:phi_ev_J} to each box in the partition.

 \section{Integrating ADHM variables}\label{sec:int_ADHM_var}

 Let us recall that the ADHM variables $(B_{1,2}, I, J)$ are the coordinates of $X = \Hom(K,K) \oplus \Hom(K,K) \oplus \Hom(N,K) \oplus \Hom(K,N)$ with $G^\vee = \rU(k)$ acting on these ADHM variables.
 The ADHM moduli space~\eqref{eq:ADHM_mod_sp_res} is given by the quotient of the level set $N = \vec{s}^{-1}(0) \subset X$ as $\mathfrak{M}^\zeta_{n,k} = N/G^\vee$, where the section is defined as $\vec{s} = \vec{\mu} - \vec{\zeta}$ with $\vec{\zeta} = (\zeta , 0, 0)$.

 Following the equivariant integral formalism presented in \S\ref{sec:equiv_int}, we define the fermionic coordinates corresponding to the anti-commuting one-forms: $(\psi_{B_{1,2}}, \psi_I, \psi_J) \in \Pi TX$.
 Furthermore, in this case, we shall apply the Mathai--Quillen formalism to incorporate the ADHM equation $\vec{s} = 0$ \index{Mathai--Quillen formalism}%
 into the equivariant integral~\cite{Moore:1997dj,Lossev:1997bz}.
 Let $(\vec{\chi}, \vec{H})$ and $(\bar{\phi},\eta)$ be the anti-ghost multiplets,%
 \footnote{%
 The latter one $(\bar{\phi},\eta)$ is also called the projection multiplet.
 }
 which take a value in $\mathfrak{g}^\vee = \operatorname{Lie} G^\vee$.%
 \footnote{%
 Not to be confused with the dual of the Lie algebra $\mathfrak{g}$ denoted by $\mathfrak{g}^*$.
 }
 Recalling the BRST transformation should be compatible with the equivariant action~\eqref{eq:equivariant_Q-trans}, we define all the transformations as follows:\\[-.5em]
 \begin{subequations}
  \begin{minipage}{.5\textwidth}
  \begin{align}
   \mathcal{Q} B_{1,2} & = \psi_{B_{1,2}} \, , \\
   \mathcal{Q} I & = \psi_I \, , \\
   \mathcal{Q} J & = \psi_J \, , \\
   \mathcal{Q} \chi_\bR & = H_\bR \, , \\
   \mathcal{Q} \chi_\bC & = H_\bC \, , \\   
   \mathcal{Q} \bar{\phi} & = \eta \, , 
  \end{align}
  \end{minipage}
  \begin{minipage}{.5\textwidth}
   \begin{align}
    \mathcal{Q} \psi_{B_{1,2}} & = [\phi, B_{1,2}] - \epsilon_{1,2} B_{1,2} \, , \\
    \mathcal{Q} \psi_I & = \phi I - I \mathsf{a} \, , \\
    \mathcal{Q} \psi_J & = - J \phi + \mathsf{a} J - \epsilon_{12} J \, , \\
    \mathcal{Q} H_\bR & = [\phi, \chi_\bR] \, , \\
    \mathcal{Q} H_\bC & = [\phi, \chi_\bC] + \epsilon_{12} \chi_\bC \, , \\
    \mathcal{Q} \eta & = [\phi,\bar{\phi}] \, ,
   \end{align}
  \end{minipage}\\[1em]
 \end{subequations}
 where $(\phi, \mathsf{a}, \epsilon_{1,2}) \in \operatorname{Lie} (\rU(k), \rU(n), \rU(1)^2)$. 
 We remark that the fixed point equation~\eqref{eq:fixed_pt1} is equivalent to the condition $\mathcal{Q} \psi_{B_{1,2}, I, J} = 0$.
 Since the ADHM variables are complex, we have the conjugate variables obeying the following BRST transformations:\\[-.5em]
 \begin{subequations}
    \begin{minipage}{.5\textwidth}
     \begin{align}
      \mathcal{Q} B_{1,2}^\dag & = \bar{\psi}_{B_{1,2}} \, , \\
      \mathcal{Q} I^\dag & = \bar{\psi}_I \, , \\
      \mathcal{Q} J^\dag & = \bar{\psi}_J \, , 
     \end{align}
   \end{minipage}
    \begin{minipage}{.5\textwidth}
     \begin{align}
      \mathcal{Q} \bar{\psi}_{B_{1,2}} & = - [\phi, B_{1,2}^\dag] + \epsilon_{1,2} B_{1,2}^\dag \, , \\
      \mathcal{Q} \bar{\psi}_I & = - I^\dag \phi + \mathsf{a} I^\dag \, , \\
      \mathcal{Q} \bar{\psi}_J & = \phi J^\dag - J^\dag \mathsf{a} + \epsilon_{12} J^\dag \, .
     \end{align}
   \end{minipage}\\
 \end{subequations}

 We turn to the integral over the ADHM moduli space $\mathfrak{M}^\zeta_{n,k} = N / G^\vee$.
 In order to perform this integral, we map the equivariant cohomology on $X$ to the ordinary cohomology on $N/G^\vee$ based on the inclusion map $i: N \hookrightarrow X$. \index{equivariant!---cohomology}
 We define the map $I \circ i^*$, consisting of the pullback $i^* : H_{G^\vee}^\bullet(X) \hookrightarrow H_{G^\vee}^\bullet(N)$, and the isomorphism $I : H_{G^\vee}^\bullet(N) \cong H^\bullet(N/G^\vee)$ (if the $G^\vee$-action is free on $N$).
 Then, we obtain the cohomology class in $N/G^\vee$, $\tilde{\alpha} = I \circ i^* \alpha(\phi)$ with $\alpha(\phi) \in \Omega_{G^\vee}^\bullet(X)$.
 The naive integral of the equivariant form over $X$ is the pushforward map $\Omega_{G^\vee}^\bullet(X) \to \Omega_{G^\vee}^\bullet(\text{pt})$, which descends to $H_{G^\vee}^\bullet(X) \to H_{G^\vee}^\bullet(\text{pt})$.
 In addition, we should fix the constant factor due to the translation invariance of the $\mathfrak{g}^\vee$-measure. \index{equivariant!---form}
 This is done as follows~\cite{Witten:1992xu}:
 First we take an arbitrary Haar measure on $G^\vee$. 
 We define the measure $d^{\operatorname{rk} \mathfrak{g}^\vee} \phi$ with the Euclidean coordinates on $\mathfrak{g}^\vee$ to be consistent with the Haar measure of $G^\vee$ at the identity.
 Then $d^{\operatorname{rk} \mathfrak{g}^\vee} \phi / \operatorname{vol} G^\vee$ is a natural measure on $\mathfrak{g}^\vee$ independent of the choice of the Haar measure on $G^\vee$.
 Hence, we define the equivariant integral $H_{G^\vee}^\bullet(X) \to \mathbb{C}$ as an integral over $X \times \mathfrak{g}^\vee$ with the measure on $\mathfrak{g}^\vee$ introduced above.

 \subsection{Path integral formalism}
 
 We now consider the equivariant volume of the moduli space $\mathfrak{M}_{n,k}$.
 The ``path integral form'' of the equivariant integral over the ADHM variables $X$ is given as follows:
 \begin{align}
  Z_{n,k}(\mathsf{a}, \epsilon_{1,2})
  := \int_{\mathfrak{M}_{n,k}} 1  
  = \int_{\mathfrak{g}^\vee} \frac{d \phi}{\operatorname{vol} G^\vee}
  \int_X \np^{-S} 
  \label{eq:ADHM_mod_sp_integral}
 \end{align}
 where the action $S = S[B_{1,2}, I, J, \psi, \vec{\chi}, \vec{H}, \bar{\phi}, \eta, \phi]$ is defined as
 \begin{align}
  & S[B_{1,2}, I, J, \psi, \vec{\chi}, \vec{H}, \bar{\phi}, \eta, \phi]
  \nonumber \\
  & \hspace{5em}
  = \mathcal{Q} \tr_K
  \left[
  \im \vec{\chi} \cdot \vec{s} + g_H \vec{\chi} \cdot \vec{H} + \frac{1}{g_V} \psi \cdot V(\bar{\phi}) + \frac{1}{g_\eta} \eta [\phi, \bar{\phi}]
  \right]
  \, .
  \label{eq:ADHM_topo_action}
 \end{align}
 The vector field $V(\bar\phi)$ is associated with the $G^\vee$-action with the transformation parameter $\bar{\phi}$: 
 \begin{align}
  \psi \cdot V(\bar\phi)
  & = \psi_{B_1} [\bar\phi, B_1] + \psi_{B_2} [\bar\phi, B_2] + \psi_I \bar{\phi} I - J \bar{\phi} \psi_J
  \nonumber \\
  & \quad
  - \bar{\psi}_{B_{1}} [\bar{\phi}, B_1^\dag] - \bar{\psi}_{B_{2}} [\bar{\phi}, B_2^\dag] - I^\dag \bar{\phi} \bar{\psi}_I + \bar{\psi}_J \bar{\phi} J^\dag
  \, .
 \end{align}
 The action $S$ is now written as the $\mathcal{Q}$-exact form, so that the path integral is independent of the formal coupling constants $(g_V, g_H, g_\eta)$.

 Let us first deal with the $H$-term:
 \begin{align}
  & g_H \mathcal{Q} \tr_K \vec{\chi} \cdot \vec{H}
  \nonumber \\
  & \hspace{3em}
  =
  g_H \tr_K
  \left[
  H_\mathbb{R} H_\mathbb{R} + H_\mathbb{C}^\dag H_\mathbb{C}
  + \chi_\mathbb{R} [\phi, \chi_\mathbb{R}]
  + \chi_\mathbb{C}^\dag \left( [\phi, \chi_\mathbb{C}] - \epsilon_{12} \chi_\mathbb{C} \right)
  \right]
  \, ,
 \end{align}
 which leads to the Gaussian terms for $(H_\mathbb{R}, H_\mathbb{C})$.
 Here, in particular, we have to take care with the fermionic bilinear term of $\chi_\mathbb{R}$, which may be the zero mode, while the parameter $\epsilon_{12}$ gives the mass for $\chi_\mathbb{C}$.
 In order to cure this issue, we introduce another $\mathcal{Q}$-exact term:
 \begin{align}
  g_{\chi_\mathbb{R}} \mathcal{Q} \tr_K \chi_\mathbb{R} \bar{\phi}
  = g_{\chi_\mathbb{R}} \tr_K \left[ H_\mathbb{R} \bar{\phi} + \chi_\mathbb{R} \eta \right]
  \, ,
 \end{align}
 which gives rise to the mass term for the anti-ghost fermions at the large $g_{\chi_\mathbb{R}}$ limit.

 In order to evaluate the $\chi_\mathbb{C}$ integral, we take the diagonal basis for $\phi$:
 \begin{align}
  \phi = \diag(\phi_1,\ldots,\phi_k)
  \, ,
  \label{eq:phi_diagonal}
 \end{align}
 and the corresponding Haar measure is given by the Vandermonde determinant:
 \begin{align}
  \frac{d\phi}{\operatorname{vol} G^\vee} =
  \frac{1}{k!} \frac{d^k \phi}{(2\pi \im)^k} \prod_{a \neq b}^k (\phi_a - \phi_b)
  \, .
  \label{eq:LMNS_Vandermonde}
 \end{align}
 Then the bilinear term of $\chi_\mathbb{C}$ is given in this basis by
 \begin{align}
  g_H \tr_K \chi_\mathbb{C}^\dag \left( [\phi, \chi_\mathbb{C}] - \epsilon_{12} \chi_\mathbb{C} \right)
  =
  g_H \sum_{1 \le a, b \le k} (\phi_{ab} - \epsilon_{12}) |\chi_{\mathbb{C},ab}|^2
 \end{align}
 with $\phi_{ab} = \phi_a - \phi_b$.
 Hence, by integrating the $\chi_\mathbb{C}$ variable, we obtain
  \begin{align}
   g_H^{k^2}  \prod_{a, b}^k (\phi_{ab} - \epsilon_{12})
   = g_H^{k^2} \, (-\epsilon_{12})^k \prod_{a \neq b}^k (\phi_{ab} - \epsilon_{12})
   \, .
   \label{eq:LMNS_ADHM}
  \end{align}
  On the other hand, the Gaussian integral of $H_\mathbb{C}$ provides the factor $g_H^{-k^2}$, so that the coupling constant $g_H$ does not appear in the integral in the end.
 Similar cancellation is found for the integral of $(H_\mathbb{R}, \chi_\mathbb{R})$.


 The next step is to introduce the ``kinetic term'' for the remaining variables, $(B_{1,2}, I, J)$ and their fermionic partners, as follows: 
 \begin{align}
  &
  \frac{g_\text{kin}}{2} \, \mathcal{Q} \tr_K
  \left[
  B_{1,2}^\dag \psi_{B_{1,2}} - \bar{\psi}_{B_{1,2}} B_{1,2} 
  + I^\dag \psi_{I} - \bar{\psi}_I I 
  + J^\dag \psi_J - \bar{\psi}_J J 
  \right]
  \nonumber \\
  & = g_\text{kin} \tr_K
  \left[
  B^\dag_{1,2} \left( [\phi, B_{1,2}] - \epsilon_{1,2} B_{1,2} \right)
  + I^\dag \left( \phi I - I \mathsf{a} \right)
  + J^\dag \left( - J \phi + \mathsf{a} J - \epsilon_{12} J \right)
  \right]
  \nonumber \\
  & \quad + g_\text{kin} \tr_K
  \left[
  \bar{\psi}_{B_{1,2}} \psi_{B_{1,2}}
  + \bar{\psi}_I \psi_I + \bar{\psi}_J \psi_J
  \right]
  \, .
 \end{align}
 This term does not affect the integral since this is $\mathcal{Q}$-exact.
 Therefore, taking the limit $g_\text{kin} \to \infty$, the mass terms (from the $V$-term) are relatively suppressed, and we may focus on the kinetic terms.
 Diagonalizing $(\phi, \mathsf{a})$ as \eqref{eq:phi_diagonal} and \eqref{eq:Coulomb_diagonal}, the bosonic part of this kinetic term is given as
 \begin{align}
  \sum_{\substack{1 \le a, b \le k \\ m = 1,2}}
  (\phi_{ab} - \epsilon_m) |B_{m,ab}|^2
  + \sum_{\substack{a = 1,\ldots,k \\ \alpha = 1, \ldots, n}}
  (\phi_a - \mathsf{a}_\alpha) |I_{a\alpha}|^2
  + \sum_{\substack{a = 1,\ldots,k \\ \alpha = 1, \ldots, n}}
  (- \phi_a + \mathsf{a}_\alpha - \epsilon_{12}) |J_{\alpha a}|^2
  \, .
 \end{align}
 Together with the overall $g_\text{kin}$ factor, the Gaussian integral of these terms yields
 \begin{align}
  g_\text{kin}^{-k^2-2nk}
  \prod_{\substack{1 \le a, b \le k \\ m = 1,2}}
  (\phi_{ab} - \epsilon_{m})^{-1}
  \prod_{\substack{a = 1,\ldots,k \\ \alpha = 1, \ldots, n}}
  (\phi_a - \mathsf{a}_\alpha)^{-1}
  (- \phi_a + \mathsf{a}_\alpha - \epsilon_{12})^{-1}
  \, .
  \label{eq:LMNS_BIJ}
 \end{align}
 We remark that the fermionic Gaussian integral cancels the $g_\text{kin}$ factor similarly to the previous case.

 \subsection{Contour integral formula}
 
 Gathering all the contributions, \eqref{eq:LMNS_Vandermonde}, \eqref{eq:LMNS_ADHM}, and \eqref{eq:LMNS_BIJ}, the path integral of the ADHM variables \eqref{eq:ADHM_mod_sp_integral} is given as the multi-variable contour integral~\cite{Losev:1997tp,Moore:1997dj,Lossev:1997bz}:
 \begin{itembox}{Losev--Moore--Nekrasov--Shatashvili (LMNS) formula}\index{LMNS formula}%
  We define the gauge polynomials and the rational function:
  \begin{subequations}
  \begin{align}
   P(\phi) = \prod_{\alpha = 1}^n (\phi - \mathsf{a}_\alpha)
   \, , \qquad
   \widetilde{P}(\phi) = \prod_{\alpha = 1}^n ( - \phi + \mathsf{a}_\alpha)
   \, ,
  \end{align}
  \begin{align}
   \mathscr{S}(\phi) = \frac{(\phi - \epsilon_1)(\phi - \epsilon_2)}{\phi (\phi - \epsilon_{12})}
   \, .
   \label{eq:S_func_4d}
  \end{align}
  \end{subequations}
  \index{S-function@$\mathscr{S}$-function}%
  Then, the equivariant integral over the instanton moduli space~\eqref{eq:ADHM_mod_sp_integral} is localized on the multi-variable contour integral, \index{equivariant!---localization}
  \begin{align}
   Z_{n,k}(\mathsf{a},\epsilon_{1,2}) = \frac{1}{k!} \frac{(-\epsilon_{12})^k}{\epsilon_{1,2}^k} \oint_{\mathsf{T}_K} \prod_{a=1}^k \frac{d \phi_a}{2\pi \im} \frac{1}{P(\phi_a) \widetilde{P}(\phi_a + \epsilon_{12})}
   \prod_{a \neq b}^k \mathscr{S}(\phi_{ab})^{-1}
   \, ,
   \label{eq:LMNS_formula}
  \end{align}
  where we denote the maximal Cartan torus of $G^\vee = \rU(k)$ by $\mathsf{T}_K = \rU(1)^k$, and $\epsilon_{1,2} = \epsilon_1 \epsilon_2$.
  The factor $k!$ is the volume of the symmetric group $\mathfrak{S}_k$, which is the Weyl group of $\rU(k)$.  
 \end{itembox}
  The function $\mathscr{S}(\phi)$ has poles at $\phi = 0, \epsilon_{12}$, and the following reflection formula holds except at these poles:
  \begin{align}
   \mathscr{S}(\epsilon_{12} - \phi) = \mathscr{S}(\phi)
   \, .
  \end{align}
  Then, the total partition function is obtained by summing up all the instanton sectors:
  \begin{align}
   Z_n(\mathsf{a},\epsilon_{1,2}) = \sum_{k=0}^\infty \mathfrak{q}^k \, Z_{n,k}(\mathsf{a},\epsilon_{1,2})
  \end{align}
  where $\mathfrak{q} \in \mathbb{C}^\times$ is the instanton fugacity given by the complexified coupling constant~\eqref{eq:inst_fugacity}. \index{fugacity (instanton)}
  We remark that the gauge polynomials are related to each other as $\widetilde{P}(\phi) = (-1)^n P(\phi)$, so that the contour integral is also written only with the polynomial $P(\phi)$:
  \begin{align}
   Z_{n,k}(\mathsf{a},\epsilon_{1,2}) = \frac{(-1)^{nk}}{k!} \frac{\epsilon_{12}^k}{\epsilon_{1,2}^k} \oint_{\mathsf{T}_K} \prod_{a=1}^k \frac{d \phi_a}{2\pi \im} \frac{1}{P(\phi_a) {P}(\phi_a + \epsilon_{12})}
   \prod_{a \neq b}^k \mathscr{S}(\phi_{ab})^{-1}
   \, .
  \end{align}
  We have a sign factor $(-1)^{nk}$, which can be absorbed by the fugacity with the redefinition, $\mathfrak{q} \to (-1)^n \mathfrak{q}$.
  Therefore, whether we use $P(\phi)$ or $\widetilde{P}(\phi)$ is a convention issue in this case.%
  \footnote{%
  As discussed in \S\ref{sec:eq_ch_formula}, we have a similar formulation for 5d $\mathcal{N} = 1$ theory and 6d $\mathcal{N} = (1,0)$ theory.
  In this case, however, an additional factor is necessary to convert $P(\phi)$ and $\widetilde{P}(\phi)$.
  }

  An interpretation of the contour integral formula from the equivariant integral point of view is as follows.
  Each term in the denominator is exactly the eigenvalue of the infinitesimal equivariant torus action $\mathsf{T}_K \times \mathsf{T}_N \times \mathsf{T}_\mathbf{Q}$ on the ADHM variables: \index{equivariant!---action}
  \begin{subequations}\label{eq:ADHM_equiv_weight}
   \begin{align}
    B_{m,ab} \ \longmapsto \ &
    (\phi_{ab} - \epsilon_m) \, B_{m,ab} \qquad (m = 1,2) \\
    I_{a \alpha} \ \longmapsto \ &    
    (\phi_a - \mathsf{a}_\alpha) \, I_{a\alpha}  \\
    J_{\alpha a} \ \longmapsto \ &
    (- \phi_a + \mathsf{a}_\alpha - \epsilon_{12}) \, J_{\alpha a} \\
    \left( \ J^\dag_{a\alpha} \ \longmapsto \ \right. &
    \left. (\phi_a - \mathsf{a}_\alpha + \epsilon_{12}) \, {J}^\dag_{a \alpha} \ \right)
   \end{align}   
  \end{subequations}
  See also \eqref{eq:fixed_pt1}.
  Comparing with the localization formula~\eqref{eq:BVAB_formula}, these factors in the denominator are interpreted as the weight contributions at the fixed point.%
  \footnote{%
  If we instead consider the contribution of $J^\dag$, it will be replaced as $\widetilde{P}(\phi + \epsilon_{12}) \to P(\phi + \epsilon_{12})$.
  }
  Then, the first factor in the numerator is the Haar measure contribution.
  Recalling the torus action on the moment map is given by
  \begin{align}
   \mathsf{T}_K \times \mathsf{T}_N \times \mathsf{T}_\mathbf{Q} : \quad
   \mu_{\mathbb{C},ab} \ \longmapsto \
   (\phi_{ab} - \epsilon_{12}) \, \mu_{\mathbb{C}, ab}
   \, ,
  \end{align}
  another factor is due to the ADHM equation $\mu_\mathbb{C} = 0$.
  Since, as mentioned above, the equivariant integral over $X$ is given as the integral over $X \times \mathfrak{g}^\vee$, we first localize the integral on $X$ with the $G^\vee$-action, then perform the remaining $\mathfrak{g}^\vee$-integral.    
  To summarize, the ADHM path integral is in general given by~\cite{Nekrasov:2004vw}
  \begin{align}
   Z_{N,K}(\mathsf{a}, \epsilon_{1,2})
   = \frac{1}{|W_{G^\vee}| \operatorname{vol} \mathsf{T}_K} \oint_{\mathsf{T}_K} d^{\operatorname{rk} \mathfrak{g}^\vee} \phi \,
   \prod_{\alpha \in \Delta} \left< \alpha, \phi \right> \,
   \frac{(\text{ADHM equation})}{(\text{ADHM variables})}
  \end{align}
  where $W_{G^\vee}$ is the Weyl group of $G^\vee$, and $\Delta$ is the set of roots for $G^\vee$.  
  The factors, (ADHM equation) and (ADHM variables), are the eigenvalues of the infinitesimal equivariant torus action on them.
  We will also discuss the case if $G$ (and also $G^\vee$) is a supergroup in Chapter~\ref{sec:super_instanton}.

  \subsection{Incorporating matter}\label{sec:LMNS_matter}

  We consider the moduli space integral in the presence of the matter fields.
  In this case, the path integral localizes on the locus of the Weyl zero mode (\S\ref{sec:topological_twist}), and such a solution is described by the additional fermionic variable $(\lambda_{af})_{a=1,\ldots,k}^{f=1,\ldots,n^\text{f}}$ and $(\bar{\lambda}_{fa})_{a=1,\ldots,k}^{f=1,\ldots,n^\text{af}}$, as shown in \S\ref{sec:ADHM_construction}.
  In order to apply the path integral formalism, we define multiplets, $(\lambda,\xi)$ and $(\bar{\lambda},\bar{\xi})$, with the BRST transformations:%
  \footnote{%
  We incorporate the $\epsilon_{12}$-shift for the antifundamental matter to be consistent with the equivariant index formula discussed in \S\ref{sec:eq_ch_formula}.}\\[-.5em]  
  \begin{subequations}
   \begin{minipage}{.5\textwidth}
    \begin{align}
     \mathcal{Q}\lambda & = \xi \\
     \mathcal{Q}\bar{\lambda} & = \bar{\xi}
    \end{align}
   \end{minipage}
    \begin{minipage}{.5\textwidth}
     \begin{align}
      \mathcal{Q} \xi & = \phi \lambda - \lambda \mathsf{m} \\
      \mathcal{Q} \bar\xi & = - \bar{\lambda} \phi + \widetilde{\mathsf{m}} \bar{\lambda} - \epsilon_{12} \bar{\lambda}
     \end{align}
   \end{minipage}
   \end{subequations}\\[1em]
  where $\mathsf{m} = (m_1,\ldots,m_{n^\text{f}}) \in \operatorname{Lie} \mathsf{T}_M$ and $\widetilde{\mathsf{m}} = (\widetilde{m}_1,\ldots,\widetilde{m}_{n^\text{af}}) \in \operatorname{Lie} \mathsf{T}_{\widetilde{M}}$
  with the maximal Cartan tori of the flavor symmetry groups, $\mathsf{T}_M = \rU(1)^{n^\text{f}} \subset \rU(n^\text{f})$ and $\mathsf{T}_{\widetilde{M}} = \rU(1)^{n^\text{af}} \subset \rU(n^\text{af})$.
  The infinitesimal equivariant torus actions on $(\lambda,\bar{\lambda})$ are given by
  \begin{subequations}
  \begin{align}
   \mathsf{T} : \quad &
   \lambda_{af} \ \longmapsto \ (\phi_a - m_f) \, \lambda_{af} \\
   \mathsf{T} : \quad &
   \bar\lambda_{fa} \ \longmapsto \ (-\phi_a + \widetilde{m}_f - \epsilon_{12}) \, \bar\lambda_{fa} 
  \end{align}
  \end{subequations}
  where we denote the total torus action by $\mathsf{T}$.
  Then we incorporate the additional contributions to the action~\eqref{eq:ADHM_topo_action}:
  \begin{subequations}  
  \begin{align}
   \frac{g_\lambda}{2} \mathcal{Q} \tr_K \left[ \lambda^\dag \xi - \xi^\dag \lambda \right]
   & = g_\lambda \tr_K \left[ \lambda^\dag (\phi \lambda - \lambda \mathsf{m}) + \xi^\dag \xi \right]
   \, , \\
   \frac{g_{\bar{\lambda}}}{2} \mathcal{Q} \tr_K \left[ \bar\lambda^\dag \bar\xi - \bar\xi^\dag \bar\lambda \right]
   & = g_\lambda \tr_K \left[ \bar\lambda^\dag (- \bar\lambda \phi + \mathsf{m} \bar{\lambda} - \epsilon_{12} \bar{\lambda}) + \bar\xi^\dag \bar\xi \right] \, ,
  \end{align}
  \end{subequations}
  which end up with the factors given by
  \begin{align}
   \prod_{\substack{a = 1,\ldots,k \\ f = 1,\ldots,n^\text{f}}} (\phi_a - m_f)
   \, , \qquad
   \prod_{\substack{a = 1,\ldots,k \\ f = 1,\ldots,n^\text{af}}} (- \phi_a + \widetilde{m}_f - \epsilon_{12})   
   \, .
  \end{align}
  Hence, the LMNS formula~\eqref{eq:LMNS_formula} in the presence of the (anti)fundamental matters is given by \index{LMNS formula!---with flavor}
  \begin{align}
   Z_{n,k}(\mathsf{a},\mathsf{m},\widetilde{\mathsf{m}},\epsilon_{1,2})
   = \frac{1}{k!} \frac{(-\epsilon_{12})^k}{\epsilon_{1,2}^k} \oint_{\mathsf{T}_{G^\vee}} \prod_{a = 1}^k \frac{d \phi_a}{2\pi \im}
   \frac{P^\text{f}(\phi_a)\widetilde{P}^\text{af}(\phi_a + \epsilon_{12})}{P(\phi_a) \widetilde{P}(\phi_a + \epsilon_{12})}
   \prod_{a \neq b}^k \mathscr{S}(\phi_{ab})^{-1}
   \, ,
   \label{eq:LMNS_formula_matter}
  \end{align}
  where the matter polynomials are defined
   \begin{align}
    P^\text{f}(\phi) = \prod_{f=1}^{n^\text{f}} (\phi - m_f)
    \, , \qquad
    \widetilde{P}^\text{af}(\phi) = \prod_{f=1}^{n^\text{af}} ( - \phi + \widetilde{m}_f)
    \, .
   \end{align}
   From the geometric point of view, the integral \eqref{eq:LMNS_formula_matter} is given by   
   \begin{align}
    Z_{n,k}(\mathsf{a}, \mathsf{m}, \widetilde{\mathsf{m}}, \epsilon_{1,2})
    := \int_{\mathfrak{M}_{n,k}} e_\mathsf{T} ( \mathbf{M}^\vee \otimes \mathbf{K} \oplus \det \mathbf{Q}^\vee \otimes \mathbf{K}^\vee \otimes \widetilde{\mathbf{M}})
    \label{eq:euler_class_insertion}
   \end{align}
   where $e_{\mathsf{T}} ( \mathbf{M}^\vee \otimes \mathbf{K} \oplus \det \mathbf{Q}^\vee \otimes \mathbf{K}^\vee \otimes \widetilde{\mathbf{M}})$ is the equivariant Euler class of the bundles over the instanton moduli space whose fibers are given by the vector spaces $(K,M,\widetilde{M})$.
   We denote the spacetime bundle by $\mathbf{Q}$ introduced in \S\ref{sec:Q_bundle}.
   In fact, these bundles are identified with the instanton part of the (anti)fundamental hypermultiplet bundles, $(\mathbf{H}^\text{f,inst},\mathbf{H}^\text{af,inst})$, over the instanton moduli space.
   See~\eqref{eq:fund_hyper_inst} in \S\ref{sec:eq_ch_formula} for details.

   \subsection{Pole analysis}

   We have derived the integral formula for the instanton partition function, as in \eqref{eq:LMNS_formula} and \eqref{eq:LMNS_formula_matter}.
   Since it is a multi-variable contour integral, we should properly indicate the integration contour as follows~\cite{Nakajima:1999,Nekrasov:2002qd}.%
   \footnote{%
   It is known in general that the Jeffrey--Kirwan (JK) residue formalism provides a correct prescription for such a contour integral~\cite{Jeffrey:1995,Jeffrey:1997}, and the argument below can be justified with this JK formalism.
   See also~\cite{Brion:1999ASENS,Szenes:2004IM} and~\cite{Benini:2013xpa,Hwang:2014uwa,Nakamura:2015zsa} for applications to gauge theory partition functions.
   }
   We first assign the ordering as 
   \begin{align}
    \frac{1}{k!} \oint \prod_{a=1}^k d \phi_a
    \ \longrightarrow \
    \oint d \phi_k \cdots \oint d \phi_1
    \, .
   \end{align}
   Then we take the integration contour, which picks up the poles at
   \begin{align}
    \phi_a - \mathsf{a}_\alpha
    \, , \qquad
    \phi_{ab} - \epsilon_1
    \, , \qquad
    \phi_{ab} - \epsilon_2
    \, .
   \end{align}
   Hence, the first pole must be $\phi_1 = \mathsf{a}_\alpha$ for $\alpha \in (1,\ldots,n)$, and the pole $\phi_{ab} - \epsilon_{1,2}$ for $a > b$ gives a relation $\phi_a = \phi_{b (< a)} + \epsilon_{1,2}$.
   Applying this procedure recursively, the poles are consistent with the eigenvalues shown in \eqref{eq:phi_ev_I2}, which are labeled by $n$-tuple partition.
   This shows that the equivariant integral over the instanton moduli space localizes on the fixed point loci under the equivariant action.
   Although we could also specify the corresponding residues to these poles from this expression, we would instead work with the equivariant character based on the localization formula in the following section.

 \section{Equivariant index formula}\label{sec:eq_ch_formula}

 The localization formula shown in \S{\ref{sec:equiv_int}} allows us to express the equivariant integral only with the fixed point contributions.
 In this process, we first clarify the fixed point, and then evaluate the weight corresponding to each fixed point.
 We here study the fixed point weight based on the equivariant character, and write down the gauge theory partition function explicitly.
  \index{equivariant!---index}

 \subsection{Spacetime bundle}\label{sec:Q_bundle}

 For the four-manifold $\mathcal{S} = \bC^2$, let $\mathbf{Q}$ be the cotangent bundle to $\mathcal{S}$ at the marked point $o$,
 \begin{align}
  \mathbf{Q} = T^*_o \mathcal{S}
  \, .
 \end{align}
 We denote the corresponding automorphism group by $\GL(\mathbf{Q})$, interpreted as complexification of the subgroup of the Lorentz group $\operatorname{Spin}(4)$, and the marked point $o$ is the fixed point under this group action.
 We split $\mathbf{Q} = \mathbf{Q}_1 \oplus \mathbf{Q}_2$ with respect to the Cartan torus
 \begin{align}
  \mathsf{T}_\mathbf{Q} = \GL(\mathbf{Q}_1) \times \GL(\mathbf{Q}_2) \subset \GL(\mathbf{Q})
  \, ,
 \end{align}
 which is the complexification of \eqref{eq:q12_torus}.
 The associated characters at the $\mathsf{T}$-fixed point are given as follows:
 \begin{align}
  \ch_\mathsf{T} \mathbf{Q}_{1,2} = q_{1,2}
  \, , \qquad
  \ch_\mathsf{T} \mathbf{Q} = q_1 + q_2
  \, , \qquad
  \ch_\mathsf{T} \wedge^2 \mathbf{Q} = q_1 q_2 =: q
  \, .
 \end{align}
 We also denote $\wedge^2 \mathbf{Q} = \det \mathbf{Q}$.
 The parameters $(q_1,q_2)$ are the equivariant parameters for $\GL(\mathbf{Q})$, $q_{1,2} = \np^{\epsilon_{1,2}} \in \bC^\times$.
 We also define
 \begin{align}
  \wedge \mathbf{Q}_{1,2} = \sum_{k} (-1)^k \wedge^k \mathbf{Q}_{1,2}
  \, , \qquad
  \wedge \mathbf{Q} = \sum_{k} (-1)^k \wedge^k \mathbf{Q}
  \, ,
 \end{align}
 and their characters
\begin{align}
 \ch_\mathsf{T} \wedge \mathbf{Q}_{1,2} = 1 - q_{1,2}
 \, , \qquad
 \ch_\mathsf{T} \wedge \mathbf{Q} = (1 - q_1)(1 - q_2)
 \, .
\end{align}
  We remark the relation
 \begin{align}
  \wedge \mathbf{Q} = \wedge \mathbf{Q}_1 \cdot \wedge \mathbf{Q}_2
  \, .
  \label{eq:Q_product}
 \end{align}

 \subsection{Framing and instanton bundles}
 \index{framing!---bundle}
 \index{instanton bundle}

 We denote the bundles on the instanton moduli space with the fibers, $N$ and $K$, by $\mathbf{N}$ and $\mathbf{K}$, which we call the framing and instanton bundles (See \S\ref{sec:ADHM_construction}).
 The corresponding automorphism groups are denoted by $\GL(\mathbf{N})$ and $\GL(\mathbf{K})$.
 Let $\mathsf{T}_N$ and $\mathsf{T}_K$ be the Cartan tori of these automorphism groups.
 They split into the direct sum:
 \begin{align}
  \mathbf{N} = \bigoplus_{\alpha = 1}^n \mathbf{N}_\alpha
  \, , \qquad
  \mathbf{K} = \bigoplus_{\alpha = 1}^n \mathbf{K}_\alpha
  \, .
 \end{align}
 Then the characters of these bundles at the $\mathsf{T}$-fixed point $\lambda \in \mathfrak{M}^\mathsf{T}$ are given as
 \begin{align}
  \ch_\mathsf{T} \mathbf{N}_\alpha = \np^{\mathsf{a}_\alpha}
  \, ,\qquad
  \ch_\mathsf{T} \mathbf{K}_\alpha = \sum_{s \in \lambda_\alpha} \np^{\mathsf{a}_\alpha} q_1^{s_1 - 1} q_2^{s_2 - 1}
  \, .
  \label{eq:NK_ch}
 \end{align}

\subsection{Universal bundle}\label{sec:univ_bundle}

Let $\mathbf{Y}_\mathcal{S}$ be the universal bundle over $\mathfrak{M}_{n,k} \times \mathcal{S}$, which splits into the direct sum of the $\mathsf{T}$-fixed ideals defined in \S\ref{sec:eq_fixed_point}:
\begin{align}
 \mathbf{Y}_\mathcal{S} = \bigoplus_{\alpha=1}^n \mathbf{N}_\alpha \otimes \mathbf{I}_{\lambda_\alpha}
 \, .
 \label{eq:univ_bundle}
\end{align}
The character of the ideal parametrized by the partition $\lambda$ is given by
\begin{subequations}
 \begin{align}
  \ch_\mathsf{T} \mathbf{I}_\lambda & = \sum_{s \not \in \lambda} q_1^{s_1 - 1} q_2^{s_2 - 1}
  \, , \\
  \ch_\mathsf{T} \mathbf{I}_\emptyset & = \sum_{s_1,s_2 = 1, \ldots, \infty} q_1^{s_1-1} q_2^{s_2-1} = \frac{1}{(1 - q_1)(1 - q_2)} 
  \, .
 \end{align}
\end{subequations}
Hence, the universal bundle character is given by
\begin{align}
 \ch_\mathsf{T} \mathbf{Y}_\mathcal{S} = \sum_{\alpha = 1}^n \sum_{s \not \in \lambda_{\alpha}} \np^{\mathsf{a}_\alpha} q_1^{s_1 - 1} q_2^{s_2 - 1}
 \, .
\end{align}

We denote the inclusion map from the fixed point $o$ to the spacetime $\mathcal{S}$ by $i_o: o \hookrightarrow \mathcal{S}$.
Then the observable bundles over $\mathfrak{M}_{n,k}$ is defined by the pullback of the universal bundle $\mathbf{Y}_o = i_o^* \mathbf{Y}_\mathcal{S}$.
The localization formula gives rise to the relation
\begin{align}
 \mathbf{Y}_\mathcal{S} = \frac{\mathbf{Y}_o}{\wedge \mathbf{Q}}
 \ \iff \
 \mathbf{Y}_o = \wedge \mathbf{Q} \cdot \mathbf{Y}_\mathcal{S}
 \, ,
 \label{eq:univ_bundle_localization} 
\end{align}
where the denominator $\wedge \mathbf{Q}$ on the left equation plays a role of the equivariant Todd class on $\mathcal{S} = \mathbb{C}^2$:
\begin{align}
 \operatorname{Td}(\mathbb{C}^2) = \frac{\epsilon_1 \epsilon_2}{(\np^{\epsilon_1} - 1)(\np^{\epsilon_2} - 1)}
 \, .
\end{align}
Thus, from \eqref{eq:univ_bundle} and \eqref{eq:univ_bundle_localization}, we obtain the observable bundle given in terms of the framing and instanton bundles:
\begin{align}
 \mathbf{Y}_o = \mathbf{N} - \wedge \mathbf{Q} \cdot \mathbf{K}
 \, .
 \label{eq:obs_bndl}
\end{align}
Since the spacetime splits into $\mathcal{S} = \mathcal{S}_1 \oplus \mathcal{S}_2$ as in \S\ref{sec:Q_bundle}, there are two compatible reductions:
\begin{equation}
 \begin{tikzcd}
  & \mathbf{Y}_\mathcal{S}
  \arrow[dd,"\wedge \mathbf{Q}"]
  \arrow[dl,"\wedge \mathbf{Q}_1",swap]
  \arrow[dr,"\wedge \mathbf{Q}_2"]
  & \\
  \mathbf{X} := \mathbf{Y}_{\mathcal{S}_1}
  \arrow[dr,"\wedge \mathbf{Q}_2",swap]
  & &
  \mathbf{Y}_{\mathcal{S}_2} =: \check{\mathbf{X}}
  \arrow[dl,"\wedge \mathbf{Q}_1"]
  \\
  & \mathbf{Y}_o &
 \end{tikzcd} 
\end{equation}
Hence we have two equivalent expressions:
\begin{subequations}
\begin{align}
 \mathbf{Y}_o
 & = \wedge \mathbf{Q}_1 \cdot \mathbf{X}
  \\
 & = \wedge \mathbf{Q}_2 \cdot \check{\mathbf{X}}
 \, .
\end{align}
\end{subequations}
The characters of $(\mathbf{X},\check{\mathbf{X}}) = (\mathbf{Y}_{\mathcal{S}_1}, \mathbf{Y}_{\mathcal{S}_2})$ are given as
\begin{align}
 \ch_\mathsf{T} \mathbf{X} = \sum_{x \in \mathcal{X}} x
 \, , \qquad
 \ch_\mathsf{T} \check{\mathbf{X}} = \sum_{\check{x} \in \check{\mathcal{X}}} \check{x}
\end{align}
where
\begin{align}
 \mathcal{X} & = \left\{ x_{\alpha,k} = \np^{\mathsf{a}_\alpha} q_1^{k-1} q_2^{\lambda_{\alpha,k}} \right\}_{\substack{\alpha = 1,\ldots n \\ k=1,\ldots,\infty}}
 \, , \quad
 \check{\mathcal{X}} = \left\{ \check{x}_{\alpha,k} = \np^{\mathsf{a}_\alpha}  q_1^{\check\lambda_{\alpha,k}} q_2^{k-1} \right\}_{\substack{\alpha = 1,\ldots n \\ k=1,\ldots,\infty}}
 \
 \in \mathfrak{M}^\mathsf{T}
 \, ,
\end{align}
with the transposed partition denoted by $\check{\lambda}$.
Namely, these two reductions are related through $q_1 \leftrightarrow q_2$ $(\epsilon_1 \leftrightarrow \epsilon_2)$.%
\footnote{%
This exchanging symmetry leads to a duality of the quantum algebras emerging from the moduli space of quiver gauge theory (Langlands duality).
See also \S\ref{sec:frac_T_op}.
}

\subsection{Index formula}

We define the Adams operation on the bundle $\mathbf{E}$ by $\mathbf{E}^{[p]}$, s.t., the character is given by $\displaystyle \ch \mathbf{E}^{[p]} = \sum_{i=1}^{\operatorname{rk} \mathbf{E}} e_i^p$ for $\displaystyle \ch \mathbf{E} = \sum_{i=1}^{\operatorname{rk} \mathbf{E}} e_i$. \index{Adams operation}
In terms of the Chern roots $e_i = \np^{x_i}$, it is given by $x_i \to p x_i$.
Then, we denote the dual of the bundle $\mathbf{E}$ by $\mathbf{E}^\vee$, s.t., the character is given by $\displaystyle \ch \mathbf{E}^\vee = \sum_{i=1}^{\operatorname{rk} \mathbf{E}} e_i^{-1}$.
It is formally written as the Adams operation, $\mathbf{E}^\vee = \mathbf{E}^{[-1]}$.

Given the character of the bundle $\displaystyle \ch \mathbf{E} = \sum_{i=1}^{\operatorname{rk} \mathbf{E}} n_i \, \np^{x_i}$ with $n_i \in \mathbb{Z}$, we define the index functor, which converts the additive class to the multiplicative class, as follows:

\begin{align}
 \mathbb{I}[\mathbf{E}] = \prod_{i=1}^{\operatorname{rk} \mathbf{E}} [x_i]^{n_i}
\end{align}
where
\begin{align}
 [x] =
 \begin{cases}
  x & (\text{Cohomology})\\
  1 - \np^{-x} & (\text{K-theory}) \\
  \theta(\np^{-x};p) & (\text{Elliptic})
 \end{cases}
 \label{eq:[x]_function}
\end{align}
We denote the theta function with the elliptic nome $p = \np^{2 \pi \im \tau} \in \mathbb{C}^\times$ by $\theta(z;p)$ defined in~\eqref{eq:theta_fn}.
The elliptic index is reduced to the K-theory in the limit $p \to 0$.
In order to obtain the cohomological index form the K-theory, rescaling the Chern roots $x_i \to \beta x_i$, then take the limit $\beta \to 0$.
The remaining factor $\beta^{\operatorname{rk} \mathbf{E}}$ should be compensated with other parameters, e.g., the coupling constant.
The index obeys the reflection formula:
\begin{align}
 \mathbb{I}[\mathbf{E}^\vee] =
 \begin{cases}
  (-1)^{\operatorname{rk} \mathbf{E}} \, \mathbb{I}[\mathbf{E}] & (\text{Cohomology}) \\
  (-1)^{\operatorname{rk} \mathbf{E}} (\det \mathbf{E}) \, \mathbb{I}[\mathbf{E}] & (\text{K-theory/Elliptic}) 
 \end{cases}
\end{align}
We may take another option for the index function instead of the previous one~\eqref{eq:[x]_function}:
\begin{align}
 [x] =
 \begin{cases}
  x & (\text{Cohomology}) \\
  \np^{x/2} - \np^{-x/2} & (\text{K-theory}) \\
  \np^{x/2} \, \theta(\np^{-x};p) & (\text{Elliptic})
 \end{cases}
 \label{eq:[x]_function_sym} 
\end{align}
which makes the reflection formulas homogeneous:
\begin{align}
 \mathbb{I}[\mathbf{E}^\vee]
 = (-1)^{\operatorname{rk} \mathbf{E}} \,
 \mathbb{I}[\mathbf{E}]
 \, .
\end{align}

From the gauge theory point of view, these index formulas are associated with 4d $\mathcal{N} = 2$ theory on $\mathcal{S}$, 5d $\mathcal{N} = 1$ theory on $\mathcal{S} \times S^1$, 6d $\mathcal{N} = (1,0)$ theory on $\mathcal{S} \times T^2$.
See \S\ref{sec:5d6d} for details.
In particular, the K-theoretic index and the elliptic index are realized as the Witten index of the supersymmetric quantum mechanics on $S^1$ and the elliptic genus on $T^2$ with the modulus $\tau$.

\subsection{Vector multiplet}

We consider the vector multiplet (gauge field) contribution to the tangent bundle of the instanton moduli space at the $\mathsf{T}$-fixed point, $T_\lambda \mathfrak{M}$ for $\lambda \in \mathfrak{M}^\mathsf{T}$, that we call the vector multiplet bundle $\mathbf{V}$.
The vector multiplet is in the adjoint representation of the gauge group $G$, which is given by the tensor product of the fundamental representation and its conjugate for $G = \rU(n)$.%
\footnote{%
For $\rO(n)$ and $\Sp(n)$ groups, the adjoint representation is given by the rank two antisymmetric and symmetric representations, respectively.
See~\cite{Marino:2004cn,Nekrasov:2004vw,Hollands:2010xa,Hwang:2014uwa} for the instanton partition function for $\rO(n)$ and $\Sp(n)$ groups.
}
Then, we define the vector multiplet bundle similarly to the localization formula~\eqref{eq:univ_bundle_localization}:
\begin{align}
 \mathbf{V} = \frac{\mathbf{Y}_o^\vee \mathbf{Y}_o}{\wedge \mathbf{Q}}
 \, .
\end{align}
In terms of the partial reduction of the universal bundle $\mathbf{Y}_{\mathcal{S}_1} = \mathbf{X}$, it is given by%
\footnote{%
There is an alternative expression based on the other reduction of the universal bundle $\mathbf{Y}_{\mathcal{S}_2} = \check{\mathbf{X}}$:
\begin{align}
 \mathbf{V} = \frac{\wedge \mathbf{Q}_2^\vee}{\wedge \mathbf{Q}_1} \, \check{\mathbf{X}}^\vee \check{\mathbf{X}}
 \, .
\end{align}
}
\begin{align}
 \mathbf{V} = \frac{\wedge \mathbf{Q}_1^\vee}{\wedge \mathbf{Q}_2} \, \mathbf{X}^\vee \mathbf{X}
 \, ,
\end{align}
and the corresponding character is
\begin{align}
 \ch_\mathsf{T} \mathbf{V} = \frac{1 - q_1^{-1}}{1 - q_2} \sum_{\substack{(x,x') \in \mathcal{X} \times \mathcal{X} \\ x \neq x'}} \frac{x'}{x}
\end{align}
where the diagonal term $x = x'$, removed here, may be interpreted as the singlet term.
Therefore the vector multiplet contribution to 4d $\mathcal{N} = 2$/5d $\mathcal{N} = 1$/6d $\mathcal{N} = (1,0)$ gauge theory partition function is given by applying the index formula to the corresponding bundle:
\begin{align}
 Z^\text{vec} := \mathbb{I}[\mathbf{V}]
 =
  \begin{cases}
  \displaystyle
  \prod_{\substack{(x,x') \in \mathcal{X} \times \mathcal{X} \\ x \neq x'}}
  \frac{\Gamma_1(\log x' - \log x - \epsilon_1;\epsilon_2)}{\Gamma_1(\log x' - \log x;\epsilon_2)}
  & (\text{4d}) \\
  \displaystyle
  \prod_{\substack{(x,x') \in \mathcal{X} \times \mathcal{X} \\ x \neq x'}}
  \frac{\Gamma_q(q_2 x/x';q_2)}{\Gamma_q(q x/x';q_2)}  
  & (\text{5d}) \\ 
  \displaystyle
  \prod_{\substack{(x,x') \in \mathcal{X} \times \mathcal{X} \\ x \neq x'}}
  \frac{\Gamma_e(q_2 x/x';p,q_2)}{\Gamma_e(q x/x';p,q_2)}
  & (\text{6d}) 
 \end{cases} 
 \label{eq:full_func_vec}
\end{align}
where $\Gamma_1(z;\epsilon)$ is the gamma function~\eqref{eq:multi_gamma}, $\Gamma_q(z;q)$ is the $q$-gamma function~\eqref{eq:q-gamma}, and $\Gamma_e(z;p,q)$ is the elliptic gamma function~\eqref{eq:e-gamma}.
The hierarchical structure of 4d, 5d, and 6d theories is obvious, which corresponds to the rational, trigonometric, and elliptic functions (See \S\ref{sec:5d6d}).
It is written more explicitly as
\begin{align}
 Z^\text{vec} = \prod_{\substack{\alpha,\alpha' = 1,\ldots,n \\ k,k' = 1,\ldots,\infty \\ (\alpha,k) \neq (\alpha',k')}}
 \frac{(\np^{\mathsf{a}_{\alpha\alpha'}} q_2^{\lambda_{\alpha,k} - \lambda_{\alpha',k'}+1} q_1^{k-k'+1};q_2)_\infty}{(\np^{\mathsf{a}_{\alpha\alpha'}} q_2^{\lambda_{\alpha,k} - \lambda_{\alpha',k'}+1} q_1^{k-k'};q_2)_\infty}
 \label{eq:full_func_vec_Kth}
\end{align}
where we define $\mathsf{a}_{\alpha\alpha'} = \mathsf{a}_\alpha - \mathsf{a}_{\alpha'}$.
This is the K-theory convention, and we have a similar expression for the cohomology (4d $\mathcal{N} = 2$) and the elliptic (6d $\mathcal{N} = (1,0)$) cases.

We remark that, if we relax the non-increasing condition of the partitions, parametrizing the fixed points in the instanton moduli space (\S\ref{sec:eq_fixed_point}), the partition function vanishes:
\begin{align}
 \lambda_{\alpha,k} < \lambda_{\alpha,k+1}
 \ \implies \ Z^\text{vec} = 0
 \, .
 \label{eq:Zvec_zero}
\end{align}
This can be seen from the expression~\eqref{eq:full_func_vec_Kth}.
Since $(z;q)_\infty = 0$ at $z q^n = 0$ $(n \in \mathbb{Z}_{\ge 0})$, we have $Z^\text{vec} = 0$ if $\lambda_{\alpha,k} - \lambda_{\alpha',k'} + 1 \le 0$ at $(\alpha',k') = (\alpha, k+1)$.

\subsection{Fundamental and antifundamental matters}
\index{matter bundle}

We denote the vector bundles over the instanton moduli space with the fibers, $M$ and $\widetilde{M}$, by $\mathbf{M}$ and $\widetilde{\mathbf{M}}$, that we call the matter bundles.
The automorphism groups are $\GL(\mathbf{M})$ and $\GL(\widetilde{\mathbf{M}})$, and the corresponding Cartan tori are denoted by $\mathsf{T}_\mathbf{M}$ and $\mathsf{T}_{\widetilde{\mathbf{M}}}$ with the characters
\begin{align}
 \ch_\mathsf{T} \mathbf{M} = \sum_{f=1}^{n^\text{f}} \np^{m_f}
 \, , \qquad
 \ch_\mathsf{T} \widetilde{\mathbf{M}} = \sum_{f=1}^{n^\text{af}} \np^{\widetilde{m}_f}
 \, .
\end{align}
We define the (anti)fundamental matter, the hypermultiplet in the (anti)fundamental representation of $G$ in $\mathcal{N} = 2$ gauge theory, contribution to the (extended) tangent bundle of the instanton moduli space:
 \begin{align}
  \mathbf{H}^\text{f} = - \frac{\mathbf{M}^\vee \mathbf{Y}_o}{\wedge \mathbf{Q}} \, , \qquad
  \mathbf{H}^\text{af} = - \frac{\mathbf{Y}_o^\vee \widetilde{\mathbf{M}}}{\wedge \mathbf{Q}}
  \, ,
 \end{align}
with the characters given by
 \begin{align}
  \ch_\mathsf{T} \mathbf{H}^\text{f} = - \frac{1}{1 - q_2}
  \sum_{(x,x') \in \mathcal{M} \times \mathcal{X}} \frac{x'}{x}
  \, , \qquad
  \ch_\mathsf{T} \mathbf{H}^\text{af} = + \frac{q_1^{-1}}{1 - q_2}
  \sum_{(x,x') \in \mathcal{X} \times \widetilde{\mathcal{M}} } \frac{x'}{x}
  \, ,
 \end{align}
where we define the set of the (multiplicative; exponentiated) mass parameters:
\begin{subequations}
\begin{align}
 \mathcal{M} & = (\mu_1,\ldots,\mu_{n^\text{f}}) := (\np^{m_1},\ldots,\np^{m_{n^\text{f}}})
 \, , \\
 \widetilde{\mathcal{M}} & = (\widetilde{\mu}_1,\ldots,\widetilde{\mu}_{n^\text{af}}) := (\np^{\widetilde{m}_1},\ldots,\np^{\widetilde{m}_{n^\text{af}}})
 \, .
\end{align}
\end{subequations}
Then the (anti)fundamental hypermultiplet contribution to the partition function is similarly expressed in terms of the gamma functions,
\begin{subequations}
\begin{align}
 Z^\text{f} & := \mathbb{I}[\mathbf{H}^\text{f}] =
  \begin{cases}
   \displaystyle
   \prod_{(x,x') \in \mathcal{M} \times \mathcal{X}}
   \Gamma_1(\log x' - \log x;\epsilon_2)^{-1}
   & (\text{4d}) \\   
  \displaystyle
   \prod_{(x,x') \in \mathcal{M} \times \mathcal{X}}
   \Gamma_q \left( q_2 x/x' ; q_2\right)^{-1}
  & (\text{5d}) \\ 
  \displaystyle
  \prod_{(x,x') \in \mathcal{M} \times \mathcal{X}}
   \Gamma_e \left( q_2 x/x' ; p, q_2\right)^{-1}
  & (\text{6d}) 
  \end{cases}
 \, , \\
 Z^\text{af} & := \mathbb{I}[\mathbf{H}^\text{af}] =
 \begin{cases}
   \displaystyle
   \prod_{(x,x') \in \mathcal{X} \times \widetilde{\mathcal{M}}}
   \Gamma_1(\log x' - \log x - \epsilon_1;\epsilon_2)
   & (\text{4d}) \\
  \displaystyle
   \prod_{(x,x') \in \mathcal{X} \times \widetilde{\mathcal{M}}}
   \Gamma_q \left( q x/x' ; q_2\right)
  & (\text{5d}) \\ 
  \displaystyle
  \prod_{(x,x') \in \mathcal{X} \times \widetilde{\mathcal{M}}} \Gamma_e \left( q x/x' ; p, q_2\right)
  & (\text{6d}) 
 \end{cases}
 \, .
\end{align}
\end{subequations}
For example, the fundamental matter contribution in the K-theory convention is explicitly given as
\begin{align}
 Z^\text{f} = \prod_{\substack{\alpha = 1,\ldots,n \\ k = 1,\ldots,\infty \\ f = 1, \ldots, n^\text{f}}}
 (\np^{m_f - \mathsf{a}_\alpha} q_2^{- \lambda_{\alpha,k} + 1} q_1^{- k + 1} ;q_2)_\infty
 \, .
\end{align}
This factor has zeros on the special locus of the parameter space.
See \S\ref{sec:inst_part_func} for details.

\subsection{Adjoint matter}\label{sec:adjoint_bundle}

We consider the contribution of the hypermultiplet in the adjoint representation.
We denote the line bundle over the instanton moduli space by $\mathbf{M}_\text{adj}$ with the character given by the adjoint mass parameter $m_\text{adj}$:
\begin{align}
 \ch_\mathsf{T} \mathbf{M}_\text{adj} = \np^{m_\text{adj}} =: \mu \in \mathbb{C}^\times
 \, .
\end{align}
We define the adjoint matter contribution to the tangent bundle:
\begin{align}
 \mathbf{H}^\text{adj} = - \mathbf{M}_\text{adj} \frac{\mathbf{Y}_o^\vee \mathbf{Y}_o}{\wedge \mathbf{Q}}
 \, ,
\end{align}
and the character is given by
\begin{align}
 \ch_\mathsf{T} \mathbf{H}^\text{adj}
 = - \mu \, \frac{1 - q_1^{-1}}{1 - q_2}
 \sum_{(x,x') \in \mathcal{X} \times \mathcal{X}} \frac{x'}{x}
 \, .
\end{align}
Then the corresponding contribution to the partition function is obtained by the index formulas:
\begin{align}
 Z^\text{adj} := \mathbb{I}[\mathbf{H}^\text{adj}]
 =
 \begin{cases}
  \displaystyle
  \prod_{(x,x') \in \mathcal{X} \times \mathcal{X}}
  \frac{\Gamma_1(\log x' - \log x + m_\text{adj};\epsilon_2)}{\Gamma_1(\log x' - \log x + m_\text{adj} - \epsilon_1;\epsilon_2)}
  & (\text{4d}) \\
  \displaystyle
  \prod_{(x,x') \in \mathcal{X} \times \mathcal{X}}
  \frac{\Gamma_q(\mu^{-1} q x/x';q_2)}  {\Gamma_q(\mu^{-1} q_2 x/x';q_2)}
  & (\text{5d}) \\ 
  \displaystyle
  \prod_{(x,x') \in \mathcal{X} \times \mathcal{X}}
  \frac{\Gamma_e(\mu^{-1} q x/x';p,q_2)}{\Gamma_e(\mu^{-1} q_2 x/x';p,q_2)}
  & (\text{6d}) 
 \end{cases}
 \label{eq:full_func_adj}
\end{align}
The gauge theory with $\mathcal{N} = 2$ supersymmetry, which consists of the vector multiplet and the adjoint hypermultiplet is called $\mathcal{N} = 2^*$ theory, and the corresponding partition function is given as
\begin{align}
 Z = Z^\text{vec} Z^\text{adj}
 \, .
\end{align}
Due to the relation
\begin{align}
 Z^\text{adj}\Big|_{m_\text{adj} = 0} =  \frac{1}{Z^\text{vec}}
 \, ,
\end{align}
the partition function becomes trivial in the limit $m_\text{adj} = 0$ because of the cancellation between $Z^\text{vec}$ and $Z^\text{adj}$.
In fact, $\mathcal{N} = 2^*$ theory is given as a mass deformation of $\mathcal{N} = 4$ theory, and the supersymmetry is restored from 4d $\mathcal{N} = 2$ to $\mathcal{N} = 4$ in the massless limit.
In addition, the adjoint mass is also interpreted as the equivariant parameter for a part of the R-symmetry, $\rU(1) \subset \SU(4) = \SO(6)$.
Realizing 4d $\mathcal{N} = 4$ theory as a stack of D3 branes in Type IIB string theory, this $\SO(6)$ symmetry is interpreted as the rotation symmetry of the transverse spacetime of D3 branes.
See also \S\ref{sec:quiver_partition_function} for a related discussion.

\section{Instanton partition function}\label{sec:inst_part_func}

\subsection{Vector multiplet}

The partition function contributions discussed in \S\ref{sec:eq_ch_formula} are given as the infinite products.
We can extract the finite product by the ratio, which is in particular called the instanton part,
\begin{align}
 Z^\text{vec,inst} = \frac{Z^\text{vec}}{\mathring{Z}^\text{vec}}
 \label{eq:vec_inst_part}
\end{align}
where $\mathring{Z}^\text{vec}$ is the perturbative part obtained with the empty configuration $\lambda = \emptyset$, and $Z^\text{vec}$ is the full partition function involving both the instanton and the perturbative contributions.
In the K-theory convention~\eqref{eq:full_func_vec_Kth}, for example, it is given by
\begin{align}
 Z^\text{vec,inst} & =
 \prod_{\substack{\alpha,\alpha' = 1,\ldots,n \\ k,k' = 1,\ldots,\infty \\ (\alpha,k) \neq (\alpha',k')}}
 \frac{(\np^{\mathsf{a}_{\alpha\alpha'}} q_2^{\lambda_{\alpha,k} - \lambda_{\alpha',k'}+1} q_1^{k-k'+1};q_2)_\infty}{{(\np^{\mathsf{a}_{\alpha\alpha'}} q_2^{\lambda_{\alpha,k} - \lambda_{\alpha',k'}+1} q_1^{k-k'};q_2)_\infty}}
 \frac{{(\np^{\mathsf{a}_{\alpha\alpha'}} q_2 \, q_1^{k-k'};q_2)_\infty}}{(\np^{\mathsf{a}_{\alpha\alpha'}} q_2 \, q_1^{k-k'+1};q_2)_\infty}
 \nonumber \\
 & =
 \prod_{\substack{\alpha,\alpha' = 1,\ldots,n \\ k,k' = 1,\ldots,\infty \\ (\alpha,k) \neq (\alpha',k')}}
 \frac{{(\np^{\mathsf{a}_{\alpha\alpha'}} q_2 \, q_1^{k-k'};q_2)_{\lambda_{\alpha,k} - \lambda_{\alpha',k'}}}}{{(\np^{\mathsf{a}_{\alpha\alpha'}} q_2 \, q_1^{k-k'+1};q_2)_{\lambda_{\alpha,k} - \lambda_{\alpha',k'}}}}
 \, ,
 \label{eq:vec_inst_part2}
\end{align}
where we use the formula~\eqref{eq:q-factorial_finite}.

We rephrase the relation~\eqref{eq:vec_inst_part} in terms of the vector bundles over the instanton moduli space.
The vector multiplet bundle is written with the framing and instanton bundles as follows:
\begin{align}
 \mathbf{V}
 & = \frac{\mathbf{N}^\vee \mathbf{N}}{\wedge \mathbf{Q}} - \det \mathbf{Q}^\vee \cdot \mathbf{K}^\vee \mathbf{N} - \mathbf{N}^\vee \mathbf{K} + \wedge \mathbf{Q}^\vee \cdot \mathbf{K}^\vee \mathbf{K}
 \nonumber \\
 & =: \mathring{\mathbf{V}} + \mathbf{V}^\text{inst}
\end{align}
where we define the perturbative and the instanton contributions:
\begin{subequations}\label{eq:V_bundle_inst}
 \begin{align}
  \mathring{\mathbf{V}} & = \frac{\mathbf{N}^\vee \mathbf{N}}{\wedge \mathbf{Q}}
  \, ,  \\
  \mathbf{V}^\text{inst} & = - \det \mathbf{Q}^\vee \cdot \mathbf{K}^\vee \mathbf{N} - \mathbf{N}^\vee \mathbf{K} + \wedge \mathbf{Q}^\vee \cdot \mathbf{K}^\vee \mathbf{K}
  \, .
 \end{align}
\end{subequations}
We remark that the dimension of the instanton part is given by
\begin{align}
 \dim \mathbf{V}^\text{inst}
 & = 2nk
 \, ,
\end{align}
which agrees with the dimension of the instanton moduli space~\eqref{eq:ADHM_dim}.

The perturbative and instanton parts are then given by
\begin{align}
 \mathring{Z}^\text{vec} = \mathbb{I}[\mathring{\mathbf{V}}]
 \, , \qquad
 Z^\text{vec,inst} = \mathbb{I}[\mathbf{V}^\text{inst}]
 \, .
\end{align}
For example, for the K-theory convention, the perturbative part is explicitly given as
\begin{align}
 \mathring{Z}^\text{vec} =
 \prod_{\alpha,\alpha' = 1,\ldots n} \Gamma_{q,2}(\np^{\mathsf{a}_{\alpha \alpha'}} q ; q_1, q_2)
\end{align}
where $\Gamma_{q,2}(z;q_1,q_2)$ is the $q$-double gamma function~\eqref{eq:q-multiple_gamma}.
Similar expressions with the double gamma function are available for the cohomology and elliptic cases as well.

The character of the instanton part is given by
\begin{align}
 \ch_\mathsf{T} \mathbf{V}^\text{inst} = - \sum_{\alpha,\alpha' = 1,\ldots,n} \np^{\mathsf{a}_{\alpha\alpha'}} \, \Xi[\lambda_\alpha,\lambda_{\alpha'}]
 \label{eq:ch_V_inst}
\end{align}
where we define
\begin{align}
 \Xi[\lambda_\alpha,\lambda_{\alpha'}]
 =
 \sum_{s \in \lambda_\alpha} q_1^{\ell_{\alpha'}(s)} q_2^{- a_\alpha(s) - 1}
 + \sum_{s \in \lambda_{\alpha'}} q_1^{- \ell_\alpha(s) - 1} q_2^{a_{\alpha'}(s)}
 \, .
\end{align}
We denote the arm and leg lengths by $(a_\alpha(s), \ell_\alpha(s))$ defined in \eqref{eq:arm_leg}.
See \S\ref{sec:comb} for details of the derivation.
Hence, applying the index formula, the vector multiplet contribution to the K-theoretic instanton partition function is given by
\begin{align}
 Z^\text{vec,inst}
 & =
 \prod_{\alpha,\alpha' = 1,\ldots,n}
 \left[
 \prod_{s \in \lambda_\alpha}
 \frac{1}{1 - \np^{\mathsf{a}_{\alpha'\alpha}} q_1^{- \ell_{\alpha}(s)} q_2^{a_\alpha(s) + 1}}
 \prod_{s \in \lambda_{\alpha'}}
 \frac{1}{1 - \np^{\mathsf{a}_{\alpha'\alpha}} q_1^{\ell_\alpha(s) + 1} q_2^{-a_{\alpha'}(s)}}
 \right]
 \, .
\end{align}
This expression is analogous to the hook length formula for the ($q$-)dimension of the irreducible representation of the symmetric group~\cite{Macdonald:1997}.
We can combinatorially show the equivalence to the previous formula~\eqref{eq:vec_inst_part2}.

\subsection{Fundamental and antifundamental matters}\label{sec:fund_instanton}

We consider the instanton part of the hypermultiplet contribution to the partition function.
We split the hypermultiplet bundle into the perturbative and instanton contributions:
\begin{subequations}\label{eq:fund_hyper_inst}
 \begin{align}
  \mathbf{H}^\text{f}
  & = - \frac{{\mathbf{M}}^\vee \mathbf{N}}{\wedge \mathbf{Q}} + {\mathbf{M}}^\vee \mathbf{K}
  \hspace{2.9em} =: \mathring{\mathbf{H}}^\text{f} + \mathbf{H}^\text{f,inst} 
  \, , \\
  \mathbf{H}^\text{af}
  & = - \frac{\mathbf{N}^\vee \widetilde{\mathbf{M}}}{\wedge \mathbf{Q}} + \det \mathbf{Q}^\vee \mathbf{K}^\vee \widetilde{\mathbf{M}}
  =: \mathring{\mathbf{H}}^\text{af} + \mathbf{H}^\text{af,inst}
  \, .
 \end{align}
\end{subequations}
Hence, the corresponding contributions are given by the index formula.
For example, the K-theory partition functions are given by
\begin{subequations}
\begin{align}
 \mathring{Z}^\text{f}
 =
 \prod_{\substack{\alpha = 1,\ldots,n \\ f = 1,\ldots,n^\text{f}}} \Gamma_{q,2}(\np^{m_f - \mathsf{a}_\alpha} q;q_1,q_2)^{-1}
 \, , \qquad
 \mathring{Z}^\text{af}
 =
 \prod_{\substack{\alpha = 1,\ldots,n \\ f = 1,\ldots,n^\text{af}}} \Gamma_{q,2}(\np^{\mathsf{a}_\alpha - \widetilde{m}_f} q;q_1,q_2)^{-1}
 \, 
 \end{align}
\begin{align}
 Z^\text{f,inst}
 =
 \prod_{\substack{\alpha = 1,\ldots,n \\ f = 1,\ldots,n^\text{f}}}
 \prod_{s \in \lambda_\alpha}
 ( 1 - \np^{m_f - \mathsf{a}_\alpha} q_1^{- s_1 + 1} q_2^{- s_2 + 1} )
 \, , \quad
 Z^\text{af,inst}
 =
 \prod_{\substack{\alpha = 1,\ldots,n \\ f = 1,\ldots,n^\text{af}}}
 \prod_{s \in \lambda_\alpha}
 ( 1 - \np^{\mathsf{a}_\alpha - \widetilde{m}_f} q_1^{s_1} q_2^{s_2} )
 \, .
 \label{eq:instanton_func_fund}
\end{align}
\end{subequations}

From the instanton partition function, we see the following property.
Tune the mass parameter as
\begin{align}
 m_f = \mathsf{a}_\alpha + (n_1 - 1) \epsilon_1 + (n_2 - 1) \epsilon_2
 \, ,
 \label{eq:Higgs_branch_locus}
\end{align}
then it obeys the pit condition~\cite{Bershtein:2018SM}:\index{pit condition}
\begin{align}
 (n_1,n_2) \in \lambda_\alpha
 \ \implies \
 Z^\text{f,inst} = 0
 \, .
\end{align}
A similar statement holds for $Z^\text{af}$ as well.
In particular, in the case $n_1 = 0$ (or $n_2 = 0$), the condition \eqref{eq:Higgs_branch_locus} provides the {\em root of Higgs branch locus}\index{root of Higgs branch} in the moduli space of the supersymmetric vacua~\cite{Dorey:1998yh,Dorey:1999zk}.
At this point, 4d $\mathcal{N} = 2$ gauge theory is reduced to 2d $\mathcal{N} = (2,2)$ theory (similarly 5d $\mathcal{N} = 1$ to 3d $\mathcal{N} = 2$, 6d $\mathcal{N} = (1,0)$ to 4d $\mathcal{N} = 1$), which can be explicitly checked by the partition function~\cite{Fujimori:2015zaa}.
See also \S\ref{sec:Higgsing} for related discussions.

\subsection{Adjoint matter}

The adjoint matter contribution is almost the same as the vector multiplet up to the adjoint mass parameter $\mu = \np^{m_\text{adj}} \in \mathbb{C}^\times$.
For the K-theory convention, the perturbative and instanton contributions are given as
\begin{subequations}
 \begin{align}
  \mathring{Z}^\text{adj} & =
  \prod_{\alpha,\alpha' = 1,\ldots n} \Gamma_{q,2}(\np^{\mathsf{a}_{\alpha \alpha'} - m_\text{adj}} q ; q_1, q_2)^{-1}
  \, ,
  \\
  Z^\text{adj,inst}
  & =
  \prod_{\alpha,\alpha' = 1,\ldots,n}
  \left[
  \prod_{s \in \lambda_\alpha}
  \left( 1 - \np^{\mathsf{a}_{\alpha'\alpha} - m_\text{adj}} q_1^{- \ell_{\alpha'}(s)} q_2^{a_\alpha(s) + 1} \right)
  \prod_{s \in \lambda_{\alpha'}}
  \left( 1 - \np^{\mathsf{a}_{\alpha'\alpha} - m_\text{adj}} q_1^{\ell_\alpha(s) + 1} q_2^{-a_{\alpha'}(s)} \right)
  \right]
  \, .
 \end{align}
\end{subequations}
Hence, we obtain a combinatorial expression for the instanton partition function of 5d $\mathcal{N} = 1^*$ theory (5d uplift of 4d $\mathcal{N} = 2^*$ theory):
\begin{align}
 Z_{n,k}^\text{inst} =
 \prod_{\alpha,\alpha'}^n
 \left[
 \prod_{s \in \lambda_\alpha}
 \frac{1 - \np^{\mathsf{a}_{\alpha'\alpha} - m_\text{adj}} q_1^{- \ell_{\alpha'}(s)} q_2^{a_\alpha(s) + 1} }{ 1 - \np^{\mathsf{a}_{\alpha'\alpha}} q_1^{- \ell_{\alpha'}(s)} q_2^{a_\alpha(s) + 1} }
 \prod_{s \in \lambda_{\alpha'}}
 \frac{ 1 - \np^{\mathsf{a}_{\alpha'\alpha} - m_\text{adj}} q_1^{\ell_\alpha(s) + 1} q_2^{-a_{\alpha'}(s)} }{ 1 - \np^{\mathsf{a}_{\alpha'\alpha}} q_1^{\ell_\alpha(s) + 1} q_2^{-a_{\alpha'}(s)} }
 \right]
 \, .
\end{align}
We see that it becomes trivial in the limit $m_\text{adj} \to 0$.
In the case of $n = 1$, in particular, the instanton partition function is given by
\begin{align}
 Z_{1,k}^\text{inst} =
 \prod_{s \in \lambda}
 \frac{\left( 1 - \mu^{-1} q_1^{- \ell(s)} q_2^{a(s) + 1} \right) \left( 1 - \mu^{-1} q_1^{\ell(s) + 1} q_2^{-a(s)} \right)}{\left( 1 - q_1^{- \ell(s)} q_2^{a(s) + 1} \right)\left( 1 - q_1^{\ell(s) + 1} q_2^{-a(s)} \right)}
 \, .
 \label{eq:A0_inst_fn}
\end{align}
See also \S\ref{sec:quiv_LMNS_formula}.

\subsection{Chern--Simons term}\label{sec:CS_partition_function}

In particular for 5d gauge theory, we can incorporate the Chern--Simons term:
\begin{align}
 Z^\text{CS} = \left( \det \mathbf{Y}_\mathcal{S} \right)^{\kappa}
 \, ,
\end{align}
which splits into the perturbative and the instanton contributions
\begin{align}
 \mathring{Z}^\text{CS}
 & = \left( \det \frac{\mathbf{N}}{\wedge \mathbf{Q}} \right)^{\kappa}
 = \prod_{\alpha = 1,\ldots,n} \prod_{i,j = 0,\ldots,\infty} \left( \np^{\mathsf{a}_\alpha} q_1^i q_2^j \right)^{\kappa}
 \, ,
 \\
 Z^\text{CS,inst}
 & = \left( \det \mathbf{K} \right)^{-\kappa}
 = \prod_{\alpha = 1,\ldots,n} \prod_{s \in \lambda_\alpha}
 \left( \np^{\mathsf{a}_\alpha} q_1^{s_1 - 1} q_2^{s_2 - 1} \right)^{-\kappa}
 \, .
\end{align}
Comparing with the (anti)fundamental contributions~\eqref{eq:instanton_func_fund}, we reproduce the Chern--Simons term in the decoupling limit $m_f, \widetilde{m}_f \to + \infty$:
The fundamental and antifundamental matters induce the positive and negative shift of the Chern--Simons level, respectively.

\subsection{Relation to the contour integral formula}

We remark the relation between the equivariant character formula and the LMNS contour integral formula.
The instanton part of the vector multiplet character is also written as
\begin{align}
  \ch_\mathsf{T} \mathbf{V}^\text{inst} & =
  - \sum_{\substack{\alpha = 1,\ldots,n \\ a = 1,\ldots, k}}
  \left(
   \np^{- \phi_a + \mathsf{a}_\alpha - \epsilon_{12}} + \np^{\phi_a - \mathsf{a}_\alpha}
  \right)
 \nonumber \\
 & \quad
 + \sum_{a,b = 1,\ldots, k}
  \left(
  \np^{\phi_{ab}} 
  - \np^{\phi_{ab} - \epsilon_1} - \np^{\phi_{ab} - \epsilon_2}
  + \np^{\phi_{ab} - \epsilon_{12}}
 \right)
 \, .
\end{align}
Here the diagonal part of $\np^{\phi_{ab}}$, namely the zero modes should be removed, and, in this expression, we do not yet impose the fixed point condition for $(\phi_i)_{i = 1,\ldots,k}$.
Then, we notice that the Chern roots of each factor coincide with the infinitesimal equivariant torus action on the ADHM variables~\eqref{eq:ADHM_equiv_weight}.
Hence, applying the index functor~\eqref{eq:[x]_function}, we obtain the integrand of the contour integral~\eqref{eq:LMNS_formula}. \index{equivariant!---action}
We can similarly reproduce the hypermultiplet contributions to the contour integral, and also obtain the K-theory/elliptic analog of the LMNS formula with the corresponding index.
\index{LMNS formula}

\subsubsection{Gauge and matter polynomials}

In general, the gauge and matter polynomials appearing in the contour integral are given as follows:
\begin{subequations}
 \begin{align}
  P(\phi) & = \prod_{\alpha = 1}^n [\phi - \mathsf{a}_\alpha]
  \, , \hspace{2.5em}
  \widetilde{P}(\phi) = \prod_{\alpha = 1}^n [- \phi + \mathsf{a}_\alpha]
  \, , \\
  P^\text{f}(\phi) & = \prod_{\alpha = 1}^{n^\text{f}} [\phi - m_f]
  \, , \qquad
  \widetilde{P}^\text{af}(\phi) = \prod_{\alpha = 1}^{n^\text{af}} [- \phi + \widetilde{m}_f ]
  \, ,
 \end{align}
\end{subequations}
where the index function $[x]$ is defined in~\eqref{eq:[x]_function}.
The polynomials, $P(\phi)$ and $\widetilde{P}(\phi)$, $P^\text{f}(\phi)$ and $\widetilde{P}^\text{af}(\phi)$, have the same zeros, so that they are equivalent up to overall factors.

For example, in the K-theory convention, we have the relation:
\begin{subequations}
\begin{align}
 \widetilde{P}(\phi)
 & = \prod_{\alpha = 1}^n \left( 1 - \np^{\phi - \mathsf{a}_\alpha} \right)
 = \left( (-1)^n \prod_{\alpha = 1}^n \np^{- \mathsf{a}_\alpha} \right) \np^{n \phi} \, P(\phi)
 \, , \\
 \widetilde{P}^\text{af}(\phi)
 & = \prod_{f = 1}^{n^\text{f}} \left( 1 - \np^{\phi - \widetilde{m}_f} \right)
 = \left( (-1)^{n^\text{af}} \prod_{f = 1}^{n^\text{af}} \np^{- \widetilde{m}_f} \right) \np^{n^\text{af} \phi} \, P^\text{af}(\phi)
 \, .
\end{align}
\end{subequations}
The polynomial part in the contour integral together with the Chern--Simons factor (level $\kappa$) is given as
\begin{align}
 \prod_{a = 1}^k \np^{-\kappa \phi_a} \,
 \frac{P^\text{f}(\phi_a) \widetilde{P}^\text{af}(\phi_a + \epsilon_{12})}{P(\phi_a) \widetilde{P}(\phi_a + \epsilon_{12})}
 & =
 \left( 
 (-1)^{n + n^\text{af}}
 \np^{- (n - n^\text{af}) \epsilon_{12} } 
 \prod_{\alpha = 1}^n \np^{\mathsf{a}_\alpha}
 \prod_{f = 1}^{n^\text{af}} \np^{- \widetilde{m}_f}
 \right)^k 
 \nonumber \\
 & \quad \times
 \prod_{a = 1}^k
 \np^{- (\kappa + n - n^\text{af}) \phi_a} \,
 \frac{P^\text{f}(\phi_a) {P}^\text{af}(\phi_a + \epsilon_{12})}{P(\phi_a) {P}(\phi_a + \epsilon_{12})}
 \, .
\end{align}
Hence, we can convert the polynomial $(\widetilde{P}(\phi), \widetilde{P}^\text{af}(\phi))$ to $(P(\phi),P^\text{af}(\phi))$ by redefinition of the coupling constant and the Chern--Simons level $(\mathfrak{q}, \kappa)$:
\begin{subequations}
 \begin{align}
 \mathfrak{q}
 & \ \longmapsto \
 \left(
 (-1)^{n + n^\text{af}}
 \np^{- (n - n^\text{af}) \epsilon_{12} } 
 \prod_{\alpha = 1}^n \np^{\mathsf{a}_\alpha}
 \prod_{f = 1}^{n^\text{af}} \np^{- \widetilde{m}_f}
 \right)
 \mathfrak{q}
  \, , \qquad
  \nonumber \\
  \kappa & \ \longmapsto \ \kappa + n - n^\text{af}
 \, .
 \end{align}
\end{subequations}
In particular, the coupling constant shift becomes simpler if we impose the special unitary condition:
$\displaystyle \sum_{\alpha = 1}^n \mathsf{a}_\alpha = \sum_{f = 1}^{n^\text{af}} \widetilde{m}_f = 0$.
In addition, one can convert the convention of the $[x]$ function from~\eqref{eq:[x]_function} to \eqref{eq:[x]_function_sym} with a similar shift of the coupling constant and the Chern--Simons level:
\begin{align}
 &
 \left.
 \prod_{a = 1}^k \np^{-\kappa \phi_a} \,
 \frac{P^\text{f}(\phi_a) \widetilde{P}^\text{af}(\phi_a + \epsilon_{12})}{P(\phi_a) \widetilde{P}(\phi_a + \epsilon_{12})}\right|_\eqref{eq:[x]_function}
 \nonumber \\
 & =
 \left.
 \left(
 \np^{- \frac{1}{2} n \epsilon_{12}}
 \prod_{f = 1}^{n^\text{f}} \np^{\frac{1}{2} m_f}
 \prod_{f = 1}^{n^\text{af}} \np^{-\frac{1}{2} \widetilde{m}_f}
 \right)^k
 \prod_{a = 1}^k \np^{-(\kappa + \frac{1}{2} n^\text{f} - \frac{1}{2} n^\text{af} )\phi_a} \,
 \frac{P^\text{f}(\phi_a) \widetilde{P}^\text{af}(\phi_a + \epsilon_{12})}{P(\phi_a) \widetilde{P}(\phi_a + \epsilon_{12})}\right|_\eqref{eq:[x]_function_sym}
 \nonumber \\
 & =
 \left.
 \left(
 (-1)^{n + n^\text{af}}
 \np^{- \frac{1}{2} n \epsilon_{12}}
 \prod_{f = 1}^{n^\text{f}} \np^{\frac{1}{2} m_f}
 \prod_{f = 1}^{n^\text{af}} \np^{-\frac{1}{2} \widetilde{m}_f}
 \right)^k
 \prod_{a = 1}^k \np^{-(\kappa + \frac{1}{2} n^\text{f} - \frac{1}{2} n^\text{af} )\phi_a} \,
 \frac{P^\text{f}(\phi_a) P^\text{af}(\phi_a + \epsilon_{12})}{P(\phi_a) {P}(\phi_a + \epsilon_{12})}\right|_\eqref{eq:[x]_function_sym} 
 \, .
\end{align}

We have a similar relation in the elliptic case.
In this case, we have to impose the condition $n = n^\text{f} = n^\text{af}$ and $\kappa = 0$, in order that the partition function shows a modular property.
Therefore, together with the special unitary condition, we can freely convert the gauge and matter functions (not polynomials anymore in the elliptic case), $(\widetilde{P}(\phi), \widetilde{P}^\text{af}(\phi))$ to $(P(\phi),P^\text{af}(\phi))$.

\subsubsection{$\mathscr{S}$-function}
\index{S-function@$\mathscr{S}$-function}

The $\mathscr{S}$-function, appearing in the contour integral~\eqref{eq:S_func_4d}, is replaced as follows for the K-theory/elliptic cases:
\begin{align}\label{eq:S_func_def}
 \mathscr{S}(\phi)
 = \frac{[\phi - \epsilon_1][\phi - \epsilon_2]}{[\phi][\phi - \epsilon_{12}]}
 =
 \begin{cases}
  \displaystyle
  \frac{(1 - q_1 z)(1 - q_2 z)}{(1 - z)(1 - q z)}
  & (\text{5d}) \\[1em]
  \displaystyle
  \frac{\theta(q_1 z;p) \theta(q_2 z;p)}{\theta(z;p) \theta(q z;p)}
  & (\text{6d})
 \end{cases}
\end{align}
where $z = \np^{-\phi}$.
This expression is also obtained from the symmetric convention of the $[x]$ function~\eqref{eq:[x]_function_sym}.
This $\mathscr{S}$-function obeys the reflection formula except at the poles $z = 1, q^{-1}$ ($\phi = 0, \epsilon_{12}$),
\begin{align}
 \mathscr{S}(q^{-1} z^{-1}) = \mathscr{S}(z)
 \quad
 \left(
 \iff \
 \mathscr{S}(\phi) = \mathscr{S}(\epsilon_{12} - \phi) 
 \right)
 \, .
 \label{eq:S_func_reflection}
\end{align}
More precisely, we have the relation
\begin{align}
 \mathscr{S}(z) - \mathscr{S}(q^{-1} z^{-1})
 =
 \begin{cases}
  \displaystyle
  \frac{(1 - q_1)(1 - q_2)}{1 - q} \left( \delta(z) - \delta(qz) \right)
  & (\text{5d}) \\[1em]
  \displaystyle
  \frac{\theta(q_1;p) \theta(q_2;p)}{(p;p)_\infty^2 \theta(q;p)} \left( \delta(z) - \delta(qz) \right)
  & (\text{6d})
 \end{cases}
 \label{eq:S_func_delta}
\end{align}
where we define the multiplicative delta function \index{delta function!multiplicative---}
\begin{align}
 \delta(z) = \sum_{n \in \mathbb{Z}} z^n
 \, ,
 \label{eq:mult_delta_fn}
\end{align}
such that
\begin{align}
 \int dz \, \delta(z) \, f(z) = f(1)
 \, .
 \label{eq:mult_delta_int}
\end{align}
Therefore, we can replace $\displaystyle \delta(z) f(z) \ \longrightarrow \ \delta(z) f(1)$ as long as concerning the integrand.

In order to show the relation~\eqref{eq:S_func_delta}, for the 5d case, we rewrite the $\mathscr{S}$-function
\begin{align}
 \mathscr{S}(z) & = \frac{(1 - q_1 z)(1 - q_2 z)}{(1 - q) z} \left( \frac{1}{1 - z} - \frac{1}{1 - qz} \right)
 \, .
 \label{eq:S_func_expansion}
\end{align}
Then we obtain
\begin{align}
 \mathscr{S}(z) - \mathscr{S}(q^{-1} z^{-1})
 & = \frac{(1 - q_1 z)(1 - q_2 z)}{(1 - q) z} \left( \frac{1}{1 - z} + \frac{1}{1 - z^{-1}} - \frac{1}{1 - qz} - \frac{1}{1 - q^{-1} z^{-1}}\right)
 \nonumber \\
 & = \frac{(1 - q_1 z)(1 - q_2 z)}{(1 - q) z} \left( \frac{1}{1 - z} + \frac{z^{-1}}{1 - z^{-1}} - \frac{1}{1 - qz} - \frac{q^{-1} z^{-1}}{1 - q^{-1} z^{-1}}\right)
 \nonumber \\
 & = \frac{(1 - q_1 z)(1 - q_2 z)}{(1 - q) z} \left( \delta(z) - \delta(qz) \right)
 \nonumber \\ 
 & = \text{RHS of \eqref{eq:S_func_delta}} 
 \, ,
\end{align}  
where we replace the prefactor $\displaystyle \frac{(1 - q_1 z)(1 - q_2 z)}{(1 - q) z} \longrightarrow \frac{(1 - q_1)(1 - q_2)}{(1 - q)} $ due to the property~\eqref{eq:mult_delta_int}.

For the 6d case, we rewrite the $\mathscr{S}$-function together with \eqref{eq:theta_reflection} and \eqref{eq:theta_id} as follows:
\begin{align}
 \mathscr{S}(z) & =
 \frac{\theta(q_1 z;p) \theta(q_2 z;p)}{(p;p)_\infty^2 \theta(q;p) z}
 \sum_{n \in \mathbb{Z}} \frac{z^n}{1 - p^n/qz}
 \nonumber \\
 & =
 \frac{\theta(q_1 z;p) \theta(q_2 z;p)}{(p;p)_\infty^2 \theta(q;p) z}
 \left(
 \sum_{0 \le n, m \le \infty} z^n (qz)^{-m} p^{nm}
 - \sum_{1 \le n, m \le \infty} z^{-n} (qz)^m p^{nm}
 \right)
 \, .
\end{align}
Then, we can show the relation~\eqref{eq:S_func_delta} similarly to the 5d case.
We remark that the $O(p^0)$ contribution to the double infinite series is given by
\begin{align}
 & \lim_{p \to 0}
 \left(
 \sum_{0 \le n, m \le \infty} z^n (qz)^{-m} p^{nm}
 - \sum_{1 \le n, m \le \infty} z^{-n} (qz)^m p^{nm} 
 \right)
 \nonumber \\
 & =
 1 + \sum_{n=1}^\infty z^n + \sum_{m=1}^\infty (qz)^{-m}
 \nonumber \\
 & =
 \frac{1}{1 - z} - \frac{1}{1 - qz}
 \, ,
\end{align}
which is consistent with the 5d expression~\eqref{eq:S_func_expansion}.

\chapter{Quiver gauge theory}
\label{chap:quiver_gauge_theory}

Quiver gauge theory is a gauge theory with a product-type gauge symmetry:
\begin{align}
 G = \prod_{i \in \Gamma_0} G_i
 \, ,
\end{align}
where $\Gamma_0$ is a set of gauge nodes (vertices).
Together with a set of edges (arrows) $\Gamma_1$, we define a quiver $\Gamma = (\Gamma_0, \Gamma_1)$. \index{quiver}
For each edge $e:i \to j$, in 4d $\mathcal{N} = 1$ gauge theory (4 SUSY) convention, we assign the chiral multiplet in the bifundamental representation of $(G_i, G_j)$.
The loop edge $e: i \to i$ denotes the adjoint chiral multiplet.

The 4d $\mathcal{N} = 2$ (8 SUSY) vector multiplet consists of the $\mathcal{N} = 1$ vector multiplet and the chiral multiplet in the adjoint representation.
Similarly, the $\mathcal{N} = 2$ bifundamental hypermultiplet consists of the chiral and anti-chiral multiplets associated with the edges, $e : i \to j$ and $e : j \to i$, so that we combine a pair of these arrows into a single unoriented edge, e.g.:
\begin{align}
  \begin{tikzpicture}[baseline=(current  bounding  box.center),thick]
   \foreach \x in {0,2,4} {
   \draw (\x,0) circle (.2);
   \draw [-stealth] (\x,0)+(135:.2) arc [x radius = .5, y radius = .7, start angle = 260, end angle = -70]; 
   }
   \draw [->-] (30:.2) -- ($(2,0)+(150:.2)$);
   \draw [-<-] (-30:.2) -- ($(2,0)+(-150:.2)$);
   \draw [->-] (2,0)+(30:.2) -- ($(4,0)+(150:.2)$);
   \draw [-<-] (2,0)+(-30:.2) -- ($(4,0)+(-150:.2)$);      
   \node at (2,-1) {\text{4 SUSY}};
   \begin{scope}[shift={(6,0)}]
    \draw (0,0) -- ++(4,0);
    \filldraw[fill=white,draw=black] (0,0) circle (.2);
    \filldraw[fill=white,draw=black] (2,0) circle (.2);
    \filldraw[fill=white,draw=black] (4,0) circle (.2);
    \node at (2,-1) {\text{8 SUSY}};    
   \end{scope}
  \end{tikzpicture}
 \label{sec:4SUSY_8SUSY_quiver}
\end{align}
For less SUSY cases, we need a different type of arrows to distinguish specific multiplets, e.g., the Fermi multiplet in 2d $\mathcal{N} = (0,2)$ theory (2 SUSY).

We remark that the quiver diagram in the 8 SUSY convention can be identified with the simply-laced Dynkin diagram.%
\footnote{%
The non-simply-laced cases would be discussed in \S\ref{sec:fractional_quiver}.
}
For example, the diagram shown in \eqref{sec:4SUSY_8SUSY_quiver} is the Dynkin diagram of $A_3$.
This is not a coincidence, and we will show the quantum algebraic structure emerging from the moduli space of quiver gauge theory in Part~\ref{part:algebra}.

\section{Instanton moduli space}\label{sec:quiver_inst_mod_sp}

Let $G_i = \rU(n_i)$ be a gauge group of the node $i \in \Gamma_0$.
We consider the instanton configuration in quiver gauge theory:\index{(A)SD YM!---equation}
\begin{align}
 F^+_i = 0
 \qquad \text{for} \quad
 ^\forall i \in \Gamma_0
 \, ,
 \label{eq:ASD_node}
\end{align}
where $F^+_i$ is the SD part of the curvature associated with the gauge node $i \in \Gamma_0$.
Then, the instanton number is defined for each node:
\begin{align}
 k_i = \frac{1}{8\pi^2} \int_\mathcal{S} \tr F_i \wedge F_i
 \, .
\end{align}
Hence, the instanton moduli space for $\Gamma$-quiver gauge theory on $\mathcal{S}$ is given by
\begin{align}
 \mathfrak{M}_{\underline{n},\underline{k}}
 = \bigsqcup_{i \in \Gamma_0} \mathfrak{M}_{n_i,k_i}
 \, ,
\end{align}
where $\mathfrak{M}_{n_i,k_i}$ is the instanton moduli space~\eqref{eq:ADHM_mod_sp_res} for the node $i \in \Gamma_0$, and we define the dimension vector:
\begin{align}
 (\underline{n},\underline{k}) = (n_i,k_i)_{i = 1,\ldots,\rk \Gamma}
 \, ,
 \label{eq:quiv_dim_vec}
\end{align}
where $\rk \Gamma = |\Gamma_0|$.
In order to obtain the instanton partition function for quiver gauge theory, we apply the equivariant character formula discussed in \S\ref{sec:eq_ch_formula} to the instanton moduli space $\mathfrak{M}_{\underline{n},\underline{k}}$.

\subsection{Vector bundles on the moduli space}\label{sec:quiver_bundles}
 \index{framing!---bundle}
 \index{instanton bundle}
 \index{matter bundle}

We define the framing, instanton bundles, and also the matter bundles on the instanton moduli space $\mathfrak{M}_{\underline{n},\underline{k}}$ for each node $i \in \Gamma_0$:
\begin{align}
 \mathbf{N} = (\mathbf{N}_i)_{i \in \Gamma_0}
 \, , \quad
 \mathbf{K} = (\mathbf{K}_i)_{i \in \Gamma_0}
 \, , \quad
 \mathbf{M} = (\mathbf{M}_i)_{i \in \Gamma_0}
 \, , \quad
 \widetilde{\mathbf{M}} = (\widetilde{\mathbf{M}}_i)_{i \in \Gamma_0}
 \, .
\end{align}
The corresponding complexified automorphism groups are given by
\begin{subequations}
 \begin{align}
  \GL(\mathbf{N}) & = \prod_{i \in \Gamma_0} \GL(\mathbf{N}_i)
  \, , \hspace{2.5em}
  \GL(\mathbf{K}) = \prod_{i \in \Gamma_0} \GL(\mathbf{K}_i)
  \, , \\
  \GL(\mathbf{M}) & = \prod_{i \in \Gamma_0} \GL(\mathbf{M}_i)
  \, , \qquad
  \GL(\widetilde{\mathbf{M}}) = \prod_{i \in \Gamma_0} \GL(\widetilde{\mathbf{M}}_i)
  \, .
 \end{align}
\end{subequations}
We denote the corresponding Cartan tori by $\mathsf{T}_K$, $\mathsf{T}_N$, etc.
The dimensions of these bundles are parametrized by the dimension vector:
\begin{align}
 (\underline{n},\underline{k},\underline{n}^\text{f}, \underline{n}^\text{af}) = (n_i,k_i,n_{i}^\text{f},n_{i}^\text{af})_{i = 1,\ldots,\rk \Gamma}
 \, ,
\end{align}
so that
\begin{align}
 \GL(\mathbf{N}_i) = \GL(\mathbb{C}^{n_i})
 \, , \qquad
 \GL(\mathbf{K}_i) = \GL(\mathbb{C}^{k_i})
 \, , \qquad \text{etc.}
\end{align}

In addition, we assign the line bundle $(\mathbf{M}_e)_{e \in \Gamma_1}$ to each edge.
The automorphism group is given by
\begin{align}
 \prod_{e \in \Gamma_1} \GL(\mathbf{M}_e)
\end{align}
with the character
\begin{align}
 \ch_\mathsf{T} \mathbf{M}_e = \np^{m_e} =: \mu_e
 \, .
\end{align}
The parameter $m_e \in \mathbb{C}$ is the bifundamental mass associated with the edge $e$, and $\mu_e \in \mathbb{C}^\times$ is the multiplicative analog.

\subsection{Equivariant fixed point and observables}\label{sec:quiv_eq_fix_pt}

Based on the vector bundles introduced above, we define the observable bundle for each gauge node $i \in \Gamma_0$ similarly to~\eqref{eq:obs_bndl}:
\begin{align}
 \mathbf{Y}_i := \mathbf{Y}_{o,i} = \mathbf{N}_i - \wedge \mathbf{Q} \cdot \mathbf{K}_i
 \, .
 \label{eq:quiver_univ_bundle}
\end{align}
Similarly to the discussion in \S\ref{sec:eq_ch_formula}, the fixed point under the equivariant action in the moduli space $\mathfrak{M}^\mathsf{T}$ is parametrized by a set of partitions:\index{equivariant!---action}
\begin{align}
 \lambda = (\lambda_{i,\alpha})_{i \in \Gamma_0, \, \alpha = 1, \ldots, n_i}
 \in
 \mathfrak{M}^\mathsf{T}
 \, .
\end{align}
Then, the partial reduction of the universal bundle gives rise to the formula:
\begin{align}
 \mathbf{Y}_i
 = \wedge \mathbf{Q}_1 \cdot \mathbf{X}_i
 = \wedge \mathbf{Q}_2 \cdot \check{\mathbf{X}}_i
 \label{eq:quiver_partial_red_univ_bundle}
\end{align}
where we define $(\mathbf{X}_i,\check{\mathbf{X}}_i) = (\mathbf{Y}_{\mathcal{S}_1,i}, \mathbf{Y}_{\mathcal{S}_2,i})$ with the characters
\begin{align}
 \ch_\mathsf{T} \mathbf{X}_i = \sum_{x \in \mathcal{X}_i} x
 \, , \qquad
 \ch_\mathsf{T} \check{\mathbf{X}}_i = \sum_{\check{x} \in \check{\mathcal{X}}_i} \check{x}
 \, ,
\end{align}
and
\begin{align}
 \mathcal{X}_i & = \left\{ x_{i,\alpha,k} = \np^{\mathsf{a}_{i,\alpha}} q_1^{k-1} q_2^{\lambda_{i,\alpha,k}} \right\}_{\substack{i \in \Gamma_0 \\ \alpha = 1,\ldots n_i \\ k=1,\ldots,\infty}}
 \, , \quad
 \check{\mathcal{X}}_i = \left\{ \check{x}_{i,\alpha,k} = \np^{\mathsf{a}_{i,\alpha}}  q_1^{\check\lambda_{i,\alpha,k}} q_2^{k-1} \right\}_{\substack{i \in \Gamma_0 \\ \alpha = 1,\ldots n_i \\ k=1,\ldots,\infty}}
 \, .
 \label{eq:X-variable_quiver}
\end{align}
We also define the union of these sets:
\begin{align}
 \mathcal{X} = \bigsqcup_{i \in \Gamma_0} \mathcal{X}_i
 \, , \qquad
 \check{\mathcal{X}} = \bigsqcup_{i \in \Gamma_0} \check{\mathcal{X}}_i
 \, ,
\end{align}
with 
\begin{align}
 \mathcal{X} \,, \, \check{\mathcal{X}}
 \in \mathfrak{M}^\mathsf{T} 
 \, .
\end{align}

\section{Instanton partition function}\label{sec:quiver_partition_function}

The quiver gauge theory partition function is obtained through the equivariant localization of the instanton moduli space.
In this Section, we derive the equivariant index formula and the contour integral formula for the instanton partition function~\cite{Shadchin:2005cc,Nekrasov:2012xe,Nekrasov:2013xda}.

\subsection{Equivariant index formula}\label{sec:quiver_index_formula}
\index{equivariant!---index}

The vector multiplet and the hypermultiplet contributions to the tangent bundle on the instanton moduli space are given as follows:
\begin{subequations}\label{eq:quiv_vect_hyp_bundles}
 \begin{align}
  \mathbf{V}_i = \frac{\mathbf{Y}_i^\vee \mathbf{Y}_i}{\wedge \mathbf{Q}}
  \, ,
 \end{align}
 \begin{align}
  \mathbf{H}_i^\text{f} = - \frac{{\mathbf{M}}_i^\vee\mathbf{Y}_i}{\wedge \mathbf{Q}}
  \, , \qquad
  \mathbf{H}_i^\text{af} = - \frac{\mathbf{Y}_i^\vee \widetilde{\mathbf{M}}_i}{\wedge \mathbf{Q}} \, ,
 \end{align}
 \begin{align}
 \mathbf{H}_{e:i \to j} = - \mathbf{M}_e \frac{\mathbf{Y}_i^\vee \mathbf{Y}_j}{\wedge \mathbf{Q}}
 \, .
\end{align} 
\end{subequations}
In this convention, the bifundamental hypermultiplet for the edge $e: i \to j$ transforms as the fundamental representation of $G_j$ and the antifundamental representation of $G_i$.
We remark that the bifundamental matter associated with the loop edge $e: i \to i$ is interpreted as the adjoint matter.
Then, the partition function is obtained by the index functor of these contributions:
\begin{align}
 Z_{\underline{n},\underline{k}}
 & = \mathbb{I} \left[ T_\lambda\mathfrak{M} \right]
 = \prod_{i \in \Gamma_0} Z^\text{vec}_i Z_i^\text{f} Z_i^\text{af} \prod_{e \in \Gamma_1} Z_e^\text{bf}
\, ,
\label{eq:Z_func_quiver1}
\end{align}
with the total tangent bundle to the instanton moduli space $\mathfrak{M} = \mathfrak{M}_{\underline{n},\underline{k}}$ at the equivariant fixed point $\lambda \in \mathfrak{M}^\mathsf{T}$ given by 
\begin{align}
 T_\lambda\mathfrak{M} = \sum_{i \in \Gamma_0} \left( \mathbf{V}_i + \mathbf{H}_i^\text{f} + \mathbf{H}_i^\text{af} \right) + \sum_{e \in \Gamma_1} \mathbf{H}_e
 \, ,
 \label{eq:TM}
\end{align}
where we formally include the matter bundles to the total tangent bundle.
Instead of this expression, similarly to the previous case~\eqref{eq:euler_class_insertion}, we can also express the total instanton partition function with the equivariant Euler class of the matter bundles, including both (anti)fundamental and bifundamental hypermultiplet bundles:
\begin{align}
 Z_{\underline{n},\underline{k}}
 & = \int_{\mathfrak{M}_{\underline{n},\underline{k}}} e_\mathsf{T}\left( \sum_{i \in \Gamma_0} \mathbf{H}_i^\text{(a)f} + \sum_{e \in \Gamma_1} \mathbf{H}_e^\text{bf} \right)
 \, .
\end{align}

In particular, for K-theory convention, corresponding to 5d gauge theory, we can also incorporate the Chern--Simons term in addition to these contributions.
The total partition function of $\Gamma$-quiver gauge theory is given as summation over all the topological sectors:
\begin{align}
 Z(\Gamma,\mathcal{S}) = \sum_{\underline{k}} \mathfrak{q}^{\underline{k}} \, Z_{\underline{n}, \underline{k}}
\end{align}
where the instanton counting parameter (fugacity) is defined \index{fugacity (instanton)}
\begin{align}
 Z^\text{top}
 = \mathfrak{q}^{\underline{k}} = \prod_{i \in \Gamma_0} \mathfrak{q}_i^{k_i}
 = \prod_{i \in \Gamma_0} Z_i^\text{top}
 \, , \qquad
 \mathfrak{q}_i = \np^{ 2 \pi \im \tau_i} \in \mathbb{C}^\times
 \, ,
 \label{eq:quiv_top}
\end{align}
with the complexified gauge coupling constants $(\tau_i)_{i \in \Gamma_0}$.

Each contribution to the (full) partition function is given by applying the index to the vector multiplet and hypermultiplet bundles similarly to \eqref{eq:full_func_vec} and \eqref{eq:full_func_adj}:
\begin{subequations}\label{eq:quiv_full_func}
\begin{align}
 Z_i^\text{vec} := \mathbb{I}[\mathbf{V}_i]
 & =
 \begin{cases}
  \displaystyle
  \prod_{\substack{(x,x') \in \mathcal{X}_i \times \mathcal{X}_i \\ x \neq x'}}
  \frac{\Gamma_1(\log x' - \log x - \epsilon_1;\epsilon_2)}{\Gamma_1(\log x' - \log x;\epsilon_2)}
  & (\text{4d}) \\
  \displaystyle
  \prod_{\substack{(x,x') \in \mathcal{X}_i \times \mathcal{X}_i \\ x \neq x'}}
  \frac{\Gamma_q(q_2 x/x';q_2)}{\Gamma_q(q x/x';q_2)}  
  & (\text{5d}) \\ 
  \displaystyle
  \prod_{\substack{(x,x') \in \mathcal{X}_i \times \mathcal{X}_i \\ x \neq x'}}
  \frac{\Gamma_e(q_2 x/x';p,q_2)}{\Gamma_e(q x/x';p,q_2)}
  & (\text{6d}) 
 \end{cases}
 \\
 Z_{e:i \to j}^\text{bf} := \mathbb{I}[\mathbf{H}_{e:i \to j}]
 & =
 \begin{cases}
  \displaystyle
  \prod_{(x,x') \in \mathcal{X}_i \times \mathcal{X}_j}
  \frac{\Gamma_1(\log x' - \log x + m_e;\epsilon_2)}{\Gamma_1(\log x' - \log x + m_e - \epsilon_1;\epsilon_2)}
  & (\text{4d}) \\
  \displaystyle
  \prod_{(x,x') \in \mathcal{X}_i \times \mathcal{X}_j}
  \frac{\Gamma_q(\mu_e^{-1} q x/x';q_2)}  {\Gamma_q(\mu_e^{-1} q_2 x/x';q_2)}
  & (\text{5d}) \\ 
  \displaystyle
  \prod_{(x,x') \in \mathcal{X}_i \times \mathcal{X}_j}
  \frac{\Gamma_e(\mu_e^{-1} q x/x';p,q_2)}{\Gamma_e(\mu_e^{-1} q_2 x/x';p,q_2)}
  & (\text{6d}) 
 \end{cases}
 \\
 Z_i^\text{f} := \mathbb{I}[\mathbf{H}_i^\text{f}] & =
  \begin{cases}
   \displaystyle
   \prod_{(x,x') \in \mathcal{M}_i \times \mathcal{X}_i}
   \Gamma_1(\log x' - \log x;\epsilon_2)^{-1}
   & (\text{4d}) \\   
  \displaystyle
   \prod_{(x,x') \in \mathcal{M}_i \times \mathcal{X}_i}
   \Gamma_q \left( q_2 x/x' ; q_2\right)^{-1}
  & (\text{5d}) \\ 
  \displaystyle
  \prod_{(x,x') \in \mathcal{M}_i \times \mathcal{X}_i}
  \Gamma_e \left( q_2 x/x' ; p, q_2\right)^{-1}
  & (\text{6d}) 
  \end{cases}
 \\
 Z^\text{af} := \mathbb{I}[\mathbf{H}_i^\text{af}] & =
 \begin{cases}
   \displaystyle
   \prod_{(x,x') \in \mathcal{X}_i \times \widetilde{\mathcal{M}}_i}
   \Gamma_1(\log x' - \log x - \epsilon_1;\epsilon_2)
   & (\text{4d}) \\
  \displaystyle
   \prod_{(x,x') \in \mathcal{X}_i \times \widetilde{\mathcal{M}}_i}
   \Gamma_q \left( q x/x' ; q_2\right)
  & (\text{5d}) \\ 
  \displaystyle
  \prod_{(x,x') \in \mathcal{X}_i \times \widetilde{\mathcal{M}}_i} \Gamma_e \left( q x/x' ; p, q_2\right)
  & (\text{6d}) 
 \end{cases}
 \, ,
\end{align}
\end{subequations}
where we define the sets of the multiplicative (anti)fundamental mass parameters
\begin{subequations}\label{eq:mass_parameters_quiv}
\begin{align}
 \mathcal{M}_i & = (\mu_{i,1},\ldots,\mu_{i,n_i^\text{f}}) := (\np^{m_{i,1}},\ldots,\np^{m_{i,n_i^\text{f}}})
 \, , \\
 \widetilde{\mathcal{M}}_i & = (\widetilde{\mu}_{i,1},\ldots,\widetilde{\mu}_{i,n_i^\text{af}}) := (\np^{\widetilde{m}_{i,1}},\ldots,\np^{\widetilde{m}_{i,n_i^\text{af}}})
 \, .
\end{align}
\end{subequations}
The gamma functions are summarized in Appendix~\ref{chap:sp_functions}.
We will discuss the associated geometric structure in Part~\ref{part2}, and the quantum algebraic structure in Part~\ref{part:algebra} based on these expressions.

\subsection{Contour integral formula}\label{sec:quiv_LMNS_formula}
\index{LMNS formula!quiver gauge theory}%

As shown in \S\ref{sec:inst_part_func}, we can also derive the contour integral formula for the instanton partition function from the equivariant character computation.
The contour integral formula for the instanton partition function of quiver gauge theory is given as follows:
\begin{align}
 Z_{\underline{n},\underline{k}}^\text{inst}
 & = \frac{Z_{\underline{n},\underline{k}}}{\mathring{Z}_{\underline{n},\underline{k}}}
 = \prod_{i \in \Gamma_0} \frac{1}{k_i!} \frac{[-\epsilon_{12}]^{k_i}}{[-\epsilon_{1,2}]^{k_i}} \oint_{\mathsf{T}_K} 
 \prod_{i \in \Gamma_0} z^\text{vec}_i z^\text{f}_i z^\text{af}_i \prod_{e \in \Gamma_1} z_e^\text{bf}
\end{align}
where each factor is given by
\begin{subequations}
 \begin{align}
  z_i^\text{vec} & =
  \prod_{a = 1}^{k_i} \frac{1}{P_i(\phi_{i,a}) \widetilde{P}_i(\phi_{i,a} + \epsilon_{12})}
  \prod_{a \neq b}^{k_i} \mathscr{S}(\phi_{i,ab})^{-1}
  \, , \\
  z_i^\text{f} & = \prod_{a=1}^{k_i} P^\text{f}_i(\phi_{i,a})
  \, , \qquad \qquad
  z_i^\text{af}  = \prod_{a=1}^{k_i} \widetilde{P}^\text{af}_i(\phi_{i,a} + \epsilon_{12})
  \, , \\
  z_{e:i \to j}^\text{bf} & =
  \prod_{a = 1}^{k_j} P_i(\phi_{j,a} + m_e)
  \prod_{a = 1}^{k_i} \widetilde{P}_j(\phi_{i,a} + \epsilon_{12} - m_e)
  \prod_{\substack{a = 1,\ldots,k_i \\ b = 1,\ldots,k_j}} \mathscr{S}(\phi_{j,b;i,a} + m_e)
  \, ,
 \end{align}
\end{subequations}
where $\phi_{i,ab} = \phi_{i,a} - \phi_{i,b}$ and $\phi_{j,b;i,a} = \phi_{j,b} - \phi_{i,a}$.
Due to the reflection formula of the $\mathscr{S}$-function~\eqref{eq:S_func_reflection}, the last factor in the bifundamental hypermultiplet contribution is also written as
\begin{align}
 \prod_{\substack{a = 1,\ldots,k_i \\ b = 1,\ldots,k_j}} \mathscr{S}(\phi_{j,b;i,a} + m_e)
 =
 \prod_{\substack{a = 1,\ldots,k_i \\ b = 1,\ldots,k_j}} \mathscr{S}(\phi_{i,a;j,b} + \epsilon_{12} - m_e)
 \, .
\end{align}
We define the gauge and matter polynomials:
\begin{subequations}
\begin{align}
 P_i (\phi) & = \prod_{\alpha = 1}^{n_i} [\phi - \mathsf{a}_{i,\alpha}]
 \, , \hspace{2.3em}
 \widetilde{P}_i (\phi) = \prod_{\alpha = 1}^{n_i} [ - \phi + \mathsf{a}_{i, \alpha}]
 \, , \\
 P^\text{f}_i (\phi) & = \prod_{f = 1}^{n_i^\text{f}} [\phi - m_{i,f}]
 \, , \qquad
 \widetilde{P}^\text{af}_i (\phi) = \prod_{f = 1}^{n_i^\text{af}} [ - \phi + \widetilde{m}_{i,f}]
 \label{eq:matter_polynomial_quiver}
 \, .
\end{align}
\end{subequations}
This formula is available for 4d $\mathcal{N} = 2$/5d $\mathcal{N} = 1$/6d $\mathcal{N} = (1,0)$ theories applying the corresponding index function $[x]$ defined in \eqref{eq:[x]_function}.
More explicitly, the contour integral is given as
\begin{align}
 Z_{\underline{n},\underline{k}}^\text{inst} & = 
 \prod_{i \in \Gamma_0} \frac{1}{k_i!} \frac{[-\epsilon_{12}]^{k_i}}{[-\epsilon_{1,2}]^{k_i}} \oint_{\mathsf{T}_K}
 \prod_{i \in \Gamma_0} \prod_{a=1}^{k_i} \frac{d \phi_{i,a}}{2 \pi \im} \,
 P^\text{f}_i(\phi_{i,a}) \widetilde{P}^\text{af}_i(\phi_{i,a} + \epsilon_{12})
 \nonumber \\
 & \qquad \times
 \prod_{i \in \Gamma_0} \prod_{a=1}^{k_i}
 \left[
 \frac{\prod_{e:i \to j} P_j(\phi_{i,a} + m_e) \prod_{e:j \to i} \widetilde{P}_j(\phi_{i,a} + \epsilon_{12} - m_e)}{P_i(\phi_{i,a}) \widetilde{P}_i(\phi_{i,a} + \epsilon_{12})}
 \right]
 \nonumber \\
 & \qquad \times
 \prod_{i \in \Gamma_0} \prod_{a \neq b}^{k_i}
 \mathscr{S}(\phi_{i,ab})^{-1}
 \prod_{\substack{e \in \Gamma_1 \\ e:i \to j}}
 \prod_{a = 1}^{k_i} \prod_{b = 1}^{k_j}
 \mathscr{S}(\phi_{j,b;i,a} + m_e)
 \, .
 \label{eq:quiver_LMNS}
\end{align}


\subsubsection{Linear quiver: $A_p$}

We consider the first nontrivial quiver gauge theory, which consists of two gauge nodes with (anti)fundamental flavors.
This is classified into $A_2$ quiver, whose quiver diagram is given as follows:
 \begin{align}
  A_2: \qquad
  \begin{tikzpicture}[baseline=(current  bounding  box.center),thick]
   \draw[-<-] (-2,0) -- ++(2,0);
   \draw[-<-] (0,0) -- ++(2,0);
   \draw[-<-] (2,0) -- ++(2,0);   
   \filldraw[fill=white,draw=black] (0,0) circle (.2);
   \filldraw[fill=white,draw=black] (2,0) circle (.2);
   \filldraw[fill=white,draw=black] ($(-2,0)+(-.2,.2)$) rectangle ++ (.4,-.4);
   \filldraw[fill=white,draw=black] ($(4,0)+(-.2,.2)$) rectangle ++ (.4,-.4); 
   \node at (0,-.7) {$\rU(n_1)$};
   \node at (2,-.7) {$\rU(n_2)$};
   \node at (-2,-.7) {$\rU(n_0)$}; 
   \node at (4,-.7) {$\rU(n_3)$}; 
  \end{tikzpicture}
 \end{align}
 where the symbol $\square$ denotes the flavor node, so that the flavor symmetries for the nodes $i = 1, 2$ are $\rU(n_0)$ and $\rU(n_3)$, respectively.
 For the edge $e: i \to j$, the hypermultiplet transforms as the fundamental representation under $G_j$, and the antifundamental representation under $G_i$.
 (This difference will be relevant in 5d theory.)
 In this case, the contour integral formula for the instanton partition function is given by 
 \begin{align}
  Z^\text{inst}_{n_{1,2},k_{1,2}}
  &
  = \frac{1}{k_{1,2}!} \frac{[-\epsilon_{12}]^{k_{1,2}}}{[-\epsilon_{1,2}]^{k_{1,2}}} 
  \oint \prod_{a=1}^{k_{1}}
  \frac{d \phi_{1,a}}{2 \pi \im} 
  \prod_{a=1}^{k_{2}}
  \frac{d \phi_{2,a}}{2 \pi \im}
  \frac{\prod_{a = 1,\ldots,k_1}^{b = 1,\ldots,k_2} \mathscr{S}(\phi_{1,a;2,b} + m)}{\prod_{a \neq b}^{k_1} \mathscr{S}(\phi_{1,ab}) \prod_{a \neq b}^{k_2} \mathscr{S}(\phi_{2,ab})}
  \nonumber \\
  & \qquad \times
  \prod_{a=1}^{k_{1}}  
  \left(
  \frac{P^\text{f}_1(\phi_{1,a})\widetilde{P}_2(\phi_{1,a} + \epsilon_{12} - m)}{P_1(\phi_{1,a}) \widetilde{P}_1(\phi_{1,a} + \epsilon_{12})}
  \right)
  \prod_{a=1}^{k_{2}}
  \left(
  \frac{P_1(\phi_{2,a} + m)\widetilde{P}^\text{af}_2(\phi_{2,a} + \epsilon_{12})}{P_2(\phi_{2,a}) \widetilde{P}_2(\phi_{2,a} + \epsilon_{12})}
  \right)
 \end{align}
 where $m = m_{2 \to 1}$ is the bifundamental mass parameter for the edge $e:1 \to 2$.

 We can similarly consider the linear quiver consisting of $p$ gauge nodes, classified into $A_p$ quiver:
 \begin{align}
  A_p: \qquad
  \begin{tikzpicture}[baseline=(current  bounding  box.center),thick]
   \draw[-<-] (-2,0) -- ++(2,0);
   \draw[-<-] (0,0) -- ++(1,0);
   \draw[-<-] (2,0) -- ++(1,0);
   \draw[dotted] (1,0) -- ++(1,0);
   \draw[-<-] (3,0) -- ++(2,0);   
   \filldraw[fill=white,draw=black] (0,0) circle (.2);
   \filldraw[fill=white,draw=black] (3,0) circle (.2);
   \filldraw[fill=white,draw=black] ($(-2,0)+(-.2,.2)$) rectangle ++ (.4,-.4);
   \filldraw[fill=white,draw=black] ($(5,0)+(-.2,.2)$) rectangle ++ (.4,-.4); 
   \node at (0,-.7) {$\rU(n_1)$};
   \node at (3,-.7) {$\rU(n_p)$};
   \node at (-2,-.7) {$\rU(n_0)$}; 
   \node at (5,-.7) {$\rU(n_{p+1})$}; 
  \end{tikzpicture}
 \end{align}
 In this case, the instanton partition function is given by
 \begin{align}
  Z_{\underline{n},\underline{k}}^\text{inst} & =
  \prod_{i=1}^p \frac{1}{k_i!} \frac{[-\epsilon_{12}]^{k_{i}}}{[-\epsilon_{1,2}]^{k_{i}}}
  \oint \prod_{i=1}^p \prod_{a = 1}^{k_i} \frac{d \phi_{i,a}}{2 \pi \im}
  \frac{P_{i-1}(\phi_{i,a} + m_{i-1,i}) \widetilde{P}_{i+1}(\phi_{i,a} + \epsilon_{12} - m_{i,i+1})}{P_i(\phi_{i,a}) \widetilde{P}_i(\phi_{i,a} + \epsilon_{12})}
  \nonumber \\
  & \hspace{7em} \times
  \prod_{i = 1}^p \prod_{a \neq b}^{k_i} \mathscr{S}(\phi_{i,ab})^{-1}
  \prod_{1 = 1}^{p-1} 
  \prod_{\substack{ a = 1,\ldots,k_i \\ b = 1,\ldots, k_{i+1} }}
  \mathscr{S}(\phi_{i+1,b;i,a} + m_{i,i+1})
 \end{align}
 where we denote $P_0(\phi) = P^\text{f}_1(\phi)$ and $\widetilde{P}_{p+1}(\phi) = \widetilde{P}^\text{af}_p(\phi)$.
 
 We remark that, in order to have asymptotically free theory in four dimensions, the gauge and flavor group ranks have to obey the condition:
  \begin{align}
   2 n_i - n_{i-1} - n_{i+1} \ge 0
   \, .
   \label{eq:AF_condition}
  \end{align}
  This condition corresponds to the convergence condition for the contour integral:
  The integrand does not diverge at $\phi_{i,a} \to \infty$ under this condition.

\subsubsection{Cyclic quiver: $\widehat{A}_{p-1}$}

We consider the cyclic quiver with $p$ gauge nodes, which is classified into $\widehat{A}_{p-1}$ quiver:
\begin{align}
 \widehat{A}_{p-1}
 \ (p = 5)
 : \qquad
 \begin{tikzpicture}[baseline=(current  bounding  box.center),thick]
  \foreach \x in {0,1,2,3,4} {
  \draw [->-] (95+\x*72:1.5) arc (95+\x*72:150+\x*72:1.5);
  \filldraw [draw=black,fill=white] (90+\x*72:1.5) circle (.3) node {$n_\x$};
  }
 \end{tikzpicture}
 \label{eq:cyclic_quiver_diagram}
\end{align}
 The node index $i$ is now defined modulo $p$, $i \equiv i + p$.
 We can at least formally incorporate the (anti)fundamental matters, but we don't consider them for simplicity.
 The instanton partition function for $\widehat{A}_{p-1}$ quiver is then given as follows:
 \begin{align}
  Z_{\underline{n},\underline{k}}^\text{inst} & =
  \prod_{i=0}^{p-1} \frac{1}{k_i!} \frac{[-\epsilon_{12}]^{k_{i}}}{[-\epsilon_{1,2}]^{k_{i}}}
  \oint \prod_{i=0}^{p-1} \prod_{a = 1}^{k_i} \frac{d \phi_{i,a}}{2 \pi \im} \, 
  \frac{
  \prod_{a = 1,\ldots, k_i}^{b = 1,\ldots,k_{i+1}}
  \mathscr{S}(\phi_{i+1,b;i,a} + m_{i,i+1})}{\prod_{a \neq b}^{k_i} \mathscr{S}(\phi_{i,ab} )}
  \nonumber \\
  & \hspace{7em} \times
  \prod_{i=0}^{p-1} \prod_{a = 1}^{k_i}
  \frac{P_{i-1}(\phi_{i,a} + m_{i-1,i}) \widetilde{P}_{i+1}(\phi_{i,a} + \epsilon_{12} - m_{i,i+1})}{P_i(\phi_{i,a}) \widetilde{P}_i(\phi_{i,a} + \epsilon_{12})}  
  \, .
  \label{eq:Z_inst_cyc_quiv}
 \end{align}
 The case with $p = 1$ is reduced to $\mathcal{N} = 2^*$ theory in four dimensions, which consists of the vector multiplet and the adjoint hypermultiplet: 
 \begin{align}
  \widehat{A}_0: \qquad
  \begin{tikzpicture}[baseline=(current  bounding  box.center),thick]
   \draw [->-] (0,0) arc (-90:270:.5);
   \filldraw[fill=white,draw=black] (0,0) circle (.2);
   \node at (0,-.7) {$\rU(n)$};
  \end{tikzpicture}
 \end{align}
 We remark that, in this case, we cannot turn on the flavor node due to the convergence of the partition function, similarly to~\eqref{eq:AF_condition}.
 The contour integral formula for the instanton partition function is given as follows:
\begin{align}
 Z_{n,k}^\text{inst}
 & = \frac{1}{k!} \frac{[-\epsilon_{12}]^k}{[-\epsilon_{1,2}]^k} 
 \oint \prod_{a=1}^k \frac{d \phi_a}{2 \pi \im} \frac{P(\phi_a + m) \widetilde{P}(\phi_a + \epsilon_{12} - m)}{P(\phi_a) \widetilde{P}(\phi_a + \epsilon_{12})}
 \frac{\prod_{1 \le a, b \le k} \mathscr{S}(\phi_{ab} + m)}{\prod_{1 \le a \neq b \le k} \mathscr{S}(\phi_{ab})}  
 \, ,
 \label{eq:LMNS_formula_2*}
\end{align}
where we denote the adjoint mass parameter by $m = m_\text{adj}$, and $\mu = \np^{m}$.
This integral becomes trivial in the limit $m \to 0$, corresponding to enhancement of supersymmetry from $\mathcal{N} = 2$ to $\mathcal{N} = 4$.

The gauge polynomial part in the integral is rewritten with 5d/6d convention as%
\begin{align}
 \frac{P(\phi + m) \widetilde{P}(\phi + \epsilon_{12} - m)}{P(\phi) \widetilde{P}(\phi + \epsilon_{12})}
 & = \mu^{-n}
 \prod_{\alpha = 1}^n \frac{[\phi - \mathsf{a}_\alpha + m][\phi - \mathsf{a}_\alpha + \epsilon_{12} - m]}{[\phi - \mathsf{a}_\alpha][\phi - \mathsf{a}_\alpha + \epsilon_{12}]}
 \nonumber \\
 & = \mu^{-n}
 \prod_{\alpha = 1}^n \mathscr{S}_{34}(\phi - \mathsf{a}_\alpha)
 \, ,
\end{align}
where we define a generic $\mathscr{S}$-function:\index{S-function@$\mathscr{S}$-function}
\begin{align}
 \mathscr{S}_{ij}(\phi) = \frac{[\phi - \epsilon_{i,j}]}{[\phi][\phi - \epsilon_{ij}]}
 \, ,
 \label{eq:Sij_fn}
\end{align}
with
\begin{subequations}
\begin{align}
 (\epsilon_1, \epsilon_2, \epsilon_3, \epsilon_4)
 = (\epsilon_1, \epsilon_2, -m, m - \epsilon_{12})
 \, ,
 \label{eq:epsilon1234}
\end{align}
\begin{align}
 \epsilon_{ij} = \epsilon_i + \epsilon_j
 \, , \qquad
 [\phi - \epsilon_{i,j}] = [\phi - \epsilon_i][\phi - \epsilon_j]
 \, , \quad \text{etc.}
\end{align}
\end{subequations}
This $\mathscr{S}$-function obeys the reflection formula similarly to~\eqref{eq:S_func_reflection}:
\begin{align}
 \mathscr{S}_{ij}(\phi) = \mathscr{S}_{ij}(\epsilon_{ij} - \phi)
\end{align}
except at the poles at $\phi = 0, \epsilon_{ij}$.
In this convention, the original $\mathscr{S}$-function~\eqref{eq:S_func_def} is denoted by $\mathscr{S}(z) = \mathscr{S}_{12}(z)$.
Then, we obtain a more symmetric form of  the contour integral~\eqref{eq:LMNS_formula_2*}:
\begin{align}
 Z_{n,k}^\text{inst}
 & =
 \frac{\mu^{-kn}}{k!}
 \frac{[-\epsilon_{12}]^k}{[-\epsilon_{1,2}]^k} \mathscr{S}_{12}(m)^k
 \oint \prod_{a=1}^k \frac{d \phi_a}{2 \pi \im}
 \prod_{\alpha = 1}^n \mathscr{S}_{34}(\phi_a - \mathsf{a}_\alpha)
 \prod_{a \neq b}^k \frac{\mathscr{S}_{12}(\phi_{ab} + m)}{\mathscr{S}_{12}(\phi_{ab})}
 \nonumber \\
 & =
 \frac{(-\mu^{-n})^k}{k!}
 \oint \prod_{a=1}^k \frac{d \phi_a}{2 \pi \im}
 \prod_{\alpha = 1}^n \mathscr{S}_{34}(\phi_a - \mathsf{a}_\alpha)
 \prod_{1 \le a, b \le k}
 \frac{[\phi_{ab} - \epsilon_{12, 23, 31}]}{[\phi_{ab} - {\epsilon_{1,2,3,4}}]}
 [\phi_{(a \neq b)}]
 \, ,
\label{eq:LMNS_formula_2*_sym}
\end{align}
where the last factor in the second line $[\phi_{(a \neq b)}]$ means the product over $1 \le a \neq b \le k$, so that it gives rise to the Haar measure of $\rU(K)$.
Due to the reflection formula for the $\mathscr{S}$-function~\eqref{eq:S_func_reflection}, we have
\begin{align}
 & \prod_{a \neq b}^k \mathscr{S}_{12}(\phi_{ab} + m)
 = \prod_{a \neq b}^k \mathscr{S}_{12}(\phi_{ab} + \epsilon_{12} - m)
 \nonumber \\
 = &
 \prod_{a \neq b}^k \mathscr{S}_{12}(\phi_{ab} - \epsilon_3)
 = \prod_{a \neq b}^k \mathscr{S}_{12}(\phi_{ab} - \epsilon_4)
 \, .
\end{align}
As mentioned in \S\ref{sec:adjoint_bundle}, the adjoint mass $m = m_\text{adj}$ is interpreted as the equivariant parameter for the transversal spacetime rotation.
For example, the partition function for $\widehat{A}_0$ quiver gauge theory with $\rU(1)$ gauge symmetry~\eqref{eq:A0_inst_fn} has the following expression in terms of the $\mathscr{S}_{34}$-function:
\begin{align}
 Z_{1,k}^\text{inst}[\lambda]
 = q_3^{|\lambda|} \prod_{s \in \lambda} \mathscr{S}_{34}(q_1^{\ell(s)+1} q_2^{-a(s)})
 = q_3^{|\lambda|} \prod_{s \in \lambda} \mathscr{S}_{34}(q_1^{-\ell(s)} q_2^{a(s)+1})
 \, ,
\end{align} 
which suggests the 34-surface $\mathbb{C}_{\epsilon_3} \times \mathbb{C}_{\epsilon_4}$ plays a role in this configuration.
It has been recently discussed that, embedding $\widehat{A}_0$ quiver gauge theory within eight-dimensional setup, which is called the {\em gauge origami} (ゲージ折り紙)~\cite{Nekrasov:2015wsu}, $(\epsilon_i)_{i = 1, 2, 3, 4}$ is interpreted as the equivariant parameter of $\Spin(8)$.
\index{gauge origami}
Then, the zero sum relation
\begin{align}
 \sum_{i = 1}^4 \epsilon_i = 0
\end{align}
is a consequence of the Calabi--Yau condition for a (complex) four-fold.
See also \S\ref{sec:origami} and \S\ref{sec:qq_ch_geom}.

\subsection{Quiver Cartan matrix}\label{sec:quiv_Cartan_matrix}

\subsubsection{Half Cartan matrix}

Given a quiver $\Gamma = (\Gamma_0, \Gamma_1)$, we define the {\em mass deformation of (half) $q$-Cartan matrix}:\index{Cartan matrix!half---}
\begin{align}
 c_{ij}^+ = \delta_{ij} - \sum_{e : i \to j} \mathbf{M}_e^\vee
 \quad \text{for} \quad
 i, j \in \Gamma_0
 \, ,
\end{align}
where $(\mathbf{M}_e)_{e \in \Gamma_1}$ is the line bundle defined in \S\ref{sec:quiver_bundles} with the character
\begin{align}
 \ch_\mathsf{T} c_{ij}^+ = \delta_{ij} - \sum_{e:i \to j} \mu_e^{-1}
\, .
\end{align}
We will use $(c_{ij}^+)_{i,j\in\Gamma_0}$ also for the character as long as no confusion.
With this $q$-Cartan matrix, we can write down the total tangent bundle~\eqref{eq:TM}
in a compact form:
\begin{align}
 T_\lambda \mathfrak{M}
 = \frac{\wedge \mathbf{Q}_1^\vee}{\wedge \mathbf{Q}_2} \sum_{(i,j) \in \Gamma_0 \times \Gamma_0} \mathbf{X}_i^\vee \, c_{ij}^{+\vee} \, \mathbf{X}_j
 \, .
 \label{eq:TM_Cartan}
\end{align}
In this expression, the (anti)fundamental matter contribution is reproduced as a bifundamental factor associated with a flavor node, which is a frozen gauge node with the zero coupling constant.

\subsubsection{Full Cartan matrix}

We also define the other half of the Cartan matrix:
\begin{align}
  c_{ij}^- & = \det \mathbf{Q}^\vee \cdot c_{ji}^{+\vee}
 \, ,
 \label{eq:half_Cartan-}
\end{align}
which is combined into the full Cartan matrix: \index{Cartan matrix!full---}
\begin{align}
 c_{ij}
 = c_{ij}^+ + c_{ij}^-
 = \left( 1 + q^{-1} \right) \, \delta_{ij}
 - \sum_{e:i \to j} \mu_e^{-1}
 - \sum_{e:j \to i} \mu_e q^{-1}
\, .
\end{align}
Roughly speaking, from the gauge theory point of view, the half Cartan matrix describes 4 SUSY quiver theory, and the full Cartan matrix corresponds to 8 SUSY quiver theory.

We denote the degree-$n$ Cartan matrix obtained through the Adams operation by \index{Adams operation}
\begin{align}
 c_{ij}^{[n]} = c_{ij}\Big|_{(q,\mu_e) \to (q^n, \mu_e^n)}
 \, .
\end{align}
Then the limit $n \to 0$ corresponds to the classical quiver Cartan matrix:
\index{Cartan matrix!classical---}
\begin{align}
 c_{ij}^{[0]} := \lim_{n \to 0} c_{ij}^{[n]} = 2 \, \delta_{ij} - \#(e:i \to j) - \#(e:j \to i)
 \, .
 \label{eq:classical_quiv_Cartan_mat}
\end{align}
The $q$-Cartan matrix is not a symmetric matrix in the current convention, but obeys the following reflection relation:
\begin{align}
 c_{ji}^{[n]} = q^{-n} c_{ij}^{[-n]}
 \, .
 \label{eq:quiv_Cartan_reflection}
\end{align}
See \S\ref{sec:fractional_quiver} for the $q$-deformation of the non-simply-laced case (fractional quiver).

\subsubsection{Classification of UV complete theory}

The quiver Cartan matrix provides a useful framework for classification of UV complete theories, which are well defined for all energy scales.
Typical examples are asymptotically free theories and conformal field theories.

The quiver gauge theories discussed here consist of the vector multiplet and the hypermultiplet in the (anti/bi)fundamental representation.
In this case, the following condition is required for UV completion:
\begin{align}
 n_i^\text{f} + n_i^\text{af} \le \sum_{j \in \Gamma_0} c_{ij}^{[0]} \, n_j
 \, .
 \label{eq:UV_completion}
\end{align}
The equality holds for superconformal theories.
The condition~\eqref{eq:UV_completion} is equivalent to the convergence condition of the contour integral~\eqref{eq:quiver_LMNS}, such that the integrand is not singular (not diverging) at $\phi_{i,a} \to \infty$.
See \cite{Nekrasov:2012xe,Bhardwaj:2013qia,Bhardwaj:2015xxa,Bhardwaj:2020gyu} for more details on the classification (even in higher dimensions).

\section{Quiver variety}\label{sec:quiver_variety}



In this Section, we provide a quiver description of the instanton moduli space, which is formulated in general as the {\em quiver variety}~\cite{Nakajima:1994qu,Nakajima:1998DM}.
See also \cite{Kirillov:2016} for the monograph on this topic.
We also discuss its relation to quiver gauge theory mentioned above.
\index{quiver!---variety}

Given a quiver $\Gamma = (\Gamma_0, \Gamma_1)$, we define the dimension vectors $(\underline{n},\underline{k}) = (n_i,k_i)_{i = 1,\ldots,\rk \Gamma}$, and the corresponding vector spaces
\begin{align}
 N = \bigoplus_{i \in \Gamma_0} N_i
 \, , \qquad
 K = \bigoplus_{i \in \Gamma_0} K_i
 \, , \qquad
 N_i = \bC^{n_i}
 \, , \qquad
 K_i = \bC^{k_i}
 \, ,
\end{align}
with the automorphism group
\begin{align}
 \rU(K) = \prod_{i \in \Gamma_0} \rU(k_i)
 \, , \qquad
 \rU(N) = \prod_{i \in \Gamma_0} \rU(n_i)
 \, .
\end{align}
We define
\begin{align}
 X_\Gamma & = \left( \bigoplus_{i \in \Gamma_0} \Hom(N_i, K_i) \oplus \Hom(K_i, N_i) \right)
 \oplus
 \left( \bigoplus_{e \in \Gamma_1} \Hom(K_i, K_j) \oplus \Hom(K_j,K_i) \right)
 \label{eq:quiv_var_X}
\end{align}
with the coordinates\\[-.7em]
\begin{subequations}\label{eq:quiv_var_coordinates}
\begin{minipage}{.5\textwidth}
 \begin{align}
  B_{e:i \to j} & \in \Hom(K_i,K_j)
  \, , \\
  \overline{B}_{e:i \to j} & \in \Hom(K_j,K_i)
  \, , 
 \end{align}
\end{minipage}
\begin{minipage}{.5\textwidth}
 \begin{align}
  I_i & \in \Hom(N_i,K_i)
  \, , \\
  J_i & \in \Hom(K_i,N_i)
  \, .
 \end{align}
\end{minipage}
\end{subequations}

\noindent
The automorphism group action on these coordinates is given by
\begin{subequations}
\begin{align}
 \rU(K) : &
 \left(
 \left(B_e, \overline{B}_e \right)_{e:i \to j},
 \left( I_i, J_i \right)_{i \in \Gamma_0}
 \right)
 \longmapsto 
 \left(
 \left( g_j B_e g_i^{-1},  g_i \overline{B}_e g_j^{-1} \right)_{e:i \to j},
 \left( g_i I_i, J_i g_i^{-1} \right)_{i \in \Gamma_0}
 \right)
 \, , \\
 \rU(N) : &
 \left(
 \left(B_e, \overline{B}_e \right)_{e:i \to j},
 \left( I_i, J_i \right)_{i \in \Gamma_0}
 \right)
 \longmapsto 
 \left(
 \left( B_e, \overline{B}_e \right)_{e:i \to j},
 \left( I_i h_i^{-1}, h_i J_i \right)_{i \in \Gamma_0}
 \right)
 \, .
\end{align}
\end{subequations}
We then define the moment maps, 
\begin{subequations}\label{eq:quiv_var_moment_map}
 \begin{align}
  \mu_{\bR} & =
  \left(
  I_i I_i^\dag - J_i^\dag J_i
  - \sum_{e:i \to j} (B_e^\dag B_e - \overline{B}_e \overline{B}_e^\dag)
  + \sum_{e:j \to i} (B_e B_e^\dag - \overline{B}_e^\dag \overline{B}_e)
  \right)_{i \in \Gamma_0}
  \, , \\
  \mu_{\bC} & =
  \left(
  I_i J_i
  - \sum_{e:i \to j} \overline{B}_e B_e + \sum_{e:j \to i} B_e \overline{B}_e
  \right)_{i \in \Gamma_0}
  \, ,
 \end{align}
\end{subequations}
such that $\mu_{\bR}, \mu_{\bC}: K \to K$.
We use the vector notation ${\mu} = (\mu_{\bR}, \real \mu_{\bC}, \imag \mu_{\bC})$, and define the parameters $\zeta \in \bR^3 \otimes \mathfrak{u}_K^*$.
Then the quiver variety is defined as a hyper-K\"ahler quotient, which is isomorphic to the GIT quotient, with the stability condition:\index{stability condition}
\begin{align}
 \mathfrak{M}_\Gamma = \mathfrak{M}_{N,K}
 = {\mu}^{-1}(\zeta) /\!\!/\!\!/ \rU(K)
 \cong \mu_\bC^{-1}(\zeta_\bC) /\!\!/ \GL(K)
 \, .
 \label{eq:quiv_var}
\end{align}
The stability is now taken into account with the remaining parameter $\zeta_\bR$.
We are in particular interested in the case with $\zeta_\bC = 0$.

\subsection{ADHM quiver}
\index{ADHM construction}

As discussed in \S\ref{sec:ADHM_construction}, the ADHM construction of instantons on $\mathcal{S} = \bC^2$ is described using the ADHM variables $(B_{1,2}, I, J)$.
From the definition \eqref{eq:ADHM_data}, $B_{1,2}$ is in the adjoint representation of $\rU(k)$, and $I$ and $J$ are in the fundamental/antifundamental representation of $\rU(k)$, and in the antifundamental/fundamental representation of $\rU(n)$.
Hence, we have a quiver diagram for the ADHM variables (ADHM quiver) in the 4 SUSY convention:
\begin{align}
  \begin{tikzpicture}[baseline=(current  bounding  box.center),thick,scale=1.2]
   \draw [->-] (.1,0) arc (0:360:.75);
   \draw [-<-] (-.05,0) arc (0:360:.6);
   \node at (-2,0) {$B_{1,2}$};
   \filldraw[fill=white,draw=black] (0,0) circle (.2) node {$K$};
   \draw [-<-] (20:.2) -- ++(2,0); 
   \draw [->-] (-20:.2) -- ++(2,0); 
   \draw (2.17,.2) rectangle node {$N$} ++(.4,-.4);
   \node at (1.2,.5) {$I$};
   \node at (1.2,-.5) {$J$};   
  \end{tikzpicture}
 \label{eq:McKay_quiver_A0}
\end{align}
This is identified with $\widehat{A}_0$ quiver with the gauge node $K$ and the flavor node $N$.
The reason of this coincidence is explained in the following.

\subsection{ADHM on ALE space}\label{sec:ADHM_ALE}
\index{ADHM construction!ALE space}

So far, we have fixed the spacetime manifold $\mathcal{S} = \bC^2$.
We now consider a family of four-manifolds, called the asymptotically locally Euclidean (ALE) spaces, and gauge theory defined on it. \index{ALE space}
Let $\Gamma$ be a finite subgroup of $\SU(2)$, to which we can apply the ADE classification, summarized in Tab.~\ref{tab:McKay_corresp}.
The ALE space is obtained through resolution of the orbifold singularity of the form $\bC^2/\Gamma$, denoted by $\mathcal{S}_\Gamma = \widetilde{\bC^2/\Gamma}$~\cite{Eguchi:1978xp,Gibbons:1979zt,Kronheimer:1989zs}.
Then, it turns out that the ADHM construction on the ALE space has a quiver description based on the McKay quiver~\cite{Kronheimer:1990MA}, which is the affine ADE quiver associated with the finite subgroup $\Gamma = ADE$, obtained through the {\em McKay correspondence}~\cite{McKay:1981}.
See also~\cite{Douglas:1996sw,Johnson:1996py} for its string theory perspective.

For example, the McKay quiver for the A-type ALE space $\mathcal{S}_{A_{p-1}} = \widetilde{\bC^2/\bZ_{p}}$ is a cyclic quiver with $p$ nodes, namely $\widehat{A}_{p-1}$ quiver (See Tab.~\ref{tab:McKay_corresp} and \eqref{eq:cyclic_quiver_diagram_ALE}).
The simplest case is $p = 1$, which gives rise to $\bC^2/\bZ_1 = \bC^2$.
In this case, the McKay quiver is $\widehat{A}_0$ quiver shown in~\eqref{eq:McKay_quiver_A0}.

\begin{table}[t]
 \begin{center}
  \renewcommand{\arraystretch}{1.5}  
  \begin{tabular}{clcc}\hline\hline
   \multicolumn{2}{c}{Finite subgroup of $\SU(2)$} & ADE class & McKay quiver $\widehat\Gamma$ \\\hline
   cyclic group & $\mathbb{Z}_{p+1}$ & $A_p$ & \dynkin{A}[1]{} \\ 
   binary dihedral group & $\mathbb{BD}_{p-2}$ & $D_p$ & \dynkin{D}[1]{} \\
   binary tetrahedral group & $\mathbb{BT}$ & $E_6$ & \dynkin{E}[1]{6} \\
   binary octahedral group & $\mathbb{BO}$ & $E_7$ & \dynkin{E}[1]{7} \\
   binary icosahedral group & $\mathbb{BI}$ & $E_8$ & \dynkin{E}[1]{8} \\
   \hline\hline
  \end{tabular}
  \renewcommand{\arraystretch}{1}  
 \end{center}
 \caption{ADE classification of finite subgroup of $\SU(2)$ and McKay quiver.}
 \label{tab:McKay_corresp}
\end{table}

\subsubsection{McKay correspondence}
\index{McKay correspondence}

Let $\rho_\mathbf{Q}$
be the doublet (fundamental) representation of $\Gamma \subset \SU(2)$, and let $c_{ij} = 2 \delta_{ij} - a_{ij}$ be the Cartan matrix of the McKay quiver denoted by $\widehat{\Gamma} = (\widehat{\Gamma}_0,\widehat{\Gamma}_1)$ with the adjacency matrix $(a_{ij})$ of the quiver.
The statement of the McKay correspondence is as follows~\cite{McKay:1981}:
\begin{align}
 \rho_\mathbf{Q} \otimes \rho_i = \bigoplus_{j \in \widehat{\Gamma}_0} a_{ij} \cdot \rho_j
\end{align}
where $(\rho_i)_{i \in \widehat{\Gamma}_0}$ is the irreducible representation of $\Gamma$ associated with the node $i \in \widehat{\Gamma}_0$.
The extended node, also called the affine node, is depicted as a white node in Tab.~\ref{tab:McKay_corresp}, which corresponds to the trivial representation of $\Gamma$.

\subsubsection{ADHM moduli space}

We consider $k$-instanton sector in $G$-gauge theory with $G = \rU(n)$ on the ALE space $\mathcal{S}_\Gamma$.
The basic idea is to consider the $\Gamma$-invariant sector of the instanton configuration in $\mathcal{S}$, which is a universal cover of $\mathcal{S}_\Gamma$.%
\footnote{%
A similar construction of vortices on the orbifold $\mathbb{C}/\mathbb{Z}_p$ is presented in \cite{Kimura:2011wh}.
}

We define the vector spaces $(N,K) = (\bC^n, \bC^k)$ as before~\eqref{eq:ADHM_vect_sp}, which are decomposed with respect to the irreducible representations of the finite group $\Gamma$:
\begin{align}
 N = \bigoplus_{i \in \widehat{\Gamma}_0} N_i \otimes \rho_i
 \, , \qquad
 K = \bigoplus_{i \in \widehat{\Gamma}_0} K_i \otimes \rho_i
 \, .
 \label{eq:Gamma_decomposition_N_K}
\end{align}
In this case, the dimension vectors are given by the multiplicity of each representation, $(\underline{n}, \underline{k}) = (n_i,k_i)_{i \in \widehat{\Gamma}_0}$.
The automorphism group is given by
\begin{align}
 \rU(N) = \prod_{i \in \widehat{\Gamma}_0} \rU(N_i)
 \, , \qquad
 \rU(K) = \prod_{i \in \widehat{\Gamma}_0} \rU(K_i)
 \, .
\end{align}
This is interpreted as gauge symmetry breaking $\rU(n) \to \prod_i \rU(n_i)$ due to the holonomy effect at the infinity of the ALE space, given as the lens space $S^3/\Gamma$.

We define the ADHM variables for the ALE space $\mathcal{S}_\Gamma$:
\begin{align}
 X_\Gamma =
 \left(
 \Hom(K,K) \oplus \Hom(K,K) \oplus \Hom(N,K) \oplus (K,N)
 \right)_\Gamma
 \, ,
\end{align}
which is a $\Gamma$-invariant sector of $X$ defined in~\eqref{eq:ADHM_var_space}.
The first two terms form a doublet $\rho_\mathbf{Q}$ in $\SU(2)$,%
\footnote{%
The ADHM variable $(B_1, B_2) \in \Hom(K,K) \otimes \Hom(K,K)$ behaves in the same way as the complex coordinate $(z_{1,2}) \in \mathbb{C}^2$ under $\SU(2)$ transformation as a doublet $\rho_\mathbf{Q}$, which is a half of the Lorentz transformation, $\operatorname{Spin}(4) = \SU(2) \times \SU(2)$.
}
so that we obtain
\begin{align}
 \left(
 \rho_\mathbf{Q} \otimes \Hom(K,K)
 \right)_\Gamma
 & =
 \rho_\mathbf{Q} \otimes
 \bigoplus_{i,i' \in \widehat{\Gamma}_0} \Hom(K_i, K_{i'}) \otimes
 \left( \Hom(\rho_i, \rho_{i'}) \right)_\Gamma
 \nonumber \\ 
 & =
 \bigoplus_{i,i' \in \widehat{\Gamma}_0} \Hom(K_i, K_{i'}) \otimes
 \bigoplus_{i,i',j \in \widehat{\Gamma}_0} a_{ij} \left( \Hom(\rho_j,\rho_{i'}) \right)_\Gamma
 \nonumber \\
 & =
 \bigoplus_{i,j} a_{ij} \Hom(K_i, K_j)
\end{align}
where we use the relation $\left( \Hom(\rho_i,\rho_{j}) \right)_\Gamma = \delta_{ij}$ as a consequence of Schur's lemma.
Similarly, we obtain
\begin{align}
 \left(
 \Hom(N,K) \oplus (K,N)
 \right)_\Gamma
 & =
 \bigoplus_{i \in \widehat{\Gamma}_0} \Hom(N_i, K_i) \oplus \Hom(K_i, N_i)
 \, .
\end{align}
This shows that the ADHM moduli space for the ALE space $\mathcal{S}_\Gamma$ is given as the quiver variety \eqref{eq:quiv_var} with respect to the McKay quiver, namely the affine ADE quiver $\widehat{\Gamma}$. \index{quiver!---variety}
In this case, the parameter $\zeta \in (\mathbb{R}^3)^{|\widehat{\Gamma}_0|}$ is incorporated with the condition $\zeta \cdot \delta = \sum_{i \in \widehat{\Gamma}_0} \zeta_i \dim \rho_i =  0$, where we denote the positive primitive imaginary root of $\widehat{\Gamma}$ by $\delta$.

\subsubsection{A-type ALE space}

We consider the A-type ALE space $\mathcal{S}_{A_{p-1}} = \widetilde{\bC^2/\mathbb{Z}_p}$ as an example.
Let $a$ be the generator of the cyclic group $\mathbb{Z}_p = \mathbb{Z}/p\mathbb{Z}$ with $a^p = 1$:
$\mathbb{Z}_p = \left\langle \, a \mid a^p = 1 \, \right\rangle = \{ 1,a,\ldots,a^{p-1} \}$.
The (one-dimensional) irreducible representation is given by $\rho_i(a) = \omega^{i}$ for $i=0,\ldots,p-1$, where we denote the primitive $p$-th root of unity by $\omega = \exp \left( 2 \pi \im / p \right)$.
The doublet representation is given by
\begin{align}
 \rho_\mathbf{Q}(a) =
 \begin{pmatrix}
  \omega & 0 \\ 0 & \omega^{-1} 
 \end{pmatrix}
 \in \SU(2)
 \, .
\end{align}
Hence, we obtain the orbifold action $\Gamma = \mathbb{Z}_p$ on the coordinate $(z_1, z_2) \in \mathbb{C}^2$ as follows:
\begin{align}
 \Gamma :
 \quad
 \begin{pmatrix}
  z_1 \\ z_2
 \end{pmatrix}
 \ \longmapsto \
 \begin{pmatrix}
  \omega & 0 \\ 0 & \omega^{-1} 
 \end{pmatrix}
 \begin{pmatrix}
  z_1 \\ z_2
 \end{pmatrix}
 \, .
 \label{eq:Zp_action_Q}
\end{align}
In this case, the McKay quiver is $\widehat{A}_{p-1}$, 
and the quiver diagram is given similarly to the cyclic quiver~\eqref{eq:cyclic_quiver_diagram}:
\begin{align}
 \widehat{A}_{p-1} \ (p = 5) : \qquad
 \begin{tikzpicture}[baseline=(current  bounding  box.center),thick]
  \foreach \x in {0,1,2,3,4} {
  \draw [->-] (95+\x*72:1.6) arc (95+\x*72:150+\x*72:1.6);
  \draw [-<-] (95+\x*72:1.4) arc (95+\x*72:149+\x*72:1.4);  
  \draw [->-] (87+\x*72:1.7) -- ++(90+\x*72:1);
  \draw [-<-] (93+\x*72:1.7) -- ++(90+\x*72:1);  
  \filldraw [draw=black,fill=white] ($(90+\x*72:2.5)+(180+\x*72:0.3)$) -- ++(90+\x*72:0.6) -- ++(\x*72:.6) -- ++(-90+\x*72:.6) -- cycle;
  \node at (90+\x*72:2.8) {$N_\x$};
  \filldraw [draw=black,fill=white] (90+\x*72:1.5) circle (.35) node {$K_\x$};
  }
 \end{tikzpicture}
 \label{eq:cyclic_quiver_diagram_ALE}
\end{align}
where edges connecting $K_i$'s are $(B_e, \overline{B}_e)$, those connecting $(N_i, K_i)$ are $(I_i, J_i)$ as in \eqref{eq:quiv_var_coordinates}.
These elements are concisely combined into the block matrix form:
\begin{subequations}\label{eq:ADHM_variables_A}
\begin{align}
 B =
 \begin{pmatrix}
  0 & & & & B_{p-1} \\
  B_0 & 0 & & & \\
  & B_1 & 0 & & \\
  & & \ddots & \ddots & \\
  & & & B_{p-2} & 0
 \end{pmatrix}
 \, , \qquad
 \overline{B} =
 \begin{pmatrix}
  0 & \overline{B}_0 & & & \\
  & 0 & \overline{B}_1 & & \\
  & & \ddots & \ddots & \\
  & & & 0 & \overline{B}_{p-2} \\
  \overline{B}_{p-1} & & & & 0
 \end{pmatrix}
\end{align}
\begin{align}
 I & =
 \begin{pmatrix}
  I_0 & & \\ & \ddots & \\ & & I_{p-1}
 \end{pmatrix}
 \, , \qquad
 J =
 \begin{pmatrix}
  J_0 & & \\ & \ddots & \\ & & J_{p-1}
 \end{pmatrix}
 \, ,
\end{align}
\end{subequations}
where each of block elements is given by $B_{i} \in \Hom(K_i,K_{i+1})$, $\overline{B}_{i} \in \Hom(K_{i+1},K_{i})$, $I_i \in \Hom(N_i, K_i)$, $J_i \in \Hom(K_i, N_i)$ with the index defined modulo $p$, $i \equiv i+p$.
Then, the moment maps~\eqref{eq:quiv_var_moment_map} are described in the same form as the ordinary ones~\eqref{eq:moment_maps}:\\[-1.5em]
\begin{subequations}
\begin{align}
 \mu_\bR & = \left( I_i I_i^\dag - J_i^\dag J_i + B_{i-1} B_{i-1}^\dag - B_i^\dag B_i + \overline{B}_i \overline{B}_i^\dag - \overline{B}_{i+1}^\dag \overline{B}_{i+1} \right)_{i \in \widehat{\Gamma}_0}
 \nonumber \\
 & = II^\dag - J^\dag J + [B, B^\dag] + [\overline{B},\overline{B}^\dag]
 \, , \\
 \mu_\bC & = \left( I_i J_i + B_{i-1} \overline{B}_{i-1} - \overline{B}_i B_i \right)_{i \in \widehat{\Gamma}_0}
 \nonumber \\
 & = IJ + [ B, \overline{B}]
 \, ,
\end{align}
\end{subequations}
with the identification $(B_1, B_2) \leftrightarrow (B, \overline{B})$.
Namely, this choice of the block matrix structure properly implements the $\mathbb{Z}_p$-projection of the ADHM variables.

\subsubsection{Equivariant integral over quiver variety}

We apply the equivariant localization to the instanton moduli space of the A-type ALE space, which is given as the quiver variety associated with the McKay quiver~\cite{Fucito:2004ry,Fujii:2005dk}. \index{equivariant!---localization}
The strategy is again to focus on the $\Gamma$ invariant sector of the instanton moduli space.

Let $\Gamma = \mathbb{Z}_p$.
Since the $\Gamma$ action on the $\SU(2)$ doublet is given as~\eqref{eq:Zp_action_Q}, it is given for the equivariant parameters as
\begin{align}
 \Gamma : \ (q_1, q_2, \np^{\mathsf{a}_\alpha})
 \ \longmapsto \
 \left( \omega q_1, \omega^{-1} q_2, \omega^{r_\alpha} \np^{\mathsf{a}_\alpha} \right)
 \label{eq:Gamma_action_equiv_parameters}
\end{align}
where we define the index map $r : (1,\ldots,n) \mapsto (0,\ldots,p-1)$. 
Recalling the decomposition of $N$ into the irreducible representations of $\Gamma$~\eqref{eq:Gamma_decomposition_N_K}, the $\Gamma$ action on $i$-th sector is given by $\Gamma: \np^{\mathsf{a}_{i,\alpha}} \to \omega^i \np^{\mathsf{a}_{i,\alpha}}$, so that the multiplicity is given by $n_i = \#( r_\alpha = i , \alpha = 1, \ldots, n)$.
Then, we obtain the $\Gamma$ invariant part of the vector multiplet character~\eqref{eq:ch_V_inst} as follows:
\begin{align}
 \ch_\mathsf{T} \mathbf{V}^\text{inst}_\Gamma
 = - \sum_{\alpha,\alpha' = 1,\ldots,n} \np^{\mathsf{a}_{\alpha\alpha'}}
 \left(
 \sum_{s \in \lambda_\alpha} q_1^{\ell_{\alpha'}(s)} q_2^{- a_\alpha(s) - 1}
 \delta_{h_{\alpha{\alpha'}}(s), r_{\alpha'\alpha}}^{(p)}
 + \sum_{s \in \lambda_{\alpha'}} q_1^{- \ell_\alpha(s) - 1} q_2^{a_{\alpha'}(s)}
 \delta_{h_{{\alpha'}\alpha}(s), r_{\alpha\alpha'}}^{(p)}
 \right)
\end{align}
where $h_{\alpha{\alpha'}}(s)$ is the relative hook length~\eqref{eq:relative_hook_length}, and we define $r_{\alpha\alpha'} = r_\alpha - r_{\alpha'}$ (mod $p$) with 
\begin{align}
 \delta_{i,j}^{(p)} =
 \begin{cases}
  1 & (i \equiv j, \text{mod} \ p) \\
  0 & (i \not \equiv j, \text{mod} \ p)
 \end{cases} 
\end{align}
The vector multiplet contribution to the instanton partition function is obtained with the index formula.
For example, in the equivariant cohomology (4d gauge theory) convention, it is given by
\begin{align}
 Z_\Gamma^\text{vec,inst}
 & =
 \prod_{\alpha,\alpha' = 1,\ldots,n}
 \left[
 \prod_{s \in \lambda_\alpha}
 \frac{1}{(\mathsf{a}_{\alpha\alpha'} + \epsilon_1 \ell_{\alpha'}(s) - \epsilon_2(a_\alpha(s) + 1))^{\delta_{h_{\alpha{\alpha'}}(s), r_{\alpha'\alpha}}^{(p)}}}
 \right.
 \nonumber \\
 & \hspace{8em} \left. \times
 \prod_{s \in \lambda_{\alpha'}}
 \frac{1}{(\mathsf{a}_{\alpha\alpha'} - \epsilon_1(\ell_\alpha(s) + 1) + \epsilon_2 a_{\alpha'}(s))^{\delta_{h_{{\alpha'}\alpha}(s), r_{\alpha\alpha'}}^{(p)}}}
 \right]
 \, .
\end{align}
We remark that this expression is also obtained through the root of unity limit of the 5d gauge theory result on $\mathbb{C}^2 \times S^1$~\cite{Kimura:2011zf,Kimura:2011gq}.%
\footnote{%
A similar reduction is discussed in~\cite{Kimura:2019xzj}.
}
The hypermultiplet contributions from the $\Gamma$ invariant sector of the instanton moduli space are similarly obtained.

We can also derive the instanton moduli space based on the ADHM variables~\eqref{eq:ADHM_variables_A}.
In this case, the complexified equivariant action is given by \index{equivariant!---action}
\begin{align}
 \GL(\mathbf{Q}) : \quad (B_i, \overline{B}_i, I_i, J_i)
 & \ \longmapsto \ (q_1^{-1} B_i, q_2^{-1} \overline{B}_i, I_i, q^{-1} J_i)
 \\
 \GL(K) : \quad (B_i, \overline{B}_i, I_i, J_i)
 & \ \longmapsto \ (g_{i+1} B_i g_i^{-1}, g_i \overline{B}_i g_{i+1}^{-1}, g_i I_i, J_i g_i^{-1})
 \\
 \GL(N) : \quad (B_i, \overline{B}_i, I_i, J_i)
 & \ \longmapsto \ (B_i, \overline{B}_i, I_i \nu_i^{-1}, \nu_i J_i)
\end{align}
where $g_i \in \GL(K_i)$, $\nu_i \in \GL(N_i)$ for $i = 0, \ldots, p-1$.
Hence, we obtain the contour integral formula for the instanton partition function as follows:
\begin{align}
 Z_{\underline{n},\underline{k}}^\text{inst}
 & =
 \prod_{i=0}^{p-1} \frac{[-\epsilon_{12}]^{k_i}}{k_i!} \oint_{\mathsf{T}_K} \prod_{i=0}^{p-1} \prod_{a=1}^{k_i} \frac{d \phi_{i,a}}{2 \pi \im} \frac{1}{P_i(\phi_{i,a}) \widetilde{P}_i(\phi_{i,a} + \epsilon_{12})}
 \prod_{a \neq b}^{k_i}
 [\phi_{i,ab}][\phi_{i,ab} - \epsilon_{12}]
 \nonumber \\
 & \hspace{8em} \times
 \prod_{\substack{a = 1,\ldots,k_i \\ b = 1,\ldots,k_{i+1}}}
 \frac{1}{[\phi_{i+1,b;i,a} - \epsilon_1][\phi_{i,a;i+1,b} - \epsilon_2]}
 \label{eq:LMNS_ALE_vec}
\end{align}
where the gauge polynomials are defined as
\begin{align}
 P_i(\phi) = \prod_{\alpha = 1}^{n_i} [\phi - \mathsf{a}_{i,\alpha}]
 \, \qquad
 \widetilde{P}_i(\phi) = \prod_{\alpha = 1}^{n_i} [ - \phi + \mathsf{a}_{i,\alpha}]
 \, .
\end{align}
In general, we can incorporate the hypermultiplets in addition to the vector multiplet contribution similarly to \S\ref{sec:LMNS_matter}.
We remark that this contour integral formula is similar, but still different from $\widehat{A}_{p-1}$ quiver gauge theory partition function~\eqref{eq:Z_inst_cyc_quiv}.

This contour integral formula is also derived from the equivariant characters.
Recalling the $\Gamma$ action on $(q_1,q_2)$~\eqref{eq:Gamma_action_equiv_parameters}, and also $\Gamma: (\np^{\mathsf{a}_{i,\alpha}}, \np^{\phi_{i,a}}) \to (\omega^i \np^{\mathsf{a}_{i,\alpha}}, \omega^i \np^{\phi_{i,a}})$, the $\Gamma$ invariant part of the vector multiplet bundle~\eqref{eq:V_bundle_inst} is given by
\begin{align}
 \mathbf{V}_\Gamma^\text{inst}
 & =
 \sum_{i \in \widehat{\Gamma}_0}
 \left[
 - \det \mathbf{Q}^\vee \cdot \mathbf{K}_i^\vee \mathbf{N}_i
 - \mathbf{N}_i^\vee \mathbf{K}_i
 \right.
 \nonumber \\
 & \hspace{3em}
 \left.
 + ( 1 + \det \mathbf{Q}^\vee) \cdot \mathbf{K}_i^\vee \mathbf{K}_i
 - \mathbf{Q}_1^\vee \cdot \mathbf{K}_i^\vee \mathbf{K}_{i+1}
 - \mathbf{Q}_2^\vee \cdot \mathbf{K}_{i+1}^\vee \mathbf{K}_{i}
 \right]
 \, .
\end{align}
Then, applying the index formula, we obtain the contour integral~\eqref{eq:LMNS_ALE_vec}.

In addition to the vector multiplet, we can similarly incorporate the hypermultiplet contributions.
For example, the adjoint representation contribution is obtained as the $\Gamma$ invariant sector of the expression in \S\ref{sec:adjoint_bundle}:
\begin{align}
 \mathbf{H}_\Gamma^\text{adj,inst}
 & =
 \mathbf{M}_\text{adj} \sum_{i \in \widehat{\Gamma}_0}
 \left[
 \det \mathbf{Q}^\vee \cdot \mathbf{K}_i^\vee \mathbf{N}_i
 + \mathbf{N}_i^\vee \mathbf{K}_i
 \right.
 \nonumber \\
 & \hspace{5.5em}
 \left.
 - ( 1 + \det \mathbf{Q}^\vee) \cdot \mathbf{K}_i^\vee \mathbf{K}_i
 + \mathbf{Q}_1^\vee \cdot \mathbf{K}_i^\vee \mathbf{K}_{i+1}
 + \mathbf{Q}_2^\vee \cdot \mathbf{K}_{i+1}^\vee \mathbf{K}_{i}
 \right]
 \, .
\end{align}
Therefore, the corresponding contour integral formula for $\widehat{A}_0$ quiver theory on $\mathcal{S}_\Gamma$ with $\Gamma = \mathbb{Z}_p$ is given by
\begin{align}
 Z_{\underline{n},\underline{k}}^\text{inst}
 & =
 \prod_{i=0}^{p-1} \frac{1}{k_i!}
 \oint_{\mathsf{T}_K} \prod_{i=0}^{p-1} \prod_{a=1}^{k_i} \frac{d \phi_{i,a}}{2 \pi \im} \frac{P_i(\phi_{i,a} + m) \widetilde{P}_i(\phi_{i,a} + \epsilon_{12} - m)}{P_i(\phi_{i,a}) \widetilde{P}_i(\phi_{i,a} + \epsilon_{12})}
 \nonumber \\
 & \hspace{8em} \times
 \prod_{1 \le a, b \le k_i}
 \frac{[\phi_{i,(a \neq b)}][\phi_{i,ab} - \epsilon_{12}]}{[\phi_{i,ab} + m][\phi_{i,ab} - \epsilon_{12} + m]} 
 \nonumber \\
 & \hspace{8em} \times
 \prod_{\substack{a = 1,\ldots,k_i \\ b = 1,\ldots,k_{i+1}}}
 \frac{[\phi_{i+1,b;i,a} - \epsilon_1 + m][\phi_{i,a;i+1,b} - \epsilon_2 + m]}{[\phi_{i+1,b;i,a} - \epsilon_1][\phi_{i,a;i+1,b} - \epsilon_2]}
 \nonumber \\
 & =
 \prod_{i=0}^{p-1}
 \frac{1}{k_i!}
 \frac{[-\epsilon_{34}]^{k_i}}{[-\epsilon_{3,4}]^{k_{i}}}
 \oint_{\mathsf{T}_K} \prod_{i=0}^{p-1} \prod_{a=1}^{k_i} \frac{d \phi_{i,a}}{2 \pi \im}
 \prod_{\alpha = 1}^{n_i} \mathscr{S}_{34}(\phi_{i,a} - \mathsf{a}_{i,\alpha})
 \nonumber \\
 & \hspace{8em} \times
 \prod_{1 \le a \neq b \le k_i} \mathscr{S}_{34}(\phi_{i,ab})^{-1}
 \prod_{\substack{a = 1,\ldots,k_i \\ b = 1,\ldots,k_{i+1}}} \mathscr{S}_{34}(\phi_{i+1,b;i,a} - \epsilon_1)
 \label{eq:quiv_var_Zp_2*}
\end{align}
where $[\phi_{i,(a \neq b)}]$ means the product for $1 \le a \neq b \le k_i$.
The last term is also written as $\mathscr{S}_{34}(\phi_{i+1,b;i,a} - \epsilon_1) = \mathscr{S}_{34}(\phi_{i,a;i+1,b} - \epsilon_2)$.
This formula is reduced to \eqref{eq:LMNS_formula_2*_sym} if $p = 1$.
See also \S\ref{sec:affine_quiv_W}.

\subsection{Gauge origami}\label{sec:origami}
\index{gauge origami}

We now discuss the relation to the instanton partition function for cyclic quiver theory~\eqref{eq:Z_inst_cyc_quiv}.
Although the gauge polynomial part is slightly different, these contour integrals look quite similar after replacement $(\epsilon_{1,2}) \longleftrightarrow (\epsilon_{3,4})$:
The roles of the bifundamental mass and the equivariant parameter for the spacetime rotation are exchanged.
In fact, $\widehat{A}_{p-1}$ quiver theory on $\mathcal{S} = \mathcal{S}_{{A}_0}$ and $\widehat{A}_0$ quiver theory on $\mathcal{S}_{{A}_{p-1}}$ have the same gauge origami realization in 8d, $\mathbb{C}^4/\mathbb{Z}_p (\times \mathbb{Z}_1)$.
Generalizing this configuration, we have the following reductions~\cite{Nekrasov:2016ydq}:%
\footnote{%
In general, we may consider the orbifold $\mathbb{C}^4/\Gamma$ with $\Gamma$ a finite subgroup of $\SU(4)$, $\Gamma \subset \SU(4)$.
}
  \begin{align}
   \begin{tikzpicture}[baseline=(current  bounding  box.center)]
    \node [rectangle, draw, fill=white,
    text width=20em, text centered, rounded corners,
    minimum height=2.5em, 
    ]
    (k-inst) at (0,1.5) {8d origami: $\mathcal{S}_{\Gamma_{12}} \times \mathcal{S}_{\Gamma_{34}}$};
%
    \node [rectangle, draw, fill=white,
    text width=10em, text centered, rounded corners,
    minimum height=2.5em, 
    ]
    (S12) at (-4,-1.5) {$\widehat{\Gamma}_{34}$ quiver on $\mathcal{S}_{\Gamma_{12}}$};
    \node [rectangle, draw, fill=white,
    text width=10em, text centered, rounded corners,
    minimum height=2.5em, 
    ]
    (S34) at (4,-1.5) {$\widehat{\Gamma}_{12}$ quiver on $\mathcal{S}_{\Gamma_{34}}$};
    \draw [-latex,blue,line width = 2pt] (-1.5,.7) -- ++(-1.5,-1.5);
    \draw [-latex,red,line width = 2pt] (1.5,.7) -- ++(1.5,-1.5);    
   \end{tikzpicture}
  \end{align}
  which explains why the relation $\Gamma_{12} \longleftrightarrow \Gamma_{34}$ is related to $(\epsilon_{1,2}) \longleftrightarrow (\epsilon_{3,4})$.

In order to realize gauge theory degrees of freedom, one may put D-branes on this 8d geometry.
For example, if one puts $n$ D3 branes on 12-surface, it realizes $\rU(n)$ gauge theory with quiver structure $\widehat{\Gamma}_{34}$ on $\mathcal{S}_{\Gamma_{12}}$ (\S\ref{sec:HW_quiver}).
If one puts $n$ D3 branes on 12-surface and $n'$ D3 branes on 34-surface simultaneously, the branes on 34-surface play a role of the codimension-four defect in 12-theory, which realizes the {\em $qq$-character} of $\widehat{\Gamma}_{34}$ quiver (\S\ref{sec:qq-ch}).
This configuration is also interpreted as $\rU(n')$ theory with $n$ defects on 34-surface obtained through $\epsilon_{1,2} \leftrightarrow \epsilon_{3,4}$ duality.
See also \S\ref{sec:qq_ch_geom}.

\section{Fractional quiver gauge theory}\label{sec:fractional_quiver}

We consider the fractional quiver gauge theory, which enables us to construct quiver gauge theory beyond the simply-laced type~\cite{Kimura:2017hez}.
The fractional quiver $\Gamma_d = (\Gamma, d)$ is a quiver with integer labels assigned to each node $d = (d_i)_{i \in \Gamma_0} \in \mathbb{Z}_{>0}^{\rk \Gamma}$. \index{quiver!fractional---}
We consider the ring $R_i = \mathbb{C}[z_1^{d_i}, z_2]$ for each node $i \in \Gamma_0$, and the equivariant gauge theory counts the associated $R_i$ ideals.

We remark that there have been several studies on the non-simply-laced type gauge theory in the context of the BPS quiver~\cite{Alexandrov:2011ac,Manschot:2013dua}, the Coulomb branch of the 3d $\mathcal{N} = 4$ gauge theory~\cite{Nakajima:2019olw}, and also the related works~\cite{Dey:2016qqp,Hanany:2019tji,Hanany:2020jzl,Grimminger:2020dmg,Bourget:2020bxh,Bourget:2020asf,Bourget:2020xdz,Bourget:2020mez}.

\subsection{Instanton moduli space}

The instanton moduli space of the fractional quiver gauge theory on $\mathcal{S}$ is defined as a disjoint union as discussed in \S\ref{sec:quiver_inst_mod_sp}:
\begin{align}
 \mathfrak{M}_{\underline{n},\underline{k}} = \bigsqcup_{i \in \Gamma_0} \mathfrak{M}_{n_i,k_i}
 \, ,
\end{align}
where $(\underline{n},\underline{k}) = (n_i,k_i)_{i \in \Gamma_0}$ are the dimension vectors as in \eqref{eq:quiv_dim_vec}.
Now each moduli space $(\mathfrak{M}_{n_i,k_i})_{i \in \Gamma_0}$ is constructed with the spacetime bundle, which is the cotangent bundle to $\mathcal{S}$ at the fixed point $o$ with the modified fiber compared with the previous case in \S\ref{sec:Q_bundle}:
\begin{align}
 \mathbf{Q}_{1^{d_i} 2} = \mathbf{Q}_{1^{d_i}} \oplus \mathbf{Q}_2
\end{align}
with the character
\begin{align}
 \ch_\mathsf{T} \mathbf{Q}_{1^{d_i}} = q_1^{d_i}
 \, .
 \label{eq:Q_1_character1}
\end{align}
We remark the relation to the previous convention $\mathbf{Q}_{1^1 2} = \mathbf{Q}_{12} = \mathbf{Q} = \mathbf{Q}_1 \oplus \mathbf{Q}_2$.
Similarly we have
\begin{align}
 \ch_\mathsf{T} \wedge \mathbf{Q}_{1^{d_i}} = 1 - q_1^{d_i}
 \, , \qquad
 \ch_\mathsf{T} \wedge \mathbf{Q}_{1^{d_i} 2} = (1 - q_1^{d_i})(1 - q_2)
 \, .
 \label{eq:Q_1_character2}
\end{align}

\subsubsection{Observable bundle}

As discussed in \S\ref{sec:univ_bundle} and \S\ref{sec:quiv_eq_fix_pt}, we define the observable bundle $\mathbf{Y}_i = \mathbf{Y}_{o,i}$ for the node $i \in \Gamma_0$ as a pullback of the universal bundle $(\mathbf{Y}_{\mathcal{S},i})_{i \in \Gamma_0}$ over the moduli space $(\mathfrak{M}_{n_i,k_i})_{i \in \Gamma_0}$:
\begin{align}
 \mathbf{Y}_{\mathcal{S},i} = \bigoplus_{\alpha = 1}^{n_i} \mathbf{N}_{i,\alpha} \otimes \mathbf{I}_{1^{d_i} 2,\lambda_{i,\alpha}}
\end{align}
where the character of the ideal $\mathbf{I}_{1^{d_i} 2,\lambda}$ for the ring $R_i = \mathbb{C}[z_1^{d_i},z_2]$ associated with the partition $\lambda$ is given by
\begin{align}
 \ch_\mathsf{T} \mathbf{I}_{1^{d_i} 2,\lambda} = \sum_{s \not \in \lambda} q_1^{d_i(s_1 - 1)} q_2^{s_2 - 1}
 \, .
\end{align}
Then, we obtain the observable bundle assigned to the node $i \in \Gamma_0$ through the pullback of the universal bundle:
\begin{align}
 \mathbf{Y}_i := \mathbf{Y}_{o,i}
 & = \mathbf{N}_i - \wedge \mathbf{Q}_{1^{d_i} 2} \cdot \mathbf{K}_i
 \, ,
\end{align}
where the bundles $(\mathbf{N}_i, \mathbf{K}_i)_{i \in \Gamma_0}$ are similarly defined as in \S\ref{sec:quiver_bundles}, whereas the character of the instanton bundle $\mathbf{K}_i = \bigoplus_{\alpha = 1}^{n_i} \mathbf{K}_{i,\alpha}$ is given by
\begin{align}
 \ch_\mathsf{T} \mathbf{K}_{i,\alpha} = \sum_{s \in \lambda_{i,\alpha}} \np^{\mathsf{a}_\alpha} q_1^{d_i(s_1 - 1)} q_2^{s_2 - 1}
 \, .
\end{align}

\subsubsection{Partial reduction of the universal bundle}

In this case, we have two distinct partial reductions of the universal bundle: 
\begin{align}
 \mathbf{Y}_i
 = \wedge \mathbf{Q}_{1^{d_i}} \cdot \mathbf{X}_i
 = \wedge \mathbf{Q}_2 \cdot \check{\mathbf{X}}_i
\end{align}
where we define $(\mathbf{X}, \check{\mathbf{X}}) = (\mathbf{X}_i,\check{\mathbf{X}}_i)_{i \in \Gamma_0} = (\mathbf{Y}_{\mathcal{S}_1,i}, \mathbf{Y}_{\mathcal{S}_2,i})_{i \in \Gamma_0}$ with the characters
\begin{subequations}
\begin{align}
 \ch_\mathsf{T} \mathbf{X}_i = \sum_{x \in \mathcal{X}_i} x
 \, , \qquad
 \ch_\mathsf{T} \check{\mathbf{X}}_i = \sum_{\check{x} \in \check{\mathcal{X}}_i} \check{x}
 \, ,
\end{align}
\begin{align}
 \mathcal{X}_i & = \left\{ x_{i,\alpha,k} = \np^{\mathsf{a}_{i,\alpha}} q_1^{d_i(k-1)} q_2^{\lambda_{i,\alpha,k}} \right\}_{\substack{i \in \Gamma_0 \\ \alpha = 1,\ldots n_i \\ k=1,\ldots,\infty}}
 \, , \quad
 \check{\mathcal{X}}_i = \left\{ \check{x}_{i,\alpha,k} = \np^{\mathsf{a}_{i,\alpha}}  q_1^{d_i \check\lambda_{i,\alpha,k}} q_2^{k-1} \right\}_{\substack{i \in \Gamma_0 \\ \alpha = 1,\ldots n_i \\ k=1,\ldots,\infty}}
 \, .
\end{align}
\end{subequations}
In fact, exchanging $q_1 \leftrightarrow q_2$ gives rise to the Langlands duality of the associated quantum algebra.
See also \S\ref{sec:frac_T_op}.

\subsubsection{Fractionalization}

The expression \eqref{eq:Q_1_character2} implies the {\em fractionalization}:
\begin{align}
 \wedge \mathbf{Q}_{1^{d_i}} = \mathbf{D}_i \cdot \wedge \mathbf{Q}_{1} 
\end{align}
where
\begin{align}
 \ch_\mathsf{T} \mathbf{D}_i = 1 + q_1 + \cdots + q_1^{d_i - 1}
 \, .
\end{align}
Similarly we see a similar fractionalization of the observable bundle:
\begin{align}
 \mathbf{Y}_i = \mathbf{D}_i \cdot \wedge \mathbf{Q}_1 \cdot \mathbf{X}_i =: \mathbf{D}_i \cdot \mathbf{y}_i
\end{align}
where $\mathbf{y}_i = \wedge \mathbf{Q}_1 \cdot \mathbf{X}_i$ corresponds to the ordinary observable bundle as defined in~\eqref{eq:quiver_partial_red_univ_bundle}.

\subsection{Instanton partition function}

We then compute the partition function of the fractional quiver gauge theory.
The vector multiplet and the hypermultiplet contributions are similarly obtained as in \S\ref{sec:quiver_index_formula} by replacing the spacetime bundle $\mathbf{Q} = \mathbf{Q}_{12}$:
\begin{subequations}\label{eq:frac_quiv_bundles}
\begin{align}
 \mathbf{V}_i
 = \frac{\mathbf{Y}_i^\vee \mathbf{Y}_i}{\wedge \mathbf{Q}_{1^{d_i} 2}}
 = \frac{\wedge \mathbf{Q}_{1^{d_i}}^\vee}{\wedge \mathbf{Q}_2} \, \mathbf{X}_i^\vee \mathbf{X}_i
 \, ,
\end{align}
 \begin{align}
  \mathbf{H}_{e:i \to j}
  = - \mathbf{M}_e \frac{\mathbf{Y}_i^\vee \mathbf{Y}_j}{\wedge \mathbf{Q}_{1^{d_{ij}} 2}}
  = - \mathbf{M}_e \frac{\wedge \mathbf{Q}_{1^{d_i}}^\vee \cdot \wedge \mathbf{Q}_{1^{d_j}}}{\wedge \mathbf{Q}_{1^{d_{ij}}} \cdot \wedge \mathbf{Q}_2} \, \mathbf{X}_i^\vee \mathbf{X}_j
 \, .
 \end{align}
  \begin{align}
  \mathbf{H}_i^\text{f} = - \frac{{\mathbf{M}}_i^\vee\mathbf{Y}_i}{\wedge \mathbf{Q}_{1^{d_i} 2}}
  \, , \qquad
  \mathbf{H}_i^\text{af} = - \frac{\mathbf{Y}_i^\vee \widetilde{\mathbf{M}}_i}{\wedge \mathbf{Q}_{1^{d_i} 2}} \, ,
 \end{align}
\end{subequations}
where we define $d_{ij} = \operatorname{gcd}(d_i,d_j)$.%
\footnote{%
We mainly consider the cases with $d_{ij} = 1$.
}
Then, these contributions are concisely combined with the quiver Cartan matrix as in \S\ref{sec:quiv_Cartan_matrix}:
\begin{align}
 \sum_{i \in \Gamma_0} \mathbf{V}_i + \sum_{e:i \to j} \mathbf{H}_{e:i \to j}
 & = \frac{\wedge \mathbf{Q}_{1^{d_i}}^\vee}{\wedge \mathbf{Q}_2} \sum_{(i,j) \in \Gamma_0 \times \Gamma_0} \mathbf{X}_i^\vee \, c_{ij}^{+\vee} \, \mathbf{X}_j
 \nonumber \\
 & = \frac{\wedge \mathbf{Q}_{1}^\vee}{\wedge \mathbf{Q}_2} \sum_{(i,j) \in \Gamma_0 \times \Gamma_0} \mathbf{X}_i^\vee \, b_{ij}^{+\vee} \, \mathbf{X}_j 
\end{align}
where the half Cartan matrix and its symmetrization are given by%
\footnote{%
As mentioned below, the symmetrized Cartan matrix is not symmetric in the strict sense, whereas we call it the symmetrization, which obeys the reflection relation~\eqref{eq:quiv_Cartan_sym_reflection}.
}
\begin{subequations}
\begin{align}
 c_{ij}^+
 & = \delta_{ij} - \sum_{e:i \to j} \mathbf{M}_e^\vee \, \frac{\wedge \mathbf{Q}_{1^{d_j}}^\vee}{\wedge \mathbf{Q}_{1^{d_{ij}}}^\vee}
 \, , \\
 b_{ij}^+
 & 
 =
 \frac{\wedge \mathbf{Q}_{1^{d_i}}}{\wedge \mathbf{Q}_{1}} \, \delta_{ij}
 - \sum_{e:i \to j} \mathbf{M}_e^\vee \, \frac{\wedge \mathbf{Q}_{1^{d_i}}}{\wedge \mathbf{Q}_{1}} \frac{\wedge \mathbf{Q}_{1^{d_j}}^\vee}{\wedge \mathbf{Q}_{1^{d_{ij}}}^\vee}
 = \frac{\wedge \mathbf{Q}_{1^{d_i}}}{\wedge \mathbf{Q}_{1}} \, c_{ij}^+
 \, .
\end{align}
\end{subequations}
The (character of) full Cartan matrix is similarly defined: \index{Cartan matrix!symmetrized---}
\begin{subequations}
\begin{align}
 c_{ij} &
 = (1 + q_{1^{d_i} 2}^{-1}) \, \delta_{ij} 
 - \sum_{e:i \to j} \mu_e^{-1} \frac{1 - q_1^{-d_j}}{1 - q_1^{-d_{ij}}}
 - \sum_{e:j \to i} \mu_e q_{1^{d_{ij}} 2}^{-1} \frac{1 - q_1^{-d_j}}{1 - q_1^{-d_{ij}}}
 \\
 b_{ij} & = 
\frac{1 - q_1^{d_i}}{1 - q_1} \, c_{ij}
 \nonumber \\
 &
 = \frac{1 - q_1^{d_i}}{1 - q_1} (1 + q_{1^{d_i} 2}^{-1}) \, \delta_{ij} 
 - \sum_{e:i \to j} \mu_e^{-1} \frac{(1 - q_1^{d_i})(1 - q_1^{-d_j})}{(1 - q_1)(1 - q_1^{-d_{ij}})}
 - \sum_{e:j \to i} \mu_e q_{1^{d_{ij}} 2}^{-1} \frac{(1 - q_1^{d_i})(1 - q_1^{-d_j})}{(1 - q_1)(1 - q_1^{-d_{ij}})}
 \label{eq:Cartan_b}
\end{align}
\end{subequations}
where we denote
\begin{align}
 q_{i^d j^{d'}} = q_i^d q_j^{d'} = \exp \left( \epsilon_{i^d j^{d'}} \right)
 \, ,
 \qquad
 \epsilon_{i^d j^{d'}} = d \epsilon_i + d' \epsilon_j
 \, .
 \label{eq:q_dd'}
\end{align}
In fact, the symmetrized Cartan matrix obeys a reflection relation analogous to \eqref{eq:quiv_Cartan_reflection}:
\begin{align}
 b_{ji}^{[n]} = q^{-n} b_{ij}^{[-n]}
 \label{eq:quiv_Cartan_sym_reflection} 
\end{align}
where we denote the $n$-th Adams operation to the symmetrized Cartan matrix by $(b_{ij}^{[n]})$. \index{Adams operation}
Namely, we can formulate the fractional quiver gauge theory just by replacing the ordinary Cartan matrix with its symmetrization.

These relations between $(c_{ij})$ and $(b_{ij})$ are reduced to the standard one in the classical limit $n \to 0$.
For the simple root $(\alpha_i)$ and the simple coroot $(\alpha_i^\vee)$, the Cartan matrix is given by
\begin{align}
 c_{ij} = (\alpha_i^\vee, \alpha_j) = \frac{(\alpha_i,\alpha_j)}{(\alpha_i, \alpha_i)}
 \, ,
\end{align}
and its symmetrization 
\begin{align}
 b_{ij} = d_i c_{ij} = (\alpha_i, \alpha_j)
\end{align}
with
\begin{align}
 d_i = (\alpha_i,\alpha_i)
 \, .
\end{align}

\subsubsection{Equivariant index formula}
\index{equivariant!---index}

The full partition function is given by applying the index functor to the vector multiplet and the hypermultiplet bundles~\eqref{eq:frac_quiv_bundles}.
For the vector multiplet contribution, it is given by replacing $q_1$ with $q_1^{d_i}$, so that we obtain
\begin{align}
 Z_i^\text{vec} := \mathbb{I}[\mathbf{V}_i]
 & =
 \begin{cases}
  \displaystyle
  \prod_{\substack{(x,x') \in \mathcal{X}_i \times \mathcal{X}_i \\ x \neq x'}}
  \frac{\Gamma_1(\log x' - \log x - d_i \epsilon_1;\epsilon_2)}{\Gamma_1(\log x' - \log x;\epsilon_2)}
  & (\text{4d}) \\
  \displaystyle
  \prod_{\substack{(x,x') \in \mathcal{X}_i \times \mathcal{X}_i \\ x \neq x'}}
  \frac{\Gamma_q(q_2 x/x';q_2)}{\Gamma_q(q_{1^{d_i} 2} x/x';q_2)}  
  & (\text{5d}) \\ 
  \displaystyle
  \prod_{\substack{(x,x') \in \mathcal{X}_i \times \mathcal{X}_i \\ x \neq x'}}
  \frac{\Gamma_e(q_2 x/x';p,q_2)}{\Gamma_e(q_{1^{d_i} 2} x/x';p,q_2)}
  & (\text{6d}) 
 \end{cases}
\end{align}
For the bifundamental hypermultiplet contribution, on the other hand, we find a peculiar structure.
Since we have
\begin{align}
 \ch_\mathsf{T} \frac{\wedge \mathbf{Q}_{1^{d_j}}}{\wedge \mathbf{Q}_{1^{d_{ij}}}}
 &
 = \frac{1 - q_1^{d_j}}{1 - q_1^{d_{ij}}}
 = 1 + q_1^{d_{ij}} + \cdots + q_1^{d_j - d_{ij}}
 \, ,
\end{align}
the character is given as
\begin{align}
 \ch_\mathsf{T} \mathbf{H}_{e:i \to j}
 = - \mu_e (1 + q_1^{d_{ij}} + \cdots + q_1^{d_j - d_{ij}}) \frac{1 - q_1^{-d_i}}{1 - q_2}
 \sum_{(x,x') \in \mathcal{X}_i \times \mathcal{X}_j} \frac{x'}{x}
 \, ,
\end{align}
which implies {\em multiplication} of the bifundamental contribution with the mass shift, $(\mu_e q_1^{r d_{ij}})_{r = 0, \ldots, d_j/d_{ij} - 1}$.
Hence, the bifundamental hypermultiplet contribution to the partition function is given by
\begin{align}
 Z_{e:i \to j}^\text{bf} := \mathbb{I}[\mathbf{H}_{e:i \to j}]
 & =
 \begin{cases}
  \displaystyle
  \prod_{(x,x') \in \mathcal{X}_i \times \mathcal{X}_j}
  \prod_{r = 0}^{d_j/d_{ij} - 1}
  \frac{\Gamma_1(\log x' - \log x + m_e + r d_{ij} \epsilon_1;\epsilon_2)}{\Gamma_1(\log x' - \log x + m_e + (r d_{ij} - d_i) \epsilon_1;\epsilon_2)}
  & (\text{4d}) \\
  \displaystyle
  \prod_{(x,x') \in \mathcal{X}_i \times \mathcal{X}_j}
  \prod_{r = 0}^{d_j/d_{ij} - 1}  
  \frac{\Gamma_q(\mu_e^{-1} q_1^{d_i - r d_{ij}} q_2 x/x';q_2)}  {\Gamma_q(\mu_e^{-1} q_1^{-r d_{ij}} q_2 x/x';q_2)}
  & (\text{5d}) \\ 
  \displaystyle
  \prod_{(x,x') \in \mathcal{X}_i \times \mathcal{X}_j}
  \prod_{r = 0}^{d_j/d_{ij} - 1}  
  \frac{\Gamma_e(\mu_e^{-1} q_1^{d_i - r d_{ij}} q_2 x/x';q_2)}  {\Gamma_e(\mu_e^{-1} q_1^{-r d_{ij}} q_2 x/x';q_2)}
  & (\text{6d}) 
 \end{cases} 
\end{align}
which is asymmetric between the source and target nodes.

\subsubsection{Contour integral formula}

We can also derive the contour integral formula for the fractional quiver gauge theory partition function similarly to \S\ref{sec:quiv_LMNS_formula}.

The instanton part of the fractional quiver bundles \eqref{eq:frac_quiv_bundles} are given by \index{LMNS formula!fractional quiver gauge theory}
\begin{subequations}
\begin{align}
 \mathbf{V}_i^\text{inst}
 & = - \mathbf{N}_i^\vee \mathbf{K}_i - \det \mathbf{Q}_{1^{d_i} 2}^\vee \cdot \mathbf{K}_i^\vee \mathbf{N}_i + \wedge \mathbf{Q}_{1^{d_i} 2}^\vee \cdot \mathbf{K}_i^\vee \mathbf{K}_i
 \, ,
 \\
 \mathbf{H}_{e:i \to j}^\text{inst}
 & =
 \mathbf{M}_e \frac{\wedge \mathbf{Q}_{1^{d_j}}}{\wedge \mathbf{Q}_{1^{d_{ij}}}} \mathbf{N}_i^\vee \mathbf{K}_j
 + \mathbf{M}_e \frac{\wedge \mathbf{Q}_{1^{d_j}}}{\wedge \mathbf{Q}_{1^{d_{ij}}}} \mathbf{K}_i^\vee \mathbf{N}_j
 - \mathbf{M}_e \frac{\wedge \mathbf{Q}_{1^{d_i} 2}^\vee \cdot \wedge \mathbf{Q}_{1^{d_j} 2}}{\wedge \mathbf{Q}_{1^{d_{ij}} 2}} \mathbf{K}_i^\vee \mathbf{K}_j
 \, .
\end{align}
\end{subequations}
Hence, we obtain the contour integral formula for the fractional quiver partition function:
\begin{align}
 Z_{\underline{n},\underline{k}}^\text{inst}
 & = \frac{Z_{\underline{n},\underline{k}}}{\mathring{Z}_{\underline{n},\underline{k}}}
 = \prod_{i \in \Gamma_0} \frac{1}{k_i!} \frac{[-\epsilon_{1^{d_i} 2}]^{k_i}}{[-\epsilon_{1^{d_i},2}]^{k_i}} \oint_{\mathsf{T}_K} 
 \prod_{i \in \Gamma_0} z^\text{vec}_i z^\text{f}_i z^\text{af}_i \prod_{e \in \Gamma_1} z_{e:i \to j}^\text{bf}
\end{align}
with each contribution
\begin{subequations}
 \begin{align}
  z_i^\text{vec} & =
  \prod_{a = 1}^{k_i} \frac{1}{P_i(\phi_{i,a}) \widetilde{P}_i(\phi_{i,a} + \epsilon_{1^{d_i}2})}
  \prod_{a \neq b}^{k_i} \mathscr{S}_{1^{d_i} 2}(\phi_{i,ab})^{-1}
  \, , \\
  z_i^\text{f} & = \prod_{a=1}^{k_i} P^\text{f}_i(\phi_{i,a})
  \, , \qquad \qquad
  z_i^\text{af}  = \prod_{a=1}^{k_i} \widetilde{P}^\text{af}_i(\phi_{i,a} + \epsilon_{1^{d_i} 2})
  \, , \\
  z_{e:i \to j}^\text{bf} & =
  \prod_{a = 1}^{k_j}
  \prod_{r = 0}^{d_j/d_{ij} - 1}
  P_i(\phi_{j,a} + m_e + r d_{ij} \epsilon_1)
  \prod_{a = 1}^{k_i}
  \prod_{r = 0}^{d_i/d_{ij} - 1}
  \widetilde{P}_j(\phi_{i,a} + (d_i - r d_{ij}) \epsilon_1 + \epsilon_2 - m_e)
  \nonumber \\
  & \quad \times
  \prod_{\substack{a = 1,\ldots,k_i \\ b = 1,\ldots,k_j}}
  \prod_{r = 0}^{d_j/d_{ij} - 1}
  \mathscr{S}_{1^{d_i} 2}(\phi_{j,b;i,a} + m_e + r d_{ij} \epsilon_1)
  \, ,
 \end{align}
\end{subequations}
and the $\mathscr{S}$-function
\index{S-function@$\mathscr{S}$-function}
\begin{align}
 \mathscr{S}_{i^d j^{d'}}(\phi)
 = \frac{[\phi - \epsilon_{i^d,j^{d'}}]}{[\phi][\phi - \epsilon_{i^d j^{d'}}]}
 = \frac{[\phi - d \epsilon_i][\phi - d' \epsilon_j]}{[\phi][\phi - d \epsilon_i - d' \epsilon_j]}
 \, .
\end{align}
Actually the multiplicity of the bifundamental contribution depends on the parameter $(d_i)_{i \in \Gamma_0}$ as before.
We remark that the instanton partition function of fractional quiver shown above has (topological) string theory realization based on its algebraic structure~\cite{Kimura:2019gon}.

\chapter{Supergroup gauge theory}\label{sec:super_instanton}

Supergroup is a natural generalization of the concept of group, which describes  symmetry of systems involving both bosonic and fermionic degrees of freedom.
Although it is typically considered as a global symmetry of QFT, there are several situations where we should consider a local supergroup symmetry, e.g., supergravity is interpreted as gauge theory with super-Poincar\'e group symmetry.
In fact, supergroup gauge theory has been studied in particular from its string theory point of view~\cite{Vafa:2001qf,Okuda:2006fb,Mikhaylov:2014aoa,Dijkgraaf:2016lym,Nekrasov:2018xsb}.
Although supergroup gauge theory is inevitably a non-unitary theory, it provides an effective description of the intersecting branes constructed in higher dimensions~\cite{Nekrasov:2017gzb,Chen:2019vvt}.

In this Chapter, we focus on gauge theory with unitary supergroup symmetry, and discuss the role of instantons to explore non-perturbative aspects of supergroup gauge theory~\cite{Kimura:2019msw}.
We will apply the equivariant localization analysis to supergroup gauge theory, and obtain the instanton partition function.
We will also discuss several relations between supergroup gauge theory and non-supergroup quiver gauge theory.

\section{Supergroup Yang--Mills theory}

In this Section, we begin with some basic aspects of supervector space and superalgebra.
Then, we discuss the Yang--Mills theory with supergroup gauge symmetry.
We will also mention how to realize such a supergroup gauge theory in terms of non-supergroup gauge theory in the unphysical parameter regime.

\subsection{Supervector space, superalgebra, and supergroup}\label{sec:sup_vec_sp}

\subsubsection{Supervector space}

Supervector space is a $\mathbb{Z}_2$-graded vector space:%
\footnote{%
See, for example,~\cite{Kac:1977em,Varadarajan:2004yz,Quella:2013oda} for the mathematical introduction to the supervector space and superalgebra.
}
\begin{align}
 V = V_0 \oplus V_1
 \, .
\end{align}
We denote the parity of an element $x \in V_\sigma$ by $|x| = \sigma$ for $\sigma = 0, 1$, which is called even/bosonic for $\sigma = 0$, and odd/fermionic for $\sigma = 1$.
We denote the parity flipped vector space of $V$ by $\Pi V$ with $(\Pi V)_0 = V_1$ and $(\Pi V)_1 = V_0$, and we define the superdimension of $V$:
\begin{align}
 \sdim V = \sum_{\sigma = 0, 1} (-1)^\sigma \dim V_\sigma = \dim V_0 - \dim V_1
 \, .
\end{align}

For supervector spaces, $V$ and $W$, a linear map $V \to W$ defines a supermatrix 
\begin{align}
 M =
 \begin{pmatrix}
  M_{00} & M_{10} \\ M_{01} & M_{11}
 \end{pmatrix}
 \in \Hom(V,W)
 \label{eq:supermatrix_block}
\end{align}
with $M_{\sigma\sigma'} \in \Hom(V_\sigma,W_{\sigma'})$.
In particular, for $V = W$, namely $M \in \End(V)$, it defines general linear supergroup $\GL(V) = \GL(V_0|V_1)$, if invertible.
We then define the supertrace operation on the supermatrix $M \in \End(V)$:
\begin{align}
 \str M = \tr_0 M - \tr_1 M = \tr M_{00} - \tr M_{11}
 \, ,
\end{align}
and the superdeterminant, also known as the Berezinian,
\begin{align}
 \sdet M
 = \frac{\det (M_{00} - M_{10} M_{11}^{-1} M_{01})}{\det M_{11}}
 = \frac{\det M_{00}}{\det(M_{11} - M_{01} M_{00}^{-1} M_{10})}
 \, .
 \label{eq:sdet_def}
\end{align}
We remark that the superdeterminant is well-defined only if $\det M_{00} \det M_{11} \neq 0$, and the following identity holds for the supertrace and the superdeterminant,
\begin{align}
 \log \str M = \sdet \log M
 \, .
\end{align}
For the supertrace, the following cyclic property holds:
\begin{align}
 \str M_1 M_2 \cdots M_n = \str M_2 \cdots M_n M_1
 \, .
 \label{eq:supertrace_cyclic}
\end{align}

\subsubsection{Superalgebra}
\label{superalgebra}

Superalgebra is a $\mathbb{Z}_2$-graded algebra:
\begin{align}
 \mathfrak{A} = \mathfrak{A}_0 \oplus \mathfrak{A}_1
 \, .
\end{align}
We denote the supercommutator for $a, b \in \mathfrak{A}$ by
\begin{align}
 [a, b] = ab - (-1)^{|a||b|} ba
 \, .
\end{align}
In particular, the Lie superalgebra is a superalgebra, obeying the Jacobi identity,
\begin{align}
 [a,[b,c]] = [[a,b],c] + (-1)^{|a||b|} [b,[a,c]]
 \, .
\end{align}
For example, the special linear Lie superalgebra $\mathfrak{sl}_{n|m}$ is defined with the supertraceless condition:
\begin{align}
 \mathfrak{sl}_{n|m} =
 \left\{
 a \in \mathfrak{A} = \mathfrak{A}_0 \oplus \mathfrak{A}_1, \dim \mathfrak{A}_0 = n, \dim \mathfrak{A}_1 = m \mid \str a = 0
 \right\}
 \, ,
\end{align}
which is classified into $A_{n-1|m-1}$ according to the classification by Kac~\cite{Kac:1977em}.

\subsection{Yang--Mills theory}

We denote a Lie supergroup and the corresponding Lie superalgebra by $G = G_0|G_1$ and $\mathfrak{g} = \mathfrak{g}_0 \oplus \mathfrak{g}_1 = \operatorname{Lie} G$.
Let $A$ be a $\mathfrak{g}$-valued one-form connection $A \in \Omega^1(\mathcal{S}) \otimes \mathbb{C}[\mathfrak{g}]$, which behaves under the $G$-gauge transformation as $G: A \longmapsto g A g^{-1} + gdg^{-1}$ for $g \in G$.
The curvature $F \in \Omega^2(\mathcal{S}) \otimes \mathbb{C}[\mathfrak{g}]$ behaves $F \longmapsto g F g^{-1}$ as well as the ordinary case shown in \S\ref{sec:YM_theory}.
Then the $G$-invariant YM action functional on the spacetime $\mathcal{S}$ is defined similarly to \eqref{eq:YM-action}:\index{Yang--Mills action!supergroup---}
\begin{align}
 S_\text{YM}[A]
 & = \frac{1}{g^2} \int_\mathcal{S} d\operatorname{vol} \, |F|^2
 \nonumber \\ 
 & = - \frac{1}{g^2} \int_\mathcal{S} \str(F \wedge \star F)
 \nonumber \\
 & = - \frac{1}{g^2} \int_\mathcal{S} \tr_0 (F \wedge \star F) - \left(-\frac{1}{g^2} \right) \int_\mathcal{S} \tr_1 (F \wedge \star F)
 \, .
 \label{eq:super_SYM}
\end{align}
In this case, we replace the inner product with the supertrace $\left< A, B \right> = - \str(AB) = - \tr_0(AB) + \tr_1(AB)$ in order to make the action gauge invariant together with the cyclic property~\eqref{eq:supertrace_cyclic}.
We remark that this action is not positive semidefinite due to the odd contribution, so that the spectrum of the supergroup YM theory is not bounded.
Indeed the supergroup YM theory violates the spin-statistics theorem because of the odd component of gauge field, which is a fermionic spin-1 field, and thus supergroup theory is non-unitary.
Even though it is unbounded, we can consider the equation of motion~\eqref{eq:eom_YM} with respect to the YM action.
It does not provide a local minimum in the configuration space in a usual sense, but we may regard it as a complex saddle point to discuss a Lefschetz thimble in the complexified configuration space~\cite{Witten:2010cx}.

\subsection{Quiver gauge theory description}\label{sec:super_quiver_realization}

The supergroup YM theory has a realization as a quiver gauge theory with the ordinary gauge group symmetry.
We consider a double copy of the YM action with the gauge groups, $G_0$ and $G_1$:
\begin{align}
 S_\text{YM}[A_0,A_1] & =
 - \frac{1}{g_0^2} \int_\mathcal{S} \tr_0 (F_0 \wedge \star F_0)
 - \frac{1}{g_1^2} \int_\mathcal{S} \tr_1 (F_1 \wedge \star F_1)
\end{align}
where $(g_\sigma)_{\sigma = 0, 1}$ are the gauge coupling constants, and $(F_\sigma = dA_\sigma + A_\sigma \wedge A_\sigma)_{\sigma = 0, 1}$ is the curvature of each gauge node with the connection $A_\sigma \in \Omega^1(\mathcal{S}) \otimes \mathbb{C}[\mathfrak{g}_\sigma]$.
Comparing with the supergroup YM action $S_\text{YM}[A]$ shown in~\eqref{eq:super_SYM}, we may assign the supergroup condition:
\begin{align}
 \frac{1}{g_0^2} + \frac{1}{g_1^2} = 0
 \, .
 \label{eq:super_cond_g}
\end{align}
This implies that one cannot make both of the couplings positive simultaneously, and thus the supergroup YM theory may be realized in a unphysical parameter regime of the quiver gauge theory, $G_0 \times G_1$.
This argument similarly applies to the Chern--Simons action, and the levels of each node $(k_\sigma)_{\sigma = 0, 1}$ should obey the condition, $k_0 + k_1 = 0$.

In addition to the gauge connections, which correspond to the diagonal blocks of the superconnection based on the expression~\eqref{eq:supermatrix_block}, the supergroup theory also incorporates the off-diagonal blocks, which transform under the bifundamental representation of $G_0 \times G_1$.
Since these off-diagonal degrees of freedom are fermions (of spin~1, so that violating the spin statistics theorem), they play a role of the bifundamental matter fields, and thus the quiver gauge theory consists of two gauge nodes and two bifundamental matters, $G_0 \times \overline{G}_1$ and $\overline{G}_0 \times G_1$.
Hence, the supergroup gauge theory with $G = G_0|G_1$ has a realization in the unphysical regime of the cyclic $\widehat{A}_1$ quiver gauge theory~\cite{Dijkgraaf:2016lym}:%
\footnote{%
We will see that other parameters (equivariant parameters, FI parameters) should be also flipped for the negative gauge node in \S\ref{sec:super_localization}.
}
 \begin{align}
  \begin{tikzpicture}[baseline=(current  bounding  box.center)]
   \draw[thick] (0,0) circle (.2) node [left=.7em] {$G_0|G_1$};
   \node at (-1,-1.) {$(g^2,k)$};
   \draw[ultra thick,blue,latex-latex] (1.5,0) -- ++(2,0);
   \begin{scope}[shift={(6,0)}]
    \draw[thick] (0,0) circle (.2) node [left=1em] {$G_0$};
    \draw[thick] (3,0) circle (.2) node [right=1em] {$G_1$};
    \node at (-1,-1.) {$(g^2,k)$};
    \node at (4,-1.) {$(-g^2,-k)$};
    \draw[thick] (45:.2) to [bend left] ++(2.7,0);
    \draw[thick] (-45:.2) to [bend right] ++(2.7,0);
   \end{scope}
  \end{tikzpicture}
 \end{align}
This argument also has a natural interpretation in the D-brane description of supergroup gauge theory shown in \S\ref{sec:brane_super}.
We remark that the ABJ(M) theory~\cite{Aharony:2008ug,Aharony:2008gk} precisely matches these conditions, which is $\rU(N_0)_k \times \rU(N_1)_{-k}$ Chern--Simons theory with bifundamental matters.
In fact, the ABJM partition function looks the form of the supergroup version of the Chern--Simons theory~\cite{Kapustin:2009kz,Drukker:2009hy,Marino:2009jd}.

\section{Decoupling trick}\label{sec:decoupling}

\subsection{Vector multiplet}

The SU$(n_0|n_1)$ vector multiplet consists of SU$(n_0)$ and SU$(n_1)$ vectors and two bifundamentals of $\mathrm{SU}(n_0) \times \mathrm{SU}(n_1)$ and $\mathrm{SU}(n_1) \times \mathrm{SU}(n_0)$, where two gauge couplings, $\mathfrak{q}_0$ and $\mathfrak{q}_1$, should obey \eqref{eq:super_cond_g}:
\begin{align}
 \mathfrak{Q} := \mathfrak{q}_0 \mathfrak{q}_1 = 1
 \, .
 \label{eq:super_cond}
\end{align}
For example, $A_1$ quiver with supergroup gauge symmetry is equivalent to $\widehat{A}_1$ quiver, which exhibits modular properties with respect to the product of exponentiated gauge couplings $\mathfrak{Q}$~\cite{Nekrasov:2012xe} in the context of the Seiberg--Witten geometry (Chapter~\ref{chap:SW_theory}).
In this sense, the supergroup condition $\mathfrak{Q} \to 1$ seems singular because it's a boundary of the convergence radius.
Instead of such a singular limit, we alternatively consider the condition $\mathfrak{Q} \to 0$ using the modular transformation, which implies decoupling the SU$(n_1)$ vector multiplet: $\mathfrak{q}_1 = 0$.%
\footnote{%
Precisely speaking, there are two possibilities: $\mathfrak{q}_0 = 0$ or $\mathfrak{q}_1 = 0$.
We assume $\mathfrak{q}_0 \neq 0$ since $\mathfrak{q}_0$ would be interpreted as the physical gauge coupling.}
In this limit, $\widehat{A}_1$ quiver is reduced to $A_1$ quiver with two SU$(n_1)$ flavor nodes:
\begin{align}
  \begin{tikzpicture}[baseline=(current bounding box.center),thick]
   \filldraw[fill=white,draw=black] (0,0) circle (.2) node [below=.7em] {$n_0|n_1$};
   \draw[-latex,very thick,blue] (1.5,0) -- ++(1,0);
     \begin{scope}[shift={(4,0)}]
      \draw (0,.75) to [bend right] (0,-.75);
      \draw (0,.75) to [bend left] (0,-.75);      
     \filldraw[fill=white,draw=black] (0,.75) circle (.2) node [right=1em] {$\mathfrak{q}_0$}; 
     \filldraw[fill=white,draw=black] (0,-.75) circle (.2) node [right=1em] {$\mathfrak{q}_1$}; 
      \draw[-latex,very thick,blue] (1.5,0) -- ++(.5,0) node [above] {\textcolor{black}{$\mathfrak{q}_1 \to 0$}} -- ++(.5,0);    
     \end{scope}
   \begin{scope}[shift={(8,0)}]
    \draw (0,.75) -- ++(-.5,-1.5) node (l) {};
    \draw (0,.75) -- ++(.5,-1.5) node (r) {};    
    \filldraw[fill=white,draw=black] (0,.75) circle (.2) node [above=.7em] {$n_0$};
    \filldraw[fill=white,draw=black] (l)++(-.2,-.2) rectangle ++(.4,.4) node [below=1em] {$n_1$};
    \filldraw[fill=white,draw=black] (r)++(-.2,-.2) rectangle ++(.4,.4) node [below=1em] {$n_1$};       
   \end{scope}   
  \end{tikzpicture}  
\end{align}
We will see this is consistent with another gauging argument from string theory point of view (\S\ref{sec:brane_super}).

\subsection{Bifundamental hypermultiplet}

In order to apply this argument to generic quiver, we consider the hypermultiplet in the bifundamental representation of $\mathrm{SU}(n_{i,0}|n_{i,1}) \times \mathrm{SU}(n_{j,0}|n_{j,1})$.
We consider $A_2$ quiver as follows:
\begin{align}
 \begin{tikzpicture}[baseline=(current bounding box.center),thick]
  \draw (0,0) -- (2.5,0);
  \filldraw[fill=white,draw=black] (0,0) circle (.2) node [below=1em] {$n_{1,0}|n_{1,1}$};
  \filldraw[fill=white,draw=black] (2.5,0) circle (.2) node [below=1em] {$n_{2,0}|n_{2,1}$};  
  \draw [blue,-latex, very thick] (3.5,0) -- ++(1.5,0);
  \begin{scope}[shift={(6.5,0)}]
   \draw (0,.75) node (i1) {} to [bend right] (0,-.75) node (i2) {};
   \draw (0,.75) to [bend left] (0,-.75);
   \draw (2.5,.75) node (j1) {} to [bend right] (2.5,-.75) node (j2) {};
   \draw (2.5,.75) to [bend left] (2.5,-.75);
   \draw (i1) -- ++(2.5,0);
   \draw (0,-.75) -- ++(2.5,0);
   \draw [dashed] (i1) -- (j2);
   \draw [dashed] (i2) -- (j1);   
     \filldraw[fill=white,draw=black] (0,.75) circle (.2) node [left=.7em] {$n_{1,0}$};
     \filldraw[fill=white,draw=black] (0,-.75) circle (.2) node [left=.7em] {$n_{1,1}$};     
   \filldraw[fill=white,draw=black] (2.5,.75) circle (.2) node [right=.7em] {$n_{2,0}$};
   \filldraw[fill=white,draw=black] (2.5,-.75) circle (.2) node [right=.7em] {$n_{2,1}$};        
  \end{scope}
 \end{tikzpicture}
\end{align}
where the solid lines are the {\em positive} bifundamental hypermultiplets, while the dashed lines are the {\em negative} (vector-like) bifundamental multiplets.
Then, turning off the negative gauge couplings, $\mathfrak{q}_{1,1}$ and $\mathfrak{q}_{2,1}$, with the assumption $n_{1,1} = n_{2,1} =: n_1$, and all the mass parameters (Coulomb moduli for the negative nodes) coincide with each other, it becomes
\begin{align}
 \begin{tikzpicture}[baseline=(current bounding box.center),thick]
  \draw (0,0) node (11) {} --++(2.5,0) node (21) {};
  \draw (11)--++(-.75,-1.5) node (12a) {};
  \draw (11)--++(0,-1.5) node (12b) {};
  \draw[dashed] (11)--++(.75,-1.5) node (12c) {};
  \draw[dashed] (21)--++(-.75,-1.5) node (22a) {};
  \draw (21)--++(0,-1.5) node (22b) {};
  \draw (21)--++(.75,-1.5) node (22c) {};
  \filldraw[fill=white,draw=black] (11) circle (.2) node [above=.7em] {$n_{1,0}$};
  \filldraw[fill=white,draw=black] (21) circle (.2) node [above=.7em] {$n_{2,0}$};    
  \filldraw[fill=white,draw=black] (12a)++(-.2,-.2) rectangle ++(.4,.4);
  \filldraw[fill=white,draw=black] (12b)++(-.2,-.2) rectangle ++(.4,.4);
  \filldraw[fill=white,draw=black] (12c)++(-.2,-.2) rectangle ++(.4,.4);
  \filldraw[fill=white,draw=black] (22a)++(-.2,-.2) rectangle ++(.4,.4);
  \filldraw[fill=white,draw=black] (22b)++(-.2,-.2) rectangle ++(.4,.4);
  \filldraw[fill=white,draw=black] (22c)++(-.2,-.2) rectangle ++(.4,.4); 
  \draw [decorate,decoration={brace,amplitude=10pt,mirror,raise=4pt},yshift=0pt]
  (-.75,-1.7) -- ++(1.5,0) node [black,midway,xshift=0cm,yshift=-2em] {$n_{1}$};
  \draw [decorate,decoration={brace,amplitude=10pt,mirror,raise=4pt},yshift=0pt]
  (1.75,-1.7) -- ++(1.5,0) node [black,midway,xshift=0cm,yshift=-2em] {$n_{1}$};
  \draw[very thick,blue,-latex] (4,-.75) --++(1.5,0);
   \begin{scope}[shift={(6.5,0)}]
  \draw (0,0) node (11) {} --++(2.5,0) node (21) {};
  \draw (11)--++(0,-1.5) node (12b) {};
  \draw (21)--++(0,-1.5) node (22b) {};
  \filldraw[fill=white,draw=black] (11) circle (.2) node [above=.7em] {$n_{1,0}$};
  \filldraw[fill=white,draw=black] (21) circle (.2) node [above=.7em] {$n_{2,0}$};    
  \filldraw[fill=white,draw=black] (12b)++(-.2,-.2) rectangle ++(.4,.4) node [below=1em] {$n_1$}; 
  \filldraw[fill=white,draw=black] (22b)++(-.2,-.2) rectangle ++(.4,.4) node [below=1em] {$n_1$};   
   \end{scope}
 \end{tikzpicture}
\end{align}
which is consistent with the brane construction discussed in \S\ref{sec:brane_super}.
If $n_{1,1} \neq n_{2,1}$, such a cancellation does not occur, and the flavor nodes become different from each other.
This procedure is naturally generalized to $A_p$ quiver theory.

\subsection{$D_p$ quiver}\label{sec:decoupling_D}

Let us apply the decoupling trick to $D_p$ quiver with $p = 4$ as an example.
Splitting the SU$(n_{i,0}|n_{i,1})$ gauge nodes into positive SU$(n_{i,0})$ and negative SU$(n_{i,1})$ nodes, we obtain the following:
\begin{align}
  \begin{tikzpicture}[baseline=(current bounding box.center),thick]   
   \draw (0,0) -- (2.,0);
   \draw (4,1) -- (2,0) -- (4,-1);
   \filldraw[fill=white,draw=black] (0,0) circle (.2) node [below=1em] {$n_{1,0}|n_{1,1}$};
   \filldraw[fill=white,draw=black] (2,0) circle (.2) node [below=1em] {$n_{2,0}|n_{2,1}$};
   \filldraw[fill=white,draw=black] (4,1) circle (.2) node [below=1em] {$n_{3,0}|n_{3,1}$};
   \filldraw[fill=white,draw=black] (4,-1) circle (.2) node [below=1em] {$n_{4,0}|n_{4,1}$};     
   \draw [blue,-latex, very thick] (5.5,0) -- ++(1.5,0);
  \begin{scope}[shift={(8.5,0)}]
   \draw (0,.75) node (i1) {} to [bend right] (0,-.75) node (i2) {};
   \draw (0,.75) to [bend left] (0,-.75);
   \draw (2.,.75) node (j1) {} to [bend right] (2.,-.75) node (j2) {};
   \draw (2.,.75) to [bend left] (2.,-.75);
   \draw (3.5,2) node (k1) {} to [bend right] ++(-60:1.5) node (k2) {};
   \draw (3.5,2) to [bend left] ++(-60:1.5);
   \draw (3.5,-2) node (l1) {} to [bend right] ++(60:1.5) node (l2) {};
   \draw (3.5,-2) to [bend left] ++(60:1.5);      
   \draw (i1) -- ++(2.,0);
   \draw (i2) -- ++(2.,0);
   \draw (j1) -- (k1);
   \draw (j1) -- (l2);
   \draw (j2) -- (k2);
   \draw (j2) -- (l1);      
   \draw [dashed] (i1) -- (j2);
   \draw [dashed] (i2) -- (j1);
   \draw [dashed] (j1) -- (k2);
   \draw [dashed] (j1) -- (l1);
   \draw [dashed] (j2) -- (k1);
   \draw [dashed] (j2) -- (l2);         
   \filldraw[fill=white,draw=black] (i1) circle (.2) node [above=.7em] {$n_{1,0}$};
   \filldraw[fill=white,draw=black] (i2) circle (.2) node [below=.7em] {$n_{1,1}$};     
   \filldraw[fill=white,draw=black] (j1) circle (.2) node [above=.7em] {$n_{2,0}$};
   \filldraw[fill=white,draw=black] (j2) circle (.2) node [below=.7em] {$n_{2,1}$};     
   \filldraw[fill=white,draw=black] (k1) circle (.2) node [above=.7em] {$n_{3,0}$};
   \filldraw[fill=white,draw=black] (k2) circle (.2) node [right=.7em] {$n_{3,1}$};
   \filldraw[fill=white,draw=black] (l2) circle (.2) node [right=.7em] {$n_{4,0}$};
   \filldraw[fill=white,draw=black] (l1) circle (.2) node [below=.7em] {$n_{4,1}$};    
  \end{scope}   
  \end{tikzpicture}
\end{align}
We then turn off the gauge couplings of the negative nodes, $\mathfrak{q}_{i,1} \to 0$.
Applying the condition $n_{1,1} = n_{2,1} = 2 n_{3,1} = 2 n_{4,1} =: 2 n_1$, and tuning the Coulomb moduli for the negative nodes, we obtain $D_4$ quiver configuration with a single SU$(2n_1)$ flavor node:
\begin{align}
 \begin{tikzpicture}[baseline=(current bounding box.center),thick]
  \draw (0,0) node (11) {} --++(2.,0) node (21) {};
  \draw (4,1) node (31) {} -- (21) -- (4,-1) node (41) {};
  \draw (11)--++(-.75,-1.5) node (12a) {};
  \draw (11)--++(0,-1.5) node (12b) {};
  \draw[dashed] (11)--++(.75,-1.5) node (12c) {};
  \draw[dashed] (21)--++(-.5,-1.5) node (22a) {};
  \draw (21)--++(.5,-1.5) node (22b) {};
  \draw (21)--++(-.75,1.5) node (22d) {};
  \draw[dashed] (21)--++(0,1.5) node (22e) {};
  \draw[dashed] (21)--++(.75,1.5) node (22f) {};  
  \draw (31) --++(1,.65) node (32a) {};
  \draw (31) --++(1,0) node (32b) {};
  \draw[dashed] (31) --++(1,-.65) node (32c) {};
  \draw[dashed] (41) --++(1,.65) node (42a) {};
  \draw (41) --++(1,0) node (42b) {};
  \draw (41) --++(1,-.65) node (42c) {};    
  \filldraw[fill=white,draw=black] (11) circle (.2); 
  \filldraw[fill=white,draw=black] (21) circle (.2); 
  \filldraw[fill=white,draw=black] (31) circle (.2); 
  \filldraw[fill=white,draw=black] (41) circle (.2); 
  \filldraw[fill=orange!50,draw=black] (12a)++(-.2,-.2) rectangle ++(.4,.4);
  \filldraw[fill=orange!50,draw=black] (12b)++(-.2,-.2) rectangle ++(.4,.4);
  \filldraw[fill=orange!50,draw=black] (12c)++(-.2,-.2) rectangle ++(.4,.4);
  \filldraw[fill=orange!50,draw=black] (22a)++(-.2,-.2) rectangle ++(.4,.4);
  \filldraw[fill=orange!50,draw=black] (22b)++(-.2,-.2) rectangle ++(.4,.4);
  \filldraw[fill=orange!50,draw=black] (22d)++(-.2,-.2) rectangle ++(.4,.4);
  \filldraw[fill=cyan!50,draw=black] (22e)++(-.2,-.2) rectangle ++(.4,.4);
  \filldraw[fill=cyan!50,draw=black] (22f)++(-.2,-.2) rectangle ++(.4,.4);  
  \filldraw[fill=cyan!50,draw=black] (32a)++(-.2,-.2) rectangle ++(.4,.4);
  \filldraw[fill=cyan!50,draw=black] (32b)++(-.2,-.2) rectangle ++(.4,.4);
  \filldraw[fill=orange!50,draw=black] (32c)++(-.2,-.2) rectangle ++(.4,.4);
  \filldraw[fill=orange!50,draw=black] (42a)++(-.2,-.2) rectangle ++(.4,.4);
  \filldraw[fill=cyan!50,draw=black] (42b)++(-.2,-.2) rectangle ++(.4,.4);
  \filldraw[fill=cyan!50,draw=black] (42c)++(-.2,-.2) rectangle ++(.4,.4);    
  \draw[very thick,blue,-latex] (6,0) --++(1.5,0);
   \begin{scope}[shift={(8.5,0)}]
    \draw (0,0) node (11) {} --++(2.,0) node (21) {};
    \draw (4,1) node (31) {} -- (21) -- (4,-1) node (41) {};
  \draw (11)--++(0,-1.) node (12b) {};
%
    \filldraw[fill=white,draw=black] (11) circle (.2); 
    \filldraw[fill=white,draw=black] (21) circle (.2);
    \filldraw[fill=white,draw=black] (31) circle (.2);
    \filldraw[fill=white,draw=black] (41) circle (.2);     
  \filldraw[fill=orange!50,draw=black] (12b)++(-.2,-.2) rectangle ++(.4,.4);  
   \end{scope}  
 \end{tikzpicture}
\end{align}
where \ \tikz[baseline=(00.base),thick] \filldraw[fill=orange!50,draw=black] (0,0)
 node (00) [above = .5em,right=1em] {= SU$(2n_1)$} rectangle ++(.4,.4); and \ \tikz[baseline=(00.base),thick] \filldraw[fill=cyan!50,draw=black] (0,0)
 node (00) [above = .5em,right=1em] {= SU$(n_1)$} rectangle ++(.4,.4); flavor nodes.
This is consistent with the brane description discussed in \S\ref{sec:brane_super}. 
 
   \subsection{\texorpdfstring{$\widehat{A}_{0}$ quiver}{Affine A0 quiver}}

Let us then consider the affine quiver $\Gamma = \widehat{A}_0$.
Here we have a parameter $\mu \in \mathbb{C}^\times$ assigned to the loop edge, which is the multiplicative adjoint mass parameter.
Applying the decoupling trick, it becomes:
\begin{align}
  \begin{tikzpicture}[baseline=(current bounding box.center),thick]
   \draw (0,0) arc (-90:270:.5) node [above right=1.3em] {$\mu$};
   \filldraw[fill=white,draw=black] (0,0) circle (.2) node [below=.7em] {$n_0|n_1$};
   \draw[-latex,very thick,blue] (2,0) -- ++(1,0);
     \begin{scope}[shift={(5,0)}]
      \draw (0,.75) arc (-90:270:.4);
      \draw (0,-.75) arc (90:450:.4);      
      \draw (0,.75) to [bend right] (0,-.75);
      \draw (0,.75) to [bend left] (0,-.75);
      \draw[dashed] (0,.75) to [bend right=70] (0,-.75);
      \draw[dashed] (0,.75) to [bend left=70] (0,-.75);            
     \filldraw[fill=white,draw=black] (0,.75) circle (.2) node [right=1em] {$\mathfrak{q}_0$}; 
     \filldraw[fill=white,draw=black] (0,-.75) circle (.2) node [right=1em] {$\mathfrak{q}_1$}; 
      \draw[-latex,very thick,blue] (2,0) -- ++(.5,0) node [above] {\textcolor{black}{$\mathfrak{q}_1 \to 0$}} -- ++(.5,0);   
     \end{scope}
   \begin{scope}[shift={(10,0)}]
    \draw (0,.75) -- ++(-.5,-1.5) node (l) {};
    \draw (0,.75) -- ++(.5,-1.5) node (r) {};
    \draw[dashed] (0,.75) -- ++(-1.5,-1.5) node (l-) {};
    \draw[dashed] (0,.75) -- ++(1.5,-1.5) node (r-) {};        
    \filldraw[fill=white,draw=black] (0,.75) circle (.2) node [above=.7em] {$n_0$};
    \filldraw[fill=white,draw=black] (l)++(-.2,-.2) rectangle ++(.4,.4) node [below=1em] {$n_1$};
    \filldraw[fill=white,draw=black] (r)++(-.2,-.2) rectangle ++(.4,.4) node [below=1em] {$n_1$};
    \filldraw[fill=white,draw=black] (l-)++(-.2,-.2) rectangle ++(.4,.4) node [below=1em] {$n_1$};
    \filldraw[fill=white,draw=black] (r-)++(-.2,-.2) rectangle ++(.4,.4) node [below=1em] {$n_1$};           
   \end{scope}   
  \end{tikzpicture}  
\end{align}
Then we have the SU$(n_0)$ gauge node, and with $2 n_1$ {\em positive} and $2 n_1$ {\em negative} fundamental matters.
In this case, due to the adjoint mass parameter, the positive and negative fundamentals cannot be canceled with each other.

\section{ADHM construction of super instanton}\label{sec:super_ADHM}
\index{ADHM construction!supergroup}

For the supergroup YM theory, we can similarly consider an instanton as a solution to the (A)SD YM equation~\eqref{eq:ASDYM_eq}, and apply the ADHM construction to the supergroup YM theory similarly to \S\ref{sec:ADHM_construction}~\cite{Kimura:2019msw}.

\subsubsection{ADHM data}

As mentioned in \S\ref{sec:ADHM_brane}, the ADHM construction is based on the duality between the instanton number $k$ and the gauge group rank $n$ (the vector spaces, $K$ and $N$).
Therefore, for the supergroup theory, the instanton number is promoted to $\mathbb{Z}_2$-graded $(k_0|k_1)$ as well as the gauge group rank $(n_0|n_1)$.
We call the corresponding solution the {\em super instanton}.
In order to describe this super instanton configuration, we define two supervector spaces:
\begin{subequations}
 \begin{align}
  N & = \mathbb{C}^{n_0|n_1} = \mathbb{C}_0^{n_0} \oplus \mathbb{C}_1^{n_1} =: N_0 \oplus N_1
  \, , \\
  K & = \mathbb{C}^{k_0|k_1} = \mathbb{C}_0^{k_0} \oplus \mathbb{C}_1^{k_1} =: K_0 \oplus K_1 
  \, .
 \end{align}
\end{subequations}
We define the base manifold of the super instanton moduli space with the linear maps between the supervector spaces:
\begin{align}
 X = \Hom(K,K) \oplus \Hom(K,K) \oplus \Hom(N,K) \oplus \Hom(K,N)
 \, ,
 \label{eq:super_ADHM_var_space}
\end{align}
and the ADHM variables
\begin{align}
  B_{1,2} \in \Hom(K,K) = \End(K)
  \, , \quad
  I \in \Hom(N,K)
  \, , \quad
  J \in \Hom(K,N)
  \, .
  \label{eq:super_ADHM_data}
\end{align}
In this case, the moment maps $(\mu_\bR,\mu_\bC): X \to \bR^3 \otimes \mathfrak{u}_{k_0|k_1}^*$ are given as the same expression as before~\eqref{eq:moment_maps}, and thus the ADHM equation is formally equivalent to the ordinary case, $\vec{\mu} = (\mu_\bR, \real\mu_\bC, \imag\mu_\bC) = (0,0,0)$.

\subsubsection{Constructing instanton}

Given the ADHM variables, the following process to construct the ASD connection is the same as before.
We start with the dual Dirac operator~\eqref{eq:dual_Dirac_op}\index{dual Dirac operator}, $D^\dag : K \otimes \mathcal{S} \oplus N \to K \otimes \mathcal{S}$, which behaves as $D^\dag D = \Delta \otimes \id_\mathcal{S}$ with $\Delta \to |z|^2 \id_K$ as $z \to \infty$.
Then, the ASD connection is constructed from the normalized zero mode of the dual Dirac operator $D^\dag \Psi = 0$ as $A = \Psi^\dag d \Psi$.
In this case, the instanton charge is similarly computed with a super-analog of Osborn's formula~\eqref{eq:Osborn_formula}:\index{(A)SD YM!---connection}
\begin{align}
 \mathsf{k} := c_2[\mathcal{S}] = \frac{1}{8 \pi^2} \int_{\mathcal{S}} \str F \wedge F = \str_K \id_K = k_0 - k_1 = \sdim_\mathbb{C} K
 \, .
\end{align}
Namely, the instanton charge is given as the superdimension of the supervector space $K$, so that it could be negative.
We call $k_0$ and $k_1$ the positive and negative instanton numbers, respectively.

\subsection{String theory perspective}\label{sec:super_ADHM_brane}

Similarly to \S\ref{sec:ADHM_brane}, the instanton configuration in the supergroup YM theory in four dimensions has a realization as a D0-D4 brane system.
In this case, we should use a peculiar brane, the negative brane (also called the ghost brane~\cite{Okuda:2006fb}).
The negative brane has negative charge and negative tension, which is different from the anti-brane having negative charge with positive tension.

\begin{table}[t]
  \begin{center}
   \renewcommand{\arraystretch}{1.2}   
  \begin{tabular}{c|cccc} \hline\hline
   & instanton & anti-instanton & negative instanton & negative anti-instanton\\ \hline
   SD or ASD & ASD & SD & ASD & SD \\
   topological \# & positive & negative & negative & positive \\
   D-brane & D0 (D0$^+$) & $\overline{\mathrm{D}0}$ ($\overline{\mathrm{D}0}^+$) & D0$^-$ & $\overline{\mathrm{D}0}^-$ \\
   \hline\hline
  \end{tabular}
  \end{center}
 \caption{Properties of instanton, anti-instanton, and negative instanton.}
 \label{tab:super_SDYM}
 \renewcommand{\arraystretch}{1}   
\end{table}

The properties of instanton, anti-instanton, and negative instanton are summarized in Tab.~\ref{tab:super_SDYM}, as a refined version of Tab.~\ref{tab:SDYM}.
It is known that the brane-anti-brane configuration is not stable, namely non-BPS state.
On the other hand, both positive and negative instantons are ASD, and hence their bound state remains BPS.
This is why we can deal with the positive and negative instantons simultaneously using the extended version of the ADHM construction as shown above.

\subsection{Instanton moduli space}\label{sec:super_inst_mod_sp}

Since the ADHM equation $\mu = 0$ has a symmetry $\rU(K) = \rU(k_0|k_1)$, the super instanton moduli space is given as a supergroup hyper-K\"ahler quotient for $(n,k) = (n_0|n_1,k_0|k_1)$:
\begin{align}
 \mathfrak{M}_{n,k} = {\mu}^{-1}(0) /\!\!/\!\!/ \rU(K)
 \, .
\end{align}
Since this moduli space is singular as before, we consider the resolution of the moduli space in order to apply the equivariant integral over the moduli space similarly to \S\ref{sec:ADHM_mod_sp}:
\begin{align}
 \mathfrak{M}_{n,k}^\zeta = \{ \mu_\mathbb{R} = (+\zeta_{\neq 0} \, \id_{K_0}) \oplus (- \zeta_{\neq 0} \, \id_{K_1}),\, \mu_\mathbb{C} = 0 \} /\!\!/\!\!/ \rU(K)
 \, .
\end{align}
Namely, we assign the resolution parameters with opposite signs for the even and odd sectors.
The physical interpretation of this assignment is explained as follows:
The resolution parameter $\zeta$ has an interpretation as the noncommutative parameter on the spacetime $\mathcal{S}$, which is also interpreted as the background flux through the Seiberg--Witten map.
Since the negative sector has an opposite charge, as mentioned in \S\ref{sec:super_ADHM_brane}, the sign of the resolution parameter is now flipped compared to the positive sector.

Hereafter we assume $\zeta>0$.%
\footnote{%
In the case with $\zeta < 0$, we should exchange $(I_0,J_1^\dag) \leftrightarrow (J_0^\dag,I_1)$ in the following argument.
}
Then, the real moment map condition $\mu_\mathbb{R} = \zeta \id_{K_0} \oplus (-\zeta) \id_{K_1}$ is replaced with the stability condition, so that we have another description of the instanton moduli space:\index{stability condition}
\begin{align}
 \widetilde{\mathfrak{M}}_{n,k} = \mu_\mathbb{C}^{-1}(0) /\!\!/ \GL(K)
 \, .
\end{align}
Since, in this case, the resolution parameters is positive $+\zeta > 0$ and negative $- \zeta < 0$ for the positive and negative sectors, we should consider both the stability and co-stability conditions simultaneously:
\begin{align}
 K_0 = \mathbb{C}[B_1, B_2] I_0(N_0)
 \, , \qquad
 K_1 = \mathbb{C}[B_1^\dag, B_2^\dag] J_1^\dag(N_1)
 \, , 
\end{align}
where we define
\begin{align}
 I_\sigma = \Hom(N_\sigma, K)
 \, , \qquad
 J_\sigma = \Hom(K, N_\sigma)
 \, ,
 \qquad (\sigma = 0, 1)
 \, .
\end{align}
The complex superdimension of the moduli space is given similarly to the ordinary case~\eqref{eq:ADHM_dim}:
\begin{align}
 \sdim \widetilde{\mathfrak{M}}_{n,k}
 & = 2 \sdim \Hom(K,K) + \sdim \Hom(N,K) + \sdim \Hom(K,N)
 \nonumber \\
 & \quad
 - \sdim \Hom(K,K) - \sdim \GL(K)
 \nonumber \\
 & = 2 (n_0 - n_1) (k_0 - k_1) = 2 \sdim N \cdot \sdim K
 \, ,
 \label{eq:super_ADHM_dim}
\end{align}
which is a super analog of the formula~\eqref{eq:ADHM_dim}.

\section{Equivariant localization}\label{sec:super_localization}
\index{equivariant!---localization}

Now we apply the equivariant localization scheme to the super instanton moduli space.
In fact, we can perform a parallel analysis as discussed in \S\ref{sec:eq_fixed_point} to the super instanton case.

\subsubsection{Framing and instanton bundles}
 \index{framing!---bundle}
 \index{instanton bundle}

We define the graded framing and instanton bundles over the super instanton moduli space:
\begin{align}
 \mathbf{N} = (\mathbf{N}_i)_{i \in \Gamma_0}
 \, , \qquad
 \mathbf{K} = (\mathbf{K}_i)_{i \in \Gamma_0} 
\end{align}
with
\begin{align}
 \mathbf{N}_i = \mathbf{N}_{i}^0 \oplus \mathbf{N}_{i}^1
 \, , \qquad
 \mathbf{K}_i = \mathbf{K}_{i}^0 \oplus \mathbf{K}_{i}^1
\end{align}
which are further decomposed as
\begin{align}
 \mathbf{N}_{i}^\sigma = \bigoplus_{\alpha = 1}^{n_{i,\sigma}} \mathbf{N}_{i,\alpha}^\sigma
 \, , \qquad
 \mathbf{K}_{i}^\sigma = \bigoplus_{\alpha = 1}^{n_{i,\sigma}} \mathbf{K}_{i,\alpha}^\sigma
 \, .
\end{align}
The automorphism groups are supergroups,
\begin{align}
 \GL(N) = \prod_{i \in \Gamma_0} \GL(n_{i,0}|n_{i,1})
 \, , \qquad
 \GL(K) = \prod_{i \in \Gamma_0} \GL(k_{i,0}|k_{i,1})
\end{align}
and the corresponding Lie superalgebras of the Cartan tori are denoted by
\begin{align}
 \mathsf{a}_i & = \diag(\mathsf{a}_{i,1}^0,\ldots,\mathsf{a}_{i,n_{i,0}}^0,\mathsf{a}_{i,1}^1,\ldots,\mathsf{a}_{i,n_{i,1}}^1)
 \in \operatorname{Lie} \mathsf{T}_{N_i} \subset \mathfrak{gl}_{N_i} 
 \, ,
 \nonumber \\
 \phi_i & = \diag(\phi_{i,1}^0,\ldots,\phi_{i,k_{i,0}}^0,\phi_{i,1}^1,\ldots,\phi_{i,k_{i,1}}^1)
 \in \operatorname{Lie} \mathsf{T}_{K_i} \subset \mathfrak{gl}_{K_i}
 \, .
\end{align}
In this case, we apply the stability and co-stability conditions to $K_0$ and $K_1$ as discussed in \S\ref{sec:super_inst_mod_sp}.
Recalling the corresponding eigenvalues are given in~\eqref{eq:phi_ev}, the equivariant character for each bundle is given by
\begin{subequations}
 \begin{align}
  \ch_\mathsf{T} \mathbf{N}_{i,\alpha}^\sigma & = \np^{\mathsf{a}_{i,\alpha}^\sigma}
  \\
  \ch_\mathsf{T} \mathbf{K}_{i,\alpha}^\sigma & =
   \begin{cases}
    \displaystyle
    \sum_{s \in \lambda_{i,\alpha}^0}
    \np^{\mathsf{a}_{i,\alpha}^0} q_1^{s_1 - 1} q_2^{s_2 - 1}
    & (\sigma = 0) \\
    \displaystyle
    \sum_{s \in \lambda_{i,\alpha}^1}
    \np^{\mathsf{a}_{i,\alpha}^1} q_1^{- s_1} q_2^{- s_2}
    & (\sigma = 1)   
   \end{cases}
 \end{align}
\end{subequations}
which are combined into the equivariant supercharacter:
\begin{align}
 \sch_\mathsf{T} \mathbf{N}_i = \ch_\mathsf{T} \mathbf{N}_i^0 - \ch_\mathsf{T} \mathbf{N}_i^1
 \, , \qquad
 \sch_\mathsf{T} \mathbf{K}_i = \ch_\mathsf{T} \mathbf{K}_i^0 - \ch_\mathsf{T} \mathbf{K}_i^1
 \, .
\end{align}

\subsubsection{Observable bundles}

We then define the graded universal bundle:
\begin{align}
 \mathbf{Y}_\mathcal{S} = (\mathbf{Y}_{\mathcal{S},i})_{i \in \Gamma_0} = (\mathbf{Y}_{\mathcal{S},i}^0 \oplus \mathbf{Y}_{\mathcal{S},i}^1)_{i \in \Gamma_0}
\end{align}
with
\begin{align}
 \mathbf{Y}_{\mathcal{S},i}^\sigma =
 \begin{cases}
  \displaystyle
  \bigoplus_{\alpha = 1}^{n_{i,0}} \mathbf{N}_{i,\alpha}^0 \otimes \mathbf{I}_{\lambda_{i,\alpha}^0}
  & (\sigma = 0) \\[1em]
  \displaystyle
  \bigoplus_{\alpha = 1}^{n_{i,1}} \mathbf{N}_{i,\alpha}^1 \otimes \mathbf{I}_{\lambda_{i,\alpha}^1}^\vee \otimes \det \mathbf{Q}^\vee
  & (\sigma = 1)
 \end{cases}
\end{align}
and also the observable bundle as the pullback at the fixed point $o \in \mathcal{S}$:
\begin{align}
 \mathbf{Y}_o = i_o^* \mathbf{Y}_\mathcal{S} = (\mathbf{N}_i - \wedge \mathbf{Q} \cdot \mathbf{K}_i)_{i \in \Gamma_0}
 \, .
\end{align}
Since $\mathbf{N}$ and $\mathbf{K}$ are graded, while $\wedge \mathbf{Q}$ is the ordinary bundle, the observable bundle is also graded.
Together with the partial reduction of the universal bundle $\mathbf{Y}_\mathcal{S}$ denoted by $\mathbf{X} = \mathbf{Y}_{\mathcal{S}_1} = (\mathbf{X}_i)_{i \in \Gamma_0}$, we obtain the expression
\begin{align}
 \mathbf{Y}_{o} = (\mathbf{Y}_{o,i})_{i \in \Gamma_0}
 = \wedge \mathbf{Q}_1 \cdot \mathbf{X}
 = \wedge \mathbf{Q}_1 \cdot \mathbf{X}^0 \oplus \wedge \mathbf{Q}_1^\vee \cdot \mathbf{X}^1
 \, ,
\end{align}
where the corresponding supercharacter is given by
\begin{align}
 \sch_\mathsf{T} \mathbf{Y}_{i}
 := \sch_\mathsf{T} \mathbf{Y}_{o,i}
 & =
 \ch_\mathsf{T} \left( \wedge \mathbf{Q}_1 \cdot \mathbf{X}^0_i \right)
 - \ch_\mathsf{T} \left( \wedge \mathbf{Q}_1^\vee \cdot \mathbf{X}^1_i \right)
 \nonumber \\
 & =
 (1 - q_1) \ch_\mathsf{T} \mathbf{X}^0_i
 - (1 - q_1^{-1}) \ch_\mathsf{T} \mathbf{X}^1_i
\end{align}
with
\begin{align}
\ch_\mathsf{T} \mathbf{X}_i^\sigma = \sum_{x \in \mathcal{X}_i^\sigma} x  
\end{align}
and
\begin{align}
 \mathcal{X}_i^\sigma =
 \left\{
 x_{i,\alpha,k}^\sigma = \np^{\mathsf{a}_{i,\alpha}^\sigma} q_1^{(-1)^\sigma (k-1)} q_2^{(-1)^\sigma \lambda_{i,\alpha,k}^\sigma}
 \right\}_{\substack{\alpha = 1,\ldots,n_{i,\sigma} \\ k = 1,\ldots,\infty}}
 \, , \quad
 \mathcal{X}^\sigma = \bigsqcup_{i \in \Gamma_0} \mathcal{X}_i^\sigma
 \, .
\end{align}
We can similarly use the expression based on the other reduction $\check{\mathbf{X}} = \mathbf{Y}_{\mathcal{S}_2}$ with the transposed partition.

\subsection{Equivariant index formula}
\index{equivariant!---index}

The graded version of the vector multiplet and the hypermultiplet bundles are given similarly to the ordinary case~\eqref{eq:quiv_vect_hyp_bundles}:
 \begin{align}
  \mathbf{V}_i
  = \frac{\mathbf{Y}_i^\vee \mathbf{Y}_i}{\wedge \mathbf{Q}}
  \, ,
  \qquad
 \mathbf{H}_{e:i \to j} = - \mathbf{M}_e \frac{\mathbf{Y}_i^\vee \mathbf{Y}_j}{\wedge \mathbf{Q}}
  \, .
  \label{eq:super_quiv_vect_hyp_bundles}
\end{align}    
These bundles have decomposition
\begin{align}
 \mathbf{V}_i = \mathbf{V}_i^0 \oplus \mathbf{V}_i^1
 \, , \qquad
 \mathbf{H}_e = \mathbf{H}_e^0 \oplus \mathbf{H}_e^1
 \, ,
\end{align}
where each part is given by
\begin{subequations}
\begin{align}
 \mathbf{V}_i^0 & =
 \frac{\wedge \mathbf{Q}_1^\vee}{\wedge \mathbf{Q}_2} \,
 \left(
 \mathbf{X}_i^{0\vee} \mathbf{X}_i^0
 + \mathbf{X}_i^{1\vee} \mathbf{X}_i^1
 \right)
 & =: \mathbf{V}_{i}^{00} + \mathbf{V}_{i}^{11} \, ,
 \\
 \mathbf{V}_i^1 & =
 \det \mathbf{Q}^\vee
 \frac{\wedge \mathbf{Q}_1^\vee}{\wedge \mathbf{Q}_2^\vee} \,
 \mathbf{X}_i^{0\vee} \mathbf{X}_i^1
 +
 \frac{\wedge \mathbf{Q}_1}{\wedge \mathbf{Q}_2} \,
 \mathbf{X}_i^{1\vee} \mathbf{X}_i^0
 & =: \mathbf{V}_{i}^{01} + \mathbf{V}_{i}^{10}
 \, , \\
 \mathbf{H}_e^0 & =
 - \mathbf{M}_e \,
 \frac{\wedge \mathbf{Q}_1^\vee}{\wedge \mathbf{Q}_2} \,
 \left(
 \mathbf{X}_i^{0\vee} \mathbf{X}_j^0
 + \mathbf{X}_i^{1\vee} \mathbf{X}_j^1
 \right)
 & =: \mathbf{H}_{e}^{00} + \mathbf{H}_{e}^{11} \, ,
 \\
 \mathbf{H}_e^1 & =
 - \mathbf{M}_e \det \mathbf{Q}^\vee \,
 \frac{\wedge \mathbf{Q}_1^\vee}{\wedge \mathbf{Q}_2^\vee} \,
 \mathbf{X}_i^{0\vee} \mathbf{X}_j^1
 - \mathbf{M}_e \,
 \frac{\wedge \mathbf{Q}_1}{\wedge \mathbf{Q}_2} \,
 \mathbf{X}_i^{1\vee} \mathbf{X}_j^0
 & =: \mathbf{H}_{e}^{01} + \mathbf{H}_{e}^{10}
 \, . 
\end{align}
\end{subequations}
Hence, the supercharacter is given by
\begin{subequations}
\begin{align}
 \sch_\mathsf{T} \mathbf{V}_i
 & = \ch_\mathsf{T} \mathbf{V}_i^0 - \ch_\mathsf{T} \mathbf{V}_i^1
 =
 \sum_{\sigma,\sigma' = 0, 1} (-1)^{\sigma + \sigma'} \ch_\mathsf{T} \mathbf{V}_i^{\sigma\sigma'}
 \, , \\
 \sch_\mathsf{T} \mathbf{H}_e
 & = \ch_\mathsf{T} \mathbf{H}_e^0 - \ch_\mathsf{T} \mathbf{H}_e^1
 =
 \sum_{\sigma,\sigma' = 0, 1} (-1)^{\sigma + \sigma'} \ch_\mathsf{T} \mathbf{H}_e^{\sigma\sigma'}
 \, .
\end{align}
\end{subequations}
The full partition function contributions are obtained by applying the equivariant index formula to these bundles:
\begin{subequations}
\begin{align}
 Z_i^\text{vec} = \prod_{\sigma,\sigma' = 0,1} Z_{i,\sigma\sigma'}^\text{vec}
 \, , \qquad &
 Z_{i,\sigma\sigma'}^\text{vec} = \mathbb{I}[(-1)^{\sigma+\sigma'}\mathbf{V}_i^{\sigma\sigma'}]
 \, , \\
 Z_e^\text{bf} = \prod_{\sigma,\sigma' = 0,1} Z_{e,\sigma\sigma'}^\text{bf}
 \, , \qquad &
 Z_{e,\sigma\sigma'}^\text{bf} = \mathbb{I}[(-1)^{\sigma+\sigma'}\mathbf{H}_e^{\sigma\sigma'}]
 \, ,
\end{align}
\end{subequations}
where each contribution in the 5d (K-theory) convention is
\begin{subequations}
\begin{align}
 Z_{i,00}^\text{vec} =
 \prod_{\substack{(x,x') \in \mathcal{X}_i^0 \times \mathcal{X}_i^0 \\ x \neq x'}}
 \frac{\Gamma_q(q_2 x/x';q_2)}{\Gamma_q(q x/x';q_2)}
 \, , \qquad &
 Z_{i,11}^\text{vec} =
 \prod_{\substack{(x,x') \in \mathcal{X}_i^1 \times \mathcal{X}_i^1 \\ x \neq x'}}
 \frac{\Gamma_q(q_2 x/x';q_2)}{\Gamma_q(q x/x';q_2)}
 \, , \\
 Z_{i,01}^\text{vec} =
 \prod_{(x,x') \in \mathcal{X}_i^0 \times \mathcal{X}_i^1}
 \frac{\Gamma_q(q x/x';q_2)}{\Gamma_q(q_1 q x/x';q_2)}
 \, , \qquad &
 Z_{i,10}^\text{vec} =
 \prod_{(x,x') \in \mathcal{X}_i^1 \times \mathcal{X}_i^0}
 \frac{\Gamma_q(q_1^{-1} q_2 x/x';q_2)}{\Gamma_q(q_2 x/x';q_2)}
 \, , \\
 Z_{e:i \to j,00}^\text{bf} =
 \prod_{(x,x') \in \mathcal{X}_i^0 \times \mathcal{X}_j^0}
 \frac{\Gamma_q(\mu_e^{-1} q x/x';q_2)}{\Gamma_q(\mu_e^{-1} q_2 x/x';q_2)}
 \, , \qquad & 
 Z_{e:i \to j,11}^\text{bf} =
 \prod_{(x,x') \in \mathcal{X}_i^1 \times \mathcal{X}_j^1}
 \frac{\Gamma_q(\mu_e^{-1} q x/x';q_2)} {\Gamma_q(\mu_e^{-1} q_2 x/x';q_2)}
 \, , \\
 Z_{e:i \to j,01}^\text{bf} =
 \prod_{(x,x') \in \mathcal{X}_i^0 \times \mathcal{X}_j^1}
 \frac{\Gamma_q(\mu_e^{-1} q_1 q x/x';q_2)}{\Gamma_q(\mu_e^{-1} q x/x';q_2)}
 \, , \qquad &
 Z_{e:i \to j,10}^\text{bf} =
 \prod_{(x,x') \in \mathcal{X}_i^1 \times \mathcal{X}_j^0}
 \frac{\Gamma_q(\mu_e^{-1} q_2 x/x';q_2)}{\Gamma_q(\mu_e^{-1} q_1^{-1} q_2 x/x';q_2)}
 \, . 
\end{align}
\end{subequations}
We can similarly obtain the 4d and 6d results with the corresponding index, replacing the $q$-gamma function $\Gamma_q(z;q_2)$ with $\Gamma_1(z;\epsilon_2)$ and $\Gamma_e(z;q_2,p)$.

\subsection{Instanton partition function}

The equivariant index formula provides the full partition function including both the instanton and the perturbative contributions, which is given as an infinite product.
As discussed in \S\ref{sec:inst_part_func}, we can extract the instanton part from the full partition function by subtracting the perturbative contributions.

The instanton part of the vector multiplet bundle and the (anti)fundamental hypermultiplet bundle (similarly obtained as~\eqref{eq:fund_hyper_inst}) are given as follows:
\begin{subequations} 
 \begin{align}
  \sch_\mathsf{T} \mathbf{V}_i^\text{inst}
  & = \ch_\mathsf{T} \mathbf{V}_{i,0}^\text{inst} - \ch_\mathsf{T} \mathbf{V}_{i,1}^{\text{inst}}
  =
  \sum_{\sigma,\sigma' = 0, 1} (-1)^{\sigma + \sigma'} \ch_\mathsf{T} \mathbf{V}_{i,\sigma\sigma'}^{\text{inst}}
  \, , \label{eq:schV_inst} \\
  \sch_\mathsf{T} \mathbf{H}_e^\text{inst}
  & = \ch_\mathsf{T} \mathbf{H}_{e,0}^{\text{inst}} - \ch_\mathsf{T} \mathbf{H}_{e,1}^{\text{inst}}
  =
  \sum_{\sigma,\sigma' = 0, 1} (-1)^{\sigma + \sigma'} \ch_\mathsf{T} \mathbf{H}_{e,\sigma\sigma'}^{\text{inst}}
  \, \\
  \sch_\mathsf{T} \mathbf{H}_i^\text{(a)f,inst}
  & = \ch_\mathsf{T} \mathbf{H}_{i,0}^\text{(a)f,inst} - \ch_\mathsf{T} \mathbf{H}_{i,1}^\text{(a)f,inst}
  =
  \sum_{\sigma,\sigma' = 0, 1} (-1)^{\sigma + \sigma'} \ch_\mathsf{T} \mathbf{H}_{i,\sigma\sigma'}^\text{(a)f,inst}
 \end{align}
\end{subequations}
where
\begin{subequations}\label{eq:super_inst_bundles}
\begin{align}
 \mathbf{V}_{i,\sigma\sigma'}^\text{inst}
 & =
 - \det \mathbf{Q}^\vee \cdot \mathbf{K}_i^{\sigma\vee} \mathbf{N}^{\sigma'}_i
 - \mathbf{N}^{\sigma\vee}_i \mathbf{K}^{\sigma'}_i
 + \wedge \mathbf{Q}^\vee \cdot \mathbf{K}^{\sigma\vee}_i \mathbf{K}^{\sigma'}_i
 \label{eq:chV_inst2}
 \\
 \mathbf{H}_{i,\sigma\sigma'}^\text{f,inst}
 & = \mathbf{M}^{\sigma\vee}_i \mathbf{K}^{\sigma'}_i
 \\
 \mathbf{H}_{i,\sigma\sigma'}^\text{af,inst}
 & = \det \mathbf{Q^\vee} \cdot \mathbf{K}^{\sigma\vee}_i \widetilde{\mathbf{M}}^{\sigma'}_i
\end{align}
\end{subequations}
with
\begin{align}
 \ch_\mathsf{T} \mathbf{M}^\sigma_i = \sum_{f = 1}^{n^\text{f}_{i,\sigma}} \np^{m_{i,f}^\sigma}
 \, , \qquad
 \ch_\mathsf{T} \widetilde{\mathbf{M}}_i^\sigma = \sum_{f = 1}^{n^\text{af}_{i,\sigma}} \np^{\widetilde{m}_{i,f}^\sigma}
 \, .
\end{align}
Then, the instanton part of the partition function is given by the index with the corresponding supercharacter~\eqref{eq:schV_inst},
\begin{align}
 Z_i^\text{vec,inst}
 = \mathbb{I}[ \mathbf{V}_i^\text{inst} ]
 = \prod_{\sigma, \sigma'=0,1} Z_{i,\sigma\sigma'}^\text{vec,inst}
 \, .
\end{align}
Applying the combinatorial formula shown in Appendix~\ref{sec:comb}, we obtain the diagonal parts of the vector multiplet contribution,
\begin{subequations} 
\begin{align}
 Z_{i,00}^{\text{vec,inst}} & =
 \prod_{\alpha,\beta}^{n_{i,0}}
 \mathcal{Z}_\text{diag}^\text{vec}(\np^{\mathsf{a}_{i,\alpha}^0},\np^{\mathsf{a}_{i,\beta}^0};\lambda_{i,\alpha}^0,\lambda_{i,\beta}^0)
 \, , \\
 Z_{i,11}^{\text{vec,inst}} & =
 \prod_{\alpha,\beta}^{n_{i,1}}
 \mathcal{Z}_\text{diag}^\text{vec}(\np^{\mathsf{a}_{i,\alpha}^1},\np^{\mathsf{a}_{i,\beta}^1};\lambda_{i,\alpha}^1,\lambda_{i,\beta}^1)
\end{align} 
\end{subequations}
where the combinatorial factor is given by
\begin{align}
 \mathcal{Z}_\text{diag}^\text{vec}(\nu,\nu';\lambda_{\alpha},\lambda_\beta)
 & = \prod_{s \in \lambda_{\alpha}}
 \left(1 - \frac{\nu}{\nu'} q_1^{-\ell_{\beta}(s)} q_2^{a_{\alpha}(s)+1}\right)^{-1}
 \prod_{s \in \lambda_{\beta}}
 \left( 1 - \frac{\nu}{\nu'} q_1^{\ell_{\alpha}(s)+1} q_2^{-a_{\beta}(s)} \right)^{-1}
\end{align}
with the arm and leg lengths defined in \eqref{eq:arm_leg}.
The off-diagonal contributions are
\begin{align}
 Z_{i,\sigma\sigma'}^{\text{vec,inst}} & =
 \prod_{\alpha=1}^{n_{i,\sigma}} \prod_{\beta=1}^{n_{i,\sigma'}}
 \mathcal{Z}_{\sigma\sigma'}^\text{vec}(\np^{\mathsf{a}_{i,\alpha}^\sigma},\np^{\mathsf{a}_{i,\alpha}^{\sigma'}};\lambda_{i,\alpha}^\sigma,\lambda_{i,\beta}^{\sigma'})
\end{align}
for $\sigma \neq \sigma'$, where
\begin{subequations} 
\begin{align}
 \mathcal{Z}_{01}^\text{vec}(\nu,\nu';\lambda_{\alpha},\lambda_\beta) & =
 \prod_{s_1=1}^{\check{\lambda}_{\alpha,1}} \prod_{s_2'=1}^{\lambda_{\beta,1}}
 \left( 1 - \frac{\nu}{\nu'} q_1^{\check{\lambda}_{\beta,s_2'}+s_1} q_2^{\lambda_{\alpha,s_1} + s_2'} \right)^{-1}
 \left( 1 - \frac{\nu}{\nu'} q_1^{s_1} q_2^{s_2'} \right)
 \nonumber \\
 & \quad \times
 \prod_{s \in \lambda_{\alpha}}
 \left(1 - \frac{\nu}{\nu'} q_1^{s_1} q_2^{\lambda_{\beta,1} + s_2} \right)
 \prod_{s' \in \lambda_{\beta}}
 \left( 1 - \frac{\nu}{\nu'} q_1^{\check{\lambda}_{\alpha,1} + s_1'} q_2^{s_2'} \right)
 \\
 \mathcal{Z}_{10}^\text{vec}(\nu,\nu';\lambda_{\alpha},\lambda_\beta) & =
 \prod_{s_1=1}^{\check{\lambda}_{\alpha,1}} \prod_{s_2'=1}^{\lambda_{\beta,1}}
 \left( 1 - \frac{\nu}{\nu'} q_1^{-\check{\lambda}_{\beta,s_2'}-s_1+1} q_2^{-\lambda_{\alpha,s_1} - s_2' + 1} \right)^{-1}
 \left( 1 - \frac{\nu}{\nu'} q_1^{-s_1+1} q_2^{-s_2+1} \right)
 \nonumber \\
 & \quad \times
 \prod_{s \in \lambda_{\alpha}}
 \left(1 - \frac{\nu}{\nu'} q_1^{-s_1 + 1} q_2^{-\lambda_{\beta,1} - s_2 + 1} \right)
 \prod_{s' \in \lambda_{\beta}}
 \left( 1 - \frac{\nu}{\nu'} q_1^{-\check{\lambda}_{\alpha,1} - s_1' + 1} q_2^{- s_2' + 1} \right)
\end{align} 
\end{subequations}
The number of factors appearing in $\mathcal{Z}_{\sigma\sigma'}$ is as follows: $|\lambda_\alpha| + |\lambda_\beta|$ factors in the denominator of $\mathcal{Z}_{00}$ and $\mathcal{Z}_{11}$, $|\lambda_\alpha| + |\lambda_\beta| + \check{\lambda}_{\alpha,1} \lambda_{\beta,1}$ factors in the numerator, $\check{\lambda}_{\alpha,1} \lambda_{\beta,1}$ factors in the denominator of $\mathcal{Z}_{01}$ and $\mathcal{Z}_{10}$, so that the numbers of factors in the numerator and the denominator are balanced in total.

We remark that the diagonal contribution is given by the well-known combinatorial formula using the arm and leg lengths of the partition.
The off-diagonal contributions are still finite products written in terms of the partition, but do not have a compact formula similar to the diagonal ones.
This situation is similar to the BCD instanton partition function, involving $\phi_a + \phi_b$ in the contour integral~\cite{Marino:2004cn,Nekrasov:2004vw}.

The bifundamental hypermultiplet contribution is similarly given as follows:
\begin{align}
 Z_{e:i \to j}^{\text{bf,inst}}
 & = \mathbb{I}[\mathbf{H}_{e:i \to j}^\text{bf,inst}]
 = \prod_{\sigma,\sigma' = 0,1} Z_{e:i \to j,\sigma\sigma'}^{\text{bf,inst}}
\end{align}
where
\begin{subequations}
\begin{align}
 Z_{e:i \to j,00}^{\text{bf,inst}}
 & = \prod_{\alpha=1}^{n_{i,0}} \prod_{\beta=1}^{n_{j,0}}
 \mathcal{Z}_\text{diag}^\text{bf}(\np^{\mathsf{a}_{i,\alpha}^0},\np^{\mathsf{a}_{j,\beta}^0},\mu_{e:i \to j};\lambda_{i,\alpha}^0,\lambda_{j,\beta}^0)
 \\
 Z_{e:i \to j,11}^{\text{bf,inst}}
 & = \prod_{\alpha=1}^{n_{i,1}} \prod_{\beta=1}^{n_{j,1}}
 \mathcal{Z}_\text{diag}^\text{bf}(\np^{\mathsf{a}_{i,\alpha}^1},\np^{\mathsf{a}_{j,\beta}^1},\mu_{e:i \to j};\lambda_{j,\beta}^1,\lambda_{i,\alpha}^1)
 \\
 Z_{e:i \to j,\sigma\sigma'}^{\text{bf,inst}}
 & = \prod_{\alpha=1}^{n_{i,\sigma}} \prod_{\beta=1}^{n_{j,\sigma'}}
 \mathcal{Z}_{\sigma\sigma'}^\text{bf}(\np^{\mathsf{a}_{i,\alpha}^\sigma},\np^{\mathsf{a}_{j,\beta}^{\sigma'}},\mu_{e:i \to j};\lambda_{i,\alpha}^\sigma,\lambda_{j,\beta}^{\sigma'})
 \qquad \text{for} \qquad \sigma \neq \sigma'
\end{align} 
\end{subequations}
with
\begin{subequations} 
\begin{align}
 \mathcal{Z}_\text{diag}^\text{bf}(\nu,\nu',\mu;\lambda_{\alpha},\lambda_\beta)
 & = \prod_{s \in \lambda_{\alpha}}
 \left(1 - \mu^{-1} \frac{\nu}{\nu'} q_1^{-\ell_{\beta}(s)} q_2^{a_{\alpha}(s)+1}\right)
 \prod_{s \in \lambda_{\beta}}
 \left( 1 - \mu^{-1} \frac{\nu}{\nu'} q_1^{\ell_{\alpha}(s)+1} q_2^{-a_{\beta}(s)} \right)
\end{align}
\begin{align}
 \mathcal{Z}_{01}^\text{bf}(\nu,\nu',\mu;\lambda_{\alpha},\lambda_\beta) & =
 \prod_{s_1=1}^{\check{\lambda}_{\alpha,1}} \prod_{s_2'=1}^{\lambda_{\beta,1}}
 \left( 1 - \mu^{-1} \frac{\nu}{\nu'} q_1^{\check{\lambda}_{\beta,s_2'}+s_1} q_2^{\lambda_{\alpha,s_1} + s_2'} \right)
 \left( 1 - \mu^{-1} \frac{\nu}{\nu'} q_1^{s_1} q_2^{s_2'} \right)^{-1}
 \nonumber \\
 & \quad \times
 \prod_{s \in \lambda_{\alpha}}
 \left(1 - \mu^{-1} \frac{\nu}{\nu'} q_1^{s_1} q_2^{\lambda_{\beta,1} + s_2} \right)^{-1}
 \prod_{s' \in \lambda_{\beta}}
 \left( 1 - \mu^{-1} \frac{\nu}{\nu'} q_1^{\check{\lambda}_{\alpha,1} + s_1'} q_2^{s_2'} \right)^{-1}
 \\
 \mathcal{Z}_{10}^\text{bf}(\nu,\nu',\mu;\lambda_{\alpha},\lambda_\beta) & =
 \prod_{s_1=1}^{\check{\lambda}_{\alpha,1}} \prod_{s_2'=1}^{\lambda_{\beta,1}}
 \left( 1 - \mu^{-1} \frac{\nu}{\nu'} q_1^{-\check{\lambda}_{\beta,s_2'}-s_1+1} q_2^{-\lambda_{\alpha,s_1} - s_2' + 1} \right)
 \left( 1 - \mu^{-1} \frac{\nu}{\nu'} q_1^{-s_1+1} q_2^{-s_2+1} \right)^{-1}
 \nonumber \\
 & \quad \times
 \prod_{s \in \lambda_{\alpha}}
 \left(1 - \mu^{-1} \frac{\nu}{\nu'} q_1^{-s_1 + 1} q_2^{-\lambda_{\beta,1} - s_2 + 1} \right)^{-1}
 \prod_{s' \in \lambda_{\beta}}
 \left( 1 - \mu^{-1} \frac{\nu}{\nu'} q_1^{-\check{\lambda}_{\alpha,1} - s_1' + 1} q_2^{- s_2' + 1} \right)^{-1}
\end{align} 
\end{subequations}
We remark that the total numbers of the factors appearing in the numerator and the denominator are the same as well as the vector multiplet.

The (anti)fundamental hypermultiplet contribution to the instanton partition function is given by
\begin{align}
 Z_{i}^{\text{(a)f,inst}}
 & = \mathbb{I}[\mathbf{H}_{i}^\text{(a)f,inst}]
 = \prod_{\sigma,\sigma' = 0,1} Z_{i,\sigma\sigma'}^{\text{(a)f,inst}}  
\end{align}
where
\begin{subequations}
\begin{align}
 Z_{i,0\sigma'}^{\text{f,inst}}
 & = \prod_{\alpha=1}^{n_{i,0}} \prod_{f=1}^{n^\text{f}_{i,\sigma'}} \prod_{s \in \lambda_{i,\alpha}^0}
 \left( 1 - \frac{\np^{\mathsf{a}_{i,\alpha}^0}}{\mu_{i,f}^{\sigma'}} q_1^{s_1} q_2^{s_2} \right)^{(-1)^{\sigma'}}
 \\
 Z_{i,1\sigma'}^{\text{f,inst}}
 & = \prod_{\alpha=1}^{n_{i,1}} \prod_{f=1}^{n^\text{f}_{i,\sigma'}} \prod_{s \in \lambda_{i,\alpha}^1}
 \left( 1 - \frac{\np^{\mathsf{a}_{i,\alpha}^1}}{\mu_{i,f}^{\sigma'}} q_1^{-s_1+1} q_2^{-s_2+1} \right)^{(-1)^{\sigma'+1}}
 \\
 Z_{i,0\sigma'}^{\text{af,inst}}
 & = \prod_{\alpha=1}^{n_{i,0}} \prod_{f=1}^{n^\text{af}_{i,\sigma'}} \prod_{s \in \lambda_{i,\alpha}^0}
 \left( 1 - \frac{\tilde{\mu}_{i,f}^{\sigma'}}{\np^{\mathsf{a}_{i,\alpha}^0}} q_1^{-s_1+1} q_2^{-s_2+1} \right)^{(-1)^{\sigma'}}
 \\
 Z_{i,1\sigma'}^{\text{af,inst}}
 & = \prod_{\alpha=1}^{n_{i,1}} \prod_{f=1}^{n^\text{af}_{i,\sigma'}} \prod_{s \in \lambda_{i,\alpha}^1}
 \left( 1 - \frac{\tilde{\mu}_{i,f}^{\sigma'}}{\np^{\mathsf{a}_{i,\alpha}^1}} q_1^{s_1} q_2^{s_2} \right)^{(-1)^{\sigma'+1}}
\end{align}
\end{subequations}

\subsection{Contour integral formula}

Now we consider the contour integral formula for the instanton partition function.
For the moment, we focus on $A_1$ quiver with the (anti)fundamental hypermultiplet for simplicity.

From the instanton part of the bundles over the moduli space \eqref{eq:super_inst_bundles}, we obtain the contour integral formula as follows:\index{LMNS formula!supergroup}
\begin{align}
 Z_{n,k}^\text{inst} = \frac{1}{k_0! k_1!} \frac{[-\epsilon_{12}]^{k_0 + k_1}}{[-\epsilon_{1,2}]^{k_0 + k_1}} \oint_{\mathsf{T}_K} \prod_{\substack{\sigma = 0, 1 \\ a = 1,\ldots,k_\sigma}} \frac{d \phi_a^\sigma}{2 \pi \im} \,
 \prod_{\sigma,\sigma' = 0, 1}
 z_{\sigma\sigma'}^\text{vec} \, z_{\sigma\sigma'}^\text{f} \, z_{\sigma\sigma'}^\text{af}
 \label{eq:super_LMNS}
\end{align}
where each contribution is
\begin{subequations}
\begin{align}
 z_{\sigma\sigma'}^\text{vec}
 & =
 \begin{cases}
  \displaystyle
   \prod_{a = 1}^{k_\sigma}
  P_\sigma(\phi_{a}^\sigma)^{-1}
  \widetilde{P}_\sigma(\phi_a^\sigma + \epsilon_{12})^{-1}
  \prod_{a \neq b}^{k_\sigma}
  \mathscr{S}(\phi_a^\sigma - \phi_b^{\sigma})^{-1}
  & (\sigma = \sigma')
  \\
  \displaystyle
 \prod_{a = 1}^{k_{\sigma'}} P_\sigma(\phi_a^{\sigma'})
 \prod_{a = 1}^{k_\sigma} \widetilde{P}_{\sigma'}(\phi_a^\sigma + \epsilon_{12})
 \prod_{\substack{a = 1,\ldots, k_\sigma \\ b = 1,\ldots, k_{\sigma'}}}
 \mathscr{S}(\phi_b^{\sigma'} - \phi_a^{\sigma})
 & (\sigma \neq \sigma')  
 \end{cases}
 \\
 z_{\sigma\sigma'}^\text{f}
 & =
 \begin{cases}
  \displaystyle
   \prod_{a = 1}^{k_\sigma} P^\text{f}_\sigma(\phi_a^\sigma)
  \\
  \displaystyle
  \prod_{a = 1}^{k_\sigma} P^\text{f}_\sigma(\phi_a^\sigma)^{-1}
 \end{cases} 
 \qquad
 z_{\sigma\sigma'}^\text{af}
 =
 \begin{cases}
  \displaystyle
   \prod_{a = 1}^{k_\sigma} \widetilde{P}^\text{af}_{\sigma'}(\phi_a^\sigma + \epsilon_{12})
  & (\sigma = \sigma')
  \\
  \displaystyle
  \prod_{a = 1}^{k_\sigma} \widetilde{P}^\text{af}_{\sigma'}(\phi_a^\sigma + \epsilon_{12})^{-1}
 & (\sigma \neq \sigma')  
 \end{cases}  
\end{align}
\end{subequations}
with the gauge and matter polynomials
\begin{subequations}
\begin{align}
 P_\sigma(\phi) = \prod_{\alpha = 1}^{n_\sigma} [\phi - \mathsf{a}_\alpha^\sigma]
 \, , \qquad &
 \widetilde{P}_\sigma(\phi) = \prod_{\alpha = 1}^{n_\sigma} [- \phi + \mathsf{a}_\alpha^\sigma]
 \, , \\
 P^\text{f}_\sigma(\phi) = \prod_{\alpha = 1}^{n^\text{f}_\sigma} [\phi - m_f^\sigma]
 \, , \qquad &
 \widetilde{P}^\text{af}_\sigma(\phi) = \prod_{\alpha = 1}^{n^\text{af}_\sigma} [- \phi + \widetilde{m}_f^\sigma]
 \, . 
\end{align}
\end{subequations}
We remark that the contour integral formula~\eqref{eq:super_LMNS} formally coincides with that for quiver gauge theory~\eqref{eq:quiver_LMNS} of $\widehat{A}_1$ quiver.
However, we should be careful about the integration contour since we impose different stability condition for the positive and negative gauge nodes.
We should take the contour to be consistent with the eigenvalues \eqref{eq:phi_ev_I} for the positive node, while \eqref{eq:phi_ev_J} for the negative node.
(We should apply the contour corresponding to \eqref{eq:phi_ev_I} to both of the gauge nodes for the ordinary $\widehat{A}_1$ quiver.)

\part{Quantum Geometry}
\label{part2}

\chapter{Seiberg--Witten geometry}\label{chap:SW_theory}

Seiberg--Witten theory is a geometric framework to describe the low energy effective theory of $\mathcal{N} = 2$ supersymmetric gauge theory in four dimensions~\cite{Seiberg:1994rs,Seiberg:1994aj}.
This geometric point of view gives rise to various interesting insights on gauge theory, including dualities in gauge theory, the brane dynamics in string/M-theory, connections with integrable system, etc.
In this Chapter, we start with description of the low energy behavior of 4d $\mathcal{N} = 2$ gauge theory, and discuss the geometric analysis based on Seiberg--Witten theory.
We will then discuss its generalization to quiver gauge theory and supergroup gauge theory, and address the string/M-theory perspective with the brane description.
We will also discuss the generalization to 5d $\mathcal{N} = 1$ theory compactified on a circle, and 6d $\mathcal{N} = (1,0)$ theory compactified on a torus.

  \section{$\mathcal{N} = 2$ gauge theory in four dimensions}\label{sec:N=2}

  In this Section, we briefly summarize the four-dimensional $\mathcal{N} = 2$ supersymmetric gauge theory, which has eight supercharges.
  See also introductory articles and references therein for details on this topic~\cite{Lerche:1996xu,AlvarezGaume:1996mv,Peskin:1997qi,DHoker:1999yni,Tachikawa:2013kta}.
  
  \subsection{Supersymmetric vacua}\label{sec:SUSY_vac}

  Let $G$ be the gauge group, and its Lie algebra $\mathfrak{g} \in \operatorname{Lie} G$.
  As mentioned in \S\ref{sec:topological_twist}, the $\mathcal{N} = 2$ vector multiplet consists of the $\mathfrak{g}$-valued components $(A_\mu, \lambda^i_\alpha, \tilde\lambda^i_{\dot{\alpha}}, \phi)$ in the adjoint representation of the gauge group $G$.
  The potential function for the complex scalar field $\phi$ is given by
  \begin{align}
   V(\phi) = \frac{1}{2 g^2} \tr [\phi,\phi^\dag]^2
   \, ,
  \end{align}
  so that the supersymmetric vacuum requires the condition
  \begin{align}
   [\phi, \phi^\dag] = 0
   \label{eq:SUSY_vac}
   \, .
  \end{align}
  The solution to this vacuum condition is immediately given by the Cartan subalgebra $\mathfrak{h} \subset \mathfrak{g}$, which is the commuting subalgebra of $\mathfrak{g}$.
  Namely, $\mathcal{N} = 2$ gauge theory has a flat direction in the potential $V(\phi)$, and the vacuum expectation value (vev)\index{vacuum expectation value (vev)} of the complex scalar plays a role of the moduli of the supersymmetric vacua.

  For the simplest example, $G = \SU(2)$, the solution to the condition~\eqref{eq:SUSY_vac} is given by a diagonal traceless matrix:
  \begin{align}
   \phi = 
   \begin{pmatrix}
    \mathsf{a} & 0 \\ 0 & - \mathsf{a}
   \end{pmatrix}
   \, .
  \end{align}
  In this case, the Cartan subalgebra of $\mathfrak{g} = \mathfrak{su}_2$ is $\mathfrak{h} = \mathfrak{u}_1$, which is parametrized by the complex parameter $\mathsf{a} \in \mathbb{C}$.
  Since the scalar field $\phi$ itself is not gauge invariant, we instead use the gauge invariant parameter to parametrize the moduli space of the supersymmetric vacua:
  \begin{align}
   u := \frac{1}{2} \tr \phi^2 = 
   \mathsf{a}^2
   \, .
   \label{eq:u_var}
  \end{align}
  We also use the same symbol for its vev $u = \frac{1}{2} \vev{\tr \phi^2}$ as long as no confusion.
  We remark that this is the second Casimir element constructed from the complex scalar $\phi$.
  In general, the chiral ring operators given by the Casimir elements of the scalar field $\phi \in \mathfrak{h}$ provide the gauge invariant coordinates of the moduli space of the vacua.

  If the scalar field has a non-zero expectation value, $u \neq 0$, the off-diagonal part of the gauge field (W-boson) obtains mass, while the diagonal part (Coulomb field) remains massless.
  Therefore, the gauge symmetry is broken into the Cartan subgroup of $G$, denoted by $H \subset G$, due to the Higgs mechanism.
  Since only the Coulomb field is massless in this case, it is called the {\em Coulomb phase}, or the {\em Coulomb branch of the moduli space of vacua}. \index{moduli space of vacua!Coulomb branch}
  Hence, the Cartan element of the complex scalar is also called the {\em Coulomb moduli}\index{Coulomb moduli} in this context.

  We have focused on the vector multiplet, but in general, we can also incorporate the hypermultiplet.
  In such a case, we may consider the vacua with $u = 0$, but with a non-zero expectation value for the scalar component in the hypermultiplet.
  This situation is instead called the {\em Higgs phase} or the {\em Higgs branch of the moduli space of the supersymmetric vacua} (\S\ref{sec:Higgsing}). \index{moduli space of vacua!Higgs branch}
  We mainly focus on the Coulomb branch of $\mathcal{N} = 2$ gauge theory in the following.

  \subsection{Low energy effective theory}

  As discussed above, we have the massive W-boson and the massless Coulomb field in the Coulomb branch of the moduli space of vacua.
  This implies that the W-boson would be decoupled and only the Coulomb field becomes relevant in the low energy regime of $\mathcal{N} = 2$ gauge theory.
  Let us thus consider the low energy effective description of the Coulomb field in this situation.
  
  We have the following field contents for $\mathcal{N} = 2$ $\rU(1)$ gauge theory:
  \begin{equation}
    \begin{tikzcd}
     & \lambda_\alpha^i
     \arrow[dl,<->]
     \arrow[r,<->]
     & A^i_\mu
     \arrow[dl,<->]
     \\
     \mathsf{a}^i
     \arrow[r,<->]
     & \tilde{\lambda}_\alpha^i
     &
    \end{tikzcd}
  \end{equation}
  for $i = 1, \ldots, \dim \mathfrak{h}$.
  In the $\mathcal{N} = 1$ superfield formalism, these fields are organized in terms of the adjoint chiral and vector multiplets:
 \begin{subequations}
   \begin{align}
    \Phi^i(\mathsf{a}^i,\tilde{\lambda}^i) & = \mathsf{a}^i + 2 \tilde{\lambda}_\alpha^i \theta^\alpha + \cdots \, , \\ 
    W_\alpha^i(A_\mu^i,\lambda^i) & = \lambda_\alpha^i + F^i_{\alpha\beta} \theta^\beta + \cdots \, ,
   \end{align}
  \end{subequations}
  with the fermionic parameter $(\theta,\bar{\theta})$.
  A generic Lagrangian of $\mathcal{N} = 2$ gauge theory is then given as
  \begin{align}
   L 
   = \frac{1}{8\pi^2} 
   \left[
   \int d^2\theta d^2 \bar{\theta} \, K(\Phi^i,\bar{\Phi}^i)
   - \int d^2 \theta \, (2 \pi \im \tau_{ij}(a)) W_{\alpha}^i W^{j\alpha} + \text{c.c.}
   \right]
  \end{align}
  where $K(\Phi,\bar{\Phi})$ is the K\"ahler potential, and $(\tau_{ij})_{i,j = 1,\ldots, \dim \mathfrak{h}}$ is a matrix analog of the complex coupling~\eqref{eq:complex_coupling}.
  The kinetic matrices for the fermions $(\lambda^i, \tilde{\lambda}^i)$ are given as follows:
\begin{subequations}
 \begin{align}
  \tilde{\lambda}^i: \quad
  &
  \frac{1}{4\pi} g_{i \bar{j}} 
  := \frac{1}{4\pi} K_{i\bar{j}} 
  = \frac{1}{4\pi} \frac{\partial^2 K}{\partial \mathsf{a}^i \partial \bar{\mathsf{a}}^j} 
  \\[.5em]
  \lambda^i: \quad
  &
  \frac{1}{2\pi} \imag \tau_{ij} = \frac{1}{4 \pi \im} \left( \tau_{ij} - \bar{\tau}_{ij} \right)
 \end{align}
\end{subequations}
  The $\mathcal{N} = 2$ supersymmetry requires the agreement of these kinetic matrices:
   \begin{align}
   \im^{-1} \left(\tau_{ij} - \bar{\tau}_{ij} \right) = K_{i\bar{j}}
    \, ,
  \end{align}
  which gives rise to the expression
  \begin{subequations}
   \begin{align}
    \tau_{ij}(\mathsf{a}) & = \mathscr{F}_{ij} := \frac{\partial^2 \mathscr{F}}{\partial \mathsf{a}^i \partial \mathsf{a}^j} \, , \\[.5em]
    K(\mathsf{a},\bar{\mathsf{a}}) & = \im \left( \bar{\mathsf{a}}_D^i \mathsf{a}_i - \bar{\mathsf{a}}_i \mathsf{a}_D^i \right)
    \, .
   \end{align}
  \end{subequations}
  We here define a (at least locally) holomorphic function $\mathscr{F}(\mathsf{a})$, which is called the {\em prepotential}\index{prepotential}, and the dual variable (the dual Coulomb moduli),\index{Coulomb moduli!dual---}
  \begin{align}
   \mathsf{a}_{D,i} 
   = \frac{\partial \mathscr{F}}{\partial \mathsf{a}^i}
   \, .
  \end{align}
  This means that there exists one-to-one correspondence between the $\mathcal{N} = 2$ Lagrangian and the holomorphic prepotential $\mathscr{F}(\mathsf{a})$.
  For example, the $\mathcal{N} = 2$ Yang--Mills action is given by the quadratic prepotential:
  \begin{align}
   \mathscr{F}(\phi) = \frac{1}{2} \tau \tr \phi^2
   \, ,
   \label{eq:prepotential_quadratic}
  \end{align}
  which is the unique case providing a renormalizable action.

  From this point of view, the moduli space of vacua is a K\"ahler manifold equipped with the K\"ahler potential $K(\mathsf{a},\bar{\mathsf{a}})$.
  Furthermore, we have shown that the Cartan part of the complex scalar provides the coordinate of the Coulomb branch, and the moduli space depends on it only through the prepotential.
  \index{moduli space of vacua!Coulomb branch}
  A K\"ahler manifold with such an additional property is called the (rigid) special K\"ahler manifold, which plays an important role in the relation to the complex algebraic integrable system~\cite{Gorsky:1995zq,Martinec:1995by,Donagi:1995cf,Seiberg:1996nz,Donagi:1998}.

  \subsection{BPS spectrum}\label{sec:BPS}

  We discuss the role of $(\mathsf{a},\mathsf{a}_D)$ from the supersymmetry algebraic point of view.
  As mentioned in \S\ref{sec:topological_twist}, the supercharges of $\mathcal{N} = 2$ algebra are given by $(Q^i_\alpha,Q^i_{\dot{\alpha}})$ with the index $(\alpha,\dot{\alpha},i)$ for $(\SU(2)_L,\SU(2)_R,\SU(2)_I)$.
  They obey the following relations:
  \begin{subequations}
   \begin{align}
    \{ Q_\alpha^i, Q^{\dag \bar{j}}_{\dot{\beta}} \} & = \delta^{i\bar{j}} \, P_\mu \, \sigma^\mu_{\alpha\dot{\beta}}
    \, , \\
    \{ Q_\alpha^i, Q^j_\beta \} & = \epsilon^{ij} \, \delta_{\alpha\beta} \, Z
    \, ,
   \end{align}
  \end{subequations}
  where $(\epsilon^{ij}, \epsilon_{\alpha\beta})$ are the invariant tensors of $\SU(2)_{I,L}$, $P_\mu$ is the translation generator, and $Z$ is a central element of the algebra, which is peculiar to the extended supersymmetry algebras.
  The center $Z$ provides a bound for the mass spectrum, that is called the BPS bound, so that $Z$ itself is also called the BPS spectrum. \index{BPS spectrum}
  In fact, the center of $\mathcal{N} = 2$ is given by a linear combination:
  \begin{align}
   Z = n \, \mathsf{a} + n_D \, \mathsf{a}_D + \sum_{f = 1}^{n^\text{f}} n_f \, m_f
   \, ,
   \label{eq:BPS_Z}
  \end{align}
  where $(m_f)_{f = 1,\ldots,n^\text{f}}$ are the fundamental mass parameters if there exists the fundamental hypermultiplet with $n,n_D,n_f \in \mathbb{Z}$.
  We will see how these BPS spectra are described from the geometric point of view.

  \section{Seiberg--Witten theory}\label{sec:SW_th}

  We have seen that the low energy effective theory of $\mathcal{N} = 2$ gauge theory is written using the prepotential $\mathscr{F}(\mathsf{a})$.
  The remaining problem is then how to determine the prepotential describing such a low energy theory.

  \subsection{Renormalization group analysis}\label{sec:RG}

  Let us start with the renormalization group analysis of the coupling constant.
  The generic form of the one-loop $\beta$-function is given by
  \begin{align}
   \beta(\mu) 
   & = \mu \frac{d}{d\mu} \left. \frac{8 \pi^2}{g^2} \right|_{\mu}
   = \frac{11}{3} C(\text{adj}) - \frac{2}{3} C(R_\text{Weyl}) - \frac{1}{3} C(R_\text{scalar})
  \end{align}
  where $\mu$ is the energy scale of the coupling constant.
  The representations of the Weyl fermion and the complex scalar field under the gauge group $G$ are denoted by $R_\text{Weyl}$ and $R_\text{scalar}$.
  The coefficient $C(R)$ is the second Casimir element of the representation $R$ under the gauge group $G$:
  \begin{align}
   - \frac{1}{2} \tr_R \left( T^a T^b \right) = C(R) \, \delta^{ab}
   \, ,
   \qquad
   a,b = 1,\ldots, \dim \mathfrak{g}
   \, ,
  \end{align}
  with the generators of the algebra $\mathfrak{g} = \operatorname{Lie} G$ denoted by $(T^a)_{a = 1,\ldots,\dim \mathfrak{g}}$.
  We apply the normalization such that the coefficient for the adjoint representation is given by the dual Coxeter number of the algebra $\mathfrak{g}$, $C(\text{adj}) = h^\vee$.
  For $G = \SU(n)$, the coefficients for the adjoint and the (anti)fundamental representations are given by
  \begin{align}
   C(\text{adj}) = n
   \, , \qquad
   C(\textbf{n}) = C(\overline{\textbf{n}})
   = \frac{1}{2}
   \, .
  \end{align}
  
  In the case of $\mathcal{N} = 2$ gauge theory, the vector multiplet consists of two Weyl fermions and a single complex scalar in addition to the gauge field in the adjoint representation.
  Incorporating the hypermultiplet contribution, the total one-loop $\beta$-function is then given by
  \begin{align}
   \mu \frac{d}{d\mu} \frac{8 \pi^2}{g^2}
   = 2 C(\text{adj}) - C(\text{hyp})
   =: b
   \, .
   \label{eq:RG_b}
  \end{align}
  For example, $\mathrm{SU}(n)$ YM theory with $n^\text{f}$ fundamental hypermultiplets yields $b = 2n - n^\text{f}$ since $\mathcal{N} = 2$ hypermultiplet consists of $\mathcal{N} = 1$ chiral and anti-chiral multiplets.
   Hence, the one-loop solution to the renormalization flow of the coupling constant is 
  \begin{align}
   \exp \left( - \frac{8\pi^2}{g^2} \right) = \left( \frac{\Lambda}{\mu} \right)^b
   \label{eq:coupling_with_scale}
  \end{align}
  where $\Lambda$ is a constant of the integration, interpreted as the dynamical scale of the system.
  As long as the coefficient $b$ is positive, the coupling constant becomes small in the high energy regime compared to the dynamical scale (asymptotic freedom).

 \subsection{One-loop exactness}

 We remark that, as mentioned earlier, it is natural to consider the complexified coupling constant \eqref{eq:complex_coupling} in supersymmetric gauge theories, so that the instanton fugacity \eqref{eq:inst_fugacity} is interpreted as a complexification of \eqref{eq:coupling_with_scale}. \index{fugacity (instanton)}
 The imaginary part of the complex coupling $\imag \tau \propto 1/g^2$ receives the quantum correction, while the real part $\real \tau \propto \theta$ does not, since it is a coefficient of the total derivative (topological) term $\tr F \wedge F$.
 Hence, requiring the holomorphy of the prepotential for $\mathcal{N} = 2$ gauge theory, the quantum correction to the coupling constant turns out to be one-loop exact:
 Higher corrections are given by positive power terms of $g^2$, which are not holomorphic any more.

 In fact, once we have a matter content with $b = 0$, it does not receive any quantum correction to the coupling constant since the corresponding $\beta$-function shall be zero at all orders.
 Such a theory may be interpreted as a marginal deformation of superconformal field theory.
 For $G = \SU(n)$, for example, we obtain such a situation, if it contains $2n$ (anti)fundamental hypermultiplets, or a single adjoint hypermultiplet.
 The latter theory is in particular given by a mass deformation of $\mathcal{N} = 4$ theory, which is called $\mathcal{N} = 2^*$ theory:
 It is enhanced to $\mathcal{N} = 4$ supersymmetric theory by turning off the adjoint mass parameter.
 See also the behavior of the partition function discussed in \S\ref{sec:adjoint_bundle}.

 \subsection{$\SU(2)$ theory}

 Let us focus on $\SU(2)$ theory for the moment.
 First of all, the bare $\rU(1)$ and $\SU(2)$ coupling constants have the relation, $\tau_{\rU(1)} = 2 \tau_{\SU(2)} =: 2 \tau_0$.
 Applying the one-loop renormalization equation, we obtain
 \begin{align}
 \tau(\mathsf{a}) 
  = 2 \tau_0 - \frac{8}{2 \pi \im} \log \frac{a}{\Lambda_0}
  = - \frac{8}{2 \pi \im} \log \frac{a}{\Lambda}
 \end{align}
 where we denote the effective $\rU(1)$ coupling by $\tau(\mathsf{a})$, and $\Lambda_0$ is the energy scale corresponding to the bare coupling constant $\tau_0$:
 \begin{align}
  \Lambda^4 = \Lambda^4_0 \, \np^{2 \pi \im \tau_0}
  \, .
 \end{align}
 Therefore, the corresponding prepotential and the dual variable are given at the one-loop level as
 \begin{subequations}
 \begin{align}
  \mathscr{F}(\mathsf{a}) & 
  = \tau_0 \mathsf{a}^2 - \frac{4 \mathsf{a}^2}{2 \pi \im} \log \frac{\mathsf{a}}{\Lambda_0} + \cdots
  = - \frac{4 \mathsf{a}^2}{2 \pi \im} \log \frac{\mathsf{a}}{\Lambda} + \cdots
  \, , \\
  \mathsf{a}_D & 
  = 2 \tau_0 \mathsf{a} - \frac{8 \mathsf{a}}{2 \pi \im} \log \frac{\mathsf{a}}{\Lambda_0} + \cdots
  = - \frac{8 \mathsf{a}}{2 \pi \im} \log \frac{\mathsf{a}}{\Lambda} + \cdots
  \, .
 \end{align}
 \end{subequations}
 There exist further corrections, but they are suppressed in the perturbative regime, $|\mathsf{a}/\Lambda| \gg 1$.
 In terms of the gauge invariant variable $u$~\eqref{eq:u_var}, they are rewritten as
 \begin{align}
  \mathscr{F}(\mathsf{a}) = - \frac{u}{2 \pi \im} \log \frac{u^2}{\Lambda^4} + \cdots
  \, , \qquad
  \mathsf{a}_D = - \frac{2 u^{\frac{1}{2}}}{2 \pi \im} \log \frac{u^2}{\Lambda^4} + \cdots
  \, .
  \label{eq:u_var_form}
 \end{align}

 Based on the expressions above, we then study the behavior of the Coulomb moduli and its dual variable $(\mathsf{a},\mathsf{a}_D)$ on the $u$-plane. \index{Coulomb moduli}
 Let $u = \np^{\im \vartheta} |u|$ and $\vartheta \to \vartheta + 2 \pi \im$, we obtain
 \begin{align}
  \begin{pmatrix}
   \mathsf{a} \\ \mathsf{a}_D
  \end{pmatrix}
  \ \xrightarrow{\vartheta \to \vartheta + 2 \pi \im} \
  \begin{pmatrix}
   - \mathsf{a} \\ - \mathsf{a}_D + 4 \mathsf{a}
  \end{pmatrix}
  = 
  M_\infty
  \begin{pmatrix}
   \mathsf{a} \\ \mathsf{a}_D
  \end{pmatrix}
 \end{align}
 where $M_\infty$ is called the monodromy matrix (of the infinity):
 \begin{align}
  M_\infty =
  \begin{pmatrix}
   -1 & 0 \\ 4 & -1
  \end{pmatrix}
  \in
  \mathrm{SL}(2,\mathbb{Z})
  \, .
 \end{align}
 Hence, there exists a singularity on the complex $u$-plane.
 In fact, the expression of the prepotential and the dual variable ~\eqref{eq:u_var_form} implies the singularities at $u = \pm \Lambda^2$, which correspond to the strong coupling limit $\tau(\mathsf{a}) \to 0$.

 \subsubsection{Seiberg--Witten curve}

 The essential part of Seiberg--Witten theory is to construct the algebraic curve, a.k.a. the {\em Seiberg--Witten curve}, which geometrically encodes all the information about the low energy behavior of $\mathcal{N} = 2$ gauge theory. 
 For this purpose, we take into account the monodromy behavior on the $u$-plane discussed above.
 We begin with the algebraic curve, that we call the Seiberg--Witten curve for $\SU(2)$ supersymmetric Yang--Mills (SYM) theory: 
 \begin{align}
  \Sigma = \{ (x, y) \in \mathbb{C} \times \mathbb{C}^\times \mid \Lambda^2 \left( y + y^{-1} \right) = x^2 - u\}
  \, .
  \label{eq:SW_curve_SU2_1}
 \end{align}
\index{Seiberg--Witten curve!SU(2)}%
 Now the mass dimensions of the variables are
 \begin{align}
  [x] = 1
  \, , \qquad
  [y] = 0
  \, , \qquad
  [u] = 2
  \, , \qquad
  [\Lambda] = 1
  \, .
 \end{align}
 We will also use another convention by the shift, $y \to y / \Lambda^2$:
 \begin{align}
  \Sigma: \ y + \frac{\Lambda^{4}}{y} = x^2 - u
  \label{eq:SW_curve_SU2_2}
 \end{align}
 with the dimension $[y] = 2$.

\begin{figure}[t]
\begin{center}
 \begin{tikzpicture}[scale = .7]

  
  \filldraw [thick,fill=white,draw=black] 
  (-3.5,-3) -- (-2,-1.5) -- (3.5,-1.5) -- (2,-3) -- cycle;
  
  
  \filldraw [fill=blue,opacity=0.4,draw=none] (-2,-2.25) rectangle (-.5,0);
  \filldraw [fill=blue,opacity=0.4,draw=none] (.5,-2.25) rectangle (2,0);  
  
  
  \filldraw [thick,fill=white,opacity=0.7,draw=black] 
  (-3.5,-.75) -- (-2,.75) -- (3.5,.75) -- (2,-.75) -- cycle;
  
  
  \draw [thick] (-2.,0) node [above] {\footnotesize{$x_1^-$}} -- (-.5,0) node [above] {\footnotesize{$x_1^+$}};
  \draw [thick] (.5,0) node [above] {\footnotesize{$x_2^-$}} -- (2.,0) node [above] {\footnotesize{$x_2^+$}};
  
  \draw [thick] (-2.,-2.25) -- (-.5,-2.25);
  \draw [thick] (.5,-2.25) -- (2.,-2.25);

  \draw [ultra thick,-latex] (4,-.75) -- ++(1,0);
  

    
  \draw [thick] (8,0) circle [x radius=2, y radius = .7];
  \draw [thick] (8,-2.25) circle [x radius=2, y radius = .7];
  
  \filldraw [fill=blue,opacity=0.4,draw=none] (6.5,-2.25) rectangle (7.5,0);
  \filldraw [fill=blue,opacity=0.4,draw=none] (8.5,-2.25) rectangle (9.5,0);  
  
  \draw [thick] (6.5,0) -- (7.5,0);
  \draw [thick] (8.5,0) -- (9.5,0);
  
  \draw [thick] (6.5,-2.25) -- (7.5,-2.25);
  \draw [thick] (8.5,-2.25) -- (9.5,-2.25);

  
  \draw [ultra thick,-latex] (11,-.75) -- ++(1,0);
  
  
  

  
  \begin{scope}[shift={(15.5,0)}]	
  
  \draw [thick] (0,-1) circle [x radius = 2.5, y radius = 1.5];
  
   \draw [thick] (-1,-.75) to [bend right = 55] (1,-.75);
  \draw [thick] (-.8,-1) .. controls (-.6,-.8) and (-.2,-.7) .. (0,-.7) .. controls (.2,-.7) and (.6,-.8) .. (.8,-1);
  

   \begin{scope}[shift={(-.5,-1.15)}]   

   \draw[rotate = 70,thick,teal] (0,0) arc [start angle = 0, end angle = 180, x radius = .65, y radius = .3] node [below = .2em,black] {$A$};

   \draw[-latex, rotate = 70,thick,teal] (0, 0) arc [start angle = 0, end angle = 120, x radius = .65, y radius = .3];

   \draw[dotted, rotate = 70,thick,teal] (0, 0) arc [start angle = 0, end angle = -180, x radius = .65, y radius = .3];  

   \end{scope}


   \draw[thick,red] (1.5,-.8) arc [start angle = 0, end angle = 360, x radius = 1.5, y radius = .8];
   \draw[thick,red,-latex] (1.5,-.8) arc [start angle = 0, end angle = 320, x radius = 1.5, y radius = .8] node  [below right,black] {$B$};
  \end{scope}

 \end{tikzpicture}

 \vspace{2em}

 \begin{tikzpicture}[scale = .8]

  
  \filldraw [thick,fill=white,draw=black] 
  (-3.5,-3) -- (-2,-1.5) -- (3.5,-1.5) -- (2,-3) -- cycle;
  
  
  \filldraw [fill=blue,opacity=0.4,draw=none] (-2,-2.25) rectangle (-.5,0);
  \filldraw [fill=blue,opacity=0.4,draw=none] (1,-2.25) rectangle (2,0);  
  
  
  \filldraw [thick,fill=white,opacity=0.7,draw=black] 
  (-3.5,-.75) -- (-2,.75) -- (3.5,.75) -- (2,-.75) -- cycle;
  
  
  \draw [thick] (-2.,0) -- (-.5,0);
  \draw [thick] (1.,0) -- (2.,0);
  
  \draw [thick] (-2.,-2.25) -- (-.5,-2.25);
  \draw [thick] (1.,-2.25) -- (2.,-2.25);
 
 
  \node at (0, 0) {$A$};
  
  
  \draw[ -latex, thick, rotate = 10, teal] (-1.3, .6) arc [start angle = -270, end angle = 90, x radius = 1, y radius = .4];


  \begin{scope}[shift={(9,0)}]	

  
  \filldraw [thick,fill=white,draw=black] 
  (-3.5,-3) -- (-2,-1.5) -- (3.5,-1.5) -- (2,-3) -- cycle;

  
   \draw[thick,->-,red] (1.5,0) -- ++(0,-2.25);
   \draw[thick,->-,rounded corners,red] (1.5,-2.25) -- (1.3,-2.45) -- (-1.45,-2.45) -- (-1.25,-2.25);
   \draw[thick,->-,red] (-1.25,-2.25) -- ++(0,2.25);

  
  \filldraw [fill=blue,opacity=0.4,draw=none] (-2,-2.25) rectangle (-.5,0);
  \filldraw [fill=blue,opacity=0.4,draw=none] (1,-2.25) rectangle (2,0);  
  
  
  
  
  \filldraw [thick,fill=white,opacity=0.7,draw=black] 
  (-3.5,-.75) -- (-2,.75) -- (3.5,.75) -- (2,-.75) -- cycle;

   
   \draw[thick,->-,rounded corners,red] (-1.25,0) -- (-1.05,.2) -- (1.7,.2) -- (1.5,0);
   
  
  \draw [thick] (-2.,0) -- (-.5,0);
  \draw [thick] (1.,0) -- (2.,0);
  
  \draw [thick] (-2.,-2.25) -- (-.5,-2.25);
  \draw [thick] (1.,-2.25) -- (2.,-2.25);
  

 
  \node at (.3, .5) {$B$};

  \end{scope}
  
 \end{tikzpicture}
\end{center}
 \caption{%
 (Top) Compactification of the Riemann surface: from two cut sheets to a torus.
 The torus has two non-contractable cycles, called $A$ and $B$ cycles.
 (Bottom) $A$ cycle and $B$ cycle on the sheets.
 There are four branch points $(x_1^\pm,x_2^\pm)$ given by~\eqref{eq:branch_points} on the sheets. 
 }
 \label{fig:Riemann_surface}
\end{figure}
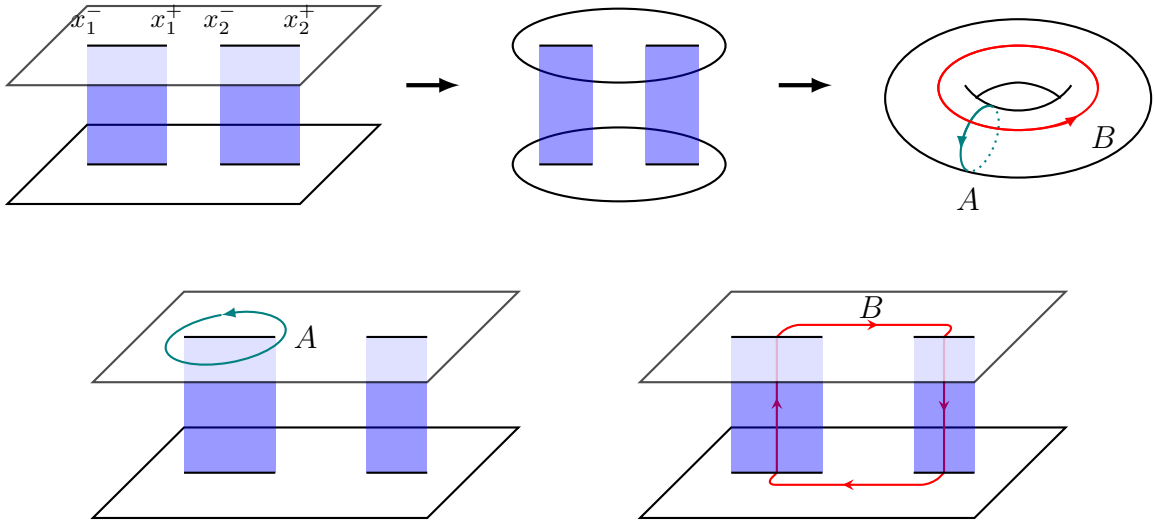

 Solving the algebraic relation \eqref{eq:SW_curve_SU2_1}, we obtain
 \begin{align}
  y = \frac{1}{2\Lambda^2} \left[ (x^2 - u) \pm \sqrt{(x^2 - u)^2 - 4 \Lambda^4} \right]
  \, ,
  \label{eq:y_solution_SU2}
 \end{align}
 where we find four branch points at
 \begin{align}
  x_1^\pm = \sqrt{u \pm 2 \Lambda^2}
  \, , \qquad
  x_2^\pm = - \sqrt{u \pm 2 \Lambda^2}
  \, .
  \label{eq:branch_points}
 \end{align}
 Due to the square root singularity, we have two branch cuts between $(x_i^-,x_i^+)_{i = 1, 2}$ on the sheets, from which we obtain a torus via compactification as shown in Fig.~\ref{fig:Riemann_surface}.
 There exist two non-contractable cycles on the torus, called $A$ and $B$ cycles as long as all the branch points are not degenerated.
 In fact, at $u = \pm 2 \Lambda^2$ and $u = \infty$,%
 \footnote{%
 We may rescale the dynamical parameter $2 \Lambda^2 \to \Lambda^2$ to obtain the agreement with the previous convention.
 }
 two of the branch points are degenerated, and this picture is not available any longer:
 \begin{subequations}
  \begin{align}
   u \to & \pm 2 \Lambda^2: \quad
   x_i^\mp \ \longrightarrow \ 0
   & (B \ \text{cycle shrinks}) \\
   u \to & \hspace{1.8em} \infty: \quad   
   (x_1^\pm,x_2^\pm) \ \longrightarrow \ (+ \infty, - \infty)
   & (A \ \text{cycle shrinks}) 
  \end{align}
 \end{subequations} 
 This is consistent with the previous argument on the singularities on the $u$-plane based on the prepotential.

 \subsubsection{Cycle integrals}

 Based on this geometric setup, we can obtain the information about the low energy effective theory of $\mathcal{N} = 2$ theory.
 We define a tautological one-form on the Seiberg--Witten curve:
  \begin{align}
   \lambda = x \frac{dy}{y} = x d \log y
   \, ,
  \end{align}
  and also the symplectic two-form
  \begin{align}
   \omega = d\lambda = dx \wedge d \log y
   \, .
  \end{align}
  Then, it is claimed that the Coulomb moduli and its dual are obtained by the contour integrals along $A$ and $B$ cycles:\index{Coulomb moduli}
  \begin{subequations}
   \begin{align}
    \mathsf{a} & = \frac{1}{2 \pi \im} \oint_A \lambda \, , \\
    \mathsf{a}_D & = \frac{1}{2 \pi \im} \oint_B \lambda \, .
   \end{align}
  \end{subequations}
  Recalling the expression of the central element in the $\mathcal{N} = 2$ algebra~\eqref{eq:BPS_Z}, the contour integral computes the mass of a BPS state.%
  %
  \footnote{%
  At this point, we do not yet include the fundamental hypermultiplet.
  See \S\ref{sec:N=2SQCD} for the case with fundamental matter.
   }
  For example, in the perturbative regime $|u| \gg |\Lambda|^2$, we see $|y| \simeq 1$ from \eqref{eq:y_solution_SU2}, and $x \simeq u^{\frac{1}{2}}$ from \eqref{eq:SW_curve_SU2_1}.
  Therefore the $A$ cycle integral is given as
  \begin{align}
   \mathsf{a} \simeq \frac{1}{2 \pi \im} \oint_{|y| = 1} u^{\frac{1}{2}} \frac{dy}{y} = u^{\frac{1}{2}}
   \, ,
  \end{align}
  which reproduces the perturbative relation.
  
  The complex coupling is then given as follows:
  \begin{align}
   \tau(\mathsf{a}) = \frac{\partial \mathsf{a}_D}{\partial \mathsf{a}} = \frac{\partial \mathsf{a}_D / \partial u}{\partial \mathsf{a} / \partial u}
   \, .
  \end{align}
  Taking the $u$-variable derivative of the one-form $\lambda$,
  \begin{align}
   \frac{\partial}{\partial u} \lambda = \left(\frac{\partial x}{\partial u}\right)\Bigg|_{y : \, \text{fixed}} \frac{dy}{y} \stackrel{\eqref{eq:SW_curve_SU2_1}}{=} \frac{1}{2} \frac{dy}{xy}
   \, ,
  \end{align}
  we obtain the expression as a ratio of the cycle integrals:
  \begin{align}
   \tau(\mathsf{a}) = \left( \oint_B \frac{dy}{xy} \right) \Big/ \left( \oint_A \frac{dy}{xy} \right)
   \, ,
  \end{align}
  which is called the complex structure of the torus. 
  Its imaginary part is always positive, $\imag \tau(\mathsf{a}) \propto 1/g^2 > 0$, which guarantees that the kinetic term of the action always has a correct sign.

  \subsection{$\SU(n)$ theory}\label{sec:SW_curve_SU(n)}

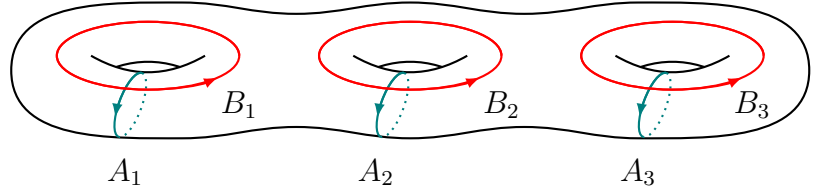
\begin{figure}[t]
 \begin{center}
  \begin{tikzpicture}[thick,scale = 1.5]

   \draw 
   (-.5,0) to [out=down,in=left]
   (1,-.6) to [out=right,in=left]   
   (2,-.5) to [out=right,in=left]
   (3,-.6) to [out=right,in=left]
   (4,-.5) to [out=right,in=left]
   (5,-.6) to [out=right,in=down]
   (6.5,0)   to [out=up,in=right]
   (5,.6)  to [out=left,in=right]
   (4,.5)  to [out=left,in=right]
   (3,.6)  to [out=left,in=right]
   (2,.5)  to [out=left,in=right]
   (1,.6)  to [out=left,in=up]  (-.5,0);

   \foreach \x in {.7,3,5.3}{
       
    \begin{scope}[shift={(\x,-.17)}]

     \draw [clip] (-.5,.3) to [bend right] (.5,.3);
     \draw (-.5,0.1) to [bend left] (.5,0.1);
    
    \end{scope}

   \begin{scope}[shift={(\x,-.17)}]

     \draw[rotate = 70,teal] (.12, .1) arc [start angle = 0, end angle = 180, x radius = .3, y radius = .1];

   \draw[-latex, rotate = 70,teal] (.12, .1) arc [start angle = 0, end angle = 120, x radius = .3, y radius = .1];

   \draw[dotted, rotate = 70,teal] (.12, .1) arc [start angle = 0, end angle = -180, x radius = .3, y radius = .1];       

    \draw[red] (.8,.3) arc [start angle = 0, end angle = 360, x radius = .8, y radius = .3];
    \draw[-latex,red] (.8,.3) arc [start angle = 0, end angle = 320, x radius = .8, y radius = .3];
    
    \end{scope}

   }
   
   \node at (.5,-.9) {$A_1$};
   \node at (2.7,-.9) {$A_2$};
   \node at (5,-.9) {$A_3$};   

   \node at (1.5,-.3) {$B_1$};
   \node at (3.8,-.3) {$B_2$};
   \node at (6.,-.3) {$B_3$};   
   
  \end{tikzpicture}
 \end{center}
  \caption{The hyperelliptic curve of genus $g = 3$ (the Seiberg--Witten curve for $\SU(4)$ SYM theory).
  There exist $2g$ non-contractable cycles denoted by $(A_\alpha,B_\alpha)_{\alpha = 1,\ldots,g}$.}
 \label{fig:cycles}
 \end{figure}

  We start with a remark that the defining relation of the curve $\Sigma$ \eqref{eq:SW_curve_SU2_1} is written in terms of the complex scalar field as%
  \footnote{%
  The Seiberg--Witten curve for generic $G$ is given by the spectral curve of the Toda integrable system associated with the affinization of the Langlands dual group of $G$, denoted by $\widehat{^LG}$~\cite{Martinec:1995by}.
  }
 \begin{align}
  x^2 - u = \det(x - \phi)
  \, .
 \end{align}
 This expression is in fact valid for generic $\SU(n)$ gauge theory.
 Hence, the Seiberg--Witten curve for $\SU(n)$ SYM theory is similarly given by
 \begin{align}
  \Sigma: \ \Lambda^n \left( y + y^{-1} \right) = \det(x - \phi) 
  \, .
  \label{eq:SW_curve_SUn}
 \end{align}\index{Seiberg--Witten curve!SU(n)@SU($n$)}%
 In this case, the right hand side is a degree-$n$ polynomial in $x$, while the degree of the variable $y$ is still two.
 Thus, there exist $n$ branch cuts on two sheets, which gives rise to the hyperelliptic curve of genus $g = n - 1$ as shown in Fig.~\ref{fig:cycles}.
 In this case, there are $2g$ non-contractable cycles $(A_\alpha,B_\alpha)_{\alpha = 1,\ldots,g}$, for which we choose the canonical basis:
 \begin{align}
  A_\alpha \cap B_\beta = \delta_{\alpha\beta}
  \, , \qquad
  A_\alpha \cap A_\beta = B_\alpha \cap B_\beta = \emptyset
  \, .
 \end{align}

 We define the one-form differential on the curve similarly to $\SU(2)$ theory
 \begin{align}
  \lambda = x \frac{dy}{y}
  \, ,
 \end{align}
 and the Coulomb moduli and the dual variables are given by the contour integrals along the non-contractable cycles:
 \begin{align}
  \mathsf{a}_\alpha = \frac{1}{2\pi \im} \oint_{A_\alpha} \lambda
  \, , \qquad
  \mathsf{a}_{D,\alpha} = \frac{1}{2\pi \im} \oint_{B_\alpha} \lambda
  \, .
 \end{align}
 In this case, the coupling matrix $(\tau_{\alpha\beta}(\mathsf{a}))_{\alpha,\beta = 1,\ldots,g}$ is given by the period matrix of the Seiberg--Witten curve $\Sigma$, whose imaginary part is positive definite, $\imag \tau_{\alpha\beta}(\mathsf{a}) > 0$.

 \subsection{$\mathcal{N} = 2$ SQCD}\label{sec:N=2SQCD}

 We discuss the geometric approach to the low energy effective theory in the presence of the hypermultiplet in the fundamental representation, namely $\mathcal{N} = 2$ supersymmetric quantum chromodynamics (SQCD).
 From the renormalization group analysis in \S\ref{sec:RG}, we may incorporate $n^\text{f}$ fundamental hypermultiplets for $n^\text{f} \le 2n$ $(b = 2n - n^\text{f} \ge 0)$ in $\SU(n)$ SYM theory.
 Let $(m_f)_{f = 1,\ldots,n^\text{f}}$ be the mass parameters for them.
 Then, the mass parameter dependence of the Seiberg--Witten curve is incorporated as the pole singularity on the curve~\cite{Hanany:1995na,Argyres:1995wt}.
 This argument is consistent with the correspondence between the BPS spectrum discussed in \S\ref{sec:BPS} and the non-trivial cycle on the curve:\index{BPS spectrum}
 Since we have the poles in addition to the $A$ and $B$ cycles, we may consider the contour surrounding the pole, which gives rise to the fundamental mass $(m_f)_{f = 1,\ldots,n^\text{f}}$.
 Hence, the Seiberg--Witten curve for $G = \SU(n)$ with $\SU(n^\text{f})$ flavor symmetry takes a form of
 \begin{align}
  \Sigma : \
  y + \Lambda^{b} \frac{P(x)}{y} = \det(x - \phi)
  \label{eq:SW_curve_SQCD}
 \end{align}
\index{Seiberg--Witten curve!SQCD}%
where 
we define the matter polynomial $P(x)$:%
\footnote{%
 This is essentially the matter polynomials~\eqref{eq:matter_polynomial_quiver} defined in \S\ref{sec:quiv_LMNS_formula} denoted by $P^\text{f}$ and $\widetilde{P}^\text{af}$.
 We do not distinguish the fundamental and antifundamental matters as long as concerning 4d theory.
}
 \begin{align}
  P(x) = \prod_{f = 1}^{n^\text{f}} ( x - m_f )
  \, .
 \end{align}
 This matter polynomial is interpreted as the characteristic polynomial with respect to the flavor symmetry group $\SU(n^\text{f})$.
 In this convention, the mass dimensions are given as $[y] = n$, $[x] = [\Lambda] = [m_f] = [\phi] = 1$.
 
\section{Quiver gauge theory}\label{sec:SW_curve_quiver}

The geometric implementation of the effective low energy theory is also possible for quiver gauge theory.
Let us briefly mention how to construct it here.
The details of the derivation will be provided in Chapter~\ref{chap:geometry}.

Let us recall the Seiberg--Witten curve for $G = \SU(n)$ theory~\eqref{eq:SW_curve_SUn}, which implies that the gauge group dependence is only found in the $x$-variable part, and the $y$-variable structure is universal: It is in general given as the hyperelliptic curve (degree two for $y$-variable).
In fact, it has been pointed out that one obtains the cubic curve for quiver gauge theory which consists of two gauge node ($A_2$ quiver)~\cite{Shadchin:2005cc}, and afterward, its representation theoretical interpretation is clarified by Nekrasov--Pestun~\cite{Nekrasov:2012xe}.
We start with the observation that the combination of $y$-variable on the curve~\eqref{eq:SW_curve_SUn}, $y + y^{-1}$, is given by the character of the two-dimensional representation of $\SL(2)$, whose Dynkin-quiver diagram is $A_1$.
This observation leads to the following statement:
\begin{itembox}{Seiberg--Witten geometry for $\Gamma$-quiver gauge theory~\cite{Nekrasov:2012xe}}
 Let $G_\Gamma$ be the simple Lie group, whose Dynkin diagram is given by the quiver $\Gamma = ADE$.
 Then, the Seiberg--Witten curve is constructed by characters of the fundamental representations of $G_\Gamma$ (affine characters for $\Gamma = \widehat{ADE}$): $\chi_i(y_1,\ldots,y_{\rk \Gamma})$ for $i \in \Gamma_0$.
 In addition, the character associated to the quiver node $i \in \Gamma_0$ is given by a polynomial in $x$, and its degree coincides with the rank of the gauge group $G_i = \rU(n_i)$ assigned to the node $i \in \Gamma_0$:
 \begin{align}
  \chi_i(y_1,\ldots,y_{\rk \Gamma}) = \mathsf{T}_{i}(x) = x^{n_i} + \cdots
  \label{eq:T_polynomiality}
 \end{align}
\end{itembox}
Here the fundamental representation is given as the highest weight representation associated with each fundamental weight assigned to the quiver node $i \in \Gamma_0$.
Let us examine this statement with several examples.

\subsection{$A_1$ quiver}

The simplest example is $A_1$ quiver, which we have already discussed.
In this case, the Seiberg--Witten curve is given by the relation:
\begin{align}
 \chi_1(y) \ : \quad y + y^{-1} = \mathsf{T}_1(x)
 \label{eq:A1_character}
\end{align}
where $\mathsf{T}_1(x)$ is a polynomial in $x$ of degree $n$ for $\SU(n)$ gauge theory assigned to the node $i = 1$.
This is consistent with the previous expression~\eqref{eq:SW_curve_SUn}, where we omit the gauge coupling factor for simplicity.
Now $\chi_1(y)$ is the two-dimensional representation character of $G_{A_1} = \SL(2)$.
In general, one may consider the higher dimensional representation character, but it would be redundant in the sense of Seiberg--Witten theory because the higher representation curve is in principle reproduced from the fundamental one by using the chiral ring relation.
From the representation theoretical point of view, it is related to the fact that any (higher) representations are constructed by the tensor product of the fundamental representations.

We remark that the relation \eqref{eq:A1_character} is equivalent to
\begin{align}
 y^2 - \mathsf{T}_1(x) \, y + 1 = 0
 \, .
 \label{eq:A1_character2}
\end{align}
We may rewrite this also in the following form:
\begin{align}
 \det(y - L(x)) = 0
\end{align}
where $L(x) \in \SL(2)$ is called the Lax matrix,\index{Lax matrix} such that
\begin{align}
 \tr L(x) = \mathsf{T}_1(x)
 \, , \qquad
 \det L(x) = 1
 \, .
\end{align}
In fact, this expression implies the correspondence between gauge theory and (classical) integrable system: 
The Seiberg--Witten curve is identified with the spectral curve associated with the $\SL(2)$ Lax matrix.

\subsection{$A_2$ quiver}

The next example is $A_2$ quiver, which consists of two gauge nodes, $\SU(n_1) \times \SU(n_2)$, with a single bifundamental hypermultiplet.
In this case, we have two fundamental characters:
\begin{subequations}\label{eq:SW_curve_A2} 
 \begin{align}
  \chi_1(y_{1,2}) : \quad & y_1 + \frac{y_2}{y_1} + \frac{1}{y_2} = \mathsf{T}_1(x) 
  \, ,
  \\
  \chi_2(y_{1,2}) : \quad & y_2 + \frac{y_1}{y_2} + \frac{1}{y_1} = \mathsf{T}_2(x) 
  \, ,
 \end{align}
\end{subequations}
where the polynomial in $x$ is given by
\begin{align}
 \mathsf{T}_i(x) = x^{n_i} + \cdots
 \, .
\end{align}
These $\chi_{1,2}(y_{1,2})$ are the characters of two three-dimensional representations of $G_{A_2} = \SL(3)$.

Since we have two relations with two $y$-variables, $y_{1,2}$, we may eliminate either $y_1$ or $y_2$, and combine two relations to a single equation:
\begin{align}\label{eq:A2_SW_curve_cl}
 y_1^3 - \mathsf{T}_1 \, y_1^2 + \mathsf{T}_2 \, y_1 - 1 = 0
 \qquad \text{or} \qquad
 y_2^3 - \mathsf{T}_2 \, y_2^2 + \mathsf{T}_1 \, y_2 - 1 = 0
 \, .
\end{align}
\index{Seiberg--Witten curve!A2 quiver@$A_2$ quiver}%
This is a cubic curve for $A_2$ quiver, which is consistent with the known result~\cite{Shadchin:2005cc}.
These cubic relations are formulated with the Lax matrix $L(x) \in \SL(3)$ as follows:\index{Lax matrix}
\begin{align}
 \det(y_1 - L(x)) = 0
 \qquad \text{or} \qquad
 \det(y_2 - \overline{L(x)}) = 0 
 \, .
\end{align}

\subsection{$A_3$ quiver}\label{sec:SW_A3}

We consider $A_3$ quiver involving three gauge nodes: 
\dynkin[mark=o,edge length=1cm,label,label macro/.code={\mathrm{SU}(n_{#1})}]{A}{3}
\
Now we have three fundamental characters,
\begin{subequations} 
 \begin{align}
  \chi_1(y_{1,2,3}) : \quad & y_1 + \frac{y_2}{y_1} + \frac{y_3}{y_2} + \frac{1}{y_3}
  & = \mathsf{T}_1(x) 
  \, , \\
  \chi_2(y_{1,2,3}) : \quad &  y_2 + \frac{y_1 y_3}{y_2} + \frac{y_1}{y_3} + \frac{y_3}{y_1} + \frac{y_2}{y_1 y_3} + \frac{1}{y_2}
  & = \mathsf{T}_2(x)
  \, , \\
  \chi_3(y_{1,2,3}) : \quad &  y_3 + \frac{y_2}{y_3} + \frac{y_1}{y_2} + \frac{1}{y_1}
  & = \mathsf{T}_3(x)
  \, .
 \end{align}
\end{subequations}
We remark the symmetry
\begin{align}
 (\chi_1,\chi_2,\chi_3)
 \ \stackrel{1 \leftrightarrow 3}{\longleftrightarrow} \
 (\chi_3,\chi_2,\chi_1)
 \, ,
\end{align}
which corresponds to the automorphism action of $A_3$ quiver:
\begin{equation}
 \begin{tikzpicture}[baseline=(current bounding box.center)]
  \tikzset{/Dynkin diagram/fold style/.style={stealth-stealth,thick, shorten <=2mm,shorten >=2mm,}}
  \dynkin[mark=o,label,root radius = .2cm, edge length = 1cm]{A}{3}
  \dynkinFold[bend left=60]{1}{3}
 \end{tikzpicture}
 \label{eq:A3_folding}
\end{equation}
See also \S\ref{sec:NS_frac} for a related argument (the folding trick).
Then, there are three possible relations to define the algebraic curve:\index{Seiberg--Witten curve!A3 quiver@$A_3$ quiver}
\begin{subequations} 
 \begin{align}
  y_1^4 - \mathsf{T}_{\tiny \yng(1)} \ y_1^3 + \mathsf{T}_{\tiny \yng(1,1)} \ y_1^2 - \mathsf{T}_{\tiny \yng(1,1,1)} \ y_1 + 1 & = 0
  \, , \\
  y_2^6 - \mathsf{T}_{\tiny \yng(1,1)} \ y_2^5 
  + \mathsf{T}_{\tiny \yng(2,1,1)} \ y_2^4 
  - \left( \mathsf{T}_{\tiny \yng(2)} + \mathsf{T}_{\tiny \yng(2,2,2)}\right) \ y_2^3 
  + \mathsf{T}_{\tiny \yng(2,1,1)} \ y_2^4 
  - \mathsf{T}_{\tiny \yng(1,1)} \ y_2 + 1
  & = 0
  \, , \label{eq:SW_curve_A3_2nd} \\
  y_3^4 - \mathsf{T}_{\tiny \yng(1,1,1)} \ y_3^3 + \mathsf{T}_{\tiny \yng(1,1)} \ y_3^2 - \mathsf{T}_{\tiny \yng(1)} \ y_3 + 1 & = 0
  \, ,
 \end{align}
\end{subequations}
where we apply the convention as follows:\\[-.5em]
\begin{subequations}
 \begin{minipage}[c]{.49\textwidth}
  \begin{align}
   \mathsf{T}_\emptyset & = 1 \\
   \mathsf{T}_{\tiny \yng(1)} & = \mathsf{T}_1
   \\
   \mathsf{T}_{\tiny \yng(1,1)} & = \mathsf{T}_2
   \\
   \mathsf{T}_{\tiny \yng(1,1,1)} & = \mathsf{T}_3
  \end{align}
 \end{minipage}
 \begin{minipage}[c]{.49\textwidth}
  \begin{align}
   \mathsf{T}_{\tiny \yng(2,1,1)} & = \mathsf{T}_{\tiny \yng(1)} \mathsf{T}_{\tiny \yng(1,1,1)} - \mathsf{T}_{\emptyset}
   \\
   \mathsf{T}_{\tiny \yng(2)} & = \mathsf{T}_{\tiny \yng(1)} \mathsf{T}_{\tiny \yng(1)} - \mathsf{T}_{\tiny \yng(1,1)}
   \\
   \mathsf{T}_{\tiny \yng(2,2,2)} & = \mathsf{T}_{\tiny \yng(1,1,1)} \mathsf{T}_{\tiny \yng(1,1,1)} - \mathsf{T}_{\tiny \yng(1,1)}   
  \end{align}
 \end{minipage}
\end{subequations}

\noindent
We remark the tensor product relations for $G_{A_3} = \SL(4)$:
\begin{subequations} 
  \begin{align}
   \wedge^2 \, \fontsize{7pt}{0pt}\selectfont \yng(1,1) 
   & = \fontsize{7pt}{0pt}\selectfont \yng(2,1,1) 
   = \left( \,  \yng(1) \otimes \fontsize{7pt}{0pt}\selectfont \yng(1,1,1) \, \right) - \emptyset
   \, , \\[.5em]
   \wedge^3 \, \fontsize{7pt}{0pt}\selectfont \yng(1,1) & = \fontsize{7pt}{0pt}\selectfont \yng(2) \oplus \yng(2,2,2) = \left(\, \yng(1) \otimes \yng(1) - \yng(1,1) \,\right) \oplus \left(\, \yng(1,1,1) \otimes \yng(1,1,1) - \yng(1,1) \,\right)
   \, .
  \end{align}
\end{subequations}
In this case, the second relation~\eqref{eq:SW_curve_A3_2nd} contains the higher representation characters, which are constructed with the fundamental ones.

   \subsection{Generic quiver}\label{sec:SW_curve_quiver_gen}

   For generic quiver $\Gamma$, we obtain the following algebraic relation for $(y_i)_{i \in \Gamma_0}$:\index{Lax matrix}
    \begin{align}
     \Sigma_i: \ \det_{R_i} (y_i - L(x)) = 0
    \end{align}
 where $L(x)$ is the $G_\Gamma$-Lax matrix, and $R_i$ is the fundamental representation associated to the node $i \in \Gamma_0$. 
 The characteristic polynomial associated with generic representation $R$ is defined
 \begin{align}
  \det_R (y - L(x)) 
  = \sum_{k = 0}^{\dim R} (-1)^k \, y^{\dim R - k} \, \tr_{\wedge^k R} L(x)
  = \sum_{k = 0}^{\dim R} (-1)^k \, y^{\dim R - k} \, \mathsf{T}_{\wedge^k R}(x)
 \end{align}
 where the $k$-th antisymmetric tensorial representation of $R$ is denoted by $\wedge^k R$ with 
 \begin{align}
  \mathsf{T}_R(x) = \tr_R L(x) 
  \, .
 \end{align}
 For any representations except for the defining (also called the standard) representation $R = {\scriptsize \yng(1)}$ (and its conjugate), the characteristic polynomial contains higher representations.
 Hence, in order to discuss the Seiberg--Witten curve for $\Gamma$-quiver gauge theory, we should use the defining representation, which ends up with the ordinary characteristic polynomial,
 \begin{align}
  \Sigma: \ \det(y - L(x)) = 0
  \, .
 \end{align}
 This implies the correspondence between $\Gamma$-quiver gauge theory and the (classical) $G_\Gamma$-integrable system.

  \section{Supergroup gauge theory}\label{sec:SW_super}

  We then consider a generalization of the geometric approach to supergroup gauge theory.
  A crucial observation is that, as shown in \S\ref{sec:SW_curve_SU(n)}, the polynomial function appearing in the Seiberg--Witten curve~\eqref{eq:SW_curve_SUn} is given as the characteristic polynomial of the corresponding gauge group.
  This is also the case for the flavor symmetry group (\S\ref{sec:N=2SQCD}).
  For supergroup theory, the corresponding characteristic function is given by the superdeterminant~\eqref{eq:sdet_def}.
  Therefore, the Seiberg--Witten curve for $\SU(n_0|n_1)$ gauge theory is given as follows~\cite{Dijkgraaf:2016lym}:\index{Seiberg--Witten curve!supergroup}
  \begin{align}
   \Sigma : \
   \Lambda^{n_0 - n_1} \left( y + y^{-1} \right) = \sdet(x - \phi) 
   \, .
   \label{eq:SW_curve_SUnm}
  \end{align}
  We remark that $n_0 - n_1$ is the superdimension of the supervector space $\mathbb{C}^{n_0|n_1}$ because the coefficient appearing in the $\beta$-function (\S\ref{sec:RG}) is replaced with the super analog of the Casimir element.

  The complex scalar transforms in the adjoint representation of the supergroup $\SU(n_0|n_1)$.
  Imposing the vacuum condition~\eqref{eq:SUSY_vac}, the scalar field $\phi$ should be diagonalized, and thus factorized, $\phi = \phi_0 \oplus \phi_1$, where each part belongs to the Cartan subalgebra of the subalgebra, $\phi_{\sigma} \in \mathfrak{h}_{\sigma} \subset \mathfrak{su}_{n_\sigma}$ for $\sigma = 0,1$.
  Hence, the supercharacteristic polynomial is given in this condition by
  \begin{align}
   \sdet(x - \phi) = \frac{\det(x - \phi_0)}{\det(x - \phi_1)} 
   = \frac{\prod_{\alpha = 1}^{n_0}(x - \mathsf{a}_\alpha^0)}{\prod_{\alpha = 1}^{n_1}(x - \mathsf{a}_\alpha^1)}
   \label{eq:sdet_Cartan}
  \end{align}
  where $(\mathsf{a}_\alpha^\sigma)_{\alpha = 1,\ldots,n_\sigma}^{\sigma = 0, 1}$ are the Coulomb moduli parameters for the supergroup gauge theory:\index{Coulomb moduli}
  \begin{align}
   \phi = 
   \begin{pmatrix}
    \phi_0 & \\ & \phi_1
   \end{pmatrix}
   \, , \qquad
   \phi_\sigma = 
   \diag(\mathsf{a}_1^\sigma,\ldots,\mathsf{a}_{n_\sigma}^\sigma)
   \, .
  \end{align}
  We can similarly impose the superflavor contribution by the supercharacteristic polynomial of the flavor supergroup $\SU(n_0^\text{f}|n_1^\text{f})$ as in \S\ref{sec:N=2SQCD}.

  Let us then see the relation to other discussions on the supergroup gauge theory.  
  We change the variable $y \to y / \Lambda^{n_0 - n_1} \det (x - \phi_1)$.
  Then the Seiberg--Witten curve~\eqref{eq:SW_curve_SUnm} together with the expression \eqref{eq:sdet_Cartan} is written in the following form:
  \begin{align}
   y + \Lambda^b \, \frac{\det(x - \phi_1)^2}{y} = \det(x - \phi_0)
   \, ,
  \end{align}
  where $b = 2 (n_0 - n_1) = 2 \sdim \mathbb{C}^{n_0|n_1}$.
  Compared with the curve for $\mathcal{N} = 2$ SQCD~\eqref{eq:SW_curve_SQCD}, it agrees with that for $\SU(n_0)$ gauge theory with $n^\text{f} = n_1 + n_1$ flavors.
  In fact, the matter polynomial is given as the square of the characteristic polynomial of $\SU(n_1)$, which means that each flavor appears as a pair, and all the mass parameters are doubled there.
  This is consistent with the argument based on the decoupling trick in \S\ref{sec:decoupling}.

  We remark that the Seiberg--Witten curve~\eqref{eq:SW_curve_SUnm} is also obtained by the unphysical limit of the curve for $\widehat{A}_1$ quiver gauge theory, as discussed in \S\ref{sec:super_quiver_realization}.
  See \cite{Dijkgraaf:2016lym} for details.

 \section{Brane dynamics and $\mathcal{N} = 2$ gauge theory}\label{sec:brane_N=2}

 In this Section, we study $\mathcal{N} = 2$ gauge theory from string/M-theory perspective.
 We will see that the Seiberg--Witten curve encodes a geometric configuration of branes from this point of view.

 \subsection{Hanany--Witten construction}\label{sec:HW_construction}

 As briefly mentioned in \S\ref{sec:ADHM_brane}, a stack of D-branes realizes non-Abelian gauge theory, where open string excitation in the parallel and transverse directions to the brane world-volume gives rise to the gauge field and (real) scalar degrees of freedom: 
 A D$p$ brane has a $(p+1)$-dimensional world-volume, and its codimension is given by $10 - (p+1) = 9 - p$.
 Let us take $p = 3$ to realize $\mathsf{d} = 4$ gauge theory.
 Then, there appear six real scalar fields, which are combined into $\mathcal{N} = 4$ vector multiplet together with the gauge fields (and also the corresponding fermionic partners).
 This is the maximally supersymmetric gauge theory in four dimensions with 16 supercharges.
 In this case, the R-symmetry $\SU(4) = \SO(6)$ has an interpretation as the rotation symmetry of the transverse directions to the D3 branes.

 \begin{figure}[t]
  \begin{center}
   \begin{tikzpicture}[scale=.8]


    \begin{scope}[shift={(-6.5,1.)}]
     
     \draw[-latex] (0,0) -- ++(1,0) node [right] {6};
     \draw[-latex] (0,0) -- ++(0,1) node [above] {45};
     \draw[-latex] (0,0) -- ++(30:1.) node [above] {789};

    \end{scope}
    
    \begin{scope}

     \node at (-2.5,2.5) [left] {(a)};
     

    \draw[thick] (-1.5,-2) -- ++(0,4) node [above] {NS5};
    \draw[thick] (+1.5,-2) -- ++(0,4) node [above] {NS5};

    \draw[latex-latex,thick,blue] (-1.5,-1.7) -- ++(1.5,0) node [below,black] {$L \propto \frac{1}{g^2}$} -- ++(1.5,0);


    \draw  (-1.5, 1.5) node [left] {$\mathsf{a}_1$} -- ++(1.5,0) node [above = .5em] {$n$ D4} -- ++(1.5,0);

    \foreach \x in {2,3} {

    \draw (-1.5, 2.1-.6*\x) node [left] {$\mathsf{a}_\x$} -- ++(3,0);
    
    }

    \node at (-1.9,-.3) {$\vdots$};
    \node at (0,-.3) {$\vdots$};    
    
    \draw (-1.5, -.9) node [left] {$\mathsf{a}_n$} -- ++(3,0);    

     
    \draw (1.5,1.2) -- ++(1.5,0) node [above = .5em] {$n^\text{f}$ D4} -- ++(1.5,0) node [right] {$m_1$};;

    \foreach \x in {2} {

    \draw (1.5, 1.8-.6*\x) -- ++(3,0) node [right] {$m_\x$};

    }

    \draw (1.5, -.6) -- ++(3,0) node [right] {$m_{n^\text{f}}$};

    \node at (3.,.1) {$\vdots$};
    \node at (4.9,.1) {$\vdots$};

    \end{scope}
    
    \begin{scope}[shift={(-4,-6)}]
     \node at (-2.5,2.5) [left] {(b)};      
     
     \draw[thick] (-1.5,-1.5) -- ++(0,3.5) node [above] {NS5};
     \draw[thick] (+1.5,-1.5) -- ++(0,3.5) node [above] {NS5};
     

    \foreach \x in {1,2,3} {

     \draw (-1.5, 2.1-.6*\x) 
     -- ++(3,0);
    
    }

    \node at (0,-.3) {$\vdots$};    
    
     \draw (-1.5, -.9) 
     -- ++(3,0);    

     

    \foreach \x in {1,2} {

     \draw (1.5, 1.8-.6*\x) -- ++(2,0); 
     \filldraw[fill=red] (3.5, 1.8-.6*\x) circle (.1);
     
    }

    \draw (1.5, -.6) -- ++(2,0); 
     \filldraw[fill=red] (3.5, -.6) circle (.1);
     
     \node at (2.5,.1) {$\vdots$};
     \node at (3.5,.1) {$\vdots$};
     
     \node at (3.5,1.5) [red,above] {D6};
    \end{scope}

    \begin{scope}[shift={(5,-6)}]
     \node at (-2.5,2.5) [left] {(c)};      
     
     \draw[thick] (-1.5,-1.5) -- ++(0,3.5) node [above] {NS5};
     \draw[thick] (+1.5,-1.5) -- ++(0,3.5) node [above] {NS5};
     

    \foreach \x in {1,2,3} {

     \draw (-1.5, 2.1-.6*\x) 
     -- ++(3,0);
    
    }

    \node at (0,-.3) {$\vdots$};    
    
     \draw (-1.5, -.9) 
     -- ++(3,0);    

     \filldraw[fill=red] (-.8, 1.2) circle (.1);
     \filldraw[fill=red] (1, .6) circle (.1);
     \filldraw[fill=red] (-.5, 0) circle (.1);     
     
    \end{scope}
    
   \end{tikzpicture}
  \end{center}
  \caption{Type IIA brane configurations of four-dimensional $\mathcal{N} = 2$ $\mathrm{SU}(n)$ gauge theory with $n^\text{f}$ fundamental hypermultiplets with (a) semi-infinite D4 branes, (b) external D4 branes terminated by D6 branes, and (c) D6 branes inside the NS5 branes via the Hanany--Witten move.}
  \label{fig:HW}
 \end{figure}
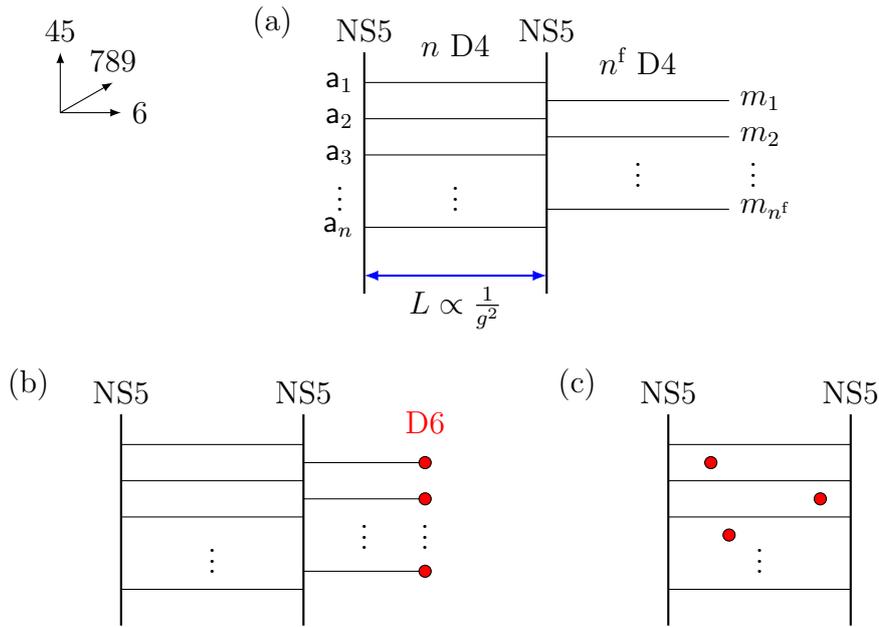

 The $\mathcal{N} = 2$ vector multiplet in four dimensions, on the other hand, contains a single complex scalar field (two real components).
 Hence, we should freeze some of the scalar fields to reduce the supersymmetry from $\mathcal{N} = 4$ to $\mathcal{N} = 2$ theory.
 A standard way to do this is to consider D4 branes suspended between NS5 branes, which is known as the Hanany--Witten construction of gauge theory with eight supercharges in string theory~\cite{Hanany:1996ie}.
 See Tab.~\ref{tab:HW}.

 \begin{table}[ht]
  \begin{center}
   \begin{tabular}{ccccccccccc}\hline\hline
    (IIA) & 0 & 1 & 2 & 3 & 4 & 5 & 6 & 7 & 8 & 9 \\\hline
    NS5 & - & - & - & - & - & - & & & & \\ 
    D4 & - & - & - & - & & & - & & & \\ 
    D6 & - & - & - & - & & & & - & - & - \\
    \hline\hline
   \end{tabular}
  \end{center}
  \caption{The brane configuration of four-dimensional $\mathcal{N} = 2$ gauge theory in Type IIA string theory.} 
  \label{tab:HW}
 \end{table}

 As shown in Fig.~\ref{fig:HW}, D4 branes are now extending in five dimensions (01236), but with a finite interval $L$ in 6-direction.
 Therefore, if the interval is sufficiently small, it would effectively behave as a four-dimensional theory.
 In fact, the interval $L$ is interpreted as the gauge coupling up to a constant factor, $L \propto 1/g^2$.
 
 In this setup, the transverse modes of D4 branes in 789-directions are fixed by the Dirichlet boundary condition with NS5 branes, which are not dynamical any longer.
 There remain two real components in the scalar field, which correspond to 45-directions, combined into a single complex scalar field in the $\mathcal{N} = 2$ vector multiplet, $\phi = \phi_4 + \im \phi_5$.
 As discussed before, the scalar field takes a value in the Cartan subalgebra of the gauge algebra, $\phi \in \mathfrak{h} \subset \mathfrak{g}$.
 In this picture, the diagonal values of the scalar field, $\phi = \diag(\mathsf{a}_1,\ldots,\mathsf{a}_n)$ for $G = \SU(n)$, are interpreted as the positions of D4 branes in 45-directions.
 
 In Fig.~\ref{fig:HW}(a), there are also external D4 branes outside the NS5 branes.
 In fact, the open strings connecting these internal and external D4 branes transform under the bifundamental representation of $\SU(n) \times \SU(n^\text{f})$.
 Since the external ones are infinitely long, their transverse modes are not dynamical.
 Similarly, we may incorporate the antifundamental matter by adding the external D4 branes going left of the NS5 brane.
 Hence, a stack of external D4 branes gives rise to the flavor symmetry $\SU(n^\text{f})$, and thus the configuration shown in Fig.~\ref{fig:HW} realizes $\mathcal{N} = 2$ gauge theory with $n^\text{f}$ fundamental hypermultiplets.
 It is also possible to terminate the external D4 branes with D6 branes (0123789), keeping the dynamics of D4 branes in 45-directions frozen by D6 branes (Fig.~\ref{fig:HW}(b)).
 Taking the D6 branes inside the NS5 branes, the suspended D4 branes are eliminated due to the Hanany--Witten effect.
 This shows that the same matter content is realized only with the internal D4 and D6 branes (Fig.~\ref{fig:HW}(c)).

 \subsection{Seiberg--Witten curve from M-theory}

 Let us remark the relation between the brane description of $\mathcal{N} = 2$ theory and the Seiberg--Witten geometry.
 First of all, since the complex scalar describes the positions of D4 branes in 45-directions, the $x$-variable in the Seiberg--Witten curve is identified with $x \simeq x_4 + \im x_5$.
 Then, the $y$-variable is instead for the positions of NS5 branes.
 (Recall the distance between NS5 branes corresponds to the gauge coupling constant.)

 Furthermore, the brane configuration discussed above, which consists of NS5 and D4 branes, is promoted to a single M5 brane after lifting up from IIA string theory to M-theory.
 Let $x_\text{M}$ be the eleventh dimension with the periodicity $x_\text{M} \simeq x_\text{M} + 2 \pi R$.
 Then, the M5 brane is extended in 0123-directions and a two-dimensional subspace of 456M, which are six dimensions in total.
 It has been shown in~\cite{Witten:1997sc} that the latter two-dimensional subspace is precisely given as the Seiberg--Witten curve of $\SU(n)$ gauge theory~\eqref{eq:SW_curve_SUn} with the identification:
 \begin{align}
  (x,y) = \left( x_4 + \im x_5, \, \exp \left( \frac{x_6 + \im x_\text{M}}{R} \right) \right) \in \mathbb{C} \times \mathbb{C}^\times
  \, .
 \end{align}
 We remark that the $y$-variable is now given as an exponential form to manifest the periodicity in M-direction.

 \subsection{Quiver gauge theory}\label{sec:HW_quiver}

 We have discussed the brane construction of a single $\SU(n)$ gauge theory.
 Let us show how to generalize this construction to quiver gauge theory involving multiple gauge nodes.

 \subsubsection{Linear quiver: $A_p$}

 We start with $A_p$ quiver, which is a linear quiver with $p$ gauge nodes.
 In this case, we consider $p+1$ parallel NS5 branes along 6-direction with suspended D4 branes.
 An example with $p = 3$ is given as
 \begin{align}
  \begin{tikzpicture}[baseline=(current  bounding  box.center)]  
   \draw[thick] (0,-2) -- ++(0,3.5) node [above] {NS5};
   \draw[thick] (3,-2) -- ++(0,3.5) node [above] {NS5};
   \draw[thick] (6,-2) -- ++(0,3.5) node [above] {NS5};
   \draw[thick] (9,-2) -- ++(0,3.5) node [above] {NS5};
   \foreach \x in {0,1,2}{
   \foreach \y in {0,2}{
   \draw (\y*3,-1.+\x) -- ++(3,0);
   }
   }
   \draw (3,-.5) -- ++(3,0);
   \draw (3,+.5) -- ++(3,0);
   \foreach \x in {-.5,+.5}{
   \draw (0,\x) -- ++(-2,0);
   \draw (9,\x) -- ++(+2,0);   
   }
   \foreach \x in {1,2,3}{
   \draw[latex-latex,thick,blue] (-3+\x*3,-1.6) -- ++(1.5,0) node [below,black] {$L_\x \propto \frac{1}{g^2_\x}$} -- ++(1.5,0);
   }
   \begin{scope}[shift={(0,-3)}]
    \draw (-1.5,0) -- (10.5,0);
    \filldraw[fill=white,draw=black] (1.5,0) circle (.2) node [below=1em] {SU$(n_1)$};
    \filldraw[fill=white,draw=black] (4.5,0) circle (.2) node [below=1em] {SU$(n_2)$};;
    \filldraw[fill=white,draw=black] (7.5,0) circle (.2) node [below=1em] {SU$(n_3)$};;
    \filldraw[fill=white,draw=black] (-1.5,-.2) rectangle ++(.4,.4) node [below=1.55em] {SU$(n_0)$};;
    \filldraw[fill=white,draw=black] (10.5,-.2) rectangle ++(.4,.4) node [below=1.55em] {SU$(n_4)$};;    
   \end{scope}
  \end{tikzpicture}
 \end{align}
 The number of D4 branes in each interval between NS5 branes provides the rank of gauge groups $(n_i)_{i = 1, \ldots, p}$. 
 In this configuration, we can add the (anti)fundamental hypermultiplet only to the first and the last nodes with the external D4 branes, where the corresponding rank of the flavor symmetries are denoted by $n_0$ and $n_{p+1}$.
 In order to add the fundamental matter to the middle nodes, we have to use internal D6 branes as in Fig.~\ref{fig:HW}.

 The open string degrees of freedom between $i$-th and $(i+1)$-st intervals give rise to the hypermultiplet in bifundamental representation of $\SU(n_i) \times \SU(n_{i+1})$.
 The corresponding bifundamental mass is given by the relative difference between the D4 brane positions in 45-directions.
 In this sense, it is natural to regard the bifundamental mass as a complex parameter, which can be compensated by the center-of-mass shift of D4 branes, namely the $\rU(1)$ gauge degrees of freedom of each gauge node.

\subsubsection{Cyclic quiver: $\widehat{A}_{p-1}$}
 
 We can similarly formulate the cyclic quiver with $p$ nodes ($\widehat{A}_{p-1}$ quiver).
 In this case, we align $p$ NS5 branes in periodic 6-direction:
 \begin{align}
  \begin{tikzpicture}[baseline=(current  bounding  box.center)]  
   \draw[thick] (0,-1.5) -- ++(0,3) node [above] {NS5};
   \draw[thick] (3,-1.5) -- ++(0,3) node [above] {NS5};
   \draw[thick] (6,-1.5) -- ++(0,3) node [above] {NS5};
   \draw[thick] (9,-1.5) -- ++(0,3) node [above] {NS5};
   \foreach \x in {0,1,2}{
   \foreach \y in {0,2}{
   \draw (\y*3,-1.+\x) -- ++(3,0);
   }
   }
   \draw (3,-.5) -- ++(3,0);
   \draw (3,+.5) -- ++(3,0);
   \foreach \x in {-.5,+.5}{
   \draw (0,\x) -- ++(-2,0);
   \draw (9,\x) -- ++(+2,0);   
   }
   \begin{scope}[shift={(0,-3)}]
    \draw (1.5,0) -- (7.5,0) -- (4.5,-1) -- cycle;
    \filldraw[fill=white,draw=black] (4.5,-1) circle (.2) node [below=1em] {SU$(n_0)$};    
    \filldraw[fill=white,draw=black] (1.5,0) circle (.2) node [above=1em] {SU$(n_1)$};
    \filldraw[fill=white,draw=black] (4.5,0) circle (.2) node [above=1em] {SU$(n_2)$};;
    \filldraw[fill=white,draw=black] (7.5,0) circle (.2) node [above=1em] {SU$(n_3)$};;
   \end{scope}
   \draw [red,very thick] (-1.7,+.5-.2) -- ++(0,+.4);
   \draw [red,very thick] (-1.7,-.5-.2) -- ++(0,+.4);
   \draw [red,very thick] (10.7,+.5-.2) -- ++(0,+.4);
   \draw [red,very thick] (10.7,-.5-.2) -- ++(0,+.4);   
  \end{tikzpicture}
 \end{align}
 Now the external D4 branes with red signs are identified with each other.
 We remark that we cannot impose the fundamental hypermultiplet to this configuration using the external D4 branes.

 Since 6-direction is periodic in this setup, we may consider T-dual transformation in this direction.
 The configuration obtained after T-duality is given as follows:\\
 \begin{table}[h]
  \begin{center}
   \begin{tabular}{ccccccccccc}\hline\hline
    (IIB) & 0 & 1 & 2 & 3 & 4 & 5 & 6 & 7 & 8 & 9 \\\hline
    TN$_{p}$ &  &  &  &  &  &  & - & - & - & - \\ 
    D3 & - & - & - & - & & & & & & \\ 
    \hline\hline
   \end{tabular}
  \end{center}
  \caption{The brane configuration of four-dimensional $\mathcal{N} = 2$ gauge theory in Type IIB string theory. We denote the $p$-centered Taub-NUT geometry by TN$_{p}$ in 6789 directions.} 
  \label{tab:HW_TN}
 \end{table}

 \noindent
  Compared to the previous one (See Tab.~\ref{tab:HW}), NS5 branes are converted to the so-called H-monopoles encoded in the Taub-NUT space (TN), which is locally asymptotic to $\mathbb{R}^3 \times S^1$ at infinity:\index{Taub--NUT space}
  The positions of NS5 branes in 789 directions correspond to the multiple centers of the Taub-NUT space.  
  See, for example,~\cite{Gregory:1997te,Tong:2002rq,Witten:2009xu} for details.
  From this point of view, the supersymmetry is reduced from $\mathcal{N} = 4$ to $\mathcal{N} = 2$ due to the H-flux of TN.
  We remark that, by taking the size of $S^1$ to infinity, we obtain the ALE space $\mathbb{R}^4/\mathbb{Z}_{p}$ from TN$_{p}$. \index{ALE space}%
  This is the shrinking limit of 6-direction on Type IIA side, and is the strong coupling limit from the gauge theory perspective, $\mathfrak{q}_\text{tot} := \mathfrak{q}_0 \mathfrak{q}_1 \cdots \mathfrak{q}_{p-1} \to 1$, where $\mathfrak{q}_i = \exp \left( 2 \pi \im \tau_i \right)$ for $i = 0, \ldots, p-1$.
  On the other hand, in the large radius limit on Type IIA side, we may lose the periodicity in 6-direction.
  This means the reduction from cyclic to linear quiver by freezing (at least) one of the gauge nodes of the quiver, $\mathfrak{q}_\text{tot} \to 0$.

  \subsubsection{D-type quiver}

  The brane construction discussed above has a natural generalization to $D_p$ and $\widehat{D}_p$ quivers, by introducing the ON$^0$ plane to the configuration~\cite{Kapustin:1998fa,Hanany:1999sj,Hayashi:2015vhy}, which is related to a combination of O5$^-$ plane and a D5 brane through the duality chain.
  For $D_p$ quiver, the brane configuration is given as follows:
  \begin{align}
   \begin{tikzpicture}[baseline=(current  bounding  box.center),scale=1.5]
   \draw[thick] (-1,-.25) -- ++(0,1.5) node [above] {NS5};
   \draw[thick] (1,-.25) -- ++(0,1.5) node [above] {NS5};
   \draw[thick] (3,-.25) -- ++(0,1.5) node [above] {NS5};
   \draw[red, very thick] (5,-.25) -- ++(0,1.5) node [above] {ON$^-$};
   \draw (-1,.2) -- ++(2,0);
   \draw (-1,.4) -- ++(2,0);
   \draw (-1,.6) -- ++(2,0);
   \draw (-1,.8) -- ++(2,0);   
   \draw (1,.1) -- ++(2,0);
   \draw (1,.3) -- ++(2,0);
   \draw (1,.5) -- ++(2,0);
   \draw (1,.7) -- ++(2,0);   
   \draw (3,.2) -- ++(2,0);
   \draw (3,.4) -- ++(2,0);
   \draw (3,.6) -- ++(2,0);
   \draw (3,.8) -- ++(2,0);      
    \begin{scope}[shift={(0,-1.5)}]
     \draw (0,0) -- (2,0);
     \draw (4,.5) -- (2,0) -- (4,-.5);
    \filldraw[fill=white,draw=black] (0,0) circle (.2) node [below=1em] {SU$(2n)$};
    \filldraw[fill=white,draw=black] (2,0) circle (.2) node [below=1em] {SU$(2n)$};;
     \filldraw[fill=white,draw=black] (4,.5) circle (.2) node [right=1em] {SU$(n)$};;
    \filldraw[fill=white,draw=black] (4,-.5) circle (.2) node [right=1em] {SU$(n)$};;     
     \end{scope}
   \end{tikzpicture}
  \end{align}
  Here the ON$^-$ plane is S-dual to an O5$^-$ plane, which would be combined into the ON$^0$ plane together with the NS5 brane.
  Furthermore, we can add the fundamental hypermultiplet, but only to the left most node.
  We can similarly deal with $\widehat{D}_p$ quiver by imposing ON$^0$ planes also on the left end of the configuration, so that we cannot incorporate the fundamental mattes for the affine case similarly to $\widehat{A}_{p-1}$ quiver.

 \subsubsection{Generic quiver}

 Although we can consider generic quiver in gauge theory, it seems not straightforward to construct such a generic quiver in string theory.
 Let us just mention a geometric realization for $\Gamma = ADE$:
 We may consider a stack of D3 branes with the transverse Taub-NUT space TN$_\Gamma$, which is reduced to the ALE space $\mathbb{C}^2/\Gamma$ in the large radius limit as well.
 Then, in this case, we would obtain the affine quiver gauge theory $\widehat{\Gamma} = \widehat{ADE}$ similarly to the discussion for the cyclic quiver.
 See also \S\ref{sec:quiver_variety} for a related discussion.

 For $\Gamma \not\in ADE$, we should apply the formalism of fractional quiver discussed in \S\ref{sec:fractional_quiver}.
 Based on the algebraic construction of the brane web formalism, we can construct arbitrary quiver gauge theory for $\Gamma = ABCDEFG$ and their affinization (including twisted versions).
See~\cite{Kimura:2019gon} for details.

 \subsection{Higgsing and vortices}\label{sec:Higgsing}

 In \S\ref{sec:fund_instanton}, we have discussed a specific property of the (anti)fundamental matter part of the instanton partition function:
 The partition function is truncated if one tunes the fundamental mass parameter related to the Coulomb moduli~\eqref{eq:Higgs_branch_locus}.
 We now discuss its brane perspective.
 
 Turning off the $\Omega$-background parameter $\epsilon_{1,2} \to 0$, the root of Higgs branch condition~\eqref{eq:Higgs_branch_locus} \index{root of Higgs branch} is simply given by equating the fundamental mass and the Coulomb moduli parameter~\cite{Dorey:1998yh,Dorey:1999zk}:
 \begin{align}
  m_f = \mathsf{a}_\alpha
  \, .
  \label{eq:Higgs_branch_locus_cl}  
 \end{align}
 Since they are interpreted as positions of the internal and external D4 branes in Type IIA configuration (Fig.~\ref{fig:HW}), under this condition, they come to the same place in 45-directions to be a single semi-infinite D4 brane.
 Furthermore, from this configuration, one can remove the right NS5 brane in 9-direction, and D2 branes suspended in 9-direction may emerge as shown below:
 \begin{align}
  \begin{tikzpicture}[baseline=(current  bounding  box.center),thick]
   \begin{scope}[shift={(0,0)}]
    \draw (-1,-1.5) -- ++(0,3) node [above] {NS5};
     \draw (+1,-1.5) -- ++(0,3) node [above] {NS5};
    \foreach \y in {0,1,2,3}{
    \draw (-1,-.9+.6*\y) -- ++(2,0);
    \draw (1,-1.2+.6*\y) -- ++(1.5,0);
    }
    \node at (.8,-2.5) {Coulomb branch}; 
    \node at (0,1.3) {D4};
    \node at (1.8,1.3) {D4};    
   \end{scope}
   \begin{scope}[shift={(5,0)}]
    \draw (-1,-1.5) -- ++(0,3) node [above] {NS5};
    \draw (+1,-1.5) -- ++(0,3) node [above] {NS5};
    \foreach \y in {0,1,2,3}{
    \draw (-1,-.9+.6*\y) -- ++(2,0);
    \draw (+1,-.9+.6*\y) -- ++(1.5,0);
    }
    \node at (1,-2.5) {$\mathsf{a}_\alpha = m_f$};
    \node at (0,1.3) {D4};    
   \end{scope}
   \begin{scope}[shift={(10,0)}]
    \draw (-1,-1.5) -- ++(0,3) node [above] {NS5};
    \draw [dotted] (.7,-1.5) -- ++(0,3); 
    \draw (+1.2,-2) -- ++(0,3.4) node [above] {NS5};
    \foreach \y in {0,1,2,3}{
    \draw (-1,-.9+.6*\y) -- ++(3,0);
    \draw [red,very thick] (.7,-.9+.6*\y) -- ++(.5,-.4);
    }
    \node at (.7,-2.5) {Higgs branch};
    \node at (1.2,.6) [right,red] {D2};
    \draw [latex-latex,blue,very thick] (-1,-1.3) -- ++(.85,0) node [below,black] {$L_6$} -- ++(.85,0);
    \draw [latex-latex,blue,very thick] (.7,-1.3) -- ++(.5,-.4) node [right,black] {$L_9$};
   \end{scope}   
  \end{tikzpicture}
  \label{fig:Higgsing}
 \end{align}
  \begin{table}[ht]
  \begin{center}
   \begin{tabular}{ccccccccccc}\hline\hline
    (IIA) & 0 & 1 & 2 & 3 & 4 & 5 & 6 & 7 & 8 & 9 \\\hline
    NS5 & - & - & - & - & - & - & & & & \\
    (D6) & - & - & - & - & & & & - & - & - \\    
    D4 & - & - & - & - & & & - & & & \\    
    D2 & - & - & & & & & & & & - \\ 
    \hline\hline
   \end{tabular}
  \end{center}
  \caption{The brane configuration of four-dimensional $\mathcal{N} = 2$ gauge theory in Higgs branch.} 
  \label{tab:HW_Higgs}
  \end{table}

  \noindent
  See also Tab.~\ref{tab:HW_Higgs}.
  A similar manipulation is possible with the internal D6 branes describing the fundamental matter related through the Hanany--Witten move.

  After removing the right NS5 brane, it is not in the Coulomb branch of the moduli space, but in the Higgs branch, where the gauge symmetry is locked with the flavor symmetry by the condition~\eqref{eq:Higgs_branch_locus_cl}. \index{moduli space of vacua!Coulomb branch} \index{moduli space of vacua!Higgs branch}
 In this case, D2 branes are interpreted as BPS vortex strings in 4d theory, and their length in 9-direction $L_9$ is identified with the Fayet--Iliopoulos (FI) parameter in four-dimensions~\cite{Hanany:2004ea}.

 This configuration has an alternative description as the world-volume theory of D2 branes, which is two-dimensional $\mathcal{N} = (2,2)$ gauge theory in 01-directions.
 From 4d $\mathcal{N} = 2$ $\SU(n)$ gauge theory with $n^\text{f} = n$ fundamental and $n^\text{af}$ antifundamental matters, we obtain the 2d $\mathcal{N} = (2,2)$ theory with $n$ fundamental, $n^\text{af}$ antifundamental, and a single adjoint chiral multiplets:%
 \footnote{%
 8 SUSY (4d $\mathcal{N} = 2$) vector multiplet splits into 4 SUSY (2d $\mathcal{N} = (2,2)$; 4d $\mathcal{N}$ = 1) vector and a adjoint chiral multiplets.}
 \begin{align}
  \begin{tikzpicture}[baseline=(current  bounding  box.center),thick,scale=.8]
       \draw (-3,0) -- (1,0);
       \filldraw [fill=white,draw=black] (-1,0) circle (.2);
       \filldraw [fill=white,draw=black] ($(-3,0)+(-.2,-.2)$) rectangle ++(.4,.4);
       \filldraw [fill=white,draw=black] ($(+1,0)+(-.2,-.2)$) rectangle ++(.4,.4);
   \draw[ultra thick,blue,-latex] (3,0) -- ++(1,0) node [black,above] {Higgsing} -- ++(1,0);
   \begin{scope}[shift={(7.5,0)}]
    \draw [->-] (2,0) -- (2,-1.5);
    \draw [->-] (2,-1.5) -- (0,0);
    \draw [->-] (2,-1.5) arc [x radius = .5, y radius = .5, start angle = 90, end angle = -270];
    \filldraw [fill=white,draw=black] ($(0,0)+(-.2,-.2)$) rectangle ++(.4,.4);
    \filldraw [fill=white,draw=black] ($(2,0)+(-.2,-.2)$) rectangle ++(.4,.4);
    \filldraw [fill=cyan!50,draw=black] (2,-1.5) circle (.2);    
   \end{scope}
  \end{tikzpicture}
 \end{align}
 where the blue circle \tikz[baseline={([yshift=-8pt]current bounding box.north)},thick] \filldraw [fill=cyan!50,draw=black] (0,0) circle (.15);
 stands for the gauge node in two dimensions, while white nodes are for four dimensions.
 The gauge group rank in two dimensions is given by the number of vortices in 4d theory in the Higgs branch.
 In addition, the gauge coupling and FI parameter for this 2d theory are identified with $L_9$ and $L_6$, respectively. 
 In this picture, the instanton in 4d realized using D0 branes is interpreted as
 an instanton also in the 2d theory, i.e., a vortex in a vortex (a trapped vortex).
 Based on such a relation between 4d and 2d theories, one can obtain the vortex partition function of 2d $\mathcal{N} = (2,2)$ theory~\cite{Dimofte:2010tz,Bonelli:2011fq}.
 See also~\cite{Fujimori:2012ab} for another, but direct derivation of the vortex partition function.

 \begin{figure}[t]
  \begin{center}
   \begin{tikzpicture}[thick,scale=.9]

     \begin{scope}
      \node at (-4,2) {(a)};

     \draw (-2,-1.5) -- ++(0,3) node [above] {NS5};
     \draw (0,-1.5) -- ++(0,3) node [above] {NS5};
     \draw (+2,-1.5) -- ++(0,3) node [above] {NS5};

     \foreach \x in {0,1,2}{

     \draw (0,.6-\x*.8) -- ++(2,0);
     \draw (2,1-\x*.8) -- ++(1.5,0);     
     \draw (0,1-\x*.8) -- ++(-2,0);
     \draw (-2,.6-\x*.8) -- ++(-1.5,0);     

     }

     \end{scope}

     \begin{scope}[shift={(9,0)}]
      \node at (-4,2) {(b)};

     \draw (-2,-1.5) -- ++(0,3) node [above] {NS5};
     \draw (0,-1.5) -- ++(0,3) node [above] {NS5};
      \draw [dotted] (1.7,-1.5) -- ++(0,3);
      \draw (+2.2,-2) -- ++(0,3) node [above] {NS5};

     \foreach \x in {0,1,2}{

     \draw (0,.6-\x*.8) -- ++(3,0);
     \draw (0,1-\x*.8) -- ++(-2,0);
     \draw (-2,.6-\x*.8) -- ++(-1.5,0);     

      \draw [red] (1.7,.6-\x*.8) -- ++(.5,-.5);
      
     }

     \end{scope}    

     \begin{scope}[shift={(0,-5)}]
      \node at (-4,2) {(c)};

      \draw (-2,-1.5) -- ++(0,3) node [above] {NS5};
      \draw (.2,-2) -- ++(0,3.2) node [above] {NS5};
      \draw [dotted] (-.3,-1.5) -- ++(0,3); 
      \draw [dotted] (1.7,-1.5) -- ++(0,3);
      \draw (+2.2,-2) -- ++(0,3.2) node [above] {NS5};

     \foreach \x in {0,1,2}{

     \draw (-2,1-\x*.8) -- ++(5,0);
     \draw (-2,.6-\x*.8) -- ++(-1.5,0);     
      
      \draw [red] (1.7,1-\x*.8) -- ++(.5,-.5);
      \draw [red] (-.3,1-\x*.8) -- ++(.5,-.5);      
      
     }
     
     \end{scope}    

     \begin{scope}[shift={(9,-5)}]
      \node at (-4,2) {(d)};

      \draw (-2,-1.5) -- ++(0,3) node [above] {NS5};
      \draw (.2,-2) -- ++(0,3.2) node [above] {NS5};
      \draw [dotted] (-.3,-1.5) -- ++(0,3); 
      \draw (.7,-2.5) -- ++(0,3.2) node [right] {NS5};

     \foreach \x in {0,1,2}{

     \draw (-2,1-\x*.8) -- ++(4.5,0);
     \draw (-2,.6-\x*.8) -- ++(-1.5,0);     
      
      \draw [red] (-.3,1-\x*.8) -- ++(.5,-.5);
      \draw [red] (.2,.7-\x*.8) -- ++(.5,-.5);      
      
     }
     
     \end{scope}        
    
   \end{tikzpicture}
  \end{center}
  \caption{Higgsing process for $A_2$ quiver. (a) Coulomb branch, (b) Higgsing the second node, (c) Higgsing both nodes, similarly described as the configuration (d). The intervals of NS5 branes in 9-direction are identified with the FI parameters.}
  \label{fig:Higgsing_quiver}
 \end{figure}
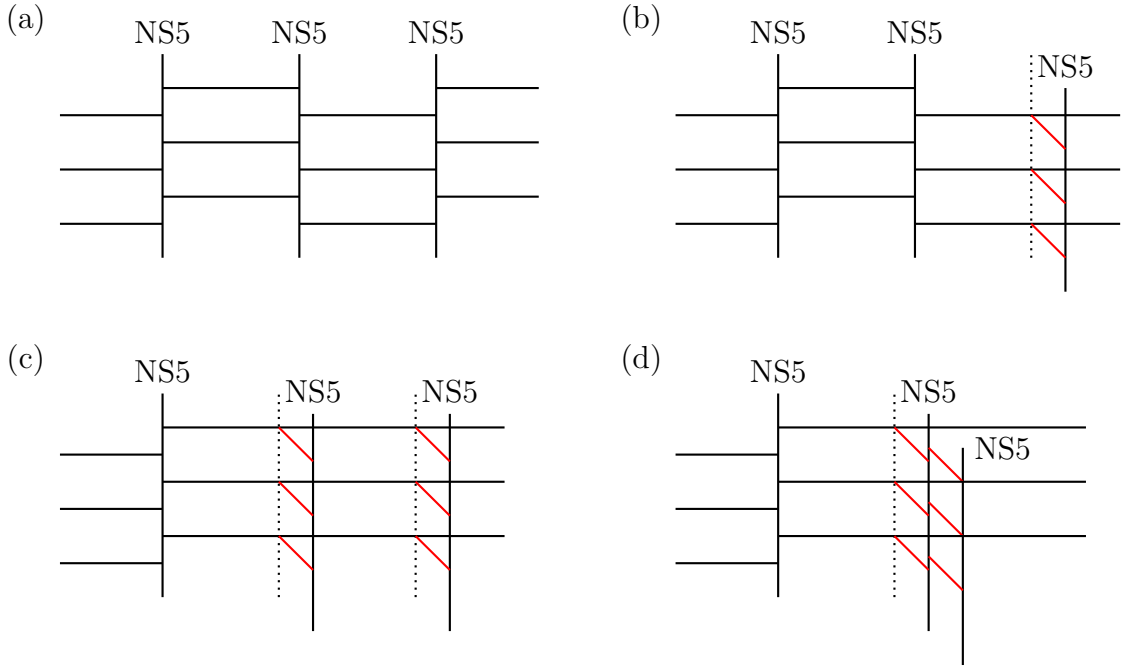

 \subsubsection{Quiver gauge theory}
 
 This manipulation is applied similarly in quiver gauge theory.
 We describe the Higgsing process for $A_2$ quiver in Fig.~\ref{fig:Higgsing_quiver}, and the corresponding quiver diagrams are given as follows:
 \begin{align}
  \begin{tikzpicture}[baseline=(current  bounding  box.center),thick,scale=.6]
      \begin{scope}[shift={(0,0)}]
       \node at (-3,1.5) {(a)};
       \draw (-3,0) -- (3,0);
      \filldraw [fill=white,draw=black] (-1,0) circle (.2);
      \filldraw [fill=white,draw=black] (+1,0) circle (.2);      
      \filldraw [fill=white,draw=black] ($(-3,0)+(-.2,-.2)$) rectangle ++(.4,.4);
      \filldraw [fill=white,draw=black] ($(+3,0)+(-.2,-.2)$) rectangle ++(.4,.4);             
      \end{scope}
      \begin{scope}[shift={(8.5,0)}]
       \node at (-3,1.5) {(b)};
       \draw (-3,0) -- (1,0);
       \draw [->-] (1,0) -- (1,-1.5);
       \draw [->-] (1,-1.5) -- (-1,0);
       \draw [->-] (1,-1.5) arc [x radius = .5, y radius = .5, start angle = 90, end angle = -270];       
     \filldraw [fill=white,draw=black] (-1,0) circle (.2);
     \filldraw [fill=white,draw=black] ($(-3,0)+(-.2,-.2)$) rectangle ++(.4,.4);
     \filldraw [fill=white,draw=black] ($(+1,0)+(-.2,-.2)$) rectangle ++(.4,.4);
     \filldraw [fill=cyan!50,draw=black] (1,-1.5) circle (.2);       
      \end{scope}
       \begin{scope}[shift={(13,0)}]
       \node at (0,1.5) {(c) \& (d)};	
       \draw [->-] (2,0) -- (2,-1.5);
       \draw [->-] (2,-1.5) -- (0,0);
       \draw [->-] (2,-1.4) -- ++(2,0);
       \draw [->-] (4,-1.6) -- ++(-2,0);       
       \draw [->-] (2,-1.65) arc [x radius = .5, y radius = .5, start angle = 90, end angle = -270];
       \draw [->-] (4,-1.65) arc [x radius = .5, y radius = .5, start angle = 90, end angle = -270];
     \filldraw [fill=white,draw=black] ($(0,0)+(-.2,-.2)$) rectangle ++(.4,.4);
     \filldraw [fill=white,draw=black] ($(2,0)+(-.2,-.2)$) rectangle ++(.4,.4);
       \filldraw [fill=cyan!50,draw=black] (2,-1.5) circle (.2);
       \filldraw [fill=cyan!50,draw=black] (4,-1.5) circle (.2);              
       \end{scope}   
  \end{tikzpicture}
 \end{align}
 In this way, we obtain the 4d-2d coupled quiver description from the Higgsing process.
 See also \cite{Gomis:2014eya} for generic linear quiver theory.

 \subsection{Higgsing in Seiberg--Witten geometry}


\subsubsection{$A_1$ quiver}

 We discuss the Seiberg--Witten geometric point of view of the Higgsing process.
 We write the Seiberg--Witten curve for $\SU(n)$ gauge theory with $(n^\text{f},n^\text{af})$ flavors \eqref{eq:SW_curve_SQCD} as follows:
 \begin{align}
  y^2 - \mathsf{T}_{1,x} \, y - P_{1,x} \widetilde{P}_{1,x} = 0
 \end{align}
 where $(P_{1,x}, \widetilde{P}_{1,x})$ are the matter polynomials of the degree $(n^\text{f},n^\text{af})$.%
\footnote{%
Compared to the previous definition~\eqref{eq:matter_polynomial_quiver}, we simply denote them by $(P_{i,x}, \widetilde{P}_{i,x})$ as long as no confusion.
}
Since, from the brane picture~\eqref{fig:Higgsing}, there remains a single NS5 brane after the Higgsing process, the algebraic curve would be given by a linear relation for the $y$-variable.
  Such a situation is realized by imposing the condition
  \begin{align}
   \mathsf{T}_{1,x} = P_{1,x} + \widetilde{P}_{1,x}
   \label{eq:T=A+D}
  \end{align}
 such that,
 \begin{align}
  y^2 - \mathsf{T}_{1,x} \, y - P_{1,x} \widetilde{P}_{1,x} =
  ( y - P_{1,x} ) ( y - \widetilde{P}_{1,x} ) = 0
  \, .
 \end{align}
 Then, the factorized curve implies the relation
 \begin{align}
  y = P_{1,x} \quad \text{or} \quad y = \widetilde{P}_{1,x}
  \, ,
 \end{align}
 which describes the Higgsed configuration with respect to the fundamental or the antifundamental matter.

  We remark that the condition~\eqref{eq:T=A+D} has a natural interpretation in the context of the integrable system:
  According to the correspondence between gauge theory and integrable system, $\mathsf{T}$-function is identified with the (eigenvalue of) transfer matrix, which is given by taking a trace of the monodromy matrix
  \begin{align}
   \mathsf{T}_x = \tr \mathcal{T}_x
   \, ,  \qquad
   \mathcal{T}_x
   =
   \begin{pmatrix}
    a(x) & b(x) \\ c(x) & d(x)
   \end{pmatrix}
   \, .
  \end{align}
  In fact, the expression~\eqref{eq:T=A+D} implies that the diagonal elements $(a(x),d(x))$ are identified with the matter polynomials $(P_{1,x},\widetilde{P}_{1,x})$ as shown in \S\ref{sec:Bethe_eq}.

   \subsubsection{$A_2$ quiver}

  This argument is straightforwardly generalized to quiver gauge theory.
 We consider $A_2$ quiver with fundamental matters.
 The Seiberg--Witten curve is described by a pair of fundamental characters:
 \begin{subequations}
  \begin{align}
   \mathsf{T}_{1,x} = y_1 + \frac{y_0 y_2}{y_1} + \frac{y_0 y_3}{y_2}
   \, , \\
   \mathsf{T}_{2,x} = y_2 + \frac{y_1 y_3}{y_2} + \frac{y_0 y_3}{y_1}
   \, ,
  \end{align}
 \end{subequations}
 where $(y_0,y_3)$ are the matter polynomials of the first and second gauge nodes.
 They are combined into a single equation to define the algebraic curve:
 \begin{align}
  y_1^3 - \mathsf{T}_{1,x} \, y_1^2 + \mathsf{T}_{2,x} \, y_0 y_1 - y_0^2 y_3 = 0
  \quad \text{or} \quad
  y_2^3 - \mathsf{T}_{2,x} \, y_2^2 + \mathsf{T}_{1,x} \, y_2 y_3 - y_0 y_3^2 = 0
  \, .
 \end{align}
 Similarly to $A_1$ quiver, let us find a factorization condition for them.
 In this case, we impose the condition
 \begin{align}
  \mathsf{T}_{1,x} - y_0 = \mathsf{T}_{2,x} - y_3 =: \widetilde{\mathsf{T}}_{1,x}
  \, ,
 \end{align}
 which leads to the factorization:
 \begin{align}
  (y_1 - y_0)(y_1^2 - \widetilde{\mathsf{T}}_{1,x} \, y_1 + y_0 y_3) = 0
  \, , \qquad
  (y_2 - y_3)(y_2^2 - \widetilde{\mathsf{T}}_{1,x} \, y_2 + y_0 y_3) = 0
  \, .
 \end{align}
 This shows that the $A_2$ curve is factorized into the linear term and the $A_1$ (quadratic) curve.
 Imposing a further condition, $\widetilde{\mathsf{T}}_{1,x} = y_0 + y_3$, the curve is completely Higgsed:
 \begin{align}
  (y_1 - y_0)^2(y_1 - y_3) = 0
  \, , \qquad
  (y_2 - y_0)(y_2 - y_3)^2 = 0
  \, .
 \end{align}
Generalization to generic quiver theory is straightforward.

 \subsection{Supergroup gauge theory}\label{sec:brane_super}

   \subsubsection{Positive and negative branes}

Following the discussion in \S\ref{sec:HW_construction}, $\mathcal{N}=2$ SU$(n_0|n_1)$ SYM theory in four dimensions is realized as a world-volume theory of positive and negative branes, D4$^+$ and D4$^-$, suspended between two separated NS5 branes~\cite{Okuda:2006fb,Dijkgraaf:2016lym}:
\begin{align}
   \begin{tikzpicture}[scale=1.5,baseline=(current  bounding  box.center)]
   \draw[thick] (-1,-1) -- ++(0,2) node [above] {NS5};
   \draw[thick] (1,-1) -- ++(0,2) node [above] {NS5};
   \draw (-1,.2) -- ++(2,0);
   \draw (-1,.4) -- ++(2,0);
   \draw (-1,.6) -- ++(2,0);
   \draw [dotted,thick] (-1,-.2) -- ++(2,0);
   \draw [dotted,thick] (-1,-.4) -- ++(2,0);
   \draw [thick,decorate,decoration={brace,amplitude=4pt,mirror,raise=4pt},yshift=0pt] (1.1,.1) -- ++(0,.6) node [black,midway,xshift=2.6em] {$n_0$ D4$^+$};
   \draw [thick,decorate,decoration={brace,amplitude=4pt,mirror,raise=4pt},yshift=0pt] (1.1,-.5) -- ++(0,.4) node [black,midway,xshift=2.6em] {$n_1$ D4$^-$};   
   \end{tikzpicture}  
\end{align}
where D4$^+$ and D4$^-$ branes are depicted as horizontal solid and dotted lines.
It has been pointed out in~\cite{Dijkgraaf:2016lym} that the negative branes are removed through gauging process:
\begin{align}
  \begin{tikzpicture}[thick,scale=1.2,baseline=(current  bounding  box.center)]
   \draw (-1,-1) -- ++(0,2) node [above] {};
   \draw (1,-1) -- ++(0,2) node [above] {};
   \draw (-1,.2) -- ++(2,0);
   \draw (-1,.4) -- ++(2,0);
   \draw (-1,.6) -- ++(2,0);
   \draw [dotted] (-1,-.1) -- ++(2,0);
   \draw [dotted] (-1,-.3) -- ++(2,0);
   \draw (-1.5,-.6) -- ++(3,0);
   \draw (-1.5,-.8) -- ++(3,0);   
   \draw[-latex,blue,very thick] (2,0) -- ++(.5,0) node [above] {} -- ++(.5,0);
   \begin{scope}[shift={(5,0)}]
   \draw (-1,-1) -- ++(0,2) node [above] {};
   \draw (1,-1) -- ++(0,2) node [above] {};
   \draw (-1,.2) -- ++(2,0);
   \draw (-1,.4) -- ++(2,0);
   \draw (-1,.6) -- ++(2,0);
   \draw [dotted] (-1,-.1) -- ++(2,0);
   \draw [dotted] (-1,-.3) -- ++(2,0);
    \draw (-1.5,-.6) -- ++(.5,0);
    \draw (-1.5,-.8) -- ++(.5,0);
    \draw (1,-.6) -- ++(.5,0);
    \draw (1,-.8) -- ++(.5,0);       
    \draw (-1,-.5) -- ++(2,0);
    \draw (-1,-.7) -- ++(2,0);    
    \draw[-latex,blue,very thick] (2,0) -- ++(.5,0) node [above] {} -- ++(.5,0);   \end{scope}
   \begin{scope}[shift={(10,0)}]
   \draw (-1,-1) -- ++(0,2) node [above] {};
   \draw (1,-1) -- ++(0,2) node [above] {};
   \draw (-1,.2) -- ++(2,0);
   \draw (-1,.4) -- ++(2,0);
   \draw (-1,.6) -- ++(2,0);
   %
   %
    \draw (-1.5,-.6) -- ++(.5,0);
    \draw (-1.5,-.8) -- ++(.5,0);
    \draw (1,-.6) -- ++(.5,0);
    \draw (1,-.8) -- ++(.5,0);           
   \end{scope}      
  \end{tikzpicture}  
\end{align}
and the resulting configuration is equivalent to SU$(n_0)$ SYM theory with $n^\text{f} = 2 n_1$ flavors.
More precisely, there are $n_1$ fundamental and $n_1$ anti-fundamental hypermultiplets having the same masses because we imposed {\em horizontal} D4$^+$ branes before gauging.
This is consistent with the decoupling trick argument in \S\ref{sec:decoupling}.
We remark that such a reduction is possible only at the special locus of the Coulomb branch of the moduli space of vacua, and this does not mean the agreement of the moduli spaces themselves.

\subsubsection{One-to-many correspondence}

The non-Abelian gauge theory is realized as a stack of D-branes.
In particular, we should use both the positive and negative branes to realize the supergroup gauge theory.
Since, in this case, there are two different branes, we should be careful of the ordering of the branes.
For example, there are three possible realizations of SU$(2|1)$ gauge theory as follows:
\begin{align}
 \begin{tikzpicture}[thick,baseline=(current bounding box.center),scale=1.]
  \node at (-.2,2.5) {(a)};
  \draw (0,0) -- ++(0,2);
  \draw (1.8,0) -- ++(0,2);
  \draw (0,1.5) -- ++(1.8,0);
  \draw (0,1) -- ++(1.8,0);
  \draw[dotted] (0,.5) -- ++(1.8,0);
  \draw (3,1.25) -- ++(0,-.5);
  \draw (-.5,1.65) -- ++(.5,0); 
  \draw (1.8,1.35) -- ++(.5,0); 
  \draw (-.5,1.15) -- ++(.5,0); 
  \draw (1.8,.85) -- ++(.5,0); 
  \draw[dotted] (-.5,.65) -- ++(.5,0);  
  \draw[dotted] (1.8,.35) -- ++(.5,0);  
  \filldraw[fill=white,draw=black] (3,1.25) circle (.15);
  \filldraw[fill=white,draw=black] (3,.75) circle (.15);
   \draw (3,.75)++(135:.15) -- ++(-45:.3);
   \draw (3,.75)++(45:.15) -- ++(-135:.3);     
  \begin{scope}[shift={(4.5,0)}]
   \node at (-.2,2.5) {(b)};
  \draw (0,0) -- ++(0,2);
  \draw (1.8,0) -- ++(0,2);
  \draw (0,1.5) -- ++(1.8,0);
  \draw[dotted] (0,1) -- ++(1.8,0);
  \draw (0,.5) -- ++(1.8,0);
  \draw (3,1.25) -- ++(0,-.5);
  \draw (-.5,1.65) -- ++(.5,0); 
  \draw (1.8,1.35) -- ++(.5,0); 
  \draw[dotted] (-.5,1.15) -- ++(.5,0); 
  \draw[dotted] (1.8,.85) -- ++(.5,0); 
  \draw (-.5,.65) -- ++(.5,0);  
  \draw (1.8,.35) -- ++(.5,0);  
   \filldraw[fill=white,draw=black] (3,1.25) circle (.15);
   \filldraw[fill=white,draw=black] (3,.75) circle (.15);
   \draw (3,1.25)++(135:.15) -- ++(-45:.3);
   \draw (3,1.25)++(45:.15) -- ++(-135:.3);
   \draw (3,.75)++(135:.15) -- ++(-45:.3);
   \draw (3,.75)++(45:.15) -- ++(-135:.3);     
  \end{scope}
  \begin{scope}[shift={(9,0)}]
   \node at (-.2,2.5) {(c)};
  \draw (0,0) -- ++(0,2);
  \draw (1.8,0) -- ++(0,2);
  \draw[dotted] (0,1.5) -- ++(1.8,0);
  \draw (0,1) -- ++(1.8,0);
  \draw (0,.5) -- ++(1.8,0);
  \draw (3,1.25) -- ++(0,-.5);
  \draw[dotted] (-.5,1.65) -- ++(.5,0); 
  \draw[dotted] (1.8,1.35) -- ++(.5,0); 
  \draw (-.5,1.15) -- ++(.5,0); 
  \draw (1.8,.85) -- ++(.5,0); 
  \draw (-.5,.65) -- ++(.5,0);  
  \draw (1.8,.35) -- ++(.5,0);  
  \filldraw[fill=white,draw=black] (3,1.25) circle (.15);
  \filldraw[fill=white,draw=black] (3,.75) circle (.15);
   \draw (3,1.25)++(135:.15) -- ++(-45:.3);
   \draw (3,1.25)++(45:.15) -- ++(-135:.3);
  \end{scope}
 \end{tikzpicture}
 \label{eq:HW_U(2|1)}
\end{align}

\noindent
This one-to-many correspondence between the gauge theory and the brane configurations is a peculiar property to the supergroup theory, which is essentially related to the ambiguity of the simple root decomposition of the supergroup.
Besides the brane configurations, we also show the corresponding Dynkin diagrams of $\mathrm{SU}(2|1)$:
(a) \dynkin[root radius=.15cm, edge length=.7cm]{A}{ot}, \
(b) \dynkin[root radius=.15cm, edge length=.7cm]{A}{tt}, and
(c) \dynkin[root radius=.15cm, edge length=.7cm]{A}{to},
where the node denoted by \dynkin[root radius=.15cm]{A}{t} is the fermionic node~\cite{Kac:1977em}.
The correspondence is given as follows:
We assign the ordinary node to the neighboring pair of D4$^+$-D4$^+$ or D4$^-$-D4$^-$ branes, and the fermionic node is assigned to the neighboring pair of D4$^+$-D4$^-$ branes.
This argument is also applicable to the external flavor branes.
Even though there are several different brane realizations, we can see that the partition function does not depend on the ordering of positive and negative branes from the topological string analysis~\cite{Kimura:2020lmc,Chen:2020rxu}.

\subsubsection{Affine quiver realization}

As discussed in \S\ref{sec:super_quiver_realization}, supergroup gauge theory has a realization in the unphysical parameter regime of quiver gauge theory ($\widehat{A}_1$ quiver).
This fact is also understood from the brane configuration.
We start with $\widehat{A}_1$ quiver gauge theory, which is a cyclic quiver with two gauge nodes, $\SU(n_1) \times \SU(n_2)$:
\begin{align}
 \begin{tikzpicture}[baseline=(current bounding box.center),scale=1.1]
  \draw [thick] (-1,-1.5) -- ++(0,3);
  \draw [thick] (+1,-1.5) -- ++(0,3);
  \begin{scope}[shift={(0,-.2)}]
  \draw (-1,.5) -- ++(2,0);
  \draw (-1,0) -- ++(2,0);
  \draw (-1,1) -- ++(2,0);
  \draw (-1,.75) arc [x radius = .75, y radius = .2, start angle = 90, end angle = 270] -- ++(2,0) arc [x radius = .75, y radius = .2, start angle = -90, end angle = 90];
   \draw (-1,.2) arc [x radius = .75, y radius = .2, start angle = 90, end angle = 270] -- ++(2,0) arc [x radius = .75, y radius = .2, start angle = -90, end angle = 90];
  \end{scope}   
  \draw [thick,blue,latex-latex] (-1,-1.2) -- ++(1,0) node [above,black] {$L_1$} -- ++(1,0);
  \draw [thick,blue,latex-latex] (-1,-1.3) arc [x radius = .75, y radius = .2, start angle = 90, end angle = 270] -- ++(1,0) node [below,black] {$L_2$} -- ++(1,0) arc [x radius = .75, y radius = .2, start angle = -90, end angle = 90];
  \draw [very thick,blue,-latex] (2.5,0) -- ++(.5,0) node [above,black] {$L_2 \to -L_1$} -- ++(.5,0);
  \begin{scope}[shift={(5.5,0)}]   
   \draw [thick] (-1,-1.5) -- ++(0,3);
   \draw [thick] (+1,-1.5) -- ++(0,3);
   \begin{scope}[shift={(0,-.2)}]
    \draw (-1,.5) -- ++(2,0);
    \draw (-1,0) -- ++(2,0);
    \draw (-1,1) -- ++(2,0);
    \draw [dotted,thick] (-1,+.25) -- ++(2,0);
    \draw [dotted,thick] (-1,.75) -- ++(2,0);    
   \end{scope}
  \draw [thick,blue,latex-latex] (-1,-1.2) -- ++(1,0) node [above,black] {$L_1$} -- ++(1,0);
  \end{scope}
  \node at (0,-2.5) {$\SU(n_1) \times \SU(n_2)$};
  \node at (5.5,-2.5) {$\SU(n_1 | n_2)$};  
  \end{tikzpicture}
\end{align}
In order to obtain the supergroup gauge theory, we apply the analytic continuation for the coupling constant $1/g_2^2 \to - 1/g_1^2$, corresponding to $L_2 \to - L_1$ in the brane configuration.
Then, the external branes are now interpreted as the internal negative branes after the analytic continuation, which is consistent with the brane configuration of the supergroup gauge theory based on the positive and negative branes.

\subsubsection{Quiver gauge theory}

\begin{figure}[ht]
\begin{center}
  \begin{tikzpicture}[thick,scale=1.2]

   \draw (-1,-1) -- ++(0,2);
   \draw (1,-1) -- ++(0,2);
   \draw (3,-1) -- ++(0,2);
   \draw (5,-1) -- ++(0,2);

   \draw (-1,.2) -- ++(2,0);
   \draw (-1,.4) -- ++(2,0);
   \draw (-1,.6) -- ++(2,0);

   \draw (1,.1) -- ++(2,0);
   \draw (1,.3) -- ++(2,0);
   \draw (1,.5) -- ++(2,0);

   \draw (3,.2) -- ++(2,0);
   \draw (3,.4) -- ++(2,0);
   \draw (3,.6) -- ++(2,0);
   
   \draw [dotted] (-1,-.2) -- ++(2,0);
   \draw [dotted] (-1,-.4) -- ++(2,0);

   \draw [dotted] (1,-.3) -- ++(2,0);
   \draw [dotted] (1,-.5) -- ++(2,0);

   \draw [dotted] (3,-.2) -- ++(2,0);
   \draw [dotted] (3,-.4) -- ++(2,0);

    \begin{scope}[shift={(0,-1.5)}]

    \draw (0,0) -- (4,0);

    \filldraw[fill=white,draw=black] (0,0) circle (.2) node [below=1em] {SU$(n_0|n_1)$};
    \filldraw[fill=white,draw=black] (2,0) circle (.2) node [below=1em] {SU$(n_0|n_1)$};;
    \filldraw[fill=white,draw=black] (4,0) circle (.2) node [below=1em] {SU$(n_0|n_1)$};;

   \draw [-latex,blue,very thick] (2,-1.1) -- ++(0,-.8);
     
    \end{scope}
   
   \begin{scope}[shift={(0,-4.5)}]

   \draw (-1,-1) -- ++(0,2);
   \draw (1,-1) -- ++(0,2);
   \draw (3,-1) -- ++(0,2);
   \draw (5,-1) -- ++(0,2);

   \draw (-1,.2) -- ++(2,0);
   \draw (-1,.4) -- ++(2,0);
   \draw (-1,.6) -- ++(2,0);

   \draw (1,.1) -- ++(2,0);
   \draw (1,.3) -- ++(2,0);
   \draw (1,.5) -- ++(2,0);

   \draw (3,.2) -- ++(2,0);
   \draw (3,.4) -- ++(2,0);
   \draw (3,.6) -- ++(2,0);
   
   \draw [dotted] (-1,-.1) -- ++(2,0);
   \draw [dotted] (-1,-.3) -- ++(2,0);

   \draw [dotted] (1,-.2) -- ++(2,0);
   \draw [dotted] (1,-.4) -- ++(2,0);

   \draw [dotted] (3,-.1) -- ++(2,0);
    \draw [dotted] (3,-.3) -- ++(2,0);

   \draw (-2,-.6) -- ++(8,0);
    \draw (-2,-.8) -- ++(8,0);

   \draw [-latex,blue,very thick] (2,-1.6) -- ++(0,-.8);    
    
   \end{scope}

   \begin{scope}[shift={(0,-8)}]

   \draw (-1,-1) -- ++(0,2);
   \draw (1,-1) -- ++(0,2);
   \draw (3,-1) -- ++(0,2);
   \draw (5,-1) -- ++(0,2);

   \draw (-1,.2) -- ++(2,0);
   \draw (-1,.4) -- ++(2,0);
   \draw (-1,.6) -- ++(2,0);

   \draw (1,.1) -- ++(2,0);
   \draw (1,.3) -- ++(2,0);
   \draw (1,.5) -- ++(2,0);

   \draw (3,.2) -- ++(2,0);
   \draw (3,.4) -- ++(2,0);
   \draw (3,.6) -- ++(2,0);
   
%
%

   \draw (-2,-.6) -- ++(1,0);
    \draw (-2,-.8) -- ++(1,0);

   \draw (5,-.6) -- ++(1,0);
   \draw (5,-.8) -- ++(1,0);    
    
   \end{scope}   

   \begin{scope}[shift={(0,-10)}]

    \draw (-2,0) -- (6,0);

    \filldraw[fill=white,draw=black] (0,0) circle (.2) node [below=1em] {SU$(n_0)$};
    \filldraw[fill=white,draw=black] (2,0) circle (.2) node [below=1em] {SU$(n_0)$};;
    \filldraw[fill=white,draw=black] (4,0) circle (.2) node [below=1em] {SU$(n_0)$};;

    \filldraw[fill=white,draw=black] (-2.2,-.2) rectangle ++(.4,.4) node [below=1.55em] {SU$(n_1)$};;
    \filldraw[fill=white,draw=black] (5.8,-.2) rectangle ++(.4,.4) node [below=1.55em] {SU$(n_1)$};;

   \end{scope}
   
  \end{tikzpicture}
\end{center}
 \caption{Reduction of supergroup quiver gauge theory via gauging.}
 \label{fig:super_quiv_red}
\end{figure}
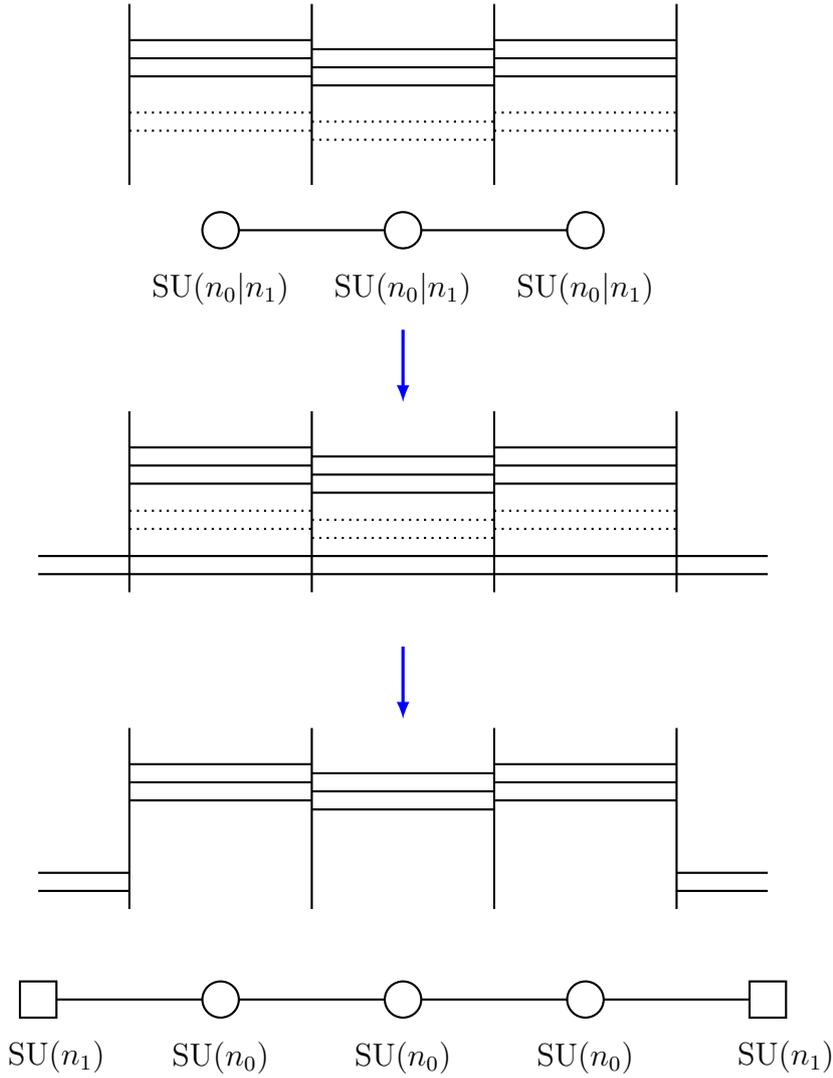

This argument is easily generalized to quiver theory.
The brane configuration of $A_p$ quiver with gauge group SU$(n_0|n_1)$ is shown in Fig.~\ref{fig:super_quiv_red}.

For the moment, we assign the same super gauge group to all the gauge node for simplicity.
The negative branes are annihilated by gauging, and the resulting theory is the linear quiver theory with gauge groups SU$(n_0)$ and flavor nodes SU$(n_1)$ attached to the left and right most nodes.
We can consider the situation with different super gauge groups assigned to each node.
However, in such a case, it is not possible to annihilate all the negative branes at the same time in general.
See also \S\ref{sec:decoupling}.

\begin{figure}[ht] 
\begin{center}
  \begin{tikzpicture}[thick,scale=1.2]

   \draw (-1,-1) -- ++(0,2) node [above] {NS5};
   \draw (1,-1) -- ++(0,2) node [above] {NS5};
   \draw (3,-1) -- ++(0,2) node [above] {NS5};
   \draw[red, very thick] (5,-1) -- ++(0,2) node [above] {ON$^-$};

   \draw (-1,.2) -- ++(2,0);
   \draw (-1,.4) -- ++(2,0);
   \draw (-1,.6) -- ++(2,0);
   \draw (-1,.8) -- ++(2,0);   

   \draw (1,.1) -- ++(2,0);
   \draw (1,.3) -- ++(2,0);
   \draw (1,.5) -- ++(2,0);
   \draw (1,.7) -- ++(2,0);   

   \draw (3,.2) -- ++(2,0);
   \draw (3,.4) -- ++(2,0);
   \draw (3,.6) -- ++(2,0);
   \draw (3,.8) -- ++(2,0);   
   
   \draw [dotted] (-1,-.2) -- ++(2,0);
   \draw [dotted] (-1,-.4) -- ++(2,0);

   \draw [dotted] (1,-.3) -- ++(2,0);
   \draw [dotted] (1,-.5) -- ++(2,0);

   \draw [dotted] (3,-.2) -- ++(2,0);
   \draw [dotted] (3,-.4) -- ++(2,0);

    \begin{scope}[shift={(0,-2)}]

     \draw (0,0) -- (2,0);
     \draw (4,.5) -- (2,0) -- (4,-.5);

    \filldraw[fill=white,draw=black] (0,0) circle (.2) node [below=1em] {SU$(2n_0|2n_1)$};
    \filldraw[fill=white,draw=black] (2,0) circle (.2) node [below=1em] {SU$(2n_0|2n_1)$};;
     \filldraw[fill=white,draw=black] (4,.5) circle (.2) node [right=1em] {SU$(n_0|n_1)$};;
    \filldraw[fill=white,draw=black] (4,-.5) circle (.2) node [right=1em] {SU$(n_0|n_1)$};;

   \draw [-latex,blue,very thick] (2,-1.4) -- ++(0,-.8);
     
    \end{scope}
   
   \begin{scope}[shift={(0,-5.5)}]

   \draw (-1,-1) -- ++(0,2);
   \draw (1,-1) -- ++(0,2);
   \draw (3,-1) -- ++(0,2);
   \draw[red,very thick] (5,-1) -- ++(0,2);

   \draw (-1,.2) -- ++(2,0);
   \draw (-1,.4) -- ++(2,0);
   \draw (-1,.6) -- ++(2,0);
   \draw (-1,.8) -- ++(2,0);    

   \draw (1,.1) -- ++(2,0);
   \draw (1,.3) -- ++(2,0);
   \draw (1,.5) -- ++(2,0);
   \draw (1,.7) -- ++(2,0);    

   \draw (3,.2) -- ++(2,0);
   \draw (3,.4) -- ++(2,0);
   \draw (3,.6) -- ++(2,0);
   \draw (3,.8) -- ++(2,0);
    
   \draw [dotted] (-1,-.1) -- ++(2,0);
   \draw [dotted] (-1,-.3) -- ++(2,0);

   \draw [dotted] (1,-.2) -- ++(2,0);
   \draw [dotted] (1,-.4) -- ++(2,0);

   \draw [dotted] (3,-.1) -- ++(2,0);
    \draw [dotted] (3,-.3) -- ++(2,0);

   \draw (-2,-.6) -- ++(7,0);
    \draw (-2,-.8) -- ++(7,0);

   \draw [-latex,blue,very thick] (2,-1.4) -- ++(0,-.8);    
    
   \end{scope}

   \begin{scope}[shift={(0,-9)}]

   \draw (-1,-1) -- ++(0,2);
   \draw (1,-1) -- ++(0,2);
   \draw (3,-1) -- ++(0,2);
   \draw[red, very thick] (5,-1) -- ++(0,2);

   \draw (-1,.2) -- ++(2,0);
   \draw (-1,.4) -- ++(2,0);
   \draw (-1,.6) -- ++(2,0);
   \draw (-1,.8) -- ++(2,0);
    
   \draw (1,.1) -- ++(2,0);
   \draw (1,.3) -- ++(2,0);
   \draw (1,.5) -- ++(2,0);
   \draw (1,.7) -- ++(2,0);
    
   \draw (3,.2) -- ++(2,0);
   \draw (3,.4) -- ++(2,0);
   \draw (3,.6) -- ++(2,0);
   \draw (3,.8) -- ++(2,0);
    
%
%

   \draw (-2,-.6) -- ++(1,0);
    \draw (-2,-.8) -- ++(1,0);

    
   \end{scope}   

   \begin{scope}[shift={(0,-11.5)}]

     \draw (-2,0) -- (2,0);
     \draw (4,.5) -- (2,0) -- (4,-.5);

    \filldraw[fill=white,draw=black] (0,0) circle (.2) node [below=1em] {SU$(2n_0)$};
    \filldraw[fill=white,draw=black] (2,0) circle (.2) node [below=1em] {SU$(2n_0)$};
    \filldraw[fill=white,draw=black] (4,.5) circle (.2) node [right=1em] {SU$(n_0)$};
    \filldraw[fill=white,draw=black] (4,-.5) circle (.2) node [right=1em] {SU$(n_0)$};    

    \filldraw[fill=white,draw=black] (-2.2,-.2) rectangle ++(.4,.4) node [below=1.55em] {SU$(2n_1)$};;

   \end{scope}
   
  \end{tikzpicture}
\end{center}
 \caption{Reduction of supergroup gauge theory via gauging for $D$-type quiver.}
 \label{fig:super_quiv_D}
\end{figure}
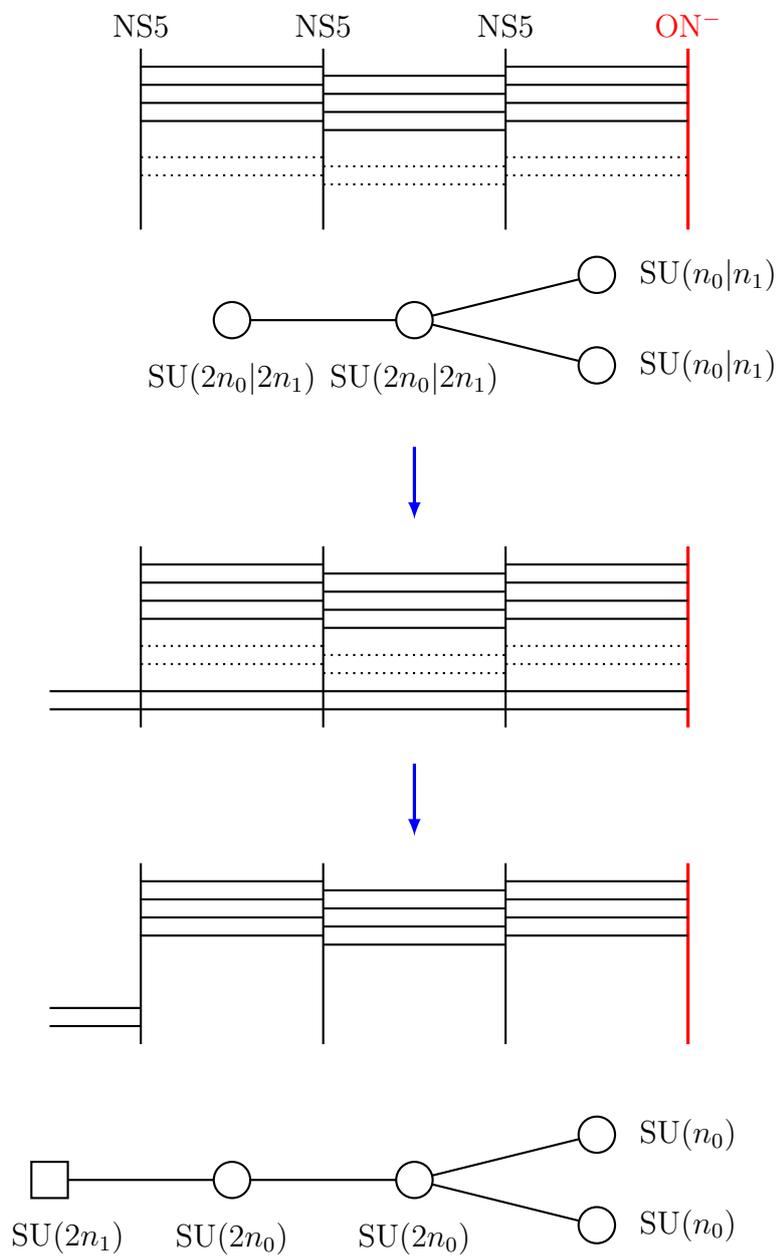

Similarly, we may consider the D-type quiver theory.
Applying the previous argument, D-type quiver gauge theory is realized using the ON$^0$ plane as in Fig.~\ref{fig:super_quiv_D}.
Thus the configuration is reduced to that for $D_r$ quiver with non-supergroup gauge nodes through the gauging process.
Such a reduction is consistent with another approach discussed in \S\ref{sec:decoupling_D}.
We can similarly deal with $\widehat{D}_r$ quiver by imposing ON$^0$ planes on the both ends of the configuration.

\section{Eight supercharge theory in higher dimensions}\label{sec:5d6d}

In this Section, we consider a higher dimensional analog of the geometric analysis discussed in this Chapter~\cite{Nekrasov:1996cz,Nekrasov:2002qd,Hollowood:2003cv}.
In fact, 6d $\mathcal{N} = (1,0)$ theory has eight supercharges, and is reduced to 4d $\mathcal{N} = 2$ theory via toroidal compactification.
The same argument is applicable to 5d $\mathcal{N} = 1$ theory.
In these cases, the corresponding R-symmetry is given by $\Sp(1) = \SU(2)$.

The bosonic components in the vector multiplet in $\mathsf{d} = 4,5,6$ are in the following:
In $\mathsf{d} = 6$, it contains the gauge field $A_{0,1,2,3,4,5}$.
In $\mathsf{d} = 5$, it contains the gauge field $A_{0,1,2,3,4}$ and also the real scalar $\phi$.
In $\mathsf{d} = 4$, it contains the gauge field $A_{0,1,2,3}$ and the complex scalar $(\phi,\bar{\phi})$.
The scalar fields in $\mathsf{d} = 4, 5$ are obtained through the dimensional reduction of the fourth and fifth components of the gauge field $A_{4, 5}$ in $\mathsf{d} = 6$:
\begin{equation}
 \begin{tikzcd}
  \text{(6d)} & \text{(5d)} & \text{(4d)}
  \\[-1em]
  A_{4,5} \arrow[r] & A_4, \phi \arrow[r] & \phi, \bar{\phi}
 \end{tikzcd}
\end{equation}
We explore the Seiberg--Witten approach to these higher dimensional theories based on the toroidal compactification.

\subsection{5d $\mathcal{N} = 1$ theory}\label{sec:5dN=1}

We now consider 5d $\mathcal{N} = 1$ gauge theory compactified on a circle $S^1$.
In this case, the fourth component of the gauge field $A_4$ and the real scalar play a similar role to the complex scalar in 4d $\mathcal{N} = 2$ gauge theory, which are combined into the Wilson loop operator along the compactification circle $S^1$:%
\footnote{%
In supersymmetric gauge theory, one should combine the gauge field and the corresponding scalar field to preserve (a part of) supersymmetry.
See, for example, \cite{Gaiotto:2007qi} for details.
}
\begin{align}
 \Phi := \operatorname{P} \exp \left( \oint_{S^1} \left( A_4 + \im \phi \, dx_4 \right) \right)
 \in G
 \, ,
\end{align}
where P stands for the path-ordering product.
We remark that $\Phi$ is a $G$-valued chiral ring operator (multiplicative), while the complex scalar in four dimensions takes a value in $\mathfrak{g} = \operatorname{Lie} G$ (additive).
Imposing the supersymmetric vacuum condition as discussed in \S\ref{sec:SUSY_vac}, $\Phi$ takes a value in the Cartan subgroup of the gauge group, $H \subset G$, which parametrizes the Coulomb branch of the moduli space of the vacua.
\index{moduli space of vacua!Coulomb branch}

For $G = \SU(n)$ with $(n^\text{f},n^\text{af})$ flavors, the Seiberg--Witten curve is given by the multiplicative analog of the curve presented in \eqref{eq:SW_curve_SQCD}:
\begin{align}
 \Sigma: \quad 
 y + \Lambda_\text{5d}^b \, \frac{x^{n - \kappa}}{y} \, P(x) \widetilde{P}(x) = \det(1 - x \, \Phi^{-1})
 \, , \qquad
 (x,y) \in \mathbb{C}^\times \times \mathbb{C}^\times
 \, ,
 \label{eq:SW_curve_5d}
\end{align}
\index{Seiberg--Witten curve!5d theory}%
where $b = 2n-n^\text{f}-n^\text{af}$, and $\Lambda_\text{5d}$ is the dynamical parameter for 5d theory.
There are several specific points to 5d theory.
$P(x)$ and $\widetilde{P}(x)$ are the multiplicative (K-theory) analogs of the matter polynomial for the fundamental and antifundamental hypermultiplets~\eqref{eq:matter_polynomial_quiver}: 
\begin{align}
 P(x) = \prod_{f = 1}^{n^\text{f}} \left( 1 - \np^{m_f} / x \right)
 \, , \qquad
 \widetilde{P}(x) = \prod_{f = 1}^{n^\text{af}} (1 - x / \np^{\widetilde{m}_f})
 \, .
\end{align}
In fact, their asymptotic behaviors are different at $x \to \infty, 0$, and we should distinguish the fundamental and antifundamental matters in five dimensions.

We denote the Chern--Simons level by $\kappa \in \mathbb{Z}$, a coefficient of the topological term $\displaystyle \int \tr \left( A \wedge F \wedge F \right)$, which is reduced to the Wilson loop on $S^1$ with the instanton background.
There is an upper bound for the level $\kappa \le n$ for $\SU(n)$ theory, concerning the asymptotic freedom.

In this case, the $x$-variable is a multiplicative variable, $x \in \mathbb{C}^\times$.
The one-form and the associated symplectic two-form on the curve are now given by
\begin{subequations}\label{eq:one_form_5d}
 \begin{align}
  \lambda & = \log x \, \frac{dy}{y} = \log x \, d \log y
  \, , \\
  \omega & = d \lambda = d \log x \wedge d \log y
  \, .
 \end{align}
\end{subequations}
Hence we can swap $x$ and $y$ as a symplectic transform.
As discussed in \S\ref{sec:SW_curve_quiver}, the $x$ and $y$ variables are corresponding to the gauge and quiver structure on the Seiberg--Witten curve.
This $x \leftrightarrow y$ symmetry implies an interesting duality between the gauge and quiver structures.
This argument will be also justified using the brane description later.

\subsubsection{Reduction to 4d theory}

Let us discuss the reduction of the 5d Seiberg--Witten curve to the 4d curve.
In order to take this scaling limit, we put the scaling parameter $R$, s.t., $x = \np^{R \, z}$, $\Phi = \diag (\np^{R \, \mathsf{a}_\alpha})_{\alpha = 1,\ldots,n}$.
In addition, we impose the special unitary condition $\displaystyle \sum_{\alpha = 1}^n \mathsf{a}_\alpha = 0$, so that $\det \Phi = 1$, for simplicity.
Then, one can rewrite the curve for 5d pure $\SU(n)$ SYM theory in the form of
\begin{align}
 y + \left( \frac{\Lambda_\text{5d}}{R} \right)^{2n} \frac{x^{-\kappa}}{y} 
 = \prod_{\alpha = 1}^{n} 
\frac{2}{R} \sinh \left( \frac{R}{2} \left( z - \mathsf{a}_\alpha \right) \right)
\end{align}
where we shift the $y$ variable as $y \to (-1)^n R^n x^{n/2} y$.
Recalling $\displaystyle \lim_{R \to 0} \frac{2}{R} \sinh \frac{R}{2} z = z$ and $\displaystyle \lim_{R \to 0} \np^{R \, z} = 1$, it is reduced to the curve for 4d theory \eqref{eq:SW_curve_SUn} in the limit $R \to 0$ under the identification
\begin{align}
 \lim_{R \to 0} \frac{\Lambda_\text{5d}}{R} = \Lambda_{\text{4d}}
 \, .
\end{align}

In fact, the scaling parameter $R$ is interpreted as the size of the compactification circle.
On the other hand, denoting $x = \np^{R \, z}$, it shows the periodicity $z \simeq z + 2 \pi \im / R$.
Namely, $x$ is a coordinate of the cylinder, $x \in \mathbb{C}^\times = \check{S}^1 \times \mathbb{R}$, where $\check{S}^1$ is the dual circle with the size $\check{R} = 1/R$.
In general, for gauge theory defined on $\mathcal{S} \times \mathcal{C}$, the $x$-variable takes a value in $\check{\mathcal{C}}$, which is dual to $\mathcal{C}$:
\begin{align}
 \mathcal{C} =
 \begin{cases}
  \text{pt} = \text{pt} \times \text{pt} & (\text{4d}) \\
  S^1 = S^1 \times \text{pt} & (\text{5d}) \\
  T^2 = S^1 \times S^1 & (\text{6d})
 \end{cases}
 \quad \longleftrightarrow \quad
 \check{\mathcal{C}} =
 \begin{cases}
  \mathbb{C} = \mathbb{R} \times \mathbb{R} & (\text{4d}) \\
  \mathbb{C}^\times = \check{S}^1 \times \mathbb{R} & (\text{5d}) \\
  \check{T}^2 = \check{S}^1 \times \check{S}^1 & (\text{6d})
 \end{cases}
 \label{eq:C_Cdual}
\end{align}
This relation is explained as follows~\cite{Seiberg:1996nz,Nekrasov:2012xe}:
We take a further compactification from 4d $\Gamma$-quiver gauge theory to 3d, $\mathcal{S} = \mathbb{R}^4 \to \mathbb{R}^3 \times S^1$.
Then the resulting three-dimensional theory is $\mathcal{N} = 4$ sigma model, and the target space is given by the moduli space of the periodic $G_\Gamma$-monopole on $S^1 \times \check{\mathcal{C}}$ as a result of the Nahm dual transformation (3d mirror symmetry).
In fact, the Seiberg--Witten curve is identified with the spectral curve of the $G_\Gamma$-monopole, and thus the $x$-variable plays a role of the spectral parameter that takes a value in $x \in \check{\mathcal{C}}$.
See also \S\ref{sec:hierarchy} for a related discussion.

\subsubsection{Brane configuration}

We can consider the brane configuration for $\mathcal{N} = 1$ gauge theory in five dimensions similarly to the discussion in \S\ref{sec:brane_N=2}.
Suppose that we compactify 4-direction on a circle $S^1$ in Type IIA setup as in Tab.~\ref{tab:HW}.
Then, we can consider the T-dual transformation in this direction to obtain Type IIB configuration~\cite{Aharony:1997bh}:

 \begin{table}[h]
  \begin{center}
   \begin{tabular}{ccccccccccc}\hline\hline
    (IIB) & 0 & 1 & 2 & 3 & 4 & 5 & 6 & 7 & 8 & 9 \\\hline
    NS5 & - & - & - & - & - & - & & & & \\ 
    D5 & - & - & - & - & - & & - & & & \\ 
    D7 & - & - & - & - & - & & & - & - & - \\
    \hline\hline
   \end{tabular}
  \end{center}
  \caption{The brane configuration of five-dimensional $\mathcal{N} = 1$ gauge theory in Type IIB string theory obtained through the T-dual transformation from the IIA setup.} 
  \label{tab:HW_5d}
 \end{table}

 \noindent
 Now NS5 and D5 branes are extended in 5 and 6 directions in addition to the common world-volume, 01234.
 In this case, these branes are converted to each other through S-duality in Type IIB string theory, and form the $(p,q)$ 5-brane web in 56-directions: \index{S-duality}
 \begin{align}
  \begin{tikzpicture}[baseline=(current  bounding  box.center)]
   \draw [thick] (-1,0) node [left] {(1,0)} -- (0,0) --++(45:1) node [right] {(1,1)};
   \draw [thick] (0,-1) node [below] {(0,1)} -- (0,0);
   \begin{scope}[shift={(-5,0)}]
    \draw [-latex] (0,0) -- ++(.75,0) node [right] {$x_6$};
    \draw [-latex] (0,0) -- ++(0,.75) node [right] {$x_5$};
   \end{scope}
  \end{tikzpicture}
 \end{align}
 where $(0,1)$ and $(1,0)$ stand for NS5 and D5 branes, and $(p,q)$-brane is a composition of $p$ D5 and $q$ NS5 branes.
 In fact, $(p,q)$-brane goes to $(p,q)$ direction in 56-plane in order to balance tensions between the 5-branes.
 In this context, D7 branes are used to terminate these 5-branes similarly to D6 branes in Type IIA setup.

 The simplest example is pure $\SU(2)$ gauge theory realized with two NS5 and two D5 branes.
 In this case, there are three possibilities providing different Chern--Simons level $\kappa = 0, 1, 2$:
 \begin{align}
  \begin{tikzpicture}[baseline=(current  bounding  box.center),thick,scale=.6]
   \node at (-2,3) {(a) $\kappa = 0$};
   \draw (-1,-1) -- (1,-1) -- (1,1) -- (-1,1) -- cycle;
   \draw (-1,-1) --++(225:1);
   \draw (1,-1) --++(-45:1);
   \draw (1,1) --++(45:1);
   \draw (-1,1) --++(135:1);
   \begin{scope}[shift={(8,0)}]
   \node at (-2,3) {(b) $\kappa = 1$};
   \draw (-3,-1) -- (1,-1) -- (1,1) -- (-1,1) -- cycle;
   \draw (-3,-1) --++(202.5:1);
   \draw (1,-1) --++(-45:1);
   \draw (1,1) --++(45:1);
   \draw (-1,1) --++(90:1);    
   \end{scope}
   \begin{scope}[shift={(16,0)}]
   \node at (-2,3) {(c) $\kappa = 2$};
   \draw (-3,-1) -- (3,-1) -- (1,1) -- (-1,1) -- cycle;
   \draw (-3,-1) --++(202.5:1);
   \draw (3,-1) --++(-22.5:1);
   \draw (1,1) --++(90:1);
   \draw (-1,1) --++(90:1);    
   \end{scope}
  \end{tikzpicture}
  \label{fig:pure_SU(2)_web}
 \end{align}
 We remark that we cannot consider higher Chern--Simons level beyond $\kappa = 2$ as mentioned in the context of the Seiberg--Witten curve.
 Furthermore, this $(p,q)$-brane web diagram is interpreted as a dual to the toric diagram of the non-compact toric Calabi--Yau three-fold.
 Such a connection between gauge theory and Calabi--Yau geometry is known as the {\em geometric engineering}~\cite{Katz:1996fh,Katz:1997eq,Dijkgraaf:2002fc}, and one can see the agreement of 5d $\mathcal{N} = 1$ instanton partition function and the topological string amplitude on the corresponding Calabi--Yau three-fold~\cite{Aganagic:2003db,Awata:2005fa,Iqbal:2007ii}.

 This construction is straightforwardly generalized to quiver gauge theory.
 Let us consider $A_3$ quiver as an example,
 \begin{equation}
  A_3: \quad
 \dynkin[mark=o,root radius=.2cm,edge length=1.5cm,label,label macro/.code={\mathrm{SU}(n_{#1})}]{A}{3}
 \, .
 \end{equation}
 The brane descriptions in Type IIA and IIB theories are given as follows:
 \begin{align}
  \begin{tikzpicture}[baseline=(current  bounding  box.center),scale=.8,thick]  
   \node at (-2,2) {(IIA)};
   \draw (0,-1.5) -- ++(0,2.5); node [above] {NS5};
   \draw (3,-1.5) -- ++(0,2.5); node [above] {NS5};
   \draw (6,-1.5) -- ++(0,2.5); node [above] {NS5};
   \draw (9,-1.5) -- ++(0,2.5); node [above] {NS5};
   \foreach \x in {0,1}{
   \foreach \y in {0,2}{
   \draw (\y*3,-1.+\x) -- ++(3,0);
   }
   }
   \draw (3,-.5) -- ++(3,0);
   \draw (3,+.5) -- ++(3,0);
   \foreach \x in {-.5,+.5}{
   \draw (0,\x) -- ++(-2,0);
   \draw (9,\x) -- ++(+2,0);   
   }
   \draw[blue,very thick,latex-latex] (4.5,-2.) -- ++(0,-.75) node [right,black] {T-dual} -- ++(0,-.75);
   \node at (-2,-3.5) {(IIB)};   
   \begin{scope}[shift={(1,-8.5)}]
    \foreach \x in {0,1,2,3}{
    \foreach \y in {0,1}{
    \draw ($\x*(1.707,.707)+\y*(.707,1.707)$) -- ++(-1,0);
    \draw ($\x*(1.707,.707)+\y*(.707,1.707)$) -- ++(0,-1);
    \draw ($\x*(1.707,.707)+\y*(.707,1.707)$) -- ++(45:1);
    }
    \draw ($\x*(1.707,.707)+2*(.707,1.707)$) -- ++(0,-1);
    }
    \foreach \y in {0,1}{
    \draw ($4*(1.707,.707)+\y*(.707,1.707)$) -- ++(-1,0);
    }
   \end{scope}
  \end{tikzpicture}
  \label{fig:brane_T-dual}
 \end{align}
 Here we also incorporate the flavor nodes to the left most and the right most gauge nodes.
 One can similarly consider the $AD$-type affine quiver by imposing the periodicity in 6-direction, and taking into account O-planes as discussed in \S\ref{sec:HW_quiver}.

 Since S-duality exchanges NS5 and D5 branes in Type IIB string theory, the IIB web diagram in \eqref{fig:brane_T-dual} also has another equivalent description:
 \index{S-duality}
 \begin{align}
  \begin{tikzpicture}[baseline=(current bounding box.center),scale=.8,thick]
   \node at (0,6) {(a)};
   \node at (11,6) {(b)};
   \begin{scope}[shift={(0,1)}]
    \foreach \x in {0,1,2,3}{
    \foreach \y in {0,1}{
    \draw ($\x*(1.707,.707)+\y*(.707,1.707)$) -- ++(-1,0);
    \draw ($\x*(1.707,.707)+\y*(.707,1.707)$) -- ++(0,-1);
    \draw ($\x*(1.707,.707)+\y*(.707,1.707)$) -- ++(45:1);
    }
    \draw ($\x*(1.707,.707)+2*(.707,1.707)$) -- ++(0,-1);
    }
    \foreach \y in {0,1}{
    \draw ($4*(1.707,.707)+\y*(.707,1.707)$) -- ++(-1,0);
    }
   \end{scope}
   \draw[blue,very thick,latex-latex] (8,3.5) -- ++(.75,0) node [above,black] {S-dual} -- ++(.75,0);
   \begin{scope}[shift={(11,0)}]
    \foreach \x in {0,1}{
    \foreach \y in {0,1,2,3}{
    \draw ($\x*(1.707,.707)+\y*(.707,1.707)$) -- ++(-1,0);
    \draw ($\x*(1.707,.707)+\y*(.707,1.707)$) -- ++(0,-1);
    \draw ($\x*(1.707,.707)+\y*(.707,1.707)$) -- ++(45:1);
    }
    \draw ($\x*(1.707,.707)+4*(.707,1.707)$) -- ++(0,-1);
    }
    \foreach \y in {0,1,2,3}{
    \draw ($2*(1.707,.707)+\y*(.707,1.707)$) -- ++(-1,0);
    }
   \end{scope}
   \begin{scope}[shift={(1.5,-2.)}]
    \draw (-2,0) -- (5.5,0);
    \foreach \x in {0,1,2}{
    \filldraw[fill=white,draw=black] (1.7*\x,0) circle (.2) node [below=1em] {SU$(2)$};
    }
    \filldraw[fill=white,draw=black] ($(-2,0)+(-.2,-.2)$) rectangle ++(.4,.4) node [below=1.4em] {SU$(2)$};
    \filldraw[fill=white,draw=black] ($(5.4,0)+(-.2,-.2)$) rectangle ++(.4,.4) node [below=1.4em] {SU$(2)$};
   \end{scope}   
   \begin{scope}[shift={(12.5,-2.)}]
    \draw (-2,0) -- (2,0);
    \filldraw[fill=white,draw=black] (0,0) circle (.2) node [below=1em] {SU$(4)$};
    \filldraw[fill=white,draw=black] ($(-2,0)+(-.2,-.2)$) rectangle ++(.4,.4) node [below=1.4em] {SU$(4)$};
    \filldraw[fill=white,draw=black] ($(2,0)+(-.2,-.2)$) rectangle ++(.4,.4) node [below=1.4em] {SU$(4)$};
   \end{scope}   
 \end{tikzpicture}
 \end{align}
 These web diagrams provide
 \begin{itemize}
  \item Theory (a): $A_3$ quiver with $G_{A_1} = \SU(2)$ gauge symmetry
  \item Theory (b): $A_1$ quiver with $G_{A_3} = \SU(4)$ gauge symmetry
 \end{itemize}
 Namely, the gauge and quiver structures are exchanged through the S-duality~\cite{Katz:1997eq,Aharony:1997bh}.
 The agreement between the partition functions of both theories are explicitly checked~\cite{Bao:2011rc}.
 This duality is specific to 5d $\mathcal{N} = 1$ gauge theory, and from geometric point of view, it is interpreted as a consequence of the symplectic transform on the Seiberg--Witten curve, $x \leftrightarrow y$, as discussed before.
 In addition, it is also possible to discuss the S-duality in the presence of the defect operators.
 See~\cite{Assel:2018rcw,Nieri:2018pev} for details.
 
 Let us briefly mention the brane description for supergroup gauge theory.
 As shown in~\cite{Okuda:2006fb}, supergroup gauge theory is engineered using the negative brane, also known as the ghost brane.
 Based on this argument, the 5-brane description of 5d $\mathcal{N} = 1$ supergroup gauge theory has been considered~\cite{Kimura:2020lmc}, and the partition function is reproduced by a generalized version of the topological vertex formalism~\cite{Aganagic:2003db,Awata:2005fa,Iqbal:2007ii}.

\subsection{6d $\mathcal{N} = (1,0)$ theory}\label{sec:6dN=1}

We then consider 6d $\mathcal{N} = (1,0)$ theory compactified on a torus $T^2$.
Since we have two periodic directions, $(x_4,x_5) \in T^2$, the chiral ring operator is given as a combination of the loop operators as follows:
\begin{equation}
 \begin{tikzcd}
  \text{(4d)} & \text{(5d)} & \text{(6d)}
  \\[-1em]
  \phi, \bar{\phi} \arrow[r] & \displaystyle \oint_{x_4} \left( A_4 + \im \phi \, dx_4 \right) \arrow[r] & \displaystyle \oint_{x_4} A_4 + \tau \oint_{x_5} A_5
 \end{tikzcd} 
\end{equation}
where $\tau$ is the modulus of the torus $T^2$.
See \cite{Benini:2013xpa} for a similar argument in two dimensions.
In this case, the Seiberg--Witten curve for $\SU(n)$ gauge theory with $n^\text{f} = 2n$ flavors is accordingly written with a doubly periodic variable~\cite{Hollowood:2003cv}:
\begin{align}
 \Sigma : \quad
 y + \frac{\mathfrak{q}}{y} \prod_{f = 1}^{2n} \theta(x/\np^{R \, m_f};p) = \prod_{\alpha = 1}^n \theta(x/\np^{R \, \mathsf{a}_\alpha};p)
 \, , \qquad
 (x,y) \in \mathbb{C}^\times \times \mathbb{C}^\times
 \, ,
 \label{eq:SW_curve_6d}
\end{align}
\index{Seiberg--Witten curve!6d theory}%
where $\theta(x;p)$ is the theta function defined in \eqref{eq:theta_fn} with the elliptic nome $p = \exp \left( 2 \pi \im \tau \right) \in \mathbb{C}^\times$, and $\mathfrak{q} \in \mathbb{C}^\times$ is the (dimensionless) coupling constant.
$R$ is the size of the circle $S^1$ in 4-direction as before.
In this case, we use the same one-form $\lambda$ and the symplectic two-form $\omega = d\lambda$ on the curve as in 5d theory~\eqref{eq:one_form_5d}.
The theta function shows a (quasi) modular property under the shift $x \to px$.
Hence, denoting $x = \np^{R \, z}$, the variable $z$ turns out to be doubly periodic under the shift:
\begin{align}
 z \ \simeq \
 z + \frac{2 \pi \im}{R} 
 \, , \
 z + \frac{2 \pi \im}{R} \tau
 \qquad
 \left(
 x \ \simeq \
 px
 \right)
 \, .
 \label{eq:elliptic_coord}
\end{align}
Namely, it takes a value in the elliptic curve
\begin{align}
 \mathcal{E} = \mathbb{C}^\times/p^\mathbb{Z} = \mathbb{C}/ \left( \mathbb{Z} \oplus \tau \mathbb{Z} \right)
 \, .
 \label{eq:elliptic_curve}
\end{align}

We remark that, since 6d $\mathcal{N} = (1,0)$ theory is a chiral theory, we should impose the anomaly free condition for the matter content.
For $\SU(n)$ gauge theory, $n^\text{af} = 2n$ is the unique possibility for the fundamental hypermultiplet.%
\footnote{%
$\SU(n)$ gauge theory with a single adjoint matter is also possible in six dimensions.
In addition, we should also introduce the tensor multiplet to cancel the anomaly, similarly to the Green--Schwarz mechanism~\cite{Green:1984bx}.
See, for example,~\cite{Ohmori:2016ahw,Tomasiello:2020} for detailed reviews.
}
From the algebraic geometric point of view, all the terms in the curve~\eqref{eq:SW_curve_6d} should be elliptic functions with the same degree.
Assuming the $y$-variable has the degree $n$, the second term on the left hand side should be also of degree $n$, which is possible only for $n^\text{f} = 2n$.

The same argument is applicable to quiver gauge theory.
In this case, the anomaly free condition is given by
\begin{align}
 n^\text{f}_i = \sum_{j \in \Gamma_0} n_j \, c_{ji}
 \label{eq:6d_conformal_cond}
\end{align}
where $(c_{ij})_{i,j \in \Gamma_0}$ is the classical quiver Cartan matrix~\eqref{eq:classical_quiv_Cartan_mat}.
We remark that this is equivalent to the conformal condition in 4d $\mathcal{N} = 2$ gauge theory, where the coupling constant becomes a dimensionless parameter.

 \subsubsection{Branes and dualities}

 Let us discuss the brane description of 6d $\mathcal{N} = (1,0)$ theory with the toroidal compactification.
 Compared to 5d $\mathcal{N} = 1$ theory, 5-direction is now periodic as well as 4-direction.
 Therefore, we can apply the T-duality along 5-direction to obtain Type IIA description~\cite{Brunner:1997gf,Hanany:1997gh}:
  \begin{table}[h]
  \begin{center}
   \begin{tabular}{ccccccccccc}\hline\hline
    (IIA) & 0 & 1 & 2 & 3 & 4 & 5 & 6 & 7 & 8 & 9 \\\hline
    NS5 & - & - & - & - & - & - & & & & \\ 
    D6 & - & - & - & - & - & - & - & & & \\ 
    D8 & - & - & - & - & - & - & & - & - & - \\
    \hline\hline
   \end{tabular}
  \end{center}
  \caption{The brane configuration of six-dimensional $\mathcal{N} = (1,0)$ gauge theory in Type IIA string theory.} 
  \label{tab:HW_6d}
  \end{table}

  \noindent
 The corresponding brane diagram is given as
 \begin{align}
  \begin{tikzpicture}[baseline=(current  bounding  box.center),thick,scale=.8]
   \foreach \x in {0,1,...,6}{
   \draw (0,\x*.1-.3) -- ++(12,0);
   }
   \foreach \y in {1,2,3}{
   \filldraw[fill=white,draw=black] (\y*3,0) circle (.4) node [above=1em] {NS5};
   }
  \draw [decorate,decoration={brace,amplitude=10pt,raise=4pt},yshift=0pt]
   (-.2,-.4) -- ++(0,.8) node [black,midway,xshift=-1.5em,yshift=0em,left] {$n$ D6};
   \draw [-latex] (11,-1.5) -- ++(1.5,0) node [right] {$x_6$};
  \end{tikzpicture}
 \end{align}
 for the chain of $\SU(n)$ gauge theories.
 Although we have such a Type IIA description, we can also describe 6d $\mathcal{N} = (1,0)$ theory using a $(p,q)$-brane configuration by imposing the periodicity in 5-direction:
 \begin{align}
  \begin{tikzpicture}[baseline=(current  bounding  box.center),scale=.7,thick]
   \begin{scope}[shift={(-4,4)}]
    \draw [-latex] (0,0) -- ++(1,0) node [right] {$x_6$};
    \draw [-latex] (0,0) -- ++(0,1) node [right] {$x_5$};
   \end{scope}
   \begin{scope}[shift={(0,0)}]
    \foreach \x in {0,1}{
    \foreach \y in {0,1,2,3}{
    \draw ($\x*(1.707,.707)+\y*(.707,1.707)$) -- ++(-1,0);
    \draw ($\x*(1.707,.707)+\y*(.707,1.707)$) -- ++(0,-1);
    \draw ($\x*(1.707,.707)+\y*(.707,1.707)$) -- ++(45:1);
    }
    \draw ($\x*(1.707,.707)+4*(.707,1.707)$) -- ++(0,-1);
    }
    \foreach \y in {0,1,2,3}{
    \draw ($2*(1.707,.707)+\y*(.707,1.707)$) -- ++(-1,0);
    }
    \draw [red,very thick] ($(0,0)+(-0.3,-.8)$) -- ++(.6,0);
    \draw [red,very thick] ($(1.707,.707)+(-0.3,-.8)$) -- ++(.6,0);
    \draw [red,very thick] ($(4*.707,3*1.707)+(-0.3,.8+.707)$) -- ++(.6,0);
    \draw [red,very thick] ($(4*.707,3*1.707)+(-0.3+1.707,.8+2*.707)$) -- ++(.6,0);
   \end{scope}
   \begin{scope}[shift={(1.5,-2.)}]
    \draw (-2,0) -- (2,0);
    \filldraw[fill=white,draw=black] (0,0) circle (.2) node [below=1em] {SU$(4)$};
    \filldraw[fill=white,draw=black] ($(-2,0)+(-.2,-.2)$) rectangle ++(.4,.4) node [below=1.4em] {SU$(4)$};
    \filldraw[fill=white,draw=black] ($(2,0)+(-.2,-.2)$) rectangle ++(.4,.4) node [below=1.4em] {SU$(4)$};
   \end{scope}     
  \end{tikzpicture}
 \end{align}
 where $(0,1)$ 5-branes with red signs are identified along the periodic 5-direction.
 We remark that other 5-branes than $(0,1)$-brane cannot be compatible with the periodicity.
 For example, the configuration for the pure $\SU(2)$ theory depicted in \eqref{fig:pure_SU(2)_web} cannot be promoted to 6d theory since the directions of external legs are not compatible with the periodicity in 5-direction.
 This is also consistent with the anomaly condition in 6d theory.

 Since our description is based on $(p,q)$ 5-branes, we can consider the S-duality for this configuration as well.
 In this case, 6d gauge theory is dual to 5d theory with affine quiver structure:
 \begin{align}
  \begin{tikzpicture}[baseline=(current bounding box.center),scale=.7,thick]
   \node at (-1,7) [right] {(a) 5d theory};
   \node at (9.5,7) [right] {(b) 6d theory};
   \begin{scope}[shift={(6,5.5)},rotate=90,xscale=-1]
    \foreach \x in {0,1}{
    \foreach \y in {0,1,2,3}{
    \draw ($\x*(1.707,.707)+\y*(.707,1.707)$) -- ++(-1,0);
    \draw ($\x*(1.707,.707)+\y*(.707,1.707)$) -- ++(0,-1);
    \draw ($\x*(1.707,.707)+\y*(.707,1.707)$) -- ++(45:1);
    }
    \draw ($\x*(1.707,.707)+4*(.707,1.707)$) -- ++(0,-1);
    }
    \foreach \y in {0,1,2,3}{
    \draw ($2*(1.707,.707)+\y*(.707,1.707)$) -- ++(-1,0);
    }
    \draw [red,very thick] ($(0,0)+(-0.2,-.8)$) -- ++(.4,0);
    \draw [red,very thick] ($(1.707,.707)+(-0.2,-.8)$) -- ++(.4,0);
    \draw [red,very thick] ($(4*.707,3*1.707)+(-0.2,.8+.707)$) -- ++(.4,0);
    \draw [red,very thick] ($(4*.707,3*1.707)+(-0.2+1.707,.8+2*.707)$) -- ++(.4,0);
   \end{scope}
   \draw[blue,very thick,latex-latex] (8,3.5) -- ++(.75,0) node [above,black] {S-dual} -- ++(.75,0);
   \begin{scope}[shift={(10.5,0)}]
    \foreach \x in {0,1}{
    \foreach \y in {0,1,2,3}{
    \draw ($\x*(1.707,.707)+\y*(.707,1.707)$) -- ++(-1,0);
    \draw ($\x*(1.707,.707)+\y*(.707,1.707)$) -- ++(0,-1);
    \draw ($\x*(1.707,.707)+\y*(.707,1.707)$) -- ++(45:1);
    }
    \draw ($\x*(1.707,.707)+4*(.707,1.707)$) -- ++(0,-1);
    }
    \foreach \y in {0,1,2,3}{
    \draw ($2*(1.707,.707)+\y*(.707,1.707)$) -- ++(-1,0);
    }
    \draw [red,very thick] ($(0,0)+(-0.2,-.8)$) -- ++(.4,0);
    \draw [red,very thick] ($(1.707,.707)+(-0.2,-.8)$) -- ++(.4,0);
    \draw [red,very thick] ($(4*.707,3*1.707)+(-0.2,.8+.707)$) -- ++(.4,0);
    \draw [red,very thick] ($(4*.707,3*1.707)+(-0.2+1.707,.8+2*.707)$) -- ++(.4,0);
   \end{scope}
   \begin{scope}[shift={(.8,-2.8)}]
    \draw (0,0) -- (3.4,0) -- (1.7,-1) -- cycle;
    \foreach \x in {0,1,2}{
    \filldraw[fill=white,draw=black] (1.7*\x,0) circle (.2) node [above=1em] {SU$(2)$};
    }
    \filldraw[fill=white,draw=black] (1.7,-1) circle (.2) node [below=1em] {SU$(2)$};
   \end{scope}   
   \begin{scope}[shift={(12.,-3.2)}]
    \draw (-2,0) -- (2,0);
    \filldraw[fill=white,draw=black] (0,0) circle (.2) node [above=1em] {SU$(4)$};
    \filldraw[fill=white,draw=black] ($(-2,0)+(-.2,-.2)$) rectangle ++(.4,.4) node [above=.6em] {SU$(4)$};
    \filldraw[fill=white,draw=black] ($(2,0)+(-.2,-.2)$) rectangle ++(.4,.4) node [above=.6em] {SU$(4)$};
   \end{scope}   
 \end{tikzpicture}
 \end{align}
 More precisely, the matter contents obtained from these web diagrams are as follows:
 \begin{itemize}
  \item Theory (a): $\widehat{A}_3$ quiver with $G_{A_1} = \SU(2)$ gauge symmetry in 5d
  \item Theory (b): $A_1$ quiver with $G_{A_3} = \SU(4)$ gauge symmetry in 6d
 \end{itemize}
 This implies that 5d theory with $\widehat{G}$ gauge symmetry is interpreted as 6d theory with $G$ gauge symmetry, and the center of mass of the bifundamental mass parameters on 5d theory side is converted to the elliptic modulus on 6d side through the duality.

 There exist similar dualities for $DE$- and $\widehat{DE}$-type quiver gauge theories.
 Let us briefly comment on it.
 The dual theory of $D$-type quiver theory in 5d is $A$-type quiver with SO-Sp gauge symmetries in 5d (also known as SO-Sp alternating quiver), and its affinization, $\widehat{D}$-type quiver theory in 5d, is dual to the SO-Sp alternating quiver in 6d~\cite{Hayashi:2015vhy}.
 For $E$-type theory, the dual theory is $(E,E)$-type conformal matter theory in 6d. 
 See \cite{DelZotto:2014hpa} for details.

 Finally, let us consider affine quiver theory in six dimensions.
 In this case, the web diagram becomes doubly periodic as follows:
 \begin{align}
  \begin{tikzpicture}[baseline=(current bounding box.center),scale=.8,thick]
   \node at (-1,7.5) [right] {(a) 6d $\widehat{A}_3$ quiver};
   \node at (9.5,7.5) [right] {(b) 6d $\widehat{A}_1$ quiver};
   \begin{scope}[shift={(6,5.5)},rotate=90,xscale=-1]
    \foreach \x in {0,1}{
    \foreach \y in {0,1,2,3}{
    \draw ($\x*(1.707,.707)+\y*(.707,1.707)$) -- ++(-1,0);
    \draw ($\x*(1.707,.707)+\y*(.707,1.707)$) -- ++(0,-1);
    \draw ($\x*(1.707,.707)+\y*(.707,1.707)$) -- ++(45:1);
    }
    \draw ($\x*(1.707,.707)+4*(.707,1.707)$) -- ++(0,-1);
    }
    \foreach \y in {0,1,2,3}{
    \draw ($2*(1.707,.707)+\y*(.707,1.707)$) -- ++(-1,0);
    }
    \draw [red,very thick] ($(0,0)+(-0.2,-.8)$) -- ++(.4,0);
    \draw [red,very thick] ($(1.707,.707)+(-0.2,-.8)$) -- ++(.4,0);
    \draw [red,very thick] ($(4*.707,3*1.707)+(-0.2,.8+.707)$) -- ++(.4,0);
    \draw [red,very thick] ($(4*.707,3*1.707)+(-0.2+1.707,.8+2*.707)$) -- ++(.4,0);
    \foreach \x in {0,1,2,3} {
    \draw [teal,ultra thick] ($\x*(.707,1.707)+(-.8,-.2)$) -- ++(0,.4);
    \draw [teal,ultra thick] ($\x*(.707,1.707)+(-.2,-.2)+2*(1.707,.707)$) -- ++(0,.4);
    }
   \end{scope}
   \draw[blue,very thick,latex-latex] (8,3.5) -- ++(.75,0) node [above,black] {S-dual} -- ++(.75,0);
   \begin{scope}[shift={(10.5,0)}]
    \foreach \x in {0,1}{
    \foreach \y in {0,1,2,3}{
    \draw ($\x*(1.707,.707)+\y*(.707,1.707)$) -- ++(-1,0);
    \draw ($\x*(1.707,.707)+\y*(.707,1.707)$) -- ++(0,-1);
    \draw ($\x*(1.707,.707)+\y*(.707,1.707)$) -- ++(45:1);
    }
    \draw ($\x*(1.707,.707)+4*(.707,1.707)$) -- ++(0,-1);
    }
    \foreach \y in {0,1,2,3}{
    \draw ($2*(1.707,.707)+\y*(.707,1.707)$) -- ++(-1,0);
    }
    \draw [red,very thick] ($(0,0)+(-0.2,-.8)$) -- ++(.4,0);
    \draw [red,very thick] ($(1.707,.707)+(-0.2,-.8)$) -- ++(.4,0);
    \draw [red,very thick] ($(4*.707,3*1.707)+(-0.2,.8+.707)$) -- ++(.4,0);
    \draw [red,very thick] ($(4*.707,3*1.707)+(-0.2+1.707,.8+2*.707)$) -- ++(.4,0);
    \foreach \x in {0,1,2,3} {
    \draw [teal,ultra thick] ($\x*(.707,1.707)+(-.8,-.2)$) -- ++(0,.4);
    \draw [teal,ultra thick] ($\x*(.707,1.707)+(-.2,-.2)+2*(1.707,.707)$) -- ++(0,.4);
    }
   \end{scope}
   \begin{scope}[shift={(.8,-2.8)}]
    \draw (0,0) -- (3.4,0) -- (1.7,-1) -- cycle;
    \foreach \x in {0,1,2}{
    \filldraw[fill=white,draw=black] (1.7*\x,0) circle (.2) node [above=1em] {SU$(2)$};
    }
    \filldraw[fill=white,draw=black] (1.7,-1) circle (.2) node [below=1em] {SU$(2)$};
   \end{scope}   
   \begin{scope}[shift={(12.,-2.7)}]
    \draw (0,0) to [bend right=45] (0,-1.5);
    \draw (0,0) to [bend left=45] (0,-1.5);
    \filldraw[fill=white,draw=black] (0,0) circle (.2) node [right=1em] {SU$(4)$};
    \filldraw[fill=white,draw=black] (0,-1.5) circle (.2) node [right=1em] {SU$(4)$};
   \end{scope}   
 \end{tikzpicture}
 \end{align}
 The matter contents obtained from these diagrams are
 \begin{itemize}
  \item Theory (a): $\widehat{A}_3$ quiver with $G_{A_1} = \SU(2)$ gauge symmetry in 6d
  \item Theory (b): $\widehat{A}_1$ quiver with $G_{A_3} = \SU(4)$ gauge symmetry in 6d
 \end{itemize}
 In this case, the duality connects two affine quiver gauge theories both in six dimensions, which are described as the little string theory.
 Under this duality, the (center of) bifundamental mass and the elliptic modulus are exchanged.
 This duality is also interpreted as the T-duality in the context of the little string theory~\cite{Bhardwaj:2015oru}.

In addition to the S-duality exchanging $(1,0)$ and $(0,1)$ 5-branes, we can also discuss the triality including $(1,1)$ 5-brane, which is specific to doubly periodic situation, namely 6d affine quiver gauge theory.
See~\cite{Hohenegger:2016yuv,Bastian:2017ing,Bastian:2017ary,Bastian:2018dfu,Bastian:2018fba} for details.
This triality is expected to be specific to the $A$-type theories~\cite{DelZotto:2020sop}.

\chapter{Quantization of geometry}\label{chap:geometry}

The low energy effective behavior of $\mathcal{N} = 2$ gauge theory has the geometric characterization due to the Seiberg--Witten theory.
In particular, the algebraic curve, called the Seiberg--Witten curve, geometrically encodes the information about the prepotential of $\mathcal{N} = 2$ theory.
In this Chapter, we show how to obtain such an algebraic object from the microscopic path integral formalism together with the instanton counting described in Part~\ref{partI}.
As discussed there, the partition function obtained from the instanton counting involves additional parameters, that we call the $\Omega$-background parameters, compared to the Seiberg--Witten geometry.
We will see that these additional parameters play a role of quantum deformation parameter in the Seiberg--Witten theory, and also point out the correspondence to quantum integrable systems.
In particular, we will see that the quantization of Seiberg--Witten curve gives rise to the TQ-relation, and the saddle point equation with respect to the instanton partition function is identified with the Bethe equation of the corresponding integrable system.

\section{Non-perturbative Schwinger--Dyson equation}

The Schwinger--Dyson equation is a functional relation between the correlation functions in QFT.
Its derivation is based on the invariance of the partition function in the path integral formalism under infinitesimal deformation of the field.
As discussed in Chapter~\ref{chap:YM_instanton}, the gauge theory partition function is given as a superposition of different topological sectors ($\theta$-vacua), so that it would be covariant under the process of adding/removing an instanton, which is similar to the coherent state in quantum mechanics.
Since changing the topological sector is a non-perturbative effect in gauge theory, the functional relation obtained from such an operation is called the non-perturbative Schwinger--Dyson equation~\cite{Nekrasov:2015wsu}.
For this purpose, we consider the adding/removing-instanton operation for the instanton partition function, and discuss its geometric interpretation.

\subsection{Add/remove instantons}\label{sec:add/remove}

The instanton partition function localizes on a fixed point locus on the instanton moduli space $\mathfrak{M}$ under the equivariant actions.
In this context, the instanton configuration is combinatorially labeled by a partition $\lambda$, so that the adding/removing-instanton operation gives rise to shift of its specific component, $\lambda_{i,\alpha,k} \to \lambda_{i,\alpha,k} \pm 1$.
We study the behavior of the instanton partition function under this partition shift.

\subsubsection{Vector multiplet}

We start with the vector multiplet bundle and its character given as%
\footnote{%
Here we omit the node index $i \in \Gamma_0$ for simplicity.
}
\begin{align}
 \mathbf{V} = \frac{\wedge \mathbf{Q}_1^\vee}{\wedge \mathbf{Q}_2} \mathbf{X}^\vee \mathbf{X}
 \ \xrightarrow{\ch_{\mathsf{T}}} \
 \frac{1 - q_1^{-1}}{1 - q_2} 
 \sum_{\substack{(x,x') \in \mathcal{X} \times \mathcal{X} \\ x \neq x'}}
 \frac{x'}{x}
 \, .
\end{align}
Then, we take the variation
\begin{align}
 \delta \mathbf{V}
 := \mathbf{V}[\mathcal{X}_{\text{ad}:(\alpha,k)}] - \mathbf{V}[\mathcal{X}]
\end{align}
where we denote the configuration obtained by adding/removing an instanton by
\begin{align}
 \mathcal{X}_{\text{ad}:(i,\alpha,k)}
 = (\mathcal{X} \backslash \{x_{i,\alpha,k}\}) \sqcup \{ q_2 x_{i,\alpha,k}\}
 \, , \qquad
 \mathcal{X}_{\text{rm}:(i,\alpha,k)}
 = (\mathcal{X} \backslash \{x_{i,\alpha,k}\}) \sqcup \{ q_2^{-1} x_{i,\alpha,k}\}
 \, .
\end{align}
We simply denote $x = x_{\alpha,k}$, hence the character of the variation is given by
\begin{align}
 \ch_{\mathsf{T}} \delta \mathbf{V}
 & = \frac{1 - q_1^{-1}}{1 - q_2} \sum_{x' \in \mathcal{X} \backslash \{x\}}
 \left(
  (q_2^{-1} - 1) \frac{x'}{x} + (q_2 - 1) \frac{x}{x'}
 \right)
 \nonumber \\
 & = (1 - q_1^{-1}) \sum_{x' \in \mathcal{X} \backslash \{x\}} \left( q_2^{-1} \frac{x'}{x} - \frac{x}{x'} \right)
 \nonumber \\
 & = (1 - q_1^{-1}) \sum_{x' \in \mathcal{X}} \left( q_2^{-1} \frac{x'}{x} - \frac{x}{x'} \right) 
 - (1 - q_1^{-1}) \lim_{x'' \to x} \left( q_2^{-1} \frac{x''}{x} - \frac{x}{x''} \right)
 \, .
 \label{eq:vec_variation}
\end{align}
We have to carefully deal with the last term since, in the limit $x'' \to x$, it gives rise to the zero mode:
\begin{align}
 \lim_{x'' \to x} \mathbb{I} \left[ \frac{x''}{x} \right]
 = \lim_{x'' \to x} [\log x'' - \log x]
 = 0
 \, .
\end{align}
In order to remove this zero mode, we replace the summation over the set $\mathcal{X}$ with the summation over $\mathcal{X}_\text{ad}$,
\begin{align}
 & (1 - q_1^{-1}) 
 \left( 
 \sum_{x' \in \mathcal{X}_\text{ad}} q_2^{-1} \frac{x'}{x} 
 - \sum_{x' \in \mathcal{X}} \frac{x}{x'} 
 \right)
 + (1 - q_1^{-1}) \lim_{x'' \to x} \left( \frac{x}{x''} - \frac{x''}{x} \right)
 \nonumber \\
 & = 
 - (1 - q_1) \sum_{x' \in \mathcal{X}_\text{ad}} q^{-1} \frac{x'}{x}
 - (1 - q_1^{-1}) \sum_{x' \in \mathcal{X}_\text{ad}} \frac{x}{x'}
 + (1 - q_1^{-1}) \lim_{x'' \to x} \left( \frac{x}{x''} - \frac{x''}{x} \right)
 \nonumber \\
 & = 
 - \ch_{\mathsf{T}} \left(
 \det \mathbf{Q}^\vee \mathbf{x}^\vee \mathbf{Y}[\mathcal{X}_\text{ad}]
 + \mathbf{Y}[\mathcal{X}]^\vee \mathbf{x}
 \right)
 + (1 - q_1^{-1}) \lim_{x'' \to x} \left( \frac{x}{x''} - \frac{x''}{x} \right)
\end{align}
where we define a bundle $\mathbf{x}$, s.t., $\ch_{\mathsf{T}} \mathbf{x} = x$, and the last term is now given by
\begin{align}
 \lim_{x'' \to x} \mathbb{I} \left[ (1 - q_1^{-1}) \left( \frac{x}{x''} - \frac{x''}{x} \right) \right]
 & = \lim_{x'' \to x} \frac{[\log x - \log x''][\log x'' - \log x - \epsilon_1]}{[\log x'' - \log x][\log x - \log x'' - \epsilon_1]} 
 \nonumber \\
 & = - 1
 \, .
\end{align}
Applying the index formula, we then obtain the ratio of the vector multiplet contributions to the partition function evaluated with $\mathcal{X}$ and $\mathcal{X}_\text{ad}$:
\begin{align}
 \mathbb{I}[\delta \mathbf{V}] 
 = \frac{Z^\text{vec}[\mathcal{X}_\text{ad}]}{Z^\text{vec}[\mathcal{X}]} 
 = - \frac{1}{\mathsf{Y}_{qx}^\vee[\mathcal{X}_\text{ad}] \mathsf{Y}_x[\mathcal{X}]}
 \label{eq:vec_reflection}
\end{align}
where we define $\mathsf{Y}$-functions:
\begin{align}
 \mathsf{Y}_{i,x} = \mathbb{I}[ \mathbf{Y}_i^\vee \mathbf{x}]
 \, , \qquad
 \mathsf{Y}^\vee_{i,x} = \mathbb{I}[ \mathbf{x}^\vee \mathbf{Y}_i]
 \, .
\end{align}
We will discuss properties of the $\mathsf{Y}$-function shortly.

\subsubsection{Vector multiplet: another derivation}

Variation of the vector multiplet bundle is similarly considered in terms of the framing and the instanton bundles.
Adding an instanton corresponds to the shift of the instanton bundle:
\begin{align}
 \mathbf{K} 
 \ \longrightarrow \
 \mathbf{K} + \mathbf{x}
 \, .
\end{align}
Recalling the expression of the observable bundle $\mathbf{Y}$ in terms of $(\mathbf{N},\mathbf{K})$ as in~\eqref{eq:obs_bndl}, and the instanton part of the vector multiplet bundle is given as~\eqref{eq:V_bundle_inst}, the variation is computed as follows:
\begin{align}
 \delta \mathbf{V} = \delta \mathbf{V}^\text{inst}
 & = 
 - \det \mathbf{Q}^\vee \mathbf{x}^\vee \mathbf{Y}
 - \mathbf{Y}^\vee \mathbf{x}
 + \wedge \mathbf{Q}^\vee
 \, .
\end{align}
The last term will give a zero mode as well as the previous computation.

In order to regularize the zero mode term, we define the shifted observable bundle:
\begin{subequations} 
 \begin{align}
  \mathbf{Y}_\text{ad} 
  & := \mathbf{N} - \wedge \mathbf{Q} \cdot \left( \mathbf{K}  + \mathbf{x} \right)
  \, ,
  \\
  \mathbf{Y}_\text{rm} 
  & := \mathbf{N} - \wedge \mathbf{Q} \cdot \left( \mathbf{K} - \mathbf{x} \right)
  \, .
 \end{align} 
\end{subequations}
Then, the variation of the vector multiplet bundle is given as
\begin{align}
 \delta \mathbf{V}
 & = 
 - \det \mathbf{Q}^\vee \mathbf{x}^\vee \mathbf{Y}_\text{ad}
 - \mathbf{Y}^\vee \mathbf{x}
 = 
 - \det \mathbf{Q}^\vee \mathbf{x}^\vee \mathbf{Y}
 - \mathbf{Y}_\text{ad}^\vee \mathbf{x}
 \, ,
\end{align}
which is consistent with the previous argument.

\subsubsection{$\mathsf{Y}$-function: observable generating function}
\index{Y-function@$\mathsf{Y}$-function}

Let us discuss properties of the $\mathsf{Y}$-function.
Since the observable bundle $(\mathbf{Y}_i)_{i \in \Gamma_0}$ has several expressions, correspondingly the $\mathsf{Y}$-function also has several forms:
\begin{subequations}\label{eq:Y_func_def}
 \begin{align}
  \mathsf{Y}_{i,x}[\mathcal{X}]
  & \stackrel{\eqref{eq:quiver_univ_bundle}}{=}
  \prod_{\alpha = 1}^{n_i} \left[
  \left( 1 - \frac{\np^{\mathsf{a}_{i,\alpha}}}{x} \right) 
  \prod_{(s_1,s_2) \in \lambda_{i,\alpha}} \mathscr{S} \left( \frac{\np^{\mathsf{a}_\alpha} q_1^{s_1 - 1} q_2^{s_2 - 1}}{x} \right)
  \right]
  \nonumber \\
  & \stackrel{\eqref{eq:quiver_partial_red_univ_bundle}}{=}
  \prod_{x' \in \mathcal{X}_i} \frac{1 - x'/x}{1 - q_1 x'/x}
  \\
  \mathsf{Y}_{i,x}[\mathcal{X}]^\vee
  & \stackrel{\eqref{eq:quiver_univ_bundle}}{=}
  \prod_{\alpha = 1}^{n_i} \left[
  \left( 1 - \frac{x}{\np^{\mathsf{a}_{i,\alpha}}} \right) 
  \prod_{(s_1,s_2) \in \lambda_{i,\alpha}} \mathscr{S} \left( \frac{\np^{\mathsf{a}_\alpha} q_1^{s_1 - 1} q_2^{s_2 - 1}}{x} \right)
  \right]
  \nonumber \\
  & \stackrel{\eqref{eq:quiver_partial_red_univ_bundle}}{=}
  \prod_{x' \in \mathcal{X}_i} \frac{1 - x/x'}{1 - q_1^{-1} x/x'}
 \end{align}
\end{subequations}
where we apply the K-theory convention, and the $\mathscr{S}$-function is defined in \eqref{eq:S_func_def}.
They also have the following expressions:
\begin{subequations} \label{eq:Y_fn_combin}
\begin{align}
 \mathsf{Y}_{i,x}[\mathcal{X}] & = \prod_{\alpha = 1}^{n_i}
 \left(
 \prod_{s \in \partial_+ \lambda_\alpha} \left( 1 - \np^{\mathsf{a}_{i,\alpha}} q_1^{s_1 - 1} q_2^{s_2 - 1} / x \right)
 \prod_{s \in \partial_- \lambda_\alpha} \left( 1 - \np^{\mathsf{a}_{i,\alpha}} q_1^{s_1} q_2^{s_2} / x \right)^{-1}
 \right)
 \, , \\
 \mathsf{Y}_{i,x}[\mathcal{X}]^\vee & = \prod_{\alpha = 1}^{n_i}
 \left(
 \prod_{s \in \partial_+ \lambda_\alpha} \left( 1 - x/\np^{\mathsf{a}_{i,\alpha}} q_1^{s_1 - 1} q_2^{s_2 - 1} \right)
 \prod_{s \in \partial_- \lambda_\alpha} \left( 1 - x/\np^{\mathsf{a}_{i,\alpha}} q_1^{s_1} q_2^{s_2} \right)^{-1}
 \right)
 \, ,
\end{align}
\end{subequations}
where $\partial_\pm \lambda$ is the outer/inner boundary of the partition:
\begin{align}
 \begin{tikzpicture}[scale=.8,baseline=(current bounding box.center)]
   \draw (0,0) rectangle ++ (1,-1);
   \draw (1,0) rectangle ++ (1,-1);
   \draw (2,0) rectangle ++ (1,-1);
   \draw (3,0) rectangle ++ (1,-1);
   \draw (4,0) rectangle ++ (1,-1);
   \draw (5,0) rectangle ++ (1,-1);
   \draw (6,0) rectangle ++ (1,-1);      
   \draw (0,-1) rectangle ++ (1,-1);
   \draw (1,-1) rectangle ++ (1,-1);
   \draw (3,-1) rectangle ++ (1,-1);
   \draw (4,-1) rectangle ++ (1,-1);   
   \draw (0,-2) rectangle ++ (1,-1);
   \draw (1,-2) rectangle ++ (1,-1);
   \draw (2,-2) rectangle ++ (1,-1);
   \draw (3,-2) rectangle ++ (1,-1);   
   \draw (0,-3) rectangle ++ (1,-1);
   \draw (1,-3) rectangle ++ (1,-1);
   \draw (2,-3) rectangle ++ (1,-1);
   \draw (0,-4) rectangle ++ (1,-1);  
  \filldraw [fill=red!50] (6.5,-.5) circle (.3);
  \filldraw [fill=red!50] (4.5,-1.5) circle (.3);
  \filldraw [fill=red!50] (3.5,-2.5) circle (.3);
  \filldraw [fill=red!50] (2.5,-3.5) circle (.3);
  \filldraw [fill=red!50] (.5,-4.5) circle (.3);
  \filldraw [fill=blue!50] (7.5,-.5) circle (.3);
  \filldraw [fill=blue!50] (5.5,-1.5) circle (.3);
  \filldraw [fill=blue!50] (4.5,-2.5) circle (.3);
  \filldraw [fill=blue!50] (3.5,-3.5) circle (.3);
  \filldraw [fill=blue!50] (1.5,-4.5) circle (.3);
  \filldraw [fill=blue!50] (.5,-5.5) circle (.3);   
 \end{tikzpicture}  
 \label{fig:outer/inner}
\end{align}
One can add a box to the outer boundary \tikz[baseline=-3pt] \filldraw[fill=blue!50] (0,0) circle (.2); $\in \partial_+ \lambda$, and one can remove a box from the inner boundary \tikz[baseline=-3pt] \filldraw[fill=red!50] (0,0) circle [radius=.2]; $\in \partial_- \lambda$ of the partition $\lambda$.
We remark $|\partial_+ \lambda| - |\partial_- \lambda| = 1$ for $^\forall \lambda$.
Furthermore, we may express the $\mathsf{Y}$-function in terms of the observable bundle:
\begin{align}
 \mathsf{Y}_{i,x} = 
 \exp \left( - \sum_{n = 1}^\infty \frac{x^{-n}}{n} \ch_\mathsf{T} \mathbf{Y}_i^{[n]} \right)
 \, .
 \label{eq:Y_func_ch_gen}
\end{align}
Since the character of the observable bundle is given as the chiral ring operator, the $\mathsf{Y}$-function plays a role of the chiral ring generating function.

Based on the finite product formula of the $\mathsf{Y}$-function, we can see its asymptotic behavior:
\begin{subequations} \label{eq:Y_asymp}
 \begin{align}
 \mathsf{Y}_{i,x} 
 & \ \longrightarrow \
 \begin{cases}
  \displaystyle
  (-x)^{-n} \prod_{\alpha = 1}^{n_i} \np^{\mathsf{a}_{i,\alpha}} & (x \to 0) \\
  1 & (x \to \infty)
 \end{cases}
 \\
 \mathsf{Y}_{i,x}^\vee
 & \ \longrightarrow \
 \begin{cases}
  1 & (x \to 0) \\
  \displaystyle
  (-x)^{+n} \prod_{\alpha = 1}^{n_i} \np^{-\mathsf{a}_{i,\alpha}} & (x \to \infty) 
 \end{cases}
\end{align}
\end{subequations}
In addition, $\mathsf{Y}_{i,x}$ and $\mathsf{Y}^\vee_{i,x}$ have the same zeros and poles depending on the instanton configuration $\mathcal{X}$:
\begin{align}
 \text{zeros}: \ x = x'
 \, ,
 \qquad
 \text{poles}: \ x = q_1 x'
 \qquad (x' \in \mathcal{X}_i)
 \, .
\end{align}
Hence they are related to each other up to an over all non-singular factor:
\begin{align}
 \mathsf{Y}_{i,x}[\mathcal{X}]
 & = \prod_{\alpha = 1}^{n_i} \left[
  \left( - \frac{\np^{\mathsf{a}_{i,\alpha}}}{x} \right)
 \left( 1 - \frac{x}{\np^{\mathsf{a}_{i,\alpha}}} \right) 
 \prod_{(s_1,s_2) \in \lambda_{i,\alpha}} \mathscr{S} \left( \frac{\np^{\mathsf{a}_\alpha} q_1^{s_1 - 1} q_2^{s_2 - 1}}{x} \right)
 \right]
 \nonumber \\
 & = 
 \left( (-x)^{-n} \prod_{\alpha = 1}^{n_i} \np^{\mathsf{a}_{i,\alpha}} \right)
 \mathsf{Y}_{i,x}[\mathcal{X}]^\vee
 \, .
 \label{eq:YtoYdual}
\end{align}
From this point of view, $\mathsf{Y}$ and $\mathsf{Y}^\vee$ are interpreted to give the expansion around $x = \infty$ and $x = 0$ of the same function, and the factor which converts from $\mathsf{Y}$ to $\mathsf{Y}^\vee$ does not depend on the instanton configuration $\mathcal{X}$.
See also its operator realization discussed in \S\ref{sec:Y_op}.

We also remark that the expression \eqref{eq:vec_reflection} may be singular because of $\mathsf{Y}_x[\mathcal{X}] = 0$ for $x \in \mathcal{X}$.
However, this singularity is cancelled by the pole of $\mathsf{Y}_{qx}[\mathcal{X}_\text{ad}]^\vee$, then the product $\mathsf{Y}_{qx}^\vee[\mathcal{X}_\text{ad}] \mathsf{Y}_x[\mathcal{X}]$ itself remains regular.
This regularity argument shall play an important role in the derivation of the Schwinger--Dyson equation.

\subsubsection{(Bi)fundamental hypermultiplet}

Let us then consider the bifundamental hypermultiplet contribution.
From \eqref{eq:quiv_vect_hyp_bundles}, the bifundamental hypermultiplet bundle is given as
\begin{align}
 \mathbf{H}_{e:i \to j} = - \mathbf{M}_e \frac{\wedge \mathbf{Q}_1^\vee}{\wedge \mathbf{Q}_2} \mathbf{X}_i^\vee \mathbf{X}_j
 \, .
\end{align}
The variation is similarly computed as follows:
\begin{subequations} 
\begin{align}
 \delta_i \mathbf{H}_{e:i \to j}
 & := 
 \mathbf{H}_{e:i \to j}[\mathcal{X}_{\text{ad}:(i,\alpha,k)}] - \mathbf{H}_{e:i \to j}[\mathcal{X}]
 = 
 \mathbf{M}_e \det \mathbf{Q}^\vee \mathbf{x}^\vee \mathbf{Y}_j[\mathcal{X}]
 \, ,
 \\
 \delta_i \mathbf{H}_{e:j \to i}
 & := 
 \mathbf{H}_{e:j \to i}[\mathcal{X}_{\text{ad}:(i,\alpha,k)}] - \mathbf{H}_{e:j \to i}[\mathcal{X}]
 =
 \mathbf{M}_e \mathbf{Y}_j[\mathcal{X}]^\vee \mathbf{x}
 \, ,
\end{align}
\end{subequations}
with $\ch_\mathsf{T} \mathbf{x} = x_{i,\alpha,k} =: x$.
In this case, we do not need to take care of the zero mode.
Hence, applying the index formula, we obtain the ratio of the bifundamental matter contributions to the partition function:
\begin{subequations} 
 \begin{align}
  \mathbb{I}[\delta_i \mathbf{H}_{e:i \to j}]
  & = \frac{Z^\text{bf}_{e:i \to j}[\mathcal{X}_{\text{ad}:(i,:a,k)}]}{Z^\text{bf}_{e:i \to j}[\mathcal{X}]}
  = \mathsf{Y}_{j,\mu_e^{-1} q x}^\vee \, , \\
  \mathbb{I}[\delta_i \mathbf{H}_{e:j \to i}]
  & = \frac{Z^\text{bf}_{e:j \to i}[\mathcal{X}_{\text{ad}:(i,:a,k)}]}{Z^\text{bf}_{e:j \to i}[\mathcal{X}]}
  = \mathsf{Y}_{j,\mu_e x}  
  \, .
 \end{align}
\end{subequations}
The (anti)fundamental hypermultiplet contribution is obtained from the bifundamental matter by freezing the gauge node:
\begin{subequations}
 \begin{align}
  \mathbb{I}[\delta \mathbf{H}_i^\text{f}]
  & = \frac{Z^\text{f}_i[\mathcal{X}_{\text{ad}:(i,\alpha,k)}]}{Z^\text{f}_i[\mathcal{X}]}
  = P_{i,x}
  \, ,
  \\
  \mathbb{I}[\delta \mathbf{H}_i^\text{af}]
  & = \frac{Z^\text{af}_i[\mathcal{X}_{\text{ad}:(i,\alpha,k)}]}{Z^\text{af}_i[\mathcal{X}]}
  = \widetilde{P}^\vee_{i,qx}
  \, ,
 \end{align}
\end{subequations}
where $(P_{i,x}, \widetilde{P}^\vee_{i,x})$ are the matter polynomials~\eqref{eq:matter_polynomial_quiver} in the K-theory convention:%
\footnote{%
We omit the symbol (f, af) for the matter polynomials as long as no confusion.
}
\begin{align}
 P_{i,x} = \prod_{f = 1}^{n_i^\text{f}} \left( 1 - x^{-1} \, \np^{m_{i,f}} \right)
 \, , \qquad
 \widetilde{P}_{i,x}^\vee = \prod_{f = 1}^{n_i^\text{af}} \left( 1 - x \, \np^{-\widetilde{m}_{i,f}} \right)
 \, .
\end{align}
These are polynomials in $x^{-1}$ and $x$, respectively, which are converted to each other as follows:
\begin{align}
 P_{i,x} =
 \left( (-x)^{-n_i^\text{f}} \prod_{f = 1}^{n_i^\text{f}} \np^{m_{i,f}} \right)
 P_{i,x}^\vee
 \, , \qquad
 \widetilde{P}_{i,x}^\vee =
 \left( (-x)^{n_i^\text{af}} \prod_{f = 1}^{n_i^\text{af}} \np^{-\widetilde{m}_{i,f}} \right)
 \widetilde{P}_{i,x}
 \, .
\end{align}

\subsubsection{Adjoint hypermultiplet}

In the case of the adjoint hypermultiplet, the computation is much similar to the vector multiplet.
For the adjoint hypermultiplet bundle given by
\begin{align}
 \mathbf{H}^\text{adj} = - \mathbf{M}_\text{adj} \frac{\wedge \mathbf{Q}_1^\vee}{\wedge \mathbf{Q}_2} \mathbf{X}^\vee \mathbf{X}
 \, ,
\end{align}
we take the variation:
\begin{align}
 \delta \mathbf{H}^\text{adj}
 & := \mathbf{H}^\text{adj}[\mathcal{X}_{\text{ad}:(\alpha,k)}] - \mathbf{H}^\text{adj}[\mathcal{X}]
 \nonumber \\
 & =
 \mathbf{M}_\text{adj} \det \mathbf{Q}^\vee \mathbf{x}^\vee \mathbf{Y}
 + \mathbf{M}_\text{adj} \mathbf{Y}^\vee \mathbf{x}
 - \mathbf{M}_\text{adj} \wedge \mathbf{Q}^\vee
 \, .
\end{align}
Then, the ratio of the adjoint matter contributions to the partition function is given by
\begin{align}
 \mathbb{I}[\delta \mathbf{H}^\text{adj}]
 = \frac{Z^\text{adj}[\mathcal{X}_{\text{ad}:(\alpha,k)}]}{Z^\text{adj}[\mathcal{X}]}
 = \mathscr{S}(\mu^{-1}) \, \mathsf{Y}_{\mu x} \mathsf{Y}^\vee_{\mu^{-1} q x}
 \label{eq:adj_add_inst}
\end{align}
where $\ch_\mathsf{T} \mathbf{M}_\text{adj} = \mu \in \mathbb{C}^\times$ is the adjoint mass parameter.
Appearance of the $\mathscr{S}$-function is specific to the adjoint matter contribution.

\subsubsection{Topological term and Chern--Simons term}

The coupling constant part, namely the topological term, simply behaves as
\begin{align}
 \frac{Z_i^\text{top}[\mathcal{X}_{\text{ad}:(i,\alpha,k)}]}{Z_i^\text{top}[\mathcal{X}]}
 = \mathfrak{q}_i
\end{align}
under the adding-instanton operation.

In addition, for 5d gauge theory, we may have the Chern--Simons term.
Based on the expression in \S\ref{sec:CS_partition_function}, we simply obtain the behavior of the partition functions under the adding-instanton operation:
\begin{align}
 \frac{Z_i^\text{CS}[\mathcal{X}_{\text{ad}:(i,\alpha,k)}]}{Z_i^\text{CS}[\mathcal{X}]}
 = x^{-\kappa_i}
 \, ,
 \qquad x = x_{i,\alpha,k}
 \, .
\end{align}

\section{$qq$-character}\label{sec:qq-ch}

Gathering all the contributions, the total partition function (except for the adjoint matter term) behaves under the adding-instanton operation as follows:
\begin{align}
 \frac{Z[\mathcal{X}_{\text{ad}:(i,\alpha,k)}]}{Z[\mathcal{X}]}
 & = - \mathfrak{q}_i \, x^{-\kappa_i}
 \frac{P_{i,x} \widetilde{P}^\vee_{i,qx}}{\mathsf{Y}_{i,qx}^\vee[\mathcal{X}_{\text{ad}:(i,\alpha,k)}] \mathsf{Y}_{i,x}[\mathcal{X}]}
 \prod_{e:i \to j} \mathsf{Y}_{j,\mu_e^{-1} q x}^\vee[\mathcal{X}]
 \prod_{e:j \to i} \mathsf{Y}_{j,\mu_e x}[\mathcal{X}]
 \label{eq:saddle_pt_finite}
\end{align}
for $x = x_{i,\alpha,k}$.
This is also written in the following form:
\begin{align}
 &
 \res_{x \, = \, x_{i,\alpha,k}}
 \Bigg[
 Z[\mathcal{X}_{\text{ad}:(i,\alpha,k)}]
 \mathsf{Y}_{i,qx}^\vee[\mathcal{X}_{\text{ad}:(i,\alpha,k)}]
 \nonumber \\
 & \hspace{5em}
 +
 Z[\mathcal{X}]
 \left(
 \mathfrak{q}_i \, x^{-\kappa_i}
 \frac{P_{i,x} \widetilde{P}^\vee_{i,qx}}{\mathsf{Y}_{i,x}[\mathcal{X}]}
 \prod_{e:i \to j} \mathsf{Y}_{j,\mu_e^{-1} q x}^\vee[\mathcal{X}]
 \prod_{e:j \to i} \mathsf{Y}_{j,\mu_e x}[\mathcal{X}]
 \right)
 \Bigg]
 = 0
 \, ,
 \label{eq:iWeyl_res} 
\end{align}
which means that the pole singularity of $\mathsf{Y}_{i,qx}^\vee[\mathcal{X}_{\text{ad}:(i,\alpha,k)}]$ at $x = x_{i,\alpha,k}$ in the first term is cancelled by the second term.
Hence, summing up all the instanton configuration, we obtain
\begin{align}
 \VEV{\mathsf{Y}_{i,qx}^\vee}
 + \mathfrak{q}_i \, x^{-\kappa_i} \, P_{i,x} \widetilde{P}^\vee_{i,qx}
 \left< \, \frac{1}{\mathsf{Y}_{i,x}}
 \prod_{e:i \to j} \mathsf{Y}_{j,\mu_e^{-1} q x}^\vee
 \prod_{e:j \to i} \mathsf{Y}_{j,\mu_e x}
 \, \right>
 \, .
 \label{eq:iWeyl_ref1}
\end{align}
In this expression, it is guaranteed that all the pole singularities in the first term are cancelled by the second contribution, which will play an important role to discuss the analytic behavior.

\subsection{iWeyl reflection}

The procedure to add a term which compensates the pole singularity of the previous term is called the {\em iWeyl reflection}.%
\footnote{%
Here $i$ is for instanton~\cite{Nekrasov:2012xe}.
}
\index{iWeyl reflection}
The reason why we can interpret this as a deformation of the Weyl reflection is explained in the following.

We denote the iWeyl reflection action:
\begin{align}
 \text{iWeyl}: \
 \mathsf{Y}_{i,qx}^\vee
 & \longmapsto \
 \mathsf{Y}_{i,qx}^\vee \times
 \left(
 \mathfrak{q}_i \, x^{-\kappa_i} 
 \frac{P_{i,x} \widetilde{P}^\vee_{i,qx}}{\mathsf{Y}_{i,qx}^\vee \mathsf{Y}_{i,x}}
 \prod_{e:i \to j} \mathsf{Y}_{j,\mu_e^{-1} q x}^\vee
 \prod_{e:j \to i} \mathsf{Y}_{j,\mu_e x}
 \right)
 \nonumber \\
 & =:
 \mathsf{Y}_{i,qx}^\vee \times
 \left(
 \mathfrak{q}_i \, x^{-\kappa_i} 
 \frac{P_{i,x} \widetilde{P}^\vee_{i,qx}}{\mathsf{A}_{i,x}}
 \right)
 \, ,
 \label{eq:iWeyl_ref}
\end{align}
where we define the $\mathsf{A}$-function
\index{A-function@$\mathsf{A}$-function}
\begin{align}
 \mathsf{A}_{i,x} =
 \mathsf{Y}_{i,qx}^\vee \mathsf{Y}_{i,x}
 \left(
 \prod_{e:i \to j} \mathsf{Y}_{j,\mu_e^{-1} q x}^\vee
 \prod_{e:j \to i} \mathsf{Y}_{j,\mu_e x}
 \right)^{-1}
 \, .
 \label{eq:A_fn_def}
\end{align}
It is also possible to write the reflection formula \eqref{eq:iWeyl_ref} only in terms of either $(\mathsf{Y}_{i,x})$ or $(\mathsf{Y}^\vee_{i,x})$ as follows:
\begin{subequations}\label{eq:iWeyl_ref_mod}
\begin{align}
 & \mathsf{Y}^\vee_{i,x}
 \ \longmapsto \
 \left(
 (-1)^{n_i + \sum_{j \to i} n_j + n_i^\text{f}}
 \np^{\sum_{j} c_{ij}^{-[\log]} n_j }
 q^{n_i^\text{f} + \kappa_i}
 \right)
 \mathfrak{q}_i \, x^{\sum_{j} c_{ij}^{-[0]} n_j - n_i^\text{f} - \kappa_i}
 \nonumber \\ &
 \hspace{13em} \times
 \frac{P_{i,q^{-1} x}^\vee \widetilde{P}^\vee_{i,x}}{\mathsf{Y}^\vee_{i,q^{-1}x}}
 \prod_{e:i \to j} \mathsf{Y}_{j,\mu_e^{-1} x}^\vee
 \prod_{e:j \to i} \mathsf{Y}_{j,\mu_e q^{-1} x}^\vee
 \, \\
 & \mathsf{Y}_{i,x}
 \ \longmapsto \
 \left(
 (-1)^{n_i + \sum_{i \to j} n_j + n_i^\text{f}}
 \np^{\sum_{j} c_{ij}^{+[\log]} n_j }
 q^{\kappa_i}
 \right)
 \mathfrak{q}_i \, x^{\sum_{j} c_{ij}^{+[0]} n_j + n_i^\text{f} - \kappa_i}
 \nonumber \\ &
 \hspace{13em} \times
 \frac{P_{i,q^{-1} x} \widetilde{P}_{i,x}}{\mathsf{Y}_{i,q^{-1}x}}
 \prod_{e:i \to j} \mathsf{Y}_{j,\mu_e^{-1} x}
 \prod_{e:j \to i} \mathsf{Y}_{j,\mu_e q^{-1} x}
 \, ,
\end{align}
\end{subequations}
where $(c_{ij}^{\pm[0]})$ is the classical version of the half Cartan matrices (See \S\ref{sec:quiv_Cartan_matrix}), and we define the logarithmic analog:
\begin{subequations}
 \begin{align}
  c_{ij}^{+[\log]} & = \delta_{ij} - \sum_{e:i \to j} \log \mu_e^{-1}
  \, , \\
  c_{ij}^{-[\log]} & =
  \log q^{-1} \delta_{ij} - \sum_{e:j \to i} \log (\mu_e q^{-1})
  \, .
 \end{align}
\end{subequations}
We also impose the special unitary condition for simplicity:
\begin{align}
 \sum_{\alpha = 1}^{n_i} \mathsf{a}_{i,\alpha} = 0
 \, \qquad
 \sum_{f=1}^{n_i^\text{f}} m_{i,f} = 0
 \, , \qquad
 \sum_{f=1}^{n_i^\text{af}} \widetilde{m}_{i,f} = 0
 \, .
 \label{eq:SU_cond}
\end{align}

In particular, we can concisely (also schematically) express the $\mathsf{A}$-function using the (classical) quiver Cartan matrix \eqref{eq:classical_quiv_Cartan_mat}:\index{A-function@$\mathsf{A}$-function}
\begin{align}
 \log \mathsf{A}_i
 =
 \log \mathsf{Y}_i^2
 - \log
 \left(
 \prod_{e:i \to j} \mathsf{Y}_j
 \prod_{e:j \to i} \mathsf{Y}_j
 \right)
 = \sum_{j \in \Gamma_0} \log \mathsf{Y}_j \, c_{ji}^{[0]} 
 \, .
 \label{eq:A_fn}
\end{align}
In order to precisely deal with the argument shift of the $\mathsf{Y}$-functions, we should use the $q$-deformation of quiver Cartan matrix.
See \S\ref{sec:A_op} for more precise relation.
Hence, identifying $\mathsf{Y}_i$ as a fundamental weight corresponding to the node $i \in \Gamma_0$, it is converted to the simple root $\mathsf{A}_i$ corresponding to the node $i$, and thus this process is interpreted as the Weyl reflection associated with a representation constructed on the quiver.

Then, in order to compensate the possible pole singularities from $\mathsf{Y}_{j,\mu_e^{-1} q x}^\vee$ and $\mathsf{Y}_{j,\mu_e x}$ in \eqref{eq:iWeyl_ref1}, we may add further terms.
In this way, the iWeyl reflection generates the highest weight module on quiver $\Gamma$ where the first term $\mathsf{Y}_{i}^\vee$ plays a role of the highest weight.
Summing up all the contributions generated by the iWeyl reflection, we shall obtain a pole-free function, which turns out to be a polynomial in the variable $x$, that is called (the vev of) the {\em $qq$-character}~\cite{Nekrasov:2015wsu} (See also \cite{Kanno:2012hk,Kanno:2013aha,Bourgine:2015szm,Bourgine:2016vsq}):\index{qq-character@$qq$-character}
\begin{align}
 \mathsf{T}_{i,x} = \mathsf{Y}_{i,x}^\vee
 + \mathfrak{q}_i \left( \frac{x}{q} \right)^{-\kappa_i}
 \frac{P_{i,q^{-1}x} \widetilde{P}^\vee_{i,x}}{\mathsf{Y}_{i,q^{-1}x}}
 \prod_{e:i \to j} \mathsf{Y}_{j,\mu_e^{-1} x}^\vee
 \prod_{e:j \to i} \mathsf{Y}_{j,\mu_e q^{-1} x}
 + \cdots
 \, .
\end{align}
Such a relation for the $\mathsf{Y}$-functions is interpreted as the non-perturbative version of the Schwinger--Dyson equation in this context.
Due to the Dynkin--Cartan classification, the highest weight module is finite dimensional for the finite-type quiver $(\det c^{[0]} > 0)$, so that the iWeyl reflection terminates within finite times.
For the affine and hyperbolic quivers $(\det c^{[0]} \le 0)$, it becomes infinite dimensional, and the corresponding $qq$-character is given by an infinite series expansion.

The $qq$-character has a physical interpretation as the codimension-four defect operator in 4d gauge theory, which generates the chiral ring operators in the full $\Omega$-background~\cite{Nekrasov:2015wsu,Nekrasov:2016ydq}.
This interpretation is also possible for 5d and 6d setups by imposing the defect branes~\cite{Kim:2016qqs,Kimura:2017auj,Agarwal:2018tso,Assel:2018rcw}.

\subsection{Supergroup gauge theory}\label{sec:iWeyl_super}

We can apply a parallel analysis for supergroup gauge theory based on the instanton partition function discussed in \S\ref{sec:super_localization}.
Since we have two sets of partitions characterizing the equivariant fixed point, we consider the partition shift for the positive and negative ones:%
\footnote{%
See \cite{Kimura:2019msw} for details of the derivation.
}
\begin{subequations}
\begin{align}
 \frac{Z[\mathcal{X}_{\text{ad:}(i,\alpha,k,0)}]}{Z[\mathcal{X}]} & =
 - \mathfrak{q}_i \, x^{-\kappa_i^0} \,
 \frac{P_{i,x} \widetilde{P}_{i,qx}^\vee}{\mathsf{Y}_{i,qx}[\mathcal{X}_\text{ad}] \mathsf{Y}_{i,x}^{\vee}[\mathcal{X}]}
 \prod_{e:i \to j} \mathsf{Y}_{j,\mu_e^{-1} q x}[\mathcal{X}]
 \prod_{e:j \to i} \mathsf{Y}_{j,\mu_e x}^\vee[\mathcal{X}]
 \Bigg|_{x = x_{i,\alpha,k}^0}
 \label{eq:ref_full+}
 \\
 \frac{Z[\mathcal{X}_{\text{ad:}(i,\alpha,k,1)}]}{Z[\mathcal{X}]} & =
 - \mathfrak{q}_i^{-1} \, (q^{-1} x)^{+\kappa_i^1} \,
 \frac{\mathsf{Y}_{i,x}[\mathcal{X}] \mathsf{Y}_{i,q^{-1}x}^{\vee}[\mathcal{X}_\text{ad}]}{P_{i,q^{-1}x} \widetilde{P}_{i,x}^\vee}
 \prod_{e:i \to j} \mathsf{Y}_{j,\mu_e^{-1} x}[\mathcal{X}]^{-1}
 \prod_{e:j \to i} \mathsf{Y}_{j,\mu_e q^{-1} x}^{\vee}[\mathcal{X}]^{-1}
 \Bigg|_{x = x_{i,\alpha,k}^1}
 \label{eq:ref_full-} 
\end{align} 
\end{subequations}
where we define the $\mathsf{Y}$-functions
\begin{align}
 \mathsf{Y}_{i,x} = \frac{\mathsf{Y}_{i,x}^0}{\mathsf{Y}_{i,x}^1}
 \, , \qquad
 \mathsf{Y}_{i,x}^\vee = \frac{\mathsf{Y}_{i,x}^{0\vee}}{\mathsf{Y}_{i,x}^{1\vee}}
 \, ,
 \label{eq:Y_fn_ratio}
\end{align}
and the matter functions
\begin{align}
 P_{i,x} = \frac{P_{i,x}^0}{P_{i,x}^1}
 \, , \qquad
 \widetilde{P}_{i,x} = \frac{\widetilde{P}_{i,x}^0}{\widetilde{P}_{i,x}^1}
 \, , \qquad
 P_{i,x}^\vee = \frac{P_{i,x}^{0\vee}}{P_{i,x}^{1\vee}}
 \, , \qquad
 \widetilde{P}_{i,x}^\vee = \frac{\widetilde{P}_{i,x}^{0\vee}}{\widetilde{P}_{i,x}^{1\vee}}
 \, ,
 \label{eq:P_ratio}
\end{align}
with
\begin{subequations}
\begin{align}
 \mathsf{Y}_{i,x}^\sigma
 = \prod_{x' \in \mathcal{X}_i^\sigma} \frac{1 - x' / x}{1 - q_1 x'/x}
 \, , \qquad
 \mathsf{Y}_{i,x}^{\sigma\vee}
 = \prod_{x' \in \mathcal{X}_i^\sigma} \frac{1 - x / x'}{1 - q_1^{-1} x/x'}
 \, ,
\end{align}
\begin{align}
 P_{i,x}^\sigma & = \prod_{\mu \in \mathcal{M}_i^\sigma} \left( 1 - \frac{\mu}{x} \right)
 \, , \qquad
 \widetilde{P}_{i,x}^\sigma = \prod_{\mu \in \widetilde{\mathcal{M}}_i^\sigma} \left( 1 - \frac{\mu}{x} \right)
 \, , \\
 P_{i,x}^{\sigma\vee} & = \prod_{\mu \in \mathcal{M}_i^\sigma} \left( 1 - \frac{x}{\mu} \right)
 \, , \qquad
 \widetilde{P}_{i,x}^{\sigma\vee} = \prod_{\mu \in \widetilde{\mathcal{M}}_i^\sigma} \left( 1 - \frac{x}{\mu} \right)
 \, .
\end{align}
\end{subequations}
Then, we see that adding/removing for the positive/negative node is equivalent to removing/adding for the negative/positive node if the two Chern--Simons levels coincide $\kappa_i^0 = \kappa_i^1$, as required for the supergroup gauge invariance.
Hence, \eqref{eq:ref_full+} and \eqref{eq:ref_full-} are essentially inverse operations.
The analysis above shows that, using the full $\mathsf{Y}$-functions~\eqref{eq:Y_fn_ratio} and the matter functions~\eqref{eq:P_ratio}, consisting of both the positive and negative ones, we can apply the same argument to construct the $qq$-character with the supergroup gauge and flavor nodes as the ordinary (non-supergroup) gauge theory.

\subsection{Higher weight current}

In general, it is possible to consider the $qq$-character starting with a generic product of the $\mathsf{Y}$-functions as the highest weight of the corresponding module:
\begin{align}
 \mathsf{Y}_{\underline{w},\underline{x}}
 = \prod_{i \in \Gamma_0} \prod_{k = 1}^{w_i} \mathsf{Y}_{i,x_{i,k}}
\end{align}
where $\underline{w} = (w_i)_{i \in \Gamma_0} \in \mathbb{Z}_{\ge 0}^{\rk \Gamma}$ are the Dynkin labels of the weight, and $\underline{x} = (x_{i,k})_{i \in \Gamma_0, k = 1,\ldots,w_i}$ are the associated parameters.
Given Dynkin labels, one can construct a highest weight representation on the quiver $\Gamma$.
In particular, we call it a fundamental representation if $w_j = \delta_{ij}$:
Only one of the Dynkin labels is one, and the others are zero, $\underline{w} = ( 0, \ldots, 0, \underbrace{1}_{i\text{-th}}, 0, \ldots, 0)$.
Then, the corresponding $qq$-character consists of the highest weight $\displaystyle \mathsf{Y}_{\underline{w},\underline{x}}$ with the remaining terms generated by the iWeyl reflections:
\begin{align}
 \mathsf{T}_{\underline{w},\underline{x}}
 = \mathsf{Y}_{\underline{w},\underline{x}} + \cdots
 \, .
 \label{eq:qq_ch_high}
\end{align}

Let us consider the iWeyl reflection for such a generic weight.
We take the highest weight,
$\mathsf{Y}^\vee_{i,qx} \mathsf{Y}^\vee_{i,qx_1} \cdots \mathsf{Y}^\vee_{i,qx_k}$.
Then, from the pole cancellation relation~\eqref{eq:iWeyl_res}, we obtain
\begin{align}
 & \res_{x \, = \, x_{i,\alpha,k}}
 \left[
 Z \cdot \mathsf{Y}^\vee_{i,qx} \mathsf{Y}^\vee_{i,qx_1} \cdots \mathsf{Y}^\vee_{i,qx_k}\Bigg|_{\mathcal{X}_{\text{ad}:(i,\alpha,k)}}
 \right.
 \nonumber \\
 & \hspace{5em}
 +
 \left.
 Z \cdot \left(
 \mathfrak{q}_i \, x^{-\kappa_i}
 P_{i,x} \widetilde{P}^\vee_{i,qx}
 \frac{\mathsf{Y}^\vee_{i,qx}}{\mathsf{A}_{i,x}} 
 \right)
 \Bigg|_{\mathcal{X}}
 \times
 \mathsf{Y}^\vee_{i,qx_1} \cdots \mathsf{Y}^\vee_{i,qx_k}
 \Bigg|_{\mathcal{X}_{\text{ad}:(i,\alpha,k)}}
 \right] = 0
 \, .
\end{align}
One may rewrite the $\mathsf{Y}$-functions evaluated with the configuration $\mathcal{X}_{\text{ad}:(i,\alpha,k)}$ using the relation
\begin{align}
 \mathsf{Y}^\vee_{i,x'}[\mathcal{X}_{\text{ad}:(i,\alpha,k)}]
 = \mathscr{S}\left( \frac{x_{i,\alpha,k}}{x'} \right) \ \mathsf{Y}^\vee_{i,x'}[\mathcal{X}]
 \stackrel{\eqref{eq:S_func_reflection}}{=} \mathscr{S}\left( \frac{q^{-1} x'}{x_{i,\alpha,k}} \right) \ \mathsf{Y}^\vee_{i,x'}[\mathcal{X}] 
 \, .
\end{align}
which follows from the expression~\eqref{eq:Y_func_def}.
Summing up the instanton configurations, we obtain the reflection with respect to the first $\mathsf{Y}$-function:
\begin{align}
 &
 \VEV{ 
 \mathsf{Y}^\vee_{i,qx} \mathsf{Y}^\vee_{i,qx_1} \cdots \mathsf{Y}^\vee_{i,qx_k} 
 }
 + \mathfrak{q}_i \, x^{-\kappa_i} \, P_{i,x} \widetilde{P}^\vee_{i,qx} \,
 \mathscr{S} \left( \frac{x_1}{x} \right) \cdots \mathscr{S} \left( \frac{x_k}{x} \right) 
 \left< \,
 \frac{\mathsf{Y}^\vee_{i,qx} \mathsf{Y}^\vee_{i,qx_1} \cdots \mathsf{Y}^\vee_{i,qx_k}}{\mathsf{A}_{i,x}}
 \, \right>
 \, .
\end{align}
Applying this process recursively, the iWeyl reflection for the higher weight is given by
\begin{align}
 \text{iWeyl}: \
 \mathsf{Y}^\vee_{i,qx_1} \cdots \mathsf{Y}^\vee_{i,qx_k} 
 \ \longmapsto \
 \sum_{I \sqcup J = \{1,\ldots,k\}}
 \prod_{\mathsf{i} \in I,\, \mathsf{j} \in J} \mathscr{S} \left( \frac{x_\mathsf{i}}{x_\mathsf{j}} \right)
 \mathsf{Y}^\vee_{i,qx_1} \cdots \mathsf{Y}^\vee_{i,qx_k} 
 \prod_{\mathsf{j} \in J} \mathsf{A}_{i,x_\mathsf{j}}^{-1}
 \label{eq:higher_ref}
\end{align}
where we only focus on the $\mathsf{Y}$ and $\mathsf{A}$-function structure, and omit the coupling constant and the matter polynomials, for simplicity.

\subsection{Collision limit}

Let us consider the collision limit of the higher weight current.
The degree-two weight has the following iWeyl reflection structure:
\begin{align}
 \mathsf{Y}^\vee_{i,qx} \mathsf{Y}^\vee_{i,qx'} 
 + \mathscr{S}\left( \frac{x'}{x} \right) \frac{\mathsf{Y}^\vee_{i,qx}\mathsf{Y}^\vee_{i,qx'}}{\mathsf{A}_{i,x}}
 + \mathscr{S}\left( \frac{x}{x'} \right) \frac{\mathsf{Y}^\vee_{i,qx}\mathsf{Y}^\vee_{i,qx'}}{\mathsf{A}_{i,x'}}
 + \frac{\mathsf{Y}^\vee_{i,qx}\mathsf{Y}^\vee_{i,qx'}}{\mathsf{A}_{i,x} \mathsf{A}_{i,x'} }
 \, .
 \label{eq:collision1} 
\end{align}
Since $\mathscr{S}(z)$ has a pole at $z = 1$ (and also at $z = q^{-1}$), we have to carefully deal with the collision limit, $x' \to x$.
In fact, in this limit, we have
\begin{align}
 \mathsf{Y}_{i,qx}^{\vee2}
 + \left( \mathfrak{c}(q_{1,2}) - \frac{(1 - q_1)(1 - q_2)}{1 - q} \partial_{\log x} \log \mathsf{A}_{i,x} \right)
 \frac{\mathsf{Y}_{i,qx}^{\vee2}}{\mathsf{A}_{i,x}}
 +  \frac{\mathsf{Y}_{i,qx}^{\vee2}}{\mathsf{A}^2_{i,x}}
 \label{eq:collision2}
\end{align}
where the constant $\mathfrak{c}(q_{1,2})$ is given by
\begin{align}
 \mathfrak{c}(q_{1,2})
 = \lim_{z \to 1} \left( \mathscr{S}(z) + \mathscr{S}(z^{-1}) \right)
 = \frac{1 - 6 q + q^2 + (q_1 + q_2)(1 + q)}{(1 - q)^2}
 \ \xrightarrow{q_{1(2)} \to 1} \ 2
 \, .
\end{align}
We can similarly consider the collision limit for the higher weight with degree greater than two.

\section{Classical limit}\label{sec:classical_lim}

\subsection{(Very) classical limit: $\epsilon_{1,2} \to 0$}

The $qq$-character has two deformation parameters $(q_1,q_2)$, and depends also on the bifundamental mass parameters $(\mu_e)_{e \in \Gamma_1}$.
In general (except for cyclic quivers), the bifundamental mass can be gauged away using the $\rU(1)$ gauge degrees of freedom of each node,%
\footnote{%
See also the argument in \S\ref{sec:HW_quiver} on the relation between the $\rU(1)$ factor and the bifundamental mass parameters.
}
 so that we just put $\mu_e \to 1$ for the moment.
 Then, in the classical limit $q_{1,2} \to 1$ ($\epsilon_{1,2} \to 0$), there is no argument shift of $\mathsf{Y}$-functions in the $qq$-character.
 It is reduced to the ordinary character of the representation associated with the quiver~\cite{Nekrasov:2012xe}, which reproduces the Seiberg--Witten geometry of $\Gamma$-quiver gauge theory as discussed in~\S\ref{sec:SW_curve_quiver}.

\subsubsection{Partition function and saddle point analysis}

As shown in \S\ref{sec:eq_fixed_point}, the $\Omega$-background parameters $(q_1,q_2)$ are introduced as the equivariant parameters with respect to the spacetime rotation, and the path integral localizes on the fixed point under the corresponding equivariant action, which regularizes the non-compactness of the spacetime manifold (IR divergence).
If the $\Omega$-background is turned off, the partition function defined through the path integral is expected to diverge as a consequence of the IR singularity.
Nekrasov's proposal was that, although it diverges, the asymptotic behavior of the partition function provides an important information, which would be identified with the Seiberg--Witten prepotential~\cite{Nekrasov:2002qd}:\index{prepotential}
\begin{align}
 Z \ \xrightarrow{\epsilon_{1,2} \to 0} \
 \exp \left( \frac{1}{\epsilon_1 \epsilon_2} \mathscr{F} + \cdots \right)
 \label{eq:Z_asymp1}
\end{align}
where the subleading terms are less singular in the limit $\epsilon_{1,2} \to 0$.
This proposal has been then confirmed by Nekrasov--Okounkov~\cite{Nekrasov:2003rj}, Nakajima--Yoshioka~\cite{Nakajima:2003pg}, and Braverman--Etingof~\cite{Braverman:2004cr}.

The asymptotic behavior of the partition function~\eqref{eq:Z_asymp1} suggests that one can apply the saddle point analysis in the limit $\epsilon_{1,2} \to 0$ as follows:
The partition function $Z$ is given as a summation over all the possible instanton configurations, but in the limit $\epsilon_{1,2} \to 0$, the saddle point configuration denoted by $\mathcal{X}_*$ dominates in the instanton sum:\index{saddle point analysis}
\begin{align}
 Z 
 & = \sum_{\mathcal{X} \in \mathfrak{M}^\mathsf{T}} Z[\mathcal{X}] 
 =: \sum_{\mathcal{X} \in \mathfrak{M}^\mathsf{T}} \exp \left( \frac{1}{\epsilon_1 \epsilon_2} \mathscr{F}[\mathcal{X}] \right) 
 \ \xrightarrow{\epsilon_{1,2} \to 0} \
 \exp \left( \frac{1}{\epsilon_1 \epsilon_2} \mathscr{F}[\mathcal{X}_*] + \cdots \right) 
 \, .
\end{align}
We see that the on-shell value of the prepotential $\mathscr{F}[\mathcal{X}_*]$ derived in this way agrees with the Seiberg--Witten prepotential.
Similarly, under this limit, the expectation value of the gauge theory observable is given by the on-shell value:
\begin{align}
 \vev{\mathcal{O}}
 & = \frac{1}{Z} \sum_{\mathcal{X} \in \mathfrak{M}^\mathsf{T}} Z[\mathcal{X}] \, \mathcal{O}[\mathcal{X}]
 \ \xrightarrow{\epsilon_{1,2} \to 0} \
 \mathcal{O}[\mathcal{X}_*]
 \, .
\end{align}
Hence, the multi-point function is simply factorized
\begin{align}
 \vev{ \mathcal{O}_1^{m_1} \cdots \mathcal{O}_n^{m_n} }
 \ \longrightarrow \
 \left( \mathcal{O}_1^{m_1} \cdots \mathcal{O}_n^{m_n} \right) [\mathcal{X}_*]
 \, .
 \label{eq:corr_fn_factorization}
\end{align}

Let us discuss how to consider the saddle point analysis associated with the instanton partition function.
The partition function is given as a discrete instanton sum, and thus the dynamical variable in this case is the discrete $x$-variable, $\mathcal{X} = (x_{i,\alpha,k})$, corresponding to two-dimensional partitions $\lambda = (\lambda_{i,\alpha,k})$.
Recalling the definition of the $x$-variable~\eqref{eq:X-variable_quiver}, it will behave as a continuous variable in the limit $\epsilon_{1,2} \to 0$.
Therefore, the saddle point configuration $\mathcal{X}_*$ is given as a solution to the saddle point equation with respect to the $x$-variable,
\begin{align}
 x \frac{\partial \mathscr{F}[\mathcal{X}]}{\partial x} = 0
 \qquad \text{for} \qquad
 ^\forall x \in \mathcal{X}
 \, .
 \label{eq:saddle_pt_eq_cl}
\end{align}
We remark that the $x$-variable is multiplicative $x \in \mathbb{C}^\times$, so that the derivative is taken with respect to $\log x$, $\partial_{\log x} = x \partial_x$.

We then point out the relation between the saddle point equation and the iWeyl reflection discussed in \S\ref{sec:qq-ch}.
The behavior of the partition function under the adding-instanton operation~\eqref{eq:saddle_pt_finite} is also written as follows:
\begin{align}
 \frac{Z[\mathcal{X}_{\text{ad}:(i,\alpha,k)}]}{Z[\mathcal{X}]}
 & = 
 \exp \left( 
 \frac{1}{\epsilon_1 \epsilon_2} \left( \mathscr{F}[\mathcal{X}_{\text{ad}:(i,\alpha,k)}] - \mathscr{F}[\mathcal{X}] \right) 
 \right)
 \nonumber \\
 & \xrightarrow{\epsilon_2 \to 0} \
 \exp \left(
 \frac{1}{\epsilon_1} x \frac{\partial \mathscr{F}[\mathcal{X}]}{\partial x} 
 \right)
 \, .
\end{align}
We remark that the configuration $\mathcal{X}_{\text{ad}:(i,\alpha,k)}$ involves the $q_2$-shift of the variable, $x \to q_2 x$.
Therefore, the saddle point equation~\eqref{eq:saddle_pt_eq_cl} is rephrased as the invariance of the partition function under the adding-instanton operation:
\begin{align}
 \frac{Z[\mathcal{X}_{\text{ad}:(i,\alpha,k)}]}{Z[\mathcal{X}]} = 1
 \, .
 \label{eq:Z/Z=1}
\end{align}
In the limit $\epsilon_{1,2} \to 0$, the pole singularities of $\mathsf{Y}$-function are promoted to the cut singularities, thus the saddle point equation provides the cut-crossing relation for $\mathsf{Y}$-functions~\cite{Nekrasov:2012xe}.

\subsection{Nekrasov--Shatashvili limit: $\epsilon_{2} \to 0$}\label{sec:NS_lim}

We can also consider a partially classical limit, $q_1 =$ finite, $q_2 \to 1$ ($\epsilon_1 =$ finite, $\epsilon_2 \to 0$), that is called the Nekrasov--Shatashvili (NS) limit in the gauge theory context~\cite{Nekrasov:2009rc}.
In this case, as shown in \cite{Nekrasov:2013xda}, the $qq$-character is reduced to the {\em $q$-character}\index{q-character@$q$-character} of the corresponding affine Yangian/quantum affine algebra~\cite{Knight:1995JA,Frenkel:1998}, which plays a central role in the connection to quantum integrable systems (\S\ref{sec:NS_integrability}).

As well as the (very) classical case $\epsilon_{1,2} \to 0$, in the NS limit, we turn off the $\Omega$-background parameter $\epsilon_2 \to 0$, and the second complex plane $\mathbb{C}_2$ becomes non-compact.
Hence we similarly expect the diverging behavior of the partition function
\begin{align}
 Z \ \xrightarrow{\epsilon_{2} \to 0} \
 \exp \left( \frac{1}{\epsilon_2} \widetilde{\mathscr{W}}(\epsilon_1) + \cdots\right)
 \, ,
\end{align}
where $\widetilde{\mathscr{W}}(\epsilon_1)$ is identified with the effective twisted superpotential of the corresponding two-dimensional $\mathcal{N} = (2,2)$ gauge theory on $\mathbb{C}_2$, and $\epsilon_1$ plays a role of the adjoint mass parameter in 2d theory.%
\footnote{%
In many cases, the adjoint mass parameter behaves as the equivariant parameter for the transverse rotation symmetry.
See \S\ref{sec:adjoint_bundle}.
}
\index{effective twisted superpotential}

Since we still have a small parameter $\epsilon_2$, we can apply the saddle point analysis as well in the NS limit:
\begin{align}
 Z =: \sum_{\mathcal{X}} \exp\left( \frac{1}{\epsilon_2} \widetilde{\mathscr{W}}[\mathcal{X}] \right)
 \ \xrightarrow{\epsilon_2 \to 0} \
 \exp \left( \frac{1}{\epsilon_2} \widetilde{\mathscr{W}}[\mathcal{X}_*] + \cdots \right)
 \, ,
\end{align}
where $\epsilon_1 \widetilde{\mathscr{W}}[\mathcal{X}] = \mathscr{F}[\mathcal{X}]$.
Since the $x$-variable is still continuous, we can apply the same argument to the limit $\epsilon_{1,2} \to 0$.
Then, the invariance of the partition function under the adding-instanton~\eqref{eq:Z/Z=1} ends up with the saddle point equation with respect to the twisted superpotential
\begin{align}
 \eqref{eq:Z/Z=1}
 \ \iff \
 x \frac{\partial \widetilde{\mathscr{W}}[\mathcal{X}]}{\partial x} = 2 \pi \im  \, \mathbb{Z}
 \qquad \text{for} \qquad
 ^\forall x \in \mathcal{X}
 \, .
 \label{eq:saddle_pt_eq_NS} 
\end{align}
From the 2d theory point of view, this equation provides the supersymmetric vacuum condition, that is called the twisted F-term condition.

\section{Examples}

\subsection{$A_1$ quiver}\label{sec:qq_ch_A1}

We then focus on a specific example, 5d $\SU(n)$ gauge theory with $(n^\text{f},n^\text{af})$ fundamental flavors, which is classified into $A_1$ quiver.
Let us discuss details of the $qq$-character in this case.

\subsubsection{Fundamental character}

Since $A_1$ quiver consists of a single gauge node, denoted by $i = 1$, the iWeyl reflection~\eqref{eq:iWeyl_ref} is reduced to
\begin{align}
 \text{iWeyl}: \
 \mathsf{Y}_{1,qx}^\vee
 & \longmapsto \
 \mathsf{Y}_{1,qx}^\vee \times
 \left(
 \mathfrak{q}_1 \, x^{-\kappa_1} 
 \frac{P_{1,x} \widetilde{P}^\vee_{1,qx}}{\mathsf{Y}_{1,qx}^\vee \mathsf{Y}_{1,x}}
 \right)
 \nonumber \\
 &
 = 
 \mathfrak{q}_1 \, x^{-\kappa_1} 
 \frac{P_{1,x} \widetilde{P}^\vee_{1,qx}}{\mathsf{Y}_{1,x}}
 \stackrel{\eqref{eq:YtoYdual}}{=}
 (-1)^n \, \mathfrak{q}_1 \, x^{n-\kappa_1} 
 \frac{P_{1,x} \widetilde{P}^\vee_{1,qx}}{\mathsf{Y}_{1,x}^\vee}
\end{align}
where we impose the special unitary condition
\begin{align}
 \sum_{\alpha = 1}^n \mathsf{a}_\alpha = 0
 \, .
\end{align}
In this case, there is no $\mathsf{Y}$-function in the numerator after the reflection:
No further pole singularities are generated, and the reflection terminates here.
Therefore, the fundamental $qq$-character for $\Gamma = A_1$ is given as follows~\cite{Nekrasov:2015wsu}:\index{qq-character@$qq$-character!A1@$A_1$}
\begin{align}
 \mathsf{T}_{1,x} 
 = 
 \mathsf{Y}_{1,x}^\vee
 + \mathfrak{q}_1 \, x^{n - \kappa_1} \,
 \frac{P_{1,q^{-1}x} \widetilde{P}^\vee_{1,x}}{\mathsf{Y}_{1,q^{-1}x}^\vee}
\end{align}
with the shift of the coupling constant%
\footnote{%
Since the coupling constant $\mathfrak{q} \in \mathbb{C}^\times$ is given by \eqref{eq:inst_fugacity} together with \eqref{eq:complex_coupling}, the sign change $\mathfrak{q} \to - \mathfrak{q}$ corresponds to the $\theta$-angle shift, $\theta \to \theta + \pi$.
This shift of the $\theta$-angle may provide a subtle difference of the global structure~\cite{Haouzi:2020zls}.
}
\begin{align}
 \frac{(-1)^n}{q^{n-\kappa_1}} \mathfrak{q}_1
 \ \longrightarrow \
 \mathfrak{q}_1
 \, .
\end{align}
Concerning the asymptotic behavior of $\mathsf{Y}^\vee$-function~\eqref{eq:Y_asymp}, the instanton average of the $qq$-character should be a polynomial in $x$ of degree $n$ (a regular function), which can be written in the form:
\begin{align}
 \VEV{\mathsf{T}_{1,x}}
 = 
 \left< \,
 \mathsf{Y}_{1,x}^\vee
 + \mathfrak{q}_1 \, x^{n - \kappa_1} \,
 \frac{P_{1,q^{-1}x} \widetilde{P}^\vee_{1,x}}{\mathsf{Y}_{1,q^{-1}x}^\vee}
 \, \right>
 =
 \det \left( 1 - x \, \Phi_1^{-1} \right)
\end{align}
for $\Phi_1 \in \SU(n)$.

Let us discuss the classical limit of this $qq$-character.
As shown in \S\ref{sec:classical_lim}, the gauge theory observable is simply replaced with its on-shell value with respect to the saddle point configuration $\mathcal{X}_*$, which we denote
\begin{align}
 y_x := \lim_{\epsilon_{1,2} \to 0} \vev{\mathsf{Y}_{1,x}^\vee} = \mathsf{Y}_{1,x}^\vee[\mathcal{X}_*]
 \, .
\end{align}
We remark that, due to the factorization property~\eqref{eq:corr_fn_factorization}, we have 
\begin{align}
 \left< \,
 \frac{1}{\mathsf{Y}_{1,q^{-1}x}^\vee}
 \, \right>
 \ \xrightarrow{\epsilon_{1,2} \to 0} \ \frac{1}{y_x}
 \, .
\end{align}
Simply denoting $y = y_x$, we reproduce the Seiberg--Witten curve for 5d $\SU(n)$ gauge theory with $(n^\text{f},n^\text{af})$ flavors~\eqref{eq:SW_curve_5d}.

\subsubsection{Higher character}

We then consider generic higher character for $A_1$ quiver.
Applying the higher reflection formula \eqref{eq:higher_ref}, the degree-$w$ character is given as follows:
\begin{align}
 \mathsf{T}_{w,\underline{x}} 
 & = \mathsf{Y}_{1,x_1}^\vee \cdots \mathsf{Y}_{1,x_w}^\vee + \cdots
 \nonumber \\ 
 & =
 \sum_{I \sqcup J = \{1,\ldots,w\}}
 \prod_{\mathsf{i} \in I,\, \mathsf{j} \in J} \mathscr{S} \left( \frac{x_\mathsf{i}}{x_\mathsf{j}} \right)
 \prod_{\mathsf{i} \in I} \mathsf{Y}_{1,x_\mathsf{i}}^\vee
 \prod_{\mathsf{j} \in J} \mathsf{Y}_{1,q^{-1}x_\mathsf{j}}^{-1}
 \, , \quad
 \underline{x} = (x_\mathsf{i})_{\mathsf{i} = 1,\ldots,w}
 \, ,
\end{align}
where we do not write the coupling constant and the matter polynomials for simplicity.
We remark that $\mathsf{A}$-function~\eqref{eq:A_fn} is given for $A_1$ quiver by
\begin{align}
 \mathsf{A}_{1,x} = \mathsf{Y}_{1,qx}^\vee \mathsf{Y}_{1,x}
 \, .
\end{align}

For example, the degree-two current $w = 2$ is given by
\begin{align}
 \mathsf{T}_{(2),(x_1,x_2)} 
 = 
 \mathsf{Y}^\vee_{1,x_1} \mathsf{Y}^\vee_{1,x_2}
 + \mathscr{S}\left(\frac{x_2}{x_1}\right) \frac{\mathsf{Y}^\vee_{1,x_2}}{\mathsf{Y}_{1,q^{-1} x_1}}
 + \mathscr{S}\left(\frac{x_1}{x_2}\right) \frac{\mathsf{Y}^\vee_{1,x_1}}{\mathsf{Y}_{1,q^{-1} x_2}}
 + \mathsf{Y}^{-1}_{1,q^{-1} x_1} \mathsf{Y}^{-1}_{1,q^{-1} x_2}
 \, .
\end{align}
This expression corresponds to the tensor product of two-dimensional representations of $G_{A_1} = \SL(2)$, which is irreducible for generic $(x_1,x_2)$ in the doubly quantum situation.
Recalling $\mathscr{S}$-function has zeros at $\mathscr{S}(z = q_{1,2}^{-1}) = 0$, one can eliminate one of the zero weight terms in the degree-two $qq$-character, for example:
\begin{align}
 \mathsf{T}_{(2),(x, q_1^{-1} x)}
 =  
 \mathsf{Y}^\vee_{1,x} \mathsf{Y}^\vee_{1,q_1^{-1} x}
 + \mathscr{S}\left( q_1 \right) \frac{\mathsf{Y}^\vee_{1,x}}{\mathsf{Y}_{1,q_1^{-2} q_2^{-1} x}}
 + \mathsf{Y}^{-1}_{1,q^{-1} x} \mathsf{Y}^{-1}_{1,q_1^{-2} q_2^{-1} x} 
 \, .
\end{align}
Let us take further limit to see the relation to the three-dimensional representation of $G_{A_1} = \SL(2)$.
Since $\mathscr{S}$-function behaves
\begin{align}
 \mathscr{S}(q_1)
 = \frac{(1 - q_1^2)(1 - q)}{(1 - q_1)(1 - q_1^2 q_2)}
 = \frac{1 - q}{1 - q_1^2 q_2} (1 + q_1)
 \ \longrightarrow \
 \begin{cases}
  2 & (q_1 \to 1) \\
  1 & (q_2 \to 1)
 \end{cases}
 \, ,
 \label{eq:S_func_NS_lim}
\end{align}
we obtain two different results in the limit:%
\footnote{%
We remark that a similar computation is also found in the twisted reduction (the root of unity limit) of the $qq$-character~\cite{Kimura:2019xzj}.
}
\begin{subequations}
 \begin{align}
  \mathsf{T}_{(2),(x, x)}^{(q_1 \to 1)}
  & =
  \mathsf{Y}^\vee_{1,x} \mathsf{Y}^\vee_{1, x}
  + 2 \, \frac{\mathsf{Y}^\vee_{1,x}}{\mathsf{Y}_{1,q_2^{-1} x}}
  + \mathsf{Y}^{-1}_{1,q_2^{-1} x} \mathsf{Y}^{-1}_{1, q_2^{-1} x} 
  =
  \left( \mathsf{T}_{1,x}^{(q_1 \to 1)} \right)^2
  \, \\
 \mathsf{T}_{(2),(x, q_1^{-1} x)}^{(q_2 \to 1)}
 & =  
 \mathsf{Y}^\vee_{1,x} \mathsf{Y}^\vee_{1,q_1^{-1} x}
 + \frac{\mathsf{Y}^\vee_{1,x}}{\mathsf{Y}_{1,q_1^{-2} x}}
 + \mathsf{Y}^{-1}_{1,q^{-1} x} \mathsf{Y}^{-1}_{1,q_1^{-2} x} 
  = \mathsf{T}_{1,x}^{(q_2 \to 1)} \mathsf{T}^{(q_2 \to 1)}_{1,q_1^{-1}} - 1
  \, ,
 \end{align}
\end{subequations}
where $\mathsf{T}_{w,\underline{x}}^{(q_{1,2} \to 1)}$ is called the $q$-character (more precisely, $q_{2,1}$-character) of $A_1$ quiver theory. 
In these expressions, the first limit corresponds to the tensor product, $\scriptsize \yng(1) \otimes \yng(1)$, and the second one is the degree-two symmetric representation, $\scriptsize \yng(2) = \yng(1) \otimes \yng(1) - \emptyset$.
In this way, the $q$-character obeys a functional relation, called the T-system,\index{T-system} that originates from the tensor product of the representations associated with quiver.%
\footnote{%
Originally the T-system was introduced as a functional relation between the T-functions obtained from the transfer matrices of the corresponding quantum integrable system.
Afterward, it has been realized that the T-function is identified with the associated $q$-character, and one can interpret it as a decomposition of the tensor product between them.
See a review article for details~\cite{Kuniba:2010ir}.
}
See also \S\ref{sec:NS_frac} for a related argument.
\index{q-character@$q$-character}

Let us briefly mention a gauge theory interpretation of the T-system.
The $q$-character is obtained in the NS limit of the $qq$-character, hence it is interpreted as a 2d reduction of the codimension-four defect, which is namely a codimension-two defect (also called the surface defect).
See \S\ref{sec:Higgsing} for a related argument.
Since the T-system is a functional relation for the $q$-characters, it is interpreted as a fusion rule for the surface defects in this context.
This would be a dual description of the bootstrap approach to the surface defect based on the class $\mathcal{S}$ perspective~\cite{Gaiotto:2012xa,Alday:2013kda,Bullimore:2014nla}.

\subsubsection{Supergroup gauge theory}

Let us briefly mention the $qq$-character for supergroup gauge theory.
As shown in \S\ref{sec:iWeyl_super}, the iWeyl reflection for supergroup theory is equivalent to the ordinary one, if we replace the $\mathsf{Y}$-function with its super analog~\eqref{eq:Y_fn_ratio}.
Recalling the super $\mathsf{Y}$-function is given as a ratio of $\mathsf{Y}^{0,1}$, the vev of the $qq$-character is not a polynomial, but a rational function:
\begin{align}
 \VEV{\mathsf{T}_{1,x}}
 = 
 \left< \,
 \mathsf{Y}_{1,x}^\vee
 + \mathfrak{q}_1 \, x^{n_0 + n_1 - \kappa_1} \,
 \frac{P_{1,q^{-1}x} \widetilde{P}^\vee_{1,x}}{\mathsf{Y}_{1,q^{-1}x}^\vee}
 \, \right>
 =
 \sdet \left( 1 - x \, \Phi_1^{-1} \right)
\end{align}
for $\Phi_1 \in \SU(n_0|n_1)$.
Taking the classical limit $\epsilon_{1,2} \to 0$, this reproduces the Seiberg--Witten curve for supergroup gauge theory discussed in \S\ref{sec:SW_super}.

\subsection{$A_2$ quiver}\label{sec:qq_A2}

Let us consider $A_2$ quiver theory.
We first rewrite the iWeyl reflection \eqref{eq:iWeyl_ref} in terms of $\mathsf{Y}^\vee$-functions,
\begin{subequations}
 \begin{align}
  \mathsf{Y}_{1,x}^\vee
  \ \longmapsto \ &
  q^{\kappa_1} \mathfrak{q}_1 \, x^{- \kappa_1} \, P_{1,q^{-1} x} \widetilde{P}^\vee_{1,x}
  \frac{\mathsf{Y}_{2,\mu^{-1} x}^\vee}{\mathsf{Y}_{1,q^{-1}x}}
  \nonumber \\ &
  =
  (-1)^{n_1}
  q^{\kappa_1 - n_1} \mathfrak{q}_1 \, x^{n_1 - \kappa_1} \, P_{1,q^{-1} x} \widetilde{P}^\vee_{1,x}
  \frac{\mathsf{Y}_{2,\mu^{-1} x}^\vee}{\mathsf{Y}^\vee_{1,q^{-1}x}}
  \nonumber \\ &
  \xrightarrow{\eqref{eq:parameter_shift_A2}} \
  \mathfrak{q}_1 \, x^{- \kappa_1} \, P_{1,q^{-1} x} \widetilde{P}^\vee_{1,x}
  \frac{\mathsf{Y}_{2,\mu^{-1} x}^\vee}{\mathsf{Y}^\vee_{1,q^{-1}x}}  
  \\
  \mathsf{Y}_{2,x}^\vee
  \ \longmapsto \ &  
  q^{\kappa_2} \mathfrak{q}_2 \, x^{- \kappa_2} \, P_{2,q^{-1} x} \widetilde{P}^\vee_{2,x}
  \frac{\mathsf{Y}_{1,\mu q^{-1} x}}{\mathsf{Y}_{2,q^{-1}x}}
  \nonumber \\ &
  =
  (-1)^{n_1 + n_2} q^{\kappa_2 - n_2 + n_1} \mu^{-n_1} \mathfrak{q}_2 \,
  x^{n_2 - n_1 - \kappa_2} \,
  P_{2,q^{-1} x} \widetilde{P}^\vee_{2,x}
  \frac{\mathsf{Y}^\vee_{1,\mu q^{-1} x}}{\mathsf{Y}^\vee_{2,q^{-1}x}}
  \nonumber \\ &
  \xrightarrow{\eqref{eq:parameter_shift_A2}} \
  \mathfrak{q}_2 \,
  x^{- \kappa_2} \,
  P_{2,q^{-1} x} \widetilde{P}^\vee_{2,x}
  \frac{\mathsf{Y}^\vee_{1,\mu q^{-1} x}}{\mathsf{Y}^\vee_{2,q^{-1}x}}  
 \end{align}
\end{subequations}
where we denote the bifundamental mass by $\mu = \mu_{1 \to 2} = \mu_{2 \to 1}^{-1} q$, and we apply the special unitary condition~\eqref{eq:SU_cond}.
We also shift the coupling constant and the Chern--Simons level $(\mathfrak{q}_i,\kappa_i)_{i = 1,2}$ as follows:
\begin{subequations}\label{eq:parameter_shift_A2}
 \begin{align}
  (-1)^{n_1}
  q^{\kappa_1 - n_1} \mathfrak{q}_1
  \ \longrightarrow \
  \mathfrak{q}_1
  \, , \qquad &
  (-1)^{n_1 + n_2} q^{\kappa_2 - n_2 + n_1} \mu^{-n_1} \mathfrak{q}_2
  \ \longrightarrow \
  \mathfrak{q}_2
  \, , \\
  n_1 - \kappa_1
  \ \longrightarrow \
  - \kappa_1
  \, , \qquad &
  n_2 - n_1 - \kappa_2
  \ \longrightarrow \
  - \kappa_2
  \, .
 \end{align}
\end{subequations}
Then, the fundamental $qq$-characters, in particular, for pure gauge theory $(n_i^\text{f},n_i^\text{af})_{i = 1,2} = 0$ with $(\kappa_i)_{i = 1,2} = 0$ are generated by the iWeyl reflections~\cite{Nekrasov:2015wsu}:\index{qq-character@$qq$-character!A2@$A_2$}
\begin{subequations}
\begin{align}
 \mathsf{T}_{1,x}
 & =
 \mathsf{Y}_{1,x}^\vee
 + \mathfrak{q}_1 \, \frac{\mathsf{Y}_{2,\mu^{-1} x}^\vee}{\mathsf{Y}^\vee_{1,q^{-1}x}}
 + \mathfrak{q}_1 \mathfrak{q}_2 \, \frac{1}{\mathsf{Y}^\vee_{2,\mu^{-1} q^{-1}x}}
 \, ,
 \\
 \mathsf{T}_{2,x}
 & =
 \mathsf{Y}_{2,x}^\vee
 + \mathfrak{q}_2 \, \frac{\mathsf{Y}^\vee_{1,\mu q^{-1} x}}{\mathsf{Y}^\vee_{2,q^{-1}x}}
 + \mathfrak{q}_1 \mathfrak{q}_2 \, \frac{1}{\mathsf{Y}^\vee_{1,\mu q^{-2} x}}
 \, .
\end{align}
\end{subequations}
These are $qq$-characters of three-dimensional representations of $G_{A_2} = \SL(3)$.

The instanton average of these $qq$-characters are polynomial functions, and their degrees are fixed by the asymptotic behaviors of the highest weight terms $(\mathsf{Y}_{1,2})$~\eqref{eq:Y_asymp}:
\begin{align}
 \VEV{\mathsf{T}_{i,x}} = \det \left(1 - x \, \Phi_i^{-1} \right)
 \, , \qquad i = 1,2
\end{align}
with $\Phi_i \in \SU(n_i)$.
Namely, they are polynomials in $x$ of degree $n_{1,2}$.
We can see that, in the classical limit $q_{1,2} \to 1$, this is consistent with the Seiberg--Witten curve for $A_2$ quiver~\eqref{eq:SW_curve_A2} discussed in \S\ref{sec:SW_curve_quiver}.

\subsection{\texorpdfstring{$\widehat{A}_{0}$ quiver}{Affine A0 quiver}}
\label{sec:qq_A0}

We next consider $\SU(n)$ gauge theory with a single adjoint hypermultiplet, which is classified into affine quiver theory $\widehat{A}_0$.
We denote the adjoint mass parameter by $m \in \mathbb{C}$, and define the multiplicative analog $\mu = \np^m \in \mathbb{C}^\times$.
Taking into account the adjoint matter contribution to the adding-instanton operation~\eqref{eq:adj_add_inst}, the iWeyl reflection for $\widehat{A}_0$ quiver is given by
\begin{align}
 \text{iWeyl}: \
 \mathsf{Y}_{1,x}^\vee
 \ \longmapsto \
 \mathfrak{q}_1 \, \mathscr{S}(\mu^{-1}) \,
 \frac{\mathsf{Y}_{1,\mu^{-1} x}^\vee \mathsf{Y}_{1,\mu q^{-1} x}}{\mathsf{Y}_{1,q^{-1} x}}
 = \mu^{-n} \mathfrak{q}_1 \, \mathscr{S}(\mu^{-1}) \, \frac{\mathsf{Y}_{1,\mu^{-1} x}^\vee \mathsf{Y}^\vee_{1,\mu q^{-1} x}}{\mathsf{Y}^\vee_{1,q^{-1} x}}
 \, .
 \label{eq:iWeyl_A0_1}
\end{align}
This is derived from the pole cancellation with the adjoint matter~\eqref{eq:iWeyl_res}:
\begin{align}
 &
 \res_{x \, = \, x_{1,\alpha,k}}
 \left[
 Z \cdot \mathsf{Y}^\vee_{1,qx}\Big|_{\mathcal{X}_{\text{ad}:(1,\alpha,k)}}
 +
 Z \cdot \left(
 \mathfrak{q}_1 \, \mathscr{S}(\mu^{-1}) \, \frac{\mathsf{Y}_{1,\mu^{-1} q x}^\vee \mathsf{Y}_{1,\mu x}}{\mathsf{Y}_{1,x}}
 \right)
 \Bigg|_{\mathcal{X}}
 \right] = 0
 \, .
\end{align}
In order to see the reflection structure more concisely, we introduce the multiplicative version of the eight-dimensional $\Omega$-background parameters~\eqref{eq:epsilon1234}:
\begin{align}
 (q_1,q_2,q_3,q_4) = (q_1,q_2,\mu^{-1},\mu q^{-1})
 \label{eq:q1234}
\end{align}
obeying the Calabi--Yau condition,
\begin{align}
 q_1 q_2 q_3 q_4 = 1
 \, .
\end{align}
Then, the iWeyl reflection~\eqref{eq:iWeyl_A0_1} is rewritten as
\begin{align}
 \text{iWeyl}: \quad
 \mathsf{Y}^\vee_{1,x}
 \ \longmapsto \
 \mathsf{Y}^\vee_{1,x}
 \times
 \left(
 q_3^n 
 \mathfrak{q}_1 \, \mathscr{S}(q_3)
 \frac{\mathsf{Y}_{1,q_3 x}^\vee \mathsf{Y}^\vee_{1,q_4 x}}{\mathsf{Y}^\vee_{1,x} \mathsf{Y}^\vee_{1,q_{34} x}}
 \right)
 \, .
 \label{eq:iWeyl_A0_2}
\end{align}
In this case, there appear another $\mathsf{Y}$-function in the numerator after the reflection, so that we should apply further reflection to cancel the poles from the new contribution.
In fact, this process does not terminate, and the $qq$-character is given as an infinite series, as a consequence of the infinite dimensionality of the affine Weyl group.
As mentioned below, it is interpreted as the character of the Fock representation of the quantum toroidal algebra of $\mathfrak{gl}_1$.

From the reflection structure~\eqref{eq:iWeyl_A0_2}, the weight appearing in $O(\mathfrak{q}_1^k)$ contribution to the $qq$-character is labeled by a partition $\lambda$ with size $k$,
\begin{align}
 \Lambda_{\lambda} & :=
 \mathsf{Y}^\vee_{1,x}
 \prod_{(s_3,s_4) \in \lambda} \frac{\mathsf{Y}_{1,q_3^{s_3} q_4^{s_4-1} x}^\vee \mathsf{Y}^\vee_{1,q_3^{s_3 - 1} q_4^{s_4} x}}{\mathsf{Y}^\vee_{q_3^{s_3 - 1} q_4^{s_4 - 1} x} \mathsf{Y}^\vee_{1,q_3^{s_3} q_4^{s_4}x}} 
 \nonumber \\ &
 = 
 \prod_{(s_3,s_4) \in \partial_+ \lambda} \mathsf{Y}^\vee_{q_3^{s_3 - 1} q_4^{s_4 - 1} x}
 \prod_{(s_3,s_4) \in \partial_- \lambda} \mathsf{Y}^{\vee-1}_{q_3^{s_3} q_4^{s_4} x}
 \label{eq:A0_module}
\end{align}
where $\partial_\pm \lambda$ is the outer/inner boundary of the partition $\lambda$ as defined in \eqref{fig:outer/inner}.
Then, the fundamental $qq$-character for $\widehat{A}_0$ quiver theory is given as a summation over the partition~\cite{Nekrasov:2015wsu}:\index{qq-character@$qq$-character!hatA0@$\widehat{A}_0$}
\begin{align}
 \mathsf{T}_{1,x}
 & = \mathsf{Y}^\vee_{1,x} + \mathfrak{q}_1 \, \mathscr{S}(\mu^{-1}) \, \frac{\mathsf{Y}_{1,q_3 x}^\vee \mathsf{Y}^\vee_{1,q_4 x}}{\mathsf{Y}^\vee_{1,q_{34} x}} + \cdots
 \nonumber \\
 & =
 \sum_{\lambda} \mathfrak{q}_1^{|\lambda|} \, Z_{34}[\lambda] \, \Lambda_\lambda
 \label{eq:qq_ch_A0}
\end{align}
where we shift the coupling constant as $q_3^n \mathfrak{q}_1 \to \mathfrak{q}_1$, and $Z_{34}[\lambda]$ is the instanton partition function of $\rU(1)$ $\widehat{A}_0$ quiver gauge theory with respect to the equivariant parameters $\epsilon_{3,4}$, evaluated with the configuration $\lambda$.

In order to show the formula~\eqref{eq:qq_ch_A0}, we again study the behavior under the adding-instanton operation, $x_{1,\alpha,k} \mapsto q_2 x_{1,\alpha,k}$ $(\mathcal{X} \to \mathcal{X}')$.
The expression~\eqref{eq:qq_ch_A0} implies the pole cancellation in the form of
\begin{align}
 \res_{q_{12}^{-1} z = x_{1,\alpha,k}}
 \left[
 Z[\mathcal{X}'] \cdot \Lambda_{\lambda}[\mathcal{X}']
 + \frac{Z_{34}[\lambda']}{Z_{34}[\lambda]} \cdot Z[\mathcal{X}] \cdot
 \Lambda_{\lambda'}[\mathcal{X}]
 \right]
 = 0
\end{align}
where we define $q_{12}^{-1} z = q_{34} z = q_3^{s_3 - 1} q_4^{s_4 - 1} x$ for $(s_3,s_4) \in \partial_+ \lambda$, and we denote a partition with an additional box $(s_3,s_4) \in \partial_+ \lambda$ by $\lambda' = \lambda \oplus (s_3,s_4)$.
Then, $Z_{34}[\lambda]$ behaves as
\begin{align}
 \frac{Z_{34}[\lambda']}{Z_{34}[\lambda]}
 & = - \frac{Z[\mathcal{X}']}{Z[\mathcal{X}]}
 \frac{\Lambda_{\lambda}[\mathcal{X}']}{\Lambda_{\lambda'}[\mathcal{X}]}
 \stackrel{\eqref{eq:iWeyl_A0_2}}{=}
 \mathfrak{q}_1 \, \mathscr{S}_{12}(q_3) \, \frac{\mathsf{Y}^\vee_{1,z}[\mathcal{X}]}{\mathsf{Y}^\vee_{1,z}[\mathcal{X}']} \,
 \frac{\Lambda_{\lambda}[\mathcal{X}']}{\Lambda_{\lambda}[\mathcal{X}]}
 \, .
\end{align}
Recalling
\begin{align}
 \frac{\mathsf{Y}_{1,x}[\mathcal{X}']}{\mathsf{Y}_{1,x}[\mathcal{X}]}
 = \mathscr{S}_{12}\left( \frac{x_{1,\alpha,k}}{x} \right)
 = \mathscr{S}_{12}\left( \frac{q_{34} z}{x} \right)
 \, ,
\end{align}
the weight factor gives rise to
\begin{align}
 \frac{\Lambda_{\lambda}[\mathcal{X}']}{\Lambda_{\lambda}[\mathcal{X}]}
 & =
 \mathscr{S}_{12} \left( q_{34} z / x \right)
 \prod_{(i,j) \in \lambda}
 \frac{\mathscr{S}_{12}(q_{34} z/q_3^{i} q_4^{j - 1} x) \mathscr{S}_{12}(q_{34} z/q_3^{i - 1} q_4^{j} x)}{\mathscr{S}_{12}(q_{34} z/q_3^{i - 1} q_4^{j - 1} x) \mathscr{S}_{12}(q_{34} z/q_3^{i} q_4^{j} x)}
 \nonumber \\
 & =
 \frac{(1 - q_{134} z/x)(1 - q_{234} z/x)}{(1 - q_{34} z/x)(1 - z/x)}
 \prod_{(i,j) \in \lambda}
 \frac{\mathscr{S}_{34}(q_3^{i-1} q_4^{j-1} x / q_{134} z) \mathscr{S}_{34}(q_3^{i-1} q_4^{j-1} x / q_{234} z)}{\mathscr{S}_{34}(q_3^{i-1} q_4^{j-1} x / q_{34} z) \mathscr{S}_{34}(q_3^{i-1} q_4^{j-1} x / z)}
 \nonumber \\
 & =
 \frac{\mathsf{Y}^{34\vee}_{1,q_{134} z} \mathsf{Y}^{34\vee}_{1,q_{234} z}}{\mathsf{Y}^{34\vee}_{1,q_{34} z} \mathsf{Y}^{34\vee}_{1,z}}\Bigg|_{\lambda}
\end{align}
where $\mathscr{S}_{34}(z)$ is defined in~\eqref{eq:Sij_fn}, and we define the 34-version of the $\mathsf{Y}$-function,
\begin{subequations}\label{eq:Y34_func_def}
 \begin{align}
  \mathsf{Y}^{34}_{i,x}[\lambda]
  & =
  \prod_{\alpha = 1}^{n^{34}_i} \left[
  \left( 1 - \frac{\np^{\mathsf{b}_{i,\alpha}}}{x} \right) 
  \prod_{(s_3,s_4) \in \lambda_{i,\alpha}} \mathscr{S}_{34} \left( \frac{\np^{\mathsf{b}_\alpha} q_3^{s_3 - 1} q_4^{s_4 - 1}}{x} \right)
  \right]
  \\
  \mathsf{Y}^{34}_{i,x}[\lambda]^{\vee}
  & =
  \prod_{\alpha = 1}^{n^{34}_i} \left[
  \left( 1 - \frac{x}{\np^{\mathsf{b}_{i,\alpha}}} \right) 
  \prod_{(s_3,s_4) \in \lambda_{i,\alpha}} \mathscr{S}_{34} \left( \frac{\np^{\mathsf{b}_\alpha} q_3^{s_3 - 1} q_3^{s_4 - 1}}{x} \right)
  \right]
 \end{align}
\end{subequations}
with the rank $(n_{i}^{34})$ and the corresponding Coulomb moduli $(\mathsf{b}_{i,\alpha})_{\alpha = 1,\ldots,n_i^{34}}$. \index{Coulomb moduli}
In this case, we have $(n_1^{34}, \mathsf{b}_{1,1}) = (1, x)$.
In fact, $\mathsf{Y}^{34\vee}_{1,q_{34} z'}[\lambda]$ has a pole, but it is cancelled as
 \begin{align}
  \lim_{z' \to z} \mathsf{Y}^{34\vee}_{1,q_{34} z'}[\lambda] \, \mathscr{S}_{12} \left( \frac{z'}{z} \right)
  = - \frac{(1 - q_1)(1 - q_2)(1 - q_{34})}{(1 - q_3)(1 - q_4)(1 - q_{12})} \,
  \mathsf{Y}^{34\vee}_{1,q_{34} z}[\lambda']
  \, .
 \end{align}
 Therefore, we obtain
 \begin{align}
  \frac{Z_{34}[\lambda']}{Z_{34}[\lambda]}
  = - q_3^{-1} \mathfrak{q}_1 \, \mathscr{S}_{34}(q_1) \,
  \frac{\mathsf{Y}^{34\vee}_{1,q_{134} z}[\lambda] \mathsf{Y}^{34\vee}_{1,q_{234} z}[\lambda]}{\mathsf{Y}^{34\vee}_{1,q_{34} z}[\lambda'] \mathsf{Y}^{34\vee}_{1,z}[\lambda]}
  = - q_3^{-1} \mathfrak{q}_1 \, \mathscr{S}_{34}(q_1) \,
  \frac{\mathsf{Y}^{34\vee}_{1,q_{1}^{-1} z}[\lambda] \mathsf{Y}^{34\vee}_{1,q_{2}^{-1} z}[\lambda]}{\mathsf{Y}^{34\vee}_{1,q_{12}^{-1} z}[\lambda'] \mathsf{Y}^{34\vee}_{1,z}[\lambda]}
  \, ,
  \label{eq:Z34_shift}
 \end{align}
 which is equivalent to the iWeyl reflection~\eqref{eq:iWeyl_A0_2} by replacing $q_{1,2} \leftrightarrow q_{3,4}$.
 This proves that $Z_{34}[\lambda]$ is the $\rU(1)$ instanton partition function of $\widehat{A}_0$ quiver theory with the equivariant parameters $\epsilon_{3,4}$.
 Such a relation between $\epsilon_{1,2} \leftrightarrow \epsilon_{3,4}$ is naturally understood from the gauge origami construction mentioned in \S\ref{sec:origami}.
 See also \S\ref{sec:qq_ch_geom}.

\subsubsection{Higher character}

We have shown that $\rU(1)$ theory on 34-surface appears from the fundamental $qq$-character of $\widehat{A}_0$ quiver.
We discuss its higher rank generalization obtained from the higher $qq$-character generated by the highest weight $\displaystyle \mathsf{Y}_{1,x_1}^\vee \cdots \mathsf{Y}_{1,x_{n'}}^\vee$ with $n' = n^{34}$.
In this case, the weight function is parametrized by $n'$-tuple partition, $\lambda^{34} = (\lambda_1^{34},\ldots,\lambda_{n'}^{34})$:
\begin{align}
 \Lambda[\lambda^{12}, \lambda^{34}]
 & = \prod_{\beta = 1}^{n'}
 \left[
 \mathsf{Y}_{1,x_\beta}^\vee
 \prod_{s \in \lambda_\beta^{34}}
 \frac{\mathsf{Y}_{1,q_3^{s_3} q_4^{s_4-1} x_\beta}^\vee \mathsf{Y}^\vee_{1,q_3^{s_3 - 1} q_4^{s_4} x_\beta}}{\mathsf{Y}^\vee_{q_3^{s_3 - 1} q_4^{s_4 - 1} x_\beta} \mathsf{Y}^\vee_{1,q_3^{s_3} q_4^{s_4}x_\beta}} 
 \right]_{\lambda^{12}}
 \nonumber \\
 & =
 \prod_{\substack{\alpha = 1,\ldots,n \\ \beta = 1,\ldots,n'}}
 \left[
 \left( 1 - \frac{x_\beta}{w_\alpha} \right)
 \prod_{s \in \lambda_{\alpha}^{12}} \mathscr{S}_{12} \left( \frac{c_\alpha^{12}(s)}{x_\beta} \right)
 \prod_{s \in \lambda_{\beta}^{34}} \mathscr{S}_{34} \left( \frac{c_\beta^{34}(s)}{w_\alpha} \right)
 \right]
 \nonumber \\
 & \quad \times
 \prod_{\substack{\alpha = 1,\ldots,n \\ \beta = 1,\ldots,n'}}
 \prod_{\substack{s \in \lambda_\alpha^{12} \\ s' \in \lambda_{\beta}^{34}}}
 \frac{1 - \left( c_\alpha^{12}(s) / c_\beta^{34}(s') \right)^\pm q_{12,23,31}}
      {1 - \left( c_\alpha^{12}(s) / c_\beta^{34}(s') \right)^\pm q_{1,2,3,4}}
 \left( 1 - \left( c_\alpha^{12}(s) / c_\beta^{34}(s') \right)^\pm \right)
\end{align}
where we apply more symmetric convention: $\lambda^{12} = (\lambda^{12}_1,\ldots,\lambda^{12}_n)$ is an $n$-tuple partition parametrizing the $\rU(n)$ instanton configuration on 12-surface.
The Coulomb moduli of $\rU(n)$ theory are denoted by $w_\alpha = \np^{\mathsf{a}_{1,\alpha}} \in \mathbb{C}^\times$, and we define the $q$-content
\begin{align}
 c_\alpha^{12}(s) = w_\alpha q_1^{s_1 - 1} q_2^{s_2 - 1}
 \, , \qquad
 c_\beta^{34}(s) = x_\beta q_3^{s_3 - 1} q_4^{s_4 - 1}
 \, .
\end{align}
We also use the convention
\begin{align}
 1 - z^\pm a = (1 - z^{+1} a)(1 - z^{-1} a)
 \, .
\end{align}
Then, we obtain exactly the same expression as \eqref{eq:Z34_shift} in this case by using the $\mathsf{Y}^{34}$-function with rank $n' = n^{34}$, so that the expansion coefficient of the degree-$n'$ $qq$-character is given by $\rU(n^{34})$ $\widehat{A}_0$ quiver theory with equivariant parameter $\epsilon_{3,4}$:
\begin{align}
 \mathsf{T}_{(n^{34}),\,\underline{x}}[\lambda^{12}]
 & =
 \left[ \mathsf{Y}^\vee_{1,x_1} \cdots \mathsf{Y}^\vee_{1,x_{n^{34}}} + \cdots \right]_{\lambda^{12}}
 \nonumber \\
 & =
 \sum_{\lambda^{34}} \mathfrak{q}_1^{|\lambda^{34}|} \, Z_{34}[\lambda^{34}] \, \Lambda[\lambda^{12},\lambda^{34}]
 \, .
\end{align}
Now the weight parameters $(x_\alpha)_{\alpha = 1,\ldots,n'}$ are identified with the Coulomb moduli for 34-theory.

From the representation theoretical point of view, the $qq$-character presented in~\eqref{eq:qq_ch_A0} is associated with the Fock module, which is parametrized by a single partition $\lambda$, and the higher weight generalization corresponds to the tensor product of the Fock spaces.
See also \S\ref{sec:affine_quiv_W} for more details on the algebraic perspectives.

\subsubsection{Classical limit: $q$-character}

In the NS limit $q_{2(1)} \to 1$, the $\mathscr{S}$-function becomes $\mathscr{S}(z) \to 1$, so that the rational function $Z_{34}[\lambda]$ in \eqref{eq:qq_ch_A0} becomes trivial, $Z_{34}[\lambda] \to 1$.
Thus, the fundamental $q$-character of $\widehat{A}_0$ quiver theory is simply given by~\cite{Nekrasov:2013xda}
\begin{align}
 \mathsf{T}_{1,x} = \sum_{\lambda} \mathfrak{q}_1^{|\lambda|} \, \Lambda_\lambda
 \, .
 \label{eq:q_ch_A0}
\end{align}
In fact, this simplification has a close relation to the argument in \S\ref{sec:adjoint_bundle}:
Taking $\mu \to 1$ ($m_\text{adj} = 0$), the partition function contribution associated with each instanton configuration becomes trivial in $\widehat{A}_0$ quiver theory.
Since there is a symmetry of $\epsilon_{1,2} \leftrightarrow \epsilon_{3,4}$ in $\widehat{A}_0$ quiver theory, the equivariant parameter $q_{2(1)}$ plays a role of the multiplicative adjoint mass for 34-theory, and the NS limit exactly corresponds to the massless limit on this side.

The $q$-character associated with $\widehat{A}_0$ quiver has been also obtained from the quantum toroidal algebra~\cite{Ginzburg:1995MRL} of $\mathfrak{gl}_1$ denoted by U$_q(\widehat{\widehat{\mathfrak{gl}}}_1)$ associated with the Fock module~\cite{Feigin:2017wnq},%
\footnote{%
Precisely speaking, the definition of the quantum toroidal algebra in~\cite{Ginzburg:1995MRL} does not apply to $\mathfrak{gl}_1$ since it is not a complex semisimple Lie algebra.
Actually the presentation of the quantum toroidal algebra of $\mathfrak{gl}_1$ is more involved than that for a complex semisimple algebra.
A relation between the quantum toroidal algebras associated with a complex semisimple Lie algebra and affine quivers (not only cyclic ones) is also discussed in~\cite{Nakajima:2002ICM}.
}
which is also known as Ding--Iohara--Miki (DIM) algebra~\cite{Ding:1996mq,Miki:2007JMP}, elliptic Hall algebra~\cite{Burban:2005DMJ,Schiffmann:2005DMJ,Schiffmann:2008CM,Schiffmann:2013DMJ}, and (spherical) double affine Hecke algebra (DAHA)~\cite{Schiffmann:2010JAC}.
See also \cite{Feigin:2009KJM,Feigin2009:JMP,Feigin:2015raa,Feigin:2017caw} and \cite{Maulik:2012wi,Braverman:2014xca} for related discussions.
This relation implies the geometric Langlands correspondence between the quantum toroidal algebra and the affine quiver W-algebra discussed in \S\ref{sec:affine_quiv_W}.
The relation between the quantum toroidal algebra and the cyclic quiver has been also pointed out in the literature~\cite{Varagnolo:1999IMRN,Schiffmann:2013DMJ,Schiffmann:2013PMIHES,Negut:2013cz,Negut:2015,Negut:2018}.

\section{Gauge origami reloaded}
\label{sec:qq_ch_geom}
\index{gauge origami}

The $qq$-character of $\widehat{A}_0$ quiver theory shows a relation between $\epsilon_{1,2} \leftrightarrow \epsilon_{3,4}$.
We now discuss it from the gauge origami point of view as in \S\ref{sec:origami}.

\subsection{8d gauge origami partition function}

We consider the gauge theory average of the $qq$-character,
\begin{align}
 Z_{n,n'; \underline{w},\underline{x} } 
 :=  \vev{ \mathsf{T}_{n'; \underline{x} } }
 = \sum_{k = 0}^\infty \mathfrak{q}_1^k \, Z_{n,n';k}
\end{align}
where
\begin{align}
 Z_{n,n';k}
 =
 \sum_{|\lambda^{12}| + |\lambda^{34}| = k} Z_{12}[\lambda^{12}] \, Z_{34}[\lambda^{34}] \, \Lambda[\lambda^{12},\lambda^{34}]
 \, .
 \label{eq:8d_combin}
\end{align}
In fact, this expression is obtained from a contour integral \index{LMNS formula!gauge origami}
\begin{align}
 Z_{n,n';k}
 &
 = \frac{1}{k!} \oint \prod_{a = 1}^k \frac{d\phi_a}{2 \pi \im}
 \prod_{\substack{\alpha = 1,\ldots,n \\ \beta = 1,\ldots,n'}}
 [\mathsf{a}_\alpha - z_\beta]
 \prod_{\beta = 1}^{n'} \mathscr{S}_{12} \left( \phi_a - z_\beta \right)
 \prod_{\alpha = 1}^{n} \mathscr{S}_{34} \left( \phi_a - \mathsf{a}_\alpha \right)
 \nonumber \\
 & \hspace{10em} \times
 \prod_{1 \le a, b \le k}
 \frac{[\phi_{ab} - \epsilon_{12, 23, 31}]}{[\phi_{ab} - {\epsilon_{1,2,3,4}}]}
 [\phi_{(a \neq b)}]
 \, ,
\end{align}
where we apply the additive convention with $x_\beta = \np^{z_\beta}$.
This expression has been also introduced as the instanton contribution in the presence of $n'$ codimension-four defects~\cite{Kim:2016qqs,Agarwal:2018tso}.
Compared to the instanton partition function of $\widehat{A}_0$ quiver~\eqref{eq:LMNS_formula_2*_sym}, there are additional contributions:
We see $\rU(n)$ and $\rU(n')$ degrees of freedom with their Cartan elements, $(w_\alpha)_{\alpha = 1,\ldots,n}$ and $(x_\beta)_{\beta = 1,\ldots,n'}$, which are coupled with $\rU(k)$ factor with the Cartan element $(\phi_a)_{a = 1,\ldots,k}$.
There is also an interaction term between $\rU(n)$ and $\rU(n')$ found in the first term of the integrand.
In the string theory language, this configuration consists of $n$ D3 branes on 12-surface and $n'$ D3 branes on 34-surface, and $k$ D(-1) branes playing a role of the instanton~\cite{Nekrasov:2016qym}.
Hence, $Z_{n,n'; \underline{w},\underline{x}}$ is interpreted as (a special case of) the partition function of the 8d gauge origami setup~\cite{Nekrasov:2016ydq}.

The 8d partition function~\eqref{eq:8d_combin} is parametrized by a pair of 2d partitions, $\lambda^{12}$ and $\lambda^{34}$, because the gauge origami configuration considered here is given by $\mathbb{C}^2 \times \mathbb{C}^2$.
In general, the partition function associated with a d-dimensional complex manifold is parametrized by d-dimensional partitions~\cite{Nekrasov:2008jjm}.
The case with d $= 3$ has been explored as the topological string amplitude associated with Calabi--Yau three-folds~\cite{Aganagic:2003db,Okounkov:2003sp,Iqbal:2007ii}.
In addition, the case with d $= 4$ has been recently proposed, and the corresponding partition function is parametrized by the solid (4d) partition~\cite{Nekrasov:2017cih,Nekrasov:2018xsb}.
See also~\cite{Cao:2017swr,Cao:2019tvv} for related works on Calabi--Yau four-folds.

\subsection{$qq$-character integral formula}\label{sec:qq_ch_int}

From 8d gauge origami point of view, the generic $qq$-character~\eqref{eq:qq_ch_high} is obtained by integrating the 34-surface degrees of freedom, namely integration over the quiver variety~\cite{Nekrasov:2015wsu},\index{qq-character@$qq$-character}\index{quiver!---variety}
\begin{align}
 \mathsf{T}_{\underline{w},\underline{x}} = \sum_{\underline{v}} \mathfrak{q}^{\underline{v}} \, \mathsf{T}_{\underline{w},\underline{v};\underline{x}}
 \label{eq:qq_ch_int_sum}
\end{align}
where $\mathfrak{q}^{\underline{v}}$ is defined in \eqref{eq:quiv_top}, and each contribution is given by 
\begin{align}
 \mathsf{T}_{\underline{w},\underline{v};\underline{x}}
 & =
 \mathsf{Y}^\vee_{\underline{w},\underline{x}}
 \prod_{i \in \Gamma_0} \frac{1}{v_i!}
 \frac{[-\epsilon_{12}]^{v_i}}{[-\epsilon_{1,2}]^{v_i}}
 \oint
 \prod_{\phi \in \underline{\phi}} \frac{d \phi}{2 \pi \im \phi} \,
 \mathsf{A}^{-1}_{\mathsf{i}(\phi),\phi}
 \prod_{i \in \Gamma_0}
 \prod_{\substack{a = 1,\ldots,v_i \\ \alpha = 1,\ldots,w_i}}
 \mathscr{S}_{12} \left( \frac{\phi_{i,a}}{x_{i,\alpha}} \right) 
 \prod_{i \in \Gamma_0} \prod_{a \neq b}^{v_i} 
 \mathscr{S}_{12} \left( \frac{\phi_{i,b}}{\phi_{i,a}} \right)^{-1}
 \nonumber \\
 & \hspace{7em} \times
 \prod_{e:i \to j} \prod_{\substack{a = 1,\ldots,v_i \\ b = 1,\ldots,v_j}}
 \mathscr{S}_{12} \left( \mu_e q^{-1} \frac{\phi_{j,b}}{\phi_{i,a}} \right)
 \prod_{e:j \to i} \prod_{\substack{a = 1,\ldots,v_i \\ b = 1,\ldots,v_j}}
 \mathscr{S}_{12} \left( \mu^{-1} \frac{\phi_{i,a}}{\phi_{j,b}} \right)
 \label{eq:qq_ch_int}
\end{align}
where $(\underline{w},\underline{v}) = (w_i,v_i)_{i \in \Gamma_0}$ are the dimension vectors used to define the quiver variety $\mathfrak{M}_{W,V}(\Gamma)$ as in \S\ref{sec:quiver_variety}.
$\underline{x} = (x_{i,\alpha})_{i \in \Gamma_0,\alpha = 1,\ldots,w_i}$ and $\underline{\phi} = (\phi_{i,a})_{i \in \Gamma_0,a = 1,\ldots,v_i}$ are the Cartan elements of $\rU(W)$ and $\rU(V)$.
We will revisit this formula based on the operator formalism in \S\ref{sec:qq_ch_VOA}.

\section{Quantization of cycle integrals}\label{sec:BS_quantum}

We have pointed out a relation between the Seiberg--Witten geometry and the classical integrable system, in particular, the correspondence between the Seiberg--Witten curve and the spectral curve of the associated classical integrable system.
On the gauge theory side, we have the $\Omega$-background parameters, and it is natural to ask what is the role of this deformation parameter on the integrable system side.
In this Section, we focus on the NS limit, $\epsilon_2 \to 0$ $(q_2 \to 1)$, and see the remaining $\Omega$-background parameter $q_1$ plays a role of the quantum deformation parameter.

\subsection{Saddle point equation}
\label{saddle point analysis}

We have discussed in \S\ref{sec:NS_lim} that, in the NS limit, we may apply the saddle point analysis with respect to the twisted superpotential $\widetilde{\mathscr{W}}$. \index{effective twisted superpotential}
Since it is also related to the prepotential $\mathscr{F}$, we rewrite the saddle point equation~\eqref{eq:saddle_pt_eq_NS} in terms of the prepotential:\index{prepotential}
\begin{align}
 x \frac{\partial \mathscr{F}}{\partial x} = 2 \pi \im \epsilon_1 \mathbb{Z}
 \, ,
 \qquad 
 x \in \mathcal{X}
 \, .
\end{align}
Let us consider $A_1$ quiver for simplicity, and recall the relation between the $x$-variable~\eqref{eq:X-variable_quiver} and the Coulomb moduli $(\mathsf{a}_\alpha)_{\alpha = 1,\ldots,n}$. \index{Coulomb moduli}
Then, this saddle point equation implies quantization of the dual Coulomb moduli parameter $(\mathsf{a}_{D}^\alpha)_{\alpha = 1,\ldots,n}$,
\begin{align}
 \mathsf{a}_{D}^\alpha = \frac{\partial \mathscr{F}}{\partial \mathsf{a}_{\alpha}} = 2 \pi \im \epsilon_1 \mathbb{Z}
 \, .
\end{align}
Since the dual variable is given by $B$-cycle integral as shown in \S\ref{sec:SW_th}, the saddle point equation in the NS limit leads to quantization of $B$-cycle integral:
\begin{align}
 \oint_{B_\alpha} \lambda = \epsilon_1 \mathbb{Z}
 \, ,
\end{align}
where we rescale the prepotential by the factor $2 \pi \im$.
In fact, such a quantization of the contour integral of the tautological one-form is interpreted as the Bohr--Sommerfeld quantization condition, and in this case, it implies that the $\Omega$-background parameter $\epsilon_1$ plays a role of the Planck constant, $\epsilon_1 \simeq \hbar$.

\subsection{$\mathsf{Y}$-function}

There is another way to see quantization of the cycle integral from $\mathsf{Y}$-function.
Let us consider the 4d (additive; cohomological) version of the $\mathsf{Y}$-function~\eqref{eq:Y_func_def}:
\begin{align}
 \mathsf{Y}_x[\mathcal{X}]
 = \prod_{x' \in \log \mathcal{X}}
 \frac{x - x'}{x - x' - \epsilon_1}
 =
 \exp
 \left(
 \sum_{n = 1}^\infty \frac{\epsilon_1^n}{n} \sum_{x' \in \log \mathcal{X}} \frac{1}{(x - x')^n}
 \right)
 \, ,
\end{align}
where we define the additive version of the $x$-variables~\eqref{eq:X-variable_quiver}:
\begin{align}
 \log \mathcal{X} = 
 \left\{
 x_{\alpha,k} = \mathsf{a}_\alpha + \epsilon_1 (k - 1) + \epsilon_2 \lambda_{\alpha,k}
 \right\}_{\alpha = 1,\ldots,n}^{k = 1,\ldots,\infty}
 \, .
 \label{eq:log_X-variable}
\end{align}
Then, $\log \mathsf{Y}_x$ shows pole singularities at $x = x'$, which contribute to the contour integral, similarly to the resolvent in the matrix model (\S\ref{sec:matrix_cycle_quantization}).
Recalling that the $\mathsf{Y}$-function is identified with the $y$-variable in the Seiberg--Witten theory, and $\lambda = x \, d \log y = - \log y \, dx$, we obtain quantization of the contour integral:
\begin{align}
 \frac{1}{2 \pi \im } \oint_\mathcal{C} \lambda = \epsilon_1 \mathbb{Z}
 \, ,
\end{align}
where the quantum number is (minus of) the number of the poles surrounded by the contour $\mathcal{C}$.
We remark that, in the classical limit $\epsilon_1 \to 0$, the poles are condensed to yield a cut singularity, and one can take the contour $\mathcal{C}$ to the non-contractable cycle on the Seiberg--Witten curve.

\section{Quantum geometry and quantum integrability}\label{sec:NS_integrability}

In addition to the quantization of the cycle integrals, there is a more direct way to see quantization of the Seiberg--Witten geometry, which shows an interesting connection to the quantum integrable system, that is called the Bethe/Gauge correspondence, as proposed by Nekrasov--Shatashvili~\cite{Nekrasov:2009rc}.
See also~\cite{Nekrasov:2009zz,Nekrasov:2009uh,Nekrasov:2009ui} and \cite{Nekrasov:2013xda} for its quiver generalization.

\subsection{Pure $\SU(n)$ Yang--Mills theory}

Let us start with the $q$-character of pure $\SU(n)$ YM theory ($A_1$ quiver) in 4d:\index{q-character@$q$-character}
\begin{align}
 \mathsf{T}_{1,x} = \mathsf{Y}_{1,x} + \frac{\mathfrak{q}_1}{\mathsf{Y}_{1,x-}}
\end{align}
where $\mathsf{T}_{1,x}$ is a polynomial in $x$ of degree $n$.%
\footnote{%
Precisely speaking, it must be considered as the vev of the $q$-character.
In the NS limit, since we can apply the saddle point analysis as discussed in \S\ref{sec:NS_lim}, it is interpreted as the on-shell value with respect to the saddle point configuration.
}
Here we use the conventions, $\mathsf{Y}_{i,x-} = \mathsf{Y}_{i,x - \epsilon_1}$, $\mathsf{Y}_{i,x--} = \mathsf{Y}_{i,x - 2 \epsilon_1}$,
$\mathsf{Y}_{i,r} = \mathsf{Y}_{i,x + r \epsilon_1}$,
etc.
We then define the $\mathsf{Q}$-function \index{Q-function@$\mathsf{Q}$-function}
\begin{align}
 \mathsf{Q}_{i,x} = \prod_{x' \in \log \mathcal{X}} ( x - x')
 \, ,
\end{align}
and thus the $\mathsf{Y}$-function is written as a ratio of $\mathsf{Q}$-functions,
\begin{align}
 \mathsf{Y}_{i,x} = \frac{\mathsf{Q}_{i,x}}{\mathsf{Q}_{i,x-}}
 \, .
\end{align}
This change of the variable is interpreted as a discrete analog of the logarithmic derivative, which is used to convert the Riccati-type differential equation to the linear differential equation.
See also \S\ref{sec:matrix_quantum_geom}.
Writing the $q$-character in terms of the $\mathsf{Q}$-function, we obtain the so-called TQ-relation for the Toda chain:\index{TQ-relation!Toda chain}
\begin{align}
 \mathsf{Q}_{1,x} + \mathfrak{q}_1 \, \mathsf{Q}_{1,x--} = \mathsf{T}_{1,x} \, \mathsf{Q}_{1,x-}
 \, .
\end{align}
In this context, the coupling constant $\mathfrak{q}_1$ plays a role of the boundary condition (twist) parameter of the periodic Toda chain.

This TQ-relation has a natural interpretation as a quantization of the Seiberg--Witten curve.
We may write the TQ-relation also in the following form:
\begin{align}
 0 = 
 \mathsf{Q}_{1,x} - \mathsf{T}_{1,x} \, \mathsf{Q}_{1,x-} + \mathfrak{q}_1 \, \mathsf{Q}_{1,x--} 
 = 
 H(\hat{x},\hat{y}) \, 
 \mathsf{Q}_{1,x--}
\end{align}
where we define the algebraic function
\begin{align}
 H(x,y) = y^2 - \mathsf{T}_{1,x} \, y + \mathfrak{q}_1
 \label{eq:SW_H_function}
\end{align}
and the operator pair:
\begin{align}
 \hat{x} = x
 \, , \qquad
 \hat{y} = \exp\left( \epsilon_1 \frac{\partial}{\partial x} \right)
 \, .
\end{align}
Compared to the discussion in \S\ref{sec:SW_curve_quiver}, we realize that the Seiberg--Witten curve for $A_1$ quiver theory~\eqref{eq:A1_character2} is given by the zero locus of the algebraic function, $\Sigma = \{\, (x,y) \in \mathbb{C} \times \mathbb{C}^\times \mid H(x,y) = 0 \,\}$, so that $H(x,y)$ is identified with the characteristic polynomial of the corresponding Lax matrix. \index{Lax matrix}
Furthermore, the operators obey the following algebraic relation
\begin{align}
 \left[ \log \hat{y} \, , \, \hat{x} \right] = \epsilon_1
 \, ,
\end{align}
which is interpreted as the canonical commutation relation with respect to the symplectic two-form on the curve, $\omega = dx \wedge d \log y$.
Therefore, identifying $\epsilon_1 \simeq \hbar$, the $q$-character provides the canonical quantization of the Seiberg--Witten geometry.

Such a quantization of an algebraic curve is called the {\em quantum curve},\index{quantum curve} which is now discussed in various research fields: matrix model~\cite{Eynard:2007kz} (See also Appendix~\ref{chap:matrix_model}), topological string~\cite{Aganagic:2003qj,Dijkgraaf:2007sw,Dijkgraaf:2008fh}, knot invariant (AJ conjecture)~\cite{Garoufalidis:2008,Garoufalidis:2004GTM}, and gauge theory~\cite{Poghossian:2010pn,Fucito:2011pn,Dorey:2011pa,Chen:2011sj,Nekrasov:2013xda}, etc.
While the classical curve is defined as the zero locus of the algebraic function as a Lagrangian submanifold of $\mathbb{C} \times \mathbb{C}^\times$, the quantum curve is given as the kernel of the quantum operator, $\operatorname{Ker} H(\hat{x},\hat{y})$, which is a subspace of the function space.
We remark that, since we may apply the symplectic transform, choice of the algebraic function $H(x,y)$ is not unique.
In order to take into account such ambiguity, it is natural to consider a ring $\mathscr{A} = \mathbb{C}[x,y]/(H(x,y))$, and from this point of view, quantization of the curve corresponds to a noncommutative ring, which is formulated as the {\em $\mathcal{D}$-module}\index{D-module@$\mathcal{D}$-module}, $\mathscr{M} = \mathbb{C}[\hat{x},\hat{y}]/(H(\hat{x},\hat{y}))$~\cite{Dijkgraaf:2008fh}.

\subsection{$\mathcal{N} = 2$ SQCD}

Let us consider $\SU(n)$ gauge theory with fundamental flavors.
The $q$-character is then given by \index{q-character@$q$-character}
\begin{align}
 \mathsf{T}_{1,x} = \mathsf{Y}_{1,x} + \mathfrak{q}_1 \, \frac{P_{1,x-} \widetilde{P}_{1,x}}{\mathsf{Y}_{1,x-}}
 \, ,
 \label{eq:q_ch_A1}
\end{align}
with the matter polynomials
\begin{align}
 P_{1,x} = \prod_{f = 1}^{n^\text{f}} (x - m_f)
 \, , \qquad
 \widetilde{P}_{1,x} = \prod_{f = 1}^{n^\text{af}} (x - \widetilde{m}_f)
 \, .
\end{align}
We remark that we deal with the fundamental and antifundamental matters separately, although in four-dimensional setup.
For the latter convenience, we shift the $\mathsf{Y}$-function, $\mathsf{Y}_{1,x} \to P_{1,x} \mathsf{Y}_{1,x}$, which leads to the shift of the $\mathsf{Q}$-function, 
\begin{align}
 \mathsf{Q}_{1,x}
 \ \longrightarrow \
 \left( \prod_{f = 1}^{n^\text{f}} \Gamma_1(x - m_f ; \epsilon_1 ) \right)
 \mathsf{Q}_{1,x}
 \, ,
 \label{eq:Q_shift}
\end{align}
where $\Gamma_1(z;\epsilon_1)$ is the (first) gamma function~\eqref{eq:multi_gamma}.
After the shift, we obtain the TQ-relation for XXX $\SL(2)$-spin chain:\index{TQ-relation!XXX spin chain}
\begin{align}
 P_{1,x-} \mathsf{Q}_{1,x} + \mathfrak{q}_1 \, \widetilde{P}_{1,x} \mathsf{Q}_{1,x--} = \mathsf{T}_{1,x} \, \mathsf{Q}_{1,x-}
 \, ,
\end{align}
which is also written in the form of quantum curve \index{quantum curve}
\begin{align}
 H(\hat{x},\hat{y}) \, \mathsf{Q}_{1,x-} = 0
\end{align}
with the algebraic function
\begin{align}
 H(x,y) = P_{1,x-} \, y^2 - \mathsf{T}_{1,x} \, y + \mathfrak{q}_1 \, \widetilde{P}_{1,x}
 \, .
\end{align}
Now the (anti)fundamental mass parameters are identified with the inhomogeneous parameters of the spin chain model.
The zero locus of this function defines the classical Seiberg--Witten curve for $\SU(n)$ SQCD.
As in the case of the pure SYM theory, We obtain the TQ-relation of the XXX spin chain as a quantization of the Seiberg--Witten curve.
See also \S\ref{sec:Higgs_truncation}.

\subsection{$A_2$ quiver}

We then discuss quantization of the Seiberg--Witten curve for quiver gauge theory.
In the NS limit, we have two fundamental $q$-characters for $A_2$ quiver:\index{q-character@$q$-character}
\begin{subequations}
 \begin{align}
  \mathsf{T}_{1,x}
  & = \mathsf{Y}_{1,x} + \frac{\mathsf{Y}_{2,x}}{\mathsf{Y}_{1,x-}} + \frac{1}{\mathsf{Y}_{2,x-}}
  = \frac{\mathsf{Q}_{1,x}}{\mathsf{Q}_{1,x-}} + \frac{\mathsf{Q}_{1,x--}}{\mathsf{Q}_{1,x-}} \frac{\mathsf{Q}_{2,x}}{\mathsf{Q}_{2,x-}} + \frac{\mathsf{Q}_{2,x--}}{\mathsf{Q}_{2,x-}}
  \, , \\
  \mathsf{T}_{2,x}
  & = \mathsf{Y}_{2,x} + \frac{\mathsf{Y}_{1,x-}}{\mathsf{Y}_{2,x-}} + \frac{1}{\mathsf{Y}_{1,x--}} 
  = \frac{\mathsf{Q}_{2,x}}{\mathsf{Q}_{2,x-}} + \frac{\mathsf{Q}_{1,x-}}{\mathsf{Q}_{1,x--}} \frac{\mathsf{Q}_{2,x--}}{\mathsf{Q}_{2,x-}} + \frac{\mathsf{Q}_{1,x---}}{\mathsf{Q}_{1,x--}}
  \, ,
 \end{align}
\end{subequations}
which are combined into a single Schr\"odinger-type difference equation (quantum curve) \index{quantum curve}
\begin{align}
 H(\hat{x}, \hat{y}) \, \mathsf{Q}_{1,x---}
 =
 \mathsf{Q}_{1,x} - \mathsf{T}_{1,x} \, \mathsf{Q}_{1,x-} + \mathsf{T}_{2,x} \, \mathsf{Q}_{1,x--} + \mathsf{Q}_{1,x---}
 = 0
 \label{eq:HQ=0_A2}
\end{align}
with the Hamiltonian defined as
\begin{align}
 H(x,y) = y^3 - \mathsf{T}_{1,x} \, y^2 + \mathsf{T}_{2,x} \, y - 1 = \det(y - L(x))
 \, .
\end{align}
for $L(x) \in \SL(3) = G_{A_2}$.
Compared to the previous expression of the curve~\eqref{eq:A2_SW_curve_cl}, the classical Seiberg--Witten curve is given as the zero locus of the Hamiltonian, $\Sigma = \{ H(x,y) = 0\}$. 
Hence, \eqref{eq:HQ=0_A2} is interpreted as a quantization of the Seiberg--Witten curve for $A_2$ quiver theory.
We can similarly consider another Hamiltonian that annihilates the $\mathsf{Q}$-function for the second node, $H(\hat{x},\hat{y}) \mathsf{Q}_{2,x} = 0$.
At this point, we do not incorporate the hypermultiplets in the fundamental representations for simplicity.
If taking into account such fundamental matters, the quantum curve~\eqref{eq:HQ=0_A2} is promoted to the TQ-relation for XXX spin chain with the symmetry $G_{A_2} = \SL(3)$.

\subsection{$A_p$ quiver}\label{sec:Ap_NSlim}

Let us move on to generic linear quiver, $\Gamma = A_p$.
The first $q$-character for $A_p$ quiver is given as follows:\index{q-character@$q$-character}
\begin{align}
 \mathsf{T}_{1,x}    
 & = \mathsf{Y}_{1,x} + \frac{\mathsf{Y}_{2,x}}{\mathsf{Y}_{1,x-}} + \cdots
 + \frac{\mathsf{Y}_{p,x}}{\mathsf{Y}_{p-1,x-}} + \frac{1}{\mathsf{Y}_{p,x-}}
 =: \sum_{i = 1}^{p+1} \Lambda_{i,x}
 \, ,
\end{align}
where the weights $(\Lambda_{i,x})_{i \in \Gamma_0}$ appearing in the highest weight module with respect to $\mathsf{Y}_{1,x}$ are given by
\begin{align}
 \Lambda_{i,x} = \frac{\mathsf{Y}_{i,x}}{\mathsf{Y}_{i-1,x-}}
 \, , \qquad
 i = 1,\ldots,p+1
 \, ,
\end{align}
with the trivial nodes, $\mathsf{Y}_{0,x} = \mathsf{Y}_{p+1,x} = 1$.
Compared to the construction shown in~\S\ref{sec:SW_curve_quiver}, these weights $(\Lambda_{i,x})_{i \in \Gamma_0}$ are identified with the eigenvalues of $G_{A_p}$-Lax matrix, $L(x) \in G_{A_p} = \SL(p+1)$, s.t., \index{Lax matrix}
\begin{align}
 \mathsf{T}_{1,x} = \tr L(x)
 \, .
\end{align}
In general, the fundamental $q$-characters have a similar expression in terms of the eigenvalues, associated with antisymmetric representation of $G_{A_p} = \SL(p+1)$,
\begin{align}
 \mathsf{T}_{i,x}
 & = \sum_{1 \le j_1 < j_2 < \cdots < j_i \le p+1} \Lambda_{j_1,x} \Lambda_{j_2,x} \cdots \Lambda_{j_i,x}
 = \tr_{\wedge^i} L(x)
 \, .
\end{align}
Then, the Hamiltonian is a generating function of these fundamental $q$-characters
\begin{align}
 H(x,y)
 = y^{p+1} - \mathsf{T}_{1,x} \, y^p + \cdots \pm \mathsf{T}_{p,x} \, y \mp 1
 = \sum_{i=0}^{p+1} (-1)^i \mathsf{T}_{i,x} \, y^{p+i-i}
\end{align}
with $\mathsf{T}_{0,x} = \mathsf{T}_{p+1,x} = 1$, which is also given as a characteristic polynomial of the Lax matrix,
\begin{align}
 H(x,y) =  \det(y - L(x)) = \prod_{i=1}^{p+1} ( y - \Lambda_{i,x} )
 \, .
\end{align}
The classical Seiberg--Witten curve is given by the zero locus of this characteristic polynomial, $\Sigma = \{ H(x,y) = 0\}$, as shown in \S\ref{sec:SW_curve_quiver}.
In the NS limit, we instead obtain the difference equation as a quantum curve:\index{quantum curve}
\begin{align}
 H( \hat{x},\hat{y}) \, \mathsf{Q}_{1,-p-1} = 0
 \, ,
\end{align}
which is promoted to the TQ-relation for the spin chain with $G_{A_p} = \SL(p+1)$ symmetry.

Since the quantum curve is a difference equation of degree $p+1$, there should be $p+1$ independent solutions, which could be obtained as follows:
\begin{align} 
 H( \hat{x},\hat{y}) \, \psi_x = \prod_{i=1}^{p+1} ( \hat{y} - \Lambda_{i,x} ) \psi_x = 0
 \quad \implies \quad
 \left( \hat{y} - \Lambda_{i,x} \right) \psi_{i,x} = 0
\end{align}
which implies
\begin{align}
 \frac{\psi_{i,x+}}{\psi_{i,x}} = \Lambda_{i,x} 
 \, .
\end{align}
The first solution $(i = 1)$ corresponds to the $\mathsf{Q}$-function, $\psi_{1,x} = \mathsf{Q}_{1,x-}$.

\section{Bethe equation}\label{sec:Bethe_eq}

We have seen that the correspondence between the gauge theory and the integrable system is promoted to the quantum level by turning on one of the $\Omega$-background parameters under the identification $\epsilon_1 = \hbar$.
In fact, the difference equation obtained through quantization of the Seiberg--Witten curve agrees with the TQ-relation of the corresponding quantum integrable system.
In this Section, we then show that the saddle point equation which determines the dominant configuration in the instanton sum also has a crucial interpretation on the quantum integrable system.

\subsection{Saddle point equation}
\index{saddle point analysis}

We revisit the saddle point equation in the NS limit.
Together with the formula~\eqref{eq:saddle_pt_finite}, the saddle point equation \eqref{eq:Z/Z=1} is given in terms of the $\mathsf{Q}$-functions as follows:
\begin{align}
 1 & = - \mathfrak{q}_i \,
 \frac{P_{i,x} \widetilde{P}_{i,x+}}{\mathsf{Y}_{i,x+} \mathsf{Y}_{i,x}}
 \prod_{e:i \to j} \mathsf{Y}_{j,x-m_e+}
 \prod_{e:j \to i} \mathsf{Y}_{j,x+m_e}
 \nonumber \\
 & = - \mathfrak{q}_i \,
 P_{i,x} \widetilde{P}_{i,x+}
 \frac{\mathsf{Q}_{i,x-}}{\mathsf{Q}_{i,x+}}
 \prod_{e:i \to j} \frac{\mathsf{Q}_{j,x-m_e+}}{\mathsf{Q}_{j,x-m_e}}
 \prod_{e:j \to i} \frac{\mathsf{Q}_{j,x+m_e}}{\mathsf{Q}_{j,x+m_e-}}
 \, ,
 \qquad
 x \in \mathcal{X}_i
 \, .
\end{align}
After several shift of the parameters, e.g., $P_{i,x+} \to P_{i,x}$, $m_e \to \frac{1}{2} \epsilon_1$, and also the redefinition of the function similar to \eqref{eq:Q_shift}, we obtain the equation, which is identified with the Bethe equation for $G_{\Gamma}$-XXX spin chain for $\Gamma = ADE$:\index{Bethe equation}
\begin{align}
 \frac{P_{i,x}}{\widetilde{P}_{i,x}}
 = - \mathfrak{q}_{i} \,
 \frac{\mathsf{Q}_{i,x-\epsilon_1}}{\mathsf{Q}_{i,x+\epsilon_1}}
 \prod_{j (\neq i)}
 \frac{\mathsf{Q}_{j,x+\frac{1}{2}\epsilon_1}}{\mathsf{Q}_{j,x-\frac{1}{2}\epsilon_1}}
 = - \mathfrak{q}_{i} \,
 \prod_{j \in \Gamma_0}
 \frac{\mathsf{Q}_{j,x-\frac{1}{2}c_{ij}^{[0]}\epsilon_1}}{\mathsf{Q}_{j,x+\frac{1}{2}c_{ij}^{[0]}\epsilon_1}}
 \, ,
 \label{eq:Bethe_eq}
\end{align}
where $(c_{ij}^{[0]})_{i,j \in \Gamma_0}$ is the classical quiver Cartan matrix~\eqref{eq:classical_quiv_Cartan_mat}.

Let us consider the simplest example, which is pure $\SU(n)$ gauge theory ($A_1$ quiver).
In this case, the saddle point equation gives rise to the Bethe equation for the Toda chain:
\begin{align}
 1 = - \mathfrak{q} \, \frac{\mathsf{Q}_{i,x-\epsilon_1}}{\mathsf{Q}_{i,x+\epsilon_1}}
 \, .
\end{align}
We remark that there is no non-trivial solution if the $\mathsf{Q}$-function is a finite polynomial, so that the $\mathsf{Q}$-function must be an infinite product.

\subsection{Higgsing and truncation}\label{sec:Higgs_truncation}

The Bethe equation obtained from the saddle point equation is still formal in the following sense:
The zeros of the $\mathsf{Q}$-function are interpreted as rapidities of magnons of the spin chain model.
Since our $\mathsf{Q}$-function constructed in gauge theory has infinitely many zeros, it describes infinite magnons on the spin chain side.
We discuss how to obtain the finite magnon configuration from gauge theory in the following.

In order to realize such a situation, let us consider the root of Higgs branch condition:\index{root of Higgs branch}
\begin{align}
 m_{i,f} = \mathsf{a}_{i,\alpha} + \tilde{n}_{i,\alpha} \, \epsilon_1
 \, .
 \label{eq:m=a+n}
\end{align}
Under this condition, as shown in \S\ref{sec:fund_instanton}, there is the restrictuion on the instanton configuration, s.t., $\ell(\lambda_{i,\alpha}) = \check{\lambda}_{i,\alpha,1} \le \tilde{n}_{i,\alpha}$.
Then, in this case, the $\mathsf{Y}$-function~\eqref{eq:Y_func_def} becomes a finite product:
\begin{align}
 \mathsf{Y}_{i,x}
 & =
 \prod_{\alpha = 1}^{n_i}
 \prod_{k = 1}^\infty
 \frac{ x - (\mathsf{a}_{i,\alpha} + (k - 1) \epsilon_1 + \lambda_{i,\alpha,k} \epsilon_2) }{ x - (\mathsf{a}_{i,\alpha} + (k - 1) \epsilon_1 + \lambda_{i,\alpha,k} \epsilon_2) - \epsilon_1}
 \nonumber \\ 
 & =
 \prod_{\alpha = 1}^{n_i}
 \left[
 \left( x - (\mathsf{a}_{i,\alpha} + \tilde{n}_{i,\alpha} \epsilon_1) \right)
 \prod_{k=1}^{\tilde{n}_{i,\alpha}}
 \frac{ x - (\mathsf{a}_{i,\alpha} + (k - 1) \epsilon_1 + \lambda_{i,\alpha,k} \epsilon_2) }{ x - (\mathsf{a}_{i,\alpha} + (k - 1) \epsilon_1 + \lambda_{i,\alpha,k} \epsilon_2) - \epsilon_1}
 \right]
 \nonumber \\
 & =:
 P_{i,x} \frac{\mathscr{Q}_{i,x}}{\mathscr{Q}_{i,x-}}
\end{align}
where we define the truncated $\mathscr{Q}$-function\index{Q-function@$\mathsf{Q}$-function}
\begin{align}
 \mathscr{Q}_{i,x} = \prod_{x' \in \log \widetilde{\mathcal{X}}_i} (x - x')
\end{align}
with the truncated version of the instanton configuration~\eqref{eq:log_X-variable},
\begin{align}
 \log \widetilde{\mathcal{X}}_i = \{ x_{i,\alpha,k} = \mathsf{a}_{i,\alpha} + \epsilon_1 (k - 1) + \epsilon_2 \lambda_{i,\alpha,k} \}_{\alpha = 1,\ldots,n_i}^{k = 1,\ldots,\tilde{n}_{i,\alpha}}
 \, .
\end{align}
As mentioned in \S\ref{sec:NS_lim}, the saddle point equation is taken with respect to the effective twisted superpotential of 2d $\mathcal{N} = (2,2)$ theory.
In this context, the $x$-variable is identified with the Cartan element of the 2d gauge algebra, $\mathfrak{h}^{\text{2d}} \subset \mathfrak{g}^\text{2d}$.
Hence, the number of the $x$-variables associated with $i$-th node is given as the rank of the gauge group $G_i^\text{2d}$.
Under the Higgs branch condition~\eqref{eq:m=a+n}, it is given by
\begin{align}
 n_i^\text{2d} = \sum_{\alpha = 1}^n \tilde{n}_{i,\alpha}
 \, .
\end{align}
In the brane configuration discussed in \S\ref{sec:Higgsing}, this is interpreted as the number of D2 branes, namely the number of vortices appearing in the Higgs phase.

\subsubsection{TQ-relation}
\index{TQ-relation!XXX spin chain}

Let us examine this prescription with $A_1$ quiver with fundamental and antifundamental matters.
Starting with the $q$-character obtained in~\eqref{eq:q_ch_A1}, we obtain
\begin{align}
 \mathsf{T}_{1,x} = P_{1,x} \frac{\mathscr{Q}_{1,x}}{\mathscr{Q}_{1,x-}}
 + \mathfrak{q}_1 \widetilde{P}_{1,x} \frac{\mathscr{Q}_{1,x--}}{\mathscr{Q}_{1,x-}}
 \, ,
\end{align}
which leads to the TQ-relation
\begin{align}
 P_{1,x} \mathscr{Q}_{1,x} - \mathsf{T}_{1,x} \, \mathscr{Q}_{1,x-} + \mathfrak{q}_1 \, \widetilde{P}_{1,x} \mathscr{Q}_{1,x--} = 0
 \, .
 \label{eq:TQ_rel_A1_truncated}
\end{align}
We remark that redefinition of the $\mathscr{Q}$-function~\eqref{eq:Q_shift} is not necessary in this case.

\subsubsection{Bethe equation}

We can similarly obtain the Bethe equation on the Higgs branch locus.
In this case, we obtain the truncated version of~\eqref{eq:Bethe_eq} from the TQ-relation by substituting a zero of the $\mathscr{Q}$-function,
\begin{align}
 \frac{P_{1,x+}}{\widetilde{P}_{1,x+}} = - \mathfrak{q}_1 \frac{\mathscr{Q}_{1,x-}}{\mathscr{Q}_{1,x+}}
 \, , \qquad
 x \in \log \widetilde{\mathcal{X}}
 \, .
\end{align}
We focus on the case with $n^\text{f} = n^{\text{af}} = L$ and parametrize $(m_f, \widetilde{m}_f) = (\nu_f - s_f \epsilon_1, \nu_f + s_f \epsilon_1)$.
Let $n^\text{2d} = n$ and $\log \widetilde{\mathcal{X}} = (x_k)_{k = 1,\ldots,n}$, then the Bethe equation for the Bethe root $x_k \in \log \widetilde{\mathcal{X}}$ is explicitly given as
\begin{align}
 \prod_{f = 1}^L
 \frac{x_k - \nu_f + s_f \epsilon_1}{x_k - \nu_f - s_f \epsilon_1}
 & = \mathfrak{q}_1
 \prod_{\substack{k' = 1 \\ (k' \neq k)}}^n
 \frac{x_k - x_{k'} - \epsilon_1}{x_k - x_{k'} + \epsilon_1}
 \, ,
\end{align}
where $(\nu_f, s_f)_{f = 1,\ldots,L}$ are identified with the inhomogeneous parameter and the spin of the site $f$ on the XXX spin chain.%
\footnote{%
This is the Bethe equation for the periodic (closed) spin chain.
Starting from 2d SO/Sp gauge theory, we obtain the spin chain with the open boundary condition~\cite{Kimura:2020bed}.
}
In addition, $\epsilon_1$ plays a role of the Planck constant, and $(n,L)$ are identified with the number of magnons and the length of the chain under this correspondence.

\subsubsection{Supergroup gauge theory}

We can similarly derive the Bethe equation from the saddle point equation of supergroup gauge theory.
In this case, we obtain two types of magnons and sites, corresponding to the positive and negative gauge and flavor nodes.
Since they have the Planck constant with opposite signs, it is interpreted as a coupled system of $n_0$ positive and $n_1$ negative magnons, $L_0$ positive and $L_1$ negative spin configurations.

Applying a similar analysis to pure SYM theory and $\widehat{A}_0$ quiver theory with supergroup gauge symmetry, we obtain the particle models (Toda, Calogero--Moser) instead of the spin chain, which is associated with the super-type root system, e.g., the double Calogero system~\cite{Sergeev:2001JNMP}.
See~\cite{Kimura:2019msw,Chen:2020rxu} for details.

\subsubsection{Surface defect in two ways}

As discussed in \S\ref{sec:Higgsing}, we shall move to the vortex theory after the Higgsing procedure, which is interpreted as the codimension-two defect of gauge theory, called the surface defect~\cite{Gukov:2008sn}.
In addition to the Higgsing, there is another way to incorporate the surface defect into gauge theory, based on the so-called ramified instanton moduli space~\cite{Feigin:2011SM,Finkelberg:2010QDZ,Kanno:2011fw}:
Instead of $\mathbb{C}^2$, one considers instantons on the (partial) orbifold $\mathbb{C} \times \left( \mathbb{C} / \mathbb{Z}_M \right)$ for $M = 1,\ldots,n$ for $\SU(n)$ gauge theory.
The case with $M = n$ is called the full defect, which is typically considered in the context of the Bethe/Gauge correspondence.
Since there is no fundamental hypermultiplet in pure $\SU(n)$ gauge theory and $\mathcal{N} = 2^*$ theory ($\widehat{A}_0$ quiver), we should utilize the ramified instanton moduli space scheme to impose the surface defect.
In this scheme, the Coulomb moduli parameter $(\mathsf{a}_\alpha)_{\alpha = 1,\ldots, n}$ and the ramified coupling constant $(\mathfrak{q}_\alpha)_{\alpha = 1,\ldots, n}$ are identified with the momenta and the particle coordinates of the quantum integrable system~\cite{Nekrasov:2017gzb,Chen:2019vvt,Chen:2020rxu,Lee:2020hfu}.
\index{Coulomb moduli}

\subsection{Dimensional hierarchy: periodicity of spectral parameter}\label{sec:hierarchy}

So far, we have focused on 4d $\mathcal{N} = 2$ gauge theory to discuss its connection with quantum integrable system through quantization of Seiberg--Witten geometry.
We can apply this quantization scheme also to 5d $\mathcal{N} = 1$ theory and 6d $\mathcal{N} = (1,0)$ theory.
Under the correspondence between gauge theory and integrable system, the $x$-variable in the Seiberg--Witten geometry is identified with the spectral parameter of the associated integrable system.
Since, in higher dimensional cases, we impose periodicity to the $x$-variable as discussed in~\S\ref{sec:5d6d}, we would obtain the integrable system with the spectral parameter with periodicity, $x \in \check{\mathcal{C}} = \mathbb{C}$ / $\mathbb{C}^\times$/ $\mathcal{E}$, as in~\eqref{eq:C_Cdual}.
This is known as the hierarchy of rational/trigonometric/elliptic integrable systems.
In fact, we would obtain the TQ-relation and the Bethe equation of XXX/XXZ/XYZ spin chain model from 4d $\mathcal{N} = 2$ / 5d $\mathcal{N} = 1$ / 6d $\mathcal{N}=(1,0)$ gauge theory on $\mathbb{C}^2 \times \mathcal{C}$.

\part{Quantum Algebra}\label{part:algebra}

\chapter{Operator formalism of gauge theory}\label{chap:operator}

We have discussed classical and quantum geometric aspects of $\mathcal{N} = 2$ gauge theory in four dimensions in the relation to various fields of physics and mathematics. 
One of the key ingredients in such aspects is the non-perturbative symmetry of gauge theory incorporated by instantons, i.e.,~covariance of path integral partition function under the adding/removing-instanton operation.
In general, symmetry of the system is rephrased as the invariance (covariance) under the corresponding transformation, which is described as a group action, and we would be able to discuss the algebraic structure from the infinitesimal version of the transformation (group action).
The purpose of this Part~\ref{part:algebra} is to explore the algebraic structure associated with the non-perturbative symmetry of gauge theory, namely the symmetry algebra of the instanton creation and annihilation.

In order to see the underlying algebraic structure of gauge theory, in this Chapter, we introduce the operator formalism of the path integral.
This formalism is in fact based on the idea of {\em BPS/CFT correspondence}~\cite{Nekrasov:2015wsu}, which claims the correspondence between the gauge theory observable (chiral ring) and the vertex operator in CFT.
We will first discuss the holomorphic deformation of the prepotential of 4d $\mathcal{N} = 2$ gauge theory to introduce the infinitely many time variables.
This allows us to construct the Fock space, and we may discuss the notion of {\em Z-state} through the operator/state correspondence.
Then, we will see that various vertex operators acting on the Fock space are used to construct the instanton partition function, and also the gauge theory observables.
We will also discuss the iWeyl reflection in the operator formalism, and see how the pole cancellation is described with the vertex operators.

\section{Holomorphic deformation}\label{sec:hol_def}

As shown in \S\ref{sec:N=2}, the low energy effective theory of $\mathcal{N} = 2$ theory is described by the prepotential.%
\footnote{%
In this Chapter, we will mostly use the K-theory (5d $\mathcal{N} = 1$ theory on a circle) convention.
} \index{prepotential}
Although the quadratic prepotential \eqref{eq:prepotential_quadratic} in particular gives rise to the renormalizable YM action, we may consider a generic holomorphic function in the context of the Seiberg--Witten geometry.
Let us consider the holomorphic deformation of the prepotential as follows~\cite{Marshakov:2006ii}:\index{holomorphic deformation}
\begin{align}
 \mathscr{F} 
 \ \longmapsto \
 \mathscr{F}(t) := \mathscr{F} + \sum_{n = 1}^\infty t_n \, \mathcal{O}_n
 \, ,
\end{align}
where the deformation parameters are called the {\em time variables}\index{time variable} from the analogy with the integrable hierarchy, and $(\mathcal{O}_n)_{n \ge 1}$ is the chiral ring operator (the Casimir elements of the complex scalar).
In particular, in the K-theory (5d $\mathcal{N} = 1$ theory) convention, it is given by the character of the observable bundle
\begin{align}
 \mathcal{O}_n = \ch_\mathsf{T} \mathbf{Y}^{[n]}
 \, .
\end{align}
See also~\cite{Nekrasov:2003rj,Nakajima:2003uh}.

Such a deformation is similarly considered for quiver gauge theory in general.
Even after the deformation, we can still apply the equivariant localization analysis to the path integral, and the resulting partition function is given as%
\footnote{%
We also denote the partition function contribution associated with the instanton configuration $\mathcal{X}$ by $Z_{\mathcal{X}} = Z[\mathcal{X}]$.
}
\begin{align}
 Z = \sum_{\mathcal{X} \in \mathfrak{M}^\mathsf{T}} Z_{\mathcal{X}}
 \ \longmapsto \
 Z(t) 
 = \sum_{\mathcal{X} \in \mathfrak{M}^\mathsf{T}} Z_{\mathcal{X}}(t)
 = \sum_{\mathcal{X} \in \mathfrak{M}^\mathsf{T}} Z_{\mathcal{X}} Z^\text{pot}_{\mathcal{X}}(t)
\end{align}
where the $t$-dependent term is called the {\em potential term}:\index{potential term}
\begin{align}
 Z_{\mathcal{X}}^\text{pot}(t)
 = \exp \left( \sum_{n = 1}^\infty t_n \, \mathcal{O}_n[\mathcal{X}] 
 \right)
 \, .
 \label{eq:por_term}
\end{align}

\subsection{Free field realization}

In the presence of the potential term, the vev of the chiral ring operator is given by the derivative with respect to the conjugate time variable,
\begin{align}
 \vev{\mathcal{O}_n} = 
 \frac{1}{Z} \sum_{\mathcal{X}} Z_\mathcal{X} \cdot \mathcal{O}_n[\mathcal{X}]
 = 
 \frac{\partial}{\partial t_n} \log Z(t)\Big|_{t=0}
 \, .
\end{align}
In this sense, the $t$-deformed partition function $Z(t)$ plays a role of the generating function of the chiral ring operators.
Then, denoting $\partial_n := \partial_{t_n}$, we obtain the identification:%
\footnote{%
This is similar to the Fourier transformation: In the presence of the factor $\np^{\im p x}$, the multiplication of the variable $p$ is equivalent to the derivative with the $x$-variable, $p \leftrightarrow -\im \partial_x$.
}
\begin{align}
 \mathcal{O}_n 
 \ \longleftrightarrow \
 \partial_n 
 \, .
\end{align}
This identification leads to the operator formalism of gauge theory with the Fock space 
\begin{align}
 \mathsf{F} = \mathbb{C}[[t_1,t_2,\ldots,\partial_1,\partial_2,\ldots]] \ket{0}
\end{align}
generated by the Heisenberg algebra $\mathscr{H} = (t_n,\partial_n)_{n \ge 1}$, s.t.,
\begin{align}
 [\partial_n, t_{n'}] = \delta_{n,n'}
 \, .
\end{align}
We define the vacuum state $\ket{0}$ as a $t$-constant annihilated by $(\partial_n)_{n \ge 1}$ and its dual,
\begin{align}
 \partial_n \ket{0} = 0
 \, , \qquad
 \bra{0} t_n = 0
 \qquad
 (n \ge 1)
 \, .
 \label{eq:vac_state}
\end{align}
For quiver theory, we define a set of the Heisenberg algebras
\begin{align}
 \mathscr{H} = \bigoplus_{i \in \Gamma_0} \mathscr{H}_i
 \, , \qquad
 \mathscr{H}_i = (t_{i,n},\partial_{i,n})_{n \ge 1} 
\end{align}
with
\begin{align}
 \left[ \partial_{i,n} \, , \, t_{j,n'} \right] = \delta_{i,j} \delta_{n,n'}
 \, .
\end{align}
We also define the zero mode to be identified with the instanton counting parameter
\begin{align}
 t_{i,0} = \log_{q_2} \mathfrak{q}_i
 \, .
\end{align}
Then, the Fock space is generated by these Heisenberg algebras
\begin{align}
 \mathsf{F} = \bigoplus_{i \in \Gamma_0} \mathsf{F}_i
 \, , \qquad
 \mathsf{F}_i = \mathbb{C}[[t_{i,n}, \partial_{i,n}]]\ket{0}
 \, .
\end{align}
We call this operator formalism based on the Heisenberg algebra the {\em free field realization} of gauge theory.

\section{$Z$-state}
\index{Z-state@$Z$-state}

In the operator formalism, since the time variables $(t_{i,n})$ are the operators, the $t$-extended partition function $Z(t)$ also behaves as an operator acting on the Fock space~$\mathsf{F}$.
This observation motivates us to explore the {\em $Z$-state} through the operator-state correspondence\index{operator-state correspondence} under the radial quantization in the operator formalism:
\begin{equation}
 \begin{tikzcd}
  \text{(Operator)} & \text{(State)} \\[-1em]
  Z(t) \arrow[r,latex-latex] & \ket{Z} 
 \end{tikzcd}
\end{equation}
where we define
\begin{align}
 \ket{Z} = Z(t) \ket{0} = \sum_{\mathcal{X}} \ket{Z_\mathcal{X}}
 \, , \qquad
 \ket{Z_\mathcal{X}} = Z_{\mathcal{X}}(t) \ket{0}
 \, .
\end{align}

Recalling the property~\eqref{eq:vac_state}, the dual vacuum plays a role of the projector onto the undeformed sector $t = 0$, $\bra{0}: \ \mathsf{F} \to \mathbb{C}$,
\begin{align}
 Z = \vev{0 \mid Z} = \bra{0} Z(t) \ket{0}
 \, .
 \label{eq:Z_projection}
\end{align}
Namely, the instanton partition function is given as the vev of the operator acting on the Fock space.%
\footnote{%
The partition function for 6d theory is instead given by the trace over the Fock space, which gives rise to the character of the corresponding module.
See \S\ref{sec:6d_trace}.
}
In the following, we will see that the operator uplift of the instanton partition function is realized using the vertex operators.

\subsection{Screening current}\label{sec:screening_current}

Compared to the index formula of the partition function~\eqref{eq:TM} together with the quiver Cartan matrix~\eqref{eq:TM_Cartan}, we arrive at the following expression of the $Z$-state for 5d $\mathcal{N} = 1$ quiver gauge theory~\cite{Kimura:2015rgi}:%
\footnote{%
A similar construction is available for 4d $\mathcal{N} = 2$ theory~\cite{Nieri:2019mdl}, and also for 6d $\mathcal{N} = (1,0)$ theory as discussed in Chapter~\ref{chap:eW}.
}
\begin{align}
 \ket{Z_\mathcal{X}}
 = Z_{\mathcal{X}}(t) \ket{0}
 = \prod_{x \in \mathcal{X}}^\succ S_{\mathsf{i}(x),x} \ket{0}
 \label{eq:Z_state_X}
\end{align}
where the product with the symbol $\succ$ is the radial ordering product with respect to the ordering in the set $\mathcal{X}$,\index{radial ordering}%
\footnote{%
For example, we take $|q_1| \ll |q_2^{-1}| < 1$ and $|\np^{\mathsf{a}}| \sim |\np^{m}| \sim 1$.
Then, define the ordering $x \succ x'$ if $|x| > |x'|$.
}
\begin{align}
 \prod_{x \in \mathcal{X}}^\succ S_x = S_{x_1} S_{x_2} S_{x_3} \cdots
 \quad \text{for} \quad
 |x_1| > |x_2| > |x_3| > \cdots
 \, ,
 \label{eq:radial_ordering_prod}
\end{align}
and we define a map $\mathsf{i}: \mathcal{X} \to \Gamma_0$, s.t.,
\begin{align} 
 \mathsf{i}(x) = i
 \ \iff \
 x \in \mathcal{X}_i
 \, , \qquad
 i \in \Gamma_0
 \, .
\end{align}
The vertex operator $(S_{i,x})_{i \in \Gamma_0}$ is called the screening current\index{screening current}
\begin{align}
 S_{i,x} 
 = 
 {: \exp \left( 
 s_{i,0} \log x + \tilde{s}_{i,0} 
 - \frac{\kappa_{i}}{2} \left( \log_{q_2} x - 1 \right) \log x
 + \sum_{n \in \mathbb{Z}_{\neq 0}} s_{i,n} \, x^{-n} 
 \right) :}
\end{align}
with the free field modes
\begin{align}
 s_{i,-n} = (1 - q_1^n) \, t_{i,n}
 \, , \qquad
 s_{i,0} = t_{i,0}
 \, , \qquad
 s_{i,n} = - \frac{1}{n} (1 - q_2^{-n})^{-1} \, c_{ji}^{[n]} \, \partial_{j,n}
 \qquad
 (n \ge 1)
 \, ,
 \label{eq:s_oscillator}
\end{align}
obeying the commutation relation
\begin{align}
 \left[ s_{i,n} \, , \, s_{j,n'} \right]
 = - \frac{1}{n} \frac{1 - q_1^n}{1 - q_2^{-n}} \, c_{ji}^{[n]} \, \delta_{n+n',0}
 \qquad
 (n \ge 1)
 \, .
 \label{eq:s_comm_rel}
\end{align}
The zero modes obey the relation
\begin{align}
 \left[ \tilde{s}_{i,0} \,,\, s_{j,n} \right] = - \beta \, c_{ji}^{[0]} \, \delta_{n,0}
 \, , \qquad
 \beta = - \frac{\epsilon_1}{\epsilon_2}
 \, .
\end{align}
The symbol $:\bullet:$ means the normal ordering, s.t., the annihilation operators $(\partial_{i,n})$ are placed on the right, while the creation operators $(t_{i,n})$ are on the left.
\index{normal ordering}

We remark that, in the construction of the screening current above, the roles of $q_1$ and $q_2$ are not on equal footing.
This is because the current convention of the partition function is based on the partial reduction of the universal bundle, $\mathbf{Y}_i = \wedge \mathbf{Q}_1 \cdot \mathbf{X}_i$, as shown in \S\ref{sec:quiv_eq_fix_pt}.
Starting with the other reduction, $\mathbf{Y}_i = \wedge \mathbf{Q}_2 \cdot \check{\mathbf{X}}_i$, we will obtain the swapped version of the screening current with $q_1 \leftrightarrow q_2$.
See \cite{Nieri:2017ntx} for a related discussion in 3d gauge theory.

In order to see the agreement between the index formula~\eqref{eq:TM} and the vertex operator representation of the partition function, we evaluate the operator product expansion (OPE) between the screening currents. \index{operator product expansion (OPE)}%
Applying the Baker--Campbell--Hausdorff formula\index{Baker--Campbell--Hausdorff formula}
\begin{align}
 \np^X \, \np^Y = \np^Z
 \, ,
\end{align}
with 
\begin{align}
 Z & = X + Y + \frac{1}{2} [X,Y] + \frac{1}{12}[X,[X,Y]] - \frac{1}{12}[Y,[X,Y]] + \cdots
 \nonumber \\
 & \longrightarrow \ 
 X + Y + \frac{1}{2} [X,Y]
 \qquad (\text{if $[X,Y]$ is a center element})
 \, ,
\end{align}
the product of the screening currents is given as follows:
\begin{align}
 S_{i,x} S_{j,x'}
 = 
 S_{ij}(x,x') \,
 {: S_{i,x} S_{j,x'} :}
\end{align}
where the pair contribution is defined
\begin{align}
 S_{ij}(x,x') = 
 \exp\left(
 - \beta \, c_{ji}^{[0]} \log x'
 - \sum_{n = 1}^\infty \frac{1}{n} \frac{1 - q_1^n}{1 - q_2^{-n}} \, c_{ji}^{[n]} \left( \frac{x'}{x} \right)^n
 \right) 
 \, .
\end{align}
Then, the $Z$-state associated with instanton configuration $\mathcal{X}$ is given by the pair contributions and the normal ordering part,
\begin{align}
 \ket{Z_\mathcal{X}} = 
 \prod_{(x \prec x') \in \mathcal{X} \times \mathcal{X} }
 S_{\mathsf{i}(x)\mathsf{i}(x')}(x,x')
 \,
 {:
 \prod_{x \in \mathcal{X}} S_{\mathsf{i}(x),x}
 :}
 \ket{0}
 \, .
\end{align}

\subsubsection{Normal ordering factor}

We first deal with the normal ordering part,
\begin{align}
 & 
 {:
 \prod_{x \in \mathcal{X}} 
 S_{\mathsf{i}(x),x}
 :}
 \ket{0} 
 \nonumber \\
 & = 
 \prod_{x \in \mathcal{X}} 
 \exp \left(
 s_{\mathsf{i}(x),0} \, \log x
 - \frac{\kappa_{\mathsf{i}(x)}}{2} \left( \log_{q_2} x - 1 \right) \log x
 + \sum_{n = 1}^\infty (1 - q_1^n) \, t_{\mathsf{i}(x),n} \, x^n
 \right)
 \ket{0}
 \nonumber \\
 & =
 \prod_{x \in \mathcal{X}}
 \mathfrak{q}_{\mathsf{i}(x)}^{\log_{q_2} x}
 \np^{
 - \frac{\kappa_{\mathsf{i}(x)}}{2} \left( \log_{q_2} x - 1 \right) \log x
 }
 \exp\left(
 \sum_{i \in \Gamma_0} \sum_{n = 1}^\infty t_{i,n} \, \ch \mathbf{Y}_{i}^{[n]}
 \right)
 \ket{0}
 \nonumber \\
 & =
 \prod_{i \in \Gamma_0}
 Z^\text{top}_i Z^\text{CS}_i Z^\text{pot}_i
 \ket{0}
 \, .
\end{align}
In order to see the identification of the first two parts, $(Z_i^\text{top}, Z_i^\text{CS})_{i \in \Gamma_0}$, it is convenient to see the behavior under the $x$-variable shift, $x \to q_2 x$ for $\mathsf{i}(x) = i$ $(x \in \mathcal{X}_i)$, and obtain the consistent behavior discussed in \S\ref{sec:add/remove}.%
\footnote{%
Precisely speaking, this agreement is up to the constant factor, which is independent of the instanton configuration, and is interpreted as the perturbative contribution.
}
The identification of the potential term is immediately obtained from the expression~\eqref{eq:por_term}. \index{potential term}
The vev of the normal ordering part gives rise to the $t$-independent part,
\begin{align}
 \bra{0}
 {:
 \prod_{x \in \mathcal{X}} 
 S_{\mathsf{i}(x),x}
 :}
 \ket{0}
 =
 \prod_{i \in \Gamma_0}
 Z^\text{top}_i Z^\text{CS}_i
 \, .
\end{align}

\subsubsection{OPE factor}
\index{operator product expansion (OPE)!S and S@$S$ and $S$}

We then evaluate the OPE factor:
\begin{align}
 \mathcal{Z}[\mathcal{X}] = 
 \prod_{(x \prec x') \in \mathcal{X} \times \mathcal{X}}
 S_{\mathsf{i}(x)\mathsf{i}(x')}(x,x')
 \, .
\end{align}
We take the $x$-variable shift, $x \to q_2 x$ $(\mathcal{X} \to \mathcal{X}')$ for $\mathsf{i}(x) = i$ $(x \in \mathcal{X}_i)$, to see the agreement with the instanton partition function,
\begin{align}
 \frac{\mathcal{Z}[\mathcal{X}']}{\mathcal{Z}[\mathcal{X}]}
 & =
 \prod_{x' (\prec x) \in \mathcal{X}}
 \exp \left(
  \sum_{n = 1}^\infty \frac{1}{n} c_{\mathsf{i}(x')i}^{[n]} (1 - q_1^n) \left( \frac{x'}{x} \right)^n
 \right)
 \nonumber \\
 & \qquad \times
 \prod_{x' (\succ x) \in \mathcal{X}}
 \exp \left(
 c_{\mathsf{i}(x')i}^{[0]} \log q_1
 +
  \sum_{n = 1}^\infty \frac{1}{n} c_{\mathsf{i}(x')i}^{[-n]} (1 - q_1^{-n}) \left( \frac{x'}{x} \right)^{-n}
 \right)
 \nonumber \\
 & =
 \prod_{x' (\neq x) \in \mathcal{X}}
 \exp \left(
  \sum_{n = 1}^\infty \frac{1}{n} c_{\mathsf{i}(x')i}^{[n]} (1 - q_1^n) \left( \frac{x'}{x} \right)^n
 \right)
\end{align}
where we apply the analytic continuation formula:
\begin{align}
 \exp \left( \log q_1 + \sum_{n = 1}^\infty \frac{z^{-n}}{n} (1 - q_1^{-n}) \right)
 & =
 q_1 \, \frac{1 - q_1^{-1} z^{-1}}{1 - z^{-1}}
 = \frac{1 - q_1 z}{1 - z}
 \nonumber \\
 & = \exp \left( \sum_{n = 1}^\infty \frac{z^n}{n} (1 - q_1^{n})  \right)
 \, .
\end{align}
Then, in terms of the $\mathsf{Y}$-function, we can rewrite this as follows:
\begin{align}
 \frac{\mathcal{Z}[\mathcal{X}']}{\mathcal{Z}[\mathcal{X}]}
 & = -
 \frac{1}{\mathsf{Y}_{i,qx}[\mathcal{X}'] \mathsf{Y}_{i,x}[\mathcal{X}]}
 \prod_{e:i \to j} \mathsf{Y}_{j,\mu_e^{-1} q x}[\mathcal{X}]
 \prod_{e:j \to i} \mathsf{Y}_{j,\mu_e x}[\mathcal{X}]
 \nonumber \\
 & = 
 \left(
 (-1)^{n_i + \sum_{i \to j} n_j}
 \np^{- \sum_{\alpha = 1}^{n_i} \mathsf{a}_{i,\alpha}}
 q^{n_i}
 \prod_{e:i \to j} \np^{\sum_{\alpha = 1}^{n_j} \mathsf{a}_{j,\alpha}} (\mu_e^{-1} q)^{-n_j}
 \right)
 x^{n_i - \sum_{i \to j} n_j}
 \nonumber \\
 & \qquad \times
 \frac{-1}{\mathsf{Y}_{i,qx}^\vee[\mathcal{X}'] \mathsf{Y}_{i,x}[\mathcal{X}]}
 \prod_{e:i \to j} \mathsf{Y}^\vee_{j,\mu_e^{-1} q x}[\mathcal{X}]
 \prod_{e:j \to i} \mathsf{Y}_{j,\mu_e x}[\mathcal{X}]
 \, ,
\end{align}
where we convert $\mathsf{Y}$ to $\mathsf{Y}^\vee$ as \eqref{eq:YtoYdual}.
Compared to the behavior under the adding-instanton operation~\eqref{eq:saddle_pt_finite}, this agrees with the equivariant index formula of the instanton partition function under the shift of the coupling constant and the Chern--Simons level:
\begin{subequations}\label{eq:parameter_shift}
\begin{align}
 \mathfrak{q}_i
 & \longleftrightarrow \
 \left(
 (-1)^{n_i + \sum_{i \to j} n_j}
 \np^{- \sum_{\alpha = 1}^{n_i} \mathsf{a}_{i,\alpha}}
 q^{n_i}
 \prod_{e:i \to j} \np^{\sum_{\alpha = 1}^{n_j} \mathsf{a}_{j,\alpha}} (\mu_e^{-1} q)^{-n_j}
 \right)
 \mathfrak{q}_i
 \, , \\
 \kappa_i
 & \longleftrightarrow \
 \kappa_i - n_i + \sum_{e:i \to j} n_j
 = \kappa_i - \sum_{j \in \Gamma_0} c_{ij}^{+[0]} n_j
 \, .
 \label{eq:CS_level_shift}
\end{align}
\end{subequations}

\subsection{Instanton sum and screening charge}\label{sec:screening_charge}

Combining the normal ordering factor and the OPE factor, we obtain the instanton partition function associated with the configuration $\mathcal{X}$ as the chiral correlator of the screening currents:
\begin{align}
 Z_\mathcal{X}
 =
 \bra{0}
 \prod_{x \in \mathcal{X}}^\succ S_{\mathsf{i}(x),x}
 \ket{0}
 \, .
 \label{eq:Z[X]_correlator}
\end{align}
Therefore, the total partition function is given by summation over $\mathcal{X}$,
\begin{align}
 Z = \sum_{\mathcal{X} \in \mathfrak{M}^\mathsf{T}}
 \bra{0}
 \prod_{x \in \mathcal{X}}^\succ S_{\mathsf{i}(x),x}
 \ket{0}
 \, .
\end{align}
In order to discuss the instanton sum, we define a set of extended configurations:
\begin{align}
 \mathfrak{M}^\mathbb{Z}
 =
 \left\{
 \np^{\mathsf{a}_{i,\alpha}} q_1^{k-1} q_2^{\mathbb{Z}}
 \right\}_{i \in \Gamma_0, \alpha = 1,\ldots,n_{i}, k = 1,\ldots,\infty}
 \, ,
\end{align}
with the asymptotics $\np^{\mathsf{a}_{i,\alpha}} q_1^{k-1} q_2^0$ at $k \to \infty$.
This is associated with an arbitrary sequence of integers with the fixed asymptotic behavior, while the partition is a non-increasing sequence of non-negative integers.
Since the configuration violating the non-increasing condition~\eqref{eq:Zvec_zero} does not contribute to the partition function, $Z_\mathcal{X} = 0$ for $\mathcal{X} \in \mathfrak{M}^\mathbb{Z} \backslash \mathfrak{M}^\mathsf{T}$, we obtain the chiral correlator expression of the partition function:
\begin{align}
 Z & =
 \sum_{\mathcal{X} \in \mathfrak{M}^\mathbb{Z}}
 \bra{0}
 \prod_{x \in \mathcal{X}}^\succ S_{\mathsf{i}(x),x}
 \ket{0} 
 =
 \bra{0}
 \prod_{x \in \mathring{\mathcal{X}}}^\succ \mathsf{S}_{\mathsf{i}(x),x}
 \ket{0}
 \, ,
 \label{eq:Z_correlator}
\end{align}
where we define the screening charge operator,%
\footnote{%
This infinite series is justified using the Jackson integral with the base $x$ denoted by $\displaystyle \oint_x dz_{q_2} \, S_{i,z}$.
See~\cite{Awata:2010yy} for a related discussion in the context of $q$-deformation of Dotsenko--Fateev integral.
}
\begin{align}
 \mathsf{S}_{i,x}
 = \sum_{k \in \mathbb{Z}}^\infty S_{i,q_2^k x}
 \, .
 \label{eq:screening_charge}
\end{align}
Then the $t$-extended partition function, which is an operator acting on the Fock space, is given as an infinite product of the screening charges
\begin{align}
 Z(t) =
 \prod_{x \in \mathring{\mathcal{X}}}^\succ \mathsf{S}_{\mathsf{i}(x),x}
 \, ,
\end{align}
and the corresponding $Z$-state is given by
\begin{align}
 \ket{Z} =
 \prod_{x \in \mathring{\mathcal{X}}}^\succ \mathsf{S}_{\mathsf{i}(x),x}
 \ket{0}
 \, .
\end{align}

We have shown that the instanton partition function has a chiral correlator expression.
Such a connection between the gauge theory, in particular, its BPS sector, and the vertex operator algebra is referred to as the {\em BPS/CFT correspondence}\index{BPS/CFT correspondence}~\cite{Nekrasov:2015wsu,Nekrasov:2016qym,Nekrasov:2016ydq,Nekrasov:2017rqy,Nekrasov:2017gzb}.
A primary example is the Alday--Gaiotto--Tachikawa (AGT) relation~\cite{Alday:2009aq,Wyllard:2009hg},\index{AGT relation} and its $q$-deformation~\cite{Awata:2009ur}, which states the equivalence between the partition function of $G$-gauge theory and the chiral conformal block of W($G$)-algebra.
Although our expression~\eqref{eq:Z_correlator} looks similar to the AGT relation, it depends only on the quiver structure, not on the gauge symmetry $G$.
We will see the underlying algebraic structure associated with the quiver structure in Chapter~\ref{chap:quiv_W}.

Another remark is that, in order to express the partition function, we have to consider infinitely many screening charges.
This is because the fixed point in the instanton moduli space is parametrized by a partition, which is an infinite sequence of non-negative integers.
Recalling the argument in \S\ref{sec:Higgs_truncation}, one can truncate the infinite product to the finite one by imposing the Higgsing condition.
The resulting chiral correlator is then interpreted as the vortex partition function in 3d quiver gauge theory~\cite{Aganagic:2013tta,Aganagic:2014oia,Aganagic:2015cta,Nedelin:2016gwu}.

\subsection{$\mathsf{V}$-operator: fundamental matter}

Although we have focused on the vector multiplet and the bifundamental hypermultiplet contributions, we can also incorporate the (anti)fundamental hypermultiplet in the operator formalism.

For this purpose, we define the $\mathsf{V}$-operator:\index{V-operator@$\mathsf{V}$-operator}
\begin{align}
 \mathsf{V}_{i,x} =
 {:
 \exp \left( \sum_{n \in \mathbb{Z}_{\neq 0}} v_{i,n} \, x^{-n} \right)
 :}
\end{align}
with
\begin{align}
 v_{i,-n} =
 - \tilde{c}_{ij}^{[n]} \, t_{j,n}
 \, , \qquad
 v_{i,n}
 = \frac{1}{n} \frac{1}{(1 - q_1^{n})(1 - q_2^{n})} \partial_{i,n}
 \label{eq:v_oscillator}
\end{align}
for $n \ge 1$, where we denote the inverse of the Cartan matrix by $(\tilde{c}_{ij})_{i,j \in \Gamma_0}$. \index{Cartan matrix!inverse of---}
Compared to the $s$-modes~\eqref{eq:s_oscillator}, we obtain
\begin{align}
 \left[ v_{i,n} \,,\, s_{j,n'} \right]
 = \frac{1}{n (1 - q_2^n)} \, \delta_{ij} \, \delta_{n+n',0}
 \, ,
 \label{eq:vs_comm_rel}
\end{align}
which gives rise to the OPE between the $\mathsf{V}$-operator and the screening current:\index{operator product expansion (OPE)!S and V@$S$ and $\mathsf{V}$}
\begin{subequations}
 \begin{align}
  S_{j,x'} \mathsf{V}_{i,x}
  & =
  {: S_{j,x'} \mathsf{V}_{i,x} :}
  \times
  \begin{cases}
   \displaystyle
   \left( q_2 \frac{x}{x'};q_2 \right)_\infty & (i = j) \\
   1 & (i \neq j)
  \end{cases}  
  \\
  \mathsf{V}_{i,x} S_{j,x'}
  & =
  {: \mathsf{V}_{i,x} S_{j,x'} :}
  \times
  \begin{cases}
   \displaystyle
   \left( \frac{x'}{x};q_2 \right)_\infty^{-1} & (i = j) \\
   1 & (i \neq j)
  \end{cases}
 \end{align}
\end{subequations}
Therefore, the fundamental and antifundamental contributions to the partition function, presented in~\eqref{eq:quiv_full_func}, are realized by the $\mathsf{V}$-operator insertion:
\begin{subequations}
\begin{align}
 Z_i^\text{f}[\mathcal{X}]
 & =
 \bra{0}
 {: \prod_{x \in \mathcal{X}}^\succ S_{\mathsf{i}(x),x} :}
 \
 {: \prod_{x \in \mathcal{M}_i} \mathsf{V}_{i, x} :}
 \ket{0}
 \, , \\
 Z_i^\text{af}[\mathcal{X}]
 & =
 \bra{0}
 {: \prod_{x \in \widetilde{ \mathcal{M} }_i } \mathsf{V}_{i,q^{-1} x} :}
 \
 {: \prod_{x \in \mathcal{X}}^\succ S_{\mathsf{i}(x),x} :}
 \ket{0}
 \, .
\end{align}
\end{subequations}
The $t$-extended partition function and the corresponding $Z$-state are then given by
\begin{subequations}\label{eq:Z_state_matter}
\begin{align}
 Z(t) & =
 {: \prod_{x \in \widetilde{ \mathcal{M} } } \mathsf{V}_{\mathsf{i}(x),q^{-1} x} :}
 \left( \prod_{x \in \mathring{\mathcal{X}}}^\succ \mathsf{S}_{\mathsf{i}(x),x} \right)
 {: \prod_{x \in \mathcal{M}} \mathsf{V}_{\mathsf{i}(x),x} :}
 \, , \\
 \ket{Z} = Z(t) \ket{0} & = 
 {: \prod_{x \in \widetilde{ \mathcal{M} } } \mathsf{V}_{\mathsf{i}(x),q^{-1} x} :}
 \left( \prod_{x \in \mathring{\mathcal{X}}}^\succ \mathsf{S}_{\mathsf{i}(x),x} \right)
 {: \prod_{x \in \mathcal{M}} \mathsf{V}_{\mathsf{i}(x),x} :}
 \ket{0}
 \, ,
\end{align}
\end{subequations}
where we similarly apply the map $\mathsf{i}(x) = i$ for $x \in \mathcal{M}_i$, $\widetilde{\mathcal{M}}_i$, and we obtain the chiral correlator representation for the instanton partition function
\begin{align}
 Z =
 \bra{0}
 {: \prod_{x \in \widetilde{ \mathcal{M} } } \mathsf{V}_{\mathsf{i}(x),q^{-1} x} :}
 \left( \prod_{x \in \mathring{\mathcal{X}}}^\succ \mathsf{S}_{\mathsf{i}(x),x} \right)
 {: \prod_{x \in \mathcal{M}} \mathsf{V}_{\mathsf{i}(x),x} :}
 \ket{0}
 =:
 \bra{\mathsf{V}}
 \prod_{x \in \mathring{\mathcal{X}}}^\succ \mathsf{S}_{\mathsf{i}(x),x}
 \ket{\mathsf{V}}
 \, ,
\end{align}
where we define the $\mathsf{V}$-states
\begin{align}
 \bra{\mathsf{V}} =
 \bra{0}
 {: \prod_{x \in \widetilde{ \mathcal{M} } } \mathsf{V}_{\mathsf{i}(x),q^{-1} x} :}
 \, , \qquad
 \ket{\mathsf{V}} =
 {: \prod_{x \in \mathcal{M}} \mathsf{V}_{\mathsf{i}(x),x} :}
 \ket{0}
 \, .
 \label{eq:V_state}
\end{align}
This expression contains a factor coming from the OPE between the $\mathsf{V}$-operators on the left and on the right, but we just omit such a factor since it is independent of the instanton configuration $\mathcal{X}$, and does not affect the expectation values.

\subsubsection{$t$-shift operator}

The expression of the (positive part of) $v$-modes~\eqref{eq:v_oscillator} implies that the $\mathsf{V}$-operator plays a role of the $t$-shift operator:
\begin{align}
 \mathsf{V}_{i,x}: \quad
 t_{i,n}
 \ \longmapsto \
 t_{i,n} + \frac{x^n}{n(1 - q_1^{n})(1 - q_2^{n})}
 \qquad
 (n \ge 1)
 \, .
\end{align}
Therefore, the (anti)fundamental matter contribution is obtained as a specific background of the $t$-variables.
A similar discussion is found in the context of topological string:
The $t$-dependent partition function corresponds to the open string amplitude, which behaves as a wavefunction, a state in the corresponding Hilbert space, and deformation of the $t$-variables are induced by the vertex operator.
See, for example, \cite{Aganagic:2003qj} for details.

\subsection{Boundary degrees of freedom}\label{sec:boundary_dof}

As discussed above, we use the $\mathsf{V}$-state $\bra{\mathsf{V}}$ to incorporate the fundamental matter, while the dual $\mathsf{V}$-state $\ket{\mathsf{V}}$ is used for the antifundamental matter.
This is related to the discrepancy between the fundamental and the antifundamental matter contributions to the partition function in 5d gauge theory (the K-theory convention).

From the vertex operator point of view, the fundamental matter contribution corresponds to the marked point around $x = 0$, while the antifundamental matter is around $x = \infty$ for $x \in \mathbb{C}^\times$.
Namely, we apply the radial quantization with the identification of the vacuum $\ket{0}$ and its dual $\bra{0}$ with $x = 0$ and $x = \infty$, respectively.
(Recall that the $Z$-state is defined with the radial ordering~\eqref{eq:radial_ordering_prod}.)\index{radial ordering}
In order to see more details of this structure, we compare the fundamental and the antifundamental contributions to the partition function~\eqref{eq:quiv_full_func} with $n_i^\text{f} = n_i^\text{af} = 1$ for simplicity:
\begin{align}
 \frac{Z_i^\text{af}}{Z_i^\text{f}}
 & =
 \prod_{x \in \mathcal{X}_i} 
 \left( q_2 \frac{\widetilde{\mu}}{x} ;q_2 \right)_\infty^{-1}
 \left( \frac{x}{\mu q^{-1}} ;q_2 \right)_\infty^{-1}
 \ \xrightarrow{\mu q^{-1} = \widetilde{\mu}} \
 \prod_{x \in \mathcal{X}_i} 
 \theta \left( \frac{x}{\mu q^{-1}} ;q_2 \right)^{-1}
 \label{eq:Zaf/Zf}
\end{align}
where $\theta(z;p)$ is the theta function defined in~\S\ref{sec:theta_fn}.
Since this is an infinite product of the theta functions, we should regularize it as follows.

\subsubsection{Boundary 4d theory on $S^3 \times S^1 = \partial(\mathbb{C}^2 \times S^1)$}

The first option is to subtract the perturbative part.
For this purpose, we focus on the trivial configuration $\mathring{\mathcal{X}}_i$, and put $n_i = 1$ again for simplicity.
Then, the infinite product becomes
\begin{align}
 \prod_{x \in \mathring{\mathcal{X}}_i} 
 \theta \left( \frac{x}{\mu q^{-1}} ;q_2 \right)^{-1}
 & =
 \prod_{k=1}^\infty 
 \theta \left( \frac{\np^{\mathsf{a}_{i,1}} q_1^{k-1}}{\mu q^{-1}} ;q_2 \right)^{-1} 
 \nonumber \\
 & =
 \left( \xi;q_1,q_2 \right)_\infty^{-1} \left( q_2 \xi^{-1};q_1^{-1},q_2 \right)_\infty^{-1}
 \nonumber \\
 & =
 \frac{\left( q \xi^{-1};q_1,q_2 \right)_\infty}{\left( \xi;q_1,q_2 \right)_\infty}
 = \Gamma_e(\xi;q_1,q_2)
 \, ,
\end{align}
where $\xi = \np^{\mathsf{a}_{i,1}}/\mu q^{-1}$.
In order to obtain the third line from the second line, we apply the analytic continuation since the multiple $q$-shifted factorial~\eqref{eq:multi_q-shift} is defined for $|q_1| < 1$.
$\Gamma_e(z;p,q)$ is the elliptic gamma function defined in \S\ref{sec:e_gamma}.
This elliptic gamma function is interpreted as 4d $\mathcal{N} = 1$ chiral multiplet contribution to the superconformal index~\cite{Romelsberger:2005eg,Kinney:2005ej} in terms of the elliptic gamma function~\cite{Dolan:2008qi}.
See also a recent review~\cite{Gadde:2020yah}.
The superconformal index is evaluated with the path integral on $S^3 \times S^1$, which is interpreted as the boundary of $\mathbb{C}^2 \times S^1$.
Hence, the fundamental and the antifundamental matter contributions on $\mathbb{C}^2 \times S^1$ are converted to each other by the boundary degrees of freedom on $S^3 \times S^1 = \partial (\mathbb{C}^2 \times S^1)$.

\subsubsection{Boundary 2d theory on $T^2 = S^1 \times S^1 = \partial(\mathbb{C} \times S^1) \subset  \partial(\mathbb{C}^2 \times S^1)$}

The other option is to impose the Higgsing condition to truncate the infinite product as discussed in \S\ref{sec:Higgs_truncation}.
Then, the instanton partition function for 5d $\mathcal{N} = 1$ theory on $\mathbb{C}^2 \times S^1$ is reduced to the vortex partition function for 3d $\mathcal{N} = 2$ theory on $\mathbb{C} \times S^1$.
In this context, the finite version of the product~\eqref{eq:Zaf/Zf} is interpreted as the $\mathcal{N} = (0,2)$ chiral multiplet contribution to the elliptic genus on $S^1 \times S^1 = \partial \left( \mathbb{C} \times S^1 \right)$~\cite{Benini:2013xpa}.
We similarly obtain the $\mathcal{N} = (0,2)$ Fermi multiplet contribution from $Z_i^\text{f}/Z_i^\text{af}$.
We can again convert the fundamental and the antifundamental matter contributions by the boundary degrees of freedom.
See \cite{Honda:2013uca,Hori:2013ika,Yoshida:2014ssa} for a similar argument.

\subsection{$\mathsf{Y}$-operator: observable generator}\label{sec:Y_op}

We discuss how to realize the $\mathsf{Y}$-function in the operator formalism.
Recalling the expression of the $\mathsf{Y}$-function in terms of the chiral ring operators~\eqref{eq:Y_func_ch_gen}, we define the $\mathsf{Y}$-operator as follows:\index{Y-operator@$\mathsf{Y}$-operator}
\begin{align}
 \mathsf{Y}_{i,x} 
 = q_1^{\tilde{\rho}_i}
 {: \exp \left( \sum_{n \in \mathbb{Z}} y_{i,n} \, x^{-n} \right) :}
\end{align}
with the component of the Weyl vector in the basis of simple roots denoted by
\begin{align}
 \tilde{\rho}_i = \sum_{j \in \Gamma_0} \tilde{c}_{ji}^{[0]}
 \, .
\end{align}
If the quiver is of affine type $(\det c^{[0]} = 0)$, we put $\tilde{\rho}_i = 0$.
The operators $(y_{i,n})_{i \in \Gamma_0, n \in \mathbb{Z}}$ are defined as
\begin{align}
 y_{i,-n} = (1 - q_1^n)(1 - q_2^n) \, \tilde{c}_{ji}^{[-n]} \, t_{j,n}
 \, , \quad
 y_{i,0} = - \log q_2 \, \tilde{c}_{ji}^{[0]} \, t_{j,0}
 \, , \quad
 y_{i,n} = - \frac{1}{n} \partial_{i,n}
 \quad
 (n \ge 1)
 \, ,
 \label{eq:y_oscillators}
\end{align}
with the commutation relation
\begin{align}
 \left[ y_{i,n} \, , y_{j,n'} \right]
 = - \frac{1}{n} (1 - q_1^n)(1 - q_2^n) \, \tilde{c}_{ji}^{[-n]} \, \delta_{n+n', 0}
 \, .
 \label{eq:y_comm_rel}
\end{align}
The relation between the $y$-mode and the $v$-mode is given as follows:
\begin{align}
 y_{i,n} = - (1 - q_1^n)(1 - q_2^n) \, v_{i,n}
 \, .
 \label{eq:y_v_relation}
\end{align}
We remark that, compared to the $\mathsf{Y}$-function~\eqref{eq:Y_func_ch_gen}, one can only fix the positive modes $(y_{i,n})_{n \ge 1}$ in the $\mathsf{Y}$-operator.
The negative modes $(y_{i,-n})_{n \ge 1}$ and the zero modes are chosen to be consistent with the commutation relations to other oscillator modes.

\subsubsection{Relation between the operators $\mathsf{Y}$ and $S$}

The commutation relation between the $y$-mode and the $s$-mode is given by
\begin{align}
 \left[ y_{i,n} \,,\, s_{j,n'} \right]
 = - \frac{1}{n} ( 1 - q_1^n) \, \delta_{ij} \, \delta_{n+n',0}
 \, , \qquad
 \left[ \tilde{s}_{i,0} \,,\, y_{j,n} \right]
 = - \log q_1 \, \delta_{ij} \, \delta_{n,0}
 \, , 
\end{align}
which gives rise to the OPE between the $\mathsf{Y}$-operator and the screening current. \index{operator product expansion (OPE)!S and Y@$S$ and $\mathsf{Y}$}
\begin{subequations}\label{eq:YS_OPE}
\begin{align}
 \mathsf{Y}_{i,x} S_{j,x'}
 & =
 {: \mathsf{Y}_{i,x} S_{j,x'} :}
 \times
 \begin{cases}
  \displaystyle
  \frac{1 - x'/x}{1 - q_1 x'/x} & (i = j) \\
  1 & (i \neq j)
 \end{cases}
 \\
 S_{j,x'} \mathsf{Y}_{i,x} 
 & =
 {: S_{j,x'} \mathsf{Y}_{i,x}  :}
 \times
 \begin{cases}
  \displaystyle
  q_1^{-1} \frac{1 - x/x'}{1 - q_1^{-1} x/x'} & (i = j) \\
  1 & (i \neq j)
 \end{cases} 
\end{align}
\end{subequations}
Naively speaking, these OPE factors coincide, $\displaystyle q_1^{-1} \frac{1 - x/x'}{1 - q_1^{-1} x/x'} = \frac{1 - x'/x}{1 - q_1 x'/x}$, but this is true except at the pole $q_1 x'/x = 1$.
In fact, we obtain a nontrivial commutation relation between the $\mathsf{Y}$-operator and the screening current, 
\begin{align}
 \left[
 \mathsf{Y}_{i,x}
 \,,\,
 S_{j,x'}
 \right]
 =
 \begin{cases}
  \displaystyle
  (1 - q_1^{-1}) \, \delta\left( q_1 \frac{x'}{x} \right) \,
  {:  \mathsf{Y}_{i,x} S_{j,x'} :}
  & ( i = j) \\
  0 & (i \neq j)
 \end{cases}
 \label{eq:YS_commutator}
\end{align}
where $\delta(z)$ is the multiplicative delta function~\eqref{eq:mult_delta_fn}.

From the OPE factors~\eqref{eq:YS_OPE}, we obtain
\begin{subequations}
\begin{align}
 \bra{0}
 \mathsf{Y}_{i,x} \,
 {: \prod_{x \in \mathcal{X}}^\succ S_{\mathsf{i}(x),x} :}
 \ket{0}
 & =
 q_1^{\tilde{\rho}_i}
 \prod_{x' \in \mathcal{X}_i}
 \frac{1 - x'/x}{1 - q_1x'/x}
 \bra{0}
 {: \prod_{x \in \mathcal{X}}^\succ S_{\mathsf{i}(x),x} :}
 \ket{0}
 \, ,
 \\
 \bra{0}
 {: \prod_{x \in \mathcal{X}}^\succ S_{\mathsf{i}(x),x} :}
 \,
 \mathsf{Y}_{i,x} 
 \ket{0}
 & =
 q_1^{\tilde{\rho}_i}
 \prod_{x' \in \mathcal{X}_i}
 q_1^{-1}
 \frac{1 - x/x'}{1 - q_1^{-1} x/x'} 
 \bra{0}
 {: \prod_{x \in \mathcal{X}}^\succ S_{\mathsf{i}(x),x} :}
 \ket{0}
 \, ,
\end{align}
\end{subequations}
which correspond to the $\mathsf{Y}$-functions, $\mathsf{Y}_{i,x}[\mathcal{X}]$ and $\mathsf{Y}_{i,x}[\mathcal{X}]^\vee$, respectively, although we have to be careful of the identification of the latter case:
The identification with the dual $\mathsf{Y}$-function, $\mathsf{Y}_{i,x}^\vee$, is up to the factor $q_1^{-1}$ inside the infinite product.

The gauge theory average of the $\mathsf{Y}$-function has a chiral correlator expression as follows:
\begin{align}
 \VEV{\mathsf{Y}_{i,x}}
 =
 \bra{\mathsf{V}}
 \mathsf{Y}_{i,x}
 \prod_{x \in \mathring{\mathcal{X}}}^\succ \mathsf{S}_{\mathsf{i}(x),x}
 \ket{\mathsf{V}}
 /
 \bra{\mathsf{V}}
 \prod_{x \in \mathring{\mathcal{X}}}^\succ \mathsf{S}_{\mathsf{i}(x),x}
 \ket{\mathsf{V}} 
 \, .
\end{align}
The expression of the vev of the $\mathsf{Y}$-function as the ratio of the correlators implies an analogy with the Baker--Akhiezer function of the integrable hierarchy under the identification of the partition function as the corresponding $\tau$-function. \index{Baker--Akhiezer function}%
See also \S\ref{sec:BA_func}.
In fact, as shown in \S\ref{sec:NS_integrability}, the $\mathsf{Y}$-function average plays a role of the wave function in the context of the quantum Seiberg--Witten curve in the NS limit.

\subsection{$\mathsf{A}$-operator: iWeyl reflection generator}\label{sec:A_op}

As discussed in \S\ref{sec:qq-ch}, the adding-instanton operator is concisely described in terms of the $\mathsf{A}$-function~\eqref{eq:A_fn_def}.
We define the operator analog of the $\mathsf{A}$-function, that we call the $\mathsf{A}$-operator:\index{A-operator@$\mathsf{A}$-operator}
\begin{align}
 \mathsf{A}_{i,x}
 = q_1
 {: \exp \left( \sum_{n \in \mathbb{Z}} a_{i,n} \, x^{-n} \right) :}
 \, .
\end{align}
The $a$-mode is defined from the $y$-mode
\begin{align}
 a_{i,n} = \sum_{j \in \Gamma_0} y_{j,n} \, c_{ji}^{[n]}
 \, .
\end{align}
From the representation theoretical point of view on the quiver, the operators $\mathsf{Y}_{i,x}$ and $\mathsf{A}_{i,x}$ correspond to the fundamental weight and the simple root associated with the node $i \in \Gamma_0$, which are converted to each other by the quiver Cartan matrix.

\subsubsection{OPE factors}

The $a$-modes are explicitly written as follows:
\begin{align}
 a_{i,-n} = (1 - q_1^n)(1 - q_2^n) \, t_{i,n}
 \, , \quad
 a_{i,0} = - \log q_2 \, t_{i,0}
 \, , \quad
 a_{i,n} = - \frac{1}{n} c_{ji}^{[n]} \, \partial_{j,n}
 \quad
 (n \ge 1)
 \, ,
\end{align}
with the commutation relation
\begin{align}
 \left[ a_{i,n} \, , a_{j,n'} \right]
 = - \frac{1}{n} (1 - q_1^n)(1 - q_2^n) \, c_{ji}^{[n]} \, \delta_{n+n', 0}
 \, .
 \label{eq:a_comm_rel}
\end{align}
The OPE between the $\mathsf{A}$-operators is then given by \index{operator product expansion (OPE)!A and A@$\mathsf{A}$ and $\mathsf{A}$}
\begin{align}
  \mathsf{A}_{i,x} \mathsf{A}_{j,x'}
  =
  {:\mathsf{A}_{i,x} \mathsf{A}_{j,x'}:} \times
  \begin{cases}
   \displaystyle
   \mathscr{S}\left(\frac{x'}{x}\right)^{-1}
   \mathscr{S}\left(q^{-1}\frac{x'}{x}\right)^{-1}
   & (i = j) \\
   \displaystyle
   \mathscr{S}\left(\mu_e q^{-1} \frac{x'}{x}\right)
   & (e: i \to j) \\
   \displaystyle
   \mathscr{S}\left(\mu_e^{-1} \frac{x'}{x}\right)
   & (e: j \to i) \\
   1 & (\text{otherwise})
  \end{cases}
 \label{eq:AA_OPE}
 \end{align}

The OPEs with other operators are similarly computed.
From the commutation relations
\begin{subequations}
 \begin{align}
  [y_{i,n} \,,\, a_{j,n'}]
  & =
  - \frac{1}{n} (1 - q_1^n)(1 - q_2^n) \, \delta_{ij} \, \delta_{n+n',0}
  \, , \\
  [v_{i,n} \,,\, a_{j,n'}]
  & = \frac{1}{n} \, \delta_{ij} \, \delta_{n+n',0}
  \, ,
 \end{align}
\end{subequations}
we obtain the OPE factors as follows:
\index{operator product expansion (OPE)!A and Y@$\mathsf{A}$ and $\mathsf{Y}$}
\index{operator product expansion (OPE)!A and V@$\mathsf{A}$ and $\mathsf{V}$}
\begin{subequations}\label{eq:AY_AV_OPE}
  \begin{align}
   \mathsf{Y}_{i,x} \mathsf{A}_{j,x'}
   =
   \mathscr{S} \left( \frac{x'}{x} \right)^{-\delta_{ij}} \,
   {:\mathsf{Y}_{i,x} \mathsf{A}_{j,x'}:}
   \, , \quad
   &
   \mathsf{A}_{j,x'} \mathsf{Y}_{i,x} 
   =
   \mathscr{S} \left( q^{-1} \frac{x}{x'} \right)^{-\delta_{ij}} \,
   {: \mathsf{A}_{j,x'} \mathsf{Y}_{i,x}:}
   \, , \\
   \mathsf{V}_{i,x} \mathsf{A}_{j,x'}
   =
   \left(1 - \frac{x'}{x} \right)^{-\delta_{ij}} \,
   {:\mathsf{V}_{i,x} \mathsf{A}_{j,x'}:}
   \, , \quad
   &
   \mathsf{A}_{j,x'} \mathsf{V}_{i,x} 
   =
   \left(1 - \frac{x}{x'} \right)^{-\delta_{ij}} \,
   {:\mathsf{A}_{j,x'} \mathsf{V}_{i,x} :}
   \, .
  \end{align}
\end{subequations}

\subsubsection{iWeyl reflection}
\index{iWeyl reflection}

Writing the $\mathsf{A}$-operator in terms of the $\mathsf{Y}$-operators, we obtain the consistent expression with the $\mathsf{A}$-function~\eqref{eq:A_fn_def}, up to the discrepancy between $\mathsf{Y}$ and $\mathsf{Y}^\vee$,
\begin{align}
 \mathsf{A}_{i,x} = {:
 \mathsf{Y}_{i,x} \mathsf{Y}_{i,q x}
 \prod_{e:i \to j} \mathsf{Y}_{j,\mu_e^{-1} q x}^{-1}
 \prod_{e:j \to i} \mathsf{Y}_{j,\mu_e x}^{-1}
 :}
 \, ,
\end{align}
which is the combination appearing in the iWeyl reflection~\eqref{eq:saddle_pt_finite}.
In addition, compared to the $s$-mode~\eqref{eq:s_oscillator}, we have 
\begin{align}
 a_{i,n} = (1 - q_2^{-n}) \, s_{i,n}
 \, ,
\end{align}
which leads to the relation between the $\mathsf{A}$-operator and the screening current,%
\footnote{%
From this relation, we may also incorporate another zero mode in the $\mathsf{A}$-operator (and also $\mathsf{Y}$-operator), $\mathsf{A}_{i,x} \to x^{\kappa_i} \, \mathsf{A}_{i,x}$, which corresponds to the Chern--Simons term.
}
\begin{align}
 \mathsf{A}_{i,x} = q_1 \, {: S_{i,x} S_{i,q_2 x}^{-1} :}
 \, .
 \label{eq:AbyS}
\end{align}
This is also consistent because, from \eqref{eq:Z[X]_correlator}, the adding-instanton schematically corresponds to inserting the $\mathsf{A}$-operator,
\begin{align}
 \frac{Z_{\mathcal{X}'}}{Z_{\mathcal{X}}}
 \ \implies \
 \bra{0}
 {: S_{i,x}^{-1} S_{i,q_2 x} :}
 \prod_{x \in \mathcal{X}}^\succ S_{\mathsf{i}(x),x}
 \ket{0}
 =
 \bra{0}
 \mathsf{A}_{i,x}^{-1}
 \prod_{x \in \mathcal{X}}^\succ S_{\mathsf{i}(x),x}
 \ket{0}
 \, .
\end{align}
More precise statement is addressed in the following.

\section{Pole cancellation mechanism}

In order to discuss the $qq$-character in the operator formalism, we discuss the pole cancellation mechanism with the vertex operators.
We start with the relation between the operators $\mathsf{A}_i$ and $S_j$ for $i = j$,
\begin{subequations}
\begin{align}
 \left[
 a_{i,n} \,,\, s_{j,n'}
 \right]
 & = - \frac{1}{n} (1 - q_1^n) \, c_{ji}^{[n]} \, \delta_{n+n',0}
 \, , \\
 \left[
 s_{i,n} \,,\, a_{j,n'}
 \right]
 & = - \frac{1}{n} (1 - q_1^{-n}) \, c_{ij}^{[-n]} \, \delta_{n+n',0} 
 \, , \\
 \left[
 \tilde{s}_{i,0} \,,\, a_{j,n}
 \right]
 & = 
 - \log q_1 \, c_{ji}^{[0]} \, \delta_{n,0}
 \, ,
\end{align}
\end{subequations}
which, for $i = j$, gives rise to \index{operator product expansion (OPE)!A and S@$\mathsf{A}$ and $S$}
\begin{subequations}
\begin{align}
 \mathsf{A}_{i,x} S_{i,x'}
 & =
 \frac{1 - x'/x}{1 - q_1 x'/x} \frac{1 - q^{-1} x'/x}{1 - q_2^{-1} x'/x} \,
 {: \mathsf{A}_{i,x} S_{i,x'} :}
 \, , \\
 S_{i,x'} \mathsf{A}_{i,x} 
 & =
 q_1^{-2}
 \frac{1 - x/x'}{1 - q_1^{-1} x/x'} \frac{1 - q x/x'}{1 - q_2 x/x'} \,
 {:  S_{i,x'} \mathsf{A}_{i,x} :}
 \, .
\end{align}
\end{subequations}
Then, we can show the pole cancellation in the following combination:
\begin{align}
 \res_{x' \to x}
 \left[
 \mathsf{Y}_{i,qx} S_{i,q_2 x'} +
 {: \mathsf{Y}_{i,qx} \mathsf{A}_{i,x}^{-1} :} \, S_{i,x'}
 \right]
 = 0
 \, ,
\end{align}
because
\begin{subequations}
\begin{align}
 \mathsf{Y}_{i,qx} S_{i,q_2 x'}
 & = \frac{1 - q_1^{-1} x'/x}{1 - x'/x} \,
 {: \mathsf{Y}_{i,qx} S_{i,q_2 x'} :}
 \, , \\
 {: \mathsf{Y}_{i,qx} \mathsf{A}_{i,x}^{-1} :} \, S_{i,x'}
 & =
 \frac{1 - q^{-1} x'/x}{1 - q_2^{-1} x'/x}
 \frac{1 - q_1 x'/x}{1 - x'/x}
 \frac{1 - q_2^{-1} x'/x}{1 - q^{-1} x'/x}
 \,
 {:\mathsf{Y}_{i,qx} \mathsf{A}_{i,x}^{-1} S_{i,x'}:}
 \nonumber \\
 & \stackrel{\eqref{eq:AbyS}}{=}
 q_1^{-1}
 \frac{1 - q_1 x'/x}{1 - x'/x} \,
 {: \mathsf{Y}_{i,qx} S_{i,q_2 x'} :}
 \, .
\end{align}
\end{subequations}
This pole cancellation is rephrased in more algebraic language.
From the commutation relations,
\begin{subequations}
 \begin{align}
  \left[
  \mathsf{Y}_{i,qx} \,,\, S_{i,x'}
  \right]
  & =
  (1 - q_1^{-1}) \, \delta\left( q_2^{-1} \frac{x'}{x} \right) \,
  {: \mathsf{Y}_{i,qx} S_{i,x'} :}
  \, , \\
  \left[
  {: \mathsf{Y}_{i,qx} \mathsf{A}_{i,x}^{-1} :}
  \,,\,
  S_{i,x'}
  \right]
  & =
  (1 - q_1) \, \delta\left( \frac{x'}{x} \right) \,
  {: \mathsf{Y}_{i,qx} \mathsf{A}_{i,x}^{-1} S_{i,x'} :}
  = - (1 - q_1^{-1}) \, \delta\left( \frac{x'}{x} \right) \,
  {: \mathsf{Y}_{i,qx} S_{i,q_2 x'} :}
  \, ,
 \end{align}
\end{subequations}
we obtain
\begin{align}
 &
 \left[
 \mathsf{Y}_{i,qx} + {: \mathsf{Y}_{i,qx} \mathsf{A}_{i,x}^{-1} :}
 \, , \,
 S_{i,x'}
 \right]
 \nonumber \\
 & = (1 - q_1^{-1}) \,
 \Bigg(
 \underbrace{
 \delta\left( q_2^{-1} \frac{x'}{x} \right) \, {: \mathsf{Y}_{i,qx} S_{i,x'} :}
 - \delta\left( \frac{x'}{x} \right) \,
  {: \mathsf{Y}_{i,qx} S_{i,q_2 x'} :}
 }_{\text{total $q_2$-difference for $x'$}}
 \Bigg)
 \, .
\end{align}
Since the right hand side is written as a total $q_2$-difference for the variable $x'$, it will vanish after the $q_2$-shifted sum, which means replacing the screening current $S_{i,x'}$ with the screening charge $\mathsf{S}_{i,x'}$.
Namely, the pole cancellation is equivalent to the commuting relation (kernel condition) to the screening charge.
After the iWeyl reflection, there may be another delta function from the second term, and we should apply another reflection to cancel the new singularity, similarly to the argument in \S\ref{sec:qq-ch}.

\chapter{Quiver W-algebra}\label{chap:quiv_W}

We have shown that the instanton partition function has a realization as a chiral correlation function of the vertex operators, whose algebraic structure depends on the quiver structure of gauge theory.
In this Chapter, we discuss the construction of the underlying vertex operator algebra, that we call the {\em quiver W-algebra}~\cite{Kimura:2015rgi}, and show that an operator uplift of the $qq$-character plays a role of the generating current of the W-algebra.
\index{quiver!---W-algebra}
We will see that, applying this formalism to the fractional quiver theory discussed in \S\ref{sec:fractional_quiver}, one can construct the quantum W-algebras associated with the non-simply-laced algebra~\cite{Kimura:2017hez}.
In addition, the formalism of quiver W-algebra is also applicable to affine quivers, which lead to a new family of W-algebras.
We will also discuss that the vertex operators introduced in this context will be utilized to express the contour integral formulas associated with integration over the instanton moduli spaces and quiver varieties.

\section{$\mathsf{T}$-operator: generating current}

Applying the pole cancellation argument recursively, we can construct the $\mathsf{T}$-operator, an operator analog of the $qq$-character, which commutes with the screening charge:\index{T-operator@$\mathsf{T}$-operator|see{$qq$-character}}
\begin{align}
 \left[
 \mathsf{T}_{i,x}
 \,,\,
 \mathsf{S}_{j,x'}
 \right]
 = 0
 \, ,
\end{align}
where the operator $\mathsf{T}_{i,x}$ is associated with the highest weight $\mathsf{Y}_{i,x}$ for each node in the quiver,
\begin{align}
 \mathsf{T}_{i,x}
 =
 \mathsf{Y}_{i,x} +
 {:\mathsf{Y}_{i,x} \mathsf{A}_{i,q^{-1} x}^{-1}:}
 + \cdots
 \, .
\end{align}
Since the operator $\mathsf{T}_{i,x}$ is a commutant of the screening charges $(\mathsf{S}_{i,x})_{i \in \Gamma_0}$, it defines a holomorphic conserved current
\begin{align}
 \partial_{\bar{x}} \mathsf{T}_{i,x} \ket{Z} = 0
 \, ,
\end{align}
with a well-defined Fourier mode expansion
\begin{align}
 \mathsf{T}_{i,x} = \sum_{n \in \mathbb{Z}} T_{i,n} \, x^{-n}
 \, .
\end{align}
Then, we arrive at the following definition:
\begin{itembox}{Quiver W-algebra~\cite{Kimura:2015rgi}}
 We define the $\mathsf{T}$-operator $(\mathsf{T}_{i,x})_{i \in \Gamma_0}$ associated with each node of quiver, $i \in \Gamma_0$, which is the operator analog of the fundamental $qq$-character.
 Then, the conserved Fourier modes of the holomorphic current $(T_{i,n})_{i\in \Gamma_0, n \in \mathbb{Z}}$ define the {\em quiver W-algebra} $\mathrm{W}_{q_{1,2}}(\mathfrak{g}_\Gamma)$ (W$(\Gamma)$ for short) as a subalgebra of the Heisenberg algebra $\mathscr{H}$.
  Namely, the $\mathsf{T}$-operator is the generating current of the quiver W-algebra.
\end{itembox}
In fact, the algebra W$_{q_{1,2}}(\Gamma)$ agrees with the $q$-deformation of the Virasoro/W-algebra for $\Gamma = ADE$~\cite{Shiraishi:1995rp,Awata:1995zk,Feigin:1995sf,Frenkel:1997}, which is reduced to the ordinary (rational; differential; additive) version of W-algebras, W$_\beta(\Gamma)$ with $\beta = - \epsilon_1/\epsilon_2$.%
\footnote{%
This reduction is obtained through the same limit as that from 5d to 4d theory discussed in \S\ref{sec:5dN=1}.
}
This formalism is also applicable beyond the finite-type quiver, affine and hyperbolic quivers, which gives rise to a new family of W-algebras.
For $\Gamma \neq ADE$, we obtain a similar statement based on the fractional quiver formalism~\cite{Kimura:2017hez}.
See \S\ref{sec:frac_W} for details.

\section{Classical limit: quantum integrability}

The vertex operators used in the operator formalism are noncommutative operators acting on the Fock space, and thus the $\mathsf{T}$-operator is also a noncommutative current in general.
Regarding the $y$- and $a$-mode commutators, \eqref{eq:y_comm_rel} and \eqref{eq:a_comm_rel} (similarly \eqref{eq:y_comm_rel_frac} and \eqref{eq:a_comm_rel_frac}), they have a factor $(1 - q_1^n)(1 - q_2^n)$, which vanishes in the NS limit, $q_2 \to 1$ (or $q_1 \to 1$).
As a result, the $\mathsf{T}$-operator becomes commutative in the NS limit, and the W-algebra is reduced to a classical Poisson algebra.
Such a commuting current can be identified with the transfer matrix of the quantum integrable system
\begin{align}
 [\mathsf{T}_{i,x} \,,\, \mathsf{T}_{j,x'}] = 0
 \, .
\end{align}
In this way, we obtain the quantum integrable system in the NS limit from the operator formalism point of view.

There is another interesting aspect of the W-algebras in the NS limit.
It has been known that there is an isomorphism between the W-algebra W$_{q_{1,2}}(\mathfrak{g}_\Gamma)$ in the classical limit, where the algebra becomes commutative algebra, and the K-theory ring of the category of representations of the quantum loop group~\cite{Frenkel:1998}.
This is interpreted as a consequence of the geometric $q$-Langlands correspondence, and is promoted to its quantum version with generic $q_{1,2}$~\cite{Frenkel:2010wd,Aganagic:2017smx}.
See also a review article~\cite{Frenkel:2005pa} on this topic.

\section{Examples}

\subsection{$A_1$ quiver}\label{sec:W_A1}

We consider $A_1$ quiver as a primary example to demonstrate the formalism of the quiver W-algebra in detail.
Since it consists of a single node, there exists a single $\mathsf{T}$-operator,\index{qq-character@$qq$-character!A1@$A_1$}
\begin{align}
 \mathsf{T}_{1,x} = \mathsf{Y}_{1,x} + \mathsf{Y}_{1,q^{-1}x}^{-1}
 \, ,
\end{align}
with the mode expansion
\begin{align}
 \mathsf{T}_{1,x} = \sum_{n \in \mathbb{Z}} T_{1,n} \, x^{-n}
 \, .
\end{align}
In order to characterize the algebraic relation of the modes $(T_{1,n})_{n \in \mathbb{Z}}$, we evaluate the OPE of the $\mathsf{T}$-operator.
The computation is in fact equivalent to that for the degree-two $qq$-character,
\begin{align}
 \mathsf{T}_{1,x} \mathsf{T}_{1,x'}
 & = f\left( \frac{x'}{x} \right)^{-1}
 \Bigg(
 {: \mathsf{Y}_{1,x} \mathsf{Y}_{1,x'} :}
 \nonumber \\
 & \hspace{8em}
 +
 \mathscr{S} \left( \frac{x'}{x} \right)
 {: \mathsf{Y}_{1,q^{-1} x}^{-1} \mathsf{Y}_{1,x'} :}
 +
 \mathscr{S} \left( \frac{x}{x'} \right)
 {: \mathsf{Y}_{1,x} \mathsf{Y}_{1,q^{-1} x'}^{-1} :}
 \nonumber \\
 & \hspace{20em}
 +
 {: \mathsf{Y}_{1,q^{-1} x}^{-1} \mathsf{Y}_{1,q^{-1} x'}^{-1} :}
 \Bigg)
\end{align}
where we define the structure function
\begin{align}
 f(z) = \exp \left(
 \sum_{n = 1}^\infty \frac{1}{n} \frac{(1 - q_1^n)(1 - q_2^n)}{1 + q^n} \, z^n
 \right)
 = \sum_{k = 0}^\infty f_k \, z^k
 \, .
\end{align}
This structure function is obtained from the OPE between the $\mathsf{Y}$-operators, so that $f(x'/x) \mathsf{T}_{1,x} \mathsf{T}_{1,x'}$ agrees with the degree-two $qq$-character $\mathsf{T}_{(2),(x,x')}$.
We also remark the relation
\begin{align}
 f(z) f(qz) = \mathscr{S}(z)
 \, .
\end{align}
Then, due to the identity of the $\mathscr{S}$-function~\eqref{eq:S_func_delta}, we obtain the OPE between the $\mathsf{T}$-operators,\index{operator product expansion (OPE)!T and T (A1)@$\mathsf{T}$ and $\mathsf{T}$ ($A_1$)}
\begin{align}
 \mathsf{T}_{(2),(x,x')} - \mathsf{T}_{(2),(x',x)}
 & =
 f\left( \frac{x'}{x} \right) \mathsf{T}_{1,x} \mathsf{T}_{1,x'}
 - f\left( \frac{x}{x'} \right) \mathsf{T}_{1,x'} \mathsf{T}_{1,x}
 \nonumber \\
 & =
 - \frac{(1 - q_1)(1 - q_2)}{1 - q}
 \left( \delta \left(q \frac{x'}{x} \right) - \delta \left(q^{-1} \frac{x'}{x} \right)\right)
 \, ,
 \label{eq:TT_OPE_A1}
\end{align}
which is equivalent to the algebraic relation for the Fourier modes,
\begin{align}
 \left[
 T_{1,n} \,,\, T_{1,n'}
 \right]
 & = - \sum_{k=1}^\infty f_k \, (T_{1,n-k} T_{1,n'+k} - T_{1,n'-k} T_{1,n+k})
 \nonumber \\
 & \hspace{10em}
 - \frac{(1 - q_1)(1 - q_2)}{1 - q} \left( q^n - q^{-n} \right) \delta_{n+n',0}
 \, .
\end{align}
We remark the commutator between $(T_{1,n})_{n \in \mathbb{Z}}$ gives rise to non-linear terms.
In this sense, it is not a Lie algebra.
It has been shown that this algebra satisfies the associativity condition, and is now known as the $q$-deformed Virasoro algebra, Vir$_{q_{1,2}}$~\cite{Shiraishi:1995rp}.
The $q$-Virasoro algebra is thought of as the $q$-deformation of W-algebra associated with $A_1$ quiver, Vir$_{q_{1,2}} = \mathrm{W}_{q_{1,2}}(A_1)$.

We remark the representation theoretical interpretation of the OPE between the $\mathsf{T}$-operators~\eqref{eq:TT_OPE_A1}.
The OPE~\eqref{eq:TT_OPE_A1} computes the antisymmetric part of the degree-two $qq$-character.
Namely it is the degree-two antisymmetric tensor product of the fundamental representation, which provides the trivial representation for $A_1$ theory, $\square \wedge \square = \emptyset$.

  \subsection{$A_2$ quiver}\label{sec:W_A2}

  Let us consider the rank 2 case, $A_2$ quiver.
  In this case, there are two fundamental $qq$-characters as discussed in \S\ref{sec:qq_A2}, so that we have two $\mathsf{T}$-operators\index{qq-character@$qq$-character!A2@$A_2$}
  \begin{subequations}
   \begin{align}
    \mathsf{T}_{1,x} & =
    \mathsf{Y}_{1,x} 
    + {: \frac{\mathsf{Y}_{2,\mu^{-1} x}}{\mathsf{Y}_{1,q^{-1} x}} :}
    + \frac{1}{\mathsf{Y}_{2,\mu^{-1} q^{-1} x}}
    \, , \\
    \mathsf{T}_{2,x} & =
    \mathsf{Y}_{2,x}
    + {: \frac{\mathsf{Y}_{1,\mu q^{-1} x}}{\mathsf{Y}_{2,q^{-1} x}} :}
    + \frac{1}{\mathsf{Y}_{1,\mu q^{-2} x}}
    \, ,
   \end{align}
  \end{subequations}
  where $\mu = \mu_{1 \to 2} = \mu_{2 \to 1}^{-1} q$ is the bifundamental mass parameter.
  Then, we can show that these $\mathsf{T}$-operators obey the following OPEs:\index{operator product expansion (OPE)!T and T (A2)@$\mathsf{T}$ and $\mathsf{T}$ ($A_2$)}
  \begin{subequations}\label{eq:A2_TT_OPE}
   \begin{align}
    &
    f_{11}\left( \frac{x'}{x} \right) \mathsf{T}_{1,x} \mathsf{T}_{1,x'}
    - f_{11}\left( \frac{x}{x'} \right) \mathsf{T}_{1,x'} \mathsf{T}_{1,x}
    \nonumber \\
    & \hspace{5em}
    =
    - \frac{(1 - q_1)(1 - q_2)}{(1 - q)}
    \left(
    \delta \left( q \frac{x'}{x} \right) \mathsf{T}_{2,\mu^{-1} x}
    - \delta \left( q^{-1} \frac{x'}{x} \right) \mathsf{T}_{2,\mu q^{-1} x}
    \right)
    \, , \\
    &
    f_{12}\left( \frac{x'}{x} \right) \mathsf{T}_{1,x} \mathsf{T}_{2,x'}
    - f_{21}\left( \frac{x}{x'} \right) \mathsf{T}_{2,x'} \mathsf{T}_{1,x}
    \nonumber \\
    & \hspace{5em}
    =
    - \frac{(1 - q_1)(1 - q_2)}{(1 - q)}
    \left(
     \delta\left( \mu q \frac{x'}{x} \right) - \delta\left( \mu q^{-2} \frac{x'}{x} \right)
    \right)
    \, , \\
    &
    f_{22}\left( \frac{x'}{x} \right) \mathsf{T}_{2,x} \mathsf{T}_{2,x'}
    - f_{22}\left( \frac{x}{x'} \right) \mathsf{T}_{2,x'} \mathsf{T}_{2,x}
    \nonumber \\
    & \hspace{5em}
    =
    - \frac{(1 - q_1)(1 - q_2)}{(1 - q)}
    \left(
    \delta \left( q \frac{x'}{x} \right) \mathsf{T}_{1,\mu q^{-1} x}
    - \delta \left( q^{-1} \frac{x'}{x} \right) \mathsf{T}_{1,\mu x}
    \right)
    \, .
   \end{align}
  \end{subequations}
  The structure function is defined as
  \begin{align}
   f_{ij}(z)
   =
   \exp \left( \sum_{n = 1}^\infty \frac{1}{n} (1 - q_1^n)(1 - q_2^n) \, \tilde{c}_{ji}^{[-n]} \, z^n \right)
   \, ,
  \end{align}
  by which the OPE between the $\mathsf{Y}$-operators is described
  \begin{align}
   \mathsf{Y}_{i,x} \mathsf{Y}_{j,x'}
   = f_{ij}\left(\frac{x'}{x}\right)^{-1} \,
   {:\mathsf{Y}_{i,x} \mathsf{Y}_{j,x'}:}
   \, .
  \end{align}
  The algebra characterized by the OPE~\eqref{eq:A2_TT_OPE} is called the $q$-deformed W$_3$ algebra, denoted by W$_{q_{1,2}}(A_2)$~\cite{Awata:1995zk}.

  As in the case of $A_1$ quiver (\S\ref{sec:W_A1}), the OPE between the $\mathsf{T}$-operators \eqref{eq:A2_TT_OPE} computes the antisymmetric tensor product of the fundamental representations of $A_2$ quiver:
  \begin{align}
   \yng(1) \ \wedge \ \yng(1) \ = \ \yng(1,1)
   \, , \qquad
   \yng(1) \ \wedge \ \yng(1,1) \ = \ \emptyset
   \, , \qquad
   \yng(1,1) \ \wedge \ \yng(1,1) \ = \ \yng(1)
   \, .
  \end{align}

  \subsection{$A_p$ quiver}

  We consider $A_p$ quiver, which is a linear quiver with $p$ gauge nodes,
\begin{equation}
 \dynkin[mark=o,scale=3.5,labels={1,2,p-1,p},
  arrow color={black,length=3mm,width=5mm}]{A}{oo.oo} 
\end{equation}
  The $\mathsf{T}$-operators for $A_p$ quiver can be constructed as the $qq$-characters of the fundamental representations of $\SL(p+1)$.
  Similarly to \S\ref{sec:Ap_NSlim}, we obtain\index{qq-character@$qq$-character!Ap@$A_p$}
  \begin{align}
   \mathsf{T}_{i,\mu_i^{-1} x}
   = \sum_{1 \le j_1 < \cdots < j_i \le p+1}
   {: \prod_{k=1}^i \Lambda_{j_k,q^{-i+k} x} :}   
  \end{align}
  where the weight operator $(\Lambda_{i,x})_{i = 1,\ldots,p_1}$ is defined
  \begin{align}
   \Lambda_{i,x}
   = {: \mathsf{Y}_{i,\mu_i^{-1} x} \mathsf{Y}_{i-1,\mu_{i-1}^{-1} q^{-1} x}^{-1} :}
  \end{align}
  with $\mathsf{Y}_{0,x} = \mathsf{Y}_{p+1,x} = 1$.
  The bifundamental mass is also defined
  \begin{align}
   \mu_i = \mu_{1 \to 2} \mu_{2 \to 3} \cdots \mu_{i-1 \to i}
   \, .
  \end{align}
  These $\mathsf{T}$-operators define the $q$-deformation of W$_{p+1}$ algebra, W$_{q_{1,2}}(A_p)$~\cite{Awata:1995zk}.
  The OPE between the $\mathsf{T}$-operators provides the antisymmetric tensor product of fundamental representations.

  \subsection{$D_p$ quiver}

We consider $D$-type quiver gauge theory with the simplest example, $\Gamma = D_4$,
  \begin{equation}
   \dynkin[mark=o,edge length=1.5cm,root radius=.2cm,label]{D}{4}
  \end{equation}
This quiver has a symmetry exchanging $1 \leftrightarrow 3 \leftrightarrow 4$, which is known as the $\SO(8)$ $(=G_{D_4})$ triality.
We denote the bifundamental mass parameters by $\mu_i := \mu_{2 \to i}$ for $i = 1, 3, 4$.
Then, the operator $\mathsf{T}_{1,x}$ is given by\index{qq-character@$qq$-character!D4@$D_4$}
\begin{align}
 \mathsf{T}_{1,x} & = \mathsf{Y}_{1,x}
 \, + : \frac{\mathsf{Y}_{2,\mu_1 q^{-1} x}}{\mathsf{Y}_{1,q^{-1} x}} :
 + : \frac{\mathsf{Y}_{3,\mu_1 \mu_3^{-1} q^{-1} x}
           \mathsf{Y}_{4,\mu_1 \mu_4^{-1} q^{-1} x}}{\mathsf{Y}_{2,\mu_1 q^{-2} x}} :
 + : \frac{\mathsf{Y}_{4,\mu_1 \mu_4^{-1} q^{-1} x}}
          {\mathsf{Y}_{3,\mu_1 \mu_3^{-1} q^{-2} x}} :
 \nonumber \\[.5em]
 & \quad 
 +
 : \frac{\mathsf{Y}_{3,\mu_1 \mu_3^{-1} q^{-1} x}}
        {\mathsf{Y}_{4,\mu_1 \mu_4^{-1} q^{-2} x}} :
 +
 : \frac{\mathsf{Y}_{2,\mu_1 q^{-2} x}}
        {\mathsf{Y}_{3,\mu_1 \mu_3^{-1} q^{-2} x}
         \mathsf{Y}_{4,\mu_1 \mu_4^{-1} q^{-2} x}}:
 +
 : \frac{\mathsf{Y}_{1,q^{-2} x}}{\mathsf{Y}_{2,\mu_1 q^{-3} x}} :
 + \, \frac{1}{\mathsf{Y}_{1,q^{-3} x}}
 \, .
\end{align}
The operators $\mathsf{T}_{3,x}$ and $\mathsf{T}_{4,x}$ are similarly obtained by permutation.
These three $\mathsf{T}$-operators correspond to three $\mathbf{8}$-representations of $\SO(8)$.
The remaining $\mathsf{T}_2$-operator, corresponding to $\mathbf{28}$-representation (adjoint representation), involves collision and derivative terms, \eqref{eq:collision1} and \eqref{eq:collision2},
\begin{align}
 \mathsf{T}_{2,x} & =
 \mathsf{T}_{2,x}^+ + \mathsf{T}_{2,x}^-
 +
 \mathscr{S}(q)
 \left(
 :\frac{\mathsf{Y}_{1,\mu_1^{-1} x}}{\mathsf{Y}_{1,\mu_1^{-1} q^{-1} x}}:
 + :\frac{\mathsf{Y}_{3,\mu_3^{-1} x}}{\mathsf{Y}_{3,\mu_3^{-1} q^{-1} x}}:
 + :\frac{\mathsf{Y}_{4,\mu_4^{-1} x}}{\mathsf{Y}_{4,\mu_4^{-1} q^{-1} x}}: 
 \right)
 \nonumber \\[.5em]
 &
 +
 \left(
 \mathfrak{c}(q_1,q_2) - \frac{(1-q_1)(1-q_2)}{1-q}
 \partial_{\log x} \log
 \left(
 \frac{\mathsf{Y}_{2,q^{-1} x} \mathsf{Y}_{2,q^{-2} x}}
       {\mathsf{Y}_{1,\mu_1^{-1} q^{-1} x}
        \mathsf{Y}_{3,\mu_3^{-1} q^{-1} x}
        \mathsf{Y}_{4,\mu_4^{-1} q^{-1} x}}
 \right)
 \right)
 :\frac{\mathsf{Y}_{2,q^{-1} x}}{\mathsf{Y}_{2,q^{-2} x}}:
\end{align}
where
\begin{subequations}
\begin{align}
 \mathsf{T}_{2,x}^{+} & =
 \mathsf{Y}_{2,x} \,
 +
 :\frac{\mathsf{Y}_{1,\mu_1^{-1}x} \mathsf{Y}_{3,\mu_3^{-1}x} \mathsf{Y}_{4,\mu_4^{-1}x}}
       {\mathsf{Y}_{2,q^{-1}x}}:
 +
 :\frac{\mathsf{Y}_{3,\mu_3^{-1} x} \mathsf{Y}_{4,\mu_4^{-1}x}}
       {\mathsf{Y}_{1,\mu_1^{-1} q^{-1} x}}:
 +
 :\frac{\mathsf{Y}_{1,\mu_1^{-1} x} \mathsf{Y}_{4,\mu_4^{-1}x}}
       {\mathsf{Y}_{3,\mu_3^{-1} q^{-1} x}}:
 \nonumber \\[.5em]
 &
 +
 :\frac{\mathsf{Y}_{1,\mu_1^{-1} x} \mathsf{Y}_{3,\mu_3^{-1}x}}
       {\mathsf{Y}_{4,\mu_4^{-1} q^{-1} x}}: 
 +
 :\frac{\mathsf{Y}_{1,\mu_1^{-1} x} \mathsf{Y}_{2,q^{-1} x}}
       {\mathsf{Y}_{3,\mu_3^{-1} q^{-1} x} \mathsf{Y}_{4,\mu_4^{-1} q^{-1} x}}: 
 +
 :\frac{\mathsf{Y}_{3,\mu_3^{-1} x} \mathsf{Y}_{2,q^{-1} x}}
       {\mathsf{Y}_{1,\mu_1^{-1} q^{-1} x} \mathsf{Y}_{4,\mu_4^{-1} q^{-1} x}}: 
 \nonumber \\[.5em]
 &
 +
 :\frac{\mathsf{Y}_{4,\mu_4^{-1} x} \mathsf{Y}_{2,q^{-1} x}}
       {\mathsf{Y}_{1,\mu_1^{-1} q^{-1} x} \mathsf{Y}_{3,\mu_3^{-1} q^{-1} x}}: 
 +
 :\frac{\mathsf{Y}_{2,q^{-1}x}^2}
       {\mathsf{Y}_{1,\mu_1^{-1} q^{-1} x}
        \mathsf{Y}_{3,\mu_3^{-1} q^{-1} x}
        \mathsf{Y}_{4,\mu_4^{-1} q^{-1} x}}:
 \nonumber \\[.5em]
 &
 +
 :\frac{\mathsf{Y}_{1,\mu_1^{-1} x}\mathsf{Y}_{1,\mu_1^{-1} q^{-1} x}}
       {\mathsf{Y}_{2,q^{-2} x}}:
 +
 :\frac{\mathsf{Y}_{3,\mu_3^{-1} x}\mathsf{Y}_{3,\mu_3^{-1} q^{-1} x}}
       {\mathsf{Y}_{2,q^{-2} x}}:
 +
 :\frac{\mathsf{Y}_{4,\mu_4^{-1} x}\mathsf{Y}_{4,\mu_4^{-1} q^{-1} x}}
       {\mathsf{Y}_{2,q^{-2} x}}:
 \, , 
\end{align}
\begin{align}
 \mathsf{T}_{2,x}^- & = \,
 :\frac{\mathsf{Y}_{1,\mu_1^{-1} q^{-1} x} \mathsf{Y}_{3,\mu_3^{-1} q^{-1} x}
        \mathsf{Y}_{4,\mu_4^{-1} q^{-1} x}}
       {\mathsf{Y}_{2,q^{-2} x}^2}:
 +
 :\frac{\mathsf{Y}_{2,q^{-1} x}}
       {\mathsf{Y}_{1,\mu_1^{-1} q^{-1} x} \mathsf{Y}_{1,\mu_1^{-1} q^{-2} x}}:
 \nonumber \\[.5em]
 &
 +
 :\frac{\mathsf{Y}_{2,q^{-1} x}}
       {\mathsf{Y}_{3,\mu_3^{-1} q^{-1} x} \mathsf{Y}_{3,\mu_3^{-1} q^{-2} x}}:
 +
 :\frac{\mathsf{Y}_{2,q^{-1} x}}
       {\mathsf{Y}_{4,\mu_4^{-1} q^{-1} x} \mathsf{Y}_{4,\mu_4^{-1} q^{-2} x}}:
 +
 :\frac{\mathsf{Y}_{3,\mu_3^{-1} q^{-1} x} \mathsf{Y}_{4,\mu_4^{-1} q^{-1} x}}
       {\mathsf{Y}_{1,\mu_1^{-1} q^{-2} x} \mathsf{Y}_{2,q^{-2} x}}:
 \nonumber \\[.5em]
 &
 +
 :\frac{\mathsf{Y}_{1,\mu_1^{-1} q^{-1} x} \mathsf{Y}_{4,\mu_4^{-1} q^{-1} x}}
       {\mathsf{Y}_{3,\mu_3^{-1} q^{-2} x} \mathsf{Y}_{2,q^{-2} x}}: 
 +
 :\frac{\mathsf{Y}_{1,\mu_1^{-1} q^{-1} x} \mathsf{Y}_{3,\mu_3^{-1} q^{-1} x}}
       {\mathsf{Y}_{4,\mu_4^{-1} q^{-2} x} \mathsf{Y}_{2,q^{-2} x}}: 
 +
 :\frac{\mathsf{Y}_{1,\mu_1^{-1} q^{-1} x}}
       {\mathsf{Y}_{3,\mu_3^{-1} q^{-2} x} \mathsf{Y}_{4,\mu_4^{-1} q^{-2} x}}:
 \nonumber \\[.5em]
 &
 +
 :\frac{\mathsf{Y}_{3,\mu_1^{-1} q^{-1} x}}
       {\mathsf{Y}_{1,\mu_1^{-1} q^{-2} x} \mathsf{Y}_{4,\mu_4^{-1} q^{-2} x}}:
 +
 :\frac{\mathsf{Y}_{4,\mu_4^{-1} q^{-1} x}}
       {\mathsf{Y}_{1,\mu_1^{-1} q^{-2} x} \mathsf{Y}_{3,\mu_3^{-1} q^{-2} x}}:
 \nonumber \\[.5em]
 &
 +
 :\frac{\mathsf{Y}_{2,q^{-2} x}}
       {\mathsf{Y}_{1,\mu_1^{-1} q^{-2} x}
        \mathsf{Y}_{3,\mu_3^{-1} q^{-2} x}
        \mathsf{Y}_{4,\mu_4^{-1} q^{-2} x}}:
 + \, \frac{1}{\mathsf{Y}_{2,q^{-3}x}}
 \, .
\end{align}
\end{subequations}
In the classical limit, this operator $\mathsf{T}_{2,x}$ is reduced to the character of 28 dimensional representation of $\SO(8)$ with extension.
The $\mathscr{S}$-factor appears at the zero weight terms, ${: \mathsf{Y}_{i,\mu_i^{-1} x}/\mathsf{Y}_{i,\mu_i^{-1} q^{-1} x} :} \xrightarrow{q \to 1} 1$, for $i = 1,3,4$.

\section{Fractional quiver W-algebra}\label{sec:frac_W}
\index{quiver!fractional---W-algebra}

The $q$-deformation of the W-algebra is given for arbitrary simple Lie algebras~\cite{Frenkel:1997}.
Applying the operator formalism to fractional quiver gauge theory introduced in \S\ref{sec:fractional_quiver}, we reproduce the $q$-deformed W-algebras beyond the simply-laced cases~\cite{Kimura:2017hez}.

\subsection{Screening current}
\index{screening current!fractional---}

We define the screening current similarly to \S\ref{sec:screening_current},
\begin{align}
 S_{i,x} 
 = 
 {: \exp \left( 
 s_{i,0} \log x + \tilde{s}_{i,0} 
 - \frac{\kappa_{i}}{2} \left( \log_{q_2} x - 1 \right) \log x
 + \sum_{n \in \mathbb{Z}_{\neq 0}} s_{i,n} \, x^{-n} 
 \right) :}
 \, .
\end{align}
In this case, the $s$-modes are slightly modified from~\eqref{eq:s_oscillator} as follows:
\begin{align}
 s_{i,-n} = (1 - q_1^{d_i n}) \, t_{i,n}
 \, , \qquad
 s_{i,0} = t_{i,0}
 \, , \qquad
 s_{i,n} = - \frac{1}{n} (1 - q_2^{-n})^{-1} \, c_{ji}^{[n]} \, \partial_{j,n}
 \qquad
 (n \ge 1)
 \, ,
\end{align}
with the commutation relation
\begin{subequations}
\begin{align}
 \left[ s_{i,n} \, , \, s_{j,n'} \right]
 & = - \frac{1}{n} \frac{1 - q_1^{d_j n}}{1 - q_2^{-n}} \, c_{ji}^{[n]} \, \delta_{n+n',0}
 = - \frac{1}{n} \frac{1 - q_1^{n}}{1 - q_2^{-n}} \, b_{ji}^{[n]} \, \delta_{n+n',0}
 \qquad
 (n \ge 1)
 \, ,
 \\
 \left[ \tilde{s}_{i,0} \,,\, s_{j,n} \right]
 & = - \beta \, c_{ji}^{[0]} \, \delta_{n,0} 
\end{align}
\end{subequations}
where $(b_{ij})_{i,j \in \Gamma_0}$ is the symmetrization of the Cartan matrix~\eqref{eq:Cartan_b}.

We remark that the $\mathsf{V}$-operator is defined to obey the same commutation relation as before~\eqref{eq:vs_comm_rel}.

\subsection{$\mathsf{Y}$-operator}
\index{Y-operator@$\mathsf{Y}$-operator!fractional---}

The $\mathsf{Y}$-operator for fractional quiver theory is modified from~\eqref{eq:y_oscillators} in \S\ref{sec:Y_op} as follows:
\begin{align}
 \mathsf{Y}_{i,x}
 = q_1^{d_i \tilde{\rho}_i} \,
 {: \exp \left( \sum_{n \in \mathbb{Z}} y_{i,n} \, x^{-n} \right) :}
 \, ,
\end{align}
with
\begin{align}
 y_{i,-n} = (1 - q_1^{d_i n})(1 - q_2^n) \, \tilde{c}_{ji}^{[-n]} \, t_{j,n}
 \, , \quad
 y_{i,0} = - \log q_2 \, \tilde{c}_{ji}^{[0]} \, t_{j,0}
 \, , \quad
 y_{i,n} = - \frac{1}{n} \partial_{i,n}
 \quad
 (n \ge 1)
 \, .
\end{align}
They obey the commutation relation
\begin{align}
 \left[ y_{i,n} \, , y_{j,n'} \right]
 = - \frac{1}{n} (1 - q_1^{d_j n})(1 - q_2^n) \, \tilde{c}_{ji}^{[-n]} \, \delta_{n+n', 0}
 \, .
 \label{eq:y_comm_rel_frac}
\end{align}
The relation to the $s$-modes is given by
\begin{align}
 \left[ y_{i,n} \,,\, s_{j,n'} \right]
 = - \frac{1}{n} ( 1 - q_1^{d_i n}) \, \delta_{ij} \, \delta_{n+n',0}
 \, , \qquad
 \left[ \tilde{s}_{i,0} \,,\, y_{j,n} \right]
 = - d_i \log q_1 \, \delta_{ij} \, \delta_{n,0}
 \, , 
\end{align}
and thus \index{operator product expansion (OPE)!S and Y@$S$ and $\mathsf{Y}$}
\begin{align}
 \left[
 \mathsf{Y}_{i,x}
 \,,\,
 S_{j,x'}
 \right]
 =
 \begin{cases}
  \displaystyle
  (1 - q_1^{-d_i}) \, \delta\left( q_1^{d_i} \frac{x'}{x} \right) \,
  {:  \mathsf{Y}_{i,x} S_{j,x'} :}
  & ( i = j) \\
  0 & (i \neq j)
 \end{cases}
\end{align}

\subsection{$\mathsf{A}$-operator}
\index{A-operator@$\mathsf{A}$-operator!fractional---}

We then consider the $\mathsf{A}$-operator for fractional quiver theory following \S\ref{sec:A_op}:
\begin{align}
 \mathsf{A}_{i,x}
 = q_1^{d_i}
 {: \exp \left( \sum_{n \in \mathbb{Z}} a_{i,n} \, x^{-n} \right) :}
 \, .
\end{align}
The $a$-modes are defined as
\begin{align}
 a_{i,-n} = (1 - q_1^{d_i n})(1 - q_2^n) \, t_{i,n}
 \, , \quad
 a_{i,0} = - \log q_2 \, t_{i,0}
 \, , \quad
 a_{i,n} = - \frac{1}{n} c_{ji}^{[n]} \, \partial_{j,n}
 \quad
 (n \ge 1)
 \, ,
\end{align}
with the commutation relation
\begin{align}
 \left[ a_{i,n} \, , a_{j,n'} \right]
 & = - \frac{1}{n} (1 - q_1^{d_j n})(1 - q_2^n) \, c_{ji}^{[n]} \, \delta_{n+n', 0}
 \nonumber \\
 & = - \frac{1}{n} (1 - q_1^{n})(1 - q_2^n) \, b_{ji}^{[n]} \, \delta_{n+n', 0} 
 \, .
 \label{eq:a_comm_rel_frac} 
\end{align}
The $a$-modes are related to other oscillators as follows:
\begin{align}
 a_{i,n} = (1 - q_2^{-n}) \, s_{i,n} = y_{j,n} \, c_{ji}^{[n]}
 \, ,
\end{align}
and thus
\begin{subequations}
 \begin{align}
  [y_{i,n} \,,\, a_{j,n'}]
  & =
  - \frac{1}{n} (1 - q_1^{d_j n})(1 - q_2^n) \, \delta_{ij} \, \delta_{n+n',0}
  \, , \\
  [v_{i,n} \,,\, a_{j,n'}]
  & = \frac{1}{n} \, \delta_{ij} \, \delta_{n+n',0}
  \, .
 \end{align}
\end{subequations}

\subsection{iWeyl reflection}
\index{iWeyl reflection}

Based on these vertex operators, we discuss the iWeyl reflection.
The $\mathsf{A}$-operator has the expression in terms of the $\mathsf{Y}$-operators
\begin{align}
 \mathsf{A}_{i,x}
 =
 {:
 \mathsf{Y}_{i,x} \mathsf{Y}_{i,q_{1^{d_i}2} x}
 \prod_{e:i \to j} \prod_{r = 0}^{d_j/d_{ij} - 1}
 \mathsf{Y}_{j,\mu_e q_1^{r d_{ij}} x}^{-1}
 \prod_{e:j \to i} \prod_{r = 0}^{d_j/d_{ij} - 1}
 \mathsf{Y}_{j,\mu_e^{-1} q_{1^{d_{ij}}2} q_1^{r d_{ij}} x}^{-1} 
 :}
\end{align}
where $q_{i^d j^{d'}}$ is defined in~\eqref{eq:q_dd'}.
Thus, the iWeyl reflection for fractional quiver is given as follows:
\begin{align}
 \text{iWeyl}: \
 \mathsf{Y}_{i,q_{1^{d_i}2} x}
 & \longmapsto \
 {: \mathsf{Y}_{i,q_{1^{d_i}2} x} \mathsf{A}_{i,x}^{-1} :}
 \nonumber \\
 & \qquad
 =
 {:
 \mathsf{Y}_{i,x}^{-1}
 \prod_{e:i \to j} \prod_{r = 0}^{d_j/d_{ij} - 1}
 \mathsf{Y}_{j,\mu_e q_1^{r d_{ij}} x}
 \prod_{e:j \to i} \prod_{r = 0}^{d_j/d_{ij} - 1}
 \mathsf{Y}_{j,\mu_e^{-1} q_{1^{d_{ij}}2} q_1^{r d_{ij}} x}
 :}
 \, ,
\end{align}
and the $\mathsf{T}$-operator is generated by the iWeyl reflection until it does not provide any further singular term,
\begin{align}
 \mathsf{T}_{i,x}
 = \mathsf{Y}_{i,x}
 + {: \mathsf{Y}_{i,x} \mathsf{A}_{i,q_{1^{d_i}2}x}^{-1}:} + \cdots
 \, .
\end{align}

\subsection{$\mathsf{T}$-operator: generating current}\label{sec:frac_T_op}

Based on the $\mathsf{T}$-operators constructed above, we define the {\em fractional quiver W-algebra} from their Fourier mode expansion:
\begin{align}
 \mathsf{T}_{i,x} = \sum_{n \in \mathbb{Z}} T_{i,n} \, x^{-n}
 \, .
\end{align}
For quivers of the finite-type, it reproduces the $q$-deformation of the W-algebras for $\Gamma = BCFG$~\cite{Frenkel:1997}, and for affine (including the twisted version) and hyperbolic-type quivers, we could construct a new family of the W-algebras.

Let us provide a comment on a duality of the W-algebras.
For the rational W-algebras, there is a duality between W$_{\beta}(\mathfrak{g})$ and W$_{\beta^{-1}}(^L\mathfrak{g})$~\cite{Feigin:1991wy}, where we denote the Langlands dual algebra of $\mathfrak{g}$ by $^L\mathfrak{g}$.
From the gauge theory point of view, this is a symmetry of exchanging $\epsilon_1 \leftrightarrow \epsilon_2$.
After the $q$-deformation, we have a similar duality, W$_{q_{1,2}}(\mathfrak{g}) \cong \mathrm{W}_{q_{2,1}}(^L\mathfrak{g})$ only if $^L\mathfrak{g} = \mathfrak{g}$, namely for the simply-laced algebras.
In fact, this duality does not hold for the non-simply-laced algebras, $^L\mathfrak{g} \neq \mathfrak{g}$, since, in this case, the corresponding gauge theory, i.e., fractional quiver gauge theory, discussed in \S\ref{sec:fractional_quiver}, is not invariant under exchange $\epsilon_1 \leftrightarrow \epsilon_2$.

   \subsection{$BC_2$ quiver}\label{sec:BC2_quiv}

   Let us consider $BC_2$ quiver 
   \dynkin[label,mark=o,root radius = .1cm, edge length = .7cm, arrow color={black,length=2mm,width=3mm}]{B}{2}    
    which is the simplest example for fractional quiver theory.%
    \footnote{%
    We simply denote $B_2$ and $C_2$ by $BC_2$ since they are indistinguishable in this context.
    Not to be confused with other $BC_n$ systems.
    }
   We assign 
   \begin{align}
    (d_1, d_2) = (2,1)
    \, ,
   \end{align}
   corresponding to the root length of each node.
   The quiver Cartan matrix is in this case given by
   \begin{align}
    (c_{ij})
    =
    \begin{pmatrix}
     1 + q_1^{-2} q_2^{-1} & - \mu^{-1} \\
     - \mu q_1^{-1} q_2^{-1} ( 1 + q_1^{-1} ) & 1 + q_1^{-1} q_2^{-1}
    \end{pmatrix}
    \ \xrightarrow{(c_{ij}^{[0]})} \
    \begin{pmatrix}
     2 & -1 \\ -2 & 2
    \end{pmatrix}
    \, ,
   \end{align}
   with the bifundamental mass parameter $\mu = \mu_{1 \to 2} = \mu_{2 \to 1}^{-1} q_1 q_2$, and its symmetrization
   \begin{align}
    (b_{ij})
    =
    \begin{pmatrix}
     (1 + q_1)(1 + q_1^{-2} q_2^{-1}) & - \mu^{-1}(1 + q_1) \\
     - \mu q_1^{-1} q_2^{-1} ( 1 + q_1^{-1}) & 1 + q_1^{-1} q_2^{-1}
    \end{pmatrix}
    \ \xrightarrow{(b_{ij}^{[0]})} \
    \begin{pmatrix}
     4 & -2 \\ -2 & 2
    \end{pmatrix}
    \, .
   \end{align}
   Then iWeyl reflection is given by
   \begin{align}
    \text{iWeyl}: \
    \left(
    \mathsf{Y}_{1,x}, \mathsf{Y}_{2,x}
    \right)
    \ \longmapsto \
    \left(
    \frac{\mathsf{Y}_{2, \mu^{-1} x} \mathsf{Y}_{2,\mu^{-1} q_1^{-1} x}}{\mathsf{Y}_{1,q_1^{-2}q_2^{-1} x}}
    \, , \,
    \frac{\mathsf{Y}_{1, \mu q_1^{-1} q_2^{-1} x}}{\mathsf{Y}_{2,q_1^{-1}q_2^{-1} x}}
    \right)
    \, ,
   \end{align}
     which generates the $\mathsf{T}$-operators\index{qq-character@$qq$-character!BC2@$BC_2$}
     \begin{subequations}\label{eq:BC2_qq_ch}
      \begin{align}
       \mathsf{T}_{1,x}
       & =
       \mathsf{Y}_{1,x}
       + \frac{\mathsf{Y}_{2, \mu^{-1} x} \mathsf{Y}_{2,\mu^{-1} q_1^{-1} x}}{\mathsf{Y}_{1,q_1^{-2}q_2^{-1} x}}
       + \mathscr{S}(q_1) \, \frac{\mathsf{Y}_{2,\mu^{-1} x}}{\mathsf{Y}_{2,\mu^{-1} q_1^{-2} q_2^{-1} x}}
       \nonumber \\
       & \hspace{10em}
       + \frac{\mathsf{Y}_{1,q_1^{-1} q_2^{-1} x}}{\mathsf{Y}_{2,\mu^{-1} q_1^{-1} q_2^{-1} x} \mathsf{Y}_{2,\mu^{-1} q_1^{-2} q_2^{-1} x} }
       + \frac{1}{\mathsf{Y}_{1,q_1^{-3} q_2^{-2} x}}
       \, , \\
       \mathsf{T}_{2,x}
       & =
       \mathsf{Y}_{2,x}
       + \frac{\mathsf{Y}_{1, \mu q_1^{-1} q_2^{-1} x}}{\mathsf{Y}_{2,q_1^{-1}q_2^{-1} x}}
       + \frac{\mathsf{Y}_{2,q_1^{-2}q_2^{-1} x}}{\mathsf{Y}_{1, \mu q_1^{-3} q_2^{-2} x}}
       + \frac{1}{\mathsf{Y}_{2,q_1^{-3} q_2^{-2} x}}
      \end{align}
     \end{subequations}
with the structure function
  \begin{align}
   f_{ij}(z) = \exp \left( \sum_{n = 1}^\infty \frac{1}{n}(1 - q_1^{d_j n})(1 - q_2^n) \, \tilde{c}_{ji}^{[-n]} \right)
   \, .
  \end{align}
  These characters correspond to the \textbf{5} (vector) and \textbf{4} (spinor) representations of $\SO(5)/\Sp(2)$.
  The OPEs of these $\mathsf{T}$-operators are given as follows:\index{operator product expansion (OPE)!T and T (BC2)@$\mathsf{T}$ and $\mathsf{T}$ ($BC_2$)}
  \begin{subequations}
   \begin{align}
    & f_{11}\left( \frac{x'}{x} \right) \mathsf{T}_{1,x} \mathsf{T}_{1,x'}
    - f_{11}\left( \frac{x}{x'} \right) \mathsf{T}_{1,x'} \mathsf{T}_{1,x}
    \nonumber \\
    & \quad =
    - \frac{(1 - q_1^2)(1 - q_2)}{1 - q_1^2 q_2}
    \Bigg(
    \delta\left( q_1^2 q_2 \frac{x'}{x} \right) f_{22}(q_1^{-1}) \mathsf{T}_{2,\mu^{-1} x} \mathsf{T}_{2,\mu^{-1} q_1^{-1} x}
    \nonumber \\
    & \hspace{13em}
    - \delta\left( q_1^{-2} q_2^{-1} \frac{x'}{x} \right) f_{22}(q_1) \mathsf{T}_{2,\mu^{-1} q_1 q_2 x} \mathsf{T}_{2,\mu^{-1} q_1^2 q_2 x}
    \Bigg)
    \nonumber \\
    & \qquad
    - \frac{(1 - q_1^2)(1 - q_2)(1 - q_1 q_2^2)(1 - q_1^3 q_2)}{(1 - q_1 q_2)(1 - q_1^2 q_2)(1 - q_1^3 q_2^2)}
    \left(
    \delta\left( q_1^3 q_2^2 \frac{x'}{x} \right)
    - \delta\left( q_1^{-3} q_2^{-2} \frac{x'}{x} \right)
    \right)
    \, ,
   \end{align}
   \begin{align}
    &
    f_{12}\left( \frac{x'}{x} \right) \mathsf{T}_{1,x} \mathsf{T}_{2,x'}
    - f_{21}\left( \frac{x}{x'} \right) \mathsf{T}_{2,x'} \mathsf{T}_{1,x}
    \nonumber \\
    & \quad =
    - \frac{(1 - q_1^2)(1 - q_2)}{1 - q_1^2 q_2}
    \left(
    \delta\left(\mu q_1^2 q_2 \frac{x'}{x} \right) \mathsf{T}_{2,\mu^{-1} x}
    - \delta\left(\mu q_1^{-3} q_2^{-2} \frac{x'}{x} \right) \mathsf{T}_{2,\mu^{-1} q_1 q_2 x}
    \right)
    \, , 
   \end{align}
   \begin{align}
    &
    f_{22}\left( \frac{x'}{x} \right) \mathsf{T}_{2,x} \mathsf{T}_{2,x'}
    - f_{22}\left( \frac{x}{x'} \right) \mathsf{T}_{2,x'} \mathsf{T}_{2,x}
    \nonumber \\
    & \quad =
    - \frac{(1 - q_1)(1 - q_2)}{1 - q_1 q_2}
    \left(
    \delta\left(\mu q_1 q_2 \frac{x'}{x} \right) \mathsf{T}_{1,\mu q_1^{-1} q_2^{-1} x}
    - \delta\left(\mu q_1^{-1} q_2^{-1} \frac{x'}{x} \right) \mathsf{T}_{1,\mu x}
    \right)
    \nonumber \\
    & \qquad
    - \frac{(1 - q_1)(1 - q_2)(1 - q_1^2 q_2^2)(1 - q_1^3 q_2)}{(1 - q_1 q_2)(1 - q_1^2 q_2)(1 - q_1^3 q_2^2)}
    \left(
    \delta\left( q_1^3 q_2^2 \frac{x'}{x} \right)
    - \delta\left( q_1^{-3} q_2^{-2} \frac{x'}{x} \right)
    \right)
   \, .
   \end{align}
  \end{subequations}
  These OPEs characterize the $q$-deformed W-algebra, W$_{q_{1,2}}(BC_2)$.
   As mentioned in \S\ref{sec:W_A1}, these OPEs correspond to the antisymmetric tensor product of the fundamental representations: $\mathbf{5} \wedge \mathbf{5} = \mathbf{10}$ (with extension), $\mathbf{5} \wedge \mathbf{4} = \mathbf{4}$, $\mathbf{4} \wedge \mathbf{4} = \mathbf{5} \oplus \mathbf{1}$.
   We remark the factor appearing in the OPE between the $\mathsf{T}_1$-operators is the degree-two $qq$-character, corresponding to the \textbf{10} representation,
   \begin{align}
    f_{22}(q_1^{-1}) \mathsf{T}_{2,\mu^{-1} x} \mathsf{T}_{2,\mu^{-1} q_1^{-1} x}
    = \mathsf{T}_{(0,2),(\mu^{-1}x, \mu^{-1} q_1^{-1}x)}
    \, .
   \end{align}

\subsection{$B_p$ quiver}\label{sec:Br}

We consider $B_p$ quiver which consists of $p$ nodes with $d_i=2$ for $i = 1,\ldots, p-1$ and $d_p = 1$,
\begin{equation}
 \dynkin[mark=o,scale=3.5,labels={1,2,p-2,p-1,p},
  arrow color={black,length=3mm,width=5mm}]{B}{oo.ooo} 
\end{equation}
In this case the local iWeyl reflection is given by
\begin{subequations}
\begin{align}
 \mathsf{Y}_{i,x}
 &
 \ \longmapsto \
 \frac{\mathsf{Y}_{i-1,\mu_{i-1 \to i}q_1^{-2}q_2^{-1}x}
       \mathsf{Y}_{i+1,\mu_{i\to i+1}^{-1}x}}
      {\mathsf{Y}_{i,q_1^{-2}q_2^{-1}x}}
 \qquad
 (i = 1,\ldots,p-2)
 \\
 \mathsf{Y}_{p-1,x}
 &
 \ \longmapsto \
 \frac{\mathsf{Y}_{p-2,\mu_{p-2 \to p-1} q_1^{-2} q_2^{-1} x}
       \mathsf{Y}_{p,\mu_{p-1 \to p}^{-1} x}
       \mathsf{Y}_{p,\mu_{p-1 \to p}^{-1} q_1^{-1} x}}
      {\mathsf{Y}_{p-1,q_1^{-2}q_2^{-1}x}}
 \\
 \mathsf{Y}_{p,x}
 &
 \ \longmapsto \
 \frac{\mathsf{Y}_{p-1,\mu_{p-1 \to p} q_1^{-1} q_2^{-1} x}}
      {\mathsf{Y}_{p,q_1^{-1} q_2^{-1} x}}
\end{align}
\end{subequations}
where we put $\mathsf{Y}_{0,x}=1$.
We introduce the weight fields
\begin{subequations}
\begin{align}
 \Lambda_{i,x}
 & =
 \frac{\mathsf{Y}_{i,\mu_i^{-1} x}}{\mathsf{Y}_{i-1,\mu_{i-1}^{-1} q_1^{-2}q_2^{-1}x}}
 \qquad (i = 1,\ldots,p-1)
 \, , \\
 \Lambda_{p,x}
 & =
 \frac{\mathsf{Y}_{p,\mu_p^{-1}x} \mathsf{Y}_{p,\mu_p^{-1}q_2^{-1}x}}
      {\mathsf{Y}_{p-1,\mu_{p-1}^{-1}q_1^{-2}q_2^{-1}x}}
 \, , \\
 \Lambda_{p+1,x}
 & =
 \frac{(1 + q_1)(1 - q_1 q_2)}{1 - q_1^2 q_2}
 \frac{\mathsf{Y}_{p,\mu_p^{-1}x}}{\mathsf{Y}_{p,\mu_p^{-1}q_1^{-2}q_2^{-1}x}}
 \, , \\
 \Lambda_{p+2,x}
 & =
 \frac{\mathsf{Y}_{p-1,\mu_{p-1}^{-1} q_1^{-1} q_2^{-1} x}}
      {\mathsf{Y}_{p,\mu_{p}^{-1} q_1^{-1} q_2^{-1} x} \mathsf{Y}_{p,\mu_{p}^{-1} q_1^{-2} q_2^{-1} x}} 
 \, , \\  
 \Lambda_{2p+2-i,x}
 & =
 \frac{\mathsf{Y}_{i-1,\mu_{i-1}^{-1} q_1^{-3} q_2^{-2} x}}
      {\mathsf{Y}_{i,\mu_{i}^{-1} q_1^{-3} q_2^{-2} x}}
 \qquad (i = 1,\ldots,p-1)
\end{align}
\end{subequations}
where we parametrize the mass parameters
\begin{align}
 \mu_i
 = \mu_{1 \to 2} \mu_{2 \to 3} \cdots \mu_{i-1 \to i}
 = \prod_{j=1}^{i-1} \mu_{j \to j+1}
 \label{eq:mass_prod_B}
\end{align}
with $\mu_1 = 1$.
Then the fundamental $qq$-character is given by
\begin{align}
 \mathsf{T}_{1,x}
 & =
 \sum_{i=1}^{2p+1}
 \Lambda_{i,x}
 \, ,
\end{align}
which corresponds to the $(2p+1)$-dimensional vector representation of $\SO(2p+1) = G_{B_p}$.

For example, we have three fundamental $qq$-characters for $B_3$ quiver,\index{qq-character@$qq$-character!B3@$B_3$}
\begin{subequations}
\begin{align}
 \mathsf{T}_{1,x}
 & =
 \mathsf{Y}_{1,x}
 + \frac{\mathsf{Y}_{2,\mu_2^{-1} x}}{\mathsf{Y}_{1,q_1^{-2} q_2^{-1} x}}
 + \frac{\mathsf{Y}_{3,\mu_3^{-1} x} \mathsf{Y}_{3,\mu_3^{-1} q_1^{-1} x}}
        {\mathsf{Y}_{2,\mu_2^{-1} q_1^{-2} q_2^{-1} x}}
 + \mathscr{S}(q_1) \frac{\mathsf{Y}_{3,\mu_3^{-1} x}}{\mathsf{Y}_{3,\mu_3^{-1} q_1^{-2} q_2^{-1} x}}
 \nonumber \\
 & \quad
 + \frac{\mathsf{Y}_{2,\mu_2^{-1} q_1^{-1} q_2^{-1} x}}
        {\mathsf{Y}_{3,\mu_3^{-1} q_1^{-1} q_2^{-1} x} \mathsf{Y}_{3,\mu_3^{-1} q_1^{-2} q_2^{-1} x}}
 + \frac{\mathsf{Y}_{1,q_1^{-3} q_2^{-2} x}}{\mathsf{Y}_{2,\mu_2^{-1} q_1^{-3} q_2^{-2} x}}
 + \frac{1}{\mathsf{Y}_{1,q_1^{-5} q_2^{-3} x}} 
 \, ,
\end{align}
\begin{align}
 &
 \mathsf{T}_{2,\mu_2^{-1} x}
 \nonumber \\
 & =
 \mathsf{Y}_{2,\mu_2^{-1} x}
 +
 \frac{\mathsf{Y}_{1,q_1^{-2} q_2^{-1} x}
       \mathsf{Y}_{3,\mu_3^{-1}} \mathsf{Y}_{3,\mu_3^{-1} q_1^{-1} x}}
      {\mathsf{Y}_{2,\mu_2^{-1} q_1^{-2} q_2^{-1} x}}
 +
 \frac{\mathsf{Y}_{3,\mu_3^{-1} x} \mathsf{Y}_{3,\mu_3^{-1} q_1^{-1} x}}
      {\mathsf{Y}_{1,q_1^{-4} q_2^{-2} x}}
 +
 \mathscr{S}(q_1)
 \frac{\mathsf{Y}_{1,q_1^{-2} q_2^{-1} x} \mathsf{Y}_{3,\mu_3^{-1} x}}
      {\mathsf{Y}_{3,\mu_3^{-1} q_1^{-2} q_2^{-1} x}}
 \nonumber \\[.5em]
 & \quad
 +
 \mathscr{S}(q_1)
 \frac{\mathsf{Y}_{3,\mu_3^{-1} x} \mathsf{Y}_{3,\mu_3^{-1} q_1^{-3} q_2^{-1} x}}
      {\mathsf{Y}_{2,\mu_2^{-1} q_1^{-4} q_2^{-2} x}}
 +
 \frac{\mathsf{Y}_{2,\mu_2^{-1} q_1^{-1} q_2^{-1} x} \mathsf{Y}_{2,\mu_2^{-1} q_1^{-2} q_2^{-1} x}}
      {\mathsf{Y}_{1,q_1^{-4} q_2^{-2} x}
       \mathsf{Y}_{3,\mu_3^{-1} q_1^{-1} q_2^{-1} x} \mathsf{Y}_{3,\mu_3^{-1} q_1^{-2} q_2^{-1} x}}
 +
 \frac{\mathsf{Y}_{1,q_1^{-2} q_2^{-1} x} \mathsf{Y}_{1,q_1^{-3} q_2^{-2} x}}
      {\mathsf{Y}_{2,\mu_2^{-1} q_1^{-3} q_2^{-2} x}}
 \nonumber \\[.5em]
 & \quad
 +
 \mathscr{S}(q_1)
 \frac{\mathsf{Y}_{2,\mu_2^{-1} q_1^{-1} q_2^{-1} x} \mathsf{Y}_{3,\mu_3^{-1} x}}{\mathsf{Y}_{1,q_1^{-4} q_2^{-2} x} \mathsf{Y}_{3,\mu_3^{-1} q_1^{-2} q_2^{-1} x}}
 +
 \frac{\mathsf{Y}_{1,q_1^{-2} q_2^{-1} x} \mathsf{Y}_{2,\mu_2^{-1} q_1^{-1} q_2^{-1} x}}
      {\mathsf{Y}_{3,\mu_3^{-1} q_1^{-1} q_2^{-1} x} \mathsf{Y}_{3,\mu_3^{-1} q_1^{-2} q_2^{-1} x}}
  \nonumber \\[.5em]
 & \quad
 +
 \mathscr{S}(q_1) \mathscr{S}(q_1^3 q_2)
 \frac{\mathsf{Y}_{3,\mu_3^{-1} x}}{\mathsf{Y}_{3,\mu_3^{-1} q_1^{-4} q_2^{-2} x}}
 +
 \mathscr{S}_{1^2 2}(q_1)
 \frac{\mathsf{Y}_{2,\mu_2^{-1} q_1^{-1} q_2^{-1} x} \mathsf{Y}_{3,\mu_3^{-1} q_1^{-3} q_2^{-1} x}}
      {\mathsf{Y}_{2,\mu_2^{-1} q_1^{-4} q_2^{-2} x} \mathsf{Y}_{3,\mu_3^{-1} q_1^{-1} q_2^{-1} x}}
 +
 \mathscr{S}_{1^2 2}(q_1^{-1})
 \frac{\mathsf{Y}_{1,q_1^{-3} q_2^{-2} x} \mathsf{Y}_{2,\mu_2^{-1} q_1^{-2} q_2^{-1} x}}
      {\mathsf{Y}_{1,q_1^{-4} q_2^{-2} x} \mathsf{Y}_{2,\mu_2^{-1} q_1^{-3} q_2^{-2} x}}
 \nonumber \\[.5em]
 & \quad
 +
 \mathscr{S}_{1^2 2}(q_1 q_2)
 \frac{\mathsf{Y}_{1,q_1^{-2} q_2^{-1} x}}{\mathsf{Y}_{1,q_1^{-5} q_2^{-3} x}}
 +
 \mathscr{S}(q_1)
 \frac{\mathsf{Y}_{2,\mu_2^{-1} q_1^{-1} q_2^{-1} x}}
      {\mathsf{Y}_{3,\mu_3^{-1} q_1^{-1} q_2^{-1} x} \mathsf{Y}_{3,\mu_3^{-1} q_1^{-4} q_2^{-2} x}}
 +
 \frac{\mathsf{Y}_{1,q_1^{-3} q_2^{-2} x}
       \mathsf{Y}_{3,\mu_3^{-1} q_1^{-2} q_2^{-1} x} \mathsf{Y}_{3,\mu_3^{-1} q_1^{-3} q_2^{-1} x}}
      {\mathsf{Y}_{2,\mu_2^{-1} q_1^{-3} q_2^{-2} x} \mathsf{Y}_{2,\mu_2^{-1} q_1^{-4} q_2^{-2} x}}
 \nonumber \\[.5em]
 & \quad
 +
 \frac{\mathsf{Y}_{2,\mu_2^{-1} q_1^{-2} q_2^{-1} x}}
      {\mathsf{Y}_{1,q_1^{-4} q_2^{-2} x} \mathsf{Y}_{1,q_1^{-5} q_2^{-3} x}}
 +
 \mathscr{S}(q_1)
 \frac{\mathsf{Y}_{1,q_1^{-3} q_2^{-2} x} \mathsf{Y}_{3,\mu_3^{-1} q_1^{-2} q_2^{-1} x}}
      {\mathsf{Y}_{2,\mu_2^{-1} q_1^{-3} q_2^{-2} x} \mathsf{Y}_{3,\mu_3^{-1} q_1^{-4} q_2^{-2} x}}
 +
 \frac{\mathsf{Y}_{3,\mu_3^{-1} q_1^{-2} q_2^{-1} x} \mathsf{Y}_{3,\mu_3^{-1} q_1^{-3} q_2^{-1} x}}
      {\mathsf{Y}_{1,q_1^{-5} q_2^{-3} x} \mathsf{Y}_{2,\mu_2^{-1} q_1^{-4} q_2^{-2} x}}
 \nonumber \\[.5em]
 & \quad
 +
 \frac{\mathsf{Y}_{1,q_1^{-3} q_2^{-2} x}}
      {\mathsf{Y}_{3,\mu_3^{-1} q_1^{-3} q_2^{-2} x} \mathsf{Y}_{3,\mu_3^{-1} q_1^{-4} q_2^{-2} x}}
 +
 \mathscr{S}(q_1)
 \frac{\mathsf{Y}_{3,\mu_3^{-1} q_1^{-2} q_2^{-1} x}}
      {\mathsf{Y}_{1,q_1^{-5} q_2^{-3} x} \mathsf{Y}_{3,\mu_3^{-1} q_1^{-4} q_2^{-2} x}}
 \nonumber \\[.5em]
 & \quad
 +
 \frac{\mathsf{Y}_{2,\mu_2^{-1} q_1^{-3} q_2^{-2} x}}
      {\mathsf{Y}_{1,q_1^{-5} q_2^{-3} x}
       \mathsf{Y}_{3,\mu_3^{-1} q_1^{-3} q_2^{-2} x} \mathsf{Y}_{3,\mu_3^{-1} q_1^{-4} q_2^{-2} x}}
 +
 \frac{1}{\mathsf{Y}_{2,\mu_2^{-1} q_1^{-5} q_2^{-3} x}}
 \, ,
\end{align}
\begin{align}
 \mathsf{T}_{3,\mu_{3}^{-1} x}
 & =
 \mathsf{Y}_{3,\mu_3^{-1} x}
 + \frac{\mathsf{Y}_{2,\mu_{2}^{-1} q_1^{-1} q_2^{-1} x}}
        {\mathsf{Y}_{3,\mu_3^{-1} q_1^{-1} q_2^{-1} x}}
 + \frac{\mathsf{Y}_{1,q_1^{-3} q_2^{-2} x} \mathsf{Y}_{3,\mu_{3}^{-1} q_1^{-2} q_2^{-1} x}}
        {\mathsf{Y}_{2,\mu_{2}^{-1} q_1^{-3} q_2^{-2} x}}
 + \frac{\mathsf{Y}_{3,\mu_3^{-1} q_1^{-2} q_2^{-1} x}}{\mathsf{Y}_{1,q_1^{-5} q_2^{-3} x}}
 \nonumber \\[.5em]
 & \quad
 + \frac{\mathsf{Y}_{1,q_1^{-3} q_2^{-2} x}}{\mathsf{Y}_{3,\mu_3^{-1} q_1^{-3} q_2^{-2} x}}
 + \frac{\mathsf{Y}_{2,\mu_2^{-1} q_1^{-3} q_2^{-2} x}}
        {\mathsf{Y}_{1,q_1^{-5} q_2^{-3} x} \mathsf{Y}_{3,\mu_{3}^{-1} q_1^{-3} q_2^{-2} x}}
 + \frac{\mathsf{Y}_{3,\mu_3^{-1} q_1^{-4} q_2^{-2} x}}{\mathsf{Y}_{2,\mu_2^{-1} q_1^{-5} q_2^{-3} x}}
 + \frac{1}{\mathsf{Y}_{3,\mu_3^{-1} q_1^{-5} q_2^{-3} x}}
 \, . 
\end{align}
\end{subequations}
They correspond to \textbf{7} (vector), \textbf{21} (adjoint), and \textbf{8} (spinor) representations of $\SO(7) = G_{B_3}$, respectively.
There are several $\mathscr{S}$-factors in the expressions which are peculiar to the $qq$-character.

\subsection{$C_p$ quiver}\label{sec:Cr}

The $C_p$ quiver consists of $p$ nodes with $d_i=1$ for $i = 1,\ldots, p-1$ and $d_p = 2$,
\begin{equation}
 \dynkin[mark=o,scale=3.5,labels={1,2,p-2,p-1,p},
  arrow color={black,length=3mm,width=5mm}]{C}{oo.ooo} 
\end{equation}
The local iWeyl reflection is
\begin{subequations}
\begin{align}
 \mathsf{Y}_{i,x}
 &
 \ \longmapsto \
 \frac{\mathsf{Y}_{i-1,\mu_{i-1 \to i} q_1^{-1} q_2^{-1} x}
       \mathsf{Y}_{i+1,\mu_{i \to i+1}^{-1} x}}
      {\mathsf{Y}_{i,q_1^{-1} q_2^{-1} x}}
 \qquad
 (i = 1, \ldots, p-1)
 \\
 \mathsf{Y}_{p,x}
 &
 \ \longmapsto \
 \frac{\mathsf{Y}_{p-1,\mu_{p-1 \to p} q_1^{-1} q_2^{-1} x}
       \mathsf{Y}_{p-1,\mu_{p-1 \to p} q_1^{-2} q_2^{-1} x}}
      {\mathsf{Y}_{p,q_1^{-2} q_2^{-1} x}}
 \, .
\end{align}
\end{subequations}
Introducing the weight fields
\begin{subequations}
\begin{align}
 \Lambda_{i,x}
 & =
 \frac{\mathsf{Y}_{i,\mu_i^{-1} x}}{\mathsf{Y}_{i-1,\mu_{i-1}^{-1} q_1^{-1}q_2^{-1}x}}
 \qquad (i = 1,\ldots,p)
 \, , \\
 \Lambda_{p+1,x}
 & =
 \frac{\mathsf{Y}_{p-1,\mu_{p-1}^{-1} q_1^{-2} q_2^{-1} x}}
      {\mathsf{Y}_{p,\mu_{p}^{-1} q_1^{-2} q_2^{-1} x}} 
 \, , \\
 \Lambda_{2p+1-i,x}
 & =
 \frac{\mathsf{Y}_{i-1,\mu_{i-1}^{-1} q_1^{-3} q_2^{-2} x}}
      {\mathsf{Y}_{i,\mu_{i}^{-1} q_1^{-3} q_2^{-2} x}}
 \qquad (i = 1,\ldots,p-1)
 \, ,
\end{align}
\end{subequations}
the fundamental $qq$-character is given by
\begin{align}
 \mathsf{T}_{1,x}
 & =
 \sum_{i=1}^{2p}
 \Lambda_{i,x}
\end{align}
which corresponds to the $2p$-dimensional representation of $\Sp(p) = G_{C_p}$.
Here we use the same notation for the mass parameter as before \eqref{eq:mass_prod_B}.

The $qq$-characters for $C_3$ quiver are explicitly given as follows:\index{qq-character@$qq$-character!C3@$C_3$}
\begin{subequations}
\begin{align}
 \mathsf{T}_{1,x}
 & =
 \mathsf{Y}_{1,x}
 +
 \frac{\mathsf{Y}_{2,\mu_2^{-1} x}}{\mathsf{Y}_{1,q_1^{-1} q_2^{-1} x}}
 +
 \frac{\mathsf{Y}_{3,\mu_3^{-1} x}}{\mathsf{Y}_{2,\mu_{2}^{-1} q_1^{-1} q_2^{-1} x}}
 +
 \frac{\mathsf{Y}_{2,\mu_2^{-1} q_1^{-2} q_2^{-1} x}}
      {\mathsf{Y}_{3,\mu_3^{-1} q_1^{-2} q_2^{-1} x}}
 +
 \frac{\mathsf{Y}_{1,q_1^{-3} q_2^{-2} x}}{\mathsf{Y}_{2,\mu_2^{-1} q_1^{-3} q_2^{-2} x}}
 +
 \frac{1}{\mathsf{Y}_{1,q_1^{-4} q_2^{-3} x}}
 \, ,
\end{align}
\begin{align}
 &
 \mathsf{T}_{2,\mu_2^{-1} x}
 \nonumber \\
 & =
 \mathsf{Y}_{2,\mu_2^{-1} x}
 +
 \frac{\mathsf{Y}_{1,q_1^{-1} q_2^{-1} x} \mathsf{Y}_{3,\mu_3^{-1} x}}
      {\mathsf{Y}_{2,\mu_2^{-1} q_1^{-1} q_2^{-1} x}}
 +
 \frac{\mathsf{Y}_{3,\mu_3^{-1} x}}{\mathsf{Y}_{1,q_1^{-2} q_2^{-2} x}}
 +
 \frac{\mathsf{Y}_{1,q_1^{-1} q_2^{-1} x} \mathsf{Y}_{2,\mu_2^{-1} q_1^{-2} q_2^{-1} x}}
      {\mathsf{Y}_{3,\mu_3^{-1} q_1^{-2} q_2^{-1} x}}
 +
 \frac{\mathsf{Y}_{2,\mu_2^{-1} q_1^{-1} q_2^{-1} x} \mathsf{Y}_{2,\mu_2^{-1} q_1^{-2} q_2^{-1} x}}
      {\mathsf{Y}_{1,q_1^{-2} q_2^{-2} x} \mathsf{Y}_{3,\mu_3^{-1} q_1^{-2} q_2^{-1} x}}
 \nonumber \\[.5em]
 & \quad
 +
 \frac{\mathsf{Y}_{1,q_1^{-1} q_2^{-1} x} \mathsf{Y}_{1,q_1^{-3} q_2^{-2} x}}
      {\mathsf{Y}_{2,\mu_2^{-1} q_1^{-3} q_2^{-2} x}}
 +
 \mathscr{S}(q_1)
 \frac{\mathsf{Y}_{1,q_1^{-3} q_2^{-2} x} \mathsf{Y}_{2,\mu_2^{-1} q_1^{-1} q_2^{-1} x}}
      {\mathsf{Y}_{1,q_1^{-2} q_2^{-2} x} \mathsf{Y}_{2,\mu_2^{-1} q_1^{-3} q_2^{-2} x}}
 +
 \mathscr{S}(q_1^2 q_2)
 \frac{\mathsf{Y}_{1,q_1^{-1} q_2^{-1} x}}{\mathsf{Y}_{1,q_1^{-4} q_2^{-3} x}}
 \nonumber \\[.5em]
 & \quad
 +
 \frac{\mathsf{Y}_{1,q_1^{-3} q_2^{-2} x} \mathsf{Y}_{3,\mu_3^{-1} q_1^{-1} q_2^{-1} x}}
      {\mathsf{Y}_{2,\mu_2^{-1} q_1^{-2} q_2^{-2} x} \mathsf{Y}_{2,\mu_2^{-1} q_1^{-3} q_2^{-2} x}}
 +
 \frac{\mathsf{Y}_{3,\mu_3^{-1} q_1^{-1} q_2^{-1} x}}
      {\mathsf{Y}_{1,q_1^{-4} q_2^{-3} x} \mathsf{Y}_{2,\mu_2^{-1} q_1^{-2} q_2^{-2} x}}
 +
 \frac{\mathsf{Y}_{1,q_1^{-3} q_2^{-2} x}}{\mathsf{Y}_{3,\mu_3^{-1} q_1^{-3} q_2^{-2} x}}
 +
 \frac{\mathsf{Y}_{2,\mu_2^{-1} q_1^{-1} q_2^{-1} x}}{\mathsf{Y}_{1,q_1^{-2} q_2^{-2} x} \mathsf{Y}_{1,q_1^{-4} q_2^{-3} x}}
 \nonumber \\[.5em]
 & \quad
 +
 \frac{\mathsf{Y}_{2,\mu_2^{-1} q_1^{-3} q_2^{-2} x}}
      {\mathsf{Y}_{1,q_1^{-4} q_2^{-3} x} \mathsf{Y}_{3,\mu_3^{-1} q_1^{-3} q_2^{-2} x}}
 +
 \frac{1}{\mathsf{Y}_{2,\mu_2^{-1} q_1^{-4} q_2^{-3} x}}
 \, ,
\end{align}
\begin{align}
 &
 \mathsf{T}_{3,\mu_3^{-1} x}
 \nonumber \\
 & =
 \mathsf{Y}_{3,\mu_3^{-1} x}
 +
 \frac{\mathsf{Y}_{2,\mu_2^{-1} q_1^{-1} q_2^{-1} x} \mathsf{Y}_{2,\mu_2^{-1} q_1^{-2} q_2^{-1} x}}
      {\mathsf{Y}_{3,\mu_3^{-1} q_1^{-2} q_2^{-1} x}}
 +
 \mathscr{S}(q_1)
 \frac{\mathsf{Y}_{1,q_1^{-3} q_2^{-2} x} \mathsf{Y}_{2,\mu_2^{-1} q_1^{-1} q_2^{-1} x}}
      {\mathsf{Y}_{2,\mu_2^{-1} q_1^{-3} q_2^{-2} x}}
 +
 \mathscr{S}(q_1)
 \frac{\mathsf{Y}_{2,\mu_2^{-1} q_1^{-1} q_2^{-1} x}}
      {\mathsf{Y}_{1,q_1^{-4} q_2^{-3} x}}
 \nonumber \\[.5em]
 & \quad
 +
 \frac{\mathsf{Y}_{1,q_1^{-2} q_2^{-2} x} \mathsf{Y}_{1,q_1^{-3} q_2^{-2} x}
       \mathsf{Y}_{3,\mu_3^{-1} q_1^{-1} q_2^{-1} x}}
      {\mathsf{Y}_{2,\mu_2^{-1} q_1^{-2} q_2^{-2} x} \mathsf{Y}_{2,\mu_2^{-1} q_1^{-3} q_2^{-2} x}}
 +
 \mathscr{S}(q_1)
 \frac{\mathsf{Y}_{1,q_1^{-2} q_2^{-2} x} \mathsf{Y}_{3,\mu_3^{-1} q_1^{-1} q_2^{-1} x}}
      {\mathsf{Y}_{1,q_1^{-4} q_2^{-3} x} \mathsf{Y}_{2,\mu_2^{-1} q_1^{-2} q_2^{-2} x}}
 +
 \frac{\mathsf{Y}_{1,q_1^{-2} q_2^{-2} x} \mathsf{Y}_{1,q_1^{-3} q_2^{-2} x}}
      {\mathsf{Y}_{3,\mu_3^{-1} q_1^{-3} q_2^{-2} x}}
 \nonumber \\[.5em]
 & \quad
 +
 \mathscr{S}(q_1)
 \frac{\mathsf{Y}_{1,q_1^{-2} q_2^{-2} x} \mathsf{Y}_{2,\mu_2^{-1} q_1^{-3} q_2^{-2} x}}
      {\mathsf{Y}_{1,q_1^{-4} q_2^{-3} x} \mathsf{Y}_{3,\mu_3^{-1} q_1^{-3} q_2^{-2} x}}
 +
 \frac{\mathsf{Y}_{3,\mu_3^{-1} q_1^{-1} q_2^{-1} x}}
      {\mathsf{Y}_{1,q_1^{-3} q_2^{-3} x} \mathsf{Y}_{1,q_1^{-4} q_2^{-3} x}}
 +
 \frac{\mathsf{Y}_{2,\mu_2^{-1} q_1^{-2} q_2^{-2} x} \mathsf{Y}_{2,\mu_2^{-1} q_1^{-3} q_2^{-2} x}}
      {\mathsf{Y}_{1,q_1^{-3} q_2^{-3} x} \mathsf{Y}_{1,q_1^{-4} q_2^{-3} x} \mathsf{Y}_{3,\mu_3^{-1} q_1^{-3} q_2^{-2} x}}
 \nonumber \\[.5em]
 & \quad
 +
 \mathscr{S}(q_1)
 \frac{\mathsf{Y}_{1,q_1^{-2} q_2^{-2} x}}{\mathsf{Y}_{2,\mu_2^{-1} q_1^{-4} q_2^{-3} x}}
 +
 \mathscr{S}(q_1)
 \frac{\mathsf{Y}_{2,\mu_2^{-1} q_1^{-2} q_2^{-2} x}}
      {\mathsf{Y}_{1,q_1^{-3} q_2^{-3} x} \mathsf{Y}_{2,\mu_2^{-1} q_1^{-4} q_2^{-3} x}}
 +
 \frac{\mathsf{Y}_{3,\mu_3^{-1} q_1^{-2} q_2^{-2} x}}
      {\mathsf{Y}_{2,\mu_2^{-1} q_1^{-3} q_2^{-3} x} \mathsf{Y}_{2,\mu_2^{-1} q_1^{-4} q_2^{-3} x}}
 +
 \frac{1}{\mathsf{Y}_{3,\mu_3^{-1} q_1^{-4} q_2^{-3} x}}
 \, .
\end{align}
\end{subequations}
They correspond to the 6, 14, and another 14 dimensional representations of $\Sp(3) = G_{C_3}$.

   \subsection{$G_2$ quiver}\label{sec:G2_quiv}

   We then consider $G_2$ quiver, which is the minimal-rank exceptional-type quiver:
   \begin{equation}
    \dynkin[mark=o,root radius = .15cm, edge length = 1.5cm, label, arrow color={black,length=3mm,width=6mm}]{G}{2}      
   \end{equation}
   with the root length parameter
   \begin{align}
    (d_1, d_2) = (3,1)
    \, .
   \end{align}
   The quiver Cartan matrix is in this case given by
   \begin{align}
    (c_{ij})
    =
    \begin{pmatrix}
     1 + q_1^{-3} q_2^{-1} & - \mu^{-1} \\
     - \mu q_1^{-1} q_2^{-1} ( 1 + q_1^{-1} + q_1^{-2} ) & 1 + q_1^{-1} q_2^{-1}
    \end{pmatrix}
    \ \xrightarrow{(c_{ij}^{[0]})} \
    \begin{pmatrix}
     2 & -1 \\ -3 & 2
    \end{pmatrix}
    \, ,
   \end{align}
   with the bifundamental mass parameter $\mu = \mu_{1 \to 2} = \mu_{2 \to 1}^{-1} q_1 q_2$, and its symmetrization
   \begin{align}
    (b_{ij})
    =
    \begin{pmatrix}
     (1 + q_1 + q_1^2)(1 + q_1^{-3} q_2^{-1}) & - \mu^{-1}(1 + q_1 + q_1^2) \\
     - \mu q_1^{-1} q_2^{-1} ( 1 + q_1^{-1} + q_1^{-2}) & 1 + q_1^{-1} q_2^{-1}
    \end{pmatrix}
    \ \xrightarrow{(b_{ij}^{[0]})} \
    \begin{pmatrix}
     6 & -3 \\ -3 & 2
    \end{pmatrix}
    \, .
   \end{align}
   Then iWeyl reflection is given by
   \begin{align}
    \text{iWeyl}: \
    \left(
    \mathsf{Y}_{1,x}, \mathsf{Y}_{2,x}
    \right)
    \ \longmapsto \
    \left(
    \frac{\mathsf{Y}_{2, \mu^{-1} x} \mathsf{Y}_{2,\mu^{-1} q_1^{-1} x} \mathsf{Y}_{2,\mu^{-1} q_1^{-2} x}}{\mathsf{Y}_{1,q_1^{-3}q_2^{-1} x}}
    \, , \,
    \frac{\mathsf{Y}_{1, \mu q_1^{-1} q_2^{-1} x}}{\mathsf{Y}_{2,q_1^{-1}q_2^{-1} x}}
    \right)
    \, ,
   \end{align}
     which generates the $\mathsf{T}$-operators\index{qq-character@$qq$-character!G2@$G_2$}
     \begin{subequations}\label{eq:G2_qq_ch}
       \begin{align}
	\mathsf{T}_{1,x}
	& =
	\mathsf{Y}_{1,x}
	+ \frac{\mathsf{Y}_{2,\mu^{-1}x} \mathsf{Y}_{2,\mu^{-1} q_1^{-1} x} \mathsf{Y}_{2,\mu^{-1} q_1^{-2} x}}{\mathsf{Y}_{1,q_1^{-3} q_2^{-1} x}}
	+ \frac{\mathsf{Y}_{1,q_1^{-1} q_2^{-1} x} \mathsf{Y}_{1,q_1^{-2} q_2^{-1} x}}{\mathsf{Y}_{2,\mu^{-1} q_1^{-1} q_2^{-1} x} \mathsf{Y}_{2,\mu^{-1} q_1^{-2} q_2^{-1} x} \mathsf{Y}_{2,\mu^{-1} q_1^{-3} q_2^{-1} x}}
	\nonumber \\
	& \quad
	+ \frac{\mathsf{Y}_{2,\mu^{-1} q_1^{-2} q_2^{-1} x} \mathsf{Y}_{2,\mu^{-1} q_1^{-3} q_2^{-1} x} \mathsf{Y}_{2,\mu^{-1} q_1^{-4} q_2^{-1} x}}{\mathsf{Y}_{1,q_1^{-4} q_2^{-2} x} \mathsf{Y}_{1,q_1^{-5} q_2^{-2} x}}
	+ \frac{\mathsf{Y}_{1,q_1^{-3} q_2^{-2} x}}{\mathsf{Y}_{2,\mu^{-1} q_1^{-3} q_2^{-2} x} \mathsf{Y}_{2,\mu^{-1} q_1^{-4} q_2^{-2} x} \mathsf{Y}_{2,\mu^{-1} q_1^{-5} q_2^{-2} x}}
	\nonumber \\
	& \quad
	+ \frac{1}{\mathsf{Y}_{1,q_1^{-6} q_2^{-3} x}}
	\nonumber \\
	& \quad
	+ \mathscr{S}_{1^2 2}(q_1)
	\Bigg[
	\frac{\mathsf{Y}_{2,\mu^{-1} x} \mathsf{Y}_{2,\mu^{-1} q_1^{-1} x}}{\mathsf{Y}_{2,\mu^{-1} q_1^{-3} q_2^{-1} x}}
	+ \frac{\mathsf{Y}_{1,q_1^{-2} q_2^{-1} x} \mathsf{Y}_{2,\mu^{-1} x}}{\mathsf{Y}_{2,\mu^{-1} q_1^{-2} q_2^{-1} x} \mathsf{Y}_{2,\mu^{-1} q_1^{-3} q_2^{-1} x}}
	+ \frac{\mathsf{Y}_{2,\mu^{-1} x} \mathsf{Y}_{2,\mu^{-1} q_1^{-4} q_2^{-1} x}}{\mathsf{Y}_{1,q_1^{-5} q_2^{-2} x}}
	\nonumber \\
	& \qquad
	+ \frac{\mathsf{Y}_{1,q_1^{-1} q_2^{-1} x}}{\mathsf{Y}_{2,\mu^{-1} q_1^{-1} q_2^{-1} x} \mathsf{Y}_{2,\mu^{-1} q_1^{-5} q_2^{-2} x}}
	+ \frac{\mathsf{Y}_{2,\mu^{-1} q_1^{-2} q_2^{-1} x} \mathsf{Y}_{2,\mu^{-1} q_1^{-3} q_2^{-1} x}}{\mathsf{Y}_{1,q_1^{-4} q_2^{-2} x} \mathsf{Y}_{2,\mu^{-1} q_1^{-5} q_2^{-2} x}}
	+ \frac{\mathsf{Y}_{2,\mu^{-1} q_1^{-4} q_2^{-1} x}}{\mathsf{Y}_{2,\mu^{-1} q_1^{-3} q_2^{-2} x} \mathsf{Y}_{2,\mu^{-1} q_1^{-4} q_2^{-2} x}}
	\Bigg]
	\nonumber \\
	& \quad
	+ \mathscr{S}_{1^3 2}(q_1)
	\frac{\mathsf{Y}_{1,q_1^{-1} q_2^{-1} x} \mathsf{Y}_{2,\mu^{-1} q_1^{-4} q_2^{-1} x}}{\mathsf{Y}_{1,q_1^{-5} q_2^{-2} x} \mathsf{Y}_{2,\mu^{-1} q_1^{-1} q_2^{-1} x}}
	+ \mathscr{S}_{1^3 2}(q_1^{-1})
	\frac{\mathsf{Y}_{1,q_1^{-2} q_2^{-1} x}}{\mathsf{Y}_{1,q_1^{-4} q_2^{-2} x}}
	\nonumber \\
	& \quad	
	+ \mathscr{S}_{1^2 2}(q_1) \mathscr{S}_{12}(q_1^4 q_2)
	\frac{\mathsf{Y}_{2,\mu^{-1} x}}{\mathsf{Y}_{2,\mu^{-1} q_1^{-5} q_2^{-2} x}}
	\, ,
       \end{align}
       \begin{align}
	\mathsf{T}_{2,x}
	& =
	\mathsf{Y}_{2,x}
	+ \frac{\mathsf{Y}_{1,\mu q_1^{-1} q_2^{-1} x}}{\mathsf{Y}_{2,q_1^{-1} q_2^{-1} x}}
	+ \frac{\mathsf{Y}_{2,q_1^{-2} q_2^{-1} x} \mathsf{Y}_{2,q_1^{-3} q_2^{-1} x}}{\mathsf{Y}_{1,\mu q_1^{-4} q_2^{-2} x}}
	\nonumber \\
	& \quad
	+ \mathscr{S}_{12}(q_1) \frac{\mathsf{Y}_{2,q_1^{-2} q_2^{-1} x}}{\mathsf{Y}_{2,q_1^{-4} q_2^{-2} x}}       
	+ \frac{\mathsf{Y}_{1,\mu q_1^{-3} q_2^{-2} x}}{\mathsf{Y}_{2,q_1^{-3} q_2^{-2} x} \mathsf{Y}_{q_1^{-4} q_2^{-2} x}}
	+ \frac{\mathsf{Y}_{2,q_1^{-5} q_2^{-2} x}}{\mathsf{Y}_{1,\mu q_1^{-6} q_2^{-3} x}}
	+ \frac{1}{\mathsf{Y}_{2,q_1^{-6} q_2^{-3} x}}
	\, .
       \end{align}
     \end{subequations}
     These $qq$-characters correspond to the \textbf{14} (with extension) and \textbf{7} representations of the exceptional Lie group $G_2$~\cite{Bouwknegt:1998da}.

     \subsection{NS$_{1,2}$ limit}\label{sec:NS_frac}

     Since the roles of $q_1$ and $q_2$ are not on equal footing for fractional quiver, there are two possible NS limits: the NS$_1$ limit $q_1 \to 1$, and the NS$_2$ limit $q_2 \to 1$.
     Then, the $qq$-character is reduced to $q_2$-character in the NS$_1$ limit, and $q_1$-character in the NS$_2$ limit.     
     We discuss these two NS limits of the $qq$-character and its relation to quantum integrable system.

     \subsubsection{$q_{1,2}$-character}
     \index{q-character@$q$-character}
     
     Let us examine $BC_2$ quiver discussed in \S\ref{sec:BC2_quiv}.
     The fundamental $qq$-characters are given in \eqref{eq:BC2_qq_ch}.
     Recalling the behavior of the $\mathscr{S}$-function in the NS limits~\eqref{eq:S_func_NS_lim}, we obtain the $q_1$-characters in the NS$_2$ limit,
     \begin{subequations}\label{eq:BC2_q1_ch}
      \begin{align}
       \mathsf{T}_{1,x}^{(q_1)}
       & =
       \mathsf{Y}_{1,x}
       + \frac{\mathsf{Y}_{2, x} \mathsf{Y}_{2,q_1^{-1} x}}{\mathsf{Y}_{1,q_1^{-2}x}}
       + \frac{\mathsf{Y}_{2,x}}{\mathsf{Y}_{2,q_1^{-2}x}}
       + \frac{\mathsf{Y}_{1,q_1^{-1}x}}{\mathsf{Y}_{2,q_1^{-1}x} \mathsf{Y}_{2,q_1^{-2}x} }
       + \frac{1}{\mathsf{Y}_{1,q_1^{-3}x}}
       \, , \\
    \mathsf{T}_{2,x}^{(q_1)}
    & =
    \mathsf{Y}_{2,x}
    + \frac{\mathsf{Y}_{1,q_1^{-1}x}}{\mathsf{Y}_{2,q_1^{-1}x}}
    + \frac{\mathsf{Y}_{2,q_1^{-2}x}}{\mathsf{Y}_{1, q_1^{-3}x}}
       + \frac{1}{\mathsf{Y}_{2,q_1^{-3}x}}
       \, ,
      \end{align}
     \end{subequations}
     and the $q_2$-characters in the NS$_1$ limit,
     \begin{subequations}
      \begin{align}
    \mathsf{T}_{1,x}^{(q_2)}
    & =
    \mathsf{Y}_{1,x}
    + \frac{\left(\mathsf{Y}_{2,x}\right)^2}{\mathsf{Y}_{1,q_2^{-1} x}}       
    + 2 \, \frac{\mathsf{Y}_{2,x}}{\mathsf{Y}_{2,q_2^{-1} x}}
       + \frac{\mathsf{Y}_{1,q_2^{-1} x}}{\left(\mathsf{Y}_{2,q_2^{-1} x}\right)^2}
       + \frac{1}{\mathsf{Y}_{1,q_2^{-2} x}}
       \, , \\
    \mathsf{T}_{2,x}^{(q_2)}
    & =
    \mathsf{Y}_{2,x}
    + \frac{\mathsf{Y}_{1,q_2^{-1} x}}{\mathsf{Y}_{2,q_2^{-1} x}}
    + \frac{\mathsf{Y}_{2,q_2^{-1} x}}{\mathsf{Y}_{1, q_2^{-2} x}}
       + \frac{1}{\mathsf{Y}_{2,q_2^{-2} x}}
       \, .
      \end{align}
     \end{subequations}
     Counting the dimension of the modules, the $q_1$-characters are still giving the \textbf{5} and \textbf{4} representations of $\SO(5)/\Sp(2)$, while the $q_2$-characters correspond to six and four dimensional representations with degenerated weight terms, in particular, in $\mathsf{T}_{1,x}^{(q_2)}$.
     This behavior of the $q_2$-characters has a natural interpretation in the relation to $AD_3$ quiver:
     They are obtained from the \textbf{6} and \textbf{4} representations of $\SU(4)/\SO(6) = G_{AD_3}$ via the folding with respect to the $\mathbb{Z}_2$-automorphism shown in \eqref{eq:A3_folding}.

     Let us then consider $G_2$ quiver.
     In this case, the fundamental $qq$-characters are given in \eqref{eq:G2_qq_ch}, and the corresponding $q_{1,2}$-characters are described as follows:
     \begin{subequations}
       \begin{align}
	\mathsf{T}_{1,x}^{(q_1)}
	& =
	\mathsf{Y}_{1,x}
	+ \frac{\mathsf{Y}_{2,\mu^{-1}x} \mathsf{Y}_{2,\mu^{-1} q_1^{-1} x} \mathsf{Y}_{2,\mu^{-1} q_1^{-2} x}}{\mathsf{Y}_{1,q_1^{-3} x}}
	+ \frac{\mathsf{Y}_{1,q_1^{-1} x} \mathsf{Y}_{1,q_1^{-2} x}}{\mathsf{Y}_{2,\mu^{-1} q_1^{-1} x} \mathsf{Y}_{2,\mu^{-1} q_1^{-2} x} \mathsf{Y}_{2,\mu^{-1} q_1^{-3} x}}
	\nonumber \\
	& \quad
	+ \frac{\mathsf{Y}_{2,\mu^{-1} q_1^{-2} x} \mathsf{Y}_{2,\mu^{-1} q_1^{-3} x} \mathsf{Y}_{2,\mu^{-1} q_1^{-4} x}}{\mathsf{Y}_{1,q_1^{-4} x} \mathsf{Y}_{1,q_1^{-5} x}}
	+ \frac{\mathsf{Y}_{1,q_1^{-3} x}}{\mathsf{Y}_{2,\mu^{-1} q_1^{-3} x} \mathsf{Y}_{2,\mu^{-1} q_1^{-4} x} \mathsf{Y}_{2,\mu^{-1} q_1^{-5} x}}
	\nonumber \\
	& \quad
	+ \frac{1}{\mathsf{Y}_{1,q_1^{-6} x}}
	+ 
	\frac{\mathsf{Y}_{2,\mu^{-1} x} \mathsf{Y}_{2,\mu^{-1} q_1^{-1} x}}{\mathsf{Y}_{2,\mu^{-1} q_1^{-3} x}}
	+ \frac{\mathsf{Y}_{1,q_1^{-2} x} \mathsf{Y}_{2,\mu^{-1} x}}{\mathsf{Y}_{2,\mu^{-1} q_1^{-2} x} \mathsf{Y}_{2,\mu^{-1} q_1^{-3} x}}
	+ \frac{\mathsf{Y}_{2,\mu^{-1} x} \mathsf{Y}_{2,\mu^{-1} q_1^{-4} x}}{\mathsf{Y}_{1,q_1^{-5} x}}
	\nonumber \\
	& \quad
	+ \frac{\mathsf{Y}_{1,q_1^{-1} x}}{\mathsf{Y}_{2,\mu^{-1} q_1^{-1}  x} \mathsf{Y}_{2,\mu^{-1} q_1^{-5} x}}
	+ \frac{\mathsf{Y}_{2,\mu^{-1} q_1^{-2} x} \mathsf{Y}_{2,\mu^{-1} q_1^{-3} x}}{\mathsf{Y}_{1,q_1^{-4} x} \mathsf{Y}_{2,\mu^{-1} q_1^{-5} x}}
	+ \frac{\mathsf{Y}_{2,\mu^{-1} q_1^{-4} x}}{\mathsf{Y}_{2,\mu^{-1} q_1^{-3} x} \mathsf{Y}_{2,\mu^{-1} q_1^{-4} x}}
	\nonumber \\
	& \quad
	+ 
	\frac{\mathsf{Y}_{1,q_1^{-1} x} \mathsf{Y}_{2,\mu^{-1} q_1^{-4} x}}{\mathsf{Y}_{1,q_1^{-5} x} \mathsf{Y}_{2,\mu^{-1} q_1^{-1} x}}
	+ 
	\frac{\mathsf{Y}_{1,q_1^{-2} x}}{\mathsf{Y}_{1,q_1^{-4} x}}
	+ 
	\frac{\mathsf{Y}_{2,\mu^{-1} x}}{\mathsf{Y}_{2,\mu^{-1} q_1^{-5} x}}
	\, ,
       \end{align}
       \begin{align}
	\mathsf{T}_{2,x}^{(q_1)}
	& =
	\mathsf{Y}_{2,x}
	+ \frac{\mathsf{Y}_{1,\mu q_1^{-1} x}}{\mathsf{Y}_{2,q_1^{-1} x}}
	+ \frac{\mathsf{Y}_{2,q_1^{-2} x} \mathsf{Y}_{2,q_1^{-3} x}}{\mathsf{Y}_{1,\mu q_1^{-4} x}}
	+ 
	\frac{\mathsf{Y}_{2,q_1^{-2} x}}{\mathsf{Y}_{2,q_1^{-4} x}}       
	+ \frac{\mathsf{Y}_{1,\mu q_1^{-3} x}}{\mathsf{Y}_{2,q_1^{-3} x} \mathsf{Y}_{q_1^{-4} x}}
	+ \frac{\mathsf{Y}_{2,q_1^{-5} x}}{\mathsf{Y}_{1,\mu q_1^{-6} x}}
	+ \frac{1}{\mathsf{Y}_{2,q_1^{-6} x}}
	\, .
       \end{align}
     \end{subequations}
     \begin{subequations}
       \begin{align}
	\mathsf{T}_{1,x}^{(q_2)}
	& =
	\mathsf{Y}_{1,x}
	+ \frac{\mathsf{Y}_{2,\mu^{-1}x}^3}{\mathsf{Y}_{1,q_2^{-1} x}}
	+ \frac{\mathsf{Y}_{1,q_2^{-1} x}^2}{\mathsf{Y}_{2,\mu^{-1} q_2^{-1} x}^3}
	+ \frac{\mathsf{Y}_{2,\mu^{-1} q_2^{-1} x}^3}{\mathsf{Y}_{1,q_2^{-2} x}^2}
	+ \frac{\mathsf{Y}_{1,q_2^{-2} x}}{\mathsf{Y}_{2,\mu^{-1} q_2^{-2} x}^3}
	+ \frac{1}{\mathsf{Y}_{1,q_2^{-3} x}}
	\nonumber \\
	& \quad
	+ 3
	\Bigg[
	\frac{\mathsf{Y}_{2,\mu^{-1} x}^2}{\mathsf{Y}_{2,\mu^{-1} q_2^{-1} x}}
	+ \frac{\mathsf{Y}_{1,q_2^{-1} x} \mathsf{Y}_{2,\mu^{-1} x}}{\mathsf{Y}_{2,\mu^{-1} q_2^{-1} x}^2}
	+ \frac{\mathsf{Y}_{2,\mu^{-1} x} \mathsf{Y}_{2,\mu^{-1} q_2^{-1} x}}{\mathsf{Y}_{1, q_2^{-2} x}}
	+ \frac{\mathsf{Y}_{1,q_2^{-1} x}}{\mathsf{Y}_{2,\mu^{-1} q_2^{-1} x} \mathsf{Y}_{2,\mu^{-1} q_2^{-2} x}}	
	\nonumber \\
	& \qquad
	+ \frac{\mathsf{Y}_{2,\mu^{-1} q_2^{-1} x}^2}{\mathsf{Y}_{1, q_2^{-2} x} \mathsf{Y}_{2,\mu^{-1} q_2^{-2} x}}
	+ \frac{\mathsf{Y}_{2,\mu^{-1} q_2^{-1} x}}{\mathsf{Y}_{2,\mu^{-1} q_2^{-2} x}^2}
	\Bigg]
	+ 4 \,
	\frac{\mathsf{Y}_{1, q_2^{-1} x} \mathsf{Y}_{2,\mu^{-1} q_2^{-1} x}}{\mathsf{Y}_{1, q_2^{-2} x} \mathsf{Y}_{2,\mu^{-1} q_2^{-1} x}}
	- 2 \,
	\frac{\mathsf{Y}_{1, q_2^{-1} x}}{\mathsf{Y}_{1, q_2^{-2} x}}
	+ 3 \,
	\frac{\mathsf{Y}_{2,\mu^{-1} x}}{\mathsf{Y}_{2,\mu^{-1} q_2^{-2} x}}
	\, ,
       \end{align}
       \begin{align}
	\mathsf{T}_{2,x}^{(q_2)}
	& =
	\mathsf{Y}_{2,x}
	+ \frac{\mathsf{Y}_{1,\mu q_2^{-1} x}}{\mathsf{Y}_{2, q_2^{-1} x}}
	+ \frac{\mathsf{Y}_{2, q_2^{-1} x}^2}{\mathsf{Y}_{1,\mu q_2^{-2} x}}
	+ 2 \, \frac{\mathsf{Y}_{2, q_2^{-1} x}}{\mathsf{Y}_{2, q_2^{-2} x}}       
	+ \frac{\mathsf{Y}_{1,\mu q_2^{-2} x}}{\mathsf{Y}_{2, q_2^{-2} x}^2}
	+ \frac{\mathsf{Y}_{2, q_2^{-2} x}}{\mathsf{Y}_{1,\mu q_2^{-3} x}}
	+ \frac{1}{\mathsf{Y}_{2, q_2^{-3} x}}
	\, .
       \end{align}
     \end{subequations}
     From these expressions, we see that the NS$_2$ limit provides the $q_1$-characters of the \textbf{14} with extension and \textbf{7} of $G_2$.
     On the other hand, the $q_2$-characters obtained in the NS$_1$ correspond to the \textbf{28} with extension and \textbf{8} representations of $\SO(8) = G_{D_4}$ via the $\mathbb{Z}_3$-folding:
	  \begin{equation}
	   \begin{tikzpicture}[baseline=(current bounding box.center)]
	    \tikzset{/Dynkin diagram/fold style/.style={stealth-stealth,thick, shorten <=2mm,shorten >=2mm,}}
	    \dynkin[mark=o,label,root radius = .2cm, edge length = 1cm,ply=3]{D}{4}
	   \end{tikzpicture}
	   \hspace{3em}
	   \tikz \draw [ultra thick,blue,-latex] (0,0) -- ++(1,0);
	   \hspace{3em}
	   \dynkin[mark=o,root radius = .15cm, edge length = 1.5cm, label, arrow color={black,length=3mm,width=6mm}]{G}{2}      
	   \label{eq:D4_folding}
	  \end{equation}	  

     \subsubsection{TQ-relation and quantization}

     We then discuss the relation to the quantum integrable systems as in \S\ref{sec:NS_integrability}.    
     We focus on the NS$_2$ limit of $BC_2$ quiver for the moment.
     In this case, we define the $\mathsf{Q}$-functions as follows:
     \begin{align}
      \mathsf{Y}_{1,x} = \frac{\mathsf{Q}_{1,0}}{\mathsf{Q}_{1,-2}}
      \, , \qquad
      \mathsf{Y}_{2,x} = \frac{\mathsf{Q}_{2,0}}{\mathsf{Q}_{2,-1}}
      \, ,
     \end{align}
     with
     \begin{align}
 \mathsf{Q}_{i,n} = \mathsf{Q}_{i,q_1^n x}
 \, .
     \end{align}     
     The $q_1$-characters~\eqref{eq:BC2_q1_ch} are written in terms of the $\mathsf{Q}$-functions in the form of
     \begin{subequations}
      \begin{align}
       \mathsf{T}_{1,0}^{(q_1)}
       & =
   \frac{\mathsf{Q}_{1,0}}{\mathsf{Q}_{1,-2}}
   + \frac{\mathsf{Q}_{1,-4}}{\mathsf{Q}_{1,-2}}
   \frac{\mathsf{Q}_{2,0}}{\mathsf{Q}_{2,-2}}
   + \frac{\mathsf{Q}_{2,0}}{\mathsf{Q}_{2,-1}}
   \frac{\mathsf{Q}_{2,-3}}{\mathsf{Q}_{2,-2}}
   + \frac{\mathsf{Q}_{1,-1}}{\mathsf{Q}_{1,-3}}
   \frac{\mathsf{Q}_{2,-3}}{\mathsf{Q}_{2,-1}}
       + \frac{\mathsf{Q}_{1,-5}}{\mathsf{Q}_{1,-3}}
       \, , \\
       \mathsf{T}_{2,0}^{(q_1)}
       & =
    \frac{\mathsf{Q}_{2,0}}{\mathsf{Q}_{2,-1}}
    + \frac{\mathsf{Q}_{1,-1}}{\mathsf{Q}_{1,-3}}
    \frac{\mathsf{Q}_{2,-2}}{\mathsf{Q}_{2,-1}}
    + \frac{\mathsf{Q}_{1,-5}}{\mathsf{Q}_{1,-3}}
    \frac{\mathsf{Q}_{2,-2}}{\mathsf{Q}_{2,-3}}
       + \frac{\mathsf{Q}_{2,-4}}{\mathsf{Q}_{2,-3}}
       \, ,
      \end{align}
     \end{subequations}
     These expressions are combined into the TQ-relation
  \begin{align}
   & \mathsf{Q}_{2,-1} \mathsf{Q}_{2,-4}
   - \mathsf{T}_{2,0} \mathsf{Q}_{2,-1} \mathsf{Q}_{2,-3}
   + \left(
     \mathsf{T}_{1,0} \mathsf{Q}_{2,-1} \mathsf{Q}_{2,-2}
   + \mathsf{Q}_{2,0} \mathsf{Q}_{2,-3}
   \right)
   \nonumber \\
   & \hspace{13em}  
   - \mathsf{T}_{2,-1} \mathsf{Q}_{2,0} \mathsf{Q}_{2,-2}
   + \mathsf{Q}_{2,1} \mathsf{Q}_{2,-2}
   = 0
   \, ,
  \end{align}
  which is schematically interpreted as quantization of the algebraic curve (quantum curve\index{quantum curve}) given as the zero locus of the algebraic function
  \begin{align}
   H(x,y) = y^4 - \mathsf{T}_{2,0} \, y^3 + (\mathsf{T}_{1,0} + 1) \, {y}^2 - \mathsf{T}_{2,-1} \, {y} + 1
   \, .
  \end{align}  
  This is the characteristic polynomial of the Lax matrix associated with the four-dimensional representation of $\SO(5)/\Sp(2)$.
  \index{Lax matrix}

 \subsubsection{Bethe equation}
 \index{Bethe equation}
 \index{saddle point analysis}

 Applying the saddle point analysis to the NS$_2$ limit of fractional quiver theory as in \S\ref{sec:Bethe_eq}, we obtain the Bethe equation for the spin chain with generic symmetry $G_\Gamma$ for $\Gamma = ABCDEFG$~\cite{Chen:2018ntf},
\begin{align}
 \frac{P_{i,x}}{\widetilde{P}_{i,x}}
 = - \mathfrak{q}_{i} \,
 \prod_{j \in \Gamma_0}
 \frac{\mathsf{Q}_{j,- \frac{1}{2} b_{ij}^{[0]}} }{\mathsf{Q}_{j,+ \frac{1}{2} b_{ij}^{[0]}}}
 \, ,
 \qquad
 x \in \mathcal{X}_i
 \, ,
\end{align}
where $(b_{ij}^{[0]})_{i,j \in \Gamma_0}$ is the (classical analog of) symmetrized Cartan matrix~\eqref{eq:Cartan_b}~\cite{Reshetikhin:1987bz}.
Compared to the simply-laced case~\eqref{eq:Bethe_eq}, the Cartan matrix is replaced by its symmetrization.

Applying the saddle point analysis to the NS$_1$ limit, on the other hand, we instead obtain a degenerated Bethe equation obtained from a naive folding trick from the corresponding simply-laced quiver.
For example, we obtain the following equation from $BC_2$ quiver in the NS$_1$ limit:
\begin{align}
 \frac{P_{1,x}}{\widetilde{P}_{1,x}}
 = - \mathfrak{q}_{1} \, \frac{\mathsf{Q}_{1,-1}}{\mathsf{Q}_{1,+1}}
 \left( \frac{\mathsf{Q}_{2,+1/2}}{\mathsf{Q}_{2,-1/2}} \right)^2
 \, , \qquad
 \frac{P_{2,x}}{\widetilde{P}_{2,x}}
 = - \mathfrak{q}_{2} \, \frac{\mathsf{Q}_{2,-1}}{\mathsf{Q}_{2,+1}}
 \frac{\mathsf{Q}_{1,+1/2}}{\mathsf{Q}_{1,-1/2}}
 \, .
\end{align}
In particular, the Bethe equation for the node $i = 1$ contains a degenerated factor $\left( \mathsf{Q}_{2,+1/2} / \mathsf{Q}_{2,-1/2} \right)^2$.
From the folding perspective, this term is obtained by identification of the nodes $i = 1, 3$ of $AD_3$ quiver.
See also~\cite{Dey:2016qqp} for a related discussion.

   \section{Affine quiver W-algebra}\label{sec:affine_quiv_W}

   The formalism of quiver W-algebra is applicable not only to the finite-type, but also affine quivers.
   We now see that a new family of W-algebras is constructed from affine quiver gauge theory.
   In this Section, we focus on the A-type affine quivers for simplicity, but the formalism discussed here is applicable to any affine quivers, including the twisted ones together with the fractional quiver formalism in \S\ref{sec:frac_W}.

   \subsection{\texorpdfstring{$\widehat{A}_0$ quiver}{Affine A0 quiver}}\label{sec:A0hat_algebra}

   We consider $\widehat{A}_0$ quiver, which is the simplest affine quiver.
   Let $\mu = \np^{m} \in \mathbb{C}^\times$ be the multiplicative adjoint mass parameter.
   The quiver Cartan matrix is given by
   \begin{align}
    c = 1 + q^{-1} - \mu^{-1} - \mu q^{-1} = (1 - \mu^{-1})(1 - \mu q^{-1}) = (1 - q_3)(1 - q_4)
   \end{align}
   where $q_{3,4}$ is defined in~\eqref{eq:q1234}.
   Then, the oscillators for the screening current~\eqref{eq:s_comm_rel}, the $\mathsf{Y}$-operator~\eqref{eq:y_comm_rel}, and the $\mathsf{A}$-operator~\eqref{eq:a_comm_rel} are given as follows:
   \begin{subequations}
   \begin{align}
  \left[ s_{1,n} \,,\, s_{1,n'} \right] 
  & = - \frac{1}{n} \frac{1 - q_1^n}{1 - q_2^{-n}} \, (1 - q_3^n)(1 - q_4^n) \, \delta_{n+n',0}
  \, , \\
  \left[ y_{1,n} \,,\, y_{1,n'} \right] 
  & =
- \frac{1}{n} \frac{(1 - q_1^n)(1 - q_2^n)}{(1 - q_3^{-n})(1 - q_4^{-n})} \, \delta_{n+n',0}
  \, , \\
  \left[ a_{1,n} \,,\, a_{1,n'} \right] 
  & =
  - \frac{1}{n} (1 - q_1^n)(1 - q_2^n)(1 - q_3^{n})(1 - q_4^n) \, \delta_{n+n',0}
  \, .
   \end{align}
   \end{subequations}
   In particular, the $a$-mode commutation relation is totally symmetric among $(q_1,q_2,q_3,q_4)$, and the OPE between the $\mathsf{A}$-operators is given by
   \begin{align}
    \mathsf{A}_{1,x} \mathsf{A}_{1,x'}
    & =
    \frac{\mathscr{S}(q_3 x'/x) \mathscr{S}(q_4 x'/x)}
         {\mathscr{S}(x'/x) \mathscr{S}(q_{34} x'/x)} \,
    {: \mathsf{A}_{1,x} \mathsf{A}_{1,x'} :}
    \nonumber \\
    & =
    \frac{1 - \left(x'/x\right)^\pm q_{12,23,31}}{1 - \left(x'/x\right)^\pm q_{1,2,3,4}}
    \left( 1 - \left(x'/x\right)^\pm \right) \,
    {: \mathsf{A}_{1,x} \mathsf{A}_{1,x'} :}
    \, .
    \label{eq:AA_OPE_A0}
   \end{align}
   We remark this OPE factor appears in the contour integral formula of the instanton partition function for $\widehat{A}_0$ quiver theory \eqref{eq:LMNS_formula_2*_sym}.
   See also the discussion in \S\ref{sec:LMNS_VOA}.
   The structure function is given in this case as
 \begin{align}
  f(z) = \exp \left( \sum_{n = 1}^\infty \frac{1}{n} \frac{(1 - q_1^n)(1 - q_2^n)}{(1 - q_3^n)(1 - q_4^n)} \, z^n \right)
 \end{align}
 obeying the relation
 \begin{align}
  \frac{f(z) f(q_{34} z)}{f(q_3 z) f(q_4 z)} = \mathscr{S}_{12}(z)
  \, .
 \end{align}
 We define an operator analog of the weight function~\eqref{eq:A0_module},
\begin{align}
 \Lambda_{\lambda} & =
 {:
\mathsf{Y}_{1,x}
 \prod_{(s_3,s_4) \in \lambda} \frac{\mathsf{Y}_{1,q_3^{s_3} q_4^{s_4-1} x} \mathsf{Y}_{1,q_3^{s_3 - 1} q_4^{s_4} x}}{\mathsf{Y}_{q_3^{s_3 - 1} q_4^{s_4 - 1} x} \mathsf{Y}_{1,q_3^{s_3} q_4^{s_4}x}}
 :} 
 \nonumber \\ &
 = 
 {:
 \prod_{(s_3,s_4) \in \partial_+ \lambda} \mathsf{Y}_{q_3^{s_3 - 1} q_4^{s_4 - 1} x}
 \prod_{(s_3,s_4) \in \partial_- \lambda} \mathsf{Y}^{-1}_{q_3^{s_3} q_4^{s_4} x}
 :}
 \, .
\end{align} 
 Then, the generating current of the affine quiver W-algebra W$_{q_{1,2},\mu}(\widehat{A}_0)$ is given by the $qq$-character derived in \S\ref{sec:qq_A0},\index{qq-character@$qq$-character!hatA0@$\widehat{A}_0$}
\begin{align}
 \mathsf{T}_{1,x}
 & =
 \sum_{\lambda} \mathfrak{q}_1^{|\lambda|} \, Z_{34}[\lambda] \, \Lambda_\lambda
 \, .
\end{align}
In contrast to the finite-type quiver W-algebras, the generating current is an infinite series of the weight operators since it is associated with the Fock module parametrized by the partition.

   \subsection{\texorpdfstring{$\widehat{A}_{p-1}$ quiver}{Affine Ap-1 quiver}}

   Let us briefly mention $\widehat{A}_{p-1}$ quiver W-algebra.
   In this case, the $qq$-character integral formula~\eqref{eq:qq_ch_int} in \S\ref{sec:qq_ch_int} agrees with the contour integral formula~\eqref{eq:quiv_var_Zp_2*} derived in \S\ref{sec:quiver_variety}.
   Hence, the generating currents $(\mathsf{T}_{i,x})_{i = 0,\ldots,p-1}$ are described as $\rU(1)$ instanton partition functions on $\widehat{A}_0$ quiver theory on $\mathbb{C}^2/\mathbb{Z}_p$ with equivariant parameters $\epsilon_{3,4}$.
   We remark that there are $p$ possible realizations of $\rU(1)$ theory with respect to the irreducible representation of $\mathbb{Z}_p$ as discussed in \S\ref{sec:ADHM_ALE}.

 \section{Integrating over quiver variety}
 \index{quiver!---variety}

 We have discussed the quantum algebraic structure of gauge theory based on the instanton partition function.
 As shown in \S\ref{sec:int_ADHM_var} and \S\ref{sec:quiver_partition_function}, the instanton partition function has the contour integral description as well.
 From this point of view, it is natural to ask what is the role of the quantum algebra in such a contour integral formula of the instanton partition function.
 In this Section, we show that the contour integral formula associated with the instanton moduli space and the quiver variety has a concise description in terms of the vertex operators used in the construction of quiver W-algebras~\cite{Kimura:2019hnw}.

 \subsection{Instanton partition function}\label{sec:LMNS_VOA}
 
 We start with the contour integral formula of the instanton partition function for quiver gauge theory~\eqref{eq:quiver_LMNS}.
 In particular, we are now interested in the 5d theory convention to show its relation to the quantum algebraic perspective.
 Using the reflection formula for the $\mathscr{S}$-function~\eqref{eq:S_func_reflection}, we have the following expression for the contour integral:\index{LMNS formula!quiver gauge theory}
\begin{align}
 Z_{\underline{n},\underline{k}}^\text{inst} & = 
 \prod_{i \in \Gamma_0} \frac{1}{k_i!} 
 \left(\frac{1 - q}{1 - q_{1,2}}\right)^{k_i}
 \oint_{\mathsf{T}_K}
 \prod_{x \in \overline{\mathcal{X}}}
 \frac{d x}{2 \pi \im x} \,
 \frac{P_{\mathsf{i}(x),x}^\text{f} \widetilde{P}_{\mathsf{i}(x),q x}^\text{af} }
 {P_{\mathsf{i}(x),x} \widetilde{P}_{\mathsf{i}(x),q x} }
 \prod_{e:\mathsf{i}(x) \to j} P_{j,\mu_e x}
 \prod_{e:j \to \mathsf{i}(x) } \widetilde{P}_{j,\mu_e^{-1} q x}
 \nonumber \\
 & \qquad \times
 \prod_{i \in \Gamma_0} \prod_{x \prec x' \in \overline{\mathcal{X}}_i \times \overline{\mathcal{X}}_i} 
 \mathscr{S} \left( \frac{x'}{x} \right)^{-1} \mathscr{S} \left( q^{-1} \frac{x'}{x} \right)^{-1} 
 \nonumber \\
 & \qquad \times
 \prod_{\substack{e \in \Gamma_1 \\ e:i \to j}}
 \prod_{(x,x') \in \overline{\mathcal{X}}_i \times \overline{\mathcal{X}}_j} 
 \mathscr{S}\left( \mu_e q^{-1} \frac{x'}{x} \right)
 \prod_{\substack{e \in \Gamma_1 \\ e:j \to i}}
 \prod_{(x,x') \in \overline{\mathcal{X}}_i \times \overline{\mathcal{X}}_j} 
 \mathscr{S}\left( \mu_e^{-1} \frac{x'}{x} \right)
 \, ,
\end{align}   
where we define a set of the integral variables,
\begin{align}
 \overline{\mathcal{X}} = \bigsqcup_{i \in \Gamma_0} \overline{\mathcal{X}}_i
 \, , \qquad
 \overline{\mathcal{X}}_i = \left\{ x_{i,\alpha,a} \right\}_{\alpha = 1,\ldots,n_i, a = 1,\ldots,k_i}
 \, .
\end{align}
Compared to the OPE between the $\mathsf{A}$-operators~\eqref{eq:AA_OPE}, all the $\mathscr{S}$-factors are reproduced from the $\mathsf{A}$-operator OPE, and the remaining factors are also obtained from the OPE between the operators $\mathsf{A}$ and $\mathsf{V}$.
Hence, we arrive at the chiral correlator representation of the contour integral formula:
\begin{align}
 Z_{\underline{n},\underline{k}}^\text{inst}
 = 
 \prod_{i \in \Gamma_0} \frac{1}{k_i!} 
 \left(\frac{1 - q}{1 - q_{1,2}}\right)^{k_i}
 \oint_{\mathsf{T}_K}
 \prod_{x \in \overline{\mathcal{X}}}
 \frac{d x}{2 \pi \im x} \,
 \vev{
 \mathsf{V}^{(\underline{n},\underline{n}^\text{af})}
 \mid
 {:\prod_{x \in \overline{\mathcal{X}}} \mathsf{A}_{\mathsf{i}(x),x}^{-1}:}
 \mid
 \mathsf{V}^{(\underline{n},\underline{n}^\text{f})} 
 }
 /
 \vev{\mathsf{V}^{(\underline{n},\underline{n}^\text{af})} \mid \mathsf{V}^{(\underline{n},\underline{n}^\text{f})}}
\end{align}
with
\begin{subequations}
\begin{align}
 {| \, \mathsf{V}^{(\underline{n},\underline{n}^\text{f})} \,\rangle}
 & =
 {:
 \prod_{x \in N} 
 \mathsf{V}_{\mathsf{i}(x),x} 
 \prod_{e:i \to j} \prod_{x \in N_j} \mathsf{V}_{i,x} 
 \prod_{x \in \mathcal{M}} \mathsf{V}_{\mathsf{i}(x), x}^{-1}
 :}\ket{0}
 \,, \\
 {\langle \, \mathsf{V}^{(\underline{n},\underline{n}^\text{af})} \,|}
 & =
 \bra{0}
 {: \
 \prod_{x \in N} 
 \mathsf{V}_{\mathsf{i}(x),q^{-1} x} 
 \prod_{e:j \to i} \prod_{x \in N_j} \mathsf{V}_{i,q^{-1} x} 
 \prod_{x \in \widetilde{\mathcal{M}}} \mathsf{V}_{\mathsf{i}(x),q^{-1} x}^{-1}
 :}
 \label{eq:V_state_contour}
\end{align}
\end{subequations}
where $(\mathcal{M},\widetilde{\mathcal{M}})$ is the set of the (anti)fundamental mass parameters~\eqref{eq:mass_parameters_quiv}, and we define a set of the Coulomb moduli parameters
\begin{align}
 N = \bigsqcup_{i \in \Gamma_0} N_i
 \, , \qquad
 N_i = \left( \np^{\mathsf{a}_{i,1}}, \ldots, \np^{\mathsf{a}_{i,n_i}} \right)
 \, .
\end{align}
We remark this definition of the $\mathsf{V}$-state~\eqref{eq:V_state_contour} is slightly different from the previous one~\eqref{eq:V_state}.

In order to express the contour integral more concisely, we define the $\mathsf{A}$-operator charge,\index{A-operator@$\mathsf{A}$-operator!---charge} denoted by $(\mathsf{W}_i)_{i \in \Gamma_0}$,
\begin{align}
 \mathsf{W}_i = \mathfrak{q}_i \,  \left( \frac{1 - q}{1 - q_{1,2}}  \right) \oint \frac{dx}{2 \pi \im x} \, \mathsf{A}_{i,x}^{-1}
 \, .
\end{align}
Then, the partition function has a formal expression
\begin{align}
 \mathfrak{q}^{\underline{k}} \, 
 Z_{\underline{n},\underline{k}}^\text{inst} 
 = 
 \prod_{i \in \Gamma_0} \frac{1}{k_i!} 
 \vev{
 \mathsf{V}^{(\underline{n},\underline{n}^\text{af})}
 \mid
 {: \prod_{i \in \Gamma_0} \mathsf{W}_i^{k_i} :}
 \mid
 \mathsf{V}^{(\underline{n},\underline{n}^\text{f})} 
 }
 /
 \vev{\mathsf{V}^{(\underline{n},\underline{n}^\text{af})} \mid \mathsf{V}^{(\underline{n},\underline{n}^\text{f})}}
 \, .
\end{align}
Furthermore, summation over all the topological sectors $k_i \in \mathbb{Z}_{\ge 0}$ leads to the exponential form,
\begin{align}
 Z & = \sum_{\underline{k}} \mathfrak{q}^{\underline{k}} \, Z_{\underline{n},\underline{k}}^\text{inst} 
 \nonumber \\
 & =
 \sum_{\underline{k}} 
 \prod_{i \in \Gamma_0} \frac{1}{k_i!} \,
 \vev{
 \mathsf{V}^{(\underline{n},\underline{n}^\text{af})}
 \mid
 {: \prod_{i \in \Gamma_0} \mathsf{W}_i^{k_i} :}
 \mid
 \mathsf{V}^{(\underline{n},\underline{n}^\text{f})} 
 }
 /
 \vev{\mathsf{V}^{(\underline{n},\underline{n}^\text{af})} \mid \mathsf{V}^{(\underline{n},\underline{n}^\text{f})}}
 \nonumber \\
 & = 
 \vev{
 \mathsf{V}^{(\underline{n},\underline{n}^\text{af})}
 \mid
 {: \prod_{i \in \Gamma_0} \np^{\mathsf{W}_i} :}
 \mid
 \mathsf{V}^{(\underline{n},\underline{n}^\text{f})} 
 }
 /
 \vev{\mathsf{V}^{(\underline{n},\underline{n}^\text{af})} \mid \mathsf{V}^{(\underline{n},\underline{n}^\text{f})}}
 \, .
\end{align}
Although it is a formal expression, we remark similarities of this expression to the several formulas in the literature, e.g., the Dotsenko--Fateev integral formula for the correlation function in 2d CFT~\cite{Dotsenko:1984nm,Dotsenko:1984ad}, the Fourier transform of the gauge theory partition function, known as the dual partition function~\cite{Nekrasov:2003rj}, the W-operator representation of the matrix integral~\cite{Morozov:2009xk}, and the partition function as the norm of the Gaiotto--Whittaker state~\cite{Gaiotto:2009ma}.
Actually such an expression is often found in the context of the integrable hierarchy as the corresponding $\tau$-function. It would be interesting to pursue the connection between the formula presented here and other similar formulas.

 \subsection{$qq$-character}\label{sec:qq_ch_VOA}

    Another application of the vertex operator formalism is the $qq$-character~\cite{Kimura:2019hnw}, which is given as the integral over the quiver variety as shown in \S\ref{sec:qq_ch_int}.
    In fact, all the factors in the integral formula~\eqref{eq:qq_ch_int} are reproduced by the OPE factors of the operators $\mathsf{Y}$ and $\mathsf{A}$ shown in~\eqref{eq:AA_OPE}, and thus we arrive at a simple integral formula
    \begin{align}
     \mathfrak{q}^{\underline{v}} \,
     \mathsf{T}_{\underline{w},\underline{v};\underline{x}}
     = 
     \prod_{i \in \Gamma_0} \frac{\mathfrak{q}_i^{v_i}}{v_i!}
     \frac{[-\epsilon_{12}]^{v_i}}{[-\epsilon_{1,2}]^{v_i}}
     \oint
     \prod_{\phi \in \underline{\phi}} \frac{d \phi}{2 \pi \im \phi} \,
     \mathsf{Y}_{\underline{w},\underline{x}}
     \mathsf{A}^{-1}_{\mathsf{i}(\phi),\phi}
     =
     \mathsf{Y}_{\underline{w},\underline{x}}
     \prod_{i \in \Gamma_0}
     \frac{1}{v_i!}
     \mathsf{W}_i^{v_i}
     \, .
    \end{align}
    Furthermore, the total $qq$-character, which is a summation over the dimension vector $\underline{v}$, is similarly expressed in the exponential form as before:\index{qq-character@$qq$-character}
    \begin{align}
     \mathsf{T}_{\underline{w},\underline{x}}
     = \sum_{\underline{v}} \mathfrak{q}^{\underline{v}} \, \mathsf{T}_{\underline{w},\underline{v};\underline{x}}
     = \mathsf{Y}_{\underline{w},\underline{x}}
     \prod_{i \in \Gamma_0} \np^{\mathsf{W}_i}
     \, .
     \label{eq:qq_ch_W}
    \end{align}
    From the representation theoretical point of view, $\mathsf{Y}_{\underline{w},\underline{x}}$ plays a role of the highest weight operator, and the summation over the dimension vector $\underline{v}$ corresponds to the summation over the Weyl orbit, and thus the operator $\displaystyle \prod_{i \in \Gamma_0} \np^{\mathsf{W}_i}$ is interpreted as the Weyl reflection generating operator, which terminates for the finite-type quiver, but generates an infinite series for the affine quiver.
    Since the $qq$-character plays a role of the generating current of the quiver W-algebra W$_{q_{1,2}}(\Gamma)$, all the generating currents are written in the form of~\eqref{eq:qq_ch_W}.
    In this sense, it provides a master formula for the generating currents of the quiver W-algebra W$_{q_{1,2}}(\Gamma)$.

  \chapter{Quiver elliptic W-algebra}\label{chap:eW}

  The quantum algebraic structure emerging from the moduli space of gauge theory is not unique to 5d $\mathcal{N} = 1$ gauge theory.
  As shown in \S\ref{sec:6dN=1}, 6d $\mathcal{N} = (1,0)$ gauge theory compactified on a torus has a similar geometric description, and it is natural to explore the underlying quantum algebraic structure of it.%
  \footnote{%
  The quantum algebra associated with 4d $\mathcal{N} = 2$ has been recently discussed~\cite{Nieri:2019mdl}.
  }
  From the correspondence between gauge theory and integrable system, 6d $\mathcal{N} = (1,0)$ theory corresponds to elliptic system. 
  In this Chapter, applying the approach used in 5d theory, we will show that an elliptic deformation of W-algebras, W$_{q_{1,2},p}(\mathfrak{g}_\Gamma)$, plays a role of the underlying algebraic structure of 6d $\mathcal{N} = (1,0)$ theory~\cite{Kimura:2016dys}.
  We will also mention a possible realization of elliptic theory based on the $q$-deformed vertex operator description, and relation to the elliptic quantum group.

  \section{Operator formalism}

  In order to see the underlying algebraic structure of 6d gauge theory compactified on a torus, we again apply the operator formalism discussed in Chapter~\ref{chap:operator}.
  We will show that the doubled Fock space is necessary to describe the elliptic correlation function.

  \subsection{Doubled Fock space}
  
  We define a pair of the Heisenberg algebras,
  \begin{align}
   \mathscr{H} = \bigoplus_{\sigma = \pm} \mathscr{H}_\sigma
   \, , \qquad
   \mathscr{H}_\sigma = (t_{i,n}^{(\sigma)}, \partial_{i,n}^{(\sigma)})_{i \in \Gamma_0, n \ge 1}
  \end{align}
  with the commutation relation
  \begin{align}
   [\partial_{i,n}^{(\sigma)} \,,\, t^{(\sigma')}_{j,n'}] = \delta_{ij} \, \delta_{n,n'} \, \delta_{\sigma,\sigma'}
   \, .
  \end{align}
  Then, we define the doubled Fock space generated by the Heisenberg algebras (the doubling trick~\cite{Clavelli:1973uk}, also related to the thermofield double):\index{doubling trick}
  \begin{align}
   \mathsf{F} = \bigoplus_{\sigma = \pm} \mathsf{F}_\sigma
   \, , \qquad
   \mathsf{F}_\sigma = \mathbb{C}[[t_{i,n}^{(\sigma)}, \partial_{i,n}^{(\sigma)}]] \ket{0}
   \, .
  \end{align}

  \subsection{Screening current}\label{sec:S_op_6d}
  \index{screening current!elliptic---}

  We define the doubled version of the screening current to construct the 6d gauge theory partition function,
  \begin{align}
   S_{i,x}
   =
   {: \exp \left(
   s_{i,0} + \tilde{s}_{i,0} - \frac{\kappa_i}{2} (\log q_2 x - 1) \log x
   + \sum_{n \in \mathbb{Z}_{\neq 0}}
   \left( s_{i,n}^{(+)} \, x^{-n} + s_{i,n}^{(-)} \, x^{+n} \right)
   \right)
   :}
  \end{align}
  where the $s$-modes are explicitly given as
  \begin{align}
   s_{i,-n}^{(\pm)} = \frac{1 - q_1^{\pm n}}{1 - p^{\pm n}} \, t_{i,n}^{(\pm)}
   \, , \qquad
   s_{i,0} = t_{i,0}
   \, , \qquad
   s_{i,n}^{(\pm)} = \mp \frac{1}{n} \frac{1}{1 - q_2^{\mp n}} \, c_{ji}^{[\pm n]} \, \partial_{j,n}^{(\pm)}
  \end{align}
  obeying the commutation relation
  \begin{align}
   \left[
   s_{i,n}^{(\pm)} \,,\, s_{j,n'}^{(\pm)}
   \right]
   = \mp \frac{1}{n} \frac{1 - q_1^{\pm n}}{(1 - p^{\pm n})(1 - q_2^{\mp n})} \, c_{ji}^{[\pm n]} \, \delta_{n+n',0}
   \, , \qquad
   \left[
   \tilde{s}_{i,0} \,,\, s_{j,n'}^{(\pm)} 
   \right]
   = - \beta \, c_{ji}^{[0]} \, \delta_{n,0}
   \, .
  \end{align}
  Then, following the same argument as in \S\ref{sec:screening_current}, the vev of the screening current pair is given by
  \index{operator product expansion (OPE)!S and S@$S$ and $S$}
  \begin{align}
   \bra{0} S_{i,x} S_{j,x'} \ket{0}
   & =
   \exp
   \left(
   - \beta \, c_{ji}^{[0]} \, \log x'
   - \sum_{m \neq 0}^\infty \frac{1-q_1^m}{m(1-p^m)(1-q_2^{-m})}
   c_{ji}^{[m]} \frac{x'^m}{x^m}
   \right)
   \, .
  \end{align}
  One can see that this pair contribution is rewritten in terms of the elliptic function summarized in \S\ref{sec:ell_fn}.  
  In this case, we have the same parameter shift as before~\eqref{eq:parameter_shift}, and we have to impose the condition for the bare Chern--Simons level
  \begin{align}
   \kappa_i = \sum_{j \in \Gamma_0} c_{ij}^{+[0]} \, n_j
   \label{eq:CS_level_6d}
  \end{align}
  to cancel the ``renormalized'' Chern--Simons level due to the shift~\eqref{eq:CS_level_shift}.

  \subsection{$Z$-state}

  The $t$-extended partition function for 6d gauge theory is given as a summation over the instanton configurations,
  \begin{align}
   Z(t) = \sum_{\mathcal{X} \in \mathfrak{M}^\mathsf{T}} Z_\mathcal{X}(t)
   \, ,
  \end{align}
  which defines the $Z$-state through the operator-state correspondence \index{operator-state correspondence}
  \begin{align}
   \ket{Z} = Z(t) \ket{0}
   \, .
  \end{align}
  We can apply the same argument as in~\S\ref{sec:screening_charge} to the current case, and we define the screening charge~\eqref{eq:screening_charge}.
  Then, we obtain the $t$-extended partition function as an infinite radial ordering product~\eqref{eq:radial_ordering_prod} over the screening charges\index{radial ordering}
  \begin{align}
   Z(t) =
   \prod_{x \in \mathring{\mathcal{X}}}^\succ \mathsf{S}_{\mathsf{i}(x),x}
   \, ,
  \end{align}
  and the corresponding $Z$-state
  \begin{align}
   \ket{Z} =
   \prod_{x \in \mathring{\mathcal{X}}}^\succ \mathsf{S}_{\mathsf{i}(x),x}
   \ket{0}
   \, .
  \end{align}
  The partition function at $t = 0$ is given as a chiral correlator of the screening charges by closing with the dual vacuum,
   \begin{align}
    Z =
    \bra{0}
    \prod_{x \in \mathring{\mathcal{X}}}^\succ \mathsf{S}_{\mathsf{i}(x),x}
    \ket{0}
    \, .
    \label{eq:Z-conf_block_6d}    
   \end{align}
   The fundamental matter contributions are similarly reproduced by the $\mathsf{V}$-operators as discussed in \S\ref{sec:V-op_6d}.

   We remark that the partition function consists of infinitely many screening charges.
   As discussed in \S\ref{sec:screening_charge}, the truncation corresponds to the Higgsing process to obtain 3d (2d) theory from 5d (4d) theory.
   Applying the same argument to 6d $\mathcal{N} = (1,0)$ theory, we obtain 4d $\mathcal{N} = 1$ theory on $\mathbb{C} \times T^2$.
   See \cite{Lodin:2017lrc} for a related discussion.

  \section{Trace formula}\label{sec:6d_trace}

  The chiral correlator for 6d theory~\eqref{eq:Z-conf_block_6d} can be written as the trace form in terms of the operators of the 5d theory (up to a normalization factor, which can be absorbed by redefinition of the gauge coupling constant),
\begin{align}
 Z & = \Tr \left[ p^{L_0}
 \prod_{x \in \mathring{\mathcal{X}}}^\succ
 \mathsf{S}_{\mathsf{i}(x),x}^\text{5d}
 \right]
 \, .
 \label{eq:Tr_formula}
\end{align}
Here the trace is taken over the Fock space $\mathsf{F}$ with respect to the 5d time variables $(t_{i,n})_{i \in \Gamma_0, n\in \mathbb{Z}_{>0}}$, and the screening charge is also defined with the oscillators used in 5d theory discussed in Chapter~\ref{chap:operator}. \index{time variable}
The energy operator $L_0$ is defined
\begin{align}
 L_0 & =
 \sum_{i \in \Gamma_0} \sum_{n=1}^\infty
 n \, t_{i,n} \, \partial_{i,n}
 \, .
\end{align}
The derivation of the trace formula~\eqref{eq:Tr_formula} is presented in the following.

Recall that the screening current correlator which gives the 5d gauge
theory partition function is
\begin{align}
 \bra{0}
 S_{i,x}^\text{5d} S_{j,x'}^\text{5d}
 \ket{0}
 & =
 \exp
 \left(
 - \sum_{m=1}^\infty \frac{1}{m} \frac{1-q_1^m}{1-q_2^{-m}}
 c_{ji}^{[m]} \frac{x'^m}{x^m}
 \right)
 \, .
 \label{eq:SS-pair-5d0}
\end{align}
Here we omit the zero mode contribution for brevity. 
There are two options to deform the 5d index to the elliptic 6d index~\eqref{eq:[x]_function}.

The first option is to modify the oscillator algebra in such a way that the normal
ordering produces the elliptic correlation function, as defined in \S\ref{sec:S_op_6d},
\begin{align}
 \bra{0}
 S_{i,x}^{\mathrm{6d}} S_{j,x'}^{\mathrm{6d}}
 \ket{0}
 & =
 \exp
 \left(
 - \sum_{m \neq 0} \frac{1-q_1^m}{m(1-p^m)(1-q_2^{-m})}
 c_{ji}^{[m]} \frac{x'^m}{x^m}
 \right)
 \, .
 \label{eq:SS-pair-5d}
\end{align}
The second option is to keep the free field oscillator commutation relations of the 5d theory, but change the definition of the correlation function to the trace as follows: 
\begin{align}
 \Big< S_{i,x}^\text{5d} S_{j,x'}^\text{5d} \Big>_\text{torus}
 =
 \Tr \Big[
 p^{L_0} S_{i,x}^\text{5d} S_{j,x'}^\text{5d}
 \Big]
 \, .
 \label{eq:SS-pair-6d}
\end{align}
The proof of the equivalence 
\begin{align}
 \bra{0} S_{i,x}^{\mathrm{6d}} S_{j,x'}^{\mathrm{6d}} \ket{0}
 &  =
 \Tr \Big[
 p^{L_0} S_{i,x}^\text{5d} S_{j,x'}^\text{5d}
 \Big]
 \, ,
 \label{eq:SS-paier-eq}
\end{align}
is shown in \S\ref{sec:torus_corr_fn}.
Then the trace formula~\eqref{eq:Tr_formula} follows.

\begin{figure}[t]
 \centering
 \begin{tikzpicture}[thick,scale=.8]

  \draw (0,0) -- (3,0)
  arc [start angle = -90, end angle = 90, x radius = .3, y radius = 1]
  -- (0,2)
  arc [start angle = 90, end angle = -270, x radius = .3, y radius = 1];


  \draw (5,0)
  arc [start angle = -90, end angle = 90, x radius = 1, y radius = 1]
  arc [start angle = 90, end angle = -270, x radius = .3, y radius = 1];

  \draw (-2,0)
  arc [start angle = -90, end angle = 90, x radius = .3, y radius = 1]
  arc [start angle = 90, end angle = 270, x radius = 1, y radius = 1];

  \draw [dotted] (-2,0)
  arc [start angle = -90, end angle = -270, x radius = .3, y radius = 1];

  \node at (-1,1) {$\times$};
  \node at (4,1) {$\times$};

  \node at (-1,-1.5) {$\times$};
  \node at (4,-1.5) {$\times$};  


  \node at (-1.6,-1.5) [left] {$\bra{0}$};
  \node at (4.6,-1.5) [right] {$\ket{0}$};  
  \node at (1.5,-1.5)
  {$\displaystyle \prod_{x \in \mathring{\mathcal{X}}}^\succ
    \mathsf{S}_{\mathsf{i}(x),x}^\text{5d} $};

  \node at (7,1) {$=$};

  \begin{scope}[shift={(11,0)}]

   \draw (0,0) -- (3,0)
   arc [start angle = -90, end angle = 90, x radius = 1, y radius = 1]
   -- (0,2)
   arc [start angle = 90, end angle = -270, x radius = .3, y radius = 1];
   
  \draw (-2,0)
  arc [start angle = -90, end angle = 90, x radius = .3, y radius = 1]
  arc [start angle = 90, end angle = 270, x radius = 1, y radius = 1];

  \draw [dotted] (-2,0)
  arc [start angle = -90, end angle = -270, x radius = .3, y radius = 1];

  \node at (-1,1) {$\times$};
  \node at (-1,-1.5) {$\times$};

   \node at (-1.6,-1.5) [left] {$\bra{0}$};
   \node at (1.7,-1.5)
   {$\displaystyle \ket{Z^\text{5d}}$};
   
  \end{scope}

  \begin{scope}[shift={(0,-6.5)}]

  \draw (-.5,0) -- (1.5,0)
  arc [start angle = -90, end angle = 90, x radius = .3, y radius = 1]
  -- (-.5,2)
  arc [start angle = 90, end angle = -270, x radius = .3, y radius = 1];

  \draw (2.25,0) -- ++(2,0)
  arc [start angle = -90, end angle = 90, x radius = .3, y radius = 1]
  -- (2.25,2)
  arc [start angle = 90, end angle = -270, x radius = .3, y radius = 1];

  \draw (5,0) -- ++(2,0)
  arc [start angle = -90, end angle = 90, x radius = .3, y radius = 1]
  -- (5,2)
  arc [start angle = 90, end angle = -270, x radius = .3, y radius = 1];
   
  \draw (8.5,0)
  arc [start angle = -90, end angle = 90, x radius = 1, y radius = 1]
  arc [start angle = 90, end angle = -270, x radius = .3, y radius = 1];

  \draw (-2,0)
  arc [start angle = -90, end angle = 90, x radius = .3, y radius = 1]
  arc [start angle = 90, end angle = 270, x radius = 1, y radius = 1];

  \draw [dotted] (-2,0)
  arc [start angle = -90, end angle = -270, x radius = .3, y radius = 1];

   \node at (-1.2,1) {$\cdots$};
   \node at (7.8,1) {$\cdots$};

   \node at (-1.2,2.5) {$\cdots$};
   \node at (7.8,2.5) {$\cdots$};   

   \node at (10.5,1) {$=$};
   \node at (10.5,-1.7) {$=$};   

   \node at (-1.2,-1.7) {$\times$};
   \node at (7.8,-1.7) {$\times$};

   \node at (-1.6,-1.7) [left] {$\bra{0}$};
   \node at (8.3,-1.7) [right] {$\ket{0}$};  
   \node at (3.25,-1.7)
   {$\displaystyle \prod_{x \in \mathring{\mathcal{X}}}^\succ
   \mathsf{S}_{\mathsf{i}(x),x}^\text{6d} $};

   \draw
   [decorate,decoration={brace,amplitude=10pt,raise=-4pt}]
   (7.8,-.5) -- (-1.3,-.5) node [black,midway,yshift=-0.6pt] {};
   

   \node at (.5,2.5) {$p^{-1}x$};
   \node at (3.25,2.5) {$x$};
   \node at (6,2.5) {$p x$};   
   
  \end{scope}

  \begin{scope}[shift={(13,-6.5)}]
   
   \draw (1,0) -- ++(1.5,0)
   arc [start angle = -90, end angle = 90, x radius = .3, y radius = 1]
   -- (1,2)
   arc [start angle = 90, end angle = -270, x radius = .3, y radius = 1];

   \draw [thick] (1,1)
   circle [x radius = .3, y radius = 1];

   \draw [thick] (2.5,0)
   arc [start angle = -90, end angle = 90, x radius = .3, y radius = 1];

   \draw [thick,dotted] (2.5,0)
   arc [start angle = -90, end angle = -270, x radius = .3, y radius = 1];

   \node at (-1.5,1) {Tr};

   \node at (-.15,1) {$p^{L_0}$ $\times$};

   \draw (-.8,-.5) -- (-1,-.5) -- (-1,2.5) -- (-.8,2.5);
   \draw (3,-.5) -- (3.2,-.5) -- (3.2,2.5) -- (3,2.5);

   \node at (1.3,-1.7) {\eqref{eq:Tr_formula}};

  \end{scope}
  
 \end{tikzpicture}
 \caption{Conformal blocks as the partition function of 5d (top) and 6d theory (bottom). The 6d block has two equivalent expressions.}
 \label{fig:conf_block}
\end{figure}

The physical meaning is as follows. For 5d gauge theory we use the cylindrical spacetime to compute the partition function, as shown in the top panel of Fig.~\ref{fig:conf_block}.
For 6d gauge theory we use the toric spacetime obtained by the identification~\eqref{eq:elliptic_coord}, illustrated in the LHS of Fig.~\ref{fig:conf_block} (bottom).
This corresponds to \eqref{eq:Z-conf_block_6d}, and is equivalent to taking the trace with the operator $p^{L_0}$ inserted.
This trace formula~\eqref{eq:Tr_formula} is also consistent with the $S$-duality in elliptic theory~\cite{Mironov:2016cyq,Iqbal:2015fvd, Nieri:2015dts} discussed in \S\ref{sec:6dN=1} because the dual theory is $\mathcal{N}=2^*$ theory (or cyclic quiver theory), whose partition function is given by the torus conformal block via the $q$-version of the AGT relation~\cite{Alday:2009aq,Awata:2009ur}.\index{AGT relation}

\subsection{Coherent state basis}

In order to obtain the torus correlation function to show the equivalence~\eqref{eq:SS-paier-eq}, we introduce the coherent state basis.
The argument in this part is essentially based on the textbook~\cite{GSW:ST1987}.

For the oscillator algebra generated by $(t, \partial_{t})$ with $[\partial_{t}, t]= 1$, we consider the coherent state basis in the Fock space
\begin{align}
 \ket{n} = \frac{t^n}{\sqrt{n!}} \ket{0}
 \, , \qquad
 \bra{n} = \bra{0} \frac{\partial^n}{\sqrt{n!}}
 \, , \qquad 
 |z) = e^{zt} \ket{0}
 \, , \qquad
 (z| = \bra{0} e^{z^* \partial}
 \, .
 \label{eq:basis}
\end{align}
The normalization is 
\begin{align}
 \vev{n \mid m} = \delta_{n,m}
 \, , \qquad
 \cvev{ z \mid w } = e^{z^* w}
 \, .
\end{align}
The states in \eqref{eq:basis} are eigenstates of the filling number operator $t \partial \ket{n} = 
n \ket{n}$ and the annihilation/creation operators $\partial |z) = z |z)$, $(z| t = (z| z^*$.
Notice that the operator $a^{t \partial_t}$ acts on the states $|z)$ and $(z|$ as,
\begin{align}
 a^{t \partial_t} |z) & = |az)
 \, , \qquad
 (z| a^{t \partial_t} = (a^* z|
 \, .
\end{align}

The identity operator can be expressed in terms of the coherent state basis:
\begin{align}
 \mathbbm{1} & =
 \frac{1}{\pi} \int d^2 z \,
 |z) e^{-|z|^2} (z|
\end{align}
where
\begin{align}
 \vev{n \mid \mathbbm{1} \mid m} = \delta_{n,m}
 \, ,
\end{align}
so that the trace of an operator over the Fock space is given by
\begin{align}
 \Tr \mathcal{O}
 & =
 \frac{1}{\pi} \int d^2 z \, e^{-|z|^2} \cvev{z \mid \mathcal{O} \mid z}
 \, .
\end{align}
Then we find the formula~\cite{Yamada:2006CFT}
\begin{align}
 \Tr \left[ a^{t \partial} e^{b \partial} e^{c t} \right]
 & =
 \frac{1}{1-a} \exp \left( \frac{abc}{1-a} \right)
 \label{eq:abc_formula}
\end{align}
since we have
\begin{align}
 \frac{1}{\pi} \int d^2 z \, e^{-|z|^2}
 \cvev{z \mid a^{t\partial_t} e^{bt} e^{c\partial_t} \mid z}
 & =
 \frac{1}{\pi} \int d^2 z \,
 e^{-(1-a)|z|^2 + ab z^* + cz}
 \, .
\end{align}

\subsection{Torus correlation function}\label{sec:torus_corr_fn}

Let us compute the torus correlation function~\eqref{eq:SS-pair-6d}.
The product of the 5d screening currents is given by
\begin{align}
 S_{i,x}^\text{5d} S_{j,x'}^\text{5d}
 & =
 \exp
 \left(
 - \sum_{m=1}^\infty \frac{1}{m} \frac{1-q_1^m}{1-q_2^{-m}}
 c_{ji}^{[m]} 
 \frac{x'^m}{x^m}
 \right)
 : S_{i,x}^\text{5d} S_{j,x'}^\text{5d} :
 \, .
\end{align}
Then we  compute the trace part
\begin{align}
 \Tr
 \left[
  p^{L_0} : S_{i,x}^\text{5d} S_{j,x'}^\text{5d} :
 \right]
 & =
 \Tr \Bigg[
 \left(
 \prod_{i' \in \Gamma_0}
 \prod_{n = 1}^\infty p^{n t_{i',n} \partial_{i',n}}
 \right)
 \exp \left(
  \sum_{n=1}^\infty (1-q_1^n) \left(x^n t_{i,n} + x'^n t_{j,n}\right)
 \right)
 \nonumber \\
 & \qquad \times
 \exp \left(
 \sum_{n=1}^\infty - \frac{1}{n(1-q_2^{-n})}
 \left(
 x^{-n} c_{ki}^{[n]} 
 \partial_{k,n}
 + x'^{-n} c_{lj}^{[n]} 
 \partial_{l,n}
 \right)
 \right) \Bigg]
 \nonumber \\ 
 & =
 \exp
 \left(
 \sum_{n=1}^\infty
 \left(
 - \frac{1-q_1^n}{n(1-q_2^{-n})}
 \frac{p^n}{1-p^n}
 c_{ji}^{[n]} 
 \frac{x'^n}{x^n} 
 +
 \frac{1-q_1^{-n}}{n(1-q_2^{n})}
 \frac{1}{1-p^{-n}}
 c_{ji}^{[-n]} 
 \frac{x^{n}}{x'^{n}}
 \right)
 \right)
 \nonumber \\
 & \qquad \times \text{const} 
\end{align}
where we have used the formulas \eqref{eq:abc_formula} and~\eqref{eq:quiv_Cartan_reflection}, and the constant term does not contain $x$ nor $x'$.
Thus we obtain the torus correlator
\begin{align}
 \Tr
 \left[
  p^{L_0} S_{i,x}^\text{5d} S_{j,x'}^\text{5d}
 \right]
 & =
 \exp
 \left(
 - \sum_{n \neq 0} \frac{1-q_1^n}{n(1-q_2^{-n})(1-p^n)}
 c_{ji}^{[n]} 
 \frac{x'^n}{x^n} 
 \right)
 \, .
\end{align}
This is equivalent to \eqref{eq:SS-pair-5d}, and proves the relation \eqref{eq:SS-paier-eq}.

\subsection{Connection to elliptic quantum group}

 It has been known that the $q$-deformation of W-algebra has a close connection with the elliptic quantum algebra U$_{q,p}(\widehat{\mathfrak{g}})$:
 The screening current of W$_{q_{1,2}}(\mathfrak{g})$ obeys essentially the same relation to the elliptic currents $e_i(z)$ and $f_i(z)$ of U$_{q,p}(\widehat{\mathfrak{g}})$~\cite{Feigin:1995sf}.%
\footnote{%
See also \cite{Konno:2020} for a recent monograph on the elliptic quantum group.
}
The relations for generic $\mathfrak{g}$ are found in~\cite{Farghly:2015ART}.
We see from \eqref{eq:SS-pair-5d0} that the 5d screening current yields
\begin{align}
 S_{i,x}^\text{5d} S_{j,x'}^\text{5d}
 & = S_{j,x'}^\text{5d} S_{i,x}^\text{5d} \times \exp
 \left(
 - \sum_{m \neq 0} \frac{1}{m} \frac{1 - q_1^m}{1 - q_2^{-m}}
 c_{ji}^{[m]} \left( \frac{x'}{x} \right)^m
 \right)
\end{align}
where we omitted the zero mode factors for simplicity.
One can rewrite the OPE factor using the theta function \eqref{eq:theta_fn} as in \S\ref{sec:boundary_dof}.
Swapping $q_1 \leftrightarrow q_2$ corresponds to swapping the currents $e_i(z) \leftrightarrow f_i(z)$.

From \eqref{eq:SS-pair-5d}, on the other hand, we obtain exactly the same relation for the 6d screening currents
\begin{align}
 S_{i,x}^\text{6d} S_{j,x'}^\text{6d}
 & = S_{j,x'}^\text{6d} S_{i,x}^\text{6d} \times \exp
 \left(
 - \sum_{m \neq 0} \frac{1}{m} \frac{1 - q_1^m}{1 - q_2^{-m}}
 c_{ji}^{[m]} \left( \frac{x'}{x} \right)^m
\right) 
\end{align}
This coincidence implies that both the $q$-deformation W$_{q_{1,2}}(\mathfrak{g})$ and the elliptic deformation W$_{q_{1,2},p}(\mathfrak{g})$ belong to the same realization of the elliptic quantum algebra U$_{q,p}(\widehat{\mathfrak{g}})$.

\section{More on elliptic vertex operators}

We then discuss the elliptic vertex operators to explore the algebraic structure for 6d gauge theory.
We will see that the previous argument in 5d is almost applicable to the present case.

\subsection{$\mathsf{V}$-operator}\label{sec:V-op_6d}
\index{V-operator@$\mathsf{V}$-operator!elliptic---}

We can incorporate the (anti)fundamental matter contribution in the operator formalism by considering another vertex operator,
\begin{align}
 \mathsf{V}_{i,x}
 & =
 \ : \exp
 \left(
  \sum_{n \neq 0}
  \left( v_{i,n}^{(+)} x^{-m} + v_{i,n}^{(-)} x^{+m} \right)
 \right):
 \, .
\end{align}
The oscillators are taken to be
\begin{align}
 v_{i,-n}^{(\pm)} & \stackrel{n > 0}{=}
 - \frac{1}{1-p^{\pm n}}
 \, \tilde{c}_{ji}^{[\pm n]} \, t_{j,n}^{(\pm)}
 \, , \qquad
 v_{i,n}^{(\pm)} \stackrel{n > 0}{=}
 \pm \frac{1}{n}
 \frac{1}{\left(1-q_1^{\pm n}\right)\left(1-q_2^{\pm n}\right)}
 \partial_{i,n}^{(\pm)}
 \, ,
\end{align}
with the commutation relation
\begin{align}
 \left[
  v_{i,n}^{(\pm)} \, , \, s_{j,n'}^{(\pm)}
 \right]
 & =
 \pm \frac{1}{n(1-p^{\pm n})(1-q_2^{\pm n})} \,
 \delta_{ij} \, \delta_{n+n',0} 
 \, .
\end{align}
The product of the $\mathsf{V}$-operator and the screening current behaves \index{operator product expansion (OPE)!S and V@$S$ and $\mathsf{V}$}
\begin{subequations}
 \begin{align}
  S_{i,x'} \mathsf{V}_{i,x} 
  & =
  \Gamma_e\left( q_2 \frac{x}{x'}; p, q_2 \right)^{-1}
  : \mathsf{V}_{i,x} S_{i,x'} :
  \, , \\
  \mathsf{V}_{i,x} S_{i,x'}
  & =
  \Gamma_e\left( \frac{x'}{x}; p, q_2 \right)
  : \mathsf{V}_{i,x} S_{i,x'} :
  \, ,
 \end{align}
\end{subequations}
which corresponds to the fundamental and antifundamental matter factors, respectively, while the OPE of the $\mathsf{V}$-operators does not yield dynamical contribution.
The $t$-extended partition function with the (anti)fundamental matter factors is given by
\begin{align}
 \ket{Z} & =
 {:
 \prod_{x \in \widetilde{\mathcal{M}}}
 \mathsf{V}_{\mathsf{i}(x),q^{-1}x}
 :}
 \left(
 \prod_{x \in \mathring{\mathcal{X}}}^\succ
 \mathsf{S}_{\mathsf{i}(x),x}
 \right)
 {:
 \prod_{x \in \mathcal{M}}
 \mathsf{V}_{\mathsf{i}(x),x}
 :}
 \ket{0}
 \label{eq:Z-state_matter}
\end{align}
which is formally equivalent to the previous expression~\eqref{eq:Z_state_matter}.

We again remark that, for the modular invariance of the non-extended partition function $Z = \vev{0 \mid Z}$, which is a conformal block of W$(\Gamma)$-algebra, we have to take into account the conformal condition~\eqref{eq:6d_conformal_cond}, although the $Z$-state \eqref{eq:Z-state_matter} is not necessarily modular invariant by itself.

\subsection{$\mathsf{Y}$-operator}
\index{Y-operator@$\mathsf{Y}$-operator!elliptic---}

In order to construct the W-algebras we define the elliptic analog of the $\mathsf{Y}$-operators, 
\begin{align}
 \mathsf{Y}_{i,x} & = \
 q_1^{\tilde{\rho}_i}
 :
 \exp
 \left(
 y_{i,0} + \left( \tilde{c}_{ji} \right)^{[0]} \kappa_j \log x
 + \sum_{n \in \mathbb{Z}_{\neq 0}}
 \left( y_{i,n}^{(+)} \, x^{-n} + y_{i,n}^{(-)} \, x^{+n} \right)
 \right)
 :
 \, ,
\end{align}
where $\tilde{\rho}_i$ is the Weyl vector defined by $\tilde{\rho}_i =
\sum_{j \in \Gamma_0} \tilde{c}_{ji}^{[0]}$, and $(\tilde{c}_{ij})$ is the
inverse of mass-deformed Cartan matrix $(c_{ij})$ as before.
In the following, we impose the condition \eqref{eq:CS_level_6d} for the Chern--Simons levels.

The oscillators $y_{i,n}^{(\pm)}$ are defined in terms of $(t_{i,n}^{(\pm)},\partial_{i,n}^{(\pm)})$ similarly to \eqref{eq:y_oscillators},
\begin{subequations}
\begin{align}
 y_{i,-n}^{(\pm)} & \stackrel{n > 0}{=}
 \frac{(1-q_1^{\pm n})(1-q_2^{\pm n})}{1-p^{\pm n}}
 \, \tilde{c}^{[\mp n]}_{ji} \, t_{j,n}^{(\pm)} 
 \, , \\
 y_{i,n}^{(\pm)} & \stackrel{n > 0}{=}
 \mp \frac{1}{n} \partial_{i,n}^{(\pm)} 
 \, , \\
 y_{i,0} & =
 - t_{j,0} \, \tilde{c}_{ji}^{[0]} \, \log q_2 
 \, ,
\end{align}
\end{subequations}
with the commutation relation:
\begin{align}
 \left[
  y_{i,n}^{(\pm)} \, , \, y_{j,n'}^{(\pm)}
 \right]
 & =
 \mp \frac{1}{n}
 \frac{(1-q_1^{\pm n})(1-q_2^{\pm n})}{1-p^{\pm n}}
 \, \tilde{c}^{[\mp n]}_{ji} \, \delta_{n+n',0} 
 \, .
 \label{eq:yy_osci_comm}
\end{align}
We remark the same relation holds to the $v$-modes as before~\eqref{eq:y_v_relation},
\begin{align}
 y_{i,n}^{(\pm)}
 = - \left(1-q_1^{\pm n}\right)\left(1-q_2^{\pm n}\right) \,
 v_{i,n}^{(\pm)}
 \, .
\end{align}
In terms of the free field $s_{i,n}^{(\pm)}$, we have 
\begin{align}
 y_{i,n}^{(\pm)} & \stackrel{n \neq 0}{=}
 (1 - q_2^{\mp n})
 \, \tilde{c}_{ji}^{[\pm n]} s_{j,n}^{(\pm)} 
 \, , \qquad
 y_{i,0} =
 - \log q_2 \, 
 \tilde{c}_{ji}^{[0]} \, s_{j,0} 
 \, ,
\end{align}
with the commutation relations between the $y$- and $s$-modes
\begin{align}
 \left[
  y_{i,n}^{(\pm)} \, , \, s_{j,n'}^{(\pm)}
 \right]
 & =
 \mp \frac{1}{n} \frac{1 - q_1^{\pm n}}{1 - p^{\pm n}}
 \delta_{n+n',0} \delta_{ij}
 \, , \qquad
 \left[
  \tilde{s}_{i,0} \, , \, y_{j,0}
 \right]
 = - \delta_{ij} \log q_1
 \, .
\end{align}
This leads to the normal ordered product (with the ordering $|x| > |x'|$) \index{operator product expansion (OPE)!S and Y@$S$ and $\mathsf{Y}$}
\begin{align}
 \mathsf{Y}_{i,x} S_{j,x'}
 & =
 {: \mathsf{Y}_{i,x} S_{j,x'} :}
 \times
 \begin{cases}
  \displaystyle
  \frac{\theta(x'/x;p)}{\theta(q_1x'/x;p)}
  & ( i = j)
  \\
  1 & (i \neq j)
 \end{cases}
 \label{eq:prod_YS}
\end{align}
The expectation value of the $\mathsf{Y}$-function has infinitely many poles at
$x = x' q_1 p^\mathbb{Z}$ for each instanton configuration
$\mathcal{X} \in \mathfrak{M}^{\mathsf{T}}$ that corresponds to the arguments of the screening currents:
\begin{align}
 \bra{0} \mathsf{Y}_{i,x} \prod_{x' \in \mathcal{X}}^\succ S_{\mathsf{i}(x'),x'} \ket{0}
 & =
 q_1^{\tilde{\rho}_i}
 \left(
 \prod_{x' \in \mathcal{X}_i}
 \frac{\theta(x'/x;p)}{\theta(q_1 x'/x;p)}
 \right)
 \bra{0} \prod_{x' \in \mathcal{X}}^\succ S_{\mathsf{i}(x'),x'} \ket{0}
 \, .
\end{align}
On the other hand, for $|x| < |x'|$, we have
\begin{align}
 S_{j,x'} \mathsf{Y}_{i,x} 
 & =
 {: S_{j,x'} \mathsf{Y}_{i,x} :}
 \times
 \begin{cases}
  \displaystyle
  q_1^{-1} \frac{\theta(x/x';p)}{\theta(q_1^{-1}x/x';p)}
  & ( i = j ) \\
  1 & (i \neq j)
 \end{cases}
 \, .
\end{align}
Therefore the commutator gives rise to
\begin{align}
 \left[
  \mathsf{Y}_{i,x} \, , \, S_{i,x'}
 \right]
 & =
 \left(
 \frac{\theta(x'/x;p)}{\theta(q_1x'/x;p)}
 - q_1^{-1} \frac{\theta(x/x';p)}{\theta(q_1^{-1}x/x';p)}
 \right)
 : \mathsf{Y}_{i,x} S_{i,x'} :
 \nonumber \\
 & =
 \frac{\theta(q_1^{-1};p)}
      {(p;p)_\infty^2} \,
 \delta(q_1 x'/x)
 : \mathsf{Y}_{i,x} S_{i,x'} :
 \, ,
\end{align}
which reproduces the previous result~\eqref{eq:YS_commutator} in the limit $p \to 0$.
The last expression is due to the relation
\begin{align}
 \frac{\theta(a z;p)}{\theta(z;p)}
 & =
 \frac{\theta(a;p)}{(p;p)_\infty^2}
 \sum_{n \in \mathbb{Z}} \frac{z^n}{1 - a p^n}
 \, ,
\end{align}
which is obtained from the identity~\eqref{eq:theta_id}.
This means that, in the limit $q_1 \to 1$, the $\mathsf{Y}$-operator commutes with the screening current, and it reproduces a commutative algebra~\cite{Nekrasov:2013xda}.

\subsection{$\mathsf{A}$-operator}
\index{A-operator@$\mathsf{A}$-operator!elliptic---}

We then define the elliptic analog of the $\mathsf{A}$-operator similarly to other vertex operators:
\begin{align}
 \mathsf{A}_{i,x} & = \
 q_1
 :
 \exp
 \left(
 a_{i,0} + \kappa_i \log x
 + \sum_{n \in \mathbb{Z}_{\neq 0}}
 \left( a_{i,n}^{(+)} \, x^{-n} + a_{i,n}^{(-)} \, x^{+n} \right)
 \right)
 :
 \, ,
\end{align}
with the free field realization
\begin{align}
 a_{i,-n}^{(\pm)} = (1 - q_1^{\pm n})(1 - q_2^{\pm n}) \, t_{i,n}^{(\pm)}
 \, , \quad
 a_{i,0} = - \log q_2 \, t_{i,0}
 \, , \quad
 a_{i,n}^{(\pm)} = \mp \frac{1}{n} c_{ji}^{[\pm n]} \, \partial_{j,n}^{(\pm)}
 \quad
 (n \ge 1)
 \, ,
\end{align}
obeying the commutation relation
\begin{align}
 \left[ a_{i,n}^{(\pm)} \, , a_{j,n'}^{\pm} \right]
 = \mp \frac{1}{n} (1 - q_1^{\pm n})(1 - q_2^{\pm n}) \, c_{ji}^{[\pm n]} \, \delta_{n+n',0}
 \, .
\end{align}
We obtain the same OPE factors between the $\mathsf{A}$-operators~\eqref{eq:AA_OPE} and with the $\mathsf{Y}$ operator~\eqref{eq:AY_AV_OPE} by replacing the $\mathscr{S}$-function with the elliptic one~\eqref{eq:S_func_def}.
The OPE with the $\mathsf{V}$ operator is given by replacing the rational factor with the theta function, \index{operator product expansion (OPE)!A and V@$\mathsf{A}$ and $\mathsf{V}$}
\begin{align}
 \mathsf{V}_{i,x} \mathsf{A}_{j,x'}
 =
 \theta\left(\frac{x'}{x};p \right)^{-\delta_{ij}} \,
 {:\mathsf{V}_{i,x} \mathsf{A}_{j,x'}:}
 \, , \quad
 &
 \mathsf{A}_{j,x'} \mathsf{V}_{i,x} 
 =
 \theta\left(\frac{x}{x'};p \right)^{-\delta_{ij}} \,
 {:\mathsf{A}_{j,x'} \mathsf{V}_{i,x} :}
   \, . 
\end{align}

\section{$\mathsf{T}$-operator}

Since we have the same relation between the operators $\mathsf{A}$, $\mathsf{Y}$, and the screening current,
\begin{align}
 \mathsf{A}_{i,x} = {:
 \mathsf{Y}_{i,x} \mathsf{Y}_{i,q x}
 \prod_{e:i \to j} \mathsf{Y}_{j,\mu_e^{-1} q x}^{-1}
 \prod_{e:j \to i} \mathsf{Y}_{j,\mu_e x}^{-1}
 :}
 = q_1 \, {: S_{i,x} S_{i,q_2 x}^{-1} :}
 \, ,
\end{align}
we can apply the iWeyl reflection and the pole cancellation mechanism to construct the holomorphic $\mathsf{T}$-operator, as a generating current of the elliptic deformation of W-algebra, W$_{q_{1,2},p}(\mathfrak{g}_{\Gamma})$:
\begin{align}
 \mathsf{T}_{i,x}
 = \mathsf{Y}_{i,x} + {:\mathsf{Y}_{i,x} \mathsf{A}_{i,q^{-1} x}^{-1}:} + \cdots
 = \sum_{n \in \mathbb{Z}} T_{i,n} \, x^{-n}
 \, .
\end{align}
Let us discuss the explicit construction with examples as follows.

\subsection{$A_1$ quiver}

The generating current for $A_1$ quiver is given as follows:\index{qq-character@$qq$-character!A1@$A_1$}
\begin{align}
 \mathsf{T}_{1,x} = \mathsf{Y}_{1,x} + \mathsf{Y}_{1,q^{-1}}^{-1}
 \, ,
\end{align}
where the expression in terms of the $\mathsf{Y}$-operators itself is formally equivalent to that discussed in \S\ref{sec:W_A1}.
This $\mathsf{T}$-operator obeys the OPE relation:\index{operator product expansion (OPE)!T and T (elliptic A1)@$\mathsf{T}$ and $\mathsf{T}$ (elliptic $A_1$)}
\begin{align}
 f\left( \frac{x'}{x} \right) \mathsf{T}_{1,x} \mathsf{T}_{1,x'}
 - f\left( \frac{x}{x'} \right) \mathsf{T}_{1,x'} \mathsf{T}_{1,x}
 = - \frac{\theta(q_1;p) \theta(q_2;p)}{(p;p)_\infty^2 \theta(q;p)}
 \left( \delta \left(q \frac{x'}{x} \right) - \delta \left(q^{-1} \frac{x'}{x} \right)\right)
 \, ,
\end{align}
where the structure function is given by
\begin{align}
 f(z) = \exp \left( \sum_{n \in \mathbb{Z}_{\neq 0}} \frac{1}{n} \frac{(1 - q_1^n)(1 - q_2^n)}{(1 - p^n)(1 + q^n)} \, z^n \right)
\end{align}
with the relation $f(z) f(qz) = \mathscr{S}(z)$.
This elliptic algebra is known to be an elliptic deformation of the Virasoro algebra, Vir$_{q_{1,2},p} = \mathrm{W}_{q_{1,2},p}(A_1)$~\cite{Nieri:2015dts}.

\subsection{$A_2$ quiver}

We then consider $A_2$ quiver.
In this case, as discussed in \S\ref{sec:W_A2}, we have two $\mathsf{T}$-operators,\index{operator product expansion (OPE)!T and T (elliptic A2)@$\mathsf{T}$ and $\mathsf{T}$ (elliptic $A_2$)}\index{qq-character@$qq$-character!A2@$A_2$}
\begin{subequations}
 \begin{align}
  &
  f_{11}\left( \frac{x'}{x} \right) \mathsf{T}_{1,x} \mathsf{T}_{1,x'}
  - f_{11}\left( \frac{x}{x'} \right) \mathsf{T}_{1,x'} \mathsf{T}_{1,x}
  \nonumber \\
  & \hspace{5em}
  =
  - \frac{\theta(q_1;p) \theta(q_2;p)}{(p;p)_\infty^2 \theta(q;p)}
  \left(
  \delta \left( q \frac{x'}{x} \right) \mathsf{T}_{2,\mu^{-1} x}
    - \delta \left( q^{-1} \frac{x'}{x} \right) \mathsf{T}_{2,\mu q^{-1} x}
  \right)
  \, , 
 \end{align}
 \begin{align}
  &
  f_{12}\left( \frac{x'}{x} \right) \mathsf{T}_{1,x} \mathsf{T}_{2,x'}
  - f_{21}\left( \frac{x}{x'} \right) \mathsf{T}_{2,x'} \mathsf{T}_{1,x}
  \nonumber \\
  & \hspace{5em}
  =
  - \frac{\theta(q_1;p) \theta(q_2;p)}{(p;p)_\infty^2 \theta(q;p)}
  \left(
  \delta\left( \mu q \frac{x'}{x} \right) - \delta\left( \mu q^{-2} \frac{x'}{x} \right)
  \right)
  \, , 
 \end{align}
 \begin{align}  
  &
  f_{22}\left( \frac{x'}{x} \right) \mathsf{T}_{2,x} \mathsf{T}_{2,x'}
  - f_{22}\left( \frac{x}{x'} \right) \mathsf{T}_{2,x'} \mathsf{T}_{2,x}
  \nonumber \\
  & \hspace{5em}
  =
  - \frac{\theta(q_1;p) \theta(q_2;p)}{(p;p)_\infty^2 \theta(q;p)}
  \left(
  \delta \left( q \frac{x'}{x} \right) \mathsf{T}_{1,\mu q^{-1} x}
  - \delta \left( q^{-1} \frac{x'}{x} \right) \mathsf{T}_{1,\mu x}
  \right)
  \, .
 \end{align}
\end{subequations}
  with the structure function
  \begin{align}
   f_{ij}(z)
   =
   \exp \left( \sum_{n \in \mathbb{Z}_{\neq 0}} \frac{1}{n} \frac{(1 - q_1^n)(1 - q_2^n)}{1 - p^n} \, \tilde{c}_{ji}^{[-n]} \, z^n \right)
   \, .
  \end{align}
  The OPE between the $\mathsf{Y}$-operators is given by
  \begin{align}
   \mathsf{Y}_{i,x} \mathsf{Y}_{j,x'}
   = f_{ij}\left(\frac{x'}{x}\right)^{-1} \,
   {:\mathsf{Y}_{i,x} \mathsf{Y}_{j,x'}:}
   \, .
  \end{align}
  The elliptic algebra generated by $(\mathsf{T}_{1,x},\mathsf{T}_{2,x})$ is an elliptic deformation of W$_3$ algebra, W$_{q_{1,2},p}(A_2)$.

\appendix
\part{Appendices}\label{part:app}

\chapter{Special functions}\label{chap:sp_functions}

We summarize the special functions used in the manuscript.

\section{Gamma functions}
\index{gamma function}
\index{gamma function!multiple---}

We define the Barnes zeta function for $(\epsilon_i)_{i=1,\ldots,k}$ for $\real(s)>k$ and $\real(\epsilon_i) > 0$:
\begin{align}
 \zeta_k(s,z;\epsilon_1,\ldots,\epsilon_k)
 = \sum_{n_1,\ldots,n_k \ge 0} \frac{1}{(z + n_1 \epsilon_1 + \cdots + n_k \epsilon_k)^s}
 \, .
\end{align}
Then, the multiple gamma function is defined as
\begin{align}
 \Gamma_k(z;\epsilon_1,\ldots,\epsilon_k)
 & = \exp \left( \frac{\partial}{\partial s}\Bigg|_{s=0} \zeta_k(s,z;\epsilon_1,\ldots,\epsilon_k) \right)
 \nonumber \\ 
 & =
 \prod_{n_1,\ldots,n_k \ge 0} \frac{1}{z + n_1 \epsilon_1 + \cdots + n_k \epsilon_k}
 \, .
 \label{eq:multi_gamma}
\end{align}
Precisely speaking, the infinite product in the second line should be a formal expression since the corresponding series expansion is available only for $\real(s)>k$.
Nevertheless, we interpret this as the zeta function regularization of the infinite product.
This gamma function is constructed to obey a functional relation
\begin{align}
 \frac{\Gamma_k(z + \epsilon_i ;\epsilon_1,\ldots,\epsilon_k)}{\Gamma_k(z;\epsilon_1,\ldots,\epsilon_k)}
 = \Gamma_{k-1}(z;\epsilon_1,\ldots,\check{\epsilon}_i,\ldots,\epsilon_k)
\end{align}
where $\check{\epsilon}_i$ means that the dependence on $\epsilon_i$ on the right hand side is removed.
We remark that the gamma function $\Gamma_k(z;\epsilon_1,\ldots,\epsilon_k)$ has poles at $z + n_1 \epsilon_1 + \cdots + n_k \epsilon_k = 0$ and no zero.

The degree-one case is related to the standard definition of the gamma function:
\begin{align}
 \Gamma_1(z;\epsilon) = \frac{\epsilon^{z/\epsilon - \frac{1}{2}}}{\sqrt{2\pi}} \Gamma(z/\epsilon)
 \, .
 \label{eq:gamma_convention}
\end{align}
Therefore, the asymptotic behavior is given by Stirling's formula:
\begin{align}
 \lim_{\epsilon \to 0} \epsilon \log \Gamma_1(z;\epsilon) = z \log z - z 
 \, .
\end{align}

\subsection{Reflection formula}\label{sec:gamma_ref}

Together with the infinite product formula of the sine function
\begin{align}
 \frac{\sin \pi z}{\pi z} = \prod_{n = 1}^\infty \left( 1 - \frac{z^2}{n^2} \right)
 \, ,
\end{align}
and the zeta function regularization $\displaystyle \prod_{n = 1}^\infty \epsilon^2 n^2 = 2 \pi / \epsilon$, we obtain the reflection formula
\begin{align}
 \Gamma_1(z;\epsilon) \Gamma_1(\epsilon - z;\epsilon)
 = \frac{1}{2 \sin \pi z / \epsilon}
 \, .
\end{align}
It is also possible to derive this formula form the standard version of the formula through the relation~\eqref{eq:gamma_convention}:
\begin{align}
 \Gamma(z) \Gamma(1 - z) = \frac{\pi}{\sin \pi z}
 \, .
\end{align}

\subsection{Multiple sine function}
\index{multiple sine function}

One may define multiple sine functions through the reflection formula for the multiple gamma functions~\cite{Kurokawa:2003},
\begin{align}
 S_k(z;\epsilon_1,\ldots,\epsilon_k) = \frac{\Gamma_k(\epsilon - z;\epsilon_1,\ldots,\epsilon_k)^{(-1)^k}}{\Gamma_k(z;\epsilon_1,\ldots,\epsilon_k)}
 \, ,
\end{align}
where
\begin{align}
 \epsilon = \sum_{i = 1}^k \epsilon_i
 \, .
\end{align}
For example, the double sine function is given as follows:
\begin{align}
 S_2(z;\epsilon_1,\epsilon_2)
 = \frac{\Gamma_2(\epsilon_1 + \epsilon_2 - z;\epsilon_1, \epsilon_2)} {\Gamma_2(z;\epsilon_1,\epsilon_2)}
 = \prod_{n,m \ge 0} \frac{z + n \epsilon_1 + m \epsilon_2}{\epsilon_1 + \epsilon_2 - z + n \epsilon_1 + m \epsilon_2}
 \, .
 \label{eq:double_sine_fn}
\end{align}
In this case, it is defined as a ratio of the double gamma functions.
We instead obtain the Upsilon function with their product,
\begin{align}
 \Upsilon(z;\epsilon_1,\epsilon_2)
 =
 \frac{1}{\Gamma_2(z;\epsilon_1,\epsilon_2) \Gamma_2(\epsilon_1 + \epsilon_2 - z;\epsilon_1,\epsilon_2)}
 \, .
\end{align}

\section{$q$-functions}

\subsection{$q$-shifted factorial}
\index{$q$-shifted factorial}

We define the $q$-shifted factorial (also known as the $q$-Pochhammer symbol):
\begin{align}
 (z;q)_n
 = \prod_{m=0}^{n-1} (1 - z q^m)
 \, .
 \label{eq:q-factorial}
\end{align}
The multivariable analog is similarly defined as
\begin{align}
 (z_1, \ldots, z_k;q)_n = (z_1;q)_n \cdots (z_k;q)_n
 \, .
\end{align}
The $q$-shifted factorial for $n \to \infty$ is given for $|q| < 1$ as
\begin{align}
 (z;q)_\infty
 = \prod_{m=0}^\infty (1 - z q^m)
 = \exp \left( - \sum_{m=1}^\infty \frac{z^m}{m (1 - q^m)} \right)
 \, .
\end{align}
For $|q| > 1$, it is given through the analytic continuation:
\begin{align}
 (z;q)_\infty = (z q^{-1};q^{-1})_\infty^{-1}
 \, .
 \label{eq:q-reflection}
\end{align}
We remark the relation
\begin{align}
 (z;q)_n
 = \frac{(z;q)_\infty}{(zq^n;q)_\infty}
 = \exp \left( - \sum_{m = 1}^\infty \frac{z^m}{m} \frac{1 - q^{mn}}{1 - q^m} \right)
 \, .
 \label{eq:q-factorial_finite}
\end{align}

\subsection{Quantum dilogarithm}
\index{quantum dilogarithm}

Let $q = \np^\epsilon$, and consider the expansion around $\epsilon = 0$ of the $q$-factorial.
We remark the expansion
\begin{align}
 \frac{1}{1 - q}
 = - \frac{1}{\epsilon} \frac{\epsilon}{\np^{\epsilon} - 1}
 = - \frac{1}{\epsilon} \sum_{n=0}^\infty B_n \, \frac{\epsilon^n}{n!}
\end{align}
where $(B_n)_{n \ge 0}$ are the Bernoulli numbers.
Then the $q$-shifted factorial is given by
\begin{align}
 (z;q)_\infty
 = \exp
 \left(
  \frac{1}{\epsilon} \sum_{n=0}^\infty \operatorname{Li}_{2-n}(z) \, B_n \, \frac{\epsilon^n}{n!}
 \right)
 = \exp \left( \frac{1}{\epsilon} \left( \operatorname{Li}_2 (z) + O(\epsilon) \right) \right)
 \label{eq:q_dilog_exp}
\end{align}
where we define the polylogarithm function for $|z| < 1$ as
\begin{align}
 \operatorname{Li}_p(z) = \sum_{m=1}^\infty \frac{z^m}{m^p}
 \, .
\end{align}
In this sense, the $q$-shifted factorial is interpreted as a $q$-deformation of the dilogarithm, which is called the quantum dilogarithm~\cite{Faddeev:1993rs}.
We remark that this quantum dilogarithm is related to, but different from Faddeev's quantum dilogarithm~\cite{Faddeev:1995nb}.

   \subsection{$q$-gamma functions}
   \index{gamma function!$q$---}   

  The formulas shown above imply that the $q$-shifted factorial is interpreted as a $q$-analog of the gamma function:%
  \footnote{%
  This is slightly different from the standard definition of the $q$-gamma function,
  \begin{align}
   \Gamma_q(x) = (1 - q)^{1 - x} \frac{(q;q)_\infty}{(q^x;q)_\infty}
   \, ,
  \end{align}
  which obeys the relation
  \begin{align}
   \Gamma_q(x + 1) = \frac{1 - q^x}{1 - q} \Gamma_q(x) = [x]_q \Gamma_q(x)
   \, .
  \end{align}
  }
\begin{align}
 \Gamma_q(z;q) := \frac{1}{(z;q)_\infty}
 \label{eq:q-gamma}
\end{align}
with poles at $z q^n = 1$ for $n \ge 0$.
In this convention, the relation~\eqref{eq:q-factorial_finite} is given by
\begin{align}
 \frac{\Gamma_q(zq^n;q)}{\Gamma_q(z;q)} = (z;q)_n = (1-z) \cdots (1 - zq^{n-1})
 \, .
 \label{eq:q-gamma_ratio}
\end{align}

A $q$-analog of the multiple gamma function is defined as
\begin{align}
 \Gamma_{q,k}(z;q_1,\ldots,q_k)
 & = (z;q_1,\ldots,q_k)_\infty^{(-1)^k}
 \nonumber \\
 & =
 \exp \left(
  (-1)^{k+1} \sum_{m=1}^\infty \frac{z^m}{m} \frac{1}{(1 - q_1^m) \cdots (1 - q_k^m)}
 \right)
 \label{eq:q-multiple_gamma}
\end{align}
with the multiple version of the $q$-shifted factorial for $|q_1|, \ldots, |q_k| < 1$:
\begin{align}
 (z;q_1,\ldots,q_k)_\infty
 & = \prod_{0 \le n_1,\ldots,n_k \le \infty} (1 - z q_1^{n_1} \cdots q_k^{n_k})
 \nonumber \\
 & =
 \exp \left(
  - \sum_{m=1}^\infty \frac{z^m}{m} \frac{1}{(1 - q_1^m) \cdots (1 - q_k^m)}
 \right)
 \, .
 \label{eq:multi_q-shift}
\end{align}
This $q$-gamma function obeys the functional relation
\begin{align}
 \frac{\Gamma_{q,k}(z q_i ;q_1,\ldots,q_k)}{\Gamma_{q,k}(z;q_1,\ldots,q_k)}
 = \Gamma_{q,k-1}(z ;q_1,\ldots,\check{q}_i,\ldots,q_k)
 \, .
 \label{eq:q-multiple_gamma_shift}
\end{align}

From this point of view, we may also consider the multiple $q$-gamma function of negative degree as follows:
\begin{align}
 \Gamma_{q,-k}(z;q_1,\ldots,q_k)
 & =
 \exp \left( (-1)^{k+1}
  \sum_{m=1}^\infty \frac{z^m}{m} (1 - q_1^m) \cdots (1 - q_k^m)
 \right)
 \, .
\end{align}
The case with $k = 2$ is the (K-theoretic) $\mathscr{S}$-function~\eqref{eq:Sij_fn}, and the case with $k = 4$ is used in the context of $\widehat{A}_0$ quiver.
See~\S\ref{sec:A0hat_algebra}.

\subsection{Partition sum}

Let $\lambda$ be a (two-dimensional) partition.
Then the summation over the partitions is given by Euler's product formula:
\begin{align}
 \sum_{\lambda} q^{|\lambda|} = \frac{1}{(q;q)_\infty}
 \, .
\end{align}
A similar result is available for the sum over the plane partitions (three-dimensional partitions) by the MacMahon function:
\begin{align}
 \sum_{\pi} q^{|\pi|} = \prod_{n=1}^\infty \frac{1}{(1 - q^n)^n}
 \, .
\end{align}

\section{Elliptic functions}\label{sec:ell_fn}

\subsection{Theta function}\label{sec:theta_fn}
\index{theta function}

The theta function with the elliptic nome $p = \np^{2 \pi \im \tau} \in \mathbb{C}^\times$ is given by
\begin{align}
 \theta(z;p)
 = (z;p)_\infty (z^{-1} p;p)_\infty
 = \exp \left( - \sum_{n \in \mathbb{Z}_{\neq 0}} \frac{z^n}{n(1 - p^n)} \right)
 \label{eq:theta_fn}
\end{align}
where $(z;p)_\infty$ is the $p$-shifted factorial~\eqref{eq:q-factorial}.
It obeys the reflection relation
\begin{align}
 \theta(z^{-1};p) = (- z^{-1}) \theta(z;p)
 \, .
 \label{eq:theta_reflection}
\end{align}

We remark that, since the $q$-shifted factorial is identified with the $q$-gamma function~\eqref{eq:q-gamma}, the relation~\eqref{eq:theta_fn} is a $q$-analog of the reflection formula of the gamma function discussed in \S\ref{sec:gamma_ref}.
In this sense, the theta function is interpreted as a $q$-analog of the sine function having zeros at
\begin{align}
 \frac{\log z}{2 \pi \im} = \mathbb{Z} + \tau \mathbb{Z}
 \, .
\end{align}

\subsubsection{An identity}

We start with Ramanujan's identity (also known as $_{1}\psi_1$ formula) for $|b/a| < |z| < 1$:
\begin{align}
 \frac{(az, p/az,p,b/z;p)_\infty}{(z,b/az,b,p/a;p)_\infty}
 = \sum_{n \in \mathbb{Z}} \frac{(a;p)_n}{(b;p)_n} \, z^n
 = {_{1}\psi_1}(a;b;z,p)
\end{align}
where we denote the bilateral basic hypergeometric series by $_r \psi_s(a_{1,\ldots,r};b_{1,\ldots,s};z,q)$.
We put $b = ap$, then we obtain
\begin{subequations}
\begin{align}
 \text{LHS}
 & = \frac{(az,p/az,p,p;p)_\infty}{(z,p/z,ap,p/a;p)_\infty}
 = \frac{\theta(az;p) (p;p)_\infty^2}{\theta(z;p) \theta(a;p)} \, (1 - a)
 \, , \\
 \text{RHS}
 & = \sum_{n \in \mathbb{Z}} \frac{1 - a}{1 - a p^n} \, z^n
 \, ,
\end{align}
\end{subequations}
which leads to the identity
\begin{align}
 \frac{\theta(az;p) (p;p)_\infty^2}{\theta(z;p) \theta(a;p)}
 = \sum_{n \in \mathbb{Z}} \frac{z^n}{1 - a p^n}
 \, .
 \label{eq:theta_id}
\end{align}

\subsection{Elliptic gamma functions}\label{sec:e_gamma}
   \index{gamma function!elliptic---}   

We define the elliptic gamma function for $|p|, |q| < 1$:
\begin{align}
 \Gamma_e(z;p,q)
 = \prod_{n,m \ge 0} \frac{(z^{-1} p q;p,q)_\infty}{(z;p,q)_\infty}
 = \exp \left( \sum_{m \in \mathbb{Z}_{\neq 0}} \frac{z^m}{m} \frac{1}{(1 - p^m)(1 - q^m)} \right)
 \, ,
 \label{eq:e-gamma}
\end{align}
which obeys the relation
\begin{align}
 \frac{\Gamma_e(zp;p,q)}{\Gamma_e(z;p,q)} = \theta(z;q)
 \, , \qquad
 \frac{\Gamma_e(zq;p,q)}{\Gamma_e(z;p,q)} = \theta(z;p)
 \, .
\end{align}
In this case, the analog of the reflection formula (\S\ref{sec:gamma_ref}) is given by
\begin{align}
 \Gamma_e(z;p,q) \Gamma_e(pq z^{-1};p,q) = 1
 \, .
\end{align}
We remark that the (inverse of) double sine function~\eqref{eq:double_sine_fn} is obtained in the scaling limit of the elliptic gamma function with $(z,p,q) = (\np^{\beta x}, \np^{\beta \epsilon_1}, \np^{\beta \epsilon_2})$ and taking $\beta \to 0$.

\subsubsection{Elliptic double gamma function}

The elliptic analog of the double gamma function is given by
\begin{align}
 \Gamma_{e,2}(z;q_1,q_2,q_3)
 & = (z;q_1,q_2,q_3)_\infty (z^{-1} q_1 q_2 q_3;q_1,q_2,q_3)_\infty
 \nonumber \\
 & =
 \exp
 \left(
  - \sum_{m \in \mathbb{Z}_{\neq 0}} \frac{z^m}{m} \frac{1}{(1 - q_1^m)(1 - q_2^m)(1 - q_3^m)}
 \right)
 \, ,
\end{align}
which obeys the relation
\begin{align}
 \frac{\Gamma_{e,2}(z q_1;q_1, q_2, q_3)}{\Gamma_{e,2}(z;q_1, q_2, q_3)} = \Gamma_{e}(z;q_2,q_3)
 \, , \qquad
 \text{etc}
 \, .
\end{align}
We can similarly construct the elliptic analog of the multiple gamma functions,
\begin{align}
 \Gamma_{e,k}(z;q_1,\ldots,q_{k+1})
 =
 \exp \left(
  (-1)^{k+1} \sum_{m \in \mathbb{Z}_{\neq 0}}^\infty \frac{z^m}{m} \frac{1}{(1 - q_1^m) \cdots (1 - q_{k+1}^m)}
 \right)
 \, ,
\end{align}
which obeys a similar shift relation to~\eqref{eq:q-multiple_gamma_shift}.
See \cite{Nishizawa:2001JPA,Narukawa:2004AM} for details.

\subsection{Elliptic analog of polylogarithm}

We define an elliptic analog of polylogarithm function:
\begin{align}
 \operatorname{Li}_k(z;p) = \sum_{n \in \mathbb{Z}_{\neq 0}} \frac{z^n}{n^k} \frac{1}{1 - p^n}
 \ \xrightarrow{p \to 0} \
 \operatorname{Li}_k(z)
 \, .
\end{align}
The first example is given by
\begin{align}
 \operatorname{Li}_1(z;p) = - \log \theta(z;p)
 \ \xrightarrow{p \to 0} \
 \operatorname{Li}_1(z) = - \log (1 - z)
 \, .
\end{align}

The elliptic gamma function has the asymptotic expansion in terms of the elliptic polylogarithm functions:
\begin{align}
 \Gamma_e(z;p,\np^{\epsilon})
 & =
 \exp
 \left(
  - \frac{1}{\epsilon} \sum_{n = 0}^\infty \operatorname{Li}_{2-n}(z;p) \, B_n \, \frac{\epsilon^n}{n!}
 \right)
 =
 \exp \left( - \frac{1}{\epsilon} \left( \operatorname{Li}_{2}(z;p) + O(\epsilon) \right)  \right)
 \, ,
\end{align}
which is analogous to the expansion~\eqref{eq:q_dilog_exp}.

\chapter{Combinatorial calculus}

\section{Partition}

The partition $\lambda$ is a sequence of non-increasing non-negative integers:
\begin{align}
 \lambda = (\lambda_1 \ge \lambda_2 \ge \cdots \ge 0) \in \mathbb{Z}_{\ge 0}^\infty
 \, .
\end{align}
We denote the transposed partition of $\lambda$ by $\check{\lambda}$.
The size of the partition is defined as
\begin{align}
 |\lambda|
 = \sum_{i=1}^\infty \lambda_i
 = \sum_{i=1}^\infty \check{\lambda}_i
 = |\check{\lambda}|
 \, .
\end{align}
For the partition $\lambda_\alpha$, we define the arm and leg lengths for $s = (s_1,s_2)$:
\begin{align}
 a_{\alpha}(s) = \lambda_{\alpha,s_1} - s_2
 \, , \qquad
 \ell_{\alpha}(s) = \check\lambda_{\alpha,s_2} - s_1
 \, .
 \label{eq:arm_leg}
\end{align}
We remark that not necessarily $s \in \lambda_\alpha$, so that $(a_\alpha(s), \ell_\alpha(s))$ may be negative.
Then the relative hook length is defined
\begin{align}
 h_{\alpha\beta}(s)
 = a_\alpha(s) + \ell_\beta(s) + 1
 = \lambda_{\alpha,s_1} + \check{\lambda}_{\beta,s_2} - s_1 - s_2 + 1
 \, .
 \label{eq:relative_hook_length}
\end{align}

\section{Instanton calculus}\label{sec:comb}

We summarize the combinatorics calculus of the partition for the instanton partition function.
Summation over the partition is expressed in the following two ways,
\begin{align}
 \sum_{s \in \lambda}
 = \sum_{s_1=1}^{\check\lambda_1} \sum_{s_2 = 1}^{\lambda_{s_1}}
 = \sum_{s_2=1}^{\lambda_1} \sum_{s_1 = 1}^{\check\lambda_{s_2}}
 \, .
\end{align}

\subsection{U($n$) theory}\label{sec:comb_U(n)}

We consider the instanton contribution to the Chern character of the bifundamental hypermultiplet (See \S\ref{sec:inst_part_func}):
\begin{align}
 \ch_\mathsf{T} \mathbf{H}_{e:i \to j}^{\text{bf,\,inst}} & =
 - \mu_e \ch_\mathsf{T} \wedge \mathbf{Q}^\vee 
 \ch_\mathsf{T} \mathbf{K}_i^\vee \ch_\mathsf{T} \mathbf{K}_j
 + \mu_e \ch_\mathsf{T} \mathbf{N}_i^\vee \ch_\mathsf{T} \mathbf{K}_j + \mu_e q^{-1} \ch_\mathsf{T} \mathbf{K}_i^\vee \ch_\mathsf{T} \mathbf{N}_j
 \nonumber \\
 & =: \sum_{\alpha=1}^{n_i} \sum_{\beta=1}^{n_j} \mu_e \frac{\nu_{j,\beta}}{\nu_{i,\alpha}} \, \Xi[\lambda_{i,\alpha},\lambda_{j,\beta}] 
\end{align}
where we define
\begin{align}
 \Xi[\lambda_{\alpha},\lambda_{\beta}]
 & =
 - (1 - q_1^{-1})(1 - q_2^{-1}) \sum_{s \in \lambda_{\alpha}} \sum_{s' \in \lambda_{\beta}} q_1^{-s_1+s_1'} q_2^{-s_2+s_2'}
 + \sum_{s \in \lambda_{\alpha}} q_1^{-s_1} q_2^{-s_2}
 + \sum_{s' \in \lambda_{\beta}} q_1^{s_1'-1} q_2^{s_2'-1}
 \label{eq:comb1}
\end{align}
From this expression, we obtain a combinatorial formula
(See, for example, \cite{Nakajima:1999})
\begin{align}
 \Xi[\lambda_{\alpha},\lambda_{\beta}]
 & =
 \sum_{s \in \lambda_{\alpha}} q_1^{\ell_{\beta}(s)} q_2^{-a_{\alpha}(s)-1}
 + \sum_{s \in \lambda_{\beta}} q_1^{-\ell_{\alpha}(s)-1} q_2^{a_{\beta}(s)}
 \, .
 \label{eq:comb2}
\end{align}
where the arm and leg lengths for each box $s = (s_1,s_2)$ in the partition are defined in \eqref{eq:arm_leg}.
We remark
\begin{align}
 q \, \Xi[\lambda_\alpha,\lambda_\beta]\Big|_{q_1,q_2}
 = \Xi[\lambda_\beta,\lambda_\alpha]\Big|_{q_1^{-1},q_2^{-1}}
 \, .
 \label{eq:comb_inv}
\end{align}
The vector multiplet contribution has a similar expression
\begin{align}
 \ch_\mathsf{T} \mathbf{V}_i^\text{inst} & =
 \ch_\mathsf{T} \wedge \mathbf{Q}^\vee 
 \ch_\mathsf{T} \mathbf{K}_i^\vee \ch_\mathsf{T} \mathbf{K}_j
 - \ch_\mathsf{T} \mathbf{N}_i^\vee \ch_\mathsf{T} \mathbf{K}_j
 - q^{-1} \ch_\mathsf{T} \mathbf{K}_i^\vee \ch_\mathsf{T} \mathbf{N}_j
 \nonumber \\
 & =
 - \sum_{\alpha,\beta=1}^{n_i}
 \frac{\nu_{i,\beta}}{\nu_{i,\alpha}} \, \Xi[\lambda_{i,\alpha},\lambda_{i,\beta}]
 \, .
\end{align}

\subsubsection{Proof of the formula \eqref{eq:comb2}}\label{sec:comb_sub}

We prove the combinatorial formula \eqref{eq:comb2}.
We partially perform the summation for the first term in \eqref{eq:comb1},
\begin{align}
 &
 - (1 - q_1^{-1})(1 - q_2^{-1}) \sum_{s \in \lambda_{\alpha}} \sum_{s' \in \lambda_{\beta}} q_1^{-s_1+s_1'} q_2^{-s_2+s_2'}
 =
 \sum_{s_1=1}^{\check\lambda_{\alpha,1}} \sum_{s_2'=1}^{\lambda_{\beta,1}}
 (1 - q_1^{\check\lambda_{\beta,s_2'}}) q_1^{-s_1}
 (1 - q_2^{-\lambda_{\alpha,s_1}}) q_2^{s_2'-1}
 \nonumber \\
 & =
 \sum_{s_1=1}^{\check\lambda_{\alpha,1}} \sum_{s_2'=1}^{\lambda_{\beta,1}}
 \left[
 q_1^{\check\lambda_{\beta,s_2'} - s_1} q_2^{-\lambda_{\alpha,s_1}+s_2'-1}
 - q_1^{- s_1} q_2^{s_2'-1}
 + (1 - q_1^{\check\lambda_{\beta,s_2'}}) q_1^{-s_1} q_2^{s_2'-1}
 + q_1^{-s_1} (1 - q_2^{-\lambda_{\alpha,s_1}}) q_2^{s_2'-1}
 \right]
 \, .
 \label{eq:comb3}
\end{align}
The third and fourth terms in \eqref{eq:comb3} are then given by
\begin{subequations}
 \begin{align}
  \sum_{s_1=1}^{\check\lambda_{\alpha,1}} \sum_{s_2'=1}^{\lambda_{\beta,1}}
  (1 - q_1^{\check\lambda_{\beta,s_2'}}) q_1^{-s_1} q_2^{s_2'-1}
  & =
  \sum_{s_1=1}^{\check\lambda_{\alpha,1}} \sum_{s_2'=1}^{\lambda_{\beta,1}}
  \sum_{s_1'=1}^{\check\lambda_{\beta,s_2'}}
  (1 - q_1) q_1^{-s_1+s_1'-1} q_2^{s_2'-1}
  \nonumber \\
  & =
  - \sum_{s' \in \lambda_\beta} (1 - q_1^{-\check\lambda_{\alpha,1}}) q_1^{s_1'-1} q_2^{s_2'-1}
  \, ,
  \\
  \sum_{s_1=1}^{\check\lambda_{\alpha,1}} \sum_{s_2'=1}^{\lambda_{\beta,1}}
  q_1^{-s_1} (1 - q_2^{-\lambda_{\alpha,s_1}}) q_2^{s_2'-1}
  & =
  \sum_{s_1=1}^{\check\lambda_{\alpha,1}} \sum_{s_2'=1}^{\lambda_{\beta,1}}
  \sum_{s_2=1}^{\lambda_{\alpha,s_1}}
  q_1^{-s_1} (1 - q_2^{-1}) q_2^{-s_2+s_2'}
  \nonumber \\
  & =
  - \sum_{s \in \lambda_\alpha}
  q_1^{-s_1} (1 - q_2^{\lambda_{\beta,1}}) q_2^{-s_2}
  \, .
 \end{align}
\end{subequations}
Combining them together, \eqref{eq:comb1} becomes
\begin{align}
 \Xi[\lambda_\alpha,\lambda_\beta]
 & =
 \sum_{s_1=1}^{\check\lambda_{\alpha,1}} \sum_{s_2'=1}^{\lambda_{\beta,1}}
 \left[
 q_1^{\check\lambda_{\beta,s_2'} - s_1} q_2^{-\lambda_{\alpha,s_1}+s_2'-1}
 - q_1^{- s_1} q_2^{s_2'-1}
 \right]
 \nonumber \\
 & \qquad
 + \sum_{s \in \lambda_\alpha} q_1^{-s_1} q_2^{\lambda_{\beta,1}-s_2}
 + \sum_{s' \in \lambda_\beta} q_1^{-\check\lambda_{\alpha,1} + s_1' - 1} q_2^{s_2'-1}
 \, .
 \label{eq:comb4} 
\end{align}

We divide it into the negative and positive parts
\begin{align}
 \Xi[\lambda_\alpha,\lambda_\beta]
 = \Xi_{q_2^{<0}}[\lambda_\alpha,\lambda_\beta] + \Xi_{q_2^{\ge 0}}[\lambda_\alpha,\lambda_\beta]
\end{align}
where $\Xi_{q_2^{<0}}$ consists of monomials with negative powers of $q_2$, while $\Xi_{q_2^{\ge 0}}$ consists of positive ones.
Let us focus on $\Xi_{q_2^{<0}}$ with \eqref{eq:comb4}.
\begin{itemize}
 \item For $\lambda_{\beta,1} > \lambda_{\alpha,s_1}$, the first term in \eqref{eq:comb4} may contribute to the negative part $\Xi_{q_2^{<0}}$.
 \item For $\lambda_{\beta,1} \le \lambda_{\alpha,s_1}$, the first and third terms may contribute to $\Xi_{q_2^{<0}}$.
\end{itemize}
In both cases, the negative part $\Xi_{q_2^{<0}}$ is given by
\begin{align}
 \Xi_{q_2^{<0}}[\lambda_\alpha,\lambda_\beta] = \sum_{s \in \lambda_\alpha}
 q_1^{\check\lambda_{\beta,s_2} - s_1} q_2^{-\lambda_{\alpha,s_1} + s_2 - 1}
 = \sum_{s \in \lambda_\alpha} q_1^{\ell_\beta(s)} q_2^{-a_\alpha(s) - 1}
 \, .
\end{align}
We can similarly obtain the positive part $\Xi_{q_2^{\ge 0}}$ by utilizing the formula \eqref{eq:comb_inv},
\begin{align}
 \Xi_{q_2^{\ge 0}}[\lambda_\alpha,\lambda_\beta] = 
 \sum_{s \in \lambda_{\beta}} q_1^{-\ell_{\alpha}(s)-1} q_2^{a_{\beta}(s)}
 \, .
\end{align}
This proves the formula \eqref{eq:comb2}.

\subsection{U($n_0|n_1$) theory}

For the supergroup theory, we consider the following contribution to the Chern character (See \S\ref{sec:super_localization}):
\begin{align}
 \ch_\mathsf{T} \mathbf{H}_{e:i \to j,\sigma\sigma'}^{\text{bf,\,inst}} & =
 - \mu_e \ch_\mathsf{T} \wedge \mathbf{Q}^\vee 
 \ch_\mathsf{T} \mathbf{K}_i^{\sigma\vee} \ch_\mathsf{T} \mathbf{K}_j^{\sigma'}
 + \mu_e \ch_\mathsf{T} \mathbf{N}_i^{\sigma\vee} \ch_\mathsf{T} \mathbf{K}_j^{\sigma'} + \mu_e q^{-1} \ch_\mathsf{T} \mathbf{K}_i^{\sigma\vee} \ch_\mathsf{T} \mathbf{N}_j^{\sigma'}
 \nonumber \\
 & =: \sum_{\alpha=1}^{n_i} \sum_{\beta=1}^{n_j} \mu_e \frac{\nu_{j,\beta}}{\nu_{i,\alpha}} \, \Xi_{\sigma\sigma'}[\lambda_{i,\alpha}^\sigma,\lambda_{j,\beta}^{\sigma'}]
\end{align}
where the diagonal factors are written using \eqref{eq:comb1} as
\begin{align}
 \Xi_{00}[\lambda_\alpha,\lambda_\beta] = \Xi[\lambda_\alpha,\lambda_\beta]
 \, , \qquad
 \Xi_{11}[\lambda_\alpha,\lambda_\beta] = \Xi[\lambda_\beta,\lambda_\alpha]
 \, .
\end{align}
The vector multiplet contribution \eqref{eq:chV_inst2} is given by
\begin{align}
 \ch_\mathsf{T} \mathbf{V}_{i,\sigma\sigma'}^\text{inst} & =
 - \sum_{\alpha=1}^{n_i} \sum_{\beta=1}^{n_i} \frac{\nu_{j,\beta}}{\nu_{i,\alpha}} \, \Xi_{\sigma\sigma'}[\lambda_{i,\alpha}^\sigma,\lambda_{i,\beta}^{\sigma'}]
 \, .
\end{align}
The off-diagonal factors are 
\begin{subequations}
\begin{align}
 \Xi_{01}[\lambda_\alpha,\lambda_\beta] & =
 - (1 - q_1^{-1})(1 - q_2^{-1}) \sum_{s \in \lambda_\alpha} \sum_{s' \in \lambda_\beta} q_1^{-s_1-s_1'+1} q_2^{-s_2-s_2'+1}
 \nonumber \\
 & \qquad 
 + \sum_{s \in \lambda_\alpha} q_1^{-s_1} q_2^{-s_2}
 + \sum_{s' \in \lambda_\beta} q_1^{-s_1'} q_2^{-s_2'}
 \label{eq:comb+-1} \\
 \Xi_{10}[\lambda_\alpha,\lambda_\beta] & =
 - (1 - q_1^{-1})(1 - q_2^{-1}) \sum_{s \in \lambda_\alpha} \sum_{s' \in \lambda_\beta} q_1^{s_1+s_1'-1} q_2^{s_2+s_2'-1}
 \nonumber \\
 & \qquad 
 + \sum_{s \in \lambda_\alpha} q_1^{s_1-1} q_2^{s_2-1}
 + \sum_{s' \in \lambda_\beta} q_1^{s_1'-1} q_2^{s_2'-1} 
\end{align}
\end{subequations}
We remark that these off-diagonal factors are symmetric under $\lambda_\alpha \leftrightarrow \lambda_\beta$,
\begin{align}
 \Xi_{01}[\lambda_\alpha,\lambda_\beta] = \Xi_{01}[\lambda_\beta,\lambda_\alpha]
 \, , \qquad
 \Xi_{10}[\lambda_\alpha,\lambda_\beta] = \Xi_{10}[\lambda_\beta,\lambda_\alpha]
 \, ,
\end{align}
and
\begin{align}
 q \, \Xi_{01}[\lambda_\alpha,\lambda_\beta]\Big|_{q_1,q_2}
 = \Xi_{10}[\lambda_\alpha,\lambda_\beta]\Big|_{q_1^{-1},q_2^{-1}}
\end{align}

We apply a similar computation to \eqref{eq:comb+-1} as discussed in \S\ref{sec:comb_U(n)}.
The first term in \eqref{eq:comb+-1} yields
\begin{align}
 &
 - (1 - q_1^{-1})(1 - q_2^{-1}) \sum_{s \in \lambda_\alpha} \sum_{s' \in \lambda_\beta} q_1^{-s_1-s_1'+1} q_2^{-s_2-s_2'+1}
 \nonumber \\
 & =
 - \sum_{s_1=1}^{\check\lambda_{\alpha,1}} \sum_{s_2'=1}^{\lambda_{\beta,1}}
 (1 - q_1^{-\check\lambda_{\beta,s_2'}}) q_1^{-s_1}
 (1 - q_2^{-\lambda_{\alpha,s_1}}) q_2^{-s_2'}
 \nonumber \\
 & =
 - \sum_{s_1=1}^{\check\lambda_{\alpha,1}} \sum_{s_2'=1}^{\lambda_{\beta,1}}
 \left[
 q_1^{-\check\lambda_{\beta,s_2'}-s_1} q_2^{-\lambda_{\alpha,s_1}-s_2'}
 - q_1^{-s_1} q_2^{-s_2'}
 + (1 - q_1^{-\check\lambda_{\beta,s_2'}}) q_1^{-s_1} q_2^{-s_2'}
 + q_1^{-s_1} (1 - q_2^{-\lambda_{\alpha,s_1}}) q_2^{-s_2'}
 \right]
 \, .
 \label{eq:comb+-2}
\end{align}
The third and fourth terms in \eqref{eq:comb+-2} are given by
\begin{subequations} 
 \begin{align}
  - \sum_{s_1=1}^{\check\lambda_{\alpha,1}} \sum_{s_2'=1}^{\lambda_{\beta,1}}
  (1 - q_1^{-\check\lambda_{\beta,s_2'}}) q_1^{-s_1} q_2^{-s_2'}
  & = - \sum_{s' \in \lambda_\beta}
  (1 - q_1^{-\check\lambda_{\alpha,1}}) q_1^{-s_1'} q_2^{-s_2'}
  \, , \\
  - \sum_{s_1=1}^{\check\lambda_{\alpha,1}} \sum_{s_2'=1}^{\lambda_{\beta,1}}
  q_1^{-s_1} (1 - q_2^{-\lambda_{\alpha,s_1}}) q_2^{-s_2'}
  & = - \sum_{s \in \lambda_\alpha} q_1^{-s_1} (1 - q_2^{-\lambda_{\beta,1}}) q_2^{-s_2}
  \, .
 \end{align}
\end{subequations}
Hence we obtain
\begin{align}
 \Xi_{01}[\lambda_\alpha,\lambda_\beta]
 & =
 - \sum_{s_1=1}^{\check\lambda_{\alpha,1}} \sum_{s_2'=1}^{\lambda_{\beta,1}}
 \left[
 q_1^{-\check\lambda_{\beta,s_2'}-s_1} q_2^{-\lambda_{\alpha,s_1}-s_2'}
 - q_1^{-s_1} q_2^{-s_2'}
 \right]
 \nonumber \\
 & \qquad
 + \sum_{s \in \lambda_\alpha} q_1^{-s_1} q_2^{-\lambda_{\beta,1}-s_2}
 + \sum_{s' \in \lambda_\beta} q_1^{-\check\lambda_{\alpha,1}-s_1'} q_2^{-s_2'}
 \, ,
\end{align}
and similarly
\begin{align}
 \Xi_{10}[\lambda_\alpha,\lambda_\beta]
 & =  - \sum_{s_1=1}^{\check\lambda_{\alpha,1}} \sum_{s_2'=1}^{\lambda_{\beta,1}}
 \left[
 q_1^{\check\lambda_{\beta,s_2'}+s_1-1} q_2^{\lambda_{\alpha,s_1}+s_2'-1}
 - q_1^{s_1-1} q_2^{s_2'-1}
 \right]
 \nonumber \\
 & \qquad 
 + \sum_{s \in \lambda_\alpha} q_1^{s_1-1} q_2^{\lambda_{\beta,1}+s_2-1}
 + \sum_{s' \in \lambda_\beta} q_1^{\check\lambda_{\alpha,1}+s_1'-1} q_2^{s_2'-1}
 \, .
\end{align}
In contrast to the diagonal part $\Xi_{00(11)}$, further simplification does not occur for these off-diagonal ones.

\chapter{Matrix model}\label{chap:matrix_model}

In this Chapter, we summarize the geometric and algebraic aspects of the matrix model, which exhibit similar perspectives to gauge theory discussed in this manuscript.
See also the manuscripts~\cite{Eynard:2015aea,Marino:2015yie} for details on this topic.

\section{Matrix integral}

The partition function of the matrix model (the path integral of zero-dimensional QFT) is given as an integral over the self-conjugate (real symmetric, complex Hermitian, quarternion self-dual) matrix:
\begin{align}
 Z = \int dH \, \np^{-\frac{1}{\hbar} \tr V(H)}
 \label{eq:matrix_integral}
\end{align}
where 
\begin{align}
 \rk H = N
 \, , \qquad
 H =
 \begin{cases}
  H^\text{T} & (\text{real symmetric})
  \\
  H^\dag & (\text{complex Hermitian})
  \\
  H^\text{D} & (\text{quaternion self-dual})
 \end{cases}
 \, ,
\end{align}
with the coupling constant $\hbar$.
We define the potential function \index{potential term}
\begin{align}
 V(x) = \sum_{k = 1}^{d+1} \frac{t_k}{k} \, x^k
 \, ,
 \label{eq:matrix_pot}
\end{align}
which is a polynomial function of degree $d+1$.%
\footnote{%
We should impose a certain condition for the coefficients $(t_k)_{k = 1, \ldots, d+1}$ for convergence of the matrix integral, which could be relaxed via complexification of the integration contour.
See~\cite{Eynard:2015aea} for details.
}
The matrix measure $dH$ is given as a product of the Lebesgue measures of all real components of the matrix $H$, denoted by $H_{ij}^{(\alpha)}$ for $\alpha = 0,\ldots,2 \beta - 1$:
\begin{align}
 dH = \prod_{i = 1}^N dH_{ii} \prod_{\substack{1 \le i < j \le N \\ \alpha = 0,\ldots,2\beta - 1}} dH_{ij}^{(\alpha)}
 \, ,
\end{align}
where the symmetry parameter $\beta$ is given by
\begin{align}
 \mathbb{R} : \ \beta = \frac{1}{2}
 \, , \qquad
 \mathbb{C} : \ \beta = 1
 \, , \qquad
 \mathbb{H} : \ \beta = 2
 \, .
\end{align}

\subsection{Eigenvalue integral representation}

The matrix measure and the potential term in the matrix integral~\eqref{eq:matrix_integral} are invariant under the similarity transformation:
\begin{align}
 H \ \longmapsto \ U H U^\dag
 \, , \qquad
 U \in (\rO(N), \rU(N), \Sp(N))
 \, .
\end{align}
Utilizing this symmetry, we choose the basis diagonalizing the matrix (gauge fixing),
\begin{align}
 H = U X U^\dag
 \,, \qquad
 X = \diag(x_1,\ldots,x_N) \in \mathbb{R}^N
 \, .
\end{align}
Then, we can write the partition function~\eqref{eq:matrix_integral} in terms of the eigenvalues:
\begin{align}
 Z
 & =
 \frac{1}{N!} \int \prod_{i = 1}^N \frac{dx_i}{2\pi} \, \np^{-\frac{\beta}{\hbar} V(x_i)} \left| \Delta_N(X) \right|^{2\beta}
 \nonumber \\
 & =:
 \frac{1}{N!} \int \prod_{i = 1}^N \frac{dx_i}{2\pi} \, \np^{ - \frac{\beta}{\hbar^2} S(X)}
 \label{eq:matrix_ev_rep}
\end{align}
where we rescale the potential function $V(x)$, and the Jacobian term is given as the Vandermonde determinant \index{Vandermonde determinant}
\begin{align}
 \Delta_N(X) = \prod_{i < j}^N (x_j - x_i) 
 \, .
\end{align}
We also define the effective action $S(X)$ as
\begin{align}
 S(X)
 =
 \hbar \sum_{i = 1}^N V(x_i)
 - 2 \hbar^2 \sum_{i<j}^N \log |x_i - x_j|
 \, .
 \label{eq:matrix_eff_action}
\end{align}
We discuss the properties of the matrix model mainly based on this eigenvalue integral representation of the partition function in the following.

\section{Saddle point analysis}
\index{saddle point analysis}

We are in particular interested in the asymptotic regime of the matrix model, which is called the 't Hooft limit:\index{'t Hooft limit}
\begin{align}
 N \ \longrightarrow \ \infty
 \, , \qquad
 \hbar \ \longrightarrow \ 0
 \, , \qquad
 t : \ \text{fixed}
 \, ,
\end{align}
where $t$ is the 't Hooft coupling defined as
\begin{align}
 t = N \hbar = O(1)
 \, .
\end{align}
In this limit, we can apply the saddle point analysis to the matrix model:
A specific configuration dominates in the integral satisfying the saddle point equation,
\begin{align}
 0 & = \frac{\partial S(X)}{\partial x_i}
 \nonumber \\
 & =
 \hbar V'(x_i) - 2 \hbar^2 \sum_{\substack{i = 1 \\ (i \neq j)}}^N \frac{1}{x_i - x_j}
 =: \hbar V_\text{eff}'(x_i)
 \label{eq:saddle_pt1}
\end{align}
for $i = 1,\ldots, N$.
We denote the effective potential by $V_\text{eff}(x)$, which consists of the one-body potential and the interaction with other eigenvalues,
\begin{align}
 V_\text{eff}(x_i)
 = V(x_i) - 2 \sum_{\substack{i = 1 \\ (i \neq j)}}^N \log |x_i - x_j|
 \, .
\end{align}

In order to solve the saddle point equation, we define an auxiliary function, called the resolvent,\index{resolvent}%
\footnote{%
Precisely speaking, we should distinguish the resolvent and its average, $\overline{W}(x) = \vev{W(x)}$, whereas, in the 't Hooft limit, the average of the observable is given by its on-shell value in general, $\displaystyle \vev{\mathcal{O}} = \frac{1}{Z} \int d H \, \np^{-\frac{1}{\hbar} \tr V(H)} \, \mathcal{O}(H) \approx \mathcal{O}(H_*)$ with the solution to the saddle point equation denoted by $H_*$.
}
\begin{align}
 W(x) = \hbar \tr \frac{1}{x - H} = \hbar \sum_{i = 1}^N \frac{1}{x - x_i}
 \ \xrightarrow{x \to \infty} \
 \frac{t}{x}
 \, .
 \label{eq:resolvent_def}
\end{align}
We see that the saddle point equation~\eqref{eq:saddle_pt1} is equivalent to the differential equation for the resolvent,
\begin{align}
 W(x)^2 + \hbar W'(x) - V'(x) W(x) + P(x) = 0
 \label{eq:saddle_pt2} 
\end{align}
where we define the polynomial function of degree $d-1$,
\begin{align}
 P(x) = \hbar \sum_{i = 1}^N \frac{V'(x) - V'(x_i)}{x - x_i}
 \, .
\end{align}
Furthermore, in the semiclassical limit $\hbar \to 0$, we may omit the derivative term in the equation~\eqref{eq:saddle_pt2}, and thus the saddle point equation is reduced to the algebraic equation
\begin{align}
 W(x)^2 - V'(x) W(x) + P(x) = 0
 \, ,
 \label{eq:saddle_pt3} 
\end{align}
which solves the resolvent
\begin{align}
 W(x) = \frac{V'(x)}{2} - \frac{1}{2} \sqrt{V'(x)^2 - 4 P(x)}
 \, .
\end{align}
We remark that the sign of the square root is fixed to be consistent with the asymptotic behavior of the resolvent~\eqref{eq:resolvent_def}.

\subsection{Eigenvalue density function}

From the saddle point equation~\eqref{eq:saddle_pt3}, we can construct the eigenvalue density function
\begin{align}
 \rho(x)
 =  \frac{1}{N} \tr \delta(x - H)
 = \frac{1}{N} \sum_{i = 1}^N \delta(x - x_i)
 \, ,
 \label{eq:DOS1}
\end{align}
satisfying the normalization condition
\begin{align}
 \int dx \, \rho(x) = 1
 \, .
\end{align}
Since the delta function is described as \index{delta function}
\begin{align}
 \delta(z)
 = \mp \frac{1}{\pi} \imag \frac{1}{z \pm \im 0}
 := \lim_{\epsilon \to 0} \mp \frac{1}{\pi} \imag \frac{1}{z \pm \im \epsilon}
 \, ,
\end{align}
the density function is given as the imaginary part of the resolvent,
\begin{align}
 \rho(x)
 & = \frac{1}{2 \pi \im} \left( W(x - \im 0) - W(x + \im 0) \right)
 \nonumber \\
 & = 
  \begin{cases}
   \displaystyle
   \frac{1}{2\pi} |M(x)| \sqrt{- \prod_{\alpha = 1}^n (x - x_\alpha^-)(x - x_\alpha^+)}
   & (x \in \mathcal{C})
   \\
   0 & (x \not\in \mathcal{C})
  \end{cases}
 \, ,
 \label{eq:DOS2} 
\end{align}
where we denote
\begin{align}
 4P(x) - V'(x)^2
 & = - M(x)^2 \prod_{\alpha = 1}^n (x - x_\alpha^-)(x - x_\alpha^+)
 \begin{cases}
  > 0 & (x \in \mathcal{C}) \\
  < 0 & (x \not\in \mathcal{C})
 \end{cases}
\end{align}
with  
\begin{align}
 \mathcal{C} = \bigsqcup_{a = 1}^n \mathcal{C}_\alpha
 \, , \qquad
 \mathcal{C}_\alpha = (x_\alpha^- , x_\alpha^+)
 \, .
\end{align}
We remark that, since $\deg V'(x) = d$, we have $1 \le n \le d$ and the degree of the polynomial function $M(x)$ is $\deg M(x) = d - n$, which is called the $n$-cut solution.
In order to characterize the $n$-cut solution profile, we define the filling fraction
\begin{align}
 \varepsilon_\alpha = \frac{N_\alpha}{N} \in \mathbb{R}_+
 \, , \qquad
 \alpha = 1, \ldots, n
\end{align}
where
\begin{align}
 N_\alpha
 = \int_{\mathcal{C}_\alpha} dx \, \rho(x)
 = \#\{ \text{eigenvalues on the cut} \ \mathcal{C}_\alpha \}
 \label{eq:filling_N}
\end{align}
together with the normalization condition
\begin{align}
 \sum_{\alpha = 1}^n \varepsilon_\alpha = 1
 \, .
\end{align}
In fact, the filling fractions are the parameters characterizing the saddle point of the matrix integral, which are interpreted as proper coordinates of the moduli space associated with the matrix model.

\subsection{Functional representation}

Using the density function~\eqref{eq:DOS1}, the potential term in the effective action~\eqref{eq:matrix_eff_action} is written as
\begin{align}
 \hbar \sum_{i = 1}^N V(x_i)
 = \frac{t}{N} \sum_{i = 1}^N \int dx \, \delta(x - x_i) \, V(x)
 = t \int dx \, \rho(x) \, V(x)
 \, .
\end{align}
Applying a similar argument to the interaction term, we obtain the functional representation of the effective action
\begin{align}
 S[\rho(x)]
 & =
 t \int dx \, \rho(x) \, V(x)
 - t^2 \, \dashint_{x \neq x'} dx dx' \, \rho(x) \rho(x') \, \log|x  - x'|
 \nonumber \\
 & \hspace{15em}
 + \sum_{\alpha = 1}^{n} \ell_\alpha \left( \varepsilon_\alpha - \int_{\mathcal{C}_\alpha} dx \, \rho(x) \right)
 \, ,
 \label{eq:eff_action_func}
\end{align}
where we denote the principal value integral by $\dashint dx$, and $(\ell_\alpha)_{\alpha = 1,\ldots,n}$ are the Lagrange multipliers to impose the condition~\eqref{eq:filling_N}.

Let us consider the saddle point analysis in this functional representation.
We take the functional derivative of the effective action,
\begin{align}
 \frac{1}{t} \frac{\delta S[\rho(x)]}{\delta \rho(x)}
 & = V(x) - 2 t \, \dashint dx' \, \rho(x') \, \log |x - x'| - t^{-1} \ell_\alpha
 \nonumber \\
 & = V_\text{eff}(x) - t^{-1} \ell_\alpha
 \, , \qquad
 x \in \mathcal{C}_\alpha
 \, ,
\end{align}
where $V_\text{eff}(x)$ is the functional version of the effective potential given in~\eqref{eq:saddle_pt1}.
Thus, from the equation of motion for the effective action, the effective potential takes a constant value for each cut,
\begin{align}
 \frac{\delta S[\rho(x)]}{\delta \rho(x)}
 = 0
 \ \implies \
 V_\text{eff}(x) = t^{-1} \ell_\alpha
 \, , \quad
 x \in \mathcal{C}_\alpha
 \, .
 \label{eq:V_eff_const}
\end{align}
Then, the (functional version of) saddle point equation~\eqref{eq:saddle_pt1} is obtained from the effective potential,
\begin{align}
 V_\text{eff}'(x) = V'(x) - 2 t \, \dashint dx' \, \frac{\rho(x')}{x - x'} = 0
 \, .
\end{align}
In this formalism, the resolvent takes a form of \index{resolvent}
\begin{align}
 W(x) = t \int dx' \, \frac{\rho(x')}{x - x'}
 \, .
\end{align}
Namely, the resolvent is given by the Hilbert transform of the density function $\rho(x)$.
Since the integrand of this expression has a singularity at $x' = x \in \mathcal{C}$, we define the regularized version of the resolvent with the principal value integral,
\begin{align}
 W^\text{reg}(x) & = t \, \dashint dx' \, \frac{\rho(x')}{x - x'}
 \qquad (x \in \mathcal{C})
 \nonumber \\
 & = \frac{1}{2} \left( W(x + \im 0) + W(x - \im 0) \right)
 = \real W(x \pm \im 0)
 \, .
\end{align}
Recalling the imaginary part of the resolvent is given by the eigenvalue density function~\eqref{eq:DOS2}, we obtain the Kramers--Kronig relation,
\begin{align}
 & W( x \pm \im 0) = \frac{1}{2} V'(x) \mp \pi \im \rho(x)
 & (x \in \mathcal{C})
 \, , \\
 \iff
 & V'_\text{eff}(x) = \pm 2 \pi \im \rho(x)
 & (x \in \mathcal{C})
 \, .
\end{align}

\section{Spectral curve}

The algebraic equation~\eqref{eq:saddle_pt3} defines the spectral curve of the matrix model, which describes the saddle point configuration of the matrix integral,
\begin{align}
 \Sigma = \{ (x, y) \in \mathbb{C} \times \mathbb{C} \mid H(x,y) = 0 \}
 \, .
\end{align}
The algebraic function $H(x,y)$ is defined as
\begin{align}
 H(x,y) = y^2 - V'(x) \, y + P(x)
 \label{eq:H_func1}
\end{align}
where we identify $y = W(x)$, and we define the one-form and the symplectic two-form on the curve,
\begin{align}
 \lambda = y \, dx
 \, , \qquad
 \omega = d\lambda = dy \wedge dx
 \, .
 \label{eq:dif_forms_matrix}
\end{align}
We also use another expression of the curve via the symplectic transform, $y \to y + V'(x)/2$,%
\footnote{%
This algebraic function is given as the characteristic polynomial of the Lax matrix, $L(x) \in \mathfrak{sl}_2$, associated with the orthogonal polynomials,
\begin{align}
 H(x,y) = \det(y - L(x))
\end{align}
with
\begin{align}
 \tr L(x) = 0
 \, , \qquad
 \det L(x) = P(x) - \frac{1}{4} V'(x)^2
 \, .
\end{align}
See~\cite{Eynard:2015aea} for details.
}
\begin{align}
 H(x, y) = y^2 - \frac{1}{4} V'(x)^2 + P(x)
 \, .
\end{align}
The degree of the spectral curve is $(2,2d)$.
From this expression, the spectral curve is written in the form of the hyperelliptic curve
\begin{align}
 \Sigma : \
 y^2 = P(x) - \frac{1}{4} V'(x)^2
 \, .
\end{align}
In particular, the $n$-cut solution~\eqref{eq:DOS2} gives rise to the spectral curve with genus $g = n - 1$.

\subsection{Cycle integrals}\label{sec:matrix_cycle_int}

For the spectral curve of the matrix model with genus $g = n - 1$, we define the $A$-cycle and $B$-cycle as in Fig.~\ref{fig:Riemann_surface} and Fig.~\ref{fig:cycles}.
Namely, the $A$-cycle is a contour surrounding the cut $\mathcal{C}_\alpha$, so that the filling fraction is given as
\begin{align}
 \varepsilon_\alpha
 = \frac{1}{N} \int_{\mathcal{C}_\alpha} dx \, \rho(x)
 = \frac{1}{2 \pi \im t} \oint_{A_\alpha} dx \, W(x)
 \, .
\end{align}
The $B$-cycle is the contour from the $\alpha$-th cut to $(\alpha+1)$-st cut, then $(\alpha+1)$-st to $\alpha$-th on the other sheet.
Recalling \eqref{eq:V_eff_const}, we obtain
\begin{align}
 \oint_{B_\alpha} \lambda = \frac{\ell_\alpha - \ell_{\alpha + 1}}{t}
 \, .
\end{align}
Thus, the contour integrals of the one-form $\lambda$ along the $A$ and $B$-cycles are given by
\begin{subequations}
 \begin{align}
  \frac{1}{2\pi \im} \oint_{A_\alpha} \lambda & = t \, \varepsilon_\alpha
  \, , \\
  \frac{1}{2\pi \im} \oint_{B_\alpha} \lambda & = \frac{1}{2 \pi \im t} \left( \ell_{\alpha} - \ell_{\alpha+1} \right)
  \, .
 \end{align}
\end{subequations}
We remark that, from the effective action~\eqref{eq:eff_action_func}, we have the relation
\begin{align}
 \ell_\alpha = \frac{\partial S}{\partial \varepsilon_\alpha}
 \, .
\end{align}
This implies the relation between the $A$ and $B$-cycle integrals through the effective action,
\begin{align}
 \oint_{B_\alpha} \lambda
 = t^{-1}
 \left(
 \frac{\partial S}{\partial \varepsilon_\alpha} -  \frac{\partial S}{\partial \varepsilon_{\alpha + 1}}
 \right)
 \, ,
\end{align}
which is analogous to the Seiberg--Witten geometry discussed in \S\ref{sec:SW_th}.

In order to obtain a closer expression to the Seiberg--Witten geometry, we take a linear combination,
\begin{align}
 \bar{\varepsilon}_\alpha = 
 \varepsilon_\alpha - \varepsilon_{\alpha + 1} 
 \, , \qquad
 \alpha = 1,\ldots,n-1,
 \, ,
\end{align}
which corresponds to the simple root of the Lie group $\SU(n)$.
Similarly define the modified $A$-cycle, $\bar{A}_\alpha = A_\alpha - A_{\alpha+1}$, then we obtain
\begin{align}
 &
 \frac{1}{2 \pi \im} \oint_{\bar{A}_\alpha} \lambda = t \, \bar{\varepsilon}_\alpha
 \, , \qquad
 \frac{1}{2 \pi \im} \oint_{B_\alpha} \lambda = \frac{1}{2 \pi \im t} \frac{\partial S}{\partial \bar{\varepsilon}_\alpha}
 \, , 
 & \alpha = 1,\ldots,n-1
 \, ,
\end{align}
where the modified filling fraction and the effective action correspond to the Coulomb moduli and the prepotential of the Seiberg--Witten theory.
\index{Coulomb moduli}
\index{prepotential}

 \section{Quantum geometry}\label{sec:matrix_quantum_geom}

 Let us discuss how to quantize the spectral curve of the matrix model.
 Recall the saddle point equation at finite $\hbar$~\eqref{eq:saddle_pt2} is a Riccati-type differential equation for the resolvent.
 In order to linearize the differential equation, we write the resolvent as \index{resolvent}
 \begin{align}
  W(x) = \hbar \frac{\psi'(x)}{\psi(x)} = \hbar \frac{d}{dx} \log \psi(x)
  \, ,
 \end{align}
 where $\psi(x)$ is the characteristic polynomial, called the wave function in this context,
 \begin{align}
  \psi(x) = \prod_{i = 1}^N (x - x_i) = \det_N(x - H)
  \, .
 \end{align}
 Then, we obtain the second order linear ODE for the wave function,
 \begin{align}
  \hbar^2 \psi''(x) - \hbar V'(x) \psi'(x) + P(x) \psi(x) = 0
  \, .
 \end{align}
 This ODE is also written in the operator form,%
 \footnote{%
 Precisely speaking, we should properly impose the ordering of the operators since the operators $(\hat{x},\hat{y})$ are noncommutative.
 }
 \begin{align}
  H(\hat{x},\hat{y}) \, \psi(x) = 0
  \, ,
 \end{align}
 where $H(x,y)$ is the algebraic function defined in \eqref{eq:H_func1}, and the operator pair is given by
 \begin{align}
  \hat{x} \, \psi(x) = x \, \psi(x)
  \, , \qquad
  \hat{y} \, \psi(x) = \hbar \frac{d}{dx} \psi(x)
  \, ,
 \end{align}
 obeying the relation
 \begin{align}
  \left[ \hat{y} \,,\, \hat{x} \right] = \hbar
  \, .
 \end{align}
 This is the canonical commutation relation with respect to the symplectic two-form defined in \eqref{eq:dif_forms_matrix}. 
 In this sense, this Schr\"odinger-type ODE is interpreted as quantization of the spectral curve (quantum curve), and $\psi(x)$ plays a role of the wave function.
 \index{quantum curve}

 \subsection{Baker--Akhiezer function}\label{sec:BA_func}
 \index{Baker--Akhiezer function}

 In order to see the connection to the integrable hierarchy, we consider the expectation value of the characteristic polynomial,
 \begin{align}
  \psi(x) & = \vev{\det(x - H)}
  \nonumber \\
  & =
  \frac{x^N}{Z(t)} \int dH \, \exp \left( - \frac{1}{\hbar} \tr V(H) - \sum_{k = 1}^\infty \frac{x^{-k}}{k} \tr H^k \right)
  \nonumber \\
  & =:
  x^N \frac{Z(t + [x])}{Z(t)}
  \, ,
 \end{align}
 where we explicitly show the $t$-dependence of the matrix integral, and we denote
 \begin{align}
  [x] = \hbar \left( x^{-1}, x^{-2}, \ldots \right)
  \, .
 \end{align}
 From this point of view, we in practice consider the situation with $d \to \infty$.
 We remark that such a shift of the $t$-variable can be imposed by the vertex operator (\S\ref{sec:VO}).
 In fact, the matrix integral $Z(t)$ is identified with the $N$-soliton solution to the $\tau$-function of the KP hierarchy, and the wave function is the corresponding Baker--Akhiezer function.
 See, for example,~\cite{Jimbo:1983if,Babelon:2003CIS} for more details.

 \subsection{Quantization of the cycle}\label{sec:matrix_cycle_quantization}

 We can also see quantization of the cycle at finite $\hbar$.
 As seen from the definition of the resolvent~\eqref{eq:resolvent_def}, it has poles with the residue $\hbar$, so that the $A$-cycle integral counts the number of eigenvalues on the cut $\mathcal{C}_\alpha$,
  \begin{align}
   \frac{1}{2 \pi \im} \oint_{A_\alpha} \lambda = \hbar \, \mathbb{Z}
   \, .
  \end{align}
  This is the Bohr--Sommerfeld quantization condition with the quantum parameter $\hbar$.
 
 \section{Quantum algebra}

 In this Section, we explore the algebraic structure of the matrix model.
 We see that the loop equation, which provides a relation for the correlation functions, is characterized using the infinite dimensional algebra.

 \subsection{Loop equation}

 Let $\mathcal{O}(H)$ be a non-singular function of the matrix $H$ (also called the observable).
 Then, as long as its expectation value $\vev{\mathcal{O}(H)}$ is finite, the following identity holds: 
 \begin{align}
  \int dH \frac{\partial}{\partial H} \left( \np^{-\frac{1}{\hbar} \tr V(H)} \, \mathcal{O}(H) \right) = 0
  \, ,
 \end{align}
 because the boundary value of the integrand is suppressed by the exponential term $\displaystyle \np^{-\frac{1}{\hbar} \tr V(H)}$.
 In the eigenvalue representation, we instead obtain the identity:
 \begin{align}
  \int dX \sum_{k = 1}^N \frac{\partial}{\partial x_k} \left( \np^{-\frac{\beta}{\hbar} \tr V(X)} \Delta_{N}(X)^{2\beta} \, \mathcal{O}(X) \right)
  = 0
  \, ,
  \label{eq:loop_id_ev}
 \end{align}
 where we can interpret $\beta$ as an arbitrary parameter.

 We consider the case with $\mathcal{O}(H) = \tr H^k = \sum_{i = 1}^N x_i^k =: p_k(X)$, where $p_k(X)$ is the $k$-th power sum polynomial of $X = (x_i)_{i = 1,\ldots,N}$.
 In this case, the identity~\eqref{eq:loop_id_ev} gives rise to
 \begin{align}
  0 & = \beta
  \left< \,
  \sum_{i = 1}^N
  \left(
  \frac{k}{\beta} x_i^{k-1} - \frac{1}{\hbar} V'(x_i) x_i^k
  + \sum_{j (\neq i)}^N \frac{2 x_i^k}{x_i - x_j}
  \right)
  \, \right>
  \nonumber \\
  & =
  \beta \sum_{m=0}^{k-1} \vev{ \tr H^m \tr H^{k-m-1}}
  + \left( 1 - \beta \right) k \vev{ \tr H^{k-1}}
  - \frac{\beta}{\hbar} \vev{\tr H^k V'(H)}  
  \, .
  \label{eq:loop_eq1}
 \end{align}
 Then, the identity with
 \begin{align}
  \mathcal{O}(H) = \tr \frac{1}{x - H} = \sum_{k = 0}^\infty x^{-k-1} \tr H^k
  \, ,
 \end{align}
 is obtained from the relation~\eqref{eq:loop_eq1} by multiplying the factor $x^{- k - 1}$ and summing over $k = 0, \ldots,\infty$,
 \begin{align}
  0 & =
  \beta
  \left< \,
  \left( \tr \frac{1}{x - H} \right)^2
  \, \right>
  + \left( 1 - \beta \right)
  \left< \,
  \tr \left( \frac{1}{x - H} \right)^2
  \, \right>
  - \frac{\beta}{\hbar}
  \left< \,
  \tr \frac{V'(H)}{x - H}
  \, \right>
  \nonumber \\
  & =
  \beta
  \left< \,
  \left( \tr \frac{1}{x - H} \right)^2
  \, \right>
  + \left( 1 - \beta \right)
  \left< \,
  \tr \left( \frac{1}{x - H} \right)^2
  \, \right>
  \nonumber \\
  & \hspace{8em}
  - \frac{\beta}{\hbar} V'(x)
  \left< \,
  \tr \frac{1}{x - H} 
  \, \right>
  + \frac{\beta}{\hbar}
  \left< \,
  \tr \frac{V'(x) - V'(H)}{x - H}
  \, \right>  
  \, ,
  \label{eq:loop_eq2}
 \end{align}
 which is called the {\em loop equation}. \index{loop equation}
 Inserting more generic operator $\mathcal{O}(H)$, we obtain various relations among the correlation functions.
 
 \subsection{Operator formalism}
  
  We then discuss the underlying algebraic structure of the loop equation obtained above.
  For this purpose, we introduce the operator formalism as follows.

  Recalling the potential function takes a form of~\eqref{eq:matrix_pot}, one may express the matrix moment as
  \begin{align}
   \vev{\tr H^n}
   & = \frac{1}{Z} \int dH \, \np^{-\frac{\beta}{\hbar} \tr V(H)} \, \tr H^n
   = - n \frac{\hbar}{\beta} \frac{\partial}{\partial t_n} \log Z
   \, ,
  \end{align}
  which implies the correspondence
  \begin{align}
   \tr H^n
   \ \iff \
   - n \frac{\hbar}{\beta} \frac{\partial}{\partial t_n}
   \label{eq:moment_op}
  \end{align}
  in the operator formalism.
  In order to apply this correspondence for $^\forall n \in \mathbb{Z}_{>0}$, we take $d \to \infty$ for the moment.


  As in \S\ref{sec:hol_def}, we introduce the Fock space $\mathsf{F} = \mathbb{C}[[t_n,\partial_n]] \ket{0}$ with the vacuum state $\partial_n \ket{0} = 0$, generated by the Heisenberg algebra $\mathscr{H} = (t_n,\partial_n)_{n \ge 1}$ with the algebraic relation $[\partial_n, t_{n'}] = \delta_{n,n'}$.
  For the latter purpose, we introduce another set of the oscillators,
  \begin{align}
   a_n = n \hbar \sqrt{\frac{2}{\beta}} \frac{\partial}{\partial t_n}
   \, , \qquad
   a_{-n} = \hbar^{-1} \sqrt{\frac{\beta}{2}} \, t_n
   \label{eq:matrix_a_mode}
  \end{align}
  obeying the algebraic relation
  \begin{align}
   \left[ a_n, a_m \right] = n \, \delta_{n+m, 0}
   \, .
  \end{align}
  With these operators, the loop equation~\eqref{eq:loop_eq1} is written as follows:
  \begin{align}
   L_{k - 1} Z = 0
   \, , \qquad
   k \ge 0
   \, ,
   \label{eq:Vir_const}
  \end{align}
  which is called the Virasoro constraint. \index{Virasoro constraint}
  The operators $(L_{k-1})_{k \ge 0}$ are given in terms of the $a$-modes,
  \begin{align}
   L_{k - 1} & =
   \frac{1}{2} \sum_{n = 0}^{k-1} a_n a_{k - n - 1} + \sum_{n = 1}^\infty a_{-n} a_{n + k - 1}
   + \frac{1}{\sqrt{2}} \left( \sqrt{\beta} - \sqrt{\beta^{-1}} \right) k \, a_{k - 1}
   \nonumber \\
   & =
   \frac{1}{2} \sum_{n \in \mathbb{Z}} {: a_n a_{k - n - 1}:}
   + \frac{1}{\sqrt{2}} \left( \sqrt{\beta} - \sqrt{\beta^{-1}} \right) k \, a_{k - 1}
   \, ,
   \label{eq:L_a_rep}
  \end{align}
  which obeys the algebraic relation for the Virasoro algebra,
  \begin{align}
   \left[ L_n, L_{n'} \right] = (n - n') L_{n+n'}
   \, .
   \label{eq:Vir_rel1}
  \end{align}
  We remark that the full Virasoro algebra $(L_n)_{n \in \mathbb{Z}}$ obeys the relation
  \begin{align}
   \left[ L_n, L_{n'} \right] = (n - n') L_{n+n'} + \frac{c}{12} n(n^2 - 1)
   \, ,
   \label{eq:Vir_rel2}
  \end{align}
  where $c$ is called the central charge.
  The current matrix model construction~\eqref{eq:Vir_rel1} does not provide the central term, because it appears from the commutation relation between the positive and negative modes, $L_n$ and $L_{-n}$; There appear only a half of the generators from the matrix model.
  Nevertheless, we can formally define the negative generators from \eqref{eq:L_a_rep} to realize the full Virasoro algebra with the central charge
  \begin{align}
   c = 1 - 6 \left( \sqrt{\beta} - \sqrt{\beta^{-1}} \right)^2
   =
   \begin{cases}
    1 & (\beta = 1) \\
    -2 & (\beta = 2^{\pm 1})
   \end{cases}
   \, .
   \label{eq:matrix_c}
  \end{align}

  \subsection{Gauge theory parameter}

  The expression of the central charge in terms of the symmetry parameter $\beta$ implies the correspondence to the $\Omega$-background parameter~\cite{Alday:2009aq},
  \begin{align}
   (\epsilon_1, \epsilon_2) = \left( \hbar , - \hbar \beta^{-1} \right)
   \ \iff \
   \left( \hbar, \beta \right) = \left( \epsilon_1, - \frac{\epsilon_1}{\epsilon_2} \right)
   \, .
  \end{align}
  Under this identification, exchanging $\epsilon_1 \leftrightarrow \epsilon_2$ implies the following symmetry on the matrix model parameters:
  \begin{align}
   \epsilon_1
   \ \longleftrightarrow \
   \epsilon_2
   \quad \iff \quad
   \left(\hbar, \beta\right)
   \ \longleftrightarrow \
   \left( - \hbar \beta^{-1}, \beta^{-1} \right)
   \, .
  \end{align}

  In fact, from the expression~\eqref{eq:matrix_ev_rep}, we obtain the effective action in the asymptotic limit
  \begin{align}
   - \frac{\hbar^2}{\beta} \log Z = \epsilon_1 \epsilon_2 \log Z
   \ \xrightarrow{\hbar \to 0} \
   S(\epsilon_\alpha,\ell_\alpha)
   \, ,
  \end{align}
  which is analogous to the asymptotic behavior of the partition function discussed in \S\ref{sec:classical_lim}, and the effective action $S(\epsilon_\alpha,\ell_\alpha)$ plays a role of the prepotential $\mathscr{F}(\mathsf{a}_\alpha,\mathsf{a}_\alpha^D)$ in gauge theory.
  See also \S\ref{sec:matrix_cycle_int}.

  \subsection{Vertex operators}\label{sec:VO}

  From the $a$-modes~\eqref{eq:matrix_a_mode}, we define the current operator, which is identified with the (derivative of) effective potential through the identification \eqref{eq:moment_op},
  \begin{align}
   J(x)
   & = \sum_{n \in \mathbb{Z}} a_n \, x^{-n - 1}
   \nonumber \\
   &
   = - \sqrt{2 \beta} \tr \frac{1}{x - H} + \hbar^{-1} \sqrt{\frac{\beta}{2}} V'(x)
   = \hbar^{-1} \sqrt{\frac{\beta}{2}} V_\text{eff}'(x)
   \, .
   \label{eq:J_op}
  \end{align}
  Similarly, we define the free boson operator
  \begin{align}
   \phi(x)
   & = - \sum_{n \in \mathbb{Z}_{\neq 0}} \frac{a_n}{n} \, x^{-n}
   + a_0 \log x + \tilde{a}_0
   \nonumber \\
   & = - \sqrt{2 \beta} \tr \log (x - H) + \hbar^{-1} \sqrt{\frac{\beta}{2}} V(x) + \tilde{a}_0
   = \hbar^{-1} \sqrt{\frac{\beta}{2}} V_\text{eff}(x) + \tilde{a}_0
  \end{align}
  with the zero mode
  \begin{align}
   [a_n , \tilde{a}_0] = \delta_{n,0}
   \, .
  \end{align}
  The free boson and the current operator have the relation
  \begin{align}
   J(x) = \partial_x \phi(x)
   \, .
  \end{align}

  We also define the vertex operator with the free boson,
  \begin{align}
   \mathsf{V}_\alpha(x) = {: \np^{\alpha \phi(x)} :}
  \end{align}
  where $\alpha$ is called the momentum parameter.
  Put $\alpha = \mp \sqrt{\beta / 2}$, then the vertex operator gives rise to the characteristic polynomial in the matrix model
  \begin{align}
   \mathsf{V}_{\mp \sqrt{\beta/2}}(x) =
   \left( \np^{- \frac{\beta}{2 \hbar} V(x)} \det(x - H)^{\beta} \right)^{\pm 1}
   \times \text{const.}
  \end{align}

  \subsubsection{Generating current}

  Let us consider the generating current of the Virasoro generators, which is called the energy-momentum tensor, and also the stress tensor,
  \begin{align}
   T(x) = \sum_{n \in \mathbb{Z}} \frac{L_n}{x^{n+2}}
   \, .
  \end{align}
  In fact, this generating current has a realization in terms of the current operator,
  \begin{align}
   T(x) = \frac{1}{2} {:JJ:}(x) - \rho J'(x)
   \label{eq:EM_tensor}
  \end{align}
  where
  \begin{align}
   \rho = \frac{1}{\sqrt{2}} ( \sqrt{\beta} - \sqrt{\beta^{-1}} )
   \, .
  \end{align}
  We remark $\rho = 0$ for $\beta = 1$.
  Since the current operator has the expression~\eqref{eq:J_op}, the generating current is given by
  \begin{align}
   T(x)
   & = \beta \left( \tr \frac{1}{x - H} \right)^2
   - \frac{\beta}{\hbar} V'(x) \tr \frac{1}{x - H}
   + \frac{\beta}{4 \hbar^2} V'(x)^2
   \nonumber \\
   & \hspace{8em}
   + (1 - \beta) \tr \left( \frac{1}{x - H} \right)^2
   + \frac{1 - \beta}{2 \hbar} V''(x)
   \, .
   \label{eq:EM_tensor_matrix}
  \end{align}
  The algebraic relation~\eqref{eq:Vir_rel2} is equivalent to the OPE between the generating currents,
  \begin{align}
   T(x) T(x') = \frac{c/2}{(x - x')^4} + \frac{2 T(x')}{(x - x')^2} + \frac{T'(x')}{x - x'} + \cdots
   \, .
  \end{align}

  \subsection{$Z$-state}
  \index{Z-state@$Z$-state}

  The OPE between the generating current and the vertex operator is given by
  \begin{align}
   T(x) V_\alpha(x') =
   \left( \frac{\Delta_\alpha}{(x - x')^2} + \frac{1}{x - x'} \frac{\partial}{\partial x'} \right) V_\alpha(x') + \cdots
  \end{align}
  where the parameter
  \begin{align}
   \Delta_\alpha = \frac{1}{2} \left( (\alpha + \rho)^2 - \rho^2 \right)
  \end{align}
  is called the conformal weight.
  We define the screening current with $\Delta_{-\sqrt{2\beta}} = 1$, \index{screening current}
  \begin{align}
   S(x) = V_{-\sqrt{2\beta}}(x)
   \, .
  \end{align}
  Then, the OPE becomes
  \begin{align}
   T(x) S(x')
   = \frac{\partial}{\partial x'} \left( \frac{1}{x - x'} \right) S(x') + \cdots
   \, ,
  \end{align}
  which is a total derivative, so that the integral of the screening current commutes with the generating current  
  \begin{align}
   \left[ T(x), \mathsf{S} \right]
   = \left[ T(x), \int dx' \, S(x') \right]
   = 0
   \, .
  \end{align}
  We define the screening charge as the integral of the screening current,
  \begin{align}
   \mathsf{S} = \int dx \, S(x)
   \, .
  \end{align}

  Then, the $Z$-state, obtained from the operator analog of the partition function through the operator/state correspondence, is constructed from the screening charge as follows:%
  \footnote{%
  Precisely speaking, we should take care of the ordering of the integral variables, $x_1 < \cdots < x_N$.
  Here we relax this condition via the analytic continuation, and multiply the factor $1/N!$.
  }
  \begin{align}
   \ket{Z}
   = \mathsf{S}^N \ket{0}
   & =
   \frac{1}{N!}
   \int \prod_{i = 1}^N dx_i \prod_{i<j}^N (x_i - x_j)^{2\beta}
   {: \prod_{i = 1}^N S(x_i) :} \ket{0}
   \nonumber \\
   & =
   \frac{1}{N!}
   \int \prod_{i = 1}^N dx_i \, \np^{-\frac{\beta}{\hbar} V(x_i)} \, \Delta_N(X)^{2\beta} \ket{0}
   \, .
  \end{align}
  In order to obtain the partition function, which takes a value in $\mathbb{C}$, we define the dual state
  \begin{align}
   \bra{d + 1} a_n
   =
   \begin{cases}
    \bra{d + 1} \left( \hbar^{-1} \sqrt{\beta/2} \, t_n \right) & (n \in (1,\ldots,d+1)) \\
    0 & (n \not\in (1,\ldots,d+1)) \\ 
   \end{cases}
  \end{align}
  having the same charge as the $Z$-state $\ket{Z}$.
  We remark that the eigenvalue of the oscillator $a_n$ (namely, the parameter $t_n$) is now a c-number.
  Then, the partition function is given as a chiral correlator of the screening charges,
  \begin{align}
   Z = \vev{ d + 1 \mid Z} = \vev{ d + 1 \mid \mathsf{S}^N \mid 0 }
   = \frac{1}{N!} \int \prod_{i = 1}^N dx_i \, \np^{-\frac{\beta}{\hbar} V(x_i)} \, \Delta_N(X)^{2\beta}
   \, ,
  \end{align}
  where the potential function is now a polynomial of degree $d+1$.

  \subsubsection{Virasoro constraint revisited}
  
  Let us discuss the Virasoro constraint from the $Z$-state perspective.
  Applying the Virasoro generating current to the vacuum, we have  
  \begin{align}
   T(x) \ket{0} = \sum_{n \in \mathbb{Z}} L_n \ket{0} x^{-n-2}
   \, .
  \end{align}
  Regularity at $x \to 0$ implies
  \begin{align}
   L_n \ket{0} = 0
   \, , \qquad
   n \ge -1
   \, .
  \end{align}
  Since the screening charge commutes with the generating current, the same argument is applied to obtain the Virasoro constraint~\eqref{eq:Vir_const}.
  Actually, compared to the expression of the generating current~\eqref{eq:EM_tensor_matrix}, the loop equation~\eqref{eq:loop_eq2} implies the negative power modes vanish in the generating current.


\newpage
\addcontentsline{toc}{chapter}{\bibname}
\bibliographystyle{amsalpha_mod}
\bibliography{/Users/k_tar/Dropbox/etc/conf}


\newpage
\addcontentsline{toc}{chapter}{\indexname}
\printindex

\end{document}